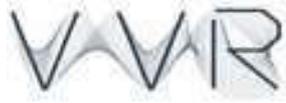
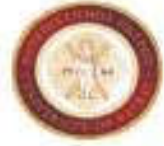

# Spectral Processing and Optimization of Static and Dynamic 3D Geometries

Doctoral Thesis

Gerasimos Arvanitis

`arvanitis@ece.upatras.gr`

Department of Electrical and Computer Engineering
University of Patras
Greece

**Supervisors:**
Professor Konstantinos Moustakas
Associate Professor Evaggelos Dermatas
Principal Researcher Aris S. Lalos

July 15, 2021

# Πιστοποίηση

Πιστοποιείται ότι η παρούσα Διδακτορική Διατριβή με θέμα:

Φασματική επεξεργασία και αλγόριθμοι βελτιστοποίησης σε στατικά και δυναμικά αντικείμενα τρισδιάστατης γεωμετρίας του Γεράσιμου Αρβανίτη του Νικολάου, Διπλωματούχου Ηλεκτρολόγου Μηχανικού & Τεχνολογίας Υπολογιστών παρουσιάστηκε δημοσίως στο Τμήμα Ηλεκτρολόγων Μηχανικών και Τεχνολογίας Υπολογιστών του Πανεπιστημίου Πατρών στις 28 Ιουλίου 2021, και εξετάσθηκε και εγκρίθηκε από την ακόλουθη Εξεταστική Επιτροπή:

- Κωνσταντίνος Μουστάκας, Καθηγητής, Τμήμα Ηλεκτρολόγων Μηχανικών & Τεχνολογίας Υπολογιστών, Πανεπιστήμιο Πατρών
- Ευάγγελος Δερματάς, Αναπλ. Καθηγητής, Τμήμα Ηλεκτρολόγων Μηχανικών & Τεχνολογίας Υπολογιστών, Πανεπιστήμιο Πατρών
- Άρης Σ. Λάλος, Κύριος Ερευνητής στο ερευνητικό ινστιτούτο 'Αθηνά'
- Ιωάννης Εμίρης, Καθηγητής, Τμήμα Πληροφορικής, Ε.Κ.Π.Α.
- Κωνσταντίνος Μπερμπερίδης, Καθηγητής, Τμήμα Μηχανικών ΗΥ και Πληροφορικής, Πανεπιστήμιο Πατρών
- Αθανάσιος Σκόδρας, Καθηγητής, Τμήμα Ηλεκτρολόγων Μηχανικών & Τεχνολογίας Υπολογιστών, Πανεπιστήμιο Πατρών
- Κατερίνα Φραγκιαδάκη, Αναπλ. Καθηγήτρια, Machine Learning Department, Carnegie Mellon University

Πάτρα, 28, 07, 2021

Ο Επιβλέπων
Καθηγητής
Κωνσταντίνος Μουστάκας

Ο Πρόεδρος του Τμήματος
Καθηγητής
Οδυσσέας Κουφοπαύλου

## Αφιέρωση

Στην σύζυγό μου και στην κορούλα μας

# Acknowledgements

First of all, I would like to thank professor Konstantinos Moustakas for his excellent supervision, collaboration and feedback throughout this thesis work. I would also like to express my gratitude to my supervisor and examiner, Dr. Aris S Lalos, for his interest, inspiration and contribution to this thesis.

# Abstract

Geometry processing of 3D objects is of primary interest in many areas of computer vision and graphics, including robot navigation, 3D object recognition, classification, feature extraction, etc.

The recent introduction of cheap range sensors has created a great interest in many new areas, driving the need for developing efficient algorithms for 3D object processing. Previously, in order to capture a 3D object, expensive specialized sensors were used, such as lasers or dedicated range images, but now this limitation has changed.

The current approaches of 3D object processing require a significant amount of manual intervention and they are still time-consuming making them unavailable for use in real-time applications.

The aim of this thesis is to present algorithms, mainly inspired by the spectral analysis, subspace tracking, etc, that can be used and facilitate many areas of low-level 3D geometry processing (i.e., reconstruction, outliers removal, denoising, compression), pattern recognition tasks (i.e., significant features extraction) and high-level applications (i.e., registration and identification of 3D objects in partially scanned and cluttered scenes), taking into consideration different types of 3D models (i.e., static and dynamic point clouds, static and dynamic 3D meshes).

# Περίληψη

Η γεωμετρική επεξεργασία τρισδιάστατων αντικειμένων έχει μεγάλο ερευνητικό ενδιαφέρον σε ένα ευρύ φάσμα εφαρμογών που σχετίζονται με την υπολογιστική όραση και τα γραφικά, όπως για παράδειγμα στην αυτόματη κίνηση robot στον χώρο, την αναγνώριση τρισδιάστατων αντικειμένων, την κατηγοριοποίηση, την εξαγωγή χαρακτηριστικών και άλλες.

Η δημιουργία νέας γενιάς οικονομικών αισθητήρων, που έχουν αναπτυχθεί τα τελευταία χρόνια, έχει παίξει σημαντικό ρόλο στην προσέλκυση ενδιαφέροντος από διάφορες περιοχές, οι οποίες έχουν σκοπό να αναπτύξουν εφαρμογές που να κάνουν χρήση της 3D πληροφορίας επωφελούμενες των πλεονεκτημάτων που προσφέρει μια πιο ολοκληρωμένη αναπαράσταση του κόσμου (3D αναπαράσταση) σε σύγκριση με την εικόνα και το ιδεο.

Οι υπάρχουσες προσεγγίσεις που σχετίζονται με την επεξεργασία 3D αντικειμένων, απαιτούν, σε σημαντικό βαθμό, ειδικές παρεμβάσεις από τους χρήστες και επίσης είναι αρκετά χρονοβόρες για να χρησιμοποιηθούν σε εφαρμογές πραγματικού χρόνου.

Στόχος της παρούσας διατριβής είναι να παρουσιάσει αλγορίθμους που μπορούν να χρησιμοποιηθούν αποτελεσματικά και ωφέλιμα σε πλήθος χαμηλού επιπέδου εφαρμογών που αφορούν την επεξεργασία 3D γεωμετρίας όπως (ανακατασκευή επιφάνειας, απομάκρυνση έκτοπων σημείων, αποθορυβοποίηση, συμπίεση) σε εργασίες που αφορούν την αναγνώριση προτύπων (εντοπισμός σημαντικών σημείων και χαρτογράφηση βάση γεωμετρικής σημαντικότητας) καθώς και σε υψηλού επιπέδου εφαρμογές (εντοπισμός 3D αντικειμένων σε σκηνές μερικής αποτύπωσης), λαμβάνοντας επίσης υπόψην διαφορετικούς τύπους 3D μοντέλων (στατικά και δυναμικά νέφη σημείων, στατικά και δυναμικά 3D αντικείμενα πλεγματικής αναπαράστασης).

# Contents

# List of Figures

# List of Tables

# Acronyms

**ADC**  Analog-to-Digital Converter

**ADMM**  Alternating Direction Method of Multipliers

**ALM**  Augmented Lagrange Multiplier

**AQNM**  Additive Quantization Model

**AR**  Augmented Reality

**BM3D**  Block-matching and 3D filtering

**bpv**  bit per vertex

**bpvf**  bit per vertex per frame

**CAD**  Computer-aided Design

**CF**  Collaborative Filtering

**CM**  Coherent Matrix

**CNN**  Convolutional Neural Network

**CT**  Computed Tomography

**DAC**  Digital-to-Analog Converter

**DAE**  Denoising AutoEncoder

**DCT**  Discrete Cosine Transform

**DNN**  Deep Neural Networks

**DOI**  Dynamic Orthogonal Iterations

**DPCS**  Dynamic Point Cloud Sequence

**EM**  Expectation-Maximization

**GFT**  Graph Fourier Transform

**GMC**  General Matrix Completion

**GNF**  Guided Normal Filtering

**GS**  Gram-Schmidt

**GSP**  Graph Spectral Processing

**HD**  Hausdorff Distance

**HR**  Householder Reflections

**ICP**  Iterative Closest Point

**IGFT**  Inverse Graph Fourier Transform

**ISVD**  Incremental Singular Value Decomposition

**k-NN**  kNearest Neighbors

**LASSO**  Least Absolute Shrinkage and Selection Operator

**LIA**  Laplacian Interpolation Approach

**LIDAR**  LIght Detection And Ranging

**LM-filt**  Limited-memory Filtering

**LSM**  Least-Square Meshes

**MBL**  Model-based Bayesian Learning

**MIMO**  Multiple-Input and Multiple-Output

**MC**  Matrix Completion

**MeshHOG**  Mesh Histogram of Oriented Gradients

**MGE**  Mean Geometric Error

**MGS**  Modified Gram Schmidt

**MHB**  Manifold Harmonics Basis

**MND**  Mean Normal Difference

**MRI**  Magnetic Resonance Imaging

**NMSVE**  Normalized Mean Square Visual Error

**OI**  Orthogonal Iterations

**PCA**  Principal Component Analysis

**PCP**  Principal Component Pursuit

**RANSAC**  RANdom SAmple Consensus

**RPCA**  Robust Principal Component Analysis

**SHOT**  Signature of Histograms of OrienTations

**SOCP**  Second-Order Cone Program

**SVD**  Singular Value Decomposition

**SVM**  Support Vector Machine

**TVPC**  Time Varying Point Cloud

**TSGSP**  Two Stage Graph Spectral Processing

**VR**  Virtual Reality

**WRLI**  Weighted Regularized Laplacian Interpolation

**3DSC**  3D shape context

**4PCS**  4-Points Congruent Sets

CHAPTER 1

# Introduction

Data, in an abstract sense, can be described as the driving force behind every action. Information is all around us, so the amount of data that we can proceed with is limitless. The 20th century was announced as the century of data. Nowadays, the one who holds the data has the power [46]. For centuries, the only way to have access to data and information was through the written word (i.e., printed books). Writing is one of the most important inventions in the history of humanity, transferring information between generations. However, this information can not be used in real-time scenarios since it is no highly interactive with the users and it mostly represents knowledge of the past or from an imagery world, making it difficult for nowadays applications to productively work only with this type of information.

From all of the human senses, the most important to perceive the information of the world around us are the senses of vision and audition. These senses and their outputs are also important for computer science since visual and audio signals can easily be digitized, using cheap sensors, in contrast to taste and olfaction. Historically, the processing of data was started by using audio signals (Fig. 1.1), then stepwise was transferred to 2D images and sequence of images (video), 3D images (Fig. 1.2) and finally to 3D geometric objects (Fig. 1.3).

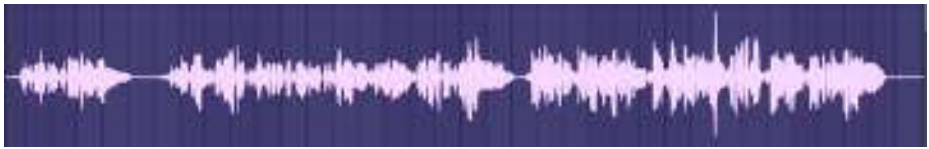

Figure 1.1: Example of an audio signal representation.

Even though the aforementioned types of signals have different forms and characteristics, they share the same needs and requirements for processing (i.e., denoising, compression, salient feature extraction). Fortunately, it has been proved through the years, that the same or very similar mathematical tools and fundamental algorithms can be used to achieve the desired processing results.

As referred above, the sense of sight plays an important role in human life.

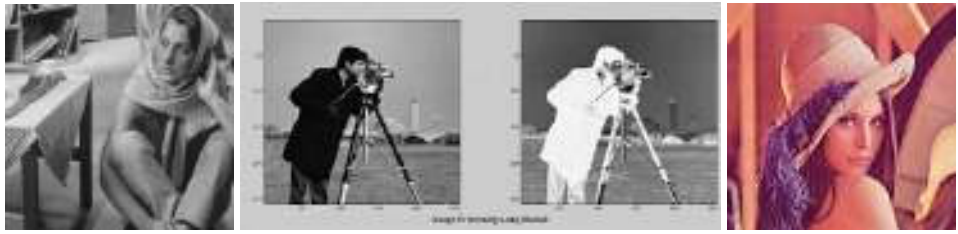

Figure 1.2: Example of 2D and 3D images.

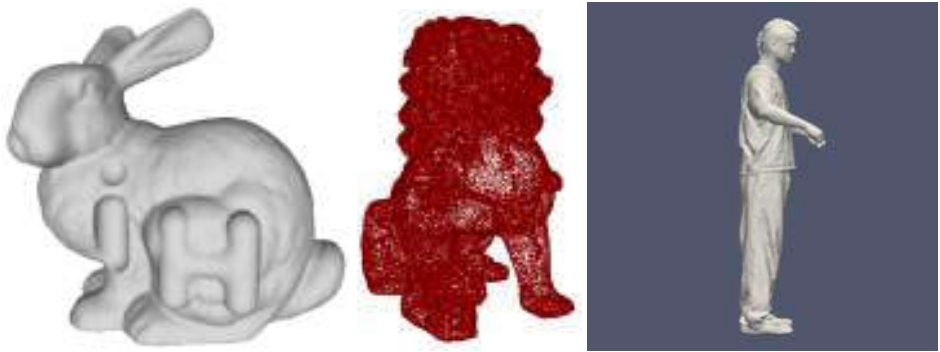

Figure 1.3: Example of 3D objects.

It allows us to connect with our surroundings, keep us safe and help us to learn more about the world. In the same way, vision is also important for robots, affecting their behavior especially when they try to interact with their environment. Robots have become smarter and their behavior is more realistic approaching humans, in many ways. One basic reason leading to this success is the learning process which is an off-line task, though. In real-time applications, robots must discover their world using only their own sensors and tools without an a priori knowledge. Undoubtedly, the type and the precision of sensors affect the robot's perception and behavior. Nowadays, the most complete and representative type of real-world information, which is closer to the human perceptual ability for the world, is the 3D models. 3D models like point clouds or 3D meshes provide a more accurate, plausible and integrated representation of reality. If a picture is worth a thousand words an interactive 3D model is worth multiple times more. 3D objects consist of a collection of vertices (i.e., point clouds), edges, and polygons in the 3D space (i.e., 3D meshes). This digitized form of 3D objects is the standard way of representing a real object of the physical world into the cyber world.

Recently, the new generation of 3D scanner devices has attracted a lot of attention due to their ability to provide more reliable results. The easiness of capturing real 3D objects has created revolutionary trends in many areas and has significantly increased the interest in graphics and dynamic 3D meshes.

Nevertheless, despite the technological evolution of 3D capturing devices, there are still some constraints that can negatively affect the quality of the produced results. In practice, 3D objects that are acquired from the 3D scanning process fail to meet the quality requirements for most practical applications. Defects like holes, missing information due to occlusions, duplicate elements, non-manifold elements, noise, outliers, etc., are introduced during the scanning process which lowers the quality of the output. Additionally, the information acquired by 3D scanners is usually huge and unorganized, creating noisy and dense 3D models that are very difficult to be efficiently handled by other high-level applications and software (e.g., 3D object recognition [47], [48], 3D matching and retrieval [49], scalable coding of static and dynamic 3D objects [50], re-meshing [51], etc.) without further processing (i.e., compression and denoising).

3D point clouds are becoming more common due to the availability of low-cost sensors. They are mainly attributed to the evolution of several 3D acquisition technologies; namely aerial laser scanning [52], [53] (large-scale buildings survey, roads, and forests), terrestrial laser scanning [54], [55], [56] (urban surveys in outdoor and indoor environments), mobile laser scanning [57] (sensors mounted on a vehicle) and more. Light detection and ranging (LIDAR), stereo, structured light, and time-of-flight are examples of sensors that capture a 3D representation of the environment. These sensors are increasingly found in mobile devices and machines such as smartphones, tablets, robots, and autonomous vehicles. As hardware technology advances, algorithms and data structures are needed to process the data generated by these sensors in innovative and meaningful ways.

Computer graphics is an important field of computer science widely used in many real-world applications. The advent of fast, inexpensive and non-commercial hardware for graphics has led to an essential demand for the implementation of new efficient algorithms that can be used in various geometric tasks. This evolution has also affected the traditional applications using image processing, until recently. These applications were adapted, according to the new trends, in order to take advantage of the advance in 3D geometry processing. The area of computer graphics attracts the attention of researchers since it can contribute and facilitate a lot of other domains from scientific applications, engineering, visualization, medical imaging to entertainment, media, game industry, etc. New technologies such as VR and AR offer a world of possibilities for task engagement neither face-to-face collaboration nor remote collaboration in a 3D environment. It creates an effective and efficient way of presentation by offering an interactive display medium in order to cater to the abilities of users' vision. This may enrich the visualization elements during decision-making.

The industry began to use 3D models very quickly and enthusiastically. Industrial design, which has a steady place in production stages, has become heavily dependent on 3D models and modeling programs. Since 3D model-ing facilitated the process during the production phase, it did not have any problems in providing funds from companies for the development of the technology. Computer animation is also an important part of computer graphics. It has attracted attention worldwide and has become one of the most successful applications of digital media technology. It has revolutionized computer animation, film world, TV and computer games industry. Many other fields such as marketing, arts, and sciences have been cooperating with computer animation which has shown us the possibilities of creating more realistic and natural scenes, comprehensive simulations of complex problems, and access to places that are difficult or impossible to discover. 3D animation simply gives life to static 3D objects, creating a sequence of static meshes each of which represents one frame. Furthermore, its applications have become more and more well-known and mostly require animated 3D models and scenes with a high degree of realism. It is therefore inevitable to compress 3D datasets.

In this part of the problem, compression algorithms come into play. Limits have always been a repressive element in developing new compression methods, codecs. Within the limits, various codecs have been developed according to the needs. Reconstruction of 3D surfaces, using data that are acquired by new generation techniques, always entails the problem of the huge mass of noisy data. This implies that numerous pre-processing steps need to be firstly implemented (i.e., denoising, outliers removal) before the reconstructed 3D model is finally available for use in a high-level application. Due to the huge amount of information, it is common that contemporary methods can not be directly applied to the initial data, so reduction methods (i.e., simplification, compression) are mandatory to be used.

This thesis deals with some of the aforementioned problems that appeared in many applications utilizing 3D models. Our main purpose is to solve them by providing novel and robust solutions using a common mathematical framework, exploiting the capabilities of spectral and geometric approaches. More specifically, in this thesis, we exhaustively discuss and present the effectiveness of our algorithms, inspired by some of the well-known and classical spectral frameworks that have been also applied in other domains like image processing. These algorithms have been carefully designed and adapted to satisfy the special requirements and restrictions of the 3D models. We also present how this framework can be used and facilitate many areas of low-level 3D geometry processing (i.e., reconstruction, outliers removal, denoising, compression), pattern recognition tasks (i.e., significant features extraction), and high-level applications (i.e., registration and identification of 3D objects in partially scanned and cluttered scenes), taking also into consideration different types of 3D models (i.e., static and dynamic point clouds, static and dynamic 3D meshes). The remainder of this manuscript is organized as follows.

In Chapter 2, we present the main areas in which the 3D models are used nowadays showing the new trends and how the use of 3D models can benefit these areas. We also present the challenges that applications utilizing 3D models mainly may face. We make an extensive discussion regarding the related state-of-the-art works of each area highlighting how they try to solve these challenges, presenting a variety of different methods and approaches, and finally, we present the contribution of this thesis in the corresponding research areas.

In Chapter 3, we introduce the fundamentals, the basic definitions and terminologies that are usually used in the 3D object processing tasks by starting from the very basic, such as the definition of the vertex, to more complex, like the definition of motion vectors. We also present the different structure representations that might have a 3D object as well as a short description of what the geometrical features and noise represents.

In Chapter 4, we present the mathematical analysis, notions and tools for 3D geometry processing that will be utilized in the next chapters of this thesis. We try to provide everything that may be useful to the readers of this thesis to be familiar with the mathematical types, definitions and assumptions that we have made, so as to be more focused on the pipeline of each approach when they will be described in the following chapters, assuming that all required mathematics are known. Additionally in this chapter, the metrics that are used for the evaluation of each method are presented, as well as the used benchmark datasets and the experimental setup.

In Chapter 5, we present in detail our contribution (i.e., assumptions and general introduction, the workflow of the methods, detailed description of the processes, experimental analysis and results) in a variety of research areas related to low-level 3D geometry processing. More specifically, the scientific areas that we have been involved in this chapter are: (i) the completion and reconstruction of incomplete static and dynamic 3D models, (ii) the outliers removal and consolidation of static and dynamic 3D meshes, (iii) the denoising of static and dynamic 3D meshes and (iv) the compression of static and dynamic 3D meshes. To mention also here that for each area, we provide several approaches taking into our consideration different aspects (e.g., structure type of the model, real-time restriction, etc.) of the problem that we intend to solve.

In Chapter 6, we continue the presentation and analysis of our contributions in other research areas related to 3D geometry processing and pattern recognition such as (i) the feature extraction from 3D objects, (ii) the saliency mapping estimation and (iii) the registration and identification of partially-observed 3D objects in cluttered scenes. We follow the same line of thought, as in the previous chapter, making an extensive analysis to each one of the proposed approaches.

Finally, in Chapter 7, we summarize the basic contribution of this thesis, we present the conclusions and we draw the future research directions. We also present the publications in journals and conferences that have contributed to the content of this thesis.

CHAPTER 2

# Trends, Challenges and Applications of 3D Geometry Processing

A 3D model is the representation of a physical object that has been created by a collection of points in the 3D coordinate system. These points are usually connected by various geometric entities such as triangles, lines or curved surfaces. Bringing objects from the real world into a virtual reality environment is an interesting but difficult task. However, the merits that it can provide are more significant than the difficulties that need to be overcome, since it can positively affect a wide variety of scientific fields, opening new directions and possibilities to these areas. Until recently, the quality of 3D models was limited by the hardware and software capabilities. Today, everyone can easily visualize and manipulate complex models through a variety of available software applications. Moreover, modern scanning technologies make it possible to generate accurate 3D models of real-life objects.

3D models are ubiquitous omnipresent and have become prevalent in many nowadays applications and research areas. They can help deep learning approaches to increase even more their accuracy since they provide a more detailed representation of the real world than the images or the videos does. While new technologies continue to evolve, the trend of using 3D models will continue to be increased. Nevertheless, there are still a lot of challenges and drawbacks that need to be addressed.

## 2.1   3D Models in Applications

The increasing interest for 3D models has brought evolution to many different scientific areas and industries, such as mobile cloud gaming and entertainment [58], heritage culture [59], medicine [60], 3D tele-immersion, communica-tion [61], [62] and more. In the following subsections, we present applications utilizing 3D objects in different domains and they can enhance the value and the expectations for future evolution in these specific areas.

### 2.1.1 Entertainment

3D models have been widely used in various entertainment applications. In the area of computer games, 3D models can be used for the creation of the game characters, background scenes, static objects and special visual effects. Game developers can easily modify these 3D models reproducing them in different poses, angles, colors, material and textures according to the requirements of the game play. 3D modeling generates realistic and immersive effects that can elevate viewers' experience to a new level.

Regarding the 3D movies, designers can easily move, translate, scale and rotate the created 3D models giving them also plausible emotions and reactions. The reproduction and reuse of the 3D models is also more feasible than the 2D drawings.

3D scanners can also be used to create digital 3D models for video games and virtual filmmaking. 3D models are quickly and accurately being created by scanning real-world objects, including actors, instead of creating models from scratch.

### 2.1.2 Immersive 3D Tele-presence Systems

Immersive tele-presentation is one of the most cutting-edge technologies in the area of direct communication and collaboration between humans. These type of systems overcome many limitations of the conventional video-based communication platforms and offers novel possibilities for an immersive remote collaboration [63].

In contrast to a VR application, where the user is transferred to a new virtual world, immersive tele-presence systems try to give the user the impression to be in another location of the real world [64] where other real people also exist and interact with them in real-time, allowing people at different locations to meet in common spatially consistent environments.

Immersive tele-presence systems can draw novel possibilities for remote collaboration to improve joint problem analysis and decision-making in context. Different users in different physical locations can be presented together inside a common area. They can explore the interior and discuss topics of interest face-to-face with speech and body language.

### 2.1.3 Industry

Visual computing technologies play an important role in several manufacturing tasks. Particularly nowadays, their role is crucial due to the new Industry 4.0 applications including: manufacturing inspection [65], quality control [66], [67], reverse engineering [68], digital twin [69] as well as autonomous repair operations.

In many manufacturing cases, the final shape of the manufactured product may differ from its originally intended design due to the irregularities encountered during the manufacturing processes. In such cases, conducting 3D shape inspection will enable quality assurance by quantifying the differences between a scanned model of the final product and a 3D CAD model of its intended design [70].

Quality control of products plays an important role in various stages of the manufacturing process. In particular, the final control of the quality of a product before being shipped to a customer is crucial for maintaining customer satisfaction and avoiding costly recalls. An automated product inspection station based on the use of a 3D scanner and related camera technology [71].

Reverse engineering is an efficient approach to significantly model the product and reduces the product development cycle. Currently, the coordinate measuring machine and the three-dimensional 3D laser scanner are widely used in the fields for shape reverse engineering and quality inspection [72].

Digital Twin is the creation of digital data from physical to collect data, data communication, simulation, predictive maintenance [73]. The digital twin is defined as a key solution for the digital transformation in manufacturing. The digital twins represent real objects or subjects with their data, functions and communication capabilities in the digital world. Digital twins are a combination of different models, data and information. Digital twins have long been pursued in the modeling and simulation community. The purpose of multi-physics simulation and system simulation is to accurately replicate and answer detailed questions about the relationship between the systems [74].

While the use of this new type of applications will expand, the number of digital 3D models will be also increased, resulting in the interest for more accurate 3D model processing.

### 2.1.4 Medicine and Healthcare

3D models can be also used in the medical industries in order to construct detailed models of organs. For example, the production of patient-specific parts directly from scan data is an obvious benefit for the medical industry. Nowadays, 3D bio-printing can print blood vessels, skin tissue, heart tissue, and ar-tificial bones for surgical therapy or transplantation. The potentials of 3D bioprinting in medical science may create many business opportunities and greatly benefits patients [75].

Complex surgical procedures can also benefit from medical 3D printing since the technology allows the creation of a relatively cheap replica of the situation that is undergoing surgery, by printing mock body parts which can be used to plan or train on difficult operations. Typically, the already available technical knowledge, i.e. MRI and CT imaging, can be used to generate the 3D model. The models require high tolerances but can be printed by non-medical companies [76].

Medical training uses detailed 3D animations of the human body to walk viewers through complex processes. When 3D animation is used in medical videos, the need for as much hands-on experience decreases. 3D animations, used in medicine, are also a step-up from the stationary, physical 3D models used in the past. Medical students can see all the movements a real human body makes.

3D modeling is also used in the healthcare industry to design medical equipment such as stethoscope, prosthetics, blood pressure machines etc. 3D models of anatomy and processes are also used to train medical staff. 3D modeling is associated with benefits such as precision and detail. The designing of medical equipment is based on accurate measurements and an attention-to-detail approach and this is the reason why it is widely deployed in the manufacturing of medical equipment.

Some phobias can also be treated: immersion therapy for arachnophobia would previously have required a collection of spiders, fear of public speaking would require a crowd, claustrophobia would need a real-life small space, and so on.

### 2.1.5 Automotive

The recent advancements in automotive electronics have resulted in safer and more convenient vehicles. An advanced driver assistance system (ADAS) is one of the core electronic systems, whose performance is strongly determined by ADAS sensors. Typical ADAS sensors include ultrasonic sensors, cameras, radio detection and ranging (Radar), and light detection and ranging (Lidar) systems. Lidar sensors are usually mounted on the roof of the vehicle and provide information regarding the objects surrounding a vehicle using laser pulses. Lidar sensors are critical to recognition algorithms of ADAS and autonomous vehicles and are essential in path planning and reactive control of autonomous vehicles [77]

Unlike cameras, 3D Lidar is robust regardless of lighting conditions, both during the day and at night and both with or without glare and shadows [78]. The lidar outputs include 3D point clouds that correspond to the scanned environments and the intensities that correspond to the reflected laser energies. [79]

Point clouds, extracted by the Lidar devices, can preserve the 3D geospatial information and the internal local structure. They can be easily analyzed for 3D semantic segmentation, 3D object detection/recognition/classification, tracking of moving objects (cars, pedestrians, etc) and localization.

Automotive AR head-up (HUD) interfaces provide a meaningful application area to examine this problem, especially given that immersive, 3D-graphics-based driving simulators are established tools to examine in-vehicle interfaces safely before testing in real vehicles [80].

In near future, the windscreen of autonomous vehicles will become an AR display, providing drivers/passengers with information, warnings and alerts [81].

### 2.1.6 Heritage

Nowadays, the preservation and maintenance of historical objects is the main priority in the area of the heritage culture. The new generation of 3D scanning devices and the new assets of technological improvements have created a fertile ground for developing tools that could facilitate challenging tasks which traditionally required a huge amount of human effort and specialized knowledge of experts (e.g., a detailed inspection of defects in a historical object due to aging). These tasks demand more human effort, especially in some special cases, such as the inspection of a large-scale or remote object (e.g., tall columns, the roof of historical buildings, etc.), where the preserver expert does not have easy access to it.

Geometric details of a 3D historical model, such as high-frequency features, and surface areas consisting of points-of-interest (e.g., cracks in an ancient vase) must be preserved, keeping their original high-resolution quality. These salient areas must be handled differently (i.e., anisotropically) from the rest 3D model (i.e., smooth or flat areas), especially in applications that deal with sensitive content such as the 3D representation of historical objects. In this type of application, any perceptual detail or even any possible defect of the object must be preserved since they represent importantly valuable information that can be used through the process of the object's maintenance and preservation.

Defects are referred to as an abnormal condition that affects the original structure of a heritage object (e.g., statue or a building). They may originate from the slow and inevitable ageing processes due to: (1) material deterioration, (2) modifications by humans (e.g., in the urban environment), and (3) climate and environmental changes [82]. The preservation and maintenance of heritage objects focus on protecting them from being destroyed and trying to preserve both their aesthetic and historical values. Conservation emphasizes the importance of preserving cultural properties and tries to extend the life of historical objects [83].

Monitoring and visualization of historical objects is a necessary process for the visual inspections, in order to constantly detect if any intervention of a professional expert is required. Therefore, it seems necessary to develop tools for implementing an effective preventive policy, which will take into account all the conservation requirements. However, visualization is not always enough for the preservation of historic buildings since the conservation professionals do not only need to navigate the 3D model but also to perform spatial and multicriteria queries in a virtual 3D environment for making decisions. This requirement is even more urgent in cases where the buildings demonstrate some critical evolution, cracks or any potential collapse [84]. Parametric tools can reduce human involvement, providing at the same time reliable inspection results in an automatic process. Such tools can facilitate the work of professionals, highlighting salient region-of-interest and providing them with other suggestions and reports.

### 2.1.7 Military

3D animations can be used for military and police training to avoid the risk of injury. In addition, the military and police might also use 3D animation to design their machining and engineering, for a better machinery performance without the need for physical manufacturing to testing.

Developed and verified a battlefield visualization framework may help in military decision-making using the technology of holograms, AR, VR and multi-touch table technology.

The application of VR and AR technologies will give a positive impact on a collaborative setting by visualizing the geography and geopolitics of the area of operation in order to get a realistic and real picture of the ensuing battlefield scenario [85].

### 2.1.8 Education

Educational models have evolved over the years, adapting to reality and the needs of society, thus involving technology in the teaching and learning processes through multimedia applications, access to information through the internet, virtual classes, communication channels, etc. Virtual reality is being used as a tool for the teaching-learning process in several research projects. For example, an implemented virtual system has been developed that allows real-time bilateral interaction between the user and the virtual environment with the aim to stimulate the abilities of children with autism spectrum disorder. Complex virtual worlds are created that can be used not only to entertain but also to educate people. In fact, the educational sector is one of the most benefited by this type of technology as a tool of teaching [86].

3D animation for educational purposes is fun and memorable for students learning a new process or idea since young people are more familiar with 3D animation because they see it so much in entertainment. 3D models from real-life objects can be used to achieve high academic standards [87]. Using 3D anatomical models in medical education are very beneficial and it is also a powerful tool in learning anatomy. Many medical instructors can use 3D models to illustrate the anatomical structure in a more effective and convenient way than using a cadaver. By using this method, students will understand the various complex anatomical structure more easily.

The innovative features of the 3D printing model can also be used to explore the ability of students' independent innovation student. In some art designing major which is closely related to 3D printing technology, the students' designing can be directly printed into a fine model, and as an effective tool, the teaching evaluation model can be improved. Geometry courses related to mathematics can abstract complex 3D geometry through 3D printing visualization, to promote students to absorb knowledge. Through the creation of 3D printing courses, and the promotion and application of 3D printing technology in the field of education, we can subvert the traditional education thought, promote innovative three-dimensional education carried out by 3D technology, to provide a solid foundation [88].

While using 3D modeling technologies in education, the following opportunities appear [89]:

- Visualization of parts and assemblies in space
- Holistic analysis of problems from related fields in relation to the same product
- Development and analysis of complex structures
- Development of abstract and logical thinking of the student

### 2.1.9 Architecture and CAD

In the architecture industries, there is the need to illustrate proposed buildings and landscapes. Building information modeling (BIM) uses a reverse modeling process that takes place from transforming point cloud into a 3D model [90]. BIM is considered a promising resource for the planned conservation of buildings thanks to its capability of archiving and organizing all the information about a building. However, a considerable amount of information has to be stored and transmitted efficiently and quickly [91], to be accessible and modifiable by many different professionals involved in the same project [92]. This need mandates the use of compressed models since the initially scanned 3D models are very dense (with millions of vertices). As a result, conventional software cannot handle them without further processing (e.g., simplification and compression).

The workflow of a preservation process starts with data acquisition, which represents the digitized information of the historical object-of-interest, collected by different sensors or devices. The point cloud is subsequently pre-processed and reduced with various systems and dedicated software. During the point cloud conversion to 3D model, several decisions have to be made, such as the range of precision, also called as Level-of-Detail (LoD) [93]. The Level-of-Detail represents the expected complexity, accuracy and change of the small-scaled features (i.e., fine details). Regarding the aspect of spatial information, the 3D models are being generated at different levels of detail and scales using methodologies based on different accurate data acquisition techniques [84]. Further research is needed in order to identify which LoD is ideal for different tasks or to easily create simplified models on demand, especially for scalable coding applications using AR/VR technologies.

With 3D animations, architects can experience their buildings in VR or AR. Architects can move through the building, experience it at all hours of the day, and identify and amend design flaws. Before this, architects had to present their designs in flat paper drawings. Now, the building can come to life before the builder has laid the first brick.

Realistic 3D models allow architects, the construction team and prospects to understand project scope well, which results in fewer surprises as the project progresses. Additionally, architects can manipulate 3D models in their preferred ways, which they cannot do with 2D drawings. They can also verify various options in the 3D scale to identify potential scenarios. This would help them to take the right steps and eliminate errors in the early stages, 3D models, a design team can gain a proper understanding of how a change in one aspect can affect the other aspects.

A building project remains in a selling process in all its stages. A detailed and realistic model of a real estate property is a crucial marketing tool that real estate businesses can use to attract prospective clients and to show them how a finished product would look like. Real estate businesses can use gorgeous and detailed 3D walkthroughs to market their properties to potential clients. Prospects find 3D real estate models to be compelling and they are more likely to feel get an immersive experience with 3D imagery compared to the level of attachment that they would have with 2D drawings.

## 2.2 State-of-the-art in the 3D Geometry Processing

In recent years, there has been increasing development of new generation sensors and 3D scanning devices, however, their precision is still limited. As a result, the acquired 3D point cloud suffers from noise, outliers, and incomplete surfaces. These abnormalities must be solved before the raw 3D scanned models are used for further processing in other high-level applications. Although the 3D static point cloud acquisition may seem straightforward, if the object moves while the 3D scanner device receives points, then the relative motion between target and object results in the capturing of false information which appears as outliers and other artifacts. Additionally, different light conditions can also affect the final results.

Range scanning is a type of 3D scanning technique that uses emission devices like light, X-ray or ultrasound. A range scanner can collect geometric information and construct a raw mesh for a target object. In practice, range scanning methods work best when a target object has a flat profile shape or a simple convex curved surface. It is capable of producing high-resolution, visually accurate models. However, it has a few limitations, since the technique uses a linear sensor, and this sensor can only detect the surface regions where light rays can reach. For concave regions where rays cannot reach (meaning that the sensor's sight is obstructed by the object itself), the scanner cannot get the geometry information. As a result, some parts of an object are left unscanned, leading to geometrical errors including holes, non-manifold, isolated elements, etc. Small cracks, deep concavities and multiple layers can cause the same problem.

To solve these problems, repairing processes must be applied to fix a raw mesh and eliminate these geometric errors. The typical 3D geometry processing stages[1] that could be applied after the scanning are:

(i) Outliers removal

(ii) Consolidation and reconstruction of the partially scanned areas

(iii) Feature extraction and pattern recognition

(iv) Denoising or smoothing

(v) Compression or simplification

These pre-processing steps alleviate many of the imperfections that are introduced by the aforementioned challenges, such as outliers, missing regions and noise, in order to faithfully reproduce the shapes of the scanned objects.

---

[1] Non necessarily with the same order or all of them to be apparent

During the surface reconstruction stage, a watertight mesh is generated from a sufficiently well sampled and processed point cloud.

In the following subsection, we will present and discuss the main challenges and the related work that try to address these limitations in different low-level 3D geometry processing applications, namely completion and reconstruction, outliers removal, denoising, compression as well as salient feature extraction and registration and identification of partially observed objects in cluttered scenes.

### 2.2.1 Completion and Reconstruction

Real-time 3D captured data play an important role in fields ranging from modeling dynamic scene geometry and holographic communication to plausible created avatars in games and others. In all these applications, the online consolidation of dynamic point cloud still remains a challenging issue since it requires the recovery of a large amount of missing data, increasing significantly the computational complexity and the execution time. Although the resolution and accuracy of 3D scanners are constantly improving, they are still unable to capture the full surface at once. The captured 3D point clouds are usually highly incomplete, stressing the need for completion approaches with low computational requirements. Even in scenarios where multiple sensors are placed around the subject, most scanned shapes are likely to exhibit large holes and these effects are attributed to occlusions, limited sensor range capabilities, high light absorption and low surface albedo [94].

**Challenges**

A lot of research has been carried out into the field of 3D mesh reconstruction and completion of static geometries, having presented excellent results applied to incomplete static meshes. However, little attention has been given to the reconstruction of animated meshes. Traditional methods usually cause temporally incoherent surfaces when they are directly applied to each frame individually. These methods do not take advantage of previous frames' knowledge, as a result, they basically deal with $n$ individual meshes instead of a sequence of temporally coherent meshes. A common approach to produce a temporal consistent dynamic mesh is to use a template prior [95], however, this approach is not ideal for real-time applications because the entire captured animated mesh is required before the execution of the process.

Another challenge that has to be addressed is the task of 3D inpainting. Inpainting has been originally used for images or videos recovering, denoising, removing undesired objects and repairing, having been widely investigated in a lot of applications. Although the problem of inpainting has been thoroughly studied in the for images and video [96], [97], [98], very few focus on dealing with incomplete surfaces in scanned 3D objects. Formerly, the majority of the state-of-the-art methods tried to solve this problem using geometrical constraints [99] in order to estimate the missing areas based on their neighbors' geometry.

**Related Works**

A fast and efficient approach for reconstructing surfaces from a set of known points has been proposed in [100] where the authors reconstruct meshes with a prescribed connectivity that approximate a set of control points in a least-squares sense. The authors in [101] focus on reconstructing watertight surfaces from unoriented point sets using a Voronoi-based variational approach, while the method in [102] tries to handle the missing points by trying to infer topological structures in the original surface at the potential expense of retaining geometric fidelity. The researchers in [103] perform reconstruction only on the available information, effectively preserving the boundaries from the scan. Recently, a new signal processing technique known as Matrix Completion (MC) [104] has been successfully applied to several computer vision problems, including the recovery of occluded faces/dynamic meshes [105], [106], [107] and the face image alignment [108]. It has been also used for the fusion of point clouds from multiview images of the same object [109]. In [110], it is applied on RGB-D data for the simultaneous tracking and reconstruction of 3D objects.

Although a large number of prior works [111] has investigated the problem of completion in static geometries, resulting in excellent filled static meshes, their direct application to every frame separately usually causes incorrect topologies and temporally incoherent surfaces. Most of the methods try to solve the problem by isolating each static point cloud of the sequence and handle it individually as a common 3D point cloud but without taking into account the temporal coherence between sequential frames [112]. Other approaches use a template prior in order to produce a temporal coherent dynamic point cloud, nonetheless, these approaches are not ideal for real-time applications because the entire captured point cloud sequence is required in advance before the execution of the process [113].

The authors in [114] assume that a Riemann surface can be uniquely determined by its conformal factor and the mean curvature, while the authors in [115] use principal curvature. The authors in [116] complete missing values based on background estimates of the unoccluded scene. In [96], local image mode filtering is used for the inpainting of a depth map created by a Kinect camera. In [117] a Markov random field-based image inpainting algorithm is presented. Recently, alternative approaches have started to gain popularity such as the exploitation of sparsity or the minimization of a quantity. The authors in [118], [97], [119] investigate the sparsity of image patches for image inpainting. In [120], [98], the researchers use low-rank matrix completion approaches for inpainting depth and light field images, correspondingly. In [121], the authors achieve surface completion by minimizing energy based on similarity of shape, while in [122] they use energy minimization of both shape and texture of a surface. The authors in [123] use sparsity constraints for inpainting highly incomplete 3D surfaces. Despite the good results, the approach requires eigen-basis decomposition which is very time-consuming, especially in the case of dense meshes. In [124], a dictionary of patches is created, trying to solve a sparsity problem. Most of the aforementioned methods use complex computational solutions or have a restricted focus only on local geometric information.

### 2.2.2 Outliers Removal

Lately, there is a surge of interest in the processing of unorganized point clouds, as they arise from the multitude and abundance of active 3D scanning technologies such as structured light and lidar systems. Based on the principle of triangulation [125] [126], 3D laser scanners typically emit laser light onto an object and their cameras capture the diffuse reflection from the object surface. However, when the scanned surfaces are highly reflective, the cameras may receive undesirable specular reflections, which can eventually record points with large measurement errors as outliers.

Outliers can be defined as clusters or sparse points, whereas the noise is usually non-isolated but exhibits different attributes from valid points. Some types of outliers that can be observed in a 3D scanned object are:

- Sparse Outliers
- Isolated Outliers
- Non-isolated Outliers
- Clusters

Specifically, the large-scaled or isolated outliers deviate so much from other observations as to arise suspicion that it is generated by a different mechanism. The outliers have to be removed in an early stage of 3D object processing since if they remain then they may be treated in the same way as the valid points decreasing the final quality of the processed object.

**Challenges**

Despite the steady improvement of the capturing sensors in both accuracy and resolution, imperfections attributed to the physical limitation of the devices and to light occlusions, still need to be addressed. One major problem of Lidar devices is the generation of outliers due to: (i) the limited quality of captured scanning devices, (ii) a relative motion between captured device and target, (iii) obstacles which can be located in front of the object of interest and/or (iv) variable lighting conditions.

Depth cameras have started to be used in a lot of robotic applications providing extra useful information (geometric and topographic) which is not able to be provided by conventional cameras. Robots need 3D visual perception in order to understand better their environment and demonstrate a more satisfying behavior. A detailed vision capability is necessary for many challenging applications especially when further processing tasks are required; like segmentation, object recognition and navigation. For example, robots need to intelligibly specify and identify their environment (borders, existing objects, obstacles) before they start to navigate inside it.

On the other hand, depth cameras are very sensitive to motion and light conditions and this may cause abnormalities to the final results, especially when the robot moves while it captures the point cloud. These abnormalities mainly attributed to (i) device limitations, (ii) surface reconstruction errors and (iii) non eliminated outliers, with the latter two having a larger influence. If a robot moves, while it receives points, then its line-of-sight focus might become unstable creating outliers to the acquired 3D point cloud. Additionally, different light conditions can affect the final results.

To overcome these limitations, it is usually formed an essential preprocessing pipeline before other operations on the point cloud (e.g. registration, surface reconstruction, estimation of normal vectors, object classification), since the removal of scanning imperfections and artifacts typically improves dramatically the quality of the subsequent procedures. More specifically, decoupling these operations can aid in avoiding premature and erroneous decisions.

**Related Works**

In literature, several works that focus on 3D static point cloud processing have yielded excellent results, but Dynamic Point Cloud Sequences (DPCSs) typically receive little attention. Although recent works [127] have started to take advantage of the temporal coherence between sequential frames, most of the existing methods approach the problem by isolating each static point cloud of the sequence and handle it individually as a common 3D point cloud [112]. Outliers removal in 3D unorganized point clouds has been addressed in many articles [26], [128]. RPCA, which has mainly been used for denoising in image applications, has started to attract a lot of attention for using it in 3D model applications. In [129] a RPCA approach has been used for the restoration of noisy point cloud data, while in [130], a simple and effective method is presented for removing noise and outliers from a point set generated by image-based 3D reconstruction techniques. The authors in [131] present a real-time approach for outliers removal using RPCA, but despite the promising results, no attention has been given to the exploitation of temporal coherence.

### 2.2.3 Denoising-Smoothing

The acquired 3D models usually come as very dense and noisy meshes that stand in need of solutions capable of distinguishing noise from local geometric features. Similarly, the surfaces extracted from volumetric data (e.g., MRI and CT devices) contain topological and geometric noise that needs to be removed before further processing. Although there are several methods in the literature for performing feature preserving mesh denoising, there are still challenges that need to be addressed, especially if we take into account that the scale of acquired data in real-time scanning operations is growing very fast. We no longer deal exclusively with individual shapes, but with entire scenes, possibly at the scale of entire cities with many objects defined as structured shapes (e.g., aerial scanning [132], [133], slam scanning [134], underwater scanning [135], scalable multi-object scanning [136], large-scale terrestrial scanning [137], large statues scanning [138]) resulting in a sequence of 3D surfaces that are affected by noise with different characteristics.

**Challenges**

Throughout the years, numerous approaches have been proposed to improve more or less some of the key feature preserving denoising characteristics [9], [10], [8]. Though, recovering the structure of large scenes is still without a doubt a stimulating challenge. To the best of our knowledge, none of the approaches is capable of identifying accurately (i) the subspace where the features of each surface part lie, (ii) the noise characteristics on the different parts of the surface. Those requirements become more essential in scenarios where large 3D models are scanned in parts, generating a sequence of 3D surfaces that arrive sequentially in time and require fast and accurate feature preserving surface denoising. In those cases, the denoising method should be also capable of mitigating dynamic noise with varying characteristics per surface part. This effect is attributed to the fact that several environmental factors that affect the surface noise pattern of each part may also change in time, e.g., variable lighting conditions and orientation of a handheld scanner, standing in need of solutions that are capable of identifying the noise and the geometry characteristics.

Despite the rapid advancements in 3D mesh and point cloud processing little attention has been given in the area of dynamic 3D mesh denoising. The main reason is that each frame of a dynamic sequence can be considered as an individual mesh which can be handled separately, providing space to researchers working in the area of adaptive processing, efficiently exploiting temporal coherence between sequential frames. However, the new trends of technology in a variety of virtual/augmented reality (VR/AR) applications, demand to be given more effort for the processing of dynamic 3D meshes denoising as a different object (3D animation). Additionally, this new approach could be proved beneficial, providing more accurate denoising results, since the dynamic sequences hide temporal information which is also could be efficiently used. 3D mesh denoising (static or dynamic) is a vital pre-processing step which must occur before other more complicated processes (e.g., transmission, segmentation, deforming, compression, etc.) take place. These processes generally require accurate inputs (fully denoised 3D models) in order to provide accurate and high-quality output results. Without a doubt, a lot of works have been presented in the area of 3D mesh denoising. However, despite the significant good results that some of them provide [9], [6], [14], [12] a lot of problems are still remain indicating the need for using more sophisticated approaches.

A big constraint of many state-of-the-art works is that they are based on the not valid assumption that the noise, affecting the surface of the 3D object, has a Gaussian distribution. This assumption does not hold in principle since in several real-life applications the type and the form of noise have different characteristics (e.g., staircase effect, outliers, device's noise, etc.). Another severe limitation is the fact that many methods utilize modified parameters for each model [139], [140], [141] and only some parameter-free approaches Wang *et al.* [10] have been proposed, which also have limitations mainly because they rely on a large dataset for the training process.

Valid identification of features gives the advantage of applying non-isotropic approaches to feature and non-feature areas addressing the aforementioned challenges. However, the identification of features, including corners and edges, in a noisy 3D mesh is not always easy, especially when spectral properties of noise and features overlap. Despite their reconstruction benefits, there are still challenges that need to be addressed. Meanwhile, several researchers, system engineers, and software developers have shown increasing interest in the application of deep learning approaches for performing several low-level information processing tasks, such as denoising, compression, etc., in image and video applications.

**Related Work**

The existing approaches that used for denoising can be classified into three core groups that treat differently noise and salient features, known as **isotropic**, **anisotropic** and **data-driven approaches** respectively.

**Isotropic Methods**. The most well-known isotropic method is the Laplacian smoothing [142] that has mesh shrinkage as a major disadvantage. In order to solve this limitation, Taubin [143] proposed a two-stage approach where two sequential filters are applied iteratively: the first one performs Laplacian smoothing while the second is used to prevent shrinkage. The accurate reconstruction depends on the shrinkage and inflation parameters, which are selected empirically and in many cases suffer from stability issues. Desbrun et al. extended this approach to irregular meshes by using the mean curvature flow analogy to rescale the mesh preventing shrinkage [144]. Several surveys that cover basic definitions and applications of the graph spectral methods have been introduced by Gotsman [145], Levy [146], Sorkine [147] and more recently by Zhang et al. [51]. All these surveys classify the spectral methods according to several criteria related to the employed operators, the application domains and the dimensionality of the spectral embeddings used. Graph spectral processing of 3D meshes relies on the singular/eigen-vectors and/or eigenspace projections derived from appropriately defined mesh operators [148]. A summary of the mesh filtering approaches that can be efficiently carried out in the spatial domain using convolution approaches is given by Taubin in [149]. Computing the truncated singular value decomposition can be extremely memory-demanding and time-consuming. To overcome these limitations, subspace tracking algorithms have been proposed as fast alternatives relying on the execution of iterative schemes for evaluating the desired eigenvectors per incoming block of floating-point data corresponding in our case, to different surface patches [150]. The Orthogonal iterations are the most widely adopted subspace tracking method, due to the fact that it results in very fast solutions when the initial input subspace is close to the subspace of interest [151].

*Spectral methods:* Spectral methods have been excessively used in the image, video, and signal processing domains trying to solve low-level problems by manipulating the eigenvalues, eigenvectors, eigenspace projections, derived from the graph Laplacian operator. In the same way, spectral methods can be utilized for the processing of 3D meshes consisting of connected vertices. However, the computational complexity and the memory requirements of these methods strongly depend on the density of the 3D model, resulting in becoming prohibitive when the number of vertices significantly increases. As it has been suggested in [152], [153], this issue can be addressed if the raw geometry data were divided and processed separately in blocks representing different overlapping parts of a mesh, namely submeshes. More specifically, the direct implementation of the SVD method on the graph Laplacian of each submesh, has an extremely high computational complexity, requiring $\mathcal{O}\left(n^3\right)$ operations, where $n$ denotes the number of vertices in a 3D mesh. Spectral methods can be classified according to several criteria related to the employed operators, the application domains and the dimensionality of the spectral embeddings used. Graph Spectral Processing of 3D meshes is based on the singular/eigenvectors and/or eigenspace projections derived from appropriately defined mesh opera-tors. There is a big variety of different tasks in which GSP has been used, such as implicit mesh fairing [154], geometry compression [147], [155] and mesh watermarking [156]. Taubin [157] was the first that treated the coordinate vertices of a 3D mesh as a 3D signal, introducing the graph Laplacian operators for discrete geometry processing. The similarities between the spectral analysis concerning the mesh Laplacian and the classical Fourier analysis motivated him for this analysis. Despite their applicability in a wide range of applications such as denoising, they require the computation of explicit eigenvector making them prohibitive for real-time scenarios. Additionally, there are a lot of applications in literature in which large-scale 3D models are scanned in parts [158], [137], [159] providing in this way a consecutive sequence of coherent 3D surfaces that need to be processed fast. Computing the truncated singular value decomposition can be extremely memory-demanding and time-consuming. To overcome these limitations, subspace tracking algorithms have been proposed as fast alternatives relying on the execution of iterative schemes for evaluating the desired eigenvectors per incoming block of floating-point data corresponding in our case, to different surface patches [160]. The most widely adopted subspace tracking method is the OI since it provides very fast solutions when the initial subspace, which is given as input, is close enough to the subspace of interest. Additionally, the size of the subspace remains at a small level [161]. The fact that both matrix multiplications and QR factorizations have been highly optimized for maximum efficiency on modern serial and parallel architectures, makes the OI approach more attractive for real-time applications.

**Anisotropic methods**. Most feature preserving mesh denoising approaches locally adjust vertex positions while respecting the underlying features and can be classified in four major categories.

*Anisotropic Geometric Diffusion:* The first one is based on anisotropic geometric diffusion [162], [163]. In [164] the new vertex position is estimated by a nonlinear weighted mean of local neighborhood vertices. The method is simple and can be easily reproduced but its computational complexity is quite high. In [165] sharp features are removed while preserving small-scale features of the geometry of the original mesh. The main limitation of these approaches is the distortion of the geometric features, which is attributed to the fact that all the vertices are handled in the same manner.

*Bilateral Filtering of Vertices and Normals:* The majority of the SoA approaches are trying to denoise a mesh taking advantage of spatial similarities of the noisy object's surface. These iterative methods use normals to estimate the weights of the filters. However, the accurate reconstruction of the geometric features strongly depends on the effect of noise in the weight estimation [4], [5]. The authors in [166] present a robust approach for denoising triangular surface meshes using a combination of bilateral filtering, feature detection, surface fitting and projection techniques. An extended version of bilateral filtering is based on normal filtering and vertex position update [9], [6], [7], [167], [168] [14]. These methods are described as two-stage iterative approaches. Firstly, the face normals are filtered and then the vertex positions are updated based on the filtered face normals. The authors in [169] propose a two-stage processing method for mesh denoising. Initially, part of the noise is removed so that the features of the mesh are preserved. A fine denoising step is applied to the points that are classified as features. Although the aforementioned methods preserve most of the sharp features, they fail to preserve features of different scales, such as medium or small-scale features. Additionally, a big issue that arises from the use of these methods is the fact that they utilize different parameters for different models.

*Sparse Optimization approaches:* The third category includes surface reconstruction and decimation approaches that regularize both vertices and normals [170]. These types of methods minimize the energy of both vertex position and normal error. The authors in [8], propose an $l_0$ minimization approach that provides accurate surface reconstruction in Gaussian noise cases, however, its performance has significantly deteriorated when considering complex noise patterns (e.g., anisotropic noise, real scan noise). Despite the good surface reconstruction results they cannot always preserve sharp features well [171] while in other cases [172] the execution time is high. In [173] a vertex classification step of noisy meshes takes place before applying the denoising process. In [174] the usage of consistent sub-neighborhoods is used for vertex classification. The method of [175] is an iterative approach combining pre-filtering, feature detection, and $l_1$-based feature recovery. The authors in [176] proposed a graph Laplacian regularization based 3D point cloud denoising algorithm. To utilize the self-similarity among surface patches, they adopted the low dimensional manifold prior, and collaboratively denoise the patches by minimizing the manifold dimension. In [177] the authors proposed to apply graph total variation to the surface normals of neighboring 3D points as regularization. This leads naturally to a $l_2$-$l_1$-norm objective function, which can be optimized elegantly using ADMM and nested gradient descent. A big drawback of these methods is their computational complexity, which results in their ability to be used in real-time applications. The main limitation of the presented approaches is the increased required computational complexity and the fact that in many cases irregularly highlights the identified sharp features.

*Tensor voting approaches:* The authors in [178] proposed an approach which exploits the synergy when facet normals and quadric surfaces are integrated to recover a piecewise smooth surface, while the existing mesh denoising techniques focus only on either the first-order features or high-order differential properties. However, they adopt a cascaded operation, which is time-consuming for large models.

**Data-driven Approaches:** Most of the aforementioned categories assume that the noise, affected the 3D object, has a Gaussian distribution. This assumption is far from real-life applications in which the type and the form of noise are much more different (staircase effect, outliers, etc.) than this simplified assumption. Only a few parameter-free approaches have appeared providing good results [179], but not without limitations mainly because of the large dataset for the training process that they require making them very time-consuming. The authors in [10] suggest a method for mesh denoising which uses training sets of noisy objects, scanned by the same devices, and extract information that is used for denoising meshes with similar noise type. While many geometric features are reconstructed adequately, geometric details are limited to those not included in the training set.

*DNN and DAE:* DNN and their applications are omnipresent in several applications (e.g., classification, compression, completion, etc.) related to the processing of 3D meshes. Nevertheless, little attention has been given to the area of feature-aware denoising. More specifically, a DAE is a type of neural network which is able to reconstruct data from corrupted input. Despite their wide adoption in the area of image denoising [180], [181], [182] to the best of our knowledge, auto-encoders have never been used for feature-aware static and/or dynamic 3D mesh denoising [183], [184]. Motivated by this fact, we introduce a novel technique for reconstructing noisy 3D surfaces using a relatively small training dataset. Nonetheless, it is widely known that despite their wide acceptance, there are still limitations that need to be addressed, which are related to the size of the training dataset and the corresponding training complexity. Additionally, in real case scenarios, the testing data may be significantly different from the training data (due to various environmental conditions, or even device limitations) degrading the performance of DNN. Despite the impressive results of recent 3D mesh denoising approaches [185], [17], [16], algorithms that are used in the area of image processing continue to inspire the area of 3D mesh processing. Motivated by this observation, we focus on converting the problem of mesh to image denoising using very robust approaches that have been used and successfully tested in this area. 3D transform-domain CF (e.g., 3D block matching - BM3D [186]) is an approach that achieves state-of-the-art image denoising performance by providing a 3D estimation that consists of jointly filtered grouped image blocks. CF is a special procedure that deals with 3D groups formed by similar 2D fragments of the image, including the following steps: the 3D transformation of a group, shrinkage of the transform spectrum, and inverse 3D transformation. The CF of the grouped blocks, reveals the details which are common between blocks since for each pixel, we obtain several different estimates that need to be combined. This method has been successfully applied in a large number of applications including image/video denoising [187], [188] deblurring, superresolution, and compression. CF approaches process the noisy images by successively extracting reference blocks and by matching blocks that are similar to the reference one, forming a 3D group matrix. After performing a 3D transform to the group formed by the overlapping blocks and attenuating the noise by hard thresholding, an aggregation step is taking place in order to form the estimate of the whole image. Recent approaches [189] suggest substituting the 3D transform-domain processing with traditional CNNs, outperforming the original BM3D approaches in terms of computational efficiency. The BM3D filtering of images and its adoption to meshes have an intuitive formulation, which leads to a simple data-driven method that addresses issues that stem from the two dimensions to manifolds in three dimensions.

### 2.2.4 Compression-Simplification

The real-time generation of dense 3D models representing real-world or synthetic objects generates massive datasets requiring efficient perception-oriented compression schemes that achieve increased compression ratios without affecting noticeably the visual quality of the object. In principle, 3D mesh compression deals with the compact representation of the connectivity and the geometry of the mesh. However, geometry encoders are much more resource-demanding than connectivity encoders [190], [191], stressing the need for novel geometry compression approaches.

Dynamic 3D meshes are becoming popular due to their capability to represent realistically the motion of real-world objects/humans, paving the road for new and more advanced immersive virtual, augmented and mixed reality experiences. However, the real-time streaming of such models introduces increasing challenges related to low cost, low-latency and scalable coding of the acquired information.

**Challenges**

Throughout the years, several approaches have been proposed to achieve reliable compression results by improving some characteristics of the compression process, like the reconstruction quality, the low computational complexity, and the compression rate.

Many augmented, virtual and extended reality applications are based on the real-time generation & streaming of static and dynamic 3D models. However, limitations in communication bandwidth, energy efficiency, and latency significantly degrade the quality of experience making these applications unattainable. Fortunately, millimeter-wave communication with large antenna arrays is a promising technology, leading to orders of magnitude increase in data rates, significantly improving spectral and energy efficiency. One of the biggest changes we have experienced so far is the possibility of moving between two different dimensions: a virtual and a physical dimension. Mixed reality technologies and bi-directional tele-presence applications, offer perceptually enriched experiences, bringing a range of benefits to business and society in various ways, introducing at the same time significant challenges, related to the acquisition and generation of reliable 3D models, the transmission data rate, the network latency, and the energy efficiency. Additionally to the aforementioned challenges, the error-prone and band-limited nature of wireless communication environments causes high packet loss/error rate, which may lead to tremendous quality degradation of real-time 3D streaming services. Most of the perception-oriented solutions build on the premise that high-frequency errors modify the appearance of the surface and thus they are much more highly noticeable as compared to the low-frequency ones [192].

Streaming of 3D animated objects can be used in applications where the geometric data is live-captured and needs to be available in real-time. Nevertheless, these types of applications demand the storage and transmission of a huge amount of 3D data. The real-time rendering of 3D models, representing either real-world or synthetic objects, generates massive datasets and it requires the use of efficient and fast algorithms for increasing the compression ratios without affecting noticeably the visual quality of the object. For this reason, various techniques on dynamic 3D mesh processing should be developed to address the growing demand and at the same time, to handle these important challenges. Geometry data are generally encoded in a lossy manner [193]. However, scalable coding seems to be the most promising approach especially in cases where the network performance is unstable, allowing practical implementations for inferring the available bandwidth and adjusting rates for chunks while balancing metrics like quality, interruptions, number of rate switches (i.e., rate stability). Additionally, a stringent latency is a vital requirement for providing a pleasant immersive VR/AR experience [194]. This means that any real-transmitted dynamic 3D mesh must be easily perceivable, at any frame of its sequence, without affecting its visual quality regardless if a decrease in the transmission rate takes place at any moment. The main purpose of low-latency applications is to avoid disturbing the user's perception because this can negatively affect his/her quality of experience.

**Related Work**

Traditional compression schemes [195], [196], suggest performing direct quantization to the three space coordinates. This form of quantization introduces high-frequency errors into the model that modify significantly the appearance of the surface, resulting in a 3D model with a blocky structure, where the reconstruction errors are highly noticeable. Hence, these algorithms are not suitable for "lossy" compression, due to their non-graceful degradation. To overcome this limitation, several works suggest working in a different domain than the spatial one, where direct quantization will result in low-frequency errors that are not easily distinguishable [147], [155], [197]. One of the first attempts in perception-oriented mesh compression is high-pass encoding [147], which builds on the idea of quantizing the differential coordinates. This approach succeeds at capturing the local relation of vertices and usually outperforms the direct quantization approaches. Similarly, the authors in [155], [197] suggest performing compression by transmitting a small number of "low frequency" components estimated by projecting the Cartesian coordinates to the graph Fourier domain. The main drawback of the graph Fourier schemes is the increased processing demands at the transmitter since they require the eigenvalue decomposition of the Laplacian matrix (e.g., evaluation of the graph Fourier basis vectors) and the projection of the vertices to a subspace, defined by these vectors.

3D tele-immersion allows the mixture of real and virtual content [198], [199]. However, these type of applications requires the real-time transmission of triangular meshes which is a very challenging issue. The authors in [200] presented a complete system for efficient 3D animated object extraction, representation, coding, and interactive rendering. In [201], the authors proposed a method for generating progressive animated models based on local feature analysis and deformation area preservation. In [202] a system is presented for real-time acquisition, transmission, and high-resolution 3D display of dynamic multiview TV content. However, this system is applied to pixels and not to point clouds with connectivity. The authors in [203] presented a linear predictive compression approach for time-consistent 3D mesh sequences supporting and exploiting scalability. The algorithm decomposes each frame of a mesh sequence in layers employing patch-based mesh simplification techniques. This layered decomposition technique is consistent in time. Following the predictive coding paradigm, local temporal and spatial dependencies between layers and frames are exploited for compression. In [204], the researches proposed a streaming compression scheme that allows encoding meshes on-the-fly by operating on a partial representation of the connectivity that is created and deleted as the mesh is fed in increments of single triangle and vertices to the compressor. The authors in [205], [206] proposed different approaches that represent 3D dynamic objects with a semi-regular mesh sequence. Then, they compress the sequence using the spatiotemporal wavelet transform. [207] used 3D affine transforms to compute the prediction errors. Additionally, they achieved to improve the coding efficiency by optimizing the prediction error quantization, using a rate control mechanism. On the other hand, in [208], [209] used an octree-based coder representing the motion between points of adjacent frames with an octree. The eight corners of each octree block are associated with eight corresponding motion vectors. The motion of each point in that block is approximated by the tri-linear interpolation of the corner motion vectors. In [210], it was exploited both the spatial correlation inside each frame (through the graph) and the temporal correlation between the frames (through the motion estimation) for color and geometry compression. In [211], a point cloud codec is used that encodes additional information in an enhancement layer, then they propose to add inter prediction to the enhancement layer in order to gain further bit rate savings. The authors in [212] proposed an algorithm traversing each mesh surface in a depth-first order. The location of each vertex can be predicted using the information of the spatial neighbor vertices as well as information of the corresponding vertices in the previous frame. In [213], the authors proposed a method for cloth compression using local cylindrical coordinates which is suitable for cloth animation compression because the dihedral component of LCC describes the main feature of the animation sequence. One limitation is that this method cannot exploit the spatial coherence, which means that the compression ratio may be low if the triangles are much smaller than the granularity of wrinkles. In [214], it was proposed a method for compression and progressive transmission of dynamic mesh sequence exploiting both temporal and spatial redundancy to effectively reduce the data size of the dynamic sequence. Although the proposed compression algorithm has many advantages, there are still problems that remain to be solved, such as the adaptive progressive transmission of the mesh sequence and the perceptual metrics focus on local relations.

In this work [215], a scheme for the compression of dynamic 3D point clouds is produced using sub-divisional meshes and graph wavelet transforms. The authors in [216] proposed an algorithm that uses a wavelet coding method for mesh sequences. A progressive mesh hierarchy and an anisotropic wavelet are built for the first frame and maintained for subsequent frames. To exploit the temporal correlation, wavelet coefficients between adjacent frames are encoded differentially. A wavelet-based transform and coding scheme is also presented by [217], in which the proposed transform preserves geometric features at lower resolutions by adaptive vertex sampling and re-triangulation. The authors in [218] discussed in detail the problem of static and dynamic 3D mesh compression, summarizing a lot of novel approaches. They presented the main categories of the literature, give comparisons and evaluate the performance of the described algorithms. Regarding the reconstruction of the dynamic 3D meshes, many works have also been presented [219]. In [220], it is presented a framework for robust geometry and motion reconstruction of complex deforming shapes. However this method requires registration of each frame making it difficult to be applied in real-time of unregistered data. In [221], it is presented a shape completion technique for creating temporally coherent watertight surfaces from real-time captured dynamic performances. A limitation of this method was that the unobserved regions in each frame have no geometric details in them. In [222], it is described an approach for the reconstruction of animated meshes from a series of time deforming point clouds, given a set of unordered point clouds that have been captured by a fast 3D scanner. Due to the high memory consumption of the MPU approximation, this algorithm was currently limited to reconstruct point clouds containing a lot of points.

### 2.2.5 Features Extraction and Saliency Map Estimation

New generation 3D scanning technologies are expected to create a revolution in Industry 4.0, facilitating a large number of virtual manufacturing tools and systems. Such applications require the accurate representation of physical objects and/or systems achieved through saliency estimation mechanisms that identify certain areas of the 3D model, leading to a meaningful and easier to analyze the representation of a 3D object. 3D saliency mapping is, therefore, guiding the selection of feature locations and is adopted in a large number of low-level 3D processing applications including denoising, compression, simplification as well as registration.

**Challenges**

The resolution and accuracy of the modern 3D scanners are constantly increasing, making them even more attractive in several vision-based manufacturing tasks, allowing the accurate generation of dynamic virtual representations of physical objects which are then used for inspection. Inspecting the parts and repairing the damages or degradations are very basic tasks for many engineering or manufacturing products. More specifically, surface defect inspection is of primary importance for engineering part quality inspection, since surface defects affect not only the appearance of parts, but also their functionality, efficiency and stability. This task mostly depends on human visual inspection by skilled inspectors. Human visual inspection is costly, labor-intensive, time-consuming, and prone to errors due to inspectors' lack of experience or fatigue, bad environmental conditions, etc. Hence automatic inspection of the surfaces using computational techniques, which is faster, more consistent and robust, is highly desired [223].

The digitized information is massive and raw, leading to the need of new essential and meaningful identification of features that will facilitate robust processing in various applications. These facts stress the need to focus on the development of computational models of visual attention, whose well-known outcomes are the saliency maps. of our method as compared to other approaches.

**Recent Works**

Visual saliency is a subjective perception cue that differentiates a region from others and immediately attracts human attention [224]. The human visual system (HVS) is evolved to automatically detect salient regions over the entire field of view [225]. It is firstly attracted by the most representative salient elements and then the visual attention is transferred to other regions [226]. Most of the existing methods try to simulate the way that the human perceptual sys-tem works, giving more emphasis to what the human brain assumes as salient information. Nevertheless, what a human assumes as a salient feature may vary from what computational methods assume as salient. On the other hand, in industrial applications, simple geometry is usually more common and useful compared to complex surfaces of high spatial frequency that would trigger human visual attention. Wei *et al.* [29] presented a 3D saliency mapping mechanism using the curvature co-occurrence histogram, following similar steps with the method proposed in [227] for extracting salient image features. Tao *et al.* [30] proposed an entropy-based saliency approach using the entropy of the normals to depict the local changes in a region. Song *et al.* [31] proposed a method which incorporates global considerations by making use of spectral attributes. An *et al.* [228] proposed a hybrid saliency taking into account both color and geometric information. Nouri *et al.* [229] proposed a saliency-based metric for the evaluation of the quality between an original and a distorted 3D mesh comparing the structural information. They also [230] proposed a saliency method, using a local vertex descriptor that is used as a basis for similarity measurement and integrated into weighted multi-scale saliency features. Zhao *et al.* [231] proposed a saliency detection method by diffusing a shape index field with a non-local means filter. Their algorithm generates a random center-surround operator to create a saliency map and use the Retinex theory to improve the saliency map. Wu *et al.* [232] proposed a 3D saliency map estimation considering both local contrast and global rarity. There are also many other methods that estimate a saliency mapping but with a completely different way. For example, the salient mapping of method [233], might be excellent for tactile-focused applications, is however, far from acceptable if applied in industrial applications like the ones demonstrated. Authors in [234] detect salient regions on the mesh based on the multi-scale Laplacian fairing results in order to use them for head pose estimation. What they assume as a salient region in their application is significantly different to we assume as salient features in our application. The authors in [235] try to automatically find landmark buildings in a city using a Context-dependency saliency mapping. In [236], the authors propose a saliency detection algorithm for large-scale colored 3D point clouds, which exploits geometric features and color features together to estimate the saliency in colored point clouds. A major drawback of the aforementioned methods is that their robustness is significantly deteriorated when applied to scanned 3D models that have been affected by noise, outliers or missing parts. Additionally, most of these works provide only visual maps to show their effectiveness and none of them have been used and evaluated in real industrial applications.

The extraction of visual saliency has been previously investigated by numerous researchers. Nevertheless, it continues to be a challenging task due to the abstract meaning of saliency. Saliency extraction represents a stimulus-driven activation process [237] where the salient part of the scene differs from its neighboring region [238] due to lack of correlation [239]. Psychophysical studies have shown that low-level salient features are processed in the early visual cortex and extract the most important information of the visual scene [240]. Saliency is mainly affected by feature type, and contrast, task difficulty, center bias, temporal variations of fixations as a recent study reports [241]. A lot of work, related to salient map extraction, has occurred in the area of image [242], [243], [244], [245], [246] and video processing [247], [248], [249]. Recently, this trend has started to attract the attention of the researchers in the area of geometry processing, where the saliency has been defined as a measure of regional importance, showing its benefits in many applications and opening a new road of processing possibilities. The term of 3D mesh saliency was coined by Lee et al. [250], that employed Gaussian weighted-averaged mean curvature-based approach to compute a perceptual metric. Several other approaches followed utilizing spectral methods [251], [252], curvature-based methods [250], [253] multiscale descriptors [254], entropy-based methods [255] and hybrid methods [256] taking into account both geometry and color. Wu et al. presented local multiscale descriptors with global rarity [254], using rotationally invariant metrics to capture local features and dissimilarity based estimators to assess global characteristics. Spectral methods for 3D meshes processing [252] were employed by Song and Liu utilizing the spectral attributes of the log-Laplacian formulation. Wei et al. [253] presented a 3D saliency mapping mechanism using the curvature co-occurrence histogram. Nouri et al. [257] proposed a saliency-based metric for the evaluation of the quality between an original and a distorted 3D mesh comparing the structural information. They also [258] proposed a saliency method, using a local vertex descriptor that is used as a basis for similarity measurement and integrated into weighted multiscale saliency features. Zhao et al. [259] proposed a saliency detection method by diffusing a shape index field with a non-local means filter. Their algorithm generates a random center-surround operator to create a saliency map and use the Retinex theory to improve it. Tao et al. [255] proposed an entropy-based saliency approach using the entropy of the normals to depict the local changes in a region. Lau et al. [260] explored the problem of tactile mesh saliency, searching for the parts of a mesh that are more likely to be grasped, pressed, or touched by humans in the real world. Song et al. [251] proposed a method that incorporates global considerations by making use of spectral attributes. An et al. [256] proposed a hybrid saliency taking into account both color and geometric information.

A new direction for the salient map extraction has started to appear by using Convolutional neural networks (CNNs). CNNs have been extensively used for the extraction of saliency maps in image processing as recent survey reports [261]. A recent study employs CNNs to extract saliency maps on 3D meshes utilizing a multi-view setup [262]. Nousias et al. [263] employed a CNN based mesh saliency extraction approach, employing a 3D geometric patch descriptor to classify faces into four classes. Greater class indices correspond to higher saliency values. Yet, our approach generates saliency values in a con-tinuous range focusing on a regression problem instread of a classification. All these CNN based approaches provide very promising results and have the clear benefit of low computational complexity in comparison with the other conventional methods, making them ideal for utilization in some very challenging situations like the processing of very dense models.

### 2.2.6 Registration and Identification of 3D Objects

The scanning and digitization of 3D objects of the real, physical world has recently attracted a lot of attention. Nowadays, there are many applications in different areas (e.g., entertainment, industry, medical visualization, military, heritage, etc.) that utilize 3D objects, either in the form of point clouds or 3D meshes. Future trends show that both this type of applications and the need for reliable 3D object representation will continue to increase. However, in practical scenarios, there are many factors that inevitably affect the quality of the acquired 3D objects, such as illumination conditions or relative motion between device and target during the scanning process, which can create random fluctuation of the data, the formation of additional and unnecessary points on the surface and points away from the surface (outliers).

**Challenges**

The device itself may also generate a pattern of systematic noise that is added to the surface of the 3D object. Additionally, due to time limitations or a random non-ideal acquisition angle, the point clouds may be incomplete or deformed, which can cause errors in matching and registration [264]. Researchers strive to overcome the existing limitations, trying to provide robust solutions that can be used in realistic circumstances and challenging scenarios.

One of the most common research problems upon digitization is the recognition of partially-observed objects in cluttered scenes, which is fundamental in numerous applications of computer vision, such as intelligent surveillance, remote manipulation of robots in manufacturing, autonomous vehicles, automatic assembly, remote sensing, retrieval, automatic object completion.

**Related Works**

Maybe the most critical step of a 3D object recognition and matching process is the feature descriptor extraction. In the literature, a great number of feature descriptors have been proposed [41], such as spin image [36], direct spacial matching [265], point's fingerprint [266], 3D shape context (3DSC) [267], snapshot [268], local shape descriptors [38], Mesh Histogram of Oriented Gradients (MeshHOG) [269], exponential map [39] and rotational projection statistics [40].

The feature descriptors can be divided into two broad categories: the global feature descriptors and the local feature descriptors. Global feature descriptors represent the geometric and topological properties of the entire 3D model, but they ignore the shape details and require an accurate segmentation of the object. Therefore, they are not usually suitable for the recognition of a partially observed object lying in cluttered scenes. The global feature-based methods are usually used in the context of 3D shape retrieval and classification. Some popular implementations include geometric 3D moments [270], shape distributions [271], viewpoint feature histogram [272], and potential well space embedding [273].

On the other hand, local descriptors focus on narrow neighborhoods, while coarse areas are still present for disambiguation [264]. They can generally handle occlusion and clutter better than the global methods [47], therefore local descriptor approaches are inherently more effective for 3D object recognition [274]. Taati and Greenspan [38] formulated the local shape descriptor for object recognition and localization in range data as an optimization problem. They presented a generalized platform for constructing local shape descriptors that subsumes a large class of existing methods, allowing for tuning to the geometry of specific models. Salti et al. [274] developed a hybrid structure between Signatures and Histograms aiming to a more favorable balance between descriptive power and robustness. Their proposed descriptor, called as Signature of Histograms of OrienTations (SHOT), attempts to leverage on the benefits of both Signatures and Histograms approaches. Buch et al. [275] introduced a method for fusing several feature matches to provide a significant increase in matching accuracy, which was consistent over all tested datasets. Lu and Wang [264] presented a novel matching algorithm of 3D point clouds based on multiple scale features and covariance matrix descriptors. They applied a combination of the curvature and eigenvalue variation, to precisely detect the key points under multiple scales. Darom and Keller [276] proposed an intrinsic scale detection scheme per interest point and utilized it to derive two scale-invariant local features for mesh models. First, they presented the Scale Invariant Spin Image local descriptor that is a scale-invariant formulation of the Spin Image descriptor, and then, they adapted the SIFT feature to mesh data by representing the vicinity of each interest point as a depth map and estimating its dominant angle using PCA to achieve rotation invariance. Lu et al. [41] presented an effective algorithm to recognize 3D objects in point clouds using multi-scale local surface features. It first detects several keypoints in each scene/model and then extracts several feature descriptors with different scales at each keypoint [277, 278].

Point cloud registration refers to the problem of aligning two or more different point clouds that do represent only partially overlapping regions. Generally, the higher the overlap, the easier the registration of the two scenes. However, there are numerous factors that can negatively affect the results [279], such as noise and outliers due to different illumination conditions or relative motion between the scanner and the scene, occlusions and tangled areas. The most well-known and widely used baseline registration method is the Iterative Closest Point (ICP) algorithm. Throughout the years, several ICP-based approaches have been presented [280–282] providing very good results; however, the registration is usually inaccurate when the two related point clouds do not have a good relative initial alignment or do not exhibit an ample overlap. Bouaziz et al. [283] introduced the sparse ICP which promises superior registration results when dealing with outliers and incomplete data. On one hand, it provides better results than the traditional ICP approach, however, it needs parameter adjustments for different use cases, limiting of application potential. Mavridis et al. [284] identified the reasons for the low efficiency of the original Sparse ICP approach and they proposed a registration pipeline that improves the convergence rate of the method by using a more efficient hybrid optimization strategy. Yuan et al. [285] try to solve this problem using a combination of ICP and Principal Component Analysis (PCA), which is used to reflect the similarity of the two point clouds. Authors of other works separate the process into a coarse and a fine registration [286, 287]. Moreover, the 4-Points Congruent Sets (4PCS) techniques are very popular, especially for global registration strategies [288–290] such as algorithms based on RANdom SAmple Consensus (RANSAC) [291–293]. Other related works try to find representative local or global features that are used later as descriptors for matching. Makadia et al. [294] used extended Gaussian images that can be approximated by spherical histograms of the surface orientations. Their algorithm uses only global information and it does not estimate any local feature. Recently, many deep learning approaches have also been presented [295–298], trying to exploit recognizable features from a training dataset, using them later for fast registration, however, they are very vulnerable (as well all the machine learning techniques) in cases where the training and the testing datasets have been created under different circumstances.

## 2.3 Contribution of this Thesis

The purpose of this thesis is to present and highlight in which way similar mathematical tools and frameworks, like spectral analysis, optimization tools, spatio-temporal analysis, subspace tracking methods, etc., can be used in different research domains of 3D geometry processing, facilitating the processes and providing exceptional results. The used mathematical tools will be presented in detail in Chapter 4. For the convenience of the readers, the contribution of this thesis has been separated into the following six subsections, equal to the number of research fields, that we have investigated.

### 2.3.1 Reconstruction and Completion of Incomplete 3D Models

A common limitation of the completion and reconstruction methods, as presented in subsection 2.2.1, is the high computational complexity that significantly affects the execution time and renders them inappropriate for real-time applications. This limitation motivated us to develop fast and effective approaches that can satisfy the reconstruction efficiency supporting at the same time real-time applications. In summary, the main contributions of this Thesis, regarding the research area of the reconstruction and completion of incomplete 3D models, are:

- We use a weighted regularized Laplacian interpolation approach that efficiently exploits the coherence between motion vectors of known vertices. To further enhance the recovery of the missing data, we suggest adding a regularization constraint that further exploits the sparse variations in the motion vector differences of a vertex and the average motion vector of its neighbors. The proposed approach is ideally suited for online settings where the point cloud sequence is not known a priori and is dynamically generated.

- We present an approach that accurately recovers highly incomplete areas of meshes consisting of a variety of small and large-scale features as well as different geometrical details. The problem of missing surfaces is solved using a fast matrix completion approach, having considered an appropriate low-rank matrix that encloses geometrical constraints.

### 2.3.2 Outliers Removal of Static and Dynamic Point Clouds

It is vital for the stability of an application, relating to point clouds, the existing outliers to have been removed before other significant processes (e.g., registration, triangulation, consolidation etc.) take place, since the removal of scanning imperfections and artifacts typically improve dramatically the quality of the subsequent procedures.

- We proposed a novel approach which takes advantage of the robustness of RPCA in order to remove outliers from unorganized static and dynamic point clouds.

- Regarding the dynamic point clouds, we presented a fast and accurate way to identify outliers from point cloud sequences taking advantage both of spatial and temporal coherence between consecutive frames.

### 2.3.3 Denoising

Mesh denoising aims to decrease or eliminate the noise, preserving all useful information such as geometric details of different scales. It is a vital preprocessing tool for improving imperfect meshes obtained by scanning devices and digitization processes. During the last years several methods have been proposed, yet, despite the significant improvement, the main need for a robust and fast algorithm that can manage large meshes with persistent noise still remains a challenge.

- We investigate for the ideal spectral subspace that can be safely removed to deteriorate the amount of noise without negatively affecting the visual quality of the reconstructed denoising model, taking also into account the removal of the highly frequent spatial features to be minimum.

- To increase the computational efficiency of the spectral decomposition we propose the separation of the 3D model into overlapped submeshes and then utilizing subspace tracking algorithms.

- We take advantage of our observation that submeshes of the same 3D models have a high coherence into the spectral domain.

- We developed an approach that operates at the spectral domain without explicitly taking into account the geometry of the mesh. Specifically, the method exploits spatial coherence and correlations in the spectral domain using by applying low-rank matrix analysis tools.

- We take advantage of coherent matrices, which are created using the GFT of each frame, in order to exploit the low-rank spectral properties of the signal subspace, in which features and noise are tangled each other.

### 2.3.4 Compression

The easiness of capturing real 3D objects and the creation of synthetic models result in the generation of a huge amount of data that are continuously increasing. This trend stresses the need for the development of novel techniques on static and dynamic 3D mesh compression, which will be capable to address the aforementioned growing demands.

- We provided a simple and easily reproducible mathematical framework, that combines both geometrical and spectral attributes. The main idea lies on the observation that, on contrary to mainstream state-of-the-art spectral coding approaches, by projecting the spatio-temporal matrix of the $\delta$ coordinates on the subspace of the covariance of the point trajectories (eigen-trajectories), we can derive superior spatio-temporal compression.

- We took advantage of the spatio-temporal coherence between consecutive differential representations and between the eigen-trajectories and graph Fourier subspaces in order to achieve very low compression rates.

- We estimated the temporal coding dictionaries by fast tracking methods based on orthogonal iterations, resulting in very fast implementation, necessary for challenging cases attributed to large 3D animation datasets and demanding execution time constraints.

- We proposed a mechanism for decomposing a mesh sequence into spatial and temporal layers that remove a single vertex at each layer. This decomposition is based on (i) topological characteristics of the mesh and (ii) the temporal behavior of each point separately as their position change through the time frames. In this way, we remove vertices that can be predicted accurately by their neighbors.

- We have created an online step for the online reconstruction of the removed frames by exploiting coherences on the subspaces corresponding to the different layered representations of the corresponding normals. This is achieved using ISVD, ending up with a significant saving in complexity, allowing a close to real-time execution using current machines.

### 2.3.5 Saliency Map and Feature Extraction

Saliency maps are compact 3D representations, generated by simplifying, annotating and/or changing the representation of a physical object/system giving more emphasis to geometrically meaningful parts. The salient features also typically satisfy important requirements such as scaling, rotation, resolution invariance that can simplify industrial processes. Our contributions to this research area are:

- We developed a method that designed to use both normals and guided normals, depending on the application. Guided normals provide more robust saliency mapping in cases of meshes affected by scanning noise and imperfections where other methods fail.

- The proposed method combines the benefits of a spectral and a geometric approach in a single unified pipeline.

- It exploits both local and global information of a model. In other words, the spectral method and the rows of the coherent matrix enclose local information while the columns of the coherent matrix encloses global information corresponding to larger patches.

- It has low-computational complexity, especially in comparison with other spectral methods that use the connectivity information of a model.

- It can be ideally used in applications related to industrial 3D models which have some special geometric characteristics (e.g., intense corners and edges) or have been affected by complex noise patterns.

- It is easily adaptive and can be efficiently adopted both by low-level applications (e.g., denoising, compression, registration, etc.), as a pre-processing step and/or by high-level industrial applications (e.g., maintenance, inspection, quality control, digital twin technologies, etc.)

### 2.3.6 Registration and Identification of 3D Objects

We assume the existence of scanned point clouds that have been acquired using low-resolution and low-cost 3D scanning devices. These noisy point clouds represent real cluttered scenes consisting of different partially-observed objects, denoted as *query* models. Additionally, we assume the existence of high-quality and complete 3D models, denoted as *target* models, which serve as the ideal representation of the *query* models. The *target* models have been acquired using high-resolution scanning devices, and have also been post-processed to remove noise and outliers. Even though the *query* and *target* models may represent the same object, they have different resolution, orientation, while the *query* object is subject to occlusion, making the processes of matching and registration an arduous task. The objective of this research is threefold;

- To identify different *query* objects partially visible in a point cloud scene.

- To match each *query* object with the corresponding original *target* model (if detected).

- To register *query* and *target* objects in the point cloud scene through 3D registration.

The successful integration of the steps is a challenging task, especially due to the presence of noise and outliers, occlusions, missing parts, different resolutions. Motivated by the need to overcome all the aforementioned challenges and limitations, inherent in each step of the process, we designed an end-to-end methodology that demonstrates the following main contributions:

- A layered broad-to-narrow registration scheme that reduces the likelihood of getting trapped in local minima, following a RANSAC-style initialization, based on density estimation in the space of rigid transformations, and a subsequent transformation refinement through a novel weighted ICP approach.

- Computational acceleration during object identification through scene segmentation acting as a data selection (i.e. reduction) step.

- Reduction of the effects of noise and object partiality, based on a novel multi-scale saliency extraction technique that allows identification of distinctive regions and reduction of ambiguity in the matching process.

- A novel point cloud descriptor combining pose information with local geometric properties that allows to identify point correspondences even in the case of extreme partiality of the *query* object. An enhanced point cloud similarity criterion is also introduced for accurate target-to-query object matching and registration.

It should be emphasized that our method does not require training data, for the matching process, since it uses only geometric descriptors of each model.

CHAPTER 3

# Fundamentals

In this chapter we discuss the basic definitions of 3D objects, their components, attributes and characteristics, and types of 3D models that are used. We also present all the corresponding notation and symbols that will be used throughout the manuscript. More details about the symbols and their short description is presented in Appendix D.

## 3.1 Basic Definitions

For the sake of completeness, we will start presenting the most basic definitions that are usually utilized in any 3D object processing application.

### 3.1.1 Vertex

Vertex **v** denotes as a point lying in to the 3D space and it is represented by a vector $\mathbf{v} = [x, y, z]$. A vertex is the basic and simplest element of a 3D object (Fig. 3.1).

$$v = [x, y, z]$$

Figure 3.1: Visual representation of a vertex.

### 3.1.2 Edges

Edge **e** represents the line segment connecting two vertices (Fig. 3.2).

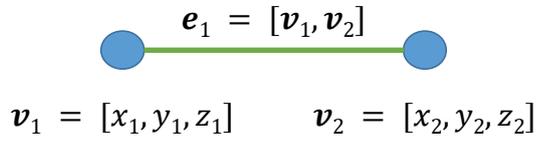

Figure 3.2: Visual representation of an edge.

### 3.1.3 Faces

A face or triangle face **f** represents the basic and the simplest surface area. It consists of 3 vertices that are connected to each other with 3 edges (Fig. 3.3).

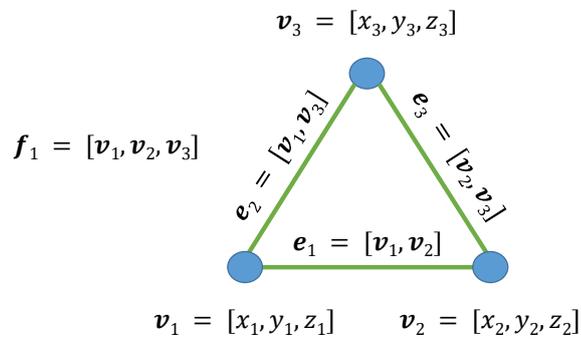

Figure 3.3: Visual representation of a triangle face.

### 3.1.4 Polygon

A polygon is a coplanar set of 2 or more faces. In Fig. 3.4, we present the simplest convex polygon consisting of 2 faces.

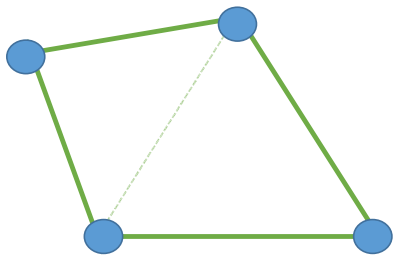

Figure 3.4: Visual representation of a simple polygon.

### 3.1.5 Centroids

Centroid **c** is the geometric center of the triangle face and it is estimated as the arithmetic mean position of all the vertex coordinates of this face (3.5). In other words, the centroid of a triangle is estimated according to:

$$\mathbf{c} = (\mathbf{v}_1 + \mathbf{v}_2 + \mathbf{v}_3)/3 \qquad (3.1)$$

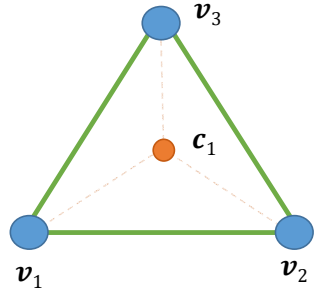

Figure 3.5: Visual representation of a centroid.

### 3.1.6 Normals

Normal or centroid normal $\mathbf{n}_c$ is a vector perpendicular to the surface of a face. The outward unit normal can be computed as:

$$\mathbf{n}_c = \frac{(\mathbf{v}_2 - \mathbf{v}_1) \times (\mathbf{v}_3 - \mathbf{v}_1)}{\|(\mathbf{v}_2 - \mathbf{v}_1) \times (\mathbf{v}_3 - \mathbf{v}_1)\|} \qquad (3.2)$$

where $\mathbf{v}_1$, $\mathbf{v}_2$ and $\mathbf{v}_3$ are the vertices that define face $\mathbf{f} = \{\mathbf{v}_1\ \mathbf{v}_2\ \mathbf{v}_3\}$ and $\times$ denotes the outer product.

On the other hand, point normals **n** are lying in a continuous but non differentiable surface, so they can only approximately estimated. The simplest way it to calculate the mean average of all $k$ centroid normals of the faces in which the point belongs, as shown in Fig. 3.6.

$$\mathbf{n} = \frac{\sum_i^k \mathbf{n}_{ci}}{k} \qquad (3.3)$$

Fig. 3.7 presents both the centroid normal $\mathbf{n}_c$ and the three point normals $\mathbf{n}_i$, $i = \{1, 2, 3\}$ of a face.

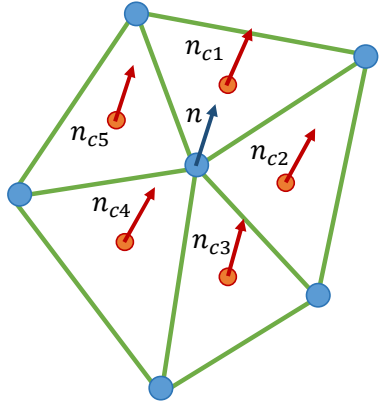

Figure 3.6: Point normal **n** estimated by the centroid normals $\mathbf{n}_c$ of the faces that belongs.

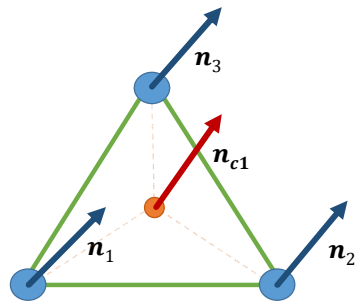

Figure 3.7: Visual representation of point normals **n** and centroid normal $\mathbf{n}_c$

### 3.1.7 Relation Between Number of Faces and Vertices

We denote as $n_f$ the number of faces and with $n$ the number of vertices. It can be proven that for 3D watertight models with mean connectivity $> 4$ the number of faces is always greater than the number of vertices.

**Proof 1**

Based on Euler's equation for the graph theory, we have:

$$n - n_e + n_f \geq 2 \tag{3.4}$$

where $n_e$ represents the number of edges. For 3D watertight models assuming that:

1. There are no vertices connected to only one or two other vertices.

2. The mean degree of each vertex is greater than 4 (i.e., the mean vertex connects via edges to 4 or more other vertices).

Taking into account the above assumptions, we can say that if the mean degree is equal to 4 then $n_e = 4n$ since each vertex creates 4 edges to connect with 4 other vertices. However, the real number of edges $n_e = 4n/2 = 2n$ because each edge has been estimated twice. Replacing this relation to Eq. (3.4) we have:

$$n - 2n + n_f \geq 2 \iff n_f \geq n + 2 \tag{3.5}$$

This result shows that for a mean connectivity $> 4$ the inequality $n_f > n$ holds. Additionally, we have to say that the mean connectivity of the used 3D models in this Thesis, and generally the most 3D models of the datasets in literature, is close to 6 (i.e. $n_e \approx 3n$), so we can easily conclude that $n_f > n$.

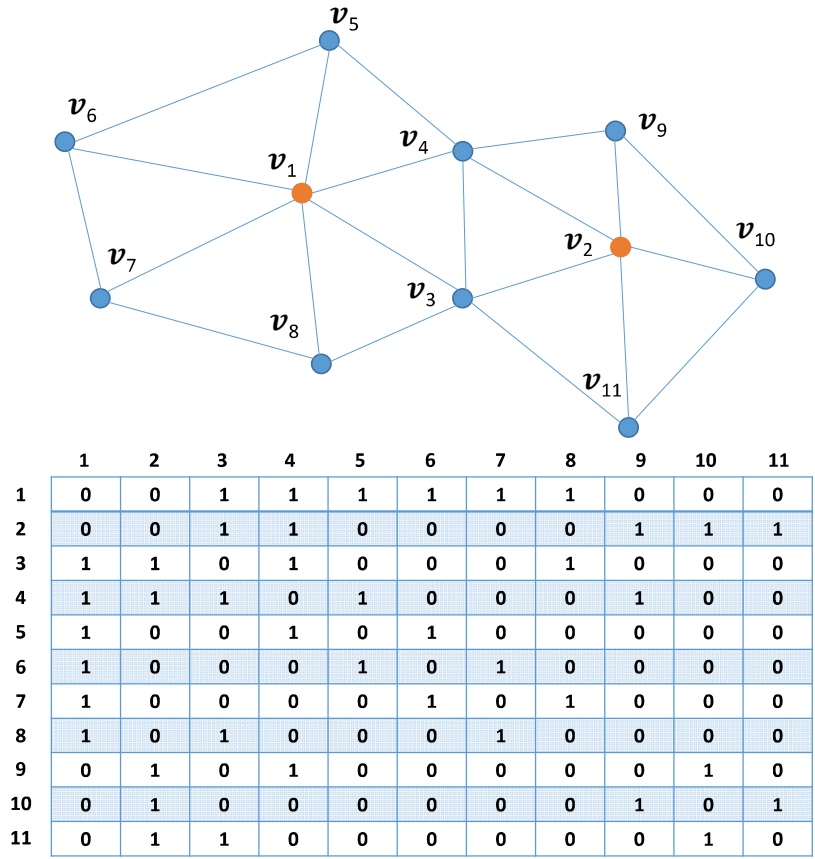

|    | 1 | 2 | 3 | 4 | 5 | 6 | 7 | 8 | 9 | 10 | 11 |
|----|---|---|---|---|---|---|---|---|---|----|----|
| 1  | 0 | 0 | 1 | 1 | 1 | 1 | 1 | 1 | 0 | 0  | 0  |
| 2  | 0 | 0 | 1 | 1 | 0 | 0 | 0 | 0 | 1 | 1  | 1  |
| 3  | 1 | 1 | 0 | 1 | 0 | 0 | 0 | 1 | 0 | 0  | 0  |
| 4  | 1 | 1 | 1 | 0 | 1 | 0 | 0 | 0 | 1 | 0  | 0  |
| 5  | 1 | 0 | 0 | 1 | 0 | 1 | 0 | 0 | 0 | 0  | 0  |
| 6  | 1 | 0 | 0 | 0 | 1 | 0 | 1 | 0 | 0 | 0  | 0  |
| 7  | 1 | 0 | 0 | 0 | 0 | 1 | 0 | 1 | 0 | 0  | 0  |
| 8  | 1 | 0 | 1 | 0 | 0 | 0 | 1 | 0 | 0 | 0  | 0  |
| 9  | 0 | 1 | 0 | 1 | 0 | 0 | 0 | 0 | 0 | 1  | 0  |
| 10 | 0 | 1 | 0 | 0 | 0 | 0 | 0 | 0 | 1 | 0  | 1  |
| 11 | 0 | 1 | 1 | 0 | 0 | 0 | 0 | 0 | 0 | 1  | 0  |

Figure 3.8: Simplified example of an adjacency matrix.

### 3.1.8 Connectivity and Adjacency Matrix

Connectivity encodes the information of all connections (edges) between vertices. A quantity that presents this information is compactly the adjacency matrix or connectivity matrix $\mathbf{C} \in \mathbb{R}^{n \times n}$ which is estimated as described below:

$$\mathbf{C}_{ij} = \begin{cases} 1 & \text{if } \mathbf{v}_i, \mathbf{v}_j \in \mathcal{E} \\ 0 & \text{otherwise} \end{cases} \quad (3.6)$$

where $\mathcal{E}$ represents the list of edges.

In Fig. 3.8, we present a simple example of a graph of connected vertices and its corresponding adjacency matrix.

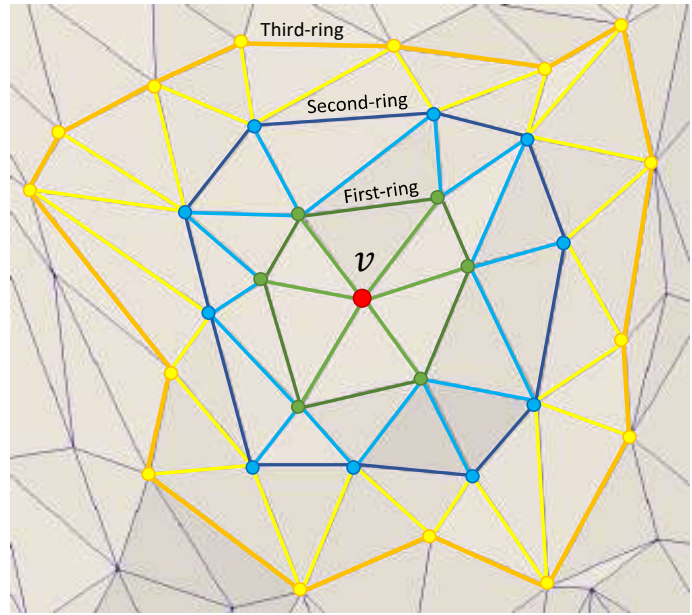

Figure 3.9: First $\Psi_{i1}$, second $\Psi_{i2}$ and third-ring $\Psi_{i3}$ of vertex $\mathbf{v}_i$.

### 3.1.9 k-ring Area

First-ring of vertex $\mathbf{v}_i$ is the area that consists of neighboring faces that share vertex $\mathbf{v}_i$. The valence $n_v$ of the first ring represents the number of vertices that constitute this area which varies even within the same 3D model. Mathematically, we can say that the first-ring neighborhood of a vertex with index $i$ is the set $\Psi_{i1}$ of vertex indices $j$ connected to $\mathbf{v}_i$ by an edge $e(i, j)$:

$$\Psi_{i1} = \{ j \,|\, (i,j) \in \mathcal{E} \} \quad (3.7)$$

Similarly to the first ring, we can define the $k$-ring where $k > 1$. To mention here that the $k$-ring encloses also the $k-1$-ring, as shown in Fig. 3.9.

### 3.1.10 k Nearest Neighbors (knn)

The knn patch is the set that is defined by the *k* nearest neighbors of the vertex $\mathbf{v}_i$. In this case, the number of the connected vertices is pre-defined and always the same for all the knn patches, on contrast to the first-ring which does not have a fixed number of vertices. $\Psi_i^k$ denotes the knn set and consists of the k nearest neighbors of vertex $\mathbf{v}_i$. In Fig. 3.10, we present an example of a k nearest neighborhood consisting of the k = 10 nearest neighbors of vertex $\mathbf{v}$.

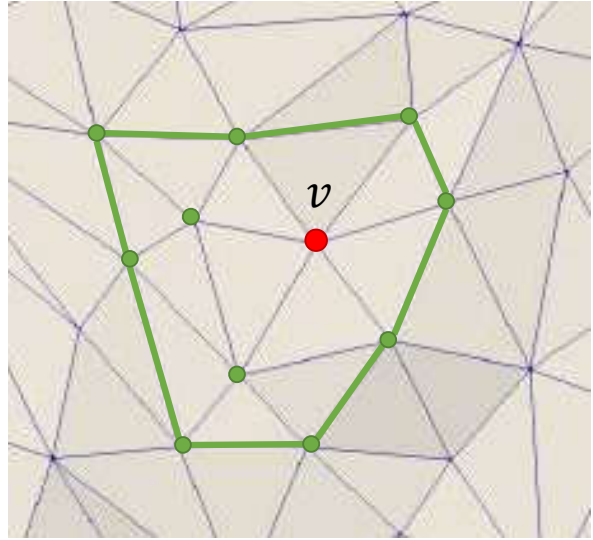

Figure 3.10: Area consisting of the k = 10 nearest neighbors of vertex **v**.

### 3.1.11 Motion vectors

Motion vectors **m** represent the motion of a vertex between two consecutive frames, denoted with the time superscript (t) and (t+1) (Fig. 3.11). In this case, the vertices must have the same index in any 3D static mesh of the dynamic sequence.

$$\mathbf{m}^{(t+1)} = \mathbf{v}^{(t+1)} - \mathbf{v}^{(t)} \qquad (3.8)$$

$$\mathbf{m}_i^{(t+1)} = [m_{xi}, m_{yi}, m_{zi}]^\top \begin{cases} m_{xi} = |v_{xi}^{(t+1)} - v_{xi}^{(t)}| \\ m_{yi} = |v_{yi}^{(t+1)} - v_{yi}^{(t)}| \quad \forall\, i = 1 \cdots n \\ m_{zi} = |v_{zi}^{(t+1)} - v_{zi}^{(t)}| \end{cases} \qquad (3.9)$$

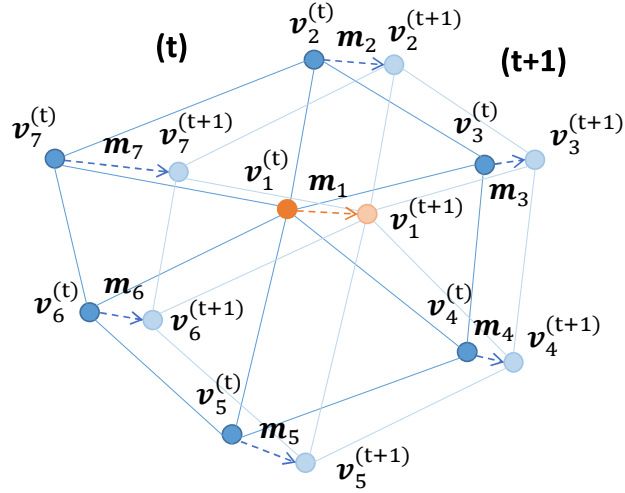

Figure 3.11: Motion vectors between two consecutive first-ring areas of frames (t) and (t+1).

## 3.2 Representation of 3D Structures

For the representation of 3D objects, different data structures can be used. However, the common element of all of these forms is the use of 3D coordinate points. In the following paragraphs, we will present some of the most popular 3D model representations.

### 3.2.1 3D Triangular Meshes

A polygon mesh $\mathcal{M}$ structure is used as a shape representation to describe the surface of 3D objects. A polygon mesh is defined as a set of vertices that are connected to each other with an edge, creating faces (Fig. 3.13). In this way, a polygon mesh consists of only three different geometric elements, namely vertices **v**, edges **e**, and polygon faces **f**. The most commonly used polygon mesh is the triangle mesh which all of their faces are triangles. The simplicity of these types of meshes results in their wide use in many applications. In this thesis, we will refer to triangle meshes using just the term meshes.

A triangle mesh $\mathcal{M}$ consists of $n$ vertices and $n_f$ faces. In this way, the 3D mesh $\mathcal{M}$ can be represented by two different sets $\mathcal{M} = \{\mathcal{V}, \mathcal{F}\}$, where the set $\mathcal{V}$ consists of vertices and the set $\mathcal{F}$ consists of the indexed faces of the mesh. The set of edges $\mathcal{E}$ is not explicitly encoded but can be directly derived from $\mathcal{V}$ and $\mathcal{F}$.

All vertices are used for the creation of a vector of vertices $\mathbf{v} = [\mathbf{x}, \mathbf{y}, \mathbf{z}]$ in the 3D coordinate space such as $\mathbf{x}, \mathbf{y}, \mathbf{z} \in \mathbb{R}^{n \times 1}$ and $\mathbf{v} \in \mathbb{R}^{n \times 3}$. This means that we have a set of $n$ points such that $\mathcal{V} = \{\mathbf{v}_1, \mathbf{v}_2, \ldots, \mathbf{v}_n\}$. Additionally, each face is defined by a set of 3 indices to vertices $f_i = [v_{i1}, v_{i2}, v_{i3}] \ \forall \ i = 1, n_f$ where $n_f > n$, so we have $n_f$ faces $\mathcal{F} = \{f_1, f_2, \ldots, f_{n_f}\}$. Examples of different 3D meshes are presented in Fig. 3.13.

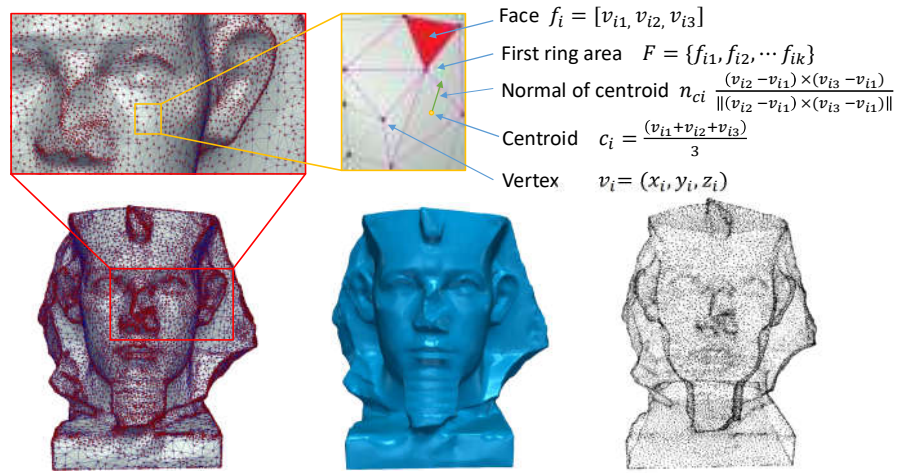

Figure 3.12: Presentation of some basic attributes that appear in 3D meshes.

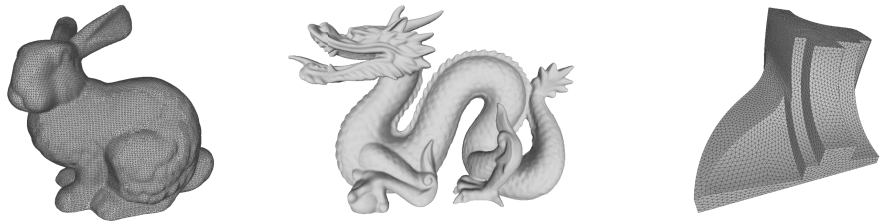

Figure 3.13: Example of 3D meshes.

### 3.2.2 Point Clouds

Another very popular 3D data representation structure is the point cloud. A point cloud is a collection of vertices $\mathbf{v} = [\mathbf{x}, \mathbf{y}, \mathbf{z}]$ in the 3D coordinate space. However, there is some additional information such as the RGB color of each specific point or the normals of each point that is usually encoded along with the raw position. Its popularity is based on the fact that 3D scanner devices estimate point positions that are only later reconstructed to meshes using triangulation algorithms. We denote the geometric part of point clouds also as $\mathcal{M}$, with the same annotation as the mesh, since it is a simplified representation of a mesh, where $\mathcal{F} = \oslash$. Some examples of 3D scanned point clouds are presented in Fig. 3.14.

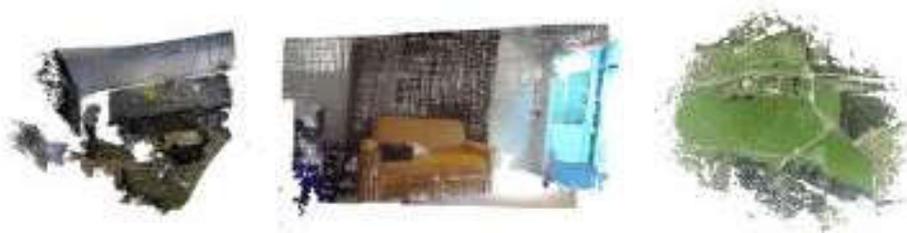

Figure 3.14: Example of scanned point clouds.

### 3.2.3 Time-varying and Dynamic Meshes

Time-varying 3D mesh is defined as a sequence of different 3D static meshes with a varying number of vertices and triangles across frames. This type can be used to represent moving objects in a multi-object scene or 3D objects in morphing. Despite the fact that this type better represents real case scenarios, other simplified forms are usually used in the literature.

An animated or dynamic mesh $\mathcal{A} = [\mathcal{M}_1; \mathcal{M}_2; \ldots \mathcal{M}_{n_s}]$ is a subcategory of time-varying mashes. It is also defined as a set of $n_s$ consecutive static 3D meshes, however in this case, each frame consists of $n$ vertices. This type can be assumed as a simplified form of a time-varying 3D mesh, where all the different frames, from $\mathcal{M}_1$ to $\mathcal{M}_{n_s}$, have the same connectivity. Additionally, the set of faces $\mathcal{F}$ remains the same, so it is easy to estimate the motion vectors between consecutive frames and to watch the state of each vertex separately during the motion of the whole 3D animation. In Fig. 3.15, we present different frames of an 3D animation model.

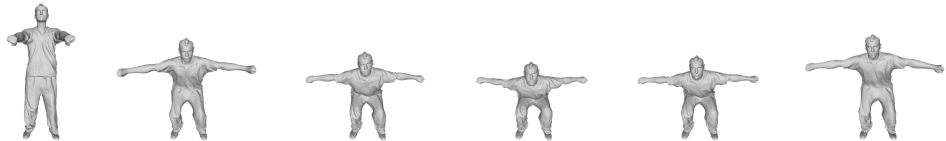

Figure 3.15: Different frames of the same 3D animation.

## 3.3 Geometrical Features

Geometric features are parts of 3D objects that can be either points, lines, curves, surfaces, etc., and represents non-trivial (non flat) parts of the 3D object.

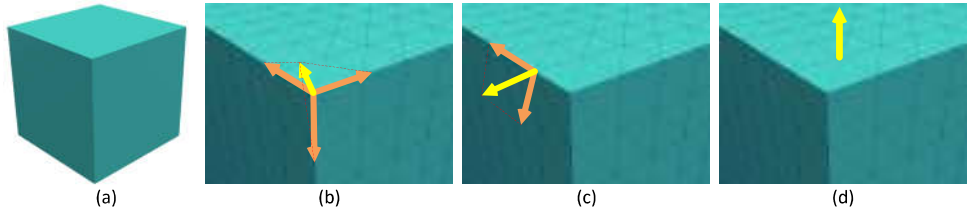

(a)  (b)  (c)  (d)

Figure 3.16: (a) Cube model, (b) vertex defined as corner feature, (c) vertex defined as edge feature, (d) vertex defined as flat area (no feature).

The categories of the primitive geometric features can be considered as:

- Corner features: Corners are high-frequency spatial features, where the normals of their neighbor faces are usually lie in three different oriented surfaces (Fig. 3.16-(a)).

- Edge features: Edges create a geometrical abnormality in the surface that are located, where the normals of their neighbor faces are usually lie in two different oriented surfaces (Fig. 3.16-(b)).

On the other hand, the compound features are usually a combination of the primitive features mentioned above. In Fig. 3.16, we present an example of primitive features in a simple cube model.

## 3.4 Noise

Noise can be described as an unpleasant amount of data that negatively affect the quality of the real signal. Usually, the vertices of a noisy mesh $\tilde{\mathcal{M}}$ satisfy the following identity:

$$\tilde{\mathbf{v}}_i = \mathbf{v}_i + \tilde{\mathbf{z}}_i, \quad \forall\, i = 1, \ldots, n \qquad (3.10)$$

where $\mathbf{v}_i$ are the noise free vertices and $\tilde{\mathbf{z}}_i$ represents a $1 \times 3$ noise vector (e.g., with distribution $\mathcal{N}(0, \sigma)$ in the case of Gaussian noise (Fig. 3.17).

The most common types of artifacts introduced in 3D meshes are:

1. Noise due to the scanning operation [10]

2. Non-uniform sampling

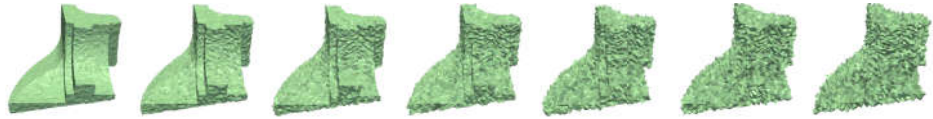

Figure 3.17: Fandisk model affected by different levels of Gaussian noise.

3. Gaussian noise [299]

4. Impulsive noise

5. Staircase effects caused by MRI and CT devices [11]

6. Surfaces with holes that can be simulated using temporal mask

7. Spatial masks

8. Complex noise patterns that may vary significantly over the different parts of the scanned surface

9. Artifact and noise caused by compression (e.g., coddyac by Vasa and Skala [1], FAMC-DCT by Mamou *et al.* [2])

10. Network errors during the transmission

In Fig. 3.18 we present an example of a 3D mesh (Armadillo) affected by different types of noise, while in Fig. 3.19, we present different types of scanning noise.

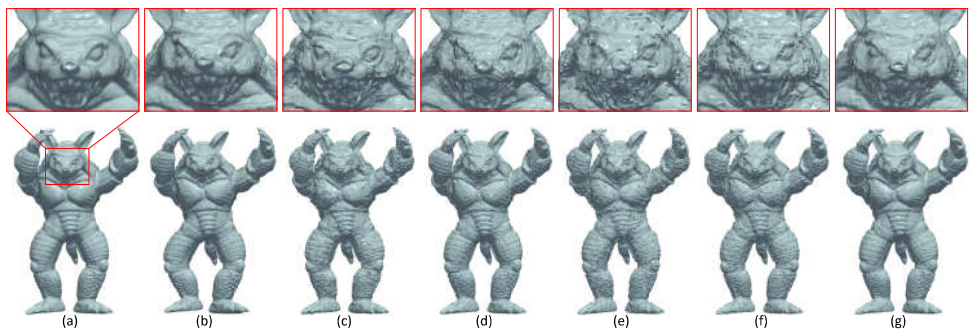

Figure 3.18: (a) Original mesh (armadillo frame 1) and the corresponding noisy meshes affected by: (b) coddyac compression by Vasa and Skala [1], (c) FAMC-DCT compression by Mamou *et al.* [2], (d) Gaussian, (e) impulsive, (f) spatial masking, (g) uniform noise.

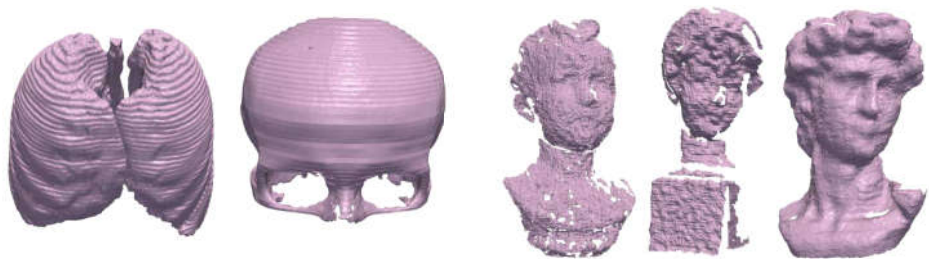

Figure 3.19: Different types of scanned noise (e.g., from MRI, kinect v1, kinect v2, kinect fusion).

CHAPTER 4

# Mathematical Notions and Tools for 3D Geometry Processing

In this chapter, we will present the mathematical tools that have been used through our research. To mention here that the proposed implementations of various approaches, which have been applied in several research areas, may have been inspired by the same theory and they may share the same mathematical principles (e.g., removing of high-frequency components is used both for denoising and compression, RPCA is used both for outliers removal and features extraction, etc., ). So the mathematical framework is defined once, in this chapter, but it can be similarly used in many other implementations. The notation table and the short description of each symbol are presented in Appendix D. To avoid readers' misunderstanding or confusion, the same notation is used in each of the following chapters.

## 4.1 Spectral Processing

Spectral processing refers to the analysis that occurs in the spectral domain space and as data quantities can be used spatial frequencies or the eigenvalues and eigenvectors after a data decomposition.

### 4.1.1 Laplacian Matrices

The Laplacian matrix of a graph $\mathcal{G}$, which correspond to the connectivity information, shows how the vertices connect each other creating triangles:

$$\mathbf{L} = \mathbf{D} - \mathbf{C} \qquad (4.1)$$

where $\mathbf{D} = diag\{d_1, \ldots, d_n\}$ is a diagonal matrix with $d_i = \sum_{j=1}^{n} C_{ij}$, being the degree of its node. The matrix $\mathbf{C}$ is used for the creation of the binary Laplacian matrix, however there are also other forms of Laplacian which are called weighted Laplacian. An example of a binary adjacency matrix has been presented in Fig. 3.8. In this case, the corresponding Laplacian matrix is presented in Fig. 4.1.

| 6 | 0 | -1 | -1 | -1 | -1 | -1 | -1 | 0 | 0 | 0 |
|---|---|----|----|----|----|----|----|----|----|----|
| 0 | 5 | -1 | -1 | 0 | 0 | 0 | 0 | -1 | -1 | -1 |
| -1 | -1 | 4 | -1 | 0 | 0 | 0 | -1 | 0 | 0 | 0 |
| -1 | -1 | -1 | 5 | -1 | 0 | 0 | 0 | -1 | 0 | 0 |
| -1 | 0 | 0 | -1 | 3 | -1 | 0 | 0 | 0 | 0 | 0 |
| -1 | 0 | 0 | 0 | -1 | 3 | -1 | 0 | 0 | 0 | 0 |
| -1 | 0 | 0 | 0 | 0 | -1 | 3 | -1 | 0 | 0 | 0 |
| -1 | 0 | -1 | 0 | 0 | 0 | -1 | 3 | 0 | 0 | 0 |
| 0 | -1 | 0 | -1 | 0 | 0 | 0 | 0 | 3 | -1 | 0 |
| 0 | -1 | 0 | 0 | 0 | 0 | 0 | 0 | -1 | 3 | -1 |
| 0 | -1 | -1 | 0 | 0 | 0 | 0 | 0 | 0 | -1 | 3 |

Figure 4.1: Simplified example of a Laplacian matrix.

However, the selection of the ideal type of weights is usually related to the application in which it will be used. For example, the weighted adjacency matrix is an ideal form for emphasizing the coherence between Laplacian matrices of different submeshes by providing geometric information; on the contrary, the binary adjacency matrix provides only connectivity information. In some cases, the binary Laplacian is perfect due to the fact that it is more universal and it remains the same for any frame of an 3D animation sequence, making it able to be used without further modification. A generalized representation of the adjacency matrix can be shown in Eq. (4.2):

$$\mathbf{C}_{ij} = \begin{cases} w_{ij} & \text{if } i,j \in \mathcal{E} \\ 0 & \text{otherwise} \end{cases} \quad (4.2)$$

The type of the connectivity matrix $\mathbf{C}$ is related to the type of the weights $w$ that it has, in Eq. (4.2). More specifically, if $w = 1$ then a binary Laplacian matrix is created, otherwise, a weighted Laplacian matrix is created. Some types of weights could be the $w = e^{-\frac{\|\mathbf{v}_i - \mathbf{v}_j\|_2^2}{2\alpha^2}}$ where $\alpha$ is the variance that sets a threshold to edge existence, or the cotangent between two vertices $w = \cot(\mathbf{v}_i, \mathbf{v}_j)$. The normalized form of the Laplacian matrix $\mathbf{L}$ is a non-negative matrix given by $\mathcal{L} = \mathbf{D}^{-1/2}\mathbf{L}\mathbf{D}^{-1/2}$

### 4.1.2 Graph Fourier Transform

The Laplacian matrix $\mathbf{L} \in \mathbb{R}^{n \times n}$, can be decomposed according to:

$$\mathbf{U}\mathbf{\Lambda}\mathbf{U}^T = \text{SVD}(\mathbf{L}) \qquad (4.3)$$

where $\mathbf{U} = [\mathbf{u}_1, \mathbf{u}_2, \ldots, \mathbf{u}_k]$ is an orthonormal matrix with the eigenvectors and $\mathbf{\Lambda} = diag\{\lambda_1, \lambda_2, \ldots, \lambda_k\}$ is a diagonal matrix with the corresponding eigenvalues.

The eigenvectors and eigenvalues of the Laplacian matrix $\mathbf{L}$ provide a spectral interpretation of the graph signals. The GFT of the a mesh $\mathcal{M}$, represented by the matrix of vertices $\mathbf{V}$, is defined as its projection onto the eigenvalues of the graph, according to:

$$\hat{\mathbf{V}} = \mathcal{T}(\mathbf{V}) = \mathbf{U}^T \mathbf{V} \qquad (4.4)$$

where $\hat{\mathbf{V}} \in \mathbb{R}^{n \times 3}$ is a matrix representing the GFT of the matrix of vertices $\mathbf{V}$ and $\mathcal{T}(.)$ represents the GFT function. We can easily observe that the number of components of a GFT matrix is equal to the number of vertices of the mesh. Correspondingly, the inverse GFT (IGFT) of each $i$ frame is given by:

$$\mathbf{V} = \mathcal{T}^{-1}(\hat{\mathbf{V}}) = \mathbf{U}\hat{\mathbf{V}} \qquad (4.5)$$

where $\mathcal{T}^{-1}(.)$ represents the IGFT function.

### 4.1.3 Subspace Decomposition and Removal of High Frequency Components

The subspace decomposition of the Laplacian matrix defined by the vertices of the mesh can be re-written as:

$$\mathbf{L} = \begin{bmatrix} \mathbf{U}_{n-n_l} & \mathbf{U}_{n_l} \end{bmatrix} \begin{bmatrix} \mathbf{\Lambda}_{n-n_l} & 0 \\ 0 & \mathbf{\Lambda}_{n_h} \end{bmatrix} \begin{bmatrix} \mathbf{U}_{n-n_l} & \mathbf{U}_{n_l} \end{bmatrix}^T \qquad (4.6)$$

where $\mathbf{U}_{n_l} = [\mathbf{u}_1, \mathbf{u}_2, \ldots, \mathbf{u}_{n_l}]$, $n_l < n$ represents the number of the remained components while $n_h = n - n_l$ the number of the removed components. Then a smoothed version of the noisy vertices can be generated by performing low pass spectral filtering of the Cartesian coordinates $\mathbf{V} = \mathbf{U}_{n_l}\hat{\mathbf{V}}$.

### 4.1.4 Orthogonal Iterations

A direct application of spectral methods to the vertices of a mesh would result in a high computational complexity which could be forbidden in applications with very dense and large meshes. A solution to this problem could be the separation of a mesh in submeshes. However, an application of a spectral method to the vertices of each submesh $\mathbf{M}[i] \; \forall \, i = 1, \cdots, n_t$, individually, would result in a complexity of the order $\mathcal{O}(\sum_{i=1}^{n_t} n_{m_i}^3)$. To minimize this complexity, an exploitation of the coherence between the spectral components of the different submeshes can be done using orthogonal iterations (OI).

The estimation of $\mathbf{U}_{n_h}[i]\ \forall\ i = 1, \cdots, n_t$ occurs according to:

$$\mathbf{U}_{n_h}[i] = Orthonorm\left\{\mathbf{T}^\zeta[i]\,\mathbf{U}_{n_h}[i-1]\right\} \quad (4.7)$$

where $\mathbf{T}[i] = (\mathbf{L}[i] + \delta\mathbf{I})^{-1}$, $\delta$ is a small positive scalar value that ensures positive definiteness of $\mathbf{T}[i]$, $\mathbf{L}[i]$ is the graph Laplacian of the vertices of the submesh $\mathbf{M}[i]$ and the power $\zeta$ plays an important role to the converge of the process. Depending on the choice of $\mathbf{T}^\zeta[i]$, we obtain alternative iterative algorithms with different convergence properties. The convergence rate of Eq. (4.7) depends on $|\lambda_{n_h+1}/\lambda_{n_h}|^\zeta$ where $\lambda_{n_h+1}$ is the $(n_h+1)$-st largest eigenvalue of $\mathbf{T}^\zeta$.

## 4.2 Guided Normals

Guided normals **g** have been successfully applied in feature-aware mesh denoising approaches [12], [17]. Fig. 4.2 presents examples of noisy normals (yellow vectors) and guided normals (orange vectors) in different surface areas (e.g., flat areas, edges, corners) showing that guided normals can be used for identifying features since they have a more robust representation under noisy conditions.

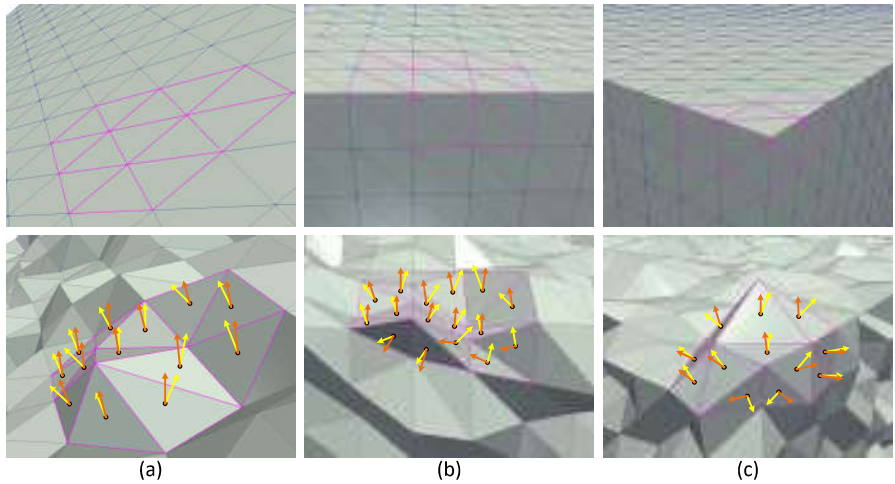

Figure 4.2: [Up] different areas of a mesh, [Down] Surface areas with noise. (a) Flat areas, (b) An edge, (c) A corner. Yellow vectors represent the noisy normals and orange vectors represent the guided normals.

For the computation of the guided normals, the estimation of the ideal patches is firstly required, as described in the following subsection 4.2.1.

### 4.2.1 Ideal Patch Estimation

For each face $f_i$, it is estimated a set $\mathcal{B}_i = \{\mathcal{P}_{i1}, \mathcal{P}_{i2}, \ldots, \mathcal{P}_{in_p}\}$ of $n_p$ candidate patches $\mathcal{P}_{ij}$. Each patch $\mathcal{P}_{ij}$ represents a small area of faces including the face $f_i$. The main purpose is to find which one of these candidate patches $\mathcal{P}_{ij}$ is the ideal representative area for each face [12], in terms of the direction of the centroid normals. Two parameters are investigated in order to identify the optimal patch for each face:

1. The difference between $\lambda_{j1}$ and $\lambda_{j2}$ eigenvalues of the covariance matrix $\mathbf{R}_{ij} = \mathbf{N}_{ij}^T \mathbf{N}_{ij} \in \mathbb{R}^{3\times 3}$ for each $\mathbf{N}_{ij}$ area, where $\mathbf{N}_{ij} = [\mathbf{n}_{c_{ij_1}}, \mathbf{n}_{c_{ij_2}}, \ldots, \mathbf{n}_{c_{ij_k}}]^T \in \mathbb{R}^{k\times 3}$ consists of the corresponding centroid normals of the candidate patch area $\mathcal{P}_{ij}$.

   The eigenvalues are estimated through the following eigenvalue decomposition: $\mathbf{R} = \mathbf{U}\mathbf{\Lambda}\mathbf{V}^T$, where the columns of $\mathbf{U}$ contain the eigenvectors of $\mathbf{R} \in \mathbb{R}^{3\times 3}$ and $\mathbf{\Lambda} = diag(\lambda_{j1}, \lambda_{j2}, \lambda_{j3})$ is a diagonal matrix with the corresponding eigenvalues.

2. Additionally to $\lambda_{j1}$ and $\lambda_{j2}$, the maximum distance between the $i$ centroid normal and the other centroid normals of the same patch is also used. Among all candidate patches, the one $\mathcal{P}^*$ with the smallest value of Eq. (5.16) is selected:
$$\mathcal{P}^* = (\mathcal{P}_{ij} \mid \min(a_j b_{ij})) \tag{4.8}$$
$$\text{where} \quad a_j = ||\lambda_{j1} - \lambda_{j2}||_2 \tag{4.9}$$
$$b_{ij} = \max(||\mathbf{n}_{ci} - \mathbf{n}_{cl}||_2) \quad \forall \, \mathbf{n}_{cl} \in \mathcal{P}_{ij} \tag{4.10}$$
$\forall \, i = 1, \cdots, n_f, \, \forall \, j = 1, \cdots, n_p$, and $\mathcal{P}^*$ represents the ideal-selected patch.

### 4.2.2 Estimation of Guided Normal

Finally, the guided normal $\mathbf{g}_i$ of the face $f_i$ is computed as the area-weighted average normal:
$$\mathbf{g}_i = \frac{\sum_{f_i \in \mathcal{P}^*} A_j \mathbf{n}_{cj}}{\left\|\sum_{f_i \in \mathcal{P}^*} A_j \mathbf{n}_{cj}\right\|_2} \quad \forall \, i = 1, \cdots, n_f \tag{4.11}$$
where $A_j$ represents the area of face $f_j$.

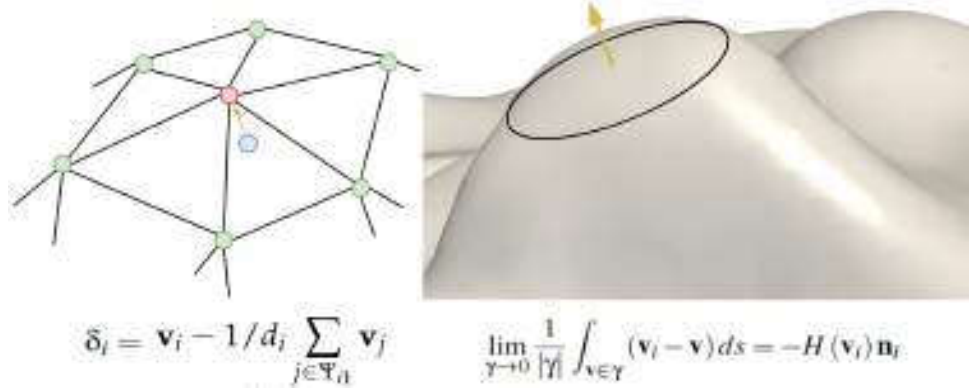

Figure 4.3: The direction of the differential coordinate vector approximates the local normal direction [3].

## 4.3 Delta Coordinates

Any vertex that belongs to the neighborhood of the first-ring area $\Psi_{i1}$ is a neighbor of $i$. The neighborhoods of undirected graphs, such as the graph of a mesh defined above are symmetric. To be more specific, if a vertex $j$ is a neighbor of vertex $i$, then also $i$ is a neighbor of $j$. The differential or $\delta$ coordinates of a mesh are calculated as the difference between the coordinates of each vertex $\mathbf{v}_i$ and the barycenter of its neighbors:

$$\boldsymbol{\delta_i} = \left[\delta_{x_i}, \delta_{y_i}, \delta_{z_i}\right] = \mathbf{v}_i - 1/d_i \sum_{j \in \Psi_{i1}} \mathbf{v}_j \quad (4.12)$$

or

$$\boldsymbol{\delta_i} = \mathbf{L}\mathbf{v} \quad (4.13)$$

where $d_i$ is the number of immediate neighbors of $i$, also known and as degree of vertex $i$. The differential coordinates correspond to the set of displacements that are produced by applying the Laplacian operator to the vertices of a mesh. The direction of $\boldsymbol{\delta}_i$ approximates the curvature flow normals as it is also shown in Fig. 4.3, i.e., $\boldsymbol{\delta}_i = -\mathcal{H}(\mathbf{v}_i)\mathbf{n}_i$ [3], where $\mathcal{H}(\mathbf{v}_i)$ represents the mean curvature at $\mathbf{v}_i$ and $\gamma$ is a closed surface curve around $\mathbf{v}_i$ and $|\gamma|$ is the length of $\gamma$.

## 4.4 Geometrical Features Extraction

Features extraction is a very important task that can be used to facilitate other high-level tasks and applications. A fast and reliable method is used for the clustering of each centroid into 3 categories taking into account the curvature of a small patch area. More specifically, for each face $f_i$ of the mesh, a patch $\mathcal{P}_i = \{f_{i_1}, f_{i_2}, \ldots, f_{i_k}\}$ is created that consists of the $k$ closest faces $f_i$ selected by the $k$-nn algorithm. These patches are used for the creation of $n_f$ matrices $\mathbf{N}$ each one of them consisting of the corresponding $k$ centroid normals:

$$\mathbf{N}_{c_i} = [\mathbf{n}_{ci_1}, \mathbf{n}_{ci_2}, \ldots, \mathbf{n}_{ci_k}]^T \ \forall \ i = 1, \ldots, n_f \tag{4.14}$$

Then, the covariance matrix for each matrix $\mathbf{N}_{c_i} \in \mathbb{R}^{k \times 3}$ is estimated, according to:

$$\mathbf{R}_{N_{c_i}} = \mathbf{N}_{c_i}^T \mathbf{N}_{c_i} \in \mathbb{R}^{3 \times 3} \tag{4.15}$$

We perform the following eigenvalue decomposition: $\mathbf{R}_{N_{c_i}} = \mathbf{U}\boldsymbol{\Lambda}\mathbf{U}^T$, where the columns of $\mathbf{U}$ are the eigenvectors of $\mathbf{R}_{N_{c_i}} \in \mathbb{R}^{3 \times 3}$ and $\boldsymbol{\Lambda} = diag(\lambda_{i1}, \lambda_{i2}, \lambda_{i3})$ is a diagonal matrix with the corresponding eigenvalues $\lambda_{ij} \ \forall \ j = 1, \cdots, 3$.

Then, three different cases [300] can be distinguished according to:

1. Corner, if ($\lambda_{i1} \cong \lambda_{i2} \cong \lambda_{i3}$)

2. Edge, if ($\lambda_{i1} \cong \lambda_{i2} < \lambda_{i3}$)

3. Flat area, if ($\lambda_{i1} < \lambda_{i2} \cong \lambda_{i3}$)

For the classification of each face as a corner, edge or flat area, the $k$-means ($k = 3$) clustering algorithm is used applied to the eigenvalues.

## 4.5 Robust Principal Component Analysis

RPCA is a tool that has been used to recover the low-rank matrix $\mathbf{E}$, corresponding to the noiseless mesh, from the observed measurements $\mathbf{M} = \mathbf{E} + \mathbf{S}$ which have been corrupted by the matrix $\mathbf{S}$ with sparse entries of large magnitude (outliers). However, the low-rank matrix $\mathbf{E}$ is not known beforehand, so we need to estimate it using the only information that we have. We start by assuming that the matrix of the observed data $\mathbf{M}$, may be decomposed as:

$$\mathbf{M} = \mathbf{E} + \mathbf{S} \tag{4.16}$$

The low-rank matrix $\mathbf{E}$ represents the space of real data while $\mathbf{S}$ is a sparse matrix representing the space where outliers lie. The dimensions of $\mathbf{E}$ and $\mathbf{S}$ matrices as well as the location of their non-zero entries are considered as unknown. The classical PCA estimates the $\mathbf{E}$ by solving:

$$\begin{aligned} \text{minimize} \quad & \|\mathbf{M} - \mathbf{E}\| \\ \text{subject to} \quad & \text{rank}(\mathbf{E}) \leq r \end{aligned} \tag{4.17}$$

Early efforts to robustify PCA have relied on robust estimates of the data covariance matrix; e.g., [301]. Recently, polynomial-time algorithms with remarkable performance guarantees have emerged for low-rank matrix recovery in the presence of sparse, but otherwise arbitrarily large errors [302] which seems to perfectly fit our problem. Their purpose is to recover the low-rank matrix $\mathbf{E}$ from the measurements $\mathbf{M} = \mathbf{E} + \mathbf{S}$ which are highly corrupted by the matrix $\mathbf{S}$ with sparse entries of large magnitude. The approach that we adopt, estimates the authentic points and identifies the outliers following the next approach.

Assuming that there are few outliers and also that the entries of $\mathbf{S}$ have an independent and identically Gaussian distribution then the above equation can be efficiently solved using the SVD. The problem of outliers in the point cloud can be considered as an idealized version of Robust PCA. Our purpose is to recover the low-rank matrix $\mathbf{E}$ from the measurements $\mathbf{M} = \mathbf{E} + \mathbf{S}$ which are highly corrupted by the matrix $\mathbf{S}$ with randomly large magnitude of entries. The approach that we use estimates the PCP by solving [303]:

$$\begin{aligned} \text{minimize} \quad & \|\mathbf{E}\|_* + \alpha \|\mathbf{S}\|_1 \\ \text{subject to} \quad & \mathbf{E} + \mathbf{S} = \mathbf{M} \end{aligned} \quad (4.18)$$

where $\|\mathbf{E}\|_* := \sum_i \sigma_i(\mathbf{E})$ denotes the nuclear norm of the matrix which is the sum of the singular values of $\mathbf{E}$. This convex PCP problem can be solved using an ALM algorithm, which works with stability across a wide range of situations without the necessity of parameters configuration. The ALM method operates on the augmented Lagrangian:

$$l(\mathbf{E}, \mathbf{S}, \mathbf{Y}) = \|\mathbf{E}\|_* + \alpha \|\mathbf{S}\|_1 + <\mathbf{Y}, \mathbf{M} - \mathbf{E} - \mathbf{S}> + \frac{\mu}{2}\|\mathbf{M} - \mathbf{E} - \mathbf{S}\|_F^2 \quad (4.19)$$

The PCP would be solved by a generic Lagrange multiplier algorithm setting $(\mathbf{E}_i, \mathbf{S}_i) = \arg\min_{\mathbf{E},\mathbf{S}} l(\mathbf{E}, \mathbf{S}, \mathbf{Y}_i)$ repeatedly, and then updating the Lagrange multiplier matrix via $\mathbf{Y}_{i+1} = \mathbf{Y}_i + \mu(\mathbf{M} - \mathbf{E}_i - \mathbf{S}_i)$.

In our case study, where a low-rank and sparse decomposition problem exists, the $min_\mathbf{E} l(\mathbf{E}, \mathbf{S}, \mathbf{Y})$ and $min_\mathbf{S} l(\mathbf{E}, \mathbf{S}, \mathbf{Y})$ both have very simple and efficient solutions. We introduce two ancillary operators $\mathcal{Q}_\tau$ and $\mathcal{D}_\tau$. $\mathcal{Q}_\tau : \mathbb{R} \longrightarrow \mathbb{R}$ denotes the shrinkage operator $\mathcal{Q}_\tau[.] = sgn(.)max(|.| - \tau, 0)$ and extends it to matrices by applying it to each element while the $\mathcal{D}_\tau(.)$ denotes the singular value thresholding operator given by $\mathcal{D}_\tau(.) = U\mathcal{Q}_\tau(\Sigma)V^T$. The estimation occurs according to:

$$\arg\min_{\mathbf{S}} l(\mathbf{E}, \mathbf{S}, \mathbf{Y}) = \mathcal{Q}_{\alpha\mu^{-1}}(\mathbf{M} - \mathbf{E} + \mu^{-1}\mathbf{Y}) \quad (4.20)$$

$$\arg\min_{\mathbf{E}} l(\mathbf{E}, \mathbf{S}, \mathbf{Y}) = \mathcal{D}_{\mu^{-1}}(\mathbf{M} - \mathbf{S} + \mu^{-1}\mathbf{Y}) \quad (4.21)$$

Firstly, the $l(.)$ is minimized with respect to $\mathbf{E}$ (fixing $\mathbf{S}$) and then the $l$ is minimized with respect to $\mathbf{S}$ (fixing $\mathbf{E}$). Finally, the Lagrange multiplier matrix $\mathbf{Y}$ is updated based on the residual $\mathbf{M} - \mathbf{E} - \mathbf{S}$.

$$\begin{aligned} \mathbf{S} &= \mathbf{Y} = 0; \mu > 0 \\ \mathbf{E}^{(t+1)} &= \mathcal{D}_{\mu^{-1}}(\mathbf{M} - \mathbf{S}^{(t)} + \mu^{-1}\mathbf{Y}^{(t)}) \\ \mathbf{S}^{(t+1)} &= \mathcal{Q}_{\alpha\mu^{-1}}(\mathbf{M} - \mathbf{E}^{(t+1)} + \mu^{-1}\mathbf{Y}^{(t)}) \\ \mathbf{Y}^{(t+1)} &= \mathbf{Y}^{(t)} + \mu(\mathbf{M} - \mathbf{E}^{(t+1)} + \mathbf{S}^{(t+1)}) \end{aligned} \quad (4.22)$$

The cost of each iteration is estimating $\mathbf{E}_{i+1}$ using singular value thresholding. This makes necessary the computation of those singular vectors of $\mathbf{M} - \mathbf{S}_i + \mu^{-1}\mathbf{Y}_i$ whose corresponding singular values exceed the threshold $\mu^{-1}$. There are two important implementation that need to be chosen; namely, the choice of $\mu$ and the stopping criterion. The value of $\mu$ is chosen as $\mu = n^2/4\|\mathbf{M}\|_1$, while the iterative process terminates when $\|\mathbf{M} - \mathbf{E} - \mathbf{S}\|_F \leq \beta\|\mathbf{M}\|_F$, with $\beta = 10^{-7}$. After this step the number of the vertices decrease, due to the removal of the outliers, so the number of the remaining (e.g., vertices) is $n_r$ where $n_r < n$ [304], [23].

Despite the effectiveness and the very good results that the aforementioned approach has provided in many applications, as have been presented by Lalos *et al.* [305], the exact decomposition of the Eq. (4.18) does not always exist especially for real noisy data (e.g., the matrix $\mathbf{M}$ of the presented problem). In this case, a more adaptive model is required, taking into account the presence of noise. So the matrix $\mathbf{M}$ can be decomposed as:

$$\mathbf{M} = \mathbf{E} + \mathbf{S} + \mathbf{G} \quad (4.23)$$

where $\mathbf{E} + \mathbf{S}$ approximates $\mathbf{M}$ and $\mathbf{G}$ is the noisy part.

## 4.6 Laplacian Interpolation to Motion Vectors

Laplacian interpolation has been used extensively in image processing [306], [307] providing extremely good results even in cases when a large amount of data is missing. In the field of dynamic point cloud consolidation, Laplacian interpolation has been used mainly for morphing [308], [309]. Despite the computational effectiveness of Laplacian interpolation, the consolidated results are smoothed.

A triangulated 3D model can be interpolated with a curved surface by putting constraints on the Laplacian matrix $\mathbf{L}$ [25]. The proposed Laplacian matrix $\mathbf{L}_w$ of Eq. (5.63), encloses all the necessary constrains in order to be efficiently used by weighted Laplacian interpolation. The process is applied to the motion vectors $\mathbf{m}$ of the vertices, represented by the matrix $\mathbf{B} = [\mathbf{m}_1 \; \mathbf{m}_2 \; \cdots \; \mathbf{m}_n] \in \mathbb{R}^{n \times 3}$. The Laplacian of $\mathbf{B}$ is written as $\Delta \mathbf{m} = \mathbf{L}_w \mathbf{B}$. Next, the matrix $\mathbf{B}$ is split into two parts: $\mathbf{B_k} \in \mathbb{R}^{n_k \times 3}$ containing the motion vectors of known vertices and

$\mathbf{B_u} \in \mathbb{R}^{n-n_k \times 3}$ containing the unspecified values. Correspondingly, the $\mathbf{L}_w$ can be partitioned into four parts:

$$\mathbf{L}_w = \begin{pmatrix} \mathbf{L}_{w_{11}} & \mathbf{L}_{w_{12}} \\ \mathbf{L}_{w_{21}} & \mathbf{L}_{w_{22}} \end{pmatrix} \quad (4.24)$$

$\mathbf{L}_{w_{11}} \in \mathbb{R}^{n_k \times n_k}$, $\mathbf{L}_{w_{12}} \in \mathbb{R}^{n_k \times (n-n_k)}$, $\mathbf{L}_{w_{21}} \in \mathbb{R}^{(n-n_k) \times n_k}$, $\mathbf{L}_{w_{22}} \in \mathbb{R}^{(n-n_k) \times (n-n_k)}$. We minimize the Euclidean norm $|\Delta \mathbf{m}|$ according to:

$$\left| \begin{pmatrix} \mathbf{L}_{w_{11}} & \mathbf{L}_{w_{12}} \\ \mathbf{L}_{w_{21}} & \mathbf{L}_{w_{22}} \end{pmatrix} \begin{pmatrix} \mathbf{B_k} \\ \mathbf{B_u} \end{pmatrix} \right| = \left| \begin{pmatrix} \mathbf{L}_{w_{11}} \\ \mathbf{L}_{w_{21}} \end{pmatrix} \mathbf{B_k} + \begin{pmatrix} \mathbf{L}_{w_{12}} \\ \mathbf{L}_{w_{22}} \end{pmatrix} \mathbf{B_u} \right| \quad (4.25)$$

This is equivalent to finding the least squares solution to:

$$\begin{pmatrix} \mathbf{L}_{w_{12}} \\ \mathbf{L}_{w_{22}} \end{pmatrix} \mathbf{B_u} = - \begin{pmatrix} \mathbf{L}_{w_{11}} \\ \mathbf{L}_{w_{21}} \end{pmatrix} \mathbf{B_k} \quad (4.26)$$

which is a system of $n$ equations with $n - n_k$ variables. The well-known least squares solution to this system of equations is given by:

$$\mathbf{B_u} = - \left( \begin{pmatrix} \mathbf{L}_{w_{12}} \\ \mathbf{L}_{w_{22}} \end{pmatrix}^\mathsf{T} \begin{pmatrix} \mathbf{L}_{w_{12}} \\ \mathbf{L}_{w_{22}} \end{pmatrix} \right)^{-1} \left( \begin{pmatrix} \mathbf{L}_{w_{12}} \\ \mathbf{L}_{w_{22}} \end{pmatrix}^\mathsf{T} \begin{pmatrix} \mathbf{L}_{w_{11}} \\ \mathbf{L}_{w_{21}} \end{pmatrix} \right) \mathbf{B_k} \quad (4.27)$$

## 4.7 Denoising of Vertices by Updating their Positions Using Denoised Normals

A lot of denoising approaches have been presented trying to denoise the normals of noisy 3D models instead of their points, directly. This type of methods have shown very good results, providing denoised normals that are very close to the original of the denoised 3D model. However, a question is how we can effectively use this useful information in order to reconstruct the geometry of the corresponding 3D denoised model. It has been proved that this can be achieved by updating the positions of the noisy vertices using the values of the denosied normals according to the following equation [6]:

$$\mathbf{v}_i^{(t+1)} = \mathbf{v}_i^{(t)} + \frac{\sum_{j \in \mathbf{\Psi}_i} \bar{\mathbf{n}}_{cj}(\langle \bar{\mathbf{n}}_{cj}, (\mathbf{c}_j^{(t)} - \mathbf{v}_i^{(t)}) \rangle)}{|\mathbf{\Psi}_{1i}|} \quad (4.28)$$

$$\mathbf{c}_j^{(t+1)} = (\mathbf{v}_{j1}^{(t+1)} + \mathbf{v}_{j2}^{(t+1)} + \mathbf{v}_{j3}^{(t+1)})/3 \ \forall \ j \in \mathbf{\Psi}_{1i} \quad (4.29)$$

where $\langle \mathbf{a}, \mathbf{b} \rangle$ represents the inner product of $\mathbf{a}$ and $\mathbf{b}$, $(t)$ represents the number of iteration and the matrix $\mathbf{\Psi}_{1i}$ is the cell of vertices that are directly connected to the vertex $\mathbf{v}_i$. This iterative process can be considered as a gradient descent process that is executed for minimizing the energy term $\sum_{j \in \mathbf{\Psi}_{1i}} \|\bar{\mathbf{n}}_{cj}(\mathbf{c}_j^{(t)} - \mathbf{v}_i^{(t)})\|^2$ across all faces.

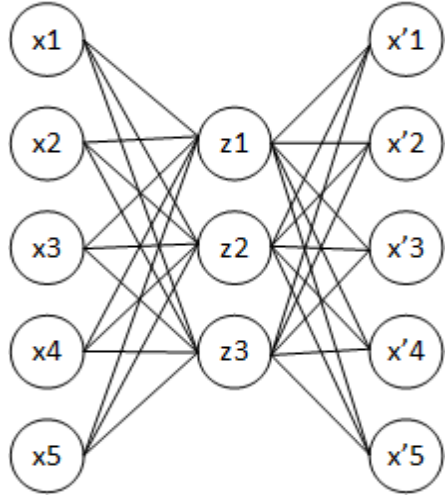

Figure 4.4: Simple architecture of an autoencoder.

## 4.8 Autoencoders

Autoencoders are part of a family of unsupervised deep learning methods, but they are also closely related to PCA (principal components analysis). Some facts about the autoencoder are:

- It is an unsupervised learning algorithm (like PCA).
- It minimizes the same objective function as PCA.
- It is a neural network.
- The neural network's target output is its input.

The architecture of an autoencoder is simply presented in Fig. 4.4.

Autoencoders compress (encode) the input $\mathbf{x} \in \mathbb{R}^d$ into a lower-dimensional code $\mathbf{z} \in \mathbb{R}^p$, where $p < d$, through a deterministic mapping $\mathbf{z} = f_\theta(\mathbf{x}) = \sigma(\mathbf{W}\mathbf{x} + \mathbf{b})$, parametrized by $\theta = \{\mathbf{W}, \mathbf{b}\}$, where $\sigma$ is an element-wise activation function (e.g., sigmoid or a rectified linear unit), $\mathbf{W} \in \mathbb{R}^{d \times p}$ is a weight matrix and $\mathbf{b} \in \mathbb{R}^{d \times 1}$ is a bias vector. Then reconstruct (decode) the output $\mathbf{x}' = h_{\theta'}(\mathbf{z}) = \sigma'(\mathbf{W}'\mathbf{z} + \mathbf{b}')$, where $\theta' = \{\mathbf{W}', \mathbf{b}'\}$, $\mathbf{W}' \in \mathbb{R}^{p \times d}$ and $\mathbf{b}' \in \mathbb{R}^{p \times 1}$. Autoencoders are trained to minimize reconstruction errors (such as squared errors):

$$\mathcal{J}(\mathbf{x}, \mathbf{x}') = \|\mathbf{x} - \mathbf{x}'\|_2 = \|\mathbf{x} - \sigma'(\mathbf{W}'(\sigma(\mathbf{W}\mathbf{x} + \mathbf{b})) + \mathbf{b}')\|_2 \quad (4.30)$$

The parameters of this model are optimized, minimizing [310]:

$$\theta^*, \theta'^* = \arg\min_{\theta, \theta'} \frac{1}{n} \sum_{i=1}^n \mathcal{J}(\mathbf{x}^{(t)}, h_{\theta'}(f_\theta(\mathbf{x}^{(t)}))) \quad (4.31)$$

## 4.9 Evaluation Metrics

The quality of the reconstructed results through the different methods and approaches that will be present in this thesis have been evaluated using a variety of different metrics that are shortly presented below:

- Dmean d is the average distance between the vertices of the reconstructed and the original 3D mesh.

- Dmax d is the maximum among the distances of the vertices of the reconstructed and the original 3D mesh.

- dist n is the average distance between the point normals of the reconstructed and the original 3D mesh.

- Dmean n is the average distance between the face normals of the reconstructed and the original 3D mesh.

- NMSVE (Normalized Mean Square Visual Error) has been shown to correlate well with perceived distortion by measuring the average error in the Laplacian and Cartesian domains [311]. NMSVE is used in order to evaluate the reconstruction quality of results, by capturing the average distortion between the original and the approximated frame:

$$NMSVE = \frac{1}{2k} \sum_{j=1}^{k} (\left\| v_i - \tilde{v}_i \right\|_2 + \left\| GL(v_i) - GL(\tilde{v}_i) \right\|_2) \quad (4.32)$$

$$GL(v_i) = v_i - \frac{\sum_{j \in \mathbf{N}_i} d_{ij}^{-1} v_j}{\sum_{j \in \mathbf{N}_i} d_{ij}^{-1}} \quad (4.33)$$

$d_{ij}$ denotes the Euclidean distance between i and j.

- $\theta$: represents the mean angle $\gamma$ (expressed in degree) between the normals of the ground truth face and the resulting face normals.

- HD: representing the average one-sided Hausdorff distance (HD) from the denoised mesh to the known ground truth mesh.

- Heatmap visualization which highlights the difference between reconstructed and original static 3D model (point cloud or mesh), described as $|\mathbf{P}_i - \bar{\mathbf{P}}_i| \ \forall \ i = 1, n$.

- KG error metric: has been introduced by Karni and Gotsman and it has been designed for the evaluation of animated triangle meshes. The detailed description of this metric is presented in [312].

- STED (Spatiotemporal edge difference): focus on the local changes of the error rather than on the absolute value of it. More details about this metric are presented in [313].

The metrics ($\theta$, NMSVE) are commonly used for the evaluation of each 3D mesh separately, while the last two (KG, STED) are used for the evaluation of the whole dynamic mesh as an object.

Additionally to the presented quantitative metrics we use also qualitative metrics like heatmap visualization highlighting, in different colors, the difference $|\mathbf{M} - \bar{\mathbf{M}}|$ between original $\mathbf{M}$ and reconstructed mesh $\bar{\mathbf{M}}$.

An index of performance of the registration task is the degree of deviation for rotation $\mathbf{R}$ and translation $\mathbf{t}$ from the ground truth rotation matrix $\mathbf{R}_g$ and translation vector $\mathbf{t}_g$. The rotation error $\epsilon_r$ and the translation error $\epsilon_t$ are defined as:

$$\epsilon_r = \arccos\left(\frac{\mathcal{S}(\mathbf{R}\mathbf{R}_g^{-1}) - 1}{2}\right)\frac{180}{\pi} \quad (4.34)$$

$$\epsilon_t = ||\mathbf{t} - \mathbf{t}_g|| \quad (4.35)$$

where $\mathcal{S}(\mathbf{A})$ determines the sum of the diagonal elements of the matrix $\mathbf{A}$ [264]. It is obvious that the smaller these errors, the better the matching results.

## 4.10 Benchmark Datasets

We have used a wide range of CAD and scanned static and dynamic 3D models. More specifically, the experiments are carried out using the following datasets.

- Scanned models [314] form Stanford dataset e.g. Stanford Bunny.

- CAD (computer-aided design) models [315] e.g. Fandisk model which is provided courtesy of MPII by the AIM@SHAPE shape Repository.

- Real scanned noisy models from 3 different devices (Kinect v1, Kinect v2, Kinect Fusion) and synthetic noise [10].

- Real scanned lungs affected by staircase effect using MRI device [11].

- A variety of CAD 3D meshes [10], affected by different levels of noise.

- A dataset consisting of artificial sequences of moving models with different types of noise by Torkhani *et al.* [316].

- A dataset consisting of motion capture animations representing humans in different kind of movement actions by Vlasic *et al.* [317].

- A dataset consisting of artificial sequences of moving animal models [318].

- A dataset, which is used for the evaluation of the outliers removal method, was taken from the "IQmulus & TerraMobilita Contest" benchmark [319]. It contains 10 highly dense point clouds ($13 \cdot 10^6 - 17 \cdot 10^6$ points) in different views of an urban environment.

To note here that the noisy models, affected by Gaussian noise, are created using the dataset of Vlasic *et al.* [317], by adding noise to the vertices of the ground truth meshes along the vertex normals, similar to Zhang *et al.* [9]. The intensity of the noise is described using a relative variance parameter:

$$\sigma_E = \frac{\sigma}{\hat{l}} \qquad (4.36)$$

where $\sigma$ denotes the variance of the Gaussian function, and $\hat{l}$ is the average edge length $l$ of the ground truth mesh.

For the experiments of registration and identification use case, we used two different datasets. The first one consists of a variety of partially scanned point clouds representing cluttered scenes of different objects, denoted as UWAOR [35], [37]. The dataset contains 50 cluttered scenes with up to 5 objects acquired with the Minolta Vivid 910 scanner in various configurations from a single viewpoint. All objects are heavily occluded (60% to 90%), as illustrated in Fig. 4.5. The second dataset, Fig. 4.6, consists of a variety of incomplete models under different viewpoints (angles), denoted as UWA3M [320] and with various percentages of occlusion. There are 22 incomplete instances of "Chef", 16 of "Chicken" and "Parasaurolophus", and 21 of "T-rex". The third dataset [321] is composed of 150 synthetic scenes, captured with a (perspective) virtual camera, and each scene contains 3 to 5 objects. The model set is composed of 20 different target objects.

### 4.10.1 Experimental Setup

The experiments were carried out on an Intel Core i7-4790HQ CPU @ 3.60GHz PC with 16 GB of RAM. The core algorithms are written in Matlab and C++.

### 4.10.2 Mathematical Framework for 3D Geometric Processing

In this thesis, the aforementioned and well-known mathematical tools, which have been mostly utilized in other domains like the image processing area, have been extended and modified to fulfill the requirements of 3D geometric processing. The purpose is to create a framework consisting of tools that can be used in different research areas of 3D geometric processing in order to facilitate their computational efficiency and the quality of their reconstructed results. In the following chapters, we will present in detail how these tools can be used in a variety of different approaches that cover a large area of 3D geometric processing applications.

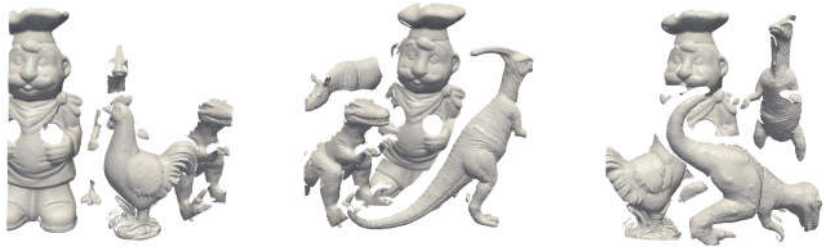

Figure 4.5: UWAOR dataset consisting of 50 partially scanned scenes from a single viewpoint.

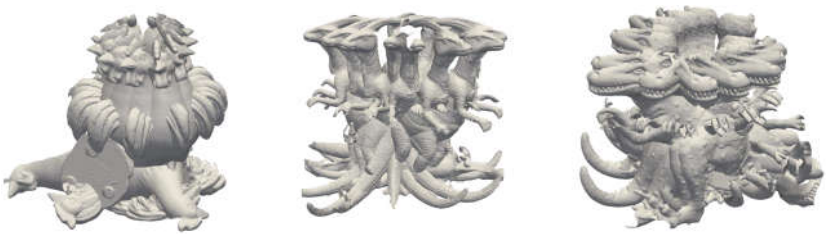

Figure 4.6: UWA3M dataset consisting of incomplete scans of 3D models in arbitrary angles.

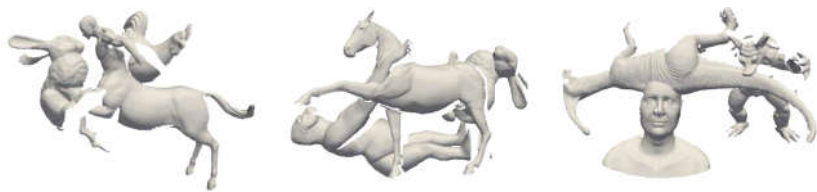

Figure 4.7: Rodolà's dataset consisting of 150 synthetic partially-observed scenes.

CHAPTER 5

# Low-level 3D Geometry Processing

In this chapter, we will present the suggested solutions trying to overcome some of the most common geometrical-related problems and challenges (as presented in Subsection 2.2), which could be appeared in an early stage of processing, using the mathematical notions and tools as presented in Chapter 4. Even though the proposed implementations use a common mathematical framework (Fig. 5.1), only for demonstration purposes and for the readers' convenience, we separate the presented applications into four sections based on the category of the low-level 3D geometry problem that each approach tries to solve.

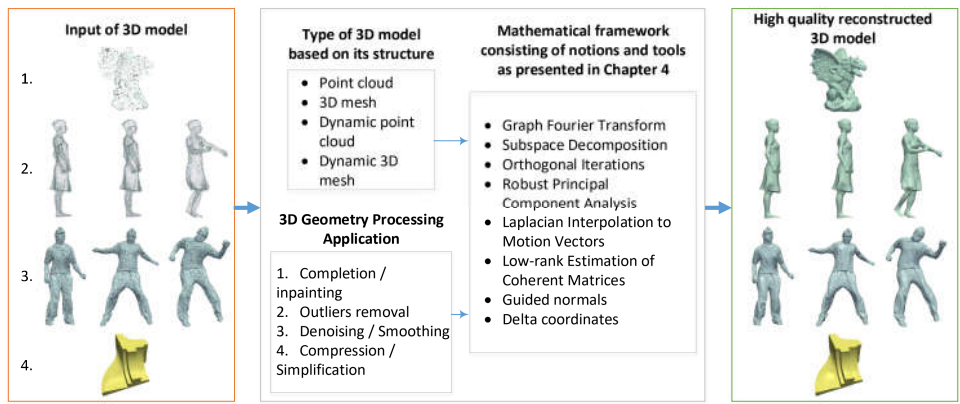

Figure 5.1: Example of different types of 3D models that need a different 3D geometry processing approach.

More specifically, the chapter is separated into the following research categories:

- Denoising (Section 5.1)
- Compression (Section 5.2)

- Completion and reconstruction of incomplete 3D models (Section 5.3)
- Outliers removal (Section 5.4)

Each one of the aforementioned categories will be discussed in detail in the following sections, showing our contribution and the suggested approaches that used to overcome the corresponding challenges.

## 5.1 Denoising of Static and Dynamic 3D Meshes

Denoising is a major problem in any signal processing application (e.g., audio, image, video, 3D object). It affects the real object of the physical world when they are digitized via a capture device. It is important, these noisy signals to be denoised before it used by other high-level applications like registration, tracking, used in immersive environments, etc. In the following subsections, we will present our contribution to the denoising process of static and dynamic meshes using spectral and deep learning approaches.

- Denoising of static 3D meshes using spectral methods (subsection 5.1.1).
- Denoising of dynamic 3D meshes (subsection 5.1.2).
- Denoising of static and dynamic 3D meshes using autoencoders (subsection 5.1.3).
- Denoising of static 3D meshes using image-based CNNs (subsection 5.1.4).

### 5.1.1 Feature Preserving Mesh Denoising Based on Graph Spectral Processing

At any noisy mesh $\tilde{M} \in \mathbb{R}^{nx3}$, the vertices and face normals that correspond to flat areas 'lie' in a low dimensional subspace of size $n_l$, while noise usually has a flat spectrum that is easily identifiable at the higher $n_h = n - n_l$ frequencies. Therefore, one way to mitigate the noise effects is to perform spectral smoothing by filtering out the higher frequencies. In mathematical terms, the subspace decomposition of the normalized Laplacian matrix defined by the vertices of the mesh can be re-written as:

$$\mathcal{L} = \begin{bmatrix} \mathbf{U}_{n_h} & \mathbf{U}_{n_l} \end{bmatrix} \begin{bmatrix} \mathbf{\Lambda}_{n_h} & 0 \\ 0 & \mathbf{\Lambda}_{n_l} \end{bmatrix} \begin{bmatrix} \mathbf{U}_{n_h} & \mathbf{U}_{n_l} \end{bmatrix}^T \tag{5.1}$$

where $\mathbf{U}_{n_l} = [\mathbf{u}_1, \mathbf{u}_2, \ldots, \mathbf{u}_{n_l}]$. Then a smoothed version of the noisy vertices can be generated by performing low pass spectral filtering of the Cartesian coordinates $\mathbf{v} = \mathbf{U}_{n_l} \mathbf{U}_{n_l}^T \tilde{\mathbf{v}}$.

Taking into account this observation, we initially decompose the original mesh in several registered 3D patches. Then, two steps of denoising are followed, namely the coarse and the fine step. In the coarse step, each part is treated individually. The coarse denoising approach is capable of identifying both i) the spectral subspace size where the features lie and ii) the level of noise, using a model-based Bayesian learning scheme. To reduce the required complexity for evaluating the spectral subspace for each part of the surface, we adopt a subspace tracking approach that exploits potential coherences between the spectral subspaces of adjacent parts. In the fine step, we eliminate the small amount of remaining noise. To that end, we apply a feature-aware GNF method. More specifically, the application of the coarse step allows the accurate identification of the geometric features, which are then used to accelerate the execution time of the conventional GNF approaches.

The coarse denoising approach suggests decomposing a mesh into smaller submeshes using a fast graph partitioning method and reducing the amount of artifacts and noise within each patch by cutting off the higher frequencies. To fine-tune the smoothing operation and preserve corners and sharp edges, we propose a novel noise level estimation scheme that dynamically evaluates the subspace size where the features 'lie' using a model-based Bayesian learning approach. One of the major limitations of GSP is the required computational complexity since it involves the evaluation of the singular vectors of a matrix (spectral coefficients) with a size equal to the number of vertices in the patch. To overcome this issue we propose a fast and efficient way for evaluating the GSP coefficients of the different patches by using an approach that is based on subspace projections. The proposed coarse denoising approach allows us to accurately cluster the face normals into features (edges, corners) and flat areas and then apply a fine denoising operation based on a graph spectral filter that averages normals that belong to a common smooth region while preserving sharp changes of normals that are indicated as features. This provides a set of face normals that are closer to the face normals of the original model, as shown in Fig. 5.2.

The processed normals are then used to update the corresponding vertex positions. In addition, this process significantly accelerates the convergence of the following process. Finally, for a given feature face we search among a set of candidate patches that contain the feature face and pick the one with the most consistent normal directions (ideal patch selection). The average normal of the chosen patch is then used as the normal of the selected face. Such guidance provides a robust estimation for the true normal in the presence of noise, enabling our denoising algorithm to handle highly noisy meshes. The proposed TSGSP architecture is illustrated in Fig. 5.3.

The main contributions of this work are:

- We provided a novel coarse denoising step that filters out a significant

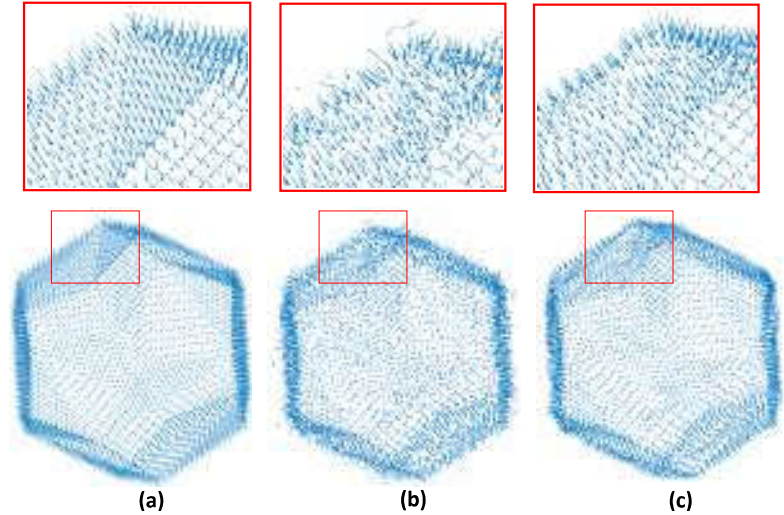

Figure 5.2: Face normals of: (**a**) The original noise-free mesh; (**b**) the noisy mesh; (**c**) Smoothed reconstructed model.

amount of noise without affecting geometric features under non stationary noise scenarios. This approach is essential for denoising large scale meshes representing complex 3D scenes, that are reconstructed in parts, affected by noise with different characteristics. To achieve that, we developed: i) An approach that automatically identifies the subspace size where sharp and small-scaled geometric features lie. Then, the subspace of interest is evaluated using orthogonal iterations, an efficient subspace tracking approach that exploits the spatial coherences between different parts of the surface, and results in very fast execution times. ii) A method which is capable of identifying the pattern of noise, using a model based Bayesian learning scheme. The identified subspace and the evaluated noise patterns are then used to mitigate variable noise patterns that are added in different surface patches that are processed individually.

- The use of the coarse step allow us to accurately identify geometric features, which are then used by the proposed feature aware GNF approach that achieves faster execution times as compared to the conventional GNF approach. More specifically, on contrary to GNF scheme, the iterative process for searching the ideal patches is not applied to the whole mesh but only in these vertices that have been classified as features.

- We proved that the fine step can be also considered as a graph spectral method and we provide extensive simulations carried out using several scanned and CAD models under a broad set of noise configurations showing that: i) The coarse scheme can be also adopted by any state-of-the-art denoising approach accelerating its execution time and allowing them to efficiently mitigate noise patterns that vary between different parts of the surface. ii) The combination of the proposed coarse and the fine graph spectral processing steps offers both enhanced robustness and low complexity requirements when dealing with anisotropic and non stationary noise patterns.

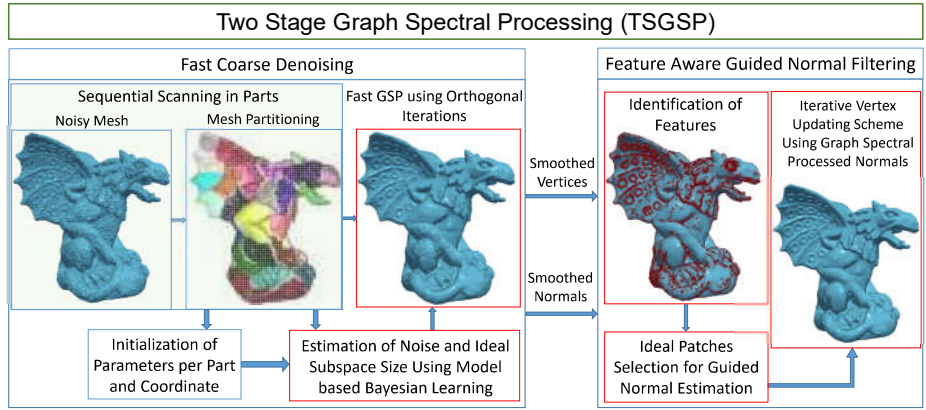

Figure 5.3: Sequential diagram of the procedure. The proposed framework is separated into a coarse denoising stage and a fine denoising stage. The proposed methods are represented with red boxes.

**Fast Sequential Processing of 3D Patches Using Orthogonal Iterations**

Calculating the graph Laplacian eigenvalues of the mesh geometry can become restrictive as the density of the models increases. To overcome this limitation, several approaches suggest processing large meshes into parts [322], [323]. In the proposed scheme, we initially separate the whole mesh into submeshes and then we replace SVD with an approximate iterative method that is much more computationally efficient.

The projection of the raw geometry data in the domain defined by the graph Laplacian eigenvalues $\mathbf{U}_{n_l}$ allows efficient representations that can be exploited by a number of applications such as, in our case, denoising. However, in order to avoid smoothing any potential feature point, we need to accurately identify the size $n_l$ of the subspace where the edges and corner points lie. The following subsection provides a dynamic scheme that can successfully identify the optimal subspace size $n_l$.

**Processing of Registered 3D Patches**

The noisy 3D mesh is partitioned into $m$ submeshes that are reconstructed individually, using the METIS method described in [323], [324]. Each submesh $\tilde{M}_i$ consists of $n_{m_i}$ points, where $\sum_{i=1}^{m} n_{m_i} = n$. To avoid the problem of edge effects, we process overlapped submeshes and estimate the weighted average of vertex positions for the overlapping nodes. The weights that are assigned to each point are proportional to the degree of the node (e.g., number of neighbors) in the corresponding submesh. For example, in Fig. 5.4 we present the weights assigned to a point (highlighted in red) that participates in three overlapped submeshes. The ideal number of submeshes depends on the total number of points of a mesh, however, an extensive evaluation study shows that submeshes with 500-800 points provide the best results in terms of both execution time and visual error.

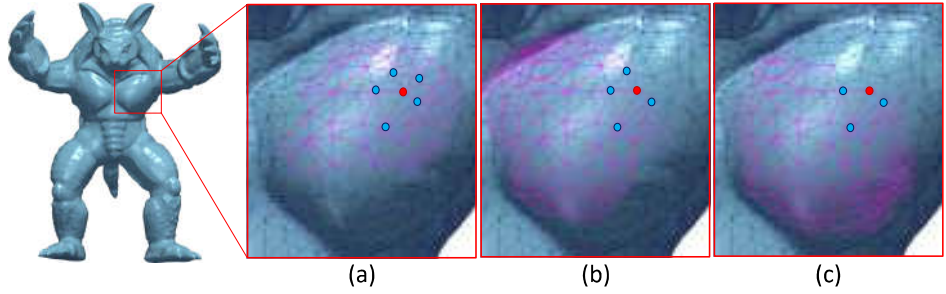

Figure 5.4: The red point has different degree in each submesh, the corresponding weights $w$ are: (a) $w = 5$, (b) $w = 4$, (c) $w = 3$.

**Dynamic Identification of $n_l$ and $\sigma_z^2$**

We start by assuming that the observed noisy vertices can be separated into two quantities, namely the clean vertices $\mathbf{v_c}$ and the external noise $\tilde{\mathbf{z}}$, as follows:

$$\mathbf{v}_j = \mathbf{v}_{c_j} + \tilde{\mathbf{z}}_j \qquad (5.2)$$
$$= \mathbf{U}_{n_{l_j}^*} \hat{\mathbf{v}}_j + \tilde{\mathbf{z}}_j \;\; j = \{x, y, z\} \qquad (5.3)$$

where $n_{l_j}^*$ represents the optimal length of features subspace size, $\hat{\mathbf{v}}_j = \mathbf{U}_{n_{l_j}^*}^T \mathbf{v}_{c_j}$ is projected clean vertices and $\tilde{\mathbf{z}}_j$ is the vector of noise coordinate $j = \{x, y, z\}$. If we assume that $n_{l_j}^*$ is known, then the $\hat{\mathbf{v}}_j$ can be modeled as a parametrized multivariate Gaussian distribution:

$$p(\hat{\mathbf{v}}_j) \sim \mathcal{N}(0, \mathbf{\Pi}_{0_j}), \;\; \mathbf{\Pi}_{0_j} = \gamma_j \mathbf{\Sigma}_j \;\; j = \{x, y, z\} \qquad (5.4)$$

where $\gamma_j$ is a scalar parameter and $\Sigma_j \in \mathbb{R}^{n_{lj}^* \times n_{lj}^*}$ is a positive definite matrix. By using the Bayes rule and assuming that the noise vector $\tilde{z}_j$ has distribution with zero mean value then $\tilde{z}_j \sim \mathcal{N}(0, \sigma_{z_j} I)$. However, in the general case, the optimal feature subspace size value of $n_lj^*$ is not known. A non optimal choice of $n_{lj}$, would lead either to the exclusion of some feature components or the inclusion of noisy components. In this case, the $l_2$ norm of the vector $\|v_j - U_{n_{lj}} \hat{v}_j\|/n_m$ will be a value outside the region $[\sigma_{z_j}^2 - \epsilon, \sigma_{z_j}^2 + \epsilon]$, where $\epsilon$ is a negligible threshold. Motivated by this observation we propose an iterative scheme that either increases or decrease the feature subspace size, so that the aforementioned termination condition is satisfied. For each choice of m the following Bayesian learning scheme is executed in order to identify the projected clean coefficients and the level of noise. In particular, the following Proposition 1 presents the Bayesian learning steps for the identification the of projected clean coefficients per $i$ submesh and per coordinate $j = \{x, y, z\}$.

*Proposition* 1. Assuming that the posterior density of $\hat{v}_{ij}$, is also Gaussian $p(\hat{v}_{ij}|v_{ij}; \sigma_{z_{ij}}, \gamma_{ij} \Sigma_{ij}) \sim \mathcal{N}(H_{ij}, \Pi_{ij})$ we can estimate its mean and covariance matrix as follows:

$$H_{ij} = \Pi_{0ij} U_{n_{lij}}^T \left( U_{n_{lij}} \Pi_{0ij} U_{n_{lij}}^T + \sigma_{z_{ij}} I_{k_i} \right)^{-1} v_{ij} \tag{5.5}$$

$$\Pi_{ij} = \Pi_{0ij} - \Pi_{0ij} U_{n_{lij}}^T \left( U_{n_{lij}} \Pi_{0ij} U_{n_{lij}}^T + \sigma_{z_{ij}} I_{k_i} \right)^{-1} U_{n_{lij}} \Pi_{0ij} \tag{5.6}$$

For finding the parameters $\sigma_{z_{ij}}, \gamma_{ij}, \Sigma_{ij}$ we employ the expectation-maximization (EM) algorithm to maximize the $p(v_{ij}; \sigma_{z_{ij}}, \gamma_{ij} \Sigma_{ij})$ per coordinate, meaning that $j = \{x, y, z\}$. For the estimation of the model parameters that maximize the aforementioned likelihood we use iteratively the following learning rules:

$$\sigma_{z_{ij}} = \frac{\|v_{ij} - U_{n_{lij}} \hat{v}_{ij}\|_2^2 + \sigma_{z_{ij}}[n_{lij} - Tr(\Pi_{ij} \Pi_{0ij}^{-1})]}{k_i} \tag{5.7}$$

$$\gamma_{ij} = \frac{Tr(\Sigma_{ij}^{-1}(\Pi_{ij} + H_{ij} H_{ij}^T))}{n_{lij}} \tag{5.8}$$

$$\Sigma_{ij} = \frac{\Pi_{ij} + H_{ij} H_{ij}^T}{\gamma_{ij}} \tag{5.9}$$

The proof of Proposition 2 is provided in the Appendices B. The performance of the proposed algorithm can be further improved by constraining the matrix $\Sigma_{ij}$ in order to have a Toeplitz symmetric structure with elements $(\Sigma_{ij})_{(q,w)} = r_{ij}^{|q-w|}$, $\forall q, w \in [1, \cdots, n_{lij}]$. This form is equivalent to modeling the elements

in the non-zero block as a first order auto-regressive process, where $r_{ij}$ can be estimated by:

$$r_{ij} = sign(q1/q0) min\{|q1/q0|, 0.99\} \quad (5.10)$$

where $q0$ is the average of the elements along the main diagonal, $q1$ is the average of elements along the main sub-diagonal of $\Sigma_{ij}$ and the value 0.99 is a bound selected by the user. The proposed scheme suggests executing the aforementioned methods iteratively, initializing the $n_{l_{ij}}$ and $\sigma^2_{z_{ij}}$ to any arbitrary value, until the following stopping criterion is satisfied $\sigma^2_{z_{ij}} - \epsilon < \|v_{ij} - U_{n_{l_{ij}}} \hat{v}_{ij}\|^2_2 / n_{mi} < \sigma^2_{z_{ij}} + \epsilon$. In each iteration $t$, if $\|v_{ij} - U_{n_{l_{ij}}} \hat{v}_{ij}\|^2_2 / n_{mi} < \sigma^2_{z_{ij}} - \epsilon$ then the length of $n_{l_{ij}}^{(t)} = n_{l_{ij}}^{(t-1)} - 1$ is decreased by one, otherwise the length of $n_{l_{ij}}^{(t)} = n_{l_{ij}}^{(t-1)} + 1$ is increased by one. For the initialization, arbitrary values of $n_{l_{ij}}^{(0)}$ and $\sigma^2_{z_{ij}}{}^{(0)}$ could be used. However, to avoid exhaustive search, we suggest an optimal initialization strategy presented in the following subsection. In Algorithm 1 we briefly describe the steps of this iterative procedure.

---

**Algorithm 1:** Identification of the optimal Subspace Size $n_{l_{ij}}^*$ per sub-mesh and per coordinate, using MBL.

---

1 Initialize the values of $n_{l_{ij}}^{(0)}$ and $\sigma^2_{z_{ij}}{}^{(0)}$, according to Eqs. (5.11) and (5.15);
2 **for** $i = 1, n_t$ **do**
3    **for** $j \in \{x, y, z\}$ **do**
4       **while** $|\|v_{ij} - U_{n_{l_{ij}}} \hat{v}_{ij}\|^2_2 / n_{mi} - \sigma^2_{z_{ij}}| < \epsilon$ **do**
5          **for** $g = 1, b$ **do**
6             Estimate the non zero values of $H_{ij}$ via Eq. (B.17) ;
7             Estimate the corresponding variances $\Pi_{ij}$ via Eq. (B.18) ;
8             Update the values of $\sigma_{z_{ij}}, \gamma_{ij}, \Sigma_{ij}, r_{ij}$ according to Eqs. (5.7) - (5.10) ;
9          **end**
10         **if** $\|v_{ij} - U_{n_{l_{ij}}} \hat{v}_{ij}\|^2_2 / n_{mi} > \sigma^2_{z_{ij}} + \epsilon$ **then** $n_{l_{ij}}^{(t)} = n_{l_{ij}}^{(t-1)} + 1$ ;
11         **else**
12            $n_{l_{ij}}^{(t)} = n_{l_{ij}}^{(t-1)} - 1$ ;
13            $t = t + 1$ ;
14       **end**
15    **end**
16 **end**

**Initialization of Parameters $n_l$ and $\sigma_z^2$**

The convergence speed of the aforementioned scheme strongly depends on the initial value of $n_{l_{ij}}^{(0)}$. An extensive simulation study using different models with geometric features of different scales showed that the subspace size where the feature lie is strongly dependent on the level of noise and the values of the geometric Laplacian. Motivated by these observations, we decided to use the following initialization strategy. An initial value that significantly accelerates this approach is the following one:

$$n_{l_{ij}}^{(0)} = n_{m_i}\left(\frac{e^{-|\widetilde{e}_{ij}|}}{\hat{\sigma}_z^{(0)}}\right) \; \forall \, i = 1, \ldots, n_t \; , \; \forall \, j \in \{x, y, z\} \quad (5.11)$$

where $\hat{\sigma}_z^{2(0)}$ is the noise variance vector, $n_t$ the total number of submeshes and $\widetilde{e}_{ij}$ is the normalized value of smoothing factor $e_{ij}$ [325], [326], expressed in mathematical terms as:

$$\mathbf{e}_i = \left\|\mathbf{L}^{1/2}[i]\mathbf{v}_i\right\|_2^2 = \mathbf{v}_i^T \mathbf{L}[i] \mathbf{v}_i, \quad \forall \, i = 1, \ldots, n_t \quad (5.12)$$

where $\mathbf{e}_i = [e_{i_x} \; e_{i_y} \; e_{i_z}]$. Small values of $\widetilde{e}_{ij}$ represent smooth areas while large values represent rough areas. For the estimation of the initial noise level $\hat{\sigma}_z^2[0]$, we start by assuming that the geometry data and the noise are uncorrelated, so the variance of the projected vertices of mesh can be expressed as [327]:

$$\sigma(\widetilde{\mathbf{v}}\vec{\mathbf{u}}) = \sigma(\mathbf{v}\vec{\mathbf{u}}) + \sigma_z^2 \quad (5.13)$$

where $\sigma(\mathbf{v}\vec{\mathbf{u}})$ represents the variance of vertices $\mathbf{v}$ in the $\vec{\mathbf{u}} \in \mathbb{R}^{3 \times 1}$ direction, $\widetilde{\mathbf{v}}$ are the affected by the noise vertices and $\sigma_z$ is the standard deviation of the noise. The direction of minimum variance can be calculated by applying the PCA on different equalized patches of the mesh. For this reason we start by separating the mesh vertices $\widetilde{\mathbf{v}}$ into $n_t$ overlapped patches. Each patch consists of $n_m$ vertices based on their relative distance which are estimated using $k$-nearest neighbor ($k$-nn) algorithm, so that the $i^{th}$ patch can be represented by the $\widetilde{\mathbf{v}}_i = [\widetilde{\mathbf{v}}_{i_1}, \widetilde{\mathbf{v}}_{i_2}, \ldots, \widetilde{\mathbf{v}}_{i_{n_m}}]^T \in \mathbb{R}^{n_m \times 3}$. The minimum variance direction can be estimated by the eigenvector associated to the minimum eigenvalue of the covariance matrix $\mathbf{R}_{\widetilde{\mathbf{v}}_{n_t}} = \frac{1}{n_t} \sum_{i=1}^{n_t} \widetilde{\mathbf{v}}_i \widetilde{\mathbf{v}}_i^T$. The variance of the vertices projected onto the minimum variance direction equals the minimum eigenvalue of the covariance matrix $\mathbf{R}_{\widetilde{\mathbf{v}}_{n_t}}$:

$$\lambda_{min}(\mathbf{R}_{\widetilde{\mathbf{v}}_{n_t}}) = \lambda_{min}(\mathbf{R}_\mathbf{v}) + \sigma_z^2 \quad (5.14)$$

where $\lambda_{min}(\mathbf{R})$ represents the minimum eigenvalue of the matrix $\mathbf{R}$. Vertices of smoothed patches, span a subspace whose dimension is smaller than $k_{n_t} \times 3$. These patches are low-rank and they can be used for the estimation of $\hat{\sigma}_z^2$ taking advantage of the fact that the minimum eigenvalue of their covariance matrix $\lambda_{min}(\mathbf{R_v})$ can be assumed as zero. Since Gaussian noise has the same power in every direction and all eigenvalues are the same, we can estimate the noise variance from the subspace spanned by the eigenvectors of the covariance matrix $\mathbf{R}_{\tilde{\mathbf{v}}_{n_t}}$ with zero eigenvalues:

$$\hat{\sigma}_z^2 = \lambda_{min}(\mathbf{R}_{\tilde{\mathbf{v}}_{n_t}}) \tag{5.15}$$

Therefore by substituting Eq. (5.15) in Eq. (5.11) we can evaluate the subspace size $n_{l_i}$ that can be used for smoothing the noisy mesh without affecting noticeably the feature points.

**Feature Preserving Surface Reconstruction**

The presented coarse denoising step generates a smooth version of the normal vectors without deteriorating features. Although, the smoothed normals of the surface provide a more accurate representation of the original normal vectors; further processing is still required in order to:

1. Better highlight specific local characteristics, like edges and corners

2. Provide a more smooth version of the flat areas

To that end, the main goal is to accurately classify face normals into features and flat areas. Then we focus on finding the most representative normals for each face, based on the identified features across different neighborhoods. After identifying the ideal neighbors and weights, we apply a two-stage graph spectral processing scheme that iteratively reconstructs the normals (e.g., each face normal is replaced by an average face normal across the ideal neighborhood) and then the vertices are updated accordingly.

**Identification of Features**

Since our focus is only on faces that represent potential features, we merge the other two cases (edge, corner) to one class (features) and another one, characterized as faces of flat areas (non-features). The described scheme is easily adaptable and can be used for the estimation of both small and large-scale features. In Fig. 5.5, we present a 3D model with geometric features of different scales, highlighted with red dots. By inspecting this figure it can be easily seen that our algorithm is capable of identifying both small and large scale features by simply modifying the patch size.

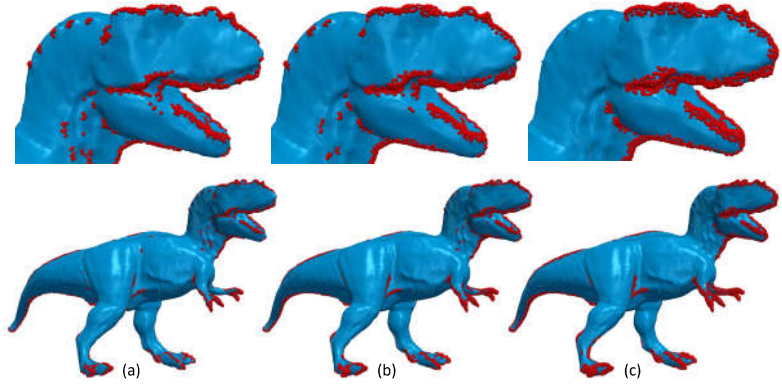

Figure 5.5: 3D model (Tyra) with a lot of small and large-scale features. Feature classification using patch area of (a) 10, (b) 20, (c) 40 neighbors.

**Selecting the Best Neighborhoods and Weights**

In this paragraph, we present a spectral method for estimating the representative face normal vectors, using an adjacency matrix of the face normals that can be considered as neighbors. To build this adjacency matrix it is important to identify the best neighborhood of each normal and then assign the appropriate bilateral weights. Our goal is to construct neighborhoods by identifying the faces that have common geometric properties. We initially execute the *k*-nn for each face of the mesh. The neighborhood of a non-feature face *i* is formed by the *k* nearest neighbors $\mathcal{P}_i$. The $n_p$ candidate neighborhoods of the a *i* feature face are all the patches, where the face *i* is participating as a nearest neighbor, defined as $\mathcal{B}_i = \{\mathcal{P}_{i1}, \mathcal{P}_{i2}, \ldots, \mathcal{P}_{in_p}\}$, where the $j^{th}$ candidate area of $i^{th}$ face is represented as $\mathcal{P}_{ij}$. Our purpose is to find which of the candidate patches $\mathcal{P}_{ij}$ is the ideal representative area for $i^{th}$ feature face.

Three parameters are investigated in order to identify the desirable areas: (a) the normalized difference $\psi_j$ between $\lambda_{j3}$ and $\lambda_{j1}$ eigenvalues (defined in the previous subsection) (b) the inner product between the face normal and the average normal of the patch and (c) the maximum distance between the face normal and the other face normals belonging to the same patch. The best neighborhood of a feature face *i* is the area that maximizes Eq. (5.16):

$$A_i = \begin{cases} (\mathcal{P}_{ij} \mid \alpha = \max(\frac{\psi_j \cdot \xi_{ij}}{\omega_{ij}})) & f_i = \text{feature} \quad \forall\, i = 1, n_f \\ \mathcal{P}_i & \text{otherwise} \quad \forall\, j = 1, n_p \end{cases} \quad (5.16)$$

$$\text{where} \quad \psi_j = \frac{|\lambda_{j1} - \lambda_{j3}|}{\lambda_{j3}} \tag{5.17}$$

$$\zeta_{ij} = \langle \hat{\mathbf{n}}_{ci}, \sum_{g \in \mathcal{P}_{ij}} \frac{\hat{\mathbf{n}}_{cg}}{|\mathcal{P}_{ij}|} \rangle \tag{5.18}$$

$$\omega_{ij} = max(|\hat{\mathbf{n}}_{ci} - \hat{\mathbf{n}}_{cg}|_2) \quad \forall g \in \mathcal{P}_{ij} \tag{5.19}$$

Finally, $A_i$ represents the area where each face has similar geometric characteristics with the face $i$. In Fig. 5.6, we illustrate different candidate areas of the selected face (highlighted in blue color). After identifying the most repre-sentative area, we focus on the estimation of the ideal values for the weighted adjacency matrix. Each connected face has a different weight, similar to the bilateral weights, which are evaluated using the following equation:

$$\mathbf{C}_w = \mathbf{W}_c \circ \mathbf{W}_s \circ \mathbf{C} \tag{5.20}$$

where $\circ$ denotes the Hadamard product, $\mathbf{C}$ is the binary adjacency matrix constructed using the k-nearest neighbors and $\mathbf{W}_c, \mathbf{W}_s \in \mathbb{R}^{n_f \times n_f}$ are estimated according to:

$$\mathbf{W}_{c_{ij}} = \begin{cases} exp(-\|\mathbf{c}_i - \mathbf{c}_j\|^2 / 2\sigma_c^2) & if \ \mathbf{C}_{ij} = 1 \\ 0 & otherwise \end{cases} \tag{5.21}$$

$$\mathbf{W}_{s_{ij}} = \begin{cases} exp(-\|\hat{\mathbf{n}}_{ci} - \hat{\mathbf{n}}_{cj}\|^2 / 2\sigma_s^2) & if \ \mathbf{C}_{ij} = 1 \\ 0 & otherwise \end{cases} \tag{5.22}$$

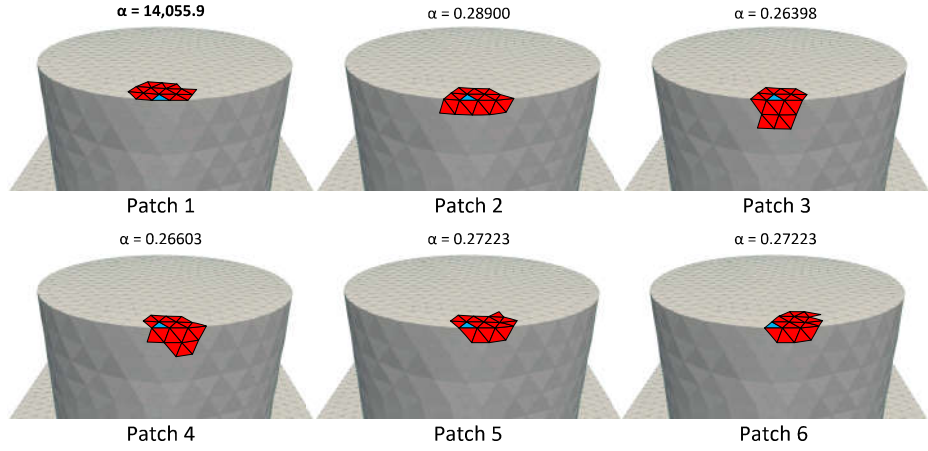

Figure 5.6: Candidate areas for the blue face.

---
**Algorithm 2:** Two Stage Graph Spectral Processing.
---
    **Function:** 3D Mesh denoising
    **Input**    : Triangle noisy mesh $\tilde{\mathbf{M}}$ corrupted by noise.
    **Output**  : Triangle denoising mesh.
    `/* Coarse reconstruction: cutting off higher frequencies  */`
1 Divide $\tilde{\mathbf{M}}$ into $s$ overlapped submeshes $\tilde{\mathbf{M}}_s$ using METIS method described in [323] [324] ;
2 Use steps of **Algorithm** 1 ;
3 Recreate the whole smoothing mesh;
    `/* Fine reconstruction: attenuating higher frequencies    */`
4 Classify each face as feature or non-feature via Eq. (4.15) ;
5 Create weighted adjacency matrix $\mathcal{A}$ via Eqs. (5.16)-(5.22) ;
6 Fine denoising of the normals $\bar{\mathbf{n}}_c$ via Eq. (5.23) ;
7 Update vertices via Eqs. (4.28)-(4.29) ;
8 **return** Triangle denoising Mesh ;
---

**Two Stage Processing of Normals and Vertices**

The ideal neighborhoods and weights are then used to fine tune the normals according to:

$$\bar{\mathbf{n}}_c = (\mathbf{D}^{-1}\mathbf{C_w})^\zeta \hat{\mathbf{n}}_c \qquad (5.23)$$

where $\hat{\mathbf{n}}_c$ represents the smoothed normals, $\zeta$ is an integer that represents a denoising factor. The fine-tuned normals are then used to update the vertices according to subsection 4.7.

**Coarse Denoising as a Pre-Processing Step**

We would like to emphasize that the coarse denoising method can be considered as an out-of-core method that could be employed by any other state-of-the-art method as a pre-processing step, optimizing both its reconstruction quality and its computational complexity. The reconstruction benefits can be easily identified by inspecting Fig. 5.7 which presents denoising results of state-of-the-art methods (first row) and the corresponding results after using the proposed coarse denoising approach as a pre-processing step (second row). Similarly, Fig. 5.7 shows the effects of the coarse denoising step in the execution time of the Guided normal approach [9]. By inspecting the number of iterations required for achieving a given reconstruction quality, the execution times of the coarse denoising step and of fine denoising iterations, it can be noted that the coarse denoising step accelerates the convergence rate of the fine denoising approach, thus significantly improving the total execution time of the whole denoising process.

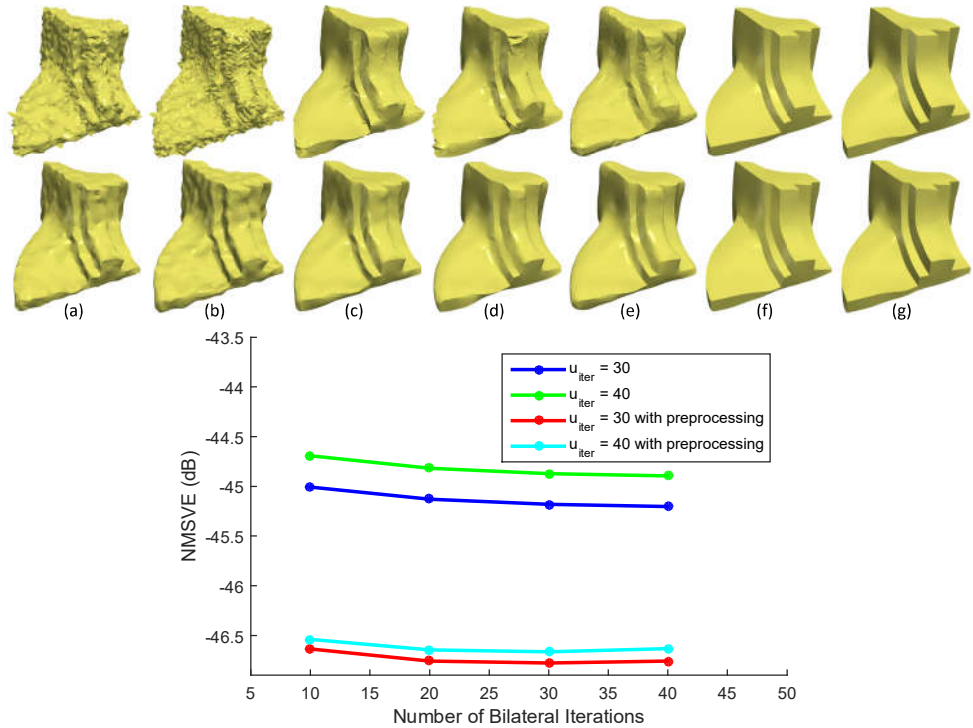

Figure 5.7: [Up] Reconstruction results with (2nd row) and without (1st row) preprocessing step using the following approaches: (a) [4], (b) [5], (c) [6], (d) [7], (e) [7], (f) [8], (g) [9]. [Down] Coarse denoising step increases the reconstruction quality. In other words, less iterations are required in order to achieve the same quality (Fandisk).

**Feature-Aware Fine Denoising Evaluation**

For the evaluation, we use a wide range of alternative approaches for selecting the best neighborhoods and the corresponding adjacency weights. More specifically, in Fig. 5.8 we present the original noisy mesh, the fine denoising output after using the $k$-nn patch directly without searching for the ideal area, the results after using Gaussian weights and finally the results after using our proposed approach.

**Computational Complexity**

Table 5.1 presents the execution times of the TSGSP approach as compared to the Guided normal algorithm [9], which is dominant among the state-of-the-art competitors. Note that while both approaches are based on a sequential update of the face normals and vertices, TSGSP results in slower execution times (68x-87x). This reduction is attributed to the application of the coarse denoising approach that filters out the high-frequency components, accelerating the convergence speed of the necessary corrections/adjustments of the vertex positions. Moreover, it should be also noted that the selection of the ideal neighborhood for each face normal correction is also a crucial parameter that contributes to the reduction of the execution time required for readjusting the mesh faces/vertices during a fine processing step.

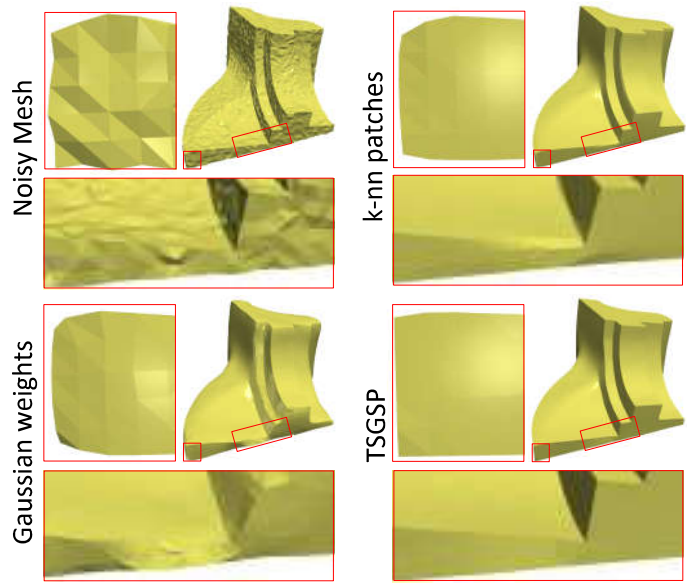

Figure 5.8: Results after using a wide range of alternative approaches for selecting the best neighborhoods and the corresponding adjacency weights.

**Results Assuming Complex Noise Patterns & Geometries**

In Figs. 5.9-5.12, we present the denoised results of our method in comparison with other state-of-the-art methods. For the experiments two different models are used which are captured by three different scanning devices (Kinect v1, Kinect-fusion, Kinect v2) as well as synthetic noise. Additionally, in each figure, the average angular difference (in degrees) is presented. The results show that the proposed scheme outperforms the other state-of-the-art concerning the reconstruction quality. Heatmap visualizations Fig. 5.13 also highlight the distortion alleviation when both components (coarse and fine) are used. The staircase effect, caused by MRI and CT devices [11], is a type of noise that requires more sophisticated schemes that differentiate staircase noise effects from surface features. In contrast to other methods, which often mistakenly consider the staircase effect as a mesh feature, our method effectively handles this case.

| | Guided mesh normal filtering | | | | | | TSGSP | | | | | | |
|---|---|---|---|---|---|---|---|---|---|---|---|---|---|
| Model | $k_{iter}$ | $t_{k_{iter}}$ | $u_{iter}$ | $t_{u_{iter}}$ | N | $t_N$ | $t_{tot}$ | $t_{cd}$ | $t_{fd}$ | $u_{iter}$ | $t_{u_{iter}}$ | N | $t_N$ | $t_{tot}$ |
| Block $\sigma$=0.4 | 40 | 200.31 | 30 | 21.88 | 17550 | 10.63 | 232.82 | 10.17 | 32.59 | 20 | 14.78 | 4457 | 2.68 | 60.32 -74% |
| Twelve $\sigma$=0.5 | 75 | 273.07 | 20 | 7.64 | 9216 | 8.11 | 289.55 | 4.72 | 18.67 | 30 | 11.54 | 2297 | 2.31 | 37.24 -87% |
| Sphere $\sigma$=0.3 | 30 | 106.91 | 20 | 17.60 | 20882 | 6.53 | 131.05 | 12.27 | 18.10 | 12 | 10.44 | 3635 | 1.22 | 42.03 -68% |
| Fandisk $\sigma$=0.7 | 50 | 257.13 | 20 | 10.61 | 12946 | 12.19 | 279.94 | 10.75 | 44.82 | 20 | 10.82 | 4464 | 4.34 | 70.73 -74% |

Table 5.1: Time performance for different models. $k_{iter}$ presents the number of iterations for bilateral filtering normal executed in $t_{k_{iter}}$, $u_{iter}$ is the number of iterations for vertex update executed in $t_{u_{iter}}$ and $N$ is the number of points searched for finding ideal patches executed in $t_N$, $t_{cd}$ corresponds to execution time for coarse denoising, feature extraction and level noise estimation, $t_{fn}$ represents the execution time for fine denoising and $t_{tot}$ is the total time expressed in seconds. The evaluation has been conducted by using the following models (V/F): Block (8771/17550), Twelve (4610/9216), Sphere (10443/20882), Fandisk (6475/12946).

The features of the mesh are successfully identified due to the application of the coarse-denoising step. By inspecting Fig. 5.14 we see that most of the state-of-the-art approaches mistakenly consider the staircase effect as geometric features that need to be preserved. The superiority of our method as compared to the presented state-of-the-art approaches is attributed to the application of the coarse step that smooths the staircase effects and allows more accurate identification of the true features. The heatmap visualization of the reconstruction error shows that the specialized method of [11] outperforms our approach, however, it should be noted that it creates irrational smoothness in the whole surface as well as a shrinking of the object, destroying also true features on the surface.

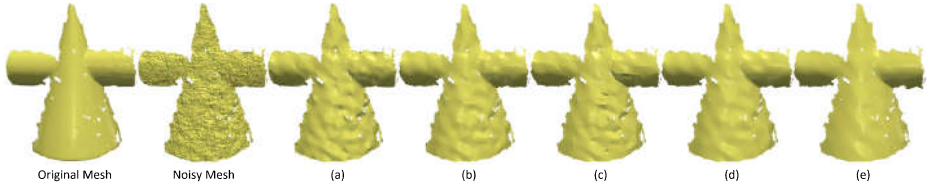

Original Mesh    Noisy Mesh    (a)    (b)    (c)    (d)    (e)

Figure 5.9: Denoising study by taking into account different approaches. The reconstruction accuracy is measured by the averaged angle $\theta$. (a) 12.98° Bilateral normal [7], (b) 10.38° Guided normals bilateral [9], (c) 10.34° $L_0$ min [8], (d) 8.91° Cascaded normal regression [10], (e) **7.98°** TSGSP. (Kinect v1-cone).

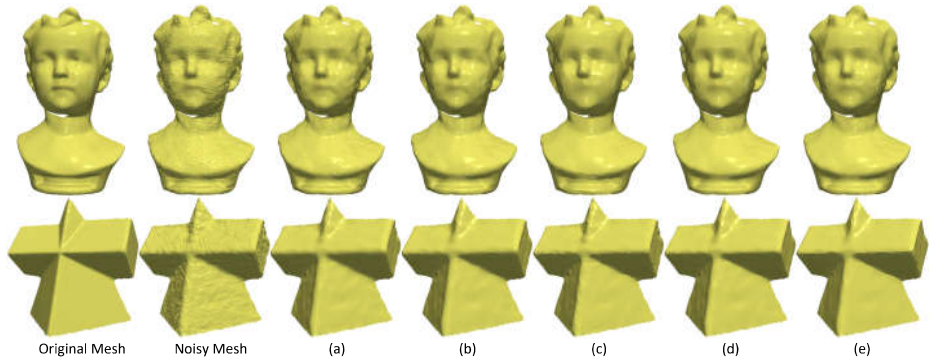

Figure 5.10: Denoising study by taking into account different approaches. The reconstruction accuracy is measured by the averaged angle $\theta$. (a) 9.39° Bilateral normal [7], (b) 7.90° Guided normals bilateral [9], (c) 7.65° $L_0$ min [8], (d) 7.79° Cascaded normal regression [10], (e) **7.59°** TSGSP. (Kinect fusion-boy), (a) 9.13° Bilateral normal [7], (b) 7.84° Guided normals bilateral [9], (c) 7.69° $L_0$ min [8], (d) 7.59° Cascaded normal regression [10], (e) **7.40°** TSGSP. (Kinect fusion-pyramid).

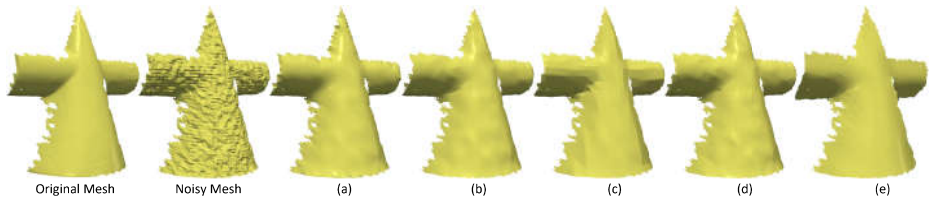

Figure 5.11: Denoising study by taking into account different approaches. The reconstruction accuracy is measured by the averaged angle $\theta$. (a) 8.50° Bilateral normal [7], (b) 7.62° Guided normals bilateral [9], (c) 7.74° $L_0$ min [8], (d) 7.68° Cascaded normal regression [10], (e) **7.41°** TSGSP. (Kinect v2-cone).

in Fig. 5.15, we present the original noisy mesh, the fine denoising output after using the k-nn patch directly without searching for the ideal area, the results after using Gaussian weights and finally the results after using our proposed approach.

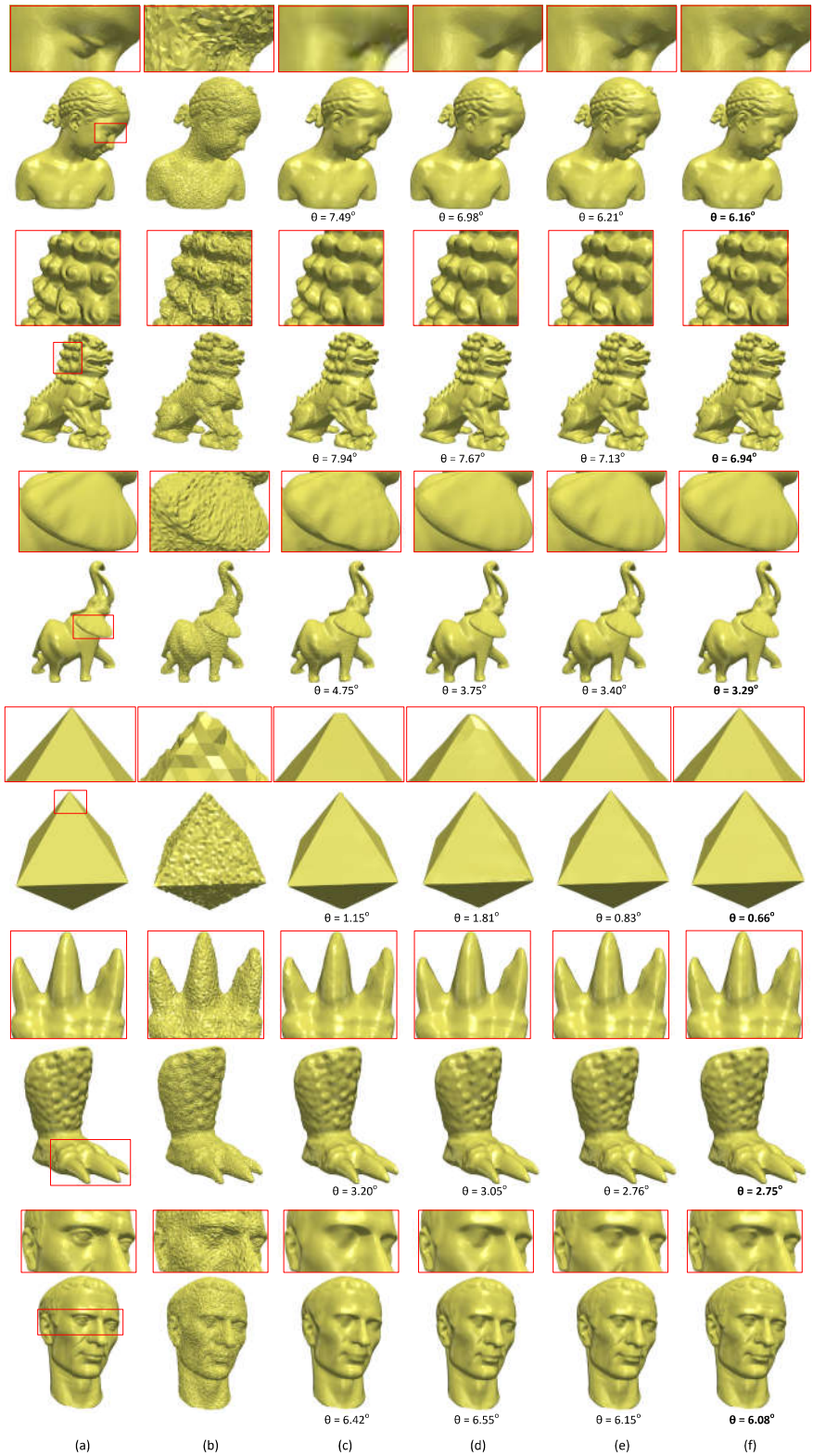

Figure 5.12: (a) Original 3D model, (b) Noisy model, and denoising results using: (c) Guided Normals Bilateral [9], (d) Cascaded Normal Regression [10], (e) parameter-free TSGSP, (f) TSGSP using ideal parameters.

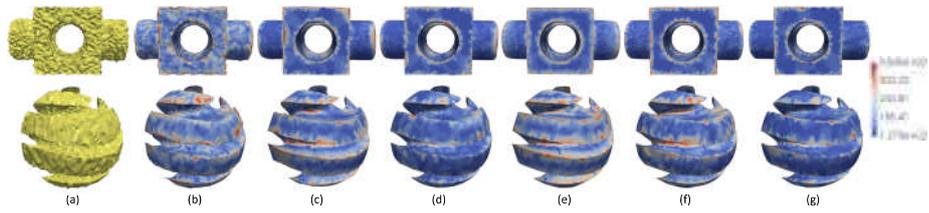

Figure 5.13: (a) Noisy models and Heatmap visualization of (b) Non Iterative [5], (c) Fast & Effective [6], (d) Bilateral Normal [7], (e) $L_0$ min [8], (f) Guided Normal Bilateral [9], (g) TSGSP.

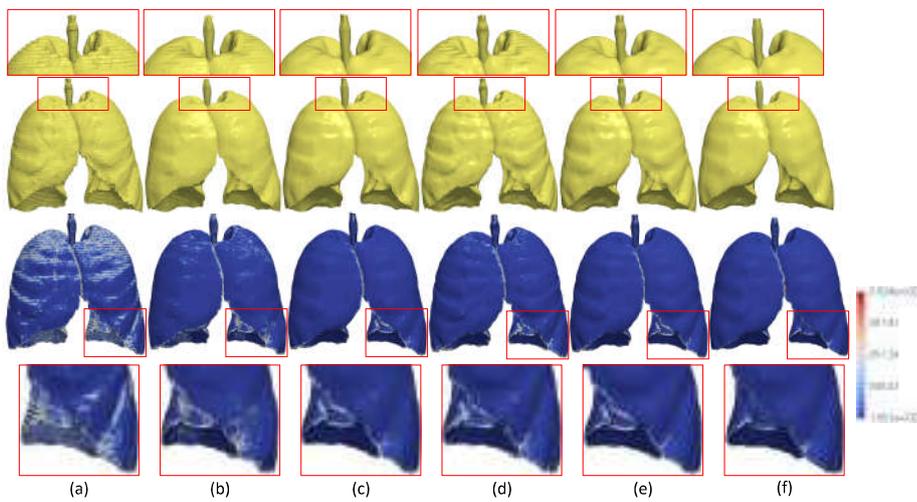

Figure 5.14: (a) Original scanned model affected by staircase effect, (b) bilateral [4], (c) fast and effective [6], (d) bilateral normal [7], (e) TSGSP, (f) Staircase-Aware Smoothing [11], (Lungs model).

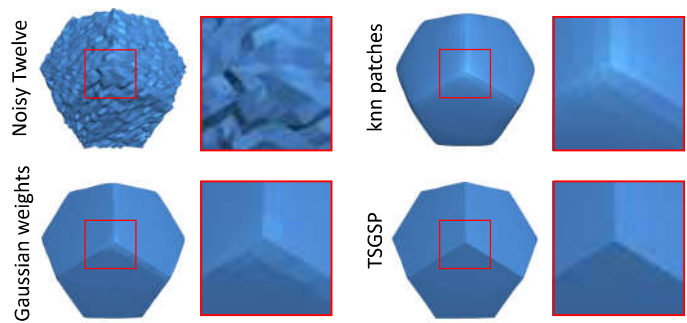

Figure 5.15: Results after using a wide range of alternative approaches for selecting the best neighborhoods and the corresponding adjacency weights.

**Data Driven Initialization Strategy**

One major advantage of data-driven methods [10] is their computational efficiency in online denoising cases. This attribute makes them capable of handling problems with tight timing restrictions in real-time. However, it should be noted that for the case of varying or unknown noise patterns that are not included in the training dataset re-training is required that is essentially a very slow and computationally costly process. More specifically, in the approach of [10] while the execution time is only 1 second, the training process, using a dataset of 63 models, takes more than 720 seconds.

We study the performance of the proposed scheme when using some fixed pre-identified initialization parameters. Those parameters have been estimated using a training set of known models and they were kept fixed for a testing set of 3D objects. This extension is based on the observation that noisy objects, created by the same scanner devices or affected by the same type of synthetic noise, require similar initialization strategies. For the experiments, we used the training set, affected by synthetic noise, provided by the benchmark of [10]. The learning process showed that besides the level of noise, geometric characteristics of the 3D object are also another factor that affects the values of the parameters. For this reason, we divided the training set into two new sets (fine geometric & CAD models). This separation comes from the observation that fine geometric models need fewer iterations and a higher value of $\sigma_s$ since they contain small-scale geometric features. On the other hand, the noise in CAD models is more easily distinguishable in flat areas stressing the need for more vertex update iterations with lower $\sigma_s$. The identified parameters are then used for evaluating the accuracy of the proposed scheme on two independent test sets consisting again by scanned and CAD models individually. The reconstruction accuracy is evaluated using the average angle difference between the normals of the original and the reconstructed object and the results are presented in Table 5.2. By inspecting this table it can be observed that the proposed scheme outperforms the method in [10] in 50% of the cases. Therefore, we conclude that our approach can be trained to work as a parameter-free method. However, a training set of objects classified in different categories (e.g., according to the geometric details or the scanning device) is required in order to identify a fixed set of parameters for each class of objects.

The training dataset which was used for the data-driven initialization strategy are illustrated in Figs. 5.16-(a) & 5.16-(b).

| Name of Model | Our Method | Cascaded Normal Regression [10] | Set of Parameters |
|---|---|---|---|
| **Block** | **2.3414°** | 2.3436° | Set 2 |
| **Bumpy torus** | **3.9688°** | 4.0464° | Set 1 |
| **Bunny hi** | 5.3268° | **5.1152°** | Set 1 |
| **Carter100K** | **6.8722°** | 7.8415° | Set 1 |
| **Child** | **6.2145°** | 6.9801° | Set 1 |
| **Chinese lion** | **7.1285°** | 7.6701° | Set 1 |
| **Cube** | **0.7747°** | 0.8656° | Set 2 |
| **Eight** | 6.0995° | **5.8017°** | Set 1 |
| **Eros100K** | **8.0866°** | 8.3486° | Set 1 |
| **Fertility** | 3.8456° | **3.6379°** | Set 1 |
| **Genus3** | **2.4576°** | 2.5751° | Set 1 |
| **Joint** | **1.6837°** | 1.697° | Set 2 |
| **Kitten** | 2.8386° | **2.8195°** | Set 1 |
| **Nicolo** | 4.5729° | **4.4868°** | Set 1 |
| **Part Lp** | **2.5046°** | 2.5422° | Set 2 |
| **Plane sphere** | 1.3816° | **1.2606°** | Set 2 |
| **Pulley** | 4.763° | **4.5899°** | Set 1 |
| **Pyramid** | **0.9446°** | 0.9912° | Set 2 |
| **Rolling stage** | 4.5187° | **4.1767°** | Set 1 |
| **Screwdriver** | 3.7645° | **2.9652°** | Set 1 |
| **Smooth feature** | **0.9847°** | 1.0085° | Set 2 |
| **Sphere** | 2.5285° | **2.3076°** | Set 2 |
| **Star** | **1.5895°** | 1.6502° | Set 2 |
| **Trim star** | 6.4181° | **4.1866°** | Set 1 |
| **Turbine Lp** | 3.7025° | **2.7707°** | Set 2 |

Table 5.2: Average angle $\theta$ using fixed pre-identified parameters on the different models included in the test set.

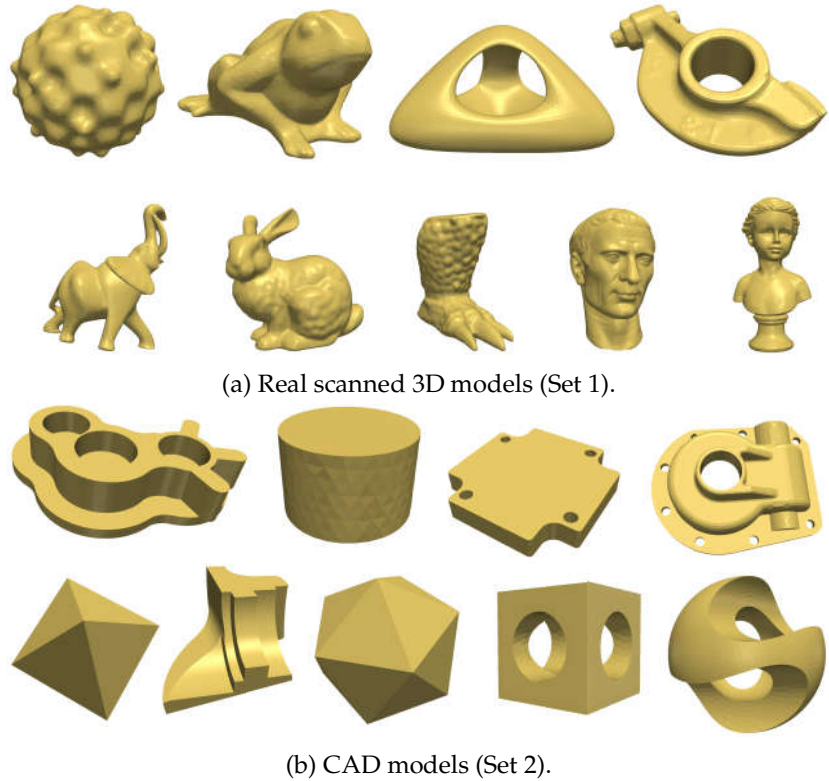

(a) Real scanned 3D models (Set 1).

(b) CAD models (Set 2).

Figure 5.16: For each dataset the pre-fixed estimated parameter are: (a) $\sigma_s = 1$, $\zeta = 3$, $u_{iter} = 5$, (b) $\sigma_s = 0.3$, $\zeta = 10$, $u_{iter} = 15$.

In Fig. 5.17, we present the denoised results of our method in comparison with the data-driven method described in [10].

**Time Varying Noise Effects**

The proposed scheme is capable of identifying both the level of noise and the spectral subspace where the features lie using a coarse step, while in the fine it identifies features and flat areas dealing efficiently with the aforementioned challenges and achieving at the same time, fast execution times. To investigate the performance of the proposed approach, as compared to the state-of-the-art methods, we provide a study, where different parts of a 3D model have been affected by different noise patterns. For this experiment, we used the dodecahedron model. We initially partitioned the 3D object into twelve different segments (corresponding to each side) and then we apply twelve different noise patterns generated assuming variable noise levels in each segment. The results of the proposed coarse to fine approach together with the other state-of-the-art approaches are presented in Fig. 5.18. The superiority of our method is more obvious when a higher level of noise appears in each area. In this case, all the other methods fail to mitigate the noise effects, creating deformations and destroying geometric features.

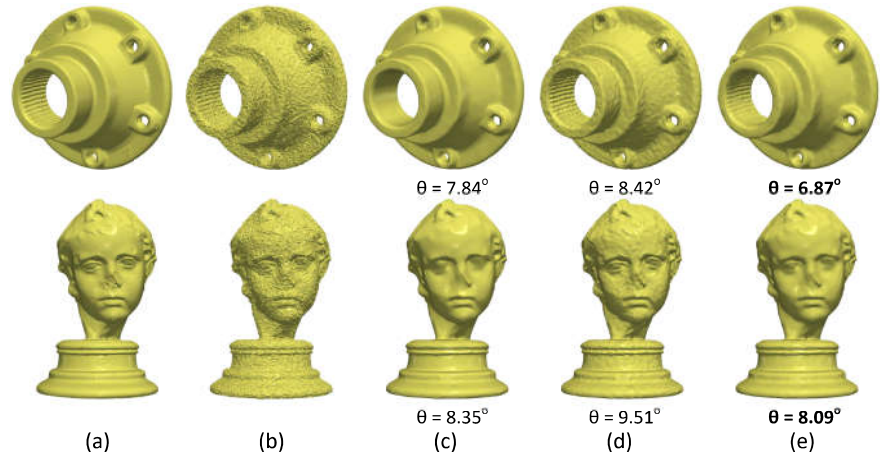

Figure 5.17: (a) Original, (b) Noisy Object, (c) Cascaded Normal Regression [10], (d) Coarse step and (e) Coarse & Fine step. (Carter100K & Eros100K models).

More specifically, Fig. 5.18 [First Row] presents the sides of Twelve model affected by lower levels of noise. In this case, the results of some methods [8], [9] and TSGSP have similarities and it seems that they effectively manage to handle the issue of varying noise patterns. The superiority of our method is more obvious in Fig. 5.18 [Second Row] where a higher level of noise appears in each area. In this case, all the other methods fail to mitigate the noise effects, creating deformations and destroying geometric features.

Additionally, we provide tables 5.3-5.4, which present the comparison results using as evaluation metrics both the HD and the angle $\theta$.

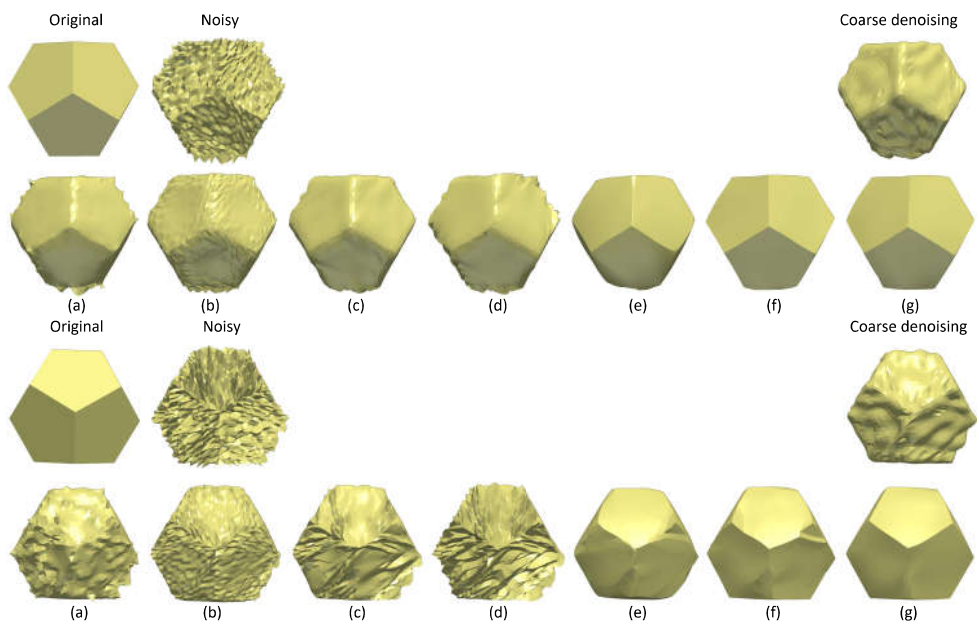

Figure 5.18: Denosing results of Twelve model which have been affected by different noise patterns in each of its twelve sides. (a) Bilateral [4], (b) Non Iterative [5], (c) Fast & Effective [6], (d) Bilateral Normal [7], (e) $L_0$ min [8], (f) Guided Normal Bilateral [9], (g) TSGSP. [First Row] Sides affected by lower levels of noise. [Second Row] Sides affected by higher levels of noise.

| Name of Model | Guided Mesh Normal | Cascade I. [10] | TSGSP using Grouped Parameters | TSGSP using Ideal Parameters |
|---|---|---|---|---|
| **Child** | $\theta = 7.4941^o$<br>HD = 0.0039 | $\theta = 6.9801^o$<br>HD = 0.0043 | $\theta = 6.2145^o$<br>**HD = 0.0038** | $\boldsymbol{\theta = 6.1615^o}$<br>**HD = 0.0038** |
| **Chinese lion** | $\theta = 7.9365^o$<br>HD = 0.5559 | $\theta = 7.6701^o$<br>HD = 0.5143 | $\theta = 7.1285^o$<br>HD = 0.4916 | $\boldsymbol{\theta = 6.939^o}$<br>**HD = 0.4893** |
| **Elephant** | $\theta = 4.7512^o$<br>HD = 0.0036 | $\theta = 3.751^o$<br>HD = 0.0028 | $\theta = 3.4015^o$<br>**HD = 0.0025** | $\boldsymbol{\theta = 3.288^o}$<br>**HD = 0.0025** |
| **Octahedron** | $\theta = 1.1468^o$<br>HD = 0.047 | $\theta = 1.8062^o$<br>HD = 0.0459 | $\theta = 0.8323^o$<br>HD = 0.0196 | $\boldsymbol{\theta = 0.6566^o}$<br>**HD = 0.0115** |
| **Leg** | $\theta = 3.1952^o$<br>HD = 0.1615 | $\theta = 3.053^o$<br>HD = 0.1774 | $\theta = 2.7617^o$<br>HD = 0.1547 | $\boldsymbol{\theta = 2.7588^o}$<br>**HD = 0.1523** |
| **Julius** | $\theta = 6.4266^o$<br>**HD = 0.0032** | $\theta = 6.5546^o$<br>HD = 0.0039 | $\theta = 6.155^o$<br>HD = 0.0086 | $\boldsymbol{\theta = 6.0766^o}$<br>HD = 0.0086 |

Table 5.3: Extended evaluation of results, which had been presented in Fig. I.14, using as metrics both the angle $\theta$ and the HD.

| Name of Model | Bilateral [4] | Non iterative [5] | Fast & Effective [6] | Bilateral Normal (l) [7] | Bilateral Normal (g) [7] | $L_0$ min [8] | Guided Normal [9] | TSGSP |
|---|---|---|---|---|---|---|---|---|
| block | $\theta = 12.7155^o$<br>HD = 0.8781 | $\theta = 13.8501^o$<br>HD = 0.653 | $\theta = 5.8023^o$<br>HD = 0.8206 | $\theta = 5.3062^o$<br>HD = 0.7257 | $\theta = 8.0165^o$<br>HD = 0.7009 | $\theta = 4.9734^o$<br>HD = 0.6698 | $\theta = 3.572^o$<br>HD = 0.71 | $\boldsymbol{\theta = 3.3026^o}$<br>**HD = 0.5869** |
| fandisk | $\theta = 22.4862^o$<br>HD = 0.2249 | $\theta = 27.9264^o$<br>HD = 0.2093 | $\theta = 13.1918^o$<br>HD = 0.1817 | $\theta = 14.2553^o$<br>HD = 0.1864 | $\theta = 15.0545^o$<br>HD = 0.1892 | $\theta = 6.2186^o$<br>HD = 0.124 | $\theta = 6.3721^o$<br>HD = 0.1388 | $\boldsymbol{\theta = 6.0669^o}$<br>**HD = 0.1208** |
| sphere | $\theta = 12.5796^o$<br>HD = 0.8106 | $\theta = 17.3618^o$<br>HD = 0.477 | $\theta = 11.892^o$<br>HD = 0.5514 | $\theta = 6.7047^o$<br>**HD = 0.3398** | $\theta = 9.6974^o$<br>HD = 0.471 | $\theta = 12.9566^o$<br>HD = 0.536 | $\theta = 10.1697^o$<br>HD = 0.4607 | $\boldsymbol{\theta = 6.2407^o}$<br>HD = 0.3917 |
| twelve | $\theta = 11.7204^o$<br>HD = 0.1357 | $\theta = 11.093^o$<br>HD = 0.1074 | $\theta = 7.4519^o$<br>HD = 0.0728 | $\theta = 7.3683^o$<br>HD = 0.0858 | $\theta = 7.271^o$<br>HD = 0.0717 | $\theta = 8.4626^o$<br>HD = 0.1301 | $\theta = 2.7542^o$<br>HD = 0.11 | $\boldsymbol{\theta = 2.6543^o}$<br>**HD = 0.0695** |

Table 5.4: Extended evaluation of results in comparison with a variety of different state-of-the-art methods using as metrics both the angle $\theta$ and the HD.

Table 5.5 presents a variety of reconstruction quality metrics for several denoising methods. By observing the quality metrics, we can verify that our method provides the best results in almost every case study.

We also present the denoising results of two real-scanned noisy 3D models (i.e., cup and wallet in Figure 5.19). Our method removes the abnormalities without over smoothing the surface of the object, preserving at the same time the high-frequency features of the objects. However, the evaluation of our method, in this case, is not feasible since the ground truth model is not known beforehand.

Table 5.5: Evaluation of the experimental results using different metrics. The lowest value per each row is highlighted in bold.

| | Metrics | Bilateral [4] | Non-Iterative [5] | Fast & Effective [6] | Bilateral (l) [7] | Bilateral (g) [7] | l0 min [8] | Guided Normal Filtering [9] | Our Approach |
|---|---|---|---|---|---|---|---|---|---|
| **Twelve (0.5)** | $\theta$ | 11.720 | 11.093 | 7.452 | 7.368 | 7.271 | 8.462 | 2.754 | **2.668** |
| | Dmean d | 0.017 | 0.0155 | 0.0115 | 0.0129 | 0.0123 | 0.0317 | 0.0128 | **0.006** |
| | Dmax d | 0.1357 | 0.1074 | 0.0728 | 0.0947 | **0.0741** | 0.1357 | 0.1594 | 0.0995 |
| | dist n | 0.1434 | 0.1301 | 0.1051 | 0.1055 | 0.1073 | 0.1518 | 0.0809 | **0.0645** |
| | NMSVE ($\times 10^{-5}$) | 6.98 | 5.4 | 4.44 | 4.26 | 4.84 | 5.43 | 5.32 | **3.55** |
| | Dmean n | 0.2113 | 0.2002 | 0.1349 | 0.1322 | 0.1331 | 0.1627 | 0.0519 | **0.0465** |
| **Block (0.4)** | $\theta$ | 12.715 | 13.850 | 5.802 | 8.016 | 5.306 | 4.973 | 3.572 | **2.9826** |
| | Dmean d | 0.1873 | 0.1425 | 0.0857 | 0.096 | 0.0744 | 0.1922 | 0.1066 | **0.0544** |
| | Dmax d | 0.8781 | 0.8684 | 0.8206 | 0.7009 | 0.8759 | 0.6836 | 0.9967 | **0.6479** |
| | dist n | 0.236 | 0.2179 | 0.1536 | 0.17 | 0.1462 | 0.1911 | 0.1443 | **0.1064** |
| | NMSVE ($\times 10^{-5}$) | 3.26 | 3.47 | 2.14 | 2.28 | 2.1 | 2.88 | 2.68 | **1.5** |
| | Dmean n | 0.3134 | 0.3106 | 0.1422 | 0.1866 | 0.128 | 0.1808 | 0.1131 | **0.0788** |
| **Fandisk (0.7)** | $\theta$ | 22.486 | 27.926 | 13.192 | 15.054 | 14.255 | 6.219 | 6.372 | **6.070** |
| | Dmean d | 0.0376 | 0.0366 | 0.0276 | 0.0321 | 0.0294 | 0.0293 | 0.0291 | **0.0155** |
| | Dmax d | 0.2412 | 0.2093 | 0.2048 | 0.188 | 0.1987 | 0.1373 | 0.1865 | **0.1207** |
| | dist n | 0.5803 | 0.6209 | 0.5447 | 0.5739 | 0.5739 | 0.4529 | 0.4495 | **0.4065** |
| | NMSVE ($\times 10^{-5}$) | 7.49 | 9.68 | 4.89 | 4.79 | 5.33 | 3.5 | 5.28 | **3.01** |
| | Dmean n | 0.403 | 0.4937 | 0.2353 | 0.2713 | 0.2535 | 0.1221 | 0.119 | **0.1104** |

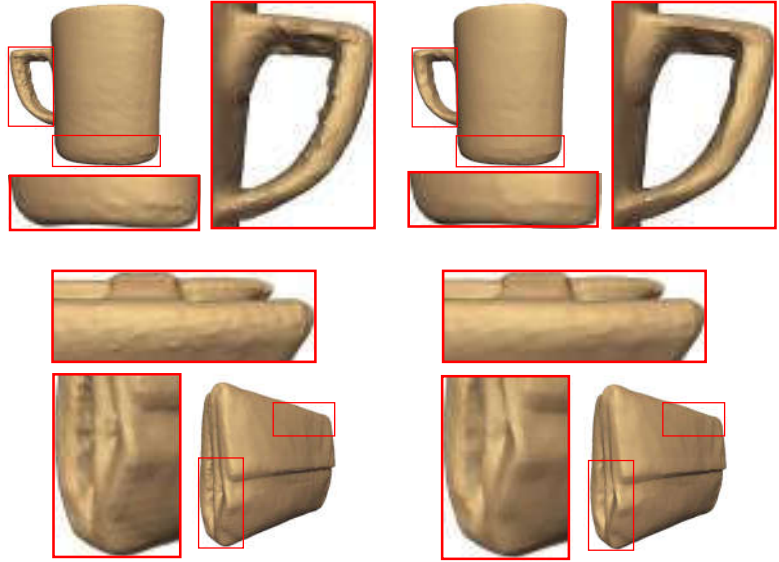

Figure 5.19: (**Left**) Three-dimensional scanned cup and wallet with abnormalities. (**Right**) Denoising results in respect of features.

### 5.1.2 Denoising of Dynamic 3D Meshes via Low-Rank Spectral Analysis

3D mesh denoising techniques could also be used for the denoising of dynamic 3D meshes, considering each frame of the sequence as a different mesh. Nevertheless, in this case, these methods do not take advantage of the temporal coherence between the consecutive meshes and they do not exploit the lower bpvf rates that methods applied directly to the dynamic 3D meshes can provide.

We followed this line of thought and we developed a simple and robust method for the denoising of a sequence of consecutive noisy frames. More specifically, the proposed approach avoids the use of complex computational solutions or focusing only on the geometric information of restricted local areas. Because of the large variety of different types of noise, we assume that a successful method must be robust and effective under any of them. In other words, we introduce a novel method which exploits similarities at the spectral frequencies of individual meshes in soft or rigid body 3D animations. The noise is mainly distributed in the high-frequencies subspace and for this reason, we use a robust principal component analysis approach in order to remove it without affecting the original components. In Fig. 5.20, the framework of the proposed approach is presented, highlighting the most important steps.

The proposed method is based on the observation that the frames of the same 3D dynamic sequence have a similar form of GFT regarding the high-frequencies components. On the other hand, the low-frequencies components, representing the main shape of a 3D object, have a different form since the shape of the moving object is changing from frame to frame. To be more specific, when a 3D object is moving its shape inevitably changes, result in the values of the low-frequencies components, while the high-frequencies components of GFT remain unchangeable. Fig. 5.21 presents an example of four different frames of the same dynamic mesh sequence and their corresponding spectral components. As we can see, only a few components (representing the low-frequencies components) take significantly different values while the rest values remain almost the same. The right values of the GFT represent the low-frequencies components (i.e. large-scale spatial features) while the left values represent the high-frequencies components (i.e. small-scale spatial features) (see Fig. 5.21-(a)). The more into the right a component is located, the more valuable is assumed for the representation of the 3D object's shape.

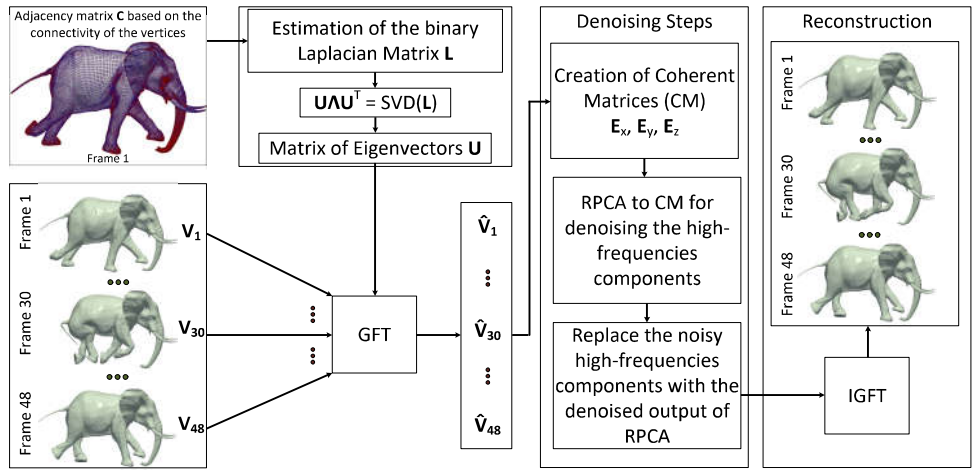

Figure 5.20: Steps of the proposed framework for denoising of dynamic 3D meshes via spectral low-rank matrix analysis.

The removal of the rightmost low-frequencies components is equal to losing any spatial information of the mesh. These components are important for the proper representation of a mesh since they are related to the basic shape. On the other hand, if the high-frequencies components are removed is not crucial and no significant changes will be observed, depending of course on the amount of the components which will be removed. In Fig. 5.22, we present an example in which different amounts of high-frequencies components have been removed from a mesh. Specifically, in Fig. 5.22-(e), we can see that even if 90% of the high-frequencies components are removed, the basic shape of the 3D object still remains and it is well recognizable.

To notice here that we use the binary Laplacian form because it is universal and exactly the same for any frame of the animated sequence, so the connectivity is estimated once since it remains the same among the frames, in contrast with the position of vertices which changes. On the other hand, the weighted adjacency matrices are different but using the binary adjacency matrix we avoid the estimation for a weighted adjacency matrix for each frame separately. Despite the fact that some previous works effectively remove the noise of a GFT representation, they also inevitably remove high-frequencies components representing small-scale features. In this work, we suggest not to remove these components but try to denoise them in order to preserve the small-scale features.

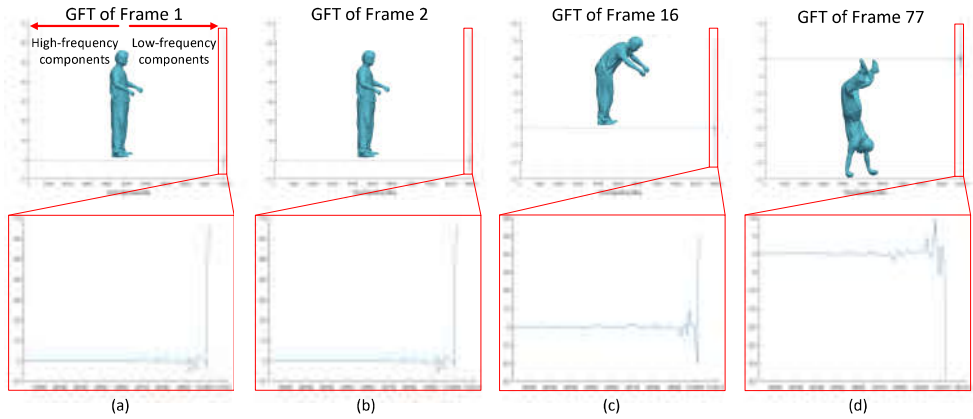

Figure 5.21: GFTs ($\hat{\mathbf{V}}_x \in \mathbb{R}^{k \times 1}$) of different frames of the same dynamic 3D mesh (Handstand model). We can easily observe that most of the components have similar magnitude values while very few components (low-frequencies components), representing the information related to shape of the object, have significant different values.

**Creating the Coherent Matrices**

The main objective is the accurate estimation, of the denoised GFT for each frame separately, in order to use it for the reconstruction of the denoised vertices by applying the IGFT according to Eq. (4.5). The whole process of the reconstruction can take place either in online mode (frame by frame) or in a parallel mode (e.g., groups of frames) increasing, even more, the denoising time processing. The matrix $\mathbf{U}$ has been estimated once, according to Eq. (4.3), but it is used for the estimation of the $n_s$ GFTs $\hat{\mathbf{V}}_j = [\ \hat{\mathbf{v}}_{1j};\ \hat{\mathbf{v}}_{2j};\ \cdots;\ \hat{\mathbf{v}}_{nj}\ ]$ of any other frame $\forall\ j = 1, \cdots, n_s$, according to Eq. (4.4), where $\hat{\mathbf{v}}_{ij} = [\hat{v}_{x_{ij}}\ \hat{v}_{y_{ij}}\ \hat{v}_{z_{ij}}]\ \forall\ i = 1, \cdots, n$. When the GFTs matrices have been estimated, we create 3 coherent matrices $\mathbf{M}_x, \mathbf{M}_y, \mathbf{M}_z$ where $\mathbf{M}_i \in \mathbb{R}^{n_s \times n} \; \forall \; i \in \{x,y,z\}$, according to:

$$\mathbf{M}_i = \begin{bmatrix} \hat{v}_{i_{11}} & \hat{v}_{i_{21}} & \cdots & \hat{v}_{i_{n_h 1}} \\ \hat{v}_{i_{12}} & \hat{v}_{i_{22}} & \cdots & \hat{v}_{i_{n_h 2}} \\ \vdots & \vdots & \ddots & \vdots \\ \hat{v}_{i_{1 n_s}} & \hat{v}_{i_{2 n_s}} & \cdots & \hat{v}_{i_{n_h n_s}} \end{bmatrix}, \; \forall \; i \in \{x,y,z\} \qquad (5.24)$$

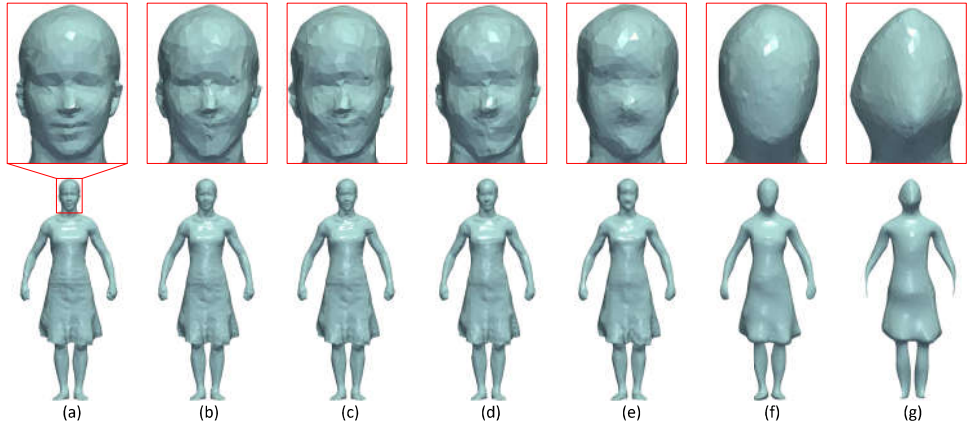

Figure 5.22: (a) Original mesh having 10002 components in total, and reconstructed models while components of the GFT have been removed: (b) 6000 components have been removed, (c) 7000 components have been removed, (d) 8000 components have been removed, (e) 9000 components have been removed, (f) 9800 components have been removed, (g) 9900 components have been removed. (Samba model frame 1).

At this point, it is worth mentioning that we do not use all the $n$ components of the GFT for the creation of the coherent matrices. The $n_l$ low-frequencies components do not have any coherence each other, among the frames, that we could take advantage of since they represent the pose/shape of the 3D object and while the object is moving, changing its pose/shape, they are also changed.

**Estimation of the Ideal Number of the Remaining Low-frequencies Components**

In this paragraph, we present the followed steps and the main assumptions related to the process of searching for the ideal number of low-frequencies components that must stay unchangeable. The main reasons why we exclude a number of low-frequencies components from taking part in the creation of the coherent matrices are:

- The low-frequencies components represent the basic information of the 3D object's shape and they must not be changed otherwise the original form of the 3D object is deformed or definitely destroyed.

- Additionally, these components do not have any coherence each other, from frame to frame, because of the different poses of the 3D object in each frame. So they can not be used for the creation of the coherent matrices.

- Finally, it has been observed that the distribution of noise, with respect to the GFT domain, mostly affect the high-frequencies components. As a result, the low-frequencies components do not need to be denoised since their denoised contribution would not be noticeable to the final reconstructed results.

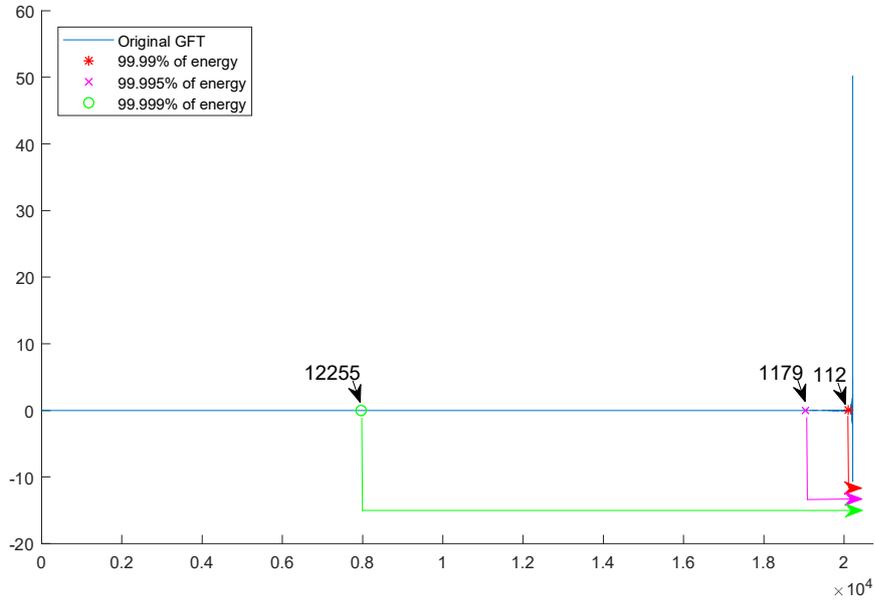

Figure 5.23: Energy component (Dinosaur model 20218 vertices).

For the estimation of the ideal value of $n_h$, which represents the number of the high-frequencies components per frame that we use for the creation of the coherent matrix, we follow the next steps. Firstly, we estimate the total energy $E_s$ of the GFT per each frame, taking into account all the $n$ components, based on the next formula:

$$E_{s_j} = \sum_{i=1}^{n} \|\hat{v}_i\|^2, \quad \forall\, j = 1, \cdots, n_s \qquad (5.25)$$

Then, we assume that the 99.99% of this energy must be kept unchangeable (low-frequencies). So we start adding high-frequencies components, from the component 1 to $n_{hj}$, until the cumulative energy of these components is equal to the 0.01% of the total energy $E_{s_j}$:

$$\sum_{i=1}^{n_{hj}} \|\hat{v}_i\|^2 = 10^{-4} E_{s_j}, \quad \forall\, j = 1, \cdots, n_s \tag{5.26}$$

where $n_{hj} < n$. The value of $n_{hj}$ may differ from frame to frame, however, each row of the **E** matrix must have an equal length. For this reason, the selected value of $n_h$ is defined as:

$$n_h = \max(n_{h1}, \cdots, n_{hn_s}) \tag{5.27}$$

In Fig. 5.23, we present an example showing how many components are required for preserving: (i) the 99.99% of the GFT's energy (112 lower frequencies components), (ii) the 99.995% of the GFT's energy (1179 lower frequencies components) and (iii) the 99.999% of the GFT's energy for a frame of the Dinosaur model (12255 lower frequencies components). In Fig. 5.24, we present an exam-ple showing how the number of the unchangeable low-frequencies components affect the quality of the denoising results. For the evaluation, we use different energy remaining approaches, in a range of [99.5% − 99.999%]. A big value of remaining energy (yellow line) means that we keep unchangeable a lot of noisy components while a small value (red line) means that we may change components that represent the main shape of the mesh and they must stay as-is.

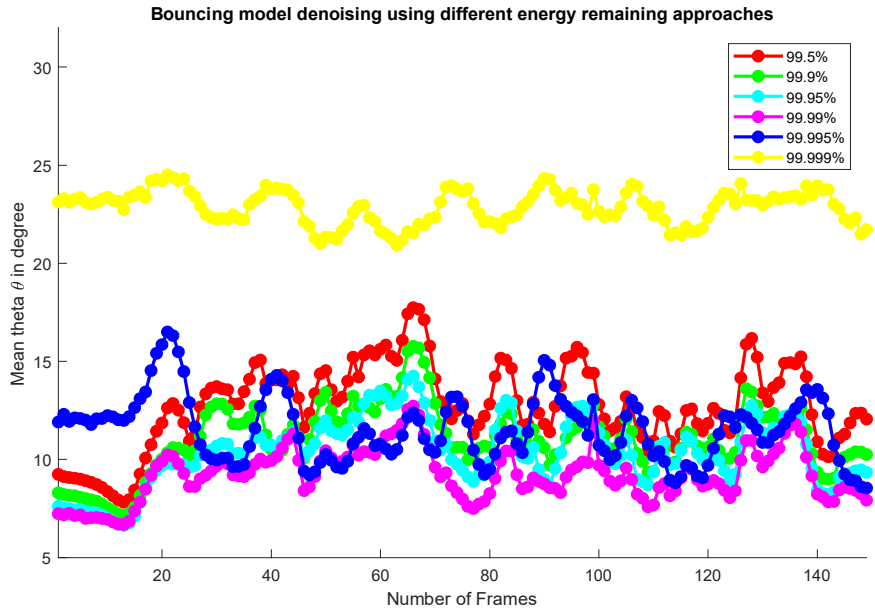

Figure 5.24: Bouncing model denoising using different energy remaining approaches.

**Properties of Robust Principal Component Analysis**

Once the coherent matrices $\mathbf{M}_i \ \forall \ i \ \in \ \{x, y, z\}$ have been created then we use RPCA in order to estimate the low-rank representation of these matrices, representing the denoising results. Generally, RPCA has been used in many applications in the area of 3D meshes processing mostly for outliers removal of unorganized point clouds [328]. A big advantage that RPCA could provide is that it can decompose a coherent matrix into a low-rank matrix $\mathbf{E}$ and a sparse matrix $\mathbf{S}$ representing the abnormalities that appear in the coherent matrix.

The motivation for using RPCA is based on the observation that GFT's values of the high-frequencies components, as appeared in different frames of the same dynamic 3D mesh, have a big coherence with respect to their form and the magnitude of their values. RPCA is specialized in finding the low-rank matrix of data providing excellent results in particular if these data have a good coherence with each other (i.e., very relevant data). Additionally, despite the fact that the noise has a unified distribution in the spatial domain (all vertices of the 3D surface can be affected equally), we observed that the noise follows a different distribution at the GFT domain in which mostly the high-frequencies components are affected. In Fig. 5.25, we present an example showing the percentage difference for each component comparing a noisy and the corresponding original mesh (Samba frame 58).

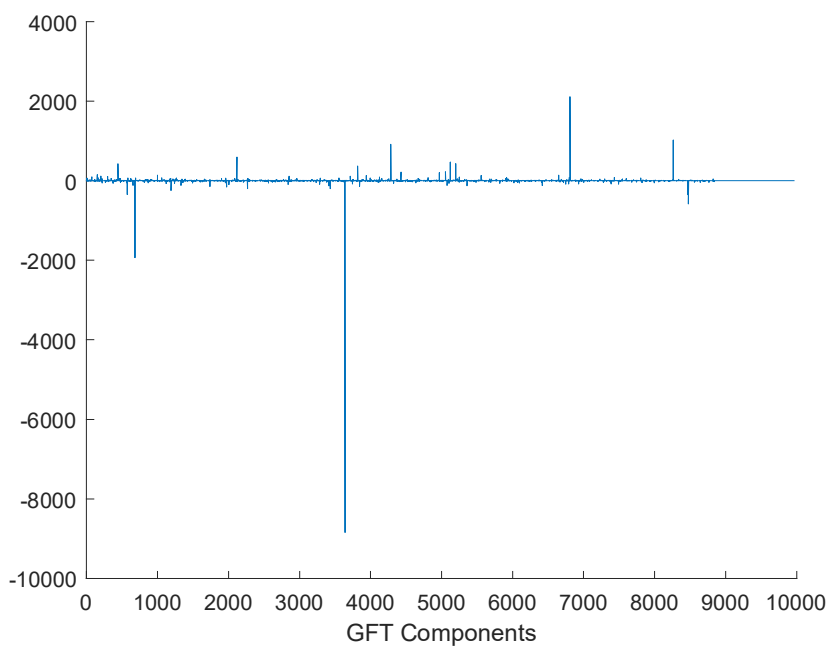

Figure 5.25: The percentage difference of each component comparing a noisy and the corresponding original mesh has a sparse representation.

**Estimation of the Low-rank Matrix Using an RPCA Approach**

For the sake of simplicity, we present here the denoising process using the coherent matrix $\mathbf{M}$, representing however all the cases of Eq. (5.66). This means that the following steps are applied three times for each one of the matrices $\mathbf{M}_x, \mathbf{M}_y, \mathbf{M}_z$. Our analysis based on the assumption that the matrix $\mathbf{M}$, which consists of the $n_h$ noisy high-frequencies components of the $n_s$ GFTs $\hat{\mathbf{V}}$ of the dynamic mesh, may be decomposed as:

$$\mathbf{M} = \mathbf{E} + \mathbf{S} \tag{5.28}$$

where $\mathbf{E}$ is a low-rank matrix representing the real (denoised) data while $\mathbf{S}$ is a sparse matrix representing the space where the outliers lie.

For this kind of problem, Zhou and Tao [329] proposed that the low-rank matrix can be modeled in a bilateral factorization form $\mathbf{WT}$ to develop an SVD free algorithm. We follow the same line of thought, by replacing $\mathbf{E}$ with its bilateral factorization:

$$\mathbf{E} = \mathbf{WT} \tag{5.29}$$

and regularizing the $l_1$ norm of the entries of the sparse matrix $\mathbf{S}$:

$$\begin{aligned} \min_{\mathbf{W},\mathbf{T},\mathbf{S}} \|\mathbf{M} - \mathbf{WT} - \mathbf{S}\|_F^2 + \lambda \|\text{vec}(\mathbf{S})\|_1 \\ \text{subject to: rank}(\mathbf{W}) = \text{rank}(\mathbf{T}) \leq r \end{aligned} \tag{5.30}$$

The $l_1$ regularization induces soft-thresholding in updating $\mathbf{S}$, which is faster than sorting, caused by cardinality constraint, as also suggested in [330] [331]. Optimizing the matrices $\mathbf{W}, \mathbf{T}$ and $\mathbf{S}$ in Eq. (5.30), we take the following updating rules:

$$\begin{aligned} \mathbf{W}^{(t+1)} &= (\mathbf{M} - \mathbf{S}^{(t)})\mathbf{T}^{(t)T}(\mathbf{T}^{(t)}\mathbf{T}^{(t)T})^+ \\ \mathbf{T}^{(t+1)} &= (\mathbf{W}^{(t+1)T}\mathbf{W}^{(t+1)})^+ \mathbf{W}^{(t+1)T}(\mathbf{M} - \mathbf{S}^{(t)}) \\ \mathbf{S}^{(t+1)} &= \mathcal{D}_\lambda(\mathbf{M} - \mathbf{W}^{(t+1)}\mathbf{T}^{(t+1)}) \end{aligned} \tag{5.31}$$

where $(t)$ denotes the $t^{th}$ iteration, $(.)^+$ is the Moore-Penrose pseudo-inverse and $\mathcal{D}_\lambda$ is an element-wise soft thresholding operator with threshold $\lambda$ such that:

$$\mathcal{D}_\lambda = \{\text{sign}(\mathbf{M}_{ij})\max(|\mathbf{M}_{ij}| - \lambda, 0) : (i,j) \in [n_s] \times [n_h]\} \tag{5.32}$$

With purpose the simplification of the updating rules in Eq. (5.31), we can say that the product $\mathbf{W}^{(t+1)}\mathbf{T}^{(t+1)}$ equals to the orthogonal projection of $\mathbf{E}$ onto the column space of $\mathbf{W}^{(t+1)}$. According to Eqs. (5.31), the column space of $\mathbf{W}^{(t+1)}$ can be represented by arbitrary orthonormal basis for the columns of $(\mathbf{M} - \mathbf{N}^{(t)})\mathbf{T}^{(t)T}$. It can be computed as $\mathbf{Q}$ via fast qr(.) decomposition:

$$\mathbf{QR} = \text{qr}((\mathbf{M} - \mathbf{S}^{(t)})\mathbf{T}^{(t)T}) \tag{5.33}$$

Then, the product $\mathbf{W}^{(t+1)}\mathbf{T}^{(t+1)}$ can be equivalently computed as:

$$\mathbf{W}^{(t+1)}\mathbf{T}^{(t+1)} = \mathbf{Q}\mathbf{Q}^T(\mathbf{M} - \mathbf{S}^{(t)}) \tag{5.34}$$

According to the above analysis, we can observe that the matrices $\mathbf{W}^{(t+1)}$ and $\mathbf{T}^{(t+1)}$ in Eqs. (5.31) can be replaced by $\mathbf{Q}$ and $\mathbf{Q}^T(\mathbf{M} - \mathbf{S}^{(t)})$ respectively, while the product $\mathbf{W}^{(t+1)}\mathbf{T}^{(t+1)}$ is kept the same. These replacements change the Eqs. (5.31), providing a faster updating procedure:

$$\begin{aligned}
\mathbf{W}^{(t+1)} &= \mathbf{Q}, \ \mathbf{QR} = \text{qr}((\mathbf{M} - \mathbf{S}^{(t)})\mathbf{T}^{(t)T}) \\
\mathbf{T}^{(t+1)} &= \mathbf{Q}^T(\mathbf{M} - \mathbf{S}^{(t)}) \\
\mathbf{S}^{(t+1)} &= \mathcal{D}_\lambda(\mathbf{M} - \mathbf{W}^{(t+1)}\mathbf{T}^{(t+1)})
\end{aligned} \tag{5.35}$$

**Reconstruction of the Model Estimating the Denoised Vertices**

Finally, the low-rank matrix $\mathbf{E} \in \mathbb{R}^{n_s \times n_h}$, constituting the denoised values, is estimated via the Eq. (5.29). More specifically, each $i$ row of this matrix consists of the $n_h$ high-frequencies denoised components of the GFT values of the $i$ frame. The total denoised GFT matrix $\mathring{\mathbf{V}}_i$ of each $i$ frame is estimated according to:

$$\mathring{\mathbf{V}}_i = [\ \overbrace{\mathbf{E}_i[1:n_{hi}]}^{n_{hi}\text{ components}}\ \overbrace{\hat{\mathbf{V}}_i[(n - n_{hi} + 1):n]}^{(n-n_{hi})\text{ components}}], \quad \forall\, i = 1, \cdots, n_s \tag{5.36}$$

keeping the $n_{hi} \leq n_h$ denoised components of the $i^{th}$ row of the matrix $\mathbf{E}$, following the assumption presented in subsection 5.1.2. Then, the final denoised vertices $\dot{\mathbf{V}}_i$ of each $i$ frame are estimated using the IGFT, applied to the denoised GFT matrix $\mathring{\mathbf{V}}_i$:

$$\dot{\mathbf{V}}_i = \mathcal{T}^{-1}(\mathring{\mathbf{V}}_i), \quad \forall\, i = 1, \cdots, n_s \tag{5.37}$$

All the aforementioned steps of the proposed approach are summarized at the Algorithm 3.

**Experimental Results and Discussion**

In this paragraph, we present the results of our approach using a variety of dynamic 3D models affected by different types of noise. The quality performance of the proposed technique is evaluated by comparing its denoising results with those of other well-known and robust techniques of the literature, such as: (i) bilateral normal denoising [7], (ii) guided mesh normal filtering [9] and (iii) two stage graph spectral processing [12].

For the experiments, we chose a variety of different models and types of noise. In table 5.6, we present information related to the noisy models that are

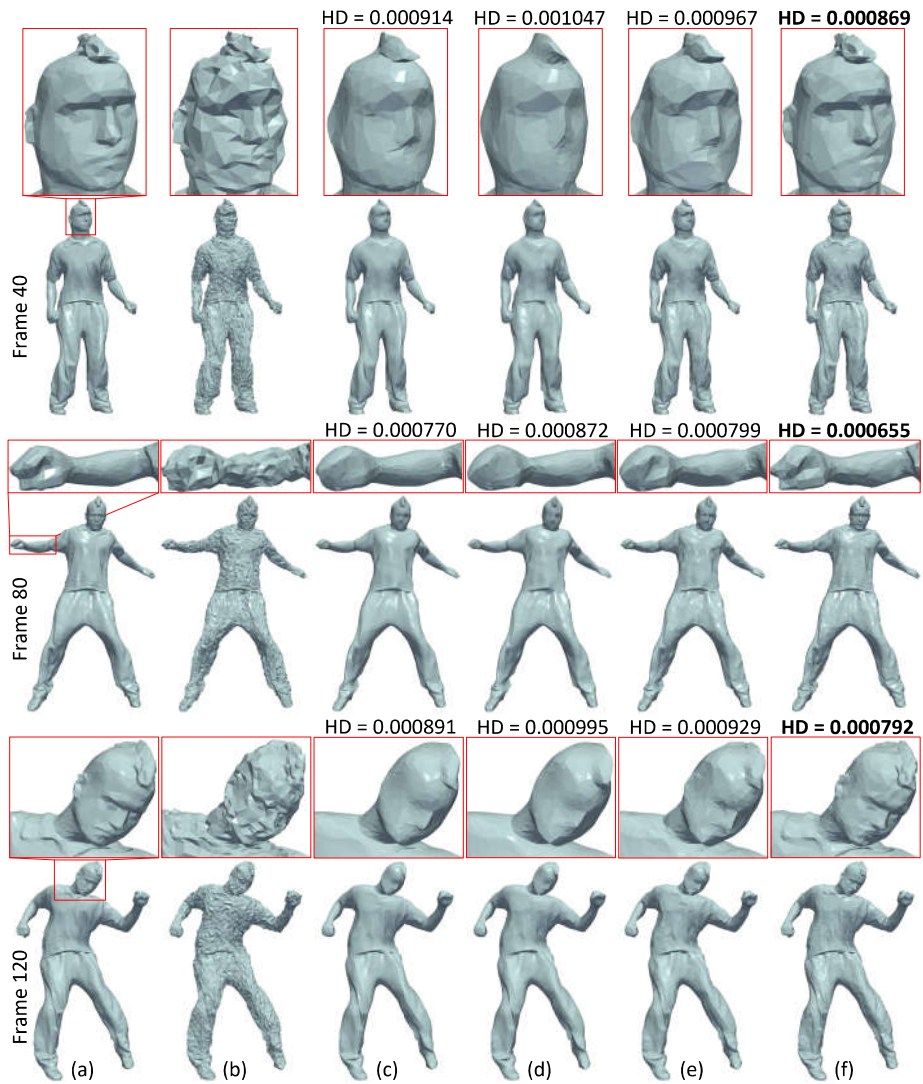

Figure 5.26: Three different frames (i.e., 40, 80 and 120) of Bouncing model. (a) Original mesh, (b) Noisy mesh, (c) Bilateral normal denoising [7], (d) Guided mesh normal filtering [9], (e) Two stage graph spectral processing [12], (f) Our approach.

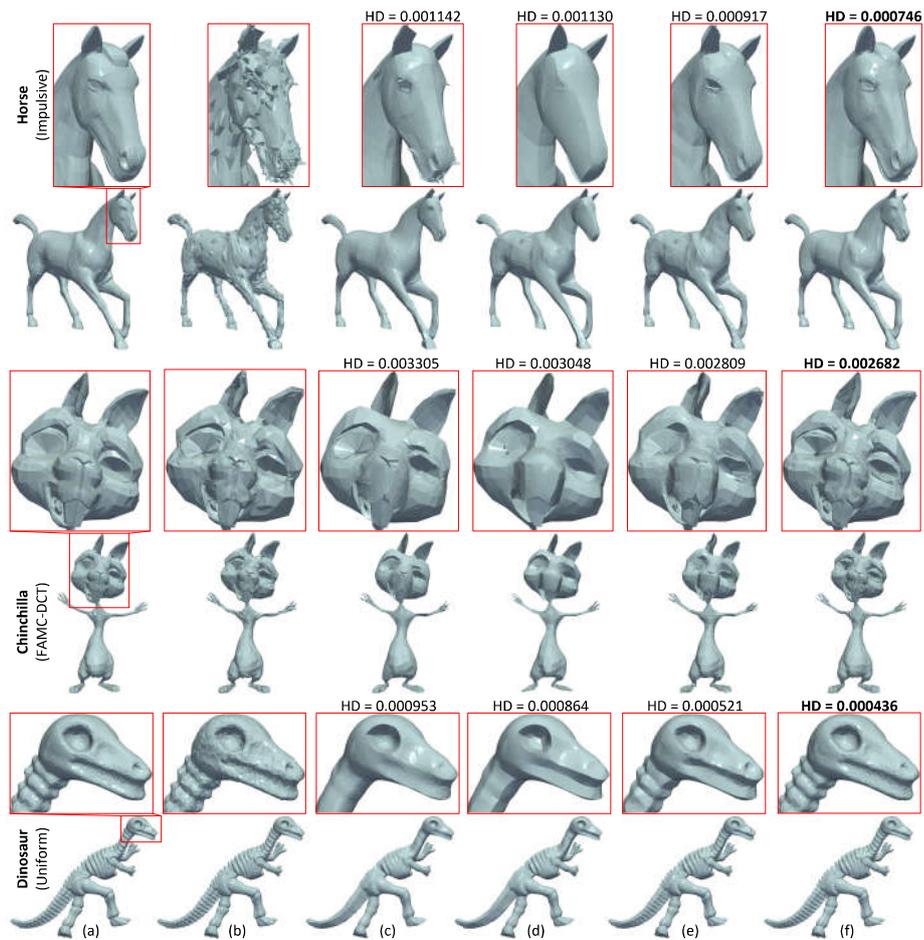

Figure 5.27: Denoising results of different noisy models (i.e., Horse, Chinchilla, Dinosaur) affected by different types of noise (i.e., Impulsive, Noise because of compression using the FAMC-DCT method and Uniform noise). (a) Original mesh, (b) Noisy mesh, (c) Bilateral normal denoising [7], (d) Guided mesh normal filtering [9], (e) Two stage graph spectral processing [12], (f) Our approach.

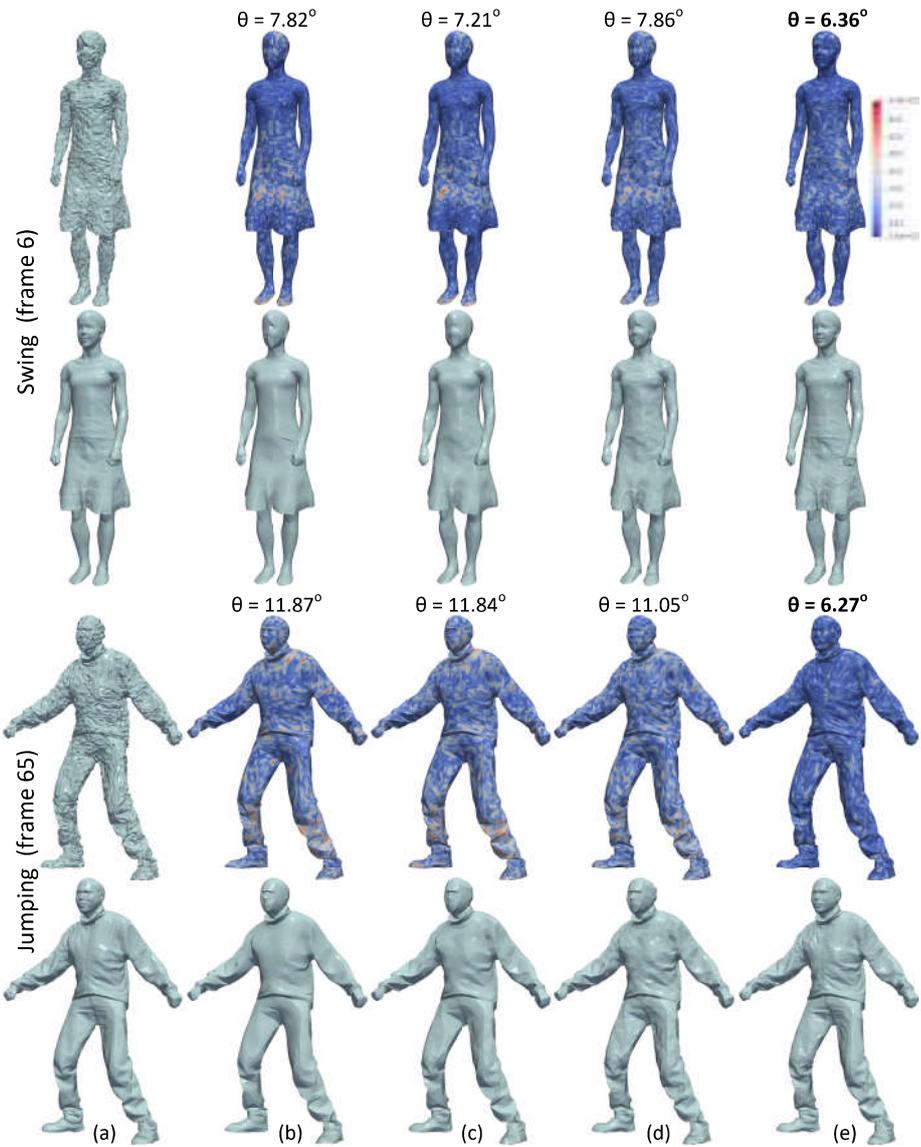

Figure 5.28: (a) Noisy [up] and Original [down] meshes of two models (i.e., Swing, Jumping). It is also provides heatmap of metric $\theta$ [up] and denoising results [down] using: (b) Bilateral normal denoising [7], (c) Guided mesh normal filtering [9], (d) Two stage graph spectral processing [12], (e) Our approach.

**Algorithm 3:** Spectral Denoising of Dynamic 3D Meshes

**Input** : Noisy sequence of meshes $\tilde{\mathbf{A}} = [\tilde{\mathcal{M}}_1; \tilde{\mathcal{M}}_2; \cdots ; \tilde{\mathcal{M}}_{n_s}]$;
**Output:** Denoised 3D animation $\dot{\mathbf{A}} = [\dot{\mathcal{M}}_1; \dot{\mathcal{M}}_2; \cdots ; \dot{\mathcal{M}}_{n_s}]$;

1 Decompose the Laplacian matrix **L** of the first frame Eq. (4.3);
2 **for** $i = 1 \cdots n_s$ **do**
3     Estimate the GFT $\hat{\mathbf{V}}_i$ of the vertices via Eq. (4.4)
4     Estimate the total energy $E_{s_i}$ of the GFT matrix via Eq. (5.25)
5     Estimate the ideal value $n_{h_i}$ of high-frequencies components via Eq. (5.26)
6 **end**
7 Select the common used value of $n_h$ via Eq. (5.27)
8 **for** $\forall j \in \{x, y, z\}$ **do**
9     Create the coherent matrix $\mathbf{M}_j$ via Eq. (5.66)
10    Estimate the low-rank matrix $\mathbf{E}_j$ via Eqs. (5.33)-(5.35)
11 **end**
12 **for** $i = 1 \cdots n_s$ **do**
13    **for** $\forall j \in \{x, y, z\}$ **do**
14        Estimate the denoised GFT $\dot{\hat{\mathbf{V}}}_{ji}$ via Eq. (5.36)
15        Estimate the denoised vertices $\dot{\mathbf{V}}_{ji}$ of the mesh via Eq. (5.37)
16    **end**
17 **end**

used, namely: (i) the number of vertices, (ii) the number of faces, (iii) the number of frames that each animation has and (iv) the type of noise which each animation has been affected with. The benefits of our method are apparent in all of the following experimental scenarios. In Fig. 5.26, we present the denoised results of three different frames (i.e., 40, 80 and 120) of the same model (Bouncing) affected by Gaussian noise. We additionally provide the values of the HD metric, for each reconstructed model, and enlarge details (in red boxes) providing an easier comparison among the meshes and techniques.

Similar results are presented in Fig. 5.27. However, in this case, each model has affected by a different type of noise. More specifically, we show the reconstructed results of three synthetic animations (i.e., Horse, Chinchilla, Dinosaur) affected by (i) impulsive, (ii) noise because of compression using the FAMC-DCT method and (iii) uniform noise, respectively. This figure highlights the effectiveness of our approach which efficiently reconstructs the denoised models without making special assumptions about the distribution and the type of noise. More importantly, the steps of the proposed method are always the same, as well as the used parameters, making it ideal for using it under different cases without requiring any special parameterization. Our method handles each model in exactly the same way regardless of the geometry of the model

| Name of Model | Vertices | Faces | Frames | Type of Noise |
|---|---|---|---|---|
| **Bouncing** | 10,002 | 20,000 | 150 | Gaussian $\sigma_E = 0.2$ |
| **Chinchilla** | 4,307 | 8,550 | 84 | FAMC-DCT |
| **Dinosaur** | 20,218 | 40,432 | 152 | Uniform |
| **Horse** | 8,431 | 16,858 | 48 | Impulsive |
| **Human** | 18,890 | 37,776 | 162 | Spacial Mask |
| **Jumping** | 9,971 | 19,938 | 150 | Gaussian $\sigma_E = 0.2$ |
| **Swing** | 10,002 | 20,000 | 150 | Gaussian $\sigma_E = 0.2$ |

Table 5.6: Brief description of the used animated noisy models.

(e.g., different geometrical features, intense edges and corners, etc.) or the type of noise.

In Fig. 5.28, we present the reconstructed results of different techniques, for one frame of two different animated models (namely, frame 6 of the Swing model and frame 65 of the Jumping model). We additionally provide the heatmap visualization of the angle $\gamma$ colorizing differently each vertex according to the value of $\gamma$ and the mean angle $\theta$. Dark blue color denotes a big similarity between two normals (practically the angle $\gamma$ goes to zero), while dark red color denotes a big difference in respect to their directions. As we can see, in any of the presented examples, the proposed method overcomes the results of the other comparison methods.

Nonetheless, despite the very good results that this method provides, we need to say that it does not perfectly work under any circumstances. For example, if the dynamic sequence is time-varying (its connectivity changes through the frames) or the meshes have been affected by a mask of noise (e.g., the noise affects each point in exactly the same way in each frame) then our approach is not able to work properly. Fig. 5.29 illustrates this unsolved problem. As we can observe, corresponding vertices of a sequence of meshes have been affected by spatial masks of noise and as a result, they maintain exactly the same form of noise per different frames.

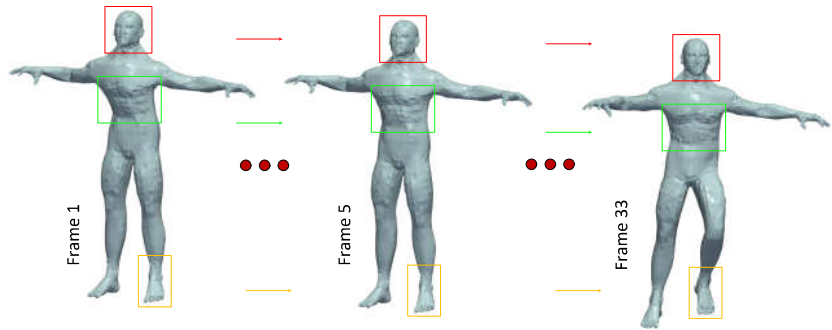

Figure 5.29: Different frames of the same animated 3D model (Human) having been affected by spacial masks of noise. This means that each frame has been affected by the same spacial mask of noise in the same areas (e.g., boxes with the same color in different frames).

### 5.1.3 Denoising of 3D Static and Dynamic Meshes using Deep Autoencoders

Recently, deep learning methods have been used in a variety of different domains showing excellent performance and pave the road for even better and robust results performing also real-time executions due to their fast computational complexity. Following these very promising trends, we investigate how these methods could be beneficially used in the denoising of 3D objects. The data-driven methods have some significant benefits as compared to the traditional mesh denoising methods [139–141], since they do not require the searching of ideal values per parameter for each model. However, their limitations mainly rely on the large dataset, required for the training process, and the corresponding training complexity. Additionally, in real case scenarios, the testing data have generally a different form from the training data (due to different light conditions, device characteristics, etc.) implementing the CNN process in real applications, to be problematic. A possible solution to this problem is to remove any objective characteristic from the data (i.e., training and testing), in order to be independent of geometry constraints (density, connectivity, etc.,), different quality of the scanner devices, or other external conditions that could affect the captured 3D model.

**Feature-aware Content-wise Denoising Approach**

We present a feature-aware content-wise denoising approach using denoising autoencoders. The proposed method has many advantages and seems very promising for future extensions. More specifically, the proposed method: (i) is totally parameter-free, (ii) it can be used for any type of noise with different patterns, and (iii) it requires a relevant small size of a dataset for the training process. The proposed method is decomposed by two main steps: (i) the feature's identification, applied both to the training and denoising process, during which we identify the features of the noisy 3D objects, and (ii) the training process, where we train different content-wise DAEs. Finally, in the denoising process, we use the already trained models for an efficient denoising process. Fig. 6.34 briefly illustrates the pipeline of the proposed schema.

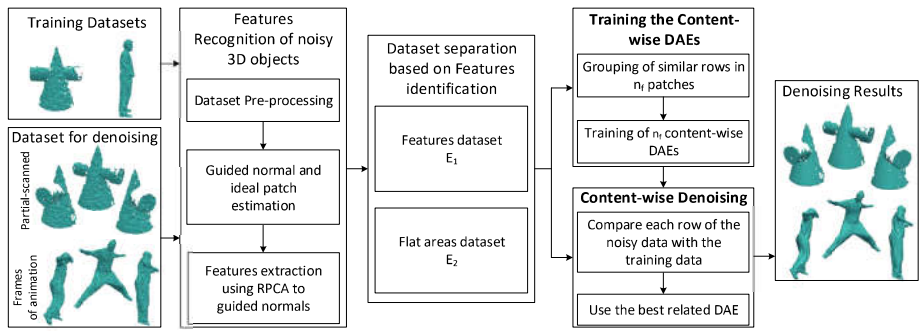

Figure 5.30: Pipeline of the proposed method for the feature-aware and content-wise 3D mesh denoising.

The contributions of this work are:

- We developed a method that is totally parameter free, giving the flexibility to be used without the need of exhaustive searching of the ideal parameters when different models are used, allowing a fully automatic runtime execution.

- We trained different deep autoencoders for flat areas and for areas with large and small-scale geometric features. This process reduces significantly the training complexity and the required training data.

- The method has no special requirements (e.g., length, connectivity) about the form of the input data, and it can be used for reconstructing several noisy 3D meshes.

**Datasets Creation Based on Type of Features**

For the generation of the training dataset, we exploit the fact that flat areas have different geometric properties and characteristics than areas with features. For this reason, we suggest splitting the whole dataset into two distinct datasets; composed of areas with features (corners, edges) and flat areas. This approach reduces the training execution time, while the two datasets encode patterns with different geometric characteristics. The classification of the different areas into the two datasets requires the identification of features which is a very challenging task, especially in the presence of noise. To overcome this problem, we use an RPCA approach applied directly to guided normals, providing robustness in various noise patterns.

**Identification of features in Noisy 3D Objects**

Nowadays, finding features of a 3D object is a straightforward process. However, finding features of a noisy 3D mesh is not always easy, especially when noise and features are tangled with each other. In many application which handle noisy objects, features need to be firstly identified, allowing us to apply different approaches (non-isotropic) to features and no-features areas providing more efficient results. Finding the features of a 3D object is a vital process since allows to apply non-isotropic approaches increasing the efficiency of the final results.

**RPCA-Based Features Identification**

We firstly estimate the guided normals [17] in order to formulate the matrix $\mathbf{M}$. Then, we apply RPCA using as input the matrix $\mathbf{M}$, exploiting the observation [332] that all the normals of flat patches have similar directions, in contrast to the normals of patches representing edges or corners (see Fig. 4.2). As a result, the entries of the matrix $\mathbf{E}$ with large values correspond to flat/smoothed areas while entries of the matrix $\mathbf{S}$ with large values correspond to geometric features of different scale. To identify features, we compare the scalar values of the entries of $\mathbf{E}$ and $\mathbf{S}$ using a threshold $e$. We assume that a centroid is a feature, if $\|\mathbf{s}_i\|_2 > \|\mathbf{e}_i\|_2 + e$. Typically, the value of the threshold $e$ that we used in our experiments is $e = 0.7$, where $\mathbf{s}$ and $\mathbf{e}$ represents the value of the sparse and low rank matrix correspondingly. After this step, two different datasets are created, defines as the dataset $\mathbf{M}_1$ representing the areas with features and $\mathbf{M}_2$ representing the flat areas, where $\mathbf{M} = \mathbf{M}_1 \cup \mathbf{M}_2$. In Fig. 5.31, we compare our results with other methods. Method [14] provides perfect recognition of features when they are used in noiseless meshes. However, they fail in noisy objects. On the other hand, the proposed method seems to provide robust results in both cases.

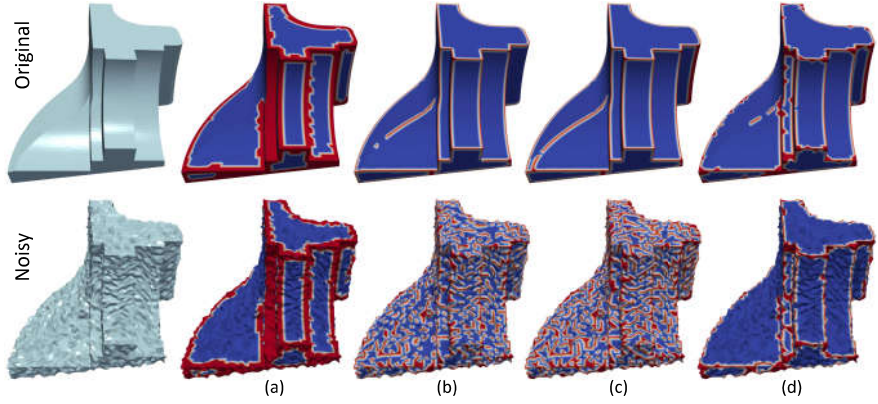

(a)         (b)         (c)         (d)

Figure 5.31: Feature detection in original and noisy 3D object (Fandisk) using different approaches. (a) The method presented in [13], (b) the method in [14], (c) the method in [14] including area standardization, (d) the proposed method.

**RPCA-Based Features Identification**

In Fig. 5.32, we present results showing that the proposed features identification algorithm, is able to identify features even in cases of noisy 3D objects where noise and features are tangled with each other. In this figure, the identified features (corners and edges) are represented with red color and the flat areas with blue color. The experimental results show that the identification process provides reliable results even when the magnitude of the noise increases.

**Training of Content-wise DAEs**

DAE is a function that receives corrupted data as input and is trained to predict the original, uncorrupted data as its output, minimizing the quantity $\mathcal{J}(\mathbf{M}, h_{\theta'}(f_{\theta}(\tilde{\mathbf{M}})))$, where $\tilde{\mathbf{M}}$ is the noisy observation of $\mathbf{M}$. In our experiments, we assume that $\sigma = \sigma'$ is the linear function. For the minimization of the cost function, we use the gradient descent algorithm while for the optimization we use the minFunc function (Quasi-Newton with Limited-Memory BFGS Updating)[1].

The guided normals are modified for ensuring unit magnitude: $\mathbf{g}_i = [\frac{g_{x_i}}{\|\mathbf{g}_i\|_2}; \frac{g_{y_i}}{\|\mathbf{g}_i\|_2}; \frac{g_{z_i}}{\|\mathbf{g}_i\|_2}]$. Additionally, we create a spatial coherent training dataset, by rotating any normal that does not lie on the first quadrant of the 3D Cartesian system. In that way all the normals lie at the positive space of the coordinate system. Thus, we create a more coherent dataset making the learning process more efficient. In

---

[1]minFunc: unconstrained differentiable multivariate optimization

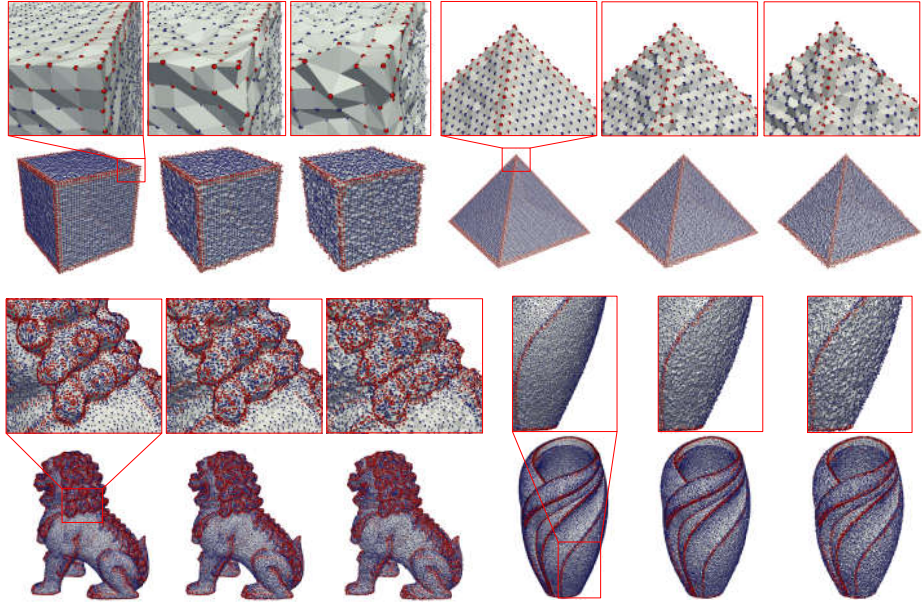

Figure 5.32: Features identification for different models (cube, pyramid, chinese lion, vase100K) affected by different levels of noise.

Fig. 5.33, we present a relative example showing the results of the aforementioned transformation in some groups of normals.

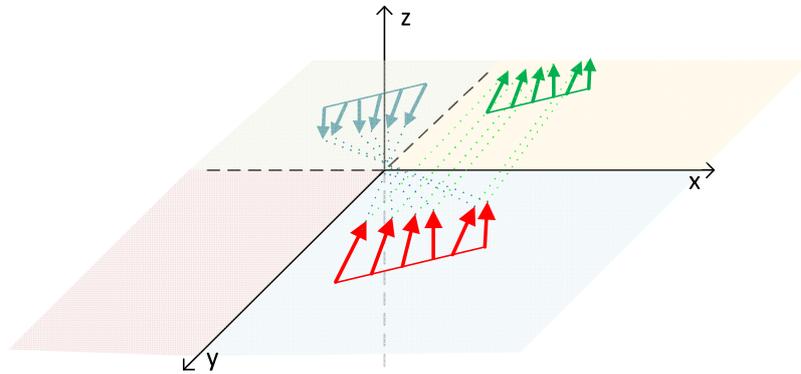

Figure 5.33: Pre-processing of dataset.

Generally, DAEs are able to find the ideal parameters $\theta$ which fit best to the original training data. Following the same approach in order to increase the effectiveness of our method, we suggest a content-wise process for learning different $\theta$s for different set data with similar properties, instead of using a universal $\theta$ for the entire dataset. In Fig. 5.34, we present a simplified example showing that the content-wise DAEs can describe better the majority of the data samples.

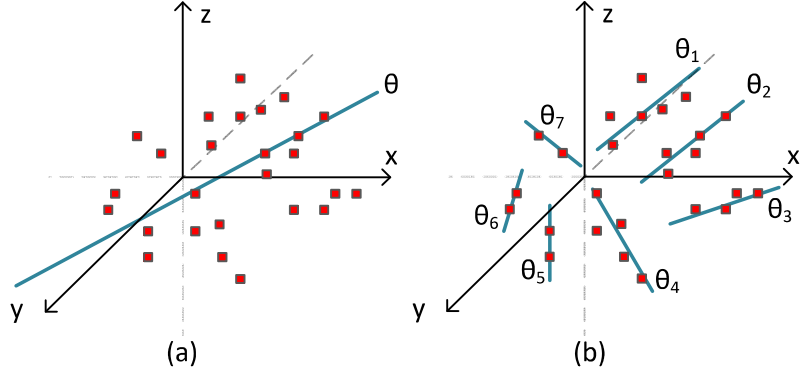

Figure 5.34: (a) Denoising autoencoder, (b) content-wise DAEs.

Regarding the training process, for each row $\mathbf{M}_i \in \mathbb{R}^{1 \times 3k}$ of the training dataset $\mathbf{M}$ in Eq. (5.66), we collect and group together any other row $\mathbf{M}_j$ that is determined by a distance metric value, lower than a small defined threshold $\epsilon$, e.g., $\|\mathbf{M}_i - \mathbf{M}_j\|_2 < \epsilon \ \forall \ i, j = 1, \cdots, n_f$. Each group of rows are then used for training different content-wise DAE. On the other hand, regarding the denoising process, for each presented noisy row $\tilde{\mathbf{M}}_i \ \forall \ i = 1, \cdots, n_f$, we select the ideal $\theta^*$ based on its similarity (distance) with the rows of the training datasets. Algorithm 4 briefly presents the steps of the proposed process.

**3D Mesh Reconstruction by Updating Vertices**

After the estimation of the reconstructed centroid normals $\bar{\mathbf{n}}_c$, we update the vertices of the output mesh, using the following robust and very well-known iterative procedure [16].

**Perfect Denoising Assuming Iterative Updating with Ideal Normals**

Fig. 5.35 presents denoising results using perfect denoised centroid normals for the iterative updating process of Eq. (11). This updating technique is able to perfectly mitigate the noise of a noisy mesh after some iterations. Specifically, after 20 iterations, the denoised results have no actually significant visual differences in comparison with the original mesh. Providing this example, we want to highlight that the presented updating method can easily converge after a finite number of iterations, especially if the used normals have a good approximation with the original (e.g., using the proposed approach).

**Algorithm 4:** Training and Content-wise Denoising

```
// Training Process
```
**Input** : Training noisy Dataset $\mathbf{M} \in \mathbb{R}^{n_f \times 3k}$, Original Dataset $\mathbf{EM'} \in \mathbb{R}^{n_f \times 3k}$;
**Output:** Denoised centroid normals $\bar{\mathbf{n}}_c$;

1. **for** $i = 1 \cdots n_f$ **do**
2.     $\mathcal{N} = \{i\}$;
3.     **for** $j = 1 \cdots n_f$ **do**
4.         **if** $\|M_i - M_j\|_2 < \epsilon$ **then**
5.             $\mathcal{N} = \mathcal{N} \cup \arg(M_j)$;
6.         **end**
7.     **end**
8.     $\theta_i = \text{train}(\mathbf{M}(q), \mathbf{M'}(q)), \ \forall q \in \mathcal{N}$;
9. **end**

```
// Content-wise Denoising Process
```
**Input** : Noisy Dataset $\tilde{\mathbf{E}} \in \mathbb{R}^{\tilde{n}_f \times 3k}$

10. **for** $i = 1 \cdots \tilde{n}_f$ **do**
11.     **for** $j = 1 \cdots n$ **do**
12.         **if** $d_j = \|\tilde{M}_i - M_j\|_2 < \epsilon$ **then**
13.             $\mathcal{K}_j = \{\theta_j, d_j\}$;
14.         **end**
15.     **end**
16.     $\theta_i^* = (\theta | \min\limits_{d \in \mathcal{K}}(\mathcal{K}(d)))$;
17.     $\bar{\mathbf{n}}_{ci} = \text{denoising}(\tilde{M}_i, \theta_i^*)$
18. **end**

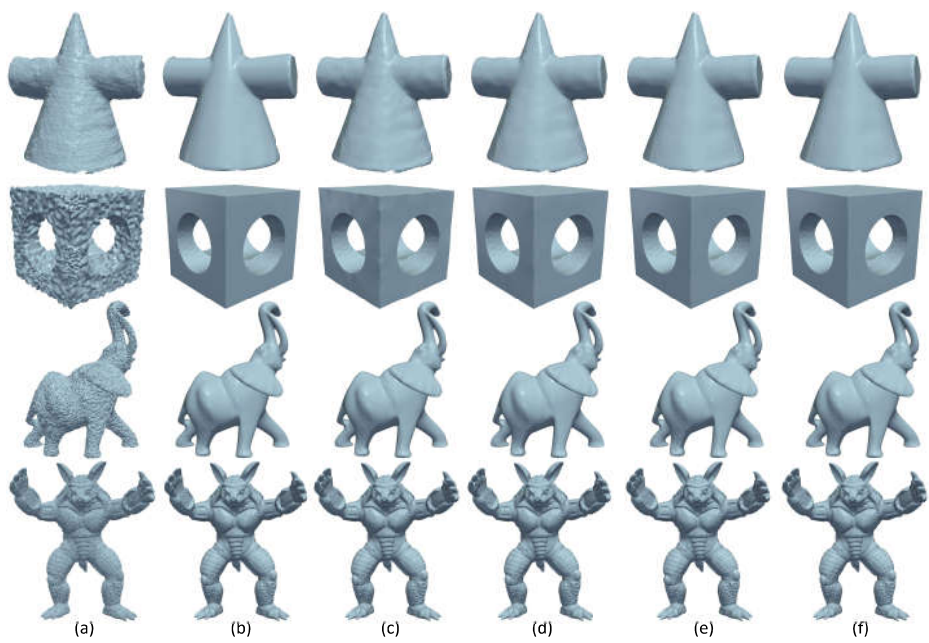

Figure 5.35: (a) Noisy object, (b) Original ground-truth object, and denoising results using perfectly denoised centroid normals after: (c) 10 iterations, (d) 20 iterations, (e) 40 iterations, (f) 60 iterations.

**Experimental Setup and Metrics**

For the experiments, we used a variety of 3D static [10] and dynamic models [317] scanned with different devices (Kinect v1, Kinect v2, synthetic) and different levels/types of noise. The quality performance of the proposed technique is compared with other well-known, recent and robust techniques in the literature, such as the (i) bilateral normal [15], (ii) fast and effective [16], (iii) guided normal [17]. The benefits of our method are apparent in the following real-case scenarios.

**Static meshes - Applied in partial-scanned areas.** Figs. 5.36, 5.37 present the denoising results from 3D partial-scanned models captured using different 3D scanning devices. The results are evaluated using two metrics ($\alpha, d_H$). We also present the first partial-scanned frame of each model which is used for the training process.

**Dynamic meshes - Applied in noisy consecutive frames.** In Fig. 5.38, we present the denoising results of animated models (crane,swing) affected by Gaussian noise with intensity $\sigma_E = \frac{\sigma}{E_{\text{mean}}} = 0.3$, where $\sigma$ is the variance of the Gaussian function, and $E_{\text{mean}}$ is the average edge length of the ground truth mesh [17]. For an easier comparison between ground-truth and reconstructed models, we provide a heatmap visualization of the error as well as the Hausdorff distance $d_H$. In both of these case studies, the results of the proposed method outperformed the results of the other methods both in visual perception and evaluation metrics. Our method avoids the over-smoothing of the models and at the same time, takes into account the special characteristics of the features, succeeding to preserve them.

### 5.1.4 Image-based 3D mesh Denoising Through a Block Matching 3D CNN Filtering Approach

Recently, Convolutional Neural Networks (CNNs) have shown their effectiveness in a variety of different processing domains (e.g., classification, segmentation, denoising, etc.,). In this subsection, we present an image-based 3D mesh denoising approach using BM3D CNN filtering applied in color images. We take advantage of the stable and robust behavior of BM3D, which has exhaustively tested for image denoising, in combination with the fast and effective behavior of CNNs. To achieve this, we create an appropriate form of images, representing patches of neighboring normals. Experimental analysis verified the correctness of our assumption while comparison with other traditional state-of-the-art methods demonstrates the potential of our approach. The proposed method has many advantages, such as (i) only fixed values of parameters are used, (ii) it requires a relatively small dataset for the training process, (iii) a general model, trained by different levels of noise, can be used for denoising,

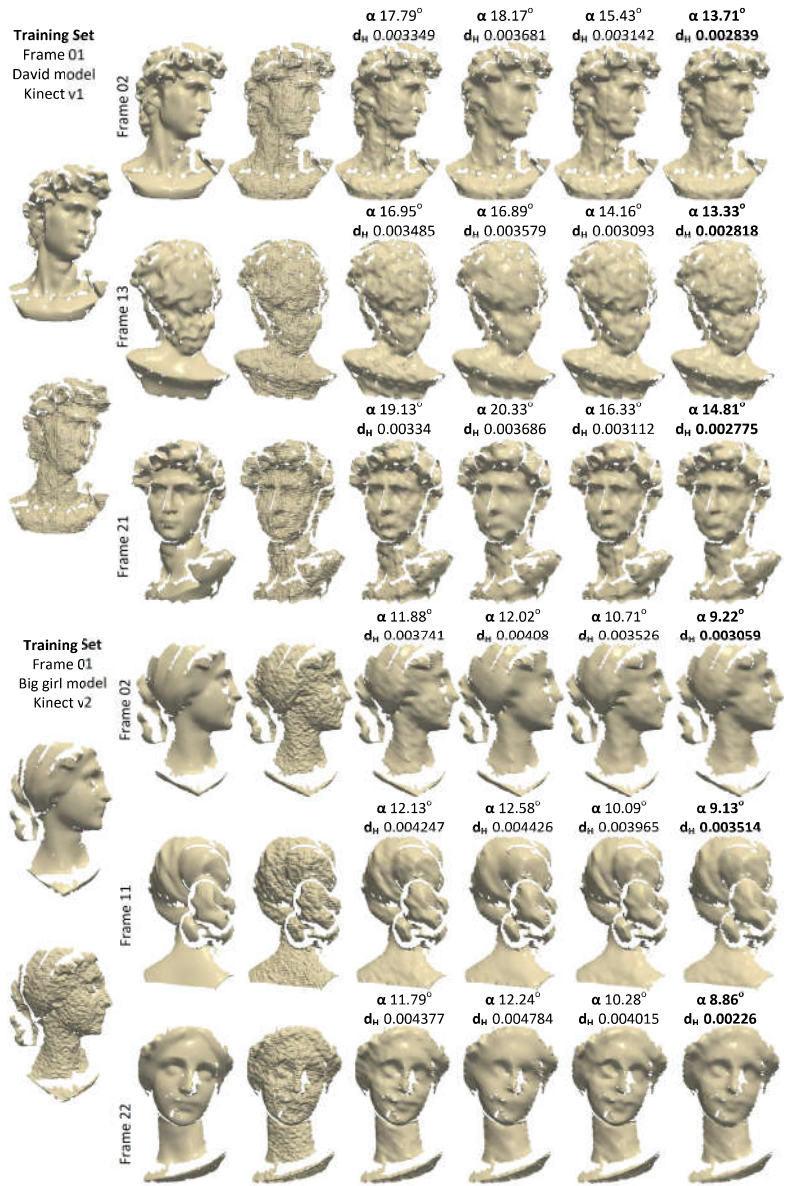

Figure 5.36: (a) Original frames, (b) noisy frames, and denoising results using: (c) bilateral normal [15], (d) fast and effective [16], (e) guided normal [17], (f) our approach.

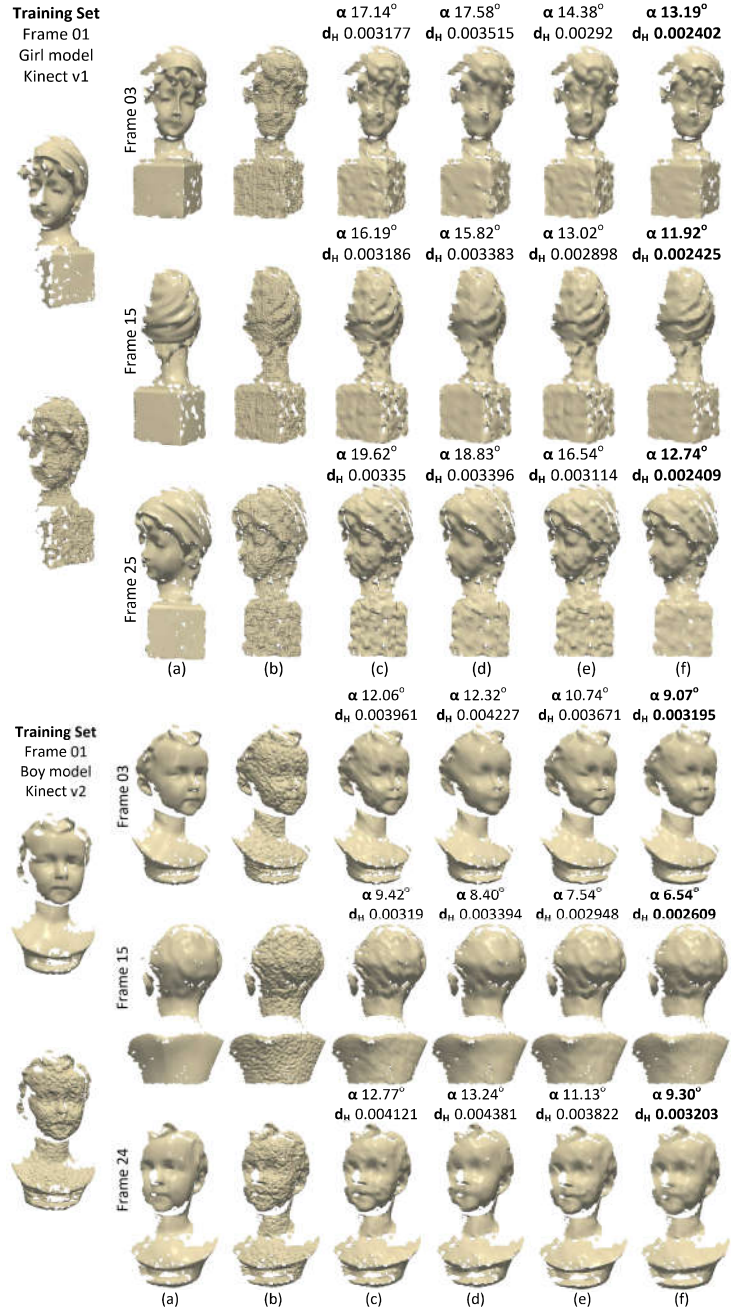

Figure 5.37: (a) Original frame, (b) noisy frame and denoising results using: (c) bilateral normal [15], (d) fast and effective [16], (e) guided normal [17], (f) our approach.

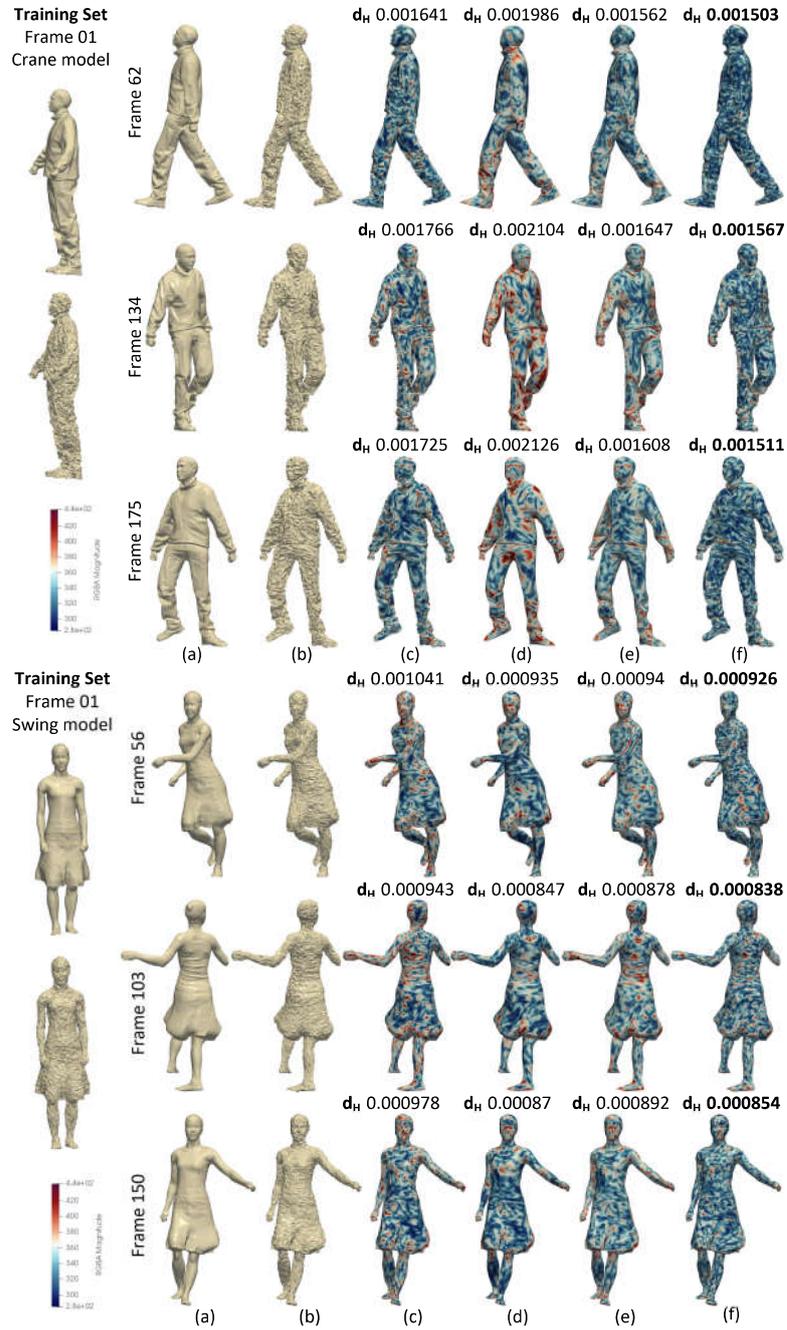

Figure 5.38: (a) Original frames, (b) noisy frames, and denoising results using: (c) bilateral normal [15], (d) fast and effective [16], (e) guided normal [17], (f) our approach.

providing acceptable results.

The method starts with the training of the CNN. After that, the training model can be used to denoise any other new noisy 3D object. However, the most vital process is a pre-processing step that creates the appropriate and uniformly-used form of the equal-sized images which efficiently encode the geometrical information of any noisy mesh. Fig. 5.39 briefly illustrates the pipeline of the proposed schema.

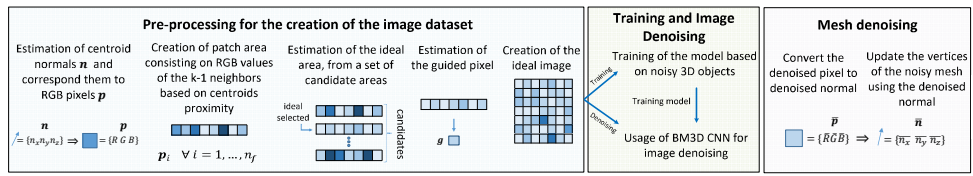

Figure 5.39: Pipeline of the proposed method for image-based 3D mesh denoising using BM3D CNN filtering.

The contributions of this work are:

- We introduced a novel pipeline for 3D mesh denoising that efficiently exploits the benefits of the 3D transform-domain CF, used for image denoising.

- The proposed data-driven method gives the flexibility for fully automatic runtime execution, without the need for exhaustive searches of ideal parameters per model.

- The proposed method has no special requirements, related to the form of the 3D mesh, since it operates with equal-sized images uniformly, encoding the useful geometrical information of any mesh.

- The proposed image-based approach results in a more efficient training process, avoiding the need for large datasets, exploiting efficiently geometrical coherence.

- We showed how a method, inspired by robust and well-known approaches used in the area of the image processing, can be also efficiently used in the area of 3D mesh processing.

**Creation of Images Based on Normals**

We start assuming that for each centroid $c_i$ there is a patch (enclosed in yellow border lines as shown in Fig. 5.40) which is created based on its proximity with the $k-1$ geometrical nearest neighboring centroids. Then, we estimate the normals of the corresponding faces and we formulate an image with size $k \times k$, where $k$ is an odd number $\{k = 2q + 1 : q \in \mathbb{Z}\}$. In the central cell $(k+1)/2$ of the first row, we place the "normal-of-interest" (in red color). The normal $\mathbf{n}_{12}$ of the closest centroid (i.e., $d_1$) is placed in the first right cell, then the normal $\mathbf{n}_{13}$ of the next closest centroid (i.e., $d_2$) is placed in the first left cell, and the completion of the row continues, in turn, using the normals of all centroids, where $d_1 < d_2 \cdots < d_{k-1}$. The next rows are completed in the same way assuming, however, that the central cell represents the normal of the closest centroid of $\mathbf{n}_1$, increasingly for each row.

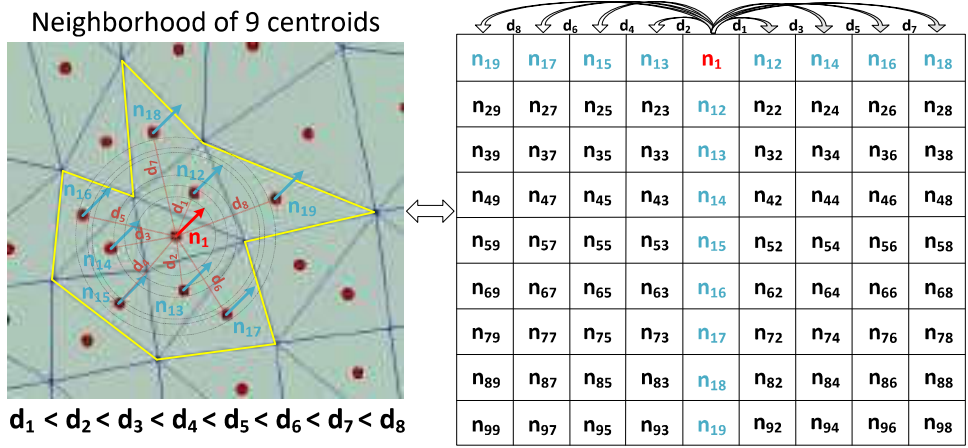

Figure 5.40: Creation of an image $9 \times 9$ using the normals of a neighborhood of 9 centroid points.

**Coherent dataset due to the orientation**

Fig. 5.41 depicts $25 \times 25$ images, representing four different patches of three different areas (i.e., flat area, edges, corners), which have been created as described in section 2.2 of the original manuscript. As we can see in Fig. 5.41-(i), images have different colors with each other, even when they belong to the same type of features (e.g., flat areas). This means that the normals of each patch have different directions in comparison with the normals of other patches even though the normals of the same patches have almost the same direction between them.

These great number of different directions, in the 3D coordinate system, resulting in the size of the training dataset that needs to be used. In this case, a large number of instances of each direction are required for an efficient learning process during the training step. However, the possible different directions are countless, so the training dataset would need to be huge.

We overcome this drawback, creating a more relevant dataset. More specif-ically, we rotate all normals of each $i$ patch about an angle $\theta_{x_i}$, so that the $i$ "normal-of-interest", to be equal to the constant vector [1 0 0]. This means that after the rotation, the corresponding "pixel-of-interest" will have red color (i.e., RGB = $\{1\ 0\ 0\}$) and the color of any other pixel will be adapted respectively, as shown in Fig. 5.41-(ii).

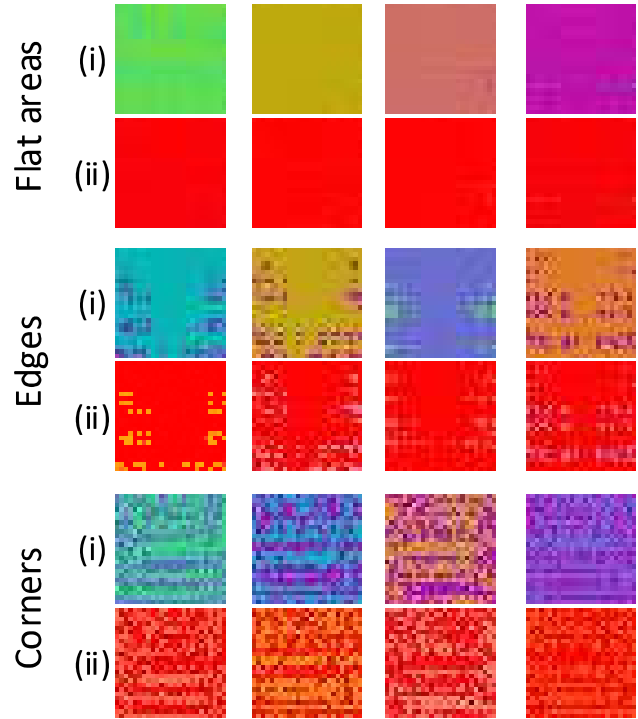

Figure 5.41: (i) original images representing patches, (ii) images after rotation.

**Creation of the Dataset Consisting of Ideal Images**

Image-guided filtering has been presented in many papers providing very good results [333] [334] [335]. Additionally, in the area of 3D mesh denoising, guided normals have demonstrated also a good denoising performance in many recent works [12] [17]. In this work, we follow the same line of thought but using guided pixels and ideal-selected images, for the filtering step of the mesh denoising.

**Estimation of the Ideal-selected Row and Guided Pixel**

For each centroid normal $\mathbf{n}_i$ of the mesh, we estimate an image $\mathcal{I}_i$ (in a similar way as presented in subsection 4.2.1), the first row $\mathcal{I}_{1i}$ of which represent the nearest neighborhood area called as patch. However, the same pixel $\mathbf{p}_i$ may be presented in more than one row/patches since each neighborhood area is created by totally overlapping with the neighboring areas of other normals. In fact, some of these rows may represent better the color of the pixel. For this reason, we collect a set $\mathcal{S}_i = \{\mathcal{I}_{1i_1}, \mathcal{I}_{1i_2}, \ldots, \mathcal{I}_{1i_{n_p}}\}$ of $n_p$ candidate rows and our main objective is to find which one of the $\mathcal{I}_{1i_j} \; \forall \; j = 1, \ldots, n_p$, is the ideal representative of the pixel $\mathbf{p}_i$ [12], in terms of the similarity of the color. For each one of these candidate rows $\mathcal{I}_{1i_j}$, we estimate the corresponding covariance matrix $\mathbf{C}_{ij} = \mathcal{I}_{1i_j}^T \mathcal{I}_{1i_j} \in \mathbb{R}^{3\times 3} \; \forall \; i = 1, \cdots, n_f, \; \forall \; j = 1, \cdots, n_p$ and we decompose it $\mathbf{C}_{ij} = \mathbf{U}\mathbf{\Lambda}\mathbf{V}$ to its eigenvectors $\mathbf{U}$ and eigenvalues $\mathbf{\Lambda} = \mathrm{diag}(\lambda_{1_{ij}}, \lambda_{2_{ij}}, \lambda_{3_{ij}})$. Then, for the estimation of the ideal-selected row $\mathcal{I}_{1i}^*$, two parameters are investigated: (a) the norm$_2$ $s_{ij} = ||\lambda_{1_{ij}} - \lambda_{2_{ij}}||_2$ of the first 2 eigenvalues and (b) the maximum color difference $h_{ij} = \max(||\mathbf{p}_i - \mathbf{p}_l||_2) \; \forall \; \mathbf{p}_l \in \mathcal{I}_{1i_j}$ between the $i$ pixel and the other pixels of the same row. Among all candidates rows, we pick this one with the smallest value of Eq. (5.38):

$$\mathcal{I}_{1i}^* = (\mathcal{I}_{1i_j} \; | \; \min(s_{ij}h_{ij})) \tag{5.38}$$

$\forall \; i = 1, \cdots, n_f, \; \forall \; j = 1, \cdots, n_p$. Finally, the guided $\mathbf{g}_i$ pixel is estimated as the average pixel of this ideal row:

$$\mathbf{g}_i = \frac{\sum_{\mathbf{p}_j \in \mathcal{I}_{1i}^*} \mathbf{p}_j}{\left\|\sum_{\mathbf{p}_j \in \mathcal{I}_{1i}^*} \mathbf{p}_j\right\|_2} \; \forall \; i = 1, \cdots, n_f \tag{5.39}$$

**Estimation of Ideal Images Based on Guided Pixels**

For the creation of the ideal images, we follow exactly the same format as this one presented in subsection 5.1.4, but using the guided pixels $\mathbf{g}_i$ instead of the normals $\mathbf{n}_i$. In Fig. 5.42, we present examples of images created by the RGB representation of: (i) the normals, and (ii) the guided pixels, for different features (i.e., flat area, edge and corner) of Fandisk model affected by different level of Gaussian noise with intensity $\sigma_E = \frac{\sigma}{E_{\mathrm{mean}}}$, where $\sigma$ is the variance of the Gaussian function, and $E_{\mathrm{mean}}$ is the average edge length [17]. As we can observe, the images, created by the guided pixels (Fig. 5.42-(ii) (b)-(d)), have a similar presentation with the original, despite the different levels of noise. On the other hand, images that have been created using the centroid normals are very noisy and differ from the original, making it difficult to be denoised, even in cases of flat areas. This observation is very important since it shows that it is not necessary to train different models for different noise patterns, as other data-driven methods do, but that a generic model would suffice. This is especially useful in cases where the level or type of noise is not known a-priori.

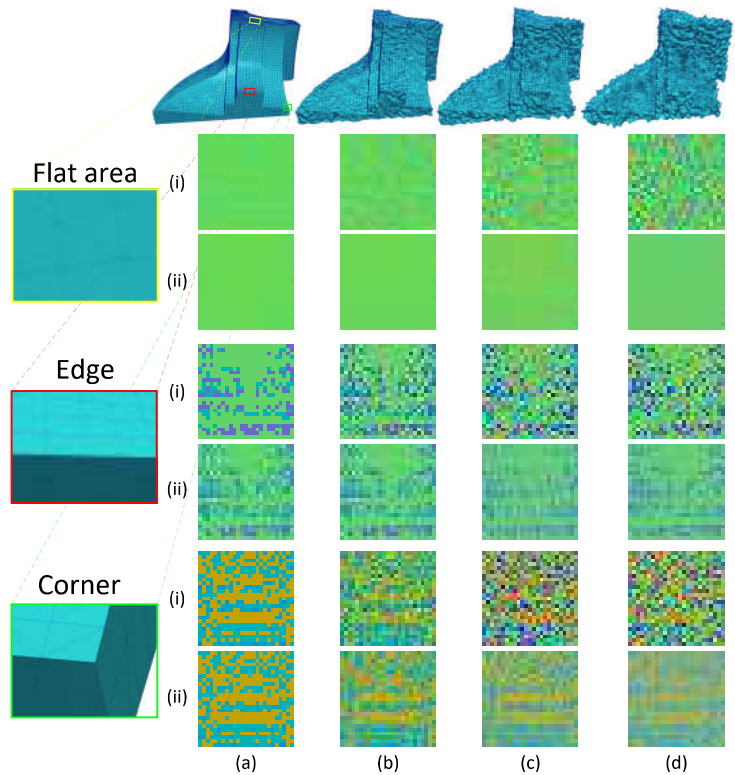

Figure 5.42: Images of 25 × 25 pixels, representing patches of different features, affected by different level of noise: (a) original model, (b) $\sigma_E = 0.2$, (c) $\sigma_E = 0.4$, (d) $\sigma_E = 0.6$. The images are created: (i) using normals, as presented in Fig. 5.40, (ii) using the guided pixels.

**BM3D CNN for Ideal Images Denoising**

In this paragraph, we will discuss, in more detail, the configuration and the characteristics of the used CNN model and how it is related to the basic layers of the BM3D approach.

BM3D uses three basic steps: (1) "Block matching" which tries to find groups of similar patches, (2) "3D wavelet shrinkage" which denoises each one of these groups in the 3D wavelet transform domain and (3) "Patch group aggregation" is an averaging procedure in which all the estimated patches return to their original positions reweighed each pixel appearing in many instances since im-ages are created by overlapping [186]. For a more efficient block matching step, we use a feature classification, similar to [12]. More specifically, we classify each $i$ centroid into three different categories of features (i.e., flat area (F), edge (E) and corner (C) ), using a $\kappa$-means ($\kappa = 3$) clustering, applied to the vector $\boldsymbol{\lambda}_i = [\lambda_{1_{i1}} \; \lambda_{2_{i1}} \; \lambda_{3_{i1}}]$ of its eigenvalues. To create a more relevant dataset, we rotate all normals of each $i$ patch about an angle $\theta_{x_i}$, as shown in Fig. 5.43, so that the $i$ "normal-of-interest" (i.e., lied in the red face), to be equal to the constant vector [1 0 0]. Additionally, in this way, a more coherent dataset is created, facilitating the learning process. Otherwise, patches with every possible normal direction would need to be provided, resulting in the creation of very large training datasets, increasing also the execution time of the learning step.

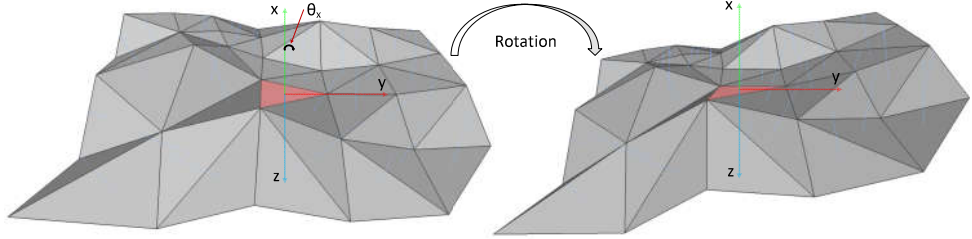

Figure 5.43: Example of patch rotation.

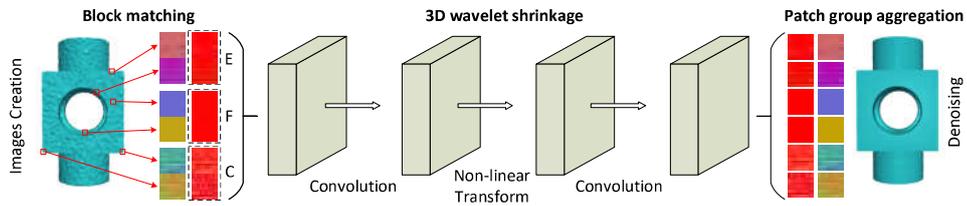

Figure 5.44: Patches of the mesh are converted to images and then they are fed into the CNN for denoising.

**Characteristics of the used BM3D CNN**

The presented CNN (Fig. 5.44), similar to [189], represents the second step of the BM3D (i.e., 3D wavelet shrinkage) and consists of two convolution layers and a nonlinear transform layer. More specifically, the first convolution layer is used as the wavelet transform step and the second convolution layer is used as the inverse wavelet transform step, they are generally presented as:

$$\mathbf{I}_{l+1} = W_l * \mathbf{I}_l \tag{5.40}$$

where $*$ indicates multidimensional convolution, $\mathbf{I}_l$ denotes the group of images for the $l \in \{1, 3\}$ layer and $W_l$ is the filter with the weights of the $l$ layer after the concatenate process. The nonlinear transform layer is used as the wavelet shrinkage step. Similar to [189], the radial basis functions (RBFs) is used instead of regular hard thresholding or Rectified Linear Unit (ReLUs), according to:

$$\mathbf{I}_3 = W_2 e^{\frac{|\mathbf{I}_2 - \mu|^2}{\sigma^2}} \quad (5.41)$$

**3D Mesh Reconstruction by Updating Vertices**

After the estimation of the denoised pixels $\bar{\mathbf{p}}$, which is the central pixel of the first row of the denoised images, we convert the RGB values to the $(n_x, n_y, n_z)$ values of the corresponding denoised centroid normals $\bar{\mathbf{n}}$. Then, we reconstruct the denoised mesh updating its vertices, using the following robust and very well-known equation [16]. Algorithm 5 summarizes the basic steps of the proposed process.

---

**Algorithm 5:** BM3D CNN for 3D Mesh Denoising

**Input**  : Noisy 3D model $\hat{\mathbf{M}} \in \mathbb{R}^{n \times 3}$;
**Output:** Denoised 3D model $\bar{\mathbf{M}} \in \mathbb{R}^{n \times 3}$;

1 Estimate the $n_f$ centroid normals **n** and convert them to corresponding pixels **p**;
2 **for** $i = 1, \cdots, n_f$ **do**
3      Estimate the ideal-selected row $\mathcal{I}_{1i}^*$ via Eq. (5.38) and the guided pixel $\mathbf{g}_i$ according to Eq. (5.39);
4      Estimate the filtered image $\mathcal{I}_i^*$ based on guided pixels;
5 **end**
6 Create 3 different types of images, based on the features classification, for a more efficient block matching;
7 Image denoising based on BM3D CNN via Eqs. (5.40)-(5.41);
8 Convert denoised pixels $\bar{\mathbf{p}}$ to denoised centroid normals $\bar{\mathbf{n}}$;
9 Reconstruct the denoised 3D mesh $\bar{\mathbf{M}}$ according to Eq. (4.28);

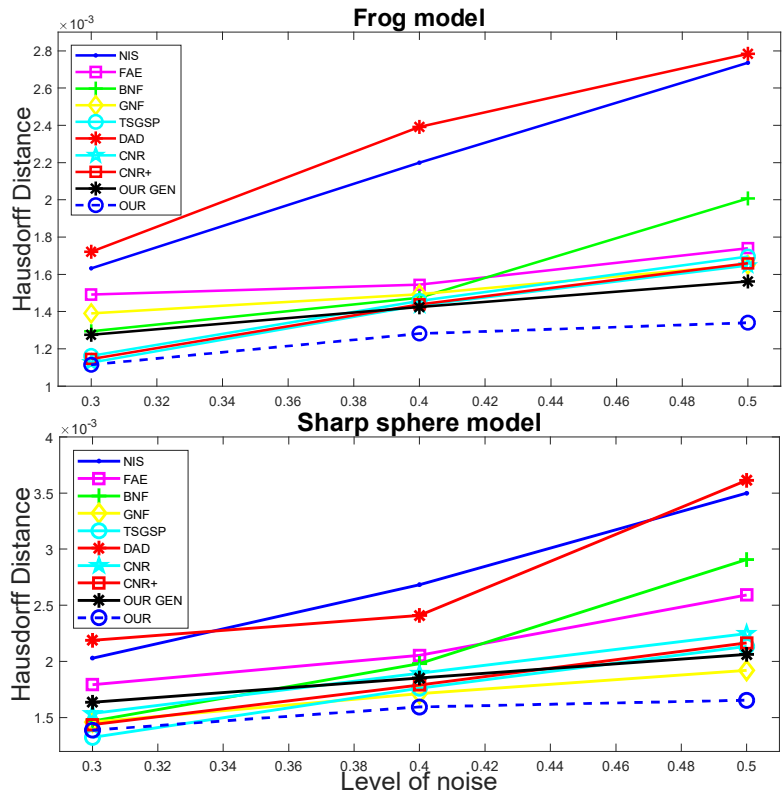

Figure 5.45: Hausdorff distance error for different levels of noise.

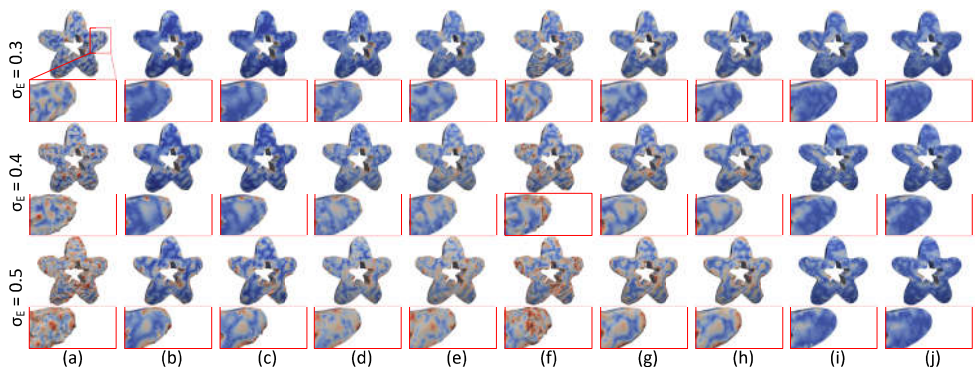

Figure 5.46: Heatmaps of (a) NIS [18], (b) FAE [16], (c) BNF [15], (d) GNF [17], (e) TSGSP [12], (f) DAD [19], (g) CNR [10], (h) CNR+, (i) GEN OUR, (j) OUR.

**Experimental Results**

For the training process, we use only 5 different models, comprised of plenty of features, consisting of $\sim$ 1.2 million (M) centroid points in total. This means that we have 1.2 M very relevant instances for the training which seem very effective, according to the experimental results.

Figure 5.47: Denoising results of: (a) NIS [18], (b) FAE [16], (c) BNF [15], (d) GNF [17], (e) TSGSP [12], (f) DAD [19], (g) CNR [10], (h) CNR+, (i) GEN OUR, (j) OUR.

The quality performance of the proposed technique is compared with other well-known and very robust techniques of the literature, namely (a) non-iterative smoothing (NIS) [18], (b) fast and effective (FAE) [16] mesh denoiing, (c) bilateral normal filtering (BNF) [15], (d) guided normal filtering (GNF) [17], (e) two stage graph spectral processing (TSGSP) [12]. We also compare the proposed approach to data-driven methods such as (f) deep autoencoders denoising (DAD) [19], and cascade normal regression [10] in two different approaches: (g) by using only a small dataset of models (CNR) (the exact same dataset used) and (h) (CNR+) by using a larger dataset. Our method is denoted as (OUR) and (GEN OUR) represents our approach but using a general training model, having been trained by different levels of noise at the same time.

In Fig. 5.47, we present original 3D models and three corresponding noisy models which have been affected by three different level of noise (i.e, $\sigma_E = 0.3, \sigma_E = 0.4, \sigma_E = 0.5$). We provide enlarged details of each reconstructed model, for easier comparison among the methods, and we also provide the metric $\theta$. As we can see from the results, the benefits of our approach are more apparent for higher levels of noise $\sigma_E > 0.3$, while for a lower level of noise, they are comparable with the results of other state-of-the-art methods. We can make similar conclusions observing the heatmaps of Fig. 5.46, which visualize in different colors the differences between the reconstructed and original mesh, in three different levels of noise. Big differences are represented by red color while small differences are represented with blue. Finally, in Fig. 5.45, we present plots of Hausdorff distance (HD) error per different level of noise, for the reconstructed models of different approaches. For a fairer comparison between data-driven and other conventional methods, we set fixed values at the parameters, which provide good results in all levels of noise, and we do not search for the ideal values per each parameter and model. Our method avoids the over-smoothing of the models and at the same time, takes into account the special characteristics of the features, succeeding to preserve them.

**Publications that have contributed to this section:**

1. Content of the subsection 5.1.1 has been presented in **J1** and **J4**

2. Content of the subsection 5.1.2 has been presented in **J2**

3. Content of the subsection 5.1.3 has been presented in **C1**, **C3** and **J3**

4. Content of the subsection 5.1.4 has been presented in **C2**

## 5.2 Compression of Static and Dynamic 3D Meshes

Spectral methods for compression have been extensively used in the processing of 3D objects. They usually take advantage of some unique properties that the eigenvalues and the eigenvectors of the decomposed Laplacian matrix have. However, despite their superior behavior and performance, they suffer from computational complexity, especially while the number of vertices of the model increases. On the other hand, isotropic compression methods handle the different geometric components of the whole 3D object in the same way without taking into account features or significant perceptual areas that must be preserved for an optative reconstruction result. In the following subsections, we will present our contribution to the compression of static and dynamic 3D meshes using spectral and feature-aware approaches.

- Compression of static 3D meshes using spectral methods (subsection 5.2.1).
- Feature aware approach for non isotropic compression using PCA (subsection 5.2.3).
- Scalable coding of dynamic 3D meshes (subsection 5.2.4).
- Spatiotemporal compression of dynamic 3D meshes (subsection 5.2.2).

### 5.2.1 Spectral Compression of 3D Meshes Using Dynamic Orthogonal Iterations

To overcome the limitations of the spectral methods, we suggest the use of a fast and efficient spectral processing approach applied to dense static and dynamic 3D meshes, which can be ideally suited for real-time compression applications. To increase the computational efficiency of the method, we exploit potential spectral coherence between adjacent parts of a mesh and then we apply an orthogonal iteration approach for the tracking of the graph Laplacian eigenspaces. Additionally, we present a dynamic version that automatically identifies the optimal subspace size that satisfies a given reconstruction quality threshold. In this way, we overcome the problem of the perceptual distortions, due to the fixed number of subspace sizes that is used for all the separated parts individually.

The decomposition of the graph Laplacian, using a direct SVD implementation, is prohibitive for very dense meshes. To overcome this drawback, several approaches have been presented in the literature, separating the 3D meshes into smaller parts [336] [153] and then they handle each one of these parts separately. Following this line of thought, we suggest the partitioning of the original large mesh into $n_t$ parts using the MeTiS algorithm described in [337]. To be able to directly apply OI, we require to process sequentially a series of matrices of the same size. To that end, we create overlapped equal-sized submeshes of $n_d$ vertices. In this case, the process for the decomposition of the $\mathbf{L}[i] \ \forall \ i = 1, \ldots, n_t$ requires $\mathcal{O}(n_t n_d^3)$ floating-point operations, which is also computational high and not acceptable for use in real-case scenarios. To overcome this problem, minimizing the computational complexity, we suggest using the processing output of a submesh as input for the orthogonal iteration process of a next submesh taking advantage of the coherence between the spectral components of the different submeshes [338], since the initialization of OI to a starting subspace close to the subspace of interest leads to a very fast solution. On the other hand, despite the better time execution performance that the orthogonal iteration approaches have, compared to the direct SVD, the careful selection of an optimal subspace size is necessary in order to simultaneously achieve both the best reconstruction quality and the fastest compression execution times. The assumption, concerning the coherence, is based on the observation that submeshes of the same mesh maintain similar geometric characteristics and connectivity information.

OI is an iterative procedure that computes the singular vectors corresponding to the dominant singular values of a symmetric, non-negative definite matrix [305]. To make the process more computational light, we suggest to preserve the $n_l$ eigenvectors corresponding to the $n_l$ lowest eigenvalue of $\mathbf{U}_{n_l}[i] = [\mathbf{u}_1, \ldots, \mathbf{u}_{n_l}] \in \mathbb{R}^{n_d \times n_l}$ for each $i$ submesh, according to Algorithm 6:

---
**Algorithm 6:** Orthogonal Iteration updating process for the $i$-th submesh

---
1  $\mathbf{U}[i]^{(0)} \leftarrow \mathbf{U}_{n_l}[i-1]$;
2  **for** $t \leftarrow 1$ **to** $t_{max}$ **do**
3  $\quad | \quad \mathbf{U}[i]^{(t)} \leftarrow \mathbf{Onorm}(\mathbf{T}[i]^\zeta \mathbf{U}[i]^{(t-1)})$;
4  **end**
5  $\mathbf{U}_{n_l}[i] \leftarrow \mathbf{U}[i]^{(t)}$;

---

where $\mathbf{T}[i] = (\mathbf{L}[i] + \delta \mathbf{I})^{-1}$ and $\delta$ denotes a very small positive scalar value ensuring the positive definiteness of the matrix $\mathbf{T}[i]$ and matrix $\mathbf{I} \ \mathbb{T}^{n_d \times n_d}$ denotes the identity matrix. The equation:

$$\mathbf{T}[i]^\zeta \mathbf{U}[i]^{(t-1)} \tag{5.42}$$

is estimated very efficiently using sparse linear system solvers, as described in [147]. The value of the power $\zeta$ plays an important role to the converge of the process that will be analyzed in following section. The convergence rate of OI depends on $|\lambda_{n_l+1} / \lambda_{n_l}|^\zeta$ where $\lambda_{n_l+1}$ is the $(n_l + 1)$-st largest eigenvalue of $\mathbf{T}[i]$ [339]. The initial subspace $\mathbf{U}_{n_l}[1]$ has to be orthonormal in order to preserve orthonormality. For this reason, $\mathbf{U}_{n_l}[1]$ is estimated by a direct SVD implementation, while all the following subspaces $\mathbf{U}_{n_l}[i]$, $i = 2, \ldots, n_t$ are estimated by adjustation, as presented in the Algorithm 6.

**Dynamic Orthogonal Iterations for Stable Reconstruction Accuracy**

In use cases where the ground truth model is known beforehand (i.e., compression), we can use this knowledge to provide a dynamic pipeline that automatically identifies the optimal subspace size $n_l$ (i.e., ideally number of remaining low-frequency components) that satisfies a specific quality requirement. This dynamic process takes into account a predefined threshold that determines the preferable perception quality of the reconstructed mesh. When we provide an initialization that is closer to the real solution, then the final results have more perceptual quality and in this way, the error between the reconstructed and the ground truth object is reduced. The method is based on the observation that the feature vectors $\mathbf{U}_{n_l}^T[i]\mathbf{v}[i]$ of each submesh $\mathbf{v}[i]$ has different subspace $\mathbf{U}_{n_l}[i]$ size and it should be carefully selected so that to have the minimum loss of information. We estimate the following mean residual vector $\mathbf{r}(t)$, in order to quantify the loss of information in each $t$ iteration:

$$\mathbf{r}^{(t)} = \sum_{j \in \{x,y,z\}} \left( \mathbf{v}_j[i] - \mathbf{U}_{n_l}[i] \, \mathbf{U}_{n_l}^T[i] \, \mathbf{v}_j[i] \right) \tag{5.43}$$

Then, we assume that when the $l_2$-norm of the metric $\mathbf{r}^{(t)}$ is lower than a given threshold $\|\mathbf{r}^{(t)}\|_2 < \epsilon_h$ then the perceptual loss is decreased and in this case the reconstructed result is assumed as acceptable. To reduce the residual error $\mathbf{r}^{(t)}$, we suggest adding one normalized column in the estimated subspace $\mathbf{U}_{n_l}[i]^{(t)} = [\mathbf{U}_{n_l}[i]^{(t-1)} \ \mathbf{r}^{(t-1)}/\|\mathbf{r}^{(t-1)}\|_2]$ and then perform orthonormalization is estimated according to:

$$\mathbf{U}_{n_l}[i]^{(t)} = \mathbf{Onorm}\left\{ \mathbf{T}^\zeta[i][\mathbf{U}_{n_l}[i]^{(t-1)} \ \frac{\mathbf{r}^{(t-1)}}{\|\mathbf{r}^{(t-1)}\|_2}] \right\} \tag{5.44}$$

On the other hand, if the $l_2$-norm of the metric $\mathbf{r}^{(t)}$ is less than a pre-defined threshold $\epsilon_l$ then the subspace size is decreased by 1 by simply selecting the first $n_{li} - 1$ columns of $\mathbf{U}_{n_l}[i]^{(t)}$. This is an iterative procedure that automatically stops when the metric $\|\mathbf{r}^{(t)}\|_2$ lies between the range of the thresholds $(\epsilon_l, \epsilon_h)$, where the threshold $\epsilon_l$ represents the lowest and $\epsilon_h$ represents the highest allowed value. This means that if the value of $\|\mathbf{r}^{(t)}\|_2$ is lower or higher of the aforementioned range then we need to increase or decrease it, respectively, according to the rules that are clearly presented in the Algorithm 7. To mention here that this process gives the flexibility to a user to easily trade his/her preference between the reconstruction quality and the computational complexity, just changing the values of the preferable thresholds. The Algorithm 7 summarizes the steps of the proposed approach.

---

**Algorithm 7:** Dynamic Orthogonal Iterations applied in the $i^{th}$ submesh

1  $\mathbf{U}[i]^{(0)} \leftarrow \mathbf{U}_{n_l}[i-1]$;
2  **for** $t = 1, \ldots, t_{max}$ **do**
3  $\quad \mathbf{U}[i]^{(t)} = \mathbf{Onorm}(\mathbf{T}[i]^{\zeta} \mathbf{U}[i]^{(t-1)})$;
4  $\quad \mathbf{r}^{(t)} = \sum_{j \in \{x,y,z\}} \left( \mathbf{v}_j[i] - \mathbf{U}_{n_l}[i] \mathbf{U}_{n_l}^T[i] \mathbf{v}_j[i] \right)$
5  $\quad$ **if** $\|\mathbf{r}^{(t)}\|_2 < \epsilon_l$ **then**
6  $\quad\quad \mathbf{U}_{n_l}[i]^{(t)} = [\mathbf{U}_{n_l}[i]^{(t-1)} \; \frac{\mathbf{r}^{(t-1)}}{\|\mathbf{r}^{(t-1)}\|_2}]$; $n_{li} \leftarrow n_{li} + 1$;
7  $\quad$ **else if** $\|\mathbf{r}^{(t)}\|_2 > \epsilon_h$ **then**
8  $\quad\quad n_{li} \leftarrow n_{li} - 1$;
9  $\quad$ **else**
10 $\quad\quad$ break;
11 $\quad$ **end**
12 **end**
13 $\mathbf{U}_{n_l}[i] \leftarrow \mathbf{U}[i]^{(t)}$;

---

**Weighted Average for Mesh Reconstruction and Guarantees of a Smooth Transition**

As we have mentioned earlier, a mesh is separated into different submeshes and then the submeshes are processed separately, using spectral techniques. However, in this case, the final reconstructed model has a loss of quality that is attributed to the dislocation of the vertices lying in the areas where two submeshes have common edges. This phenomenon is known as "edge" effect (see Fig. 5.48) and it requires special treatment in order to be mitigated or eliminated. To overcome this problem, we create overlapped submeshes [336] [153] [12] extending each submesh using also neighboring vertices of the boundary nodes of adjacent submeshes until fulfilling a predefined number of $n_d$ vertices, in total, for all submeshes of the mesh. This operation reduces the error introduced and additionally creates equal-sized submeshes which are necessary for the proceeding of the OI. In Fig. 5.48, we present different segmentation scenarios using MeTis algorithm. Inspecting the second line of this figure, which presents the reconstructed model highlighting the edges of the triangles, it is apparent that the more the parts of the segmentation are, the more apparent the edge effect is.

The edge effect is attributed to missing neighbors inevitably caused by the mesh segmentation. Missing neighbors means missing connectivity which re-sulting in missing entries in the graph Laplacian matrix. However, an efficient way to deal effectively with this limitation is to combine the reconstructed geometry of the overlapped parts. The weights that are assigned to each point are proportional to the degree of the node (e.g., number of neighbors) in the corresponding submesh. Overlapping ensures that each vertex will participate in more than one submesh, and thus the probability of having the same degree (in at least one of them) significantly increases. In Fig. 5.49, we present an example showing the weights assigned to a point (highlighted in red) that participates in three overlapped submeshes. The steps that are followed for the estimation of the weighted average coordinates of the overlapped points are presented in Algorithm 8.

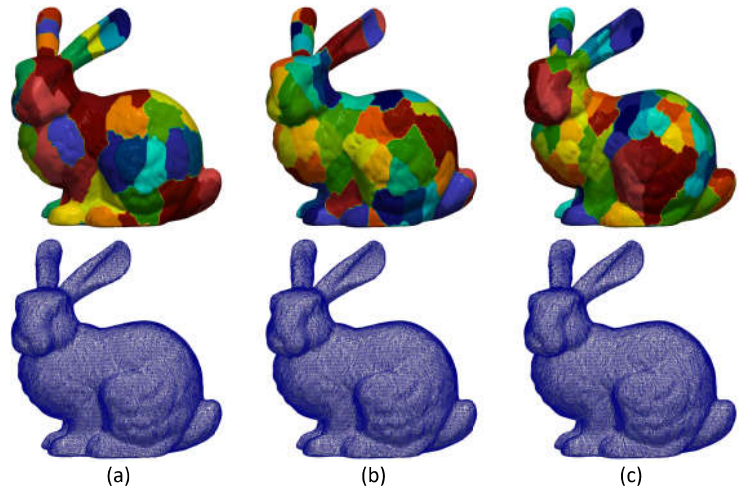

Figure 5.48: (**First line**) Segmentation of bunny model using MeTis algorithm in (**a**) 70, (**b**) 100 and (**c**) 200 parts. (**Second line**) The corresponding reconstructed models without applying overlapping process (edge effect is apparent).

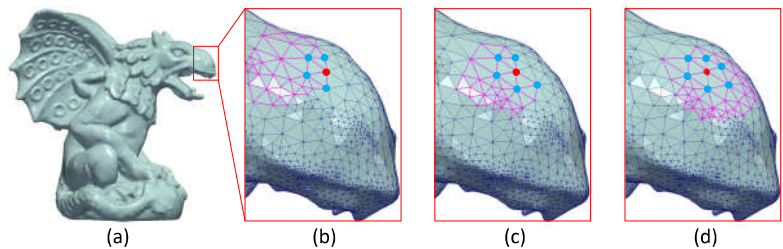

Figure 5.49: The red point has different degree in each submesh of the (**a**) original model (Gargoyle model), the corresponding weights are: (**b**) w = 4, (**c**) w = 5, (**d**) w = 6.

Additionally, we investigate whether the segmentation and the processing

**Algorithm 8:** Weighted average process for the reconstruction of a mesh

1  **for** $i = 1,..,n$ **do**
2     Find the $p_i \geq 1$ overlapped submeshes in which the $i^{th}$ point appears;
3     Set the indices of these $p_i$ submeshes in a vector $\mathbf{q}_i \in \mathbb{R}^{p_i \times 1}$;
4     $\mathbf{sumv}_i = [sumx_i, sumy_i, sumz_i] = [0, 0, 0]$;
5     $sumw_i = 0$;
6     **for** $j \in \mathbf{q}_i$ **do**
7        Find the degree $w_{ij}$ of $i^{th}$ point $\mathbf{v}_i = [x_i, y_i, z_i]$ in the $j^{th}$ submesh;
8        $\mathbf{sumv}_i = \mathbf{sumv}_i + w_{ij}\mathbf{v}_{ij}$;
9        $sumw_i = sumw_i + w_{ij}$;
10    **end**
11    $\tilde{\mathbf{v}}_{ij} = \frac{\mathbf{sumv}_i}{sumw_i}$;
12 **end**

of the overlapped patches guarantee the smooth transition in different cases where edge points belong to flat or sharp areas. At this point, it should be mentioned that the edge points could be part of edges, corners or flat areas. In the following, we present results showing that the way we treat the edge points guarantees, in all the aforementioned cases, a smooth transition successfully mitigating the edge effects.

The process starts using the MeTis algorithm for the identification of the initial parts. Then, each part is extended, using the neighbors of the boundary nodes that belong to adjacent parts until all of them have the same predefined size. Consequently, each boundary point participates in more than one segment. The weights that are assigned to each point, which participates in more than one part, represent its degree (i.e, the number of connected neighbors) in the specific part (see Fig. 5.50). The final position of an edge point is evaluated using the weighted average approach as mentioned above.

We show the distribution of error in the internal and the boundary points of each submesh. For this specific study we consider three different cases that are described below:

- **Non-Overlapping case**, where each node participates in only one part.

- **Overlapping case**, where each part is extended using the neighbors of the boundary nodes that belong to adjacent parts. Thus, each boundary point participates in more than one part, which are reconstructed individually. The final position of a boundary point is evaluated using the simple average of the reconstructed positions.

- **Weighted Overlapping case**, where each part is extended using the neighbors of the boundary nodes that belong to adjacent parts, and the final position of a boundary point is evaluated using a weighted average. The weights assigned to each point that participates in more than one part, represent its degree (i.e, the number of its neighbors) in the specific part.

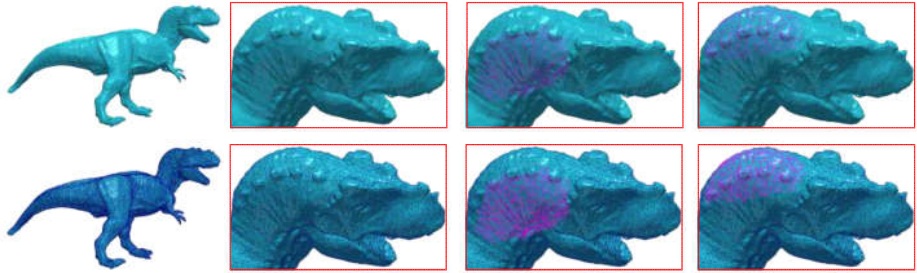

Figure 5.50: Overlapped parts means that each boundary point belongs to more than one part and its degree may vary significantly between different parts.

The standard deviation of the reconstructed error in the internal and the boundary points of each submesh for each one of the aforementioned cases is provided in Fig. 5.51. For the creation of this figure, we used eight models in total (fandisk, armadillo, block, tyra, twelve, bunny, gargoyle, Julio) and we took into account the reconstructed error per each patch of all models. On each box, the central mark is the median, the edges of the box are the 25*th* and 75*th* percentiles, and the whiskers extend to the most extreme data points that are not considered outliers. By inspecting this figure, it can be clearly shown that the weighting scheme guarantees a smooth transition, since the distribution of error in the internal and boundary points has almost identical characteristics, significantly outperforming the other two cases.

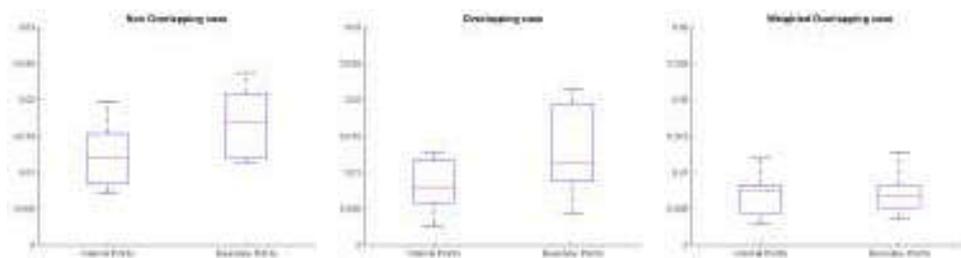

Figure 5.51: Standard deviation of the reconstructed error in the internal and the boundary points of each submesh for each one of the aforementioned cases.

Similar conclusions can be also perceived by observing the Fig. 5.52. In this figure, the results of a denoising step are presented after partitioning Fandisk model in a different number of submeshes (10 , 15 and 20, respectively). It is obvious that the error on the boundary nodes is minimized in the weighted average case, while the segmentation effects are very noticeable in the other two cases.

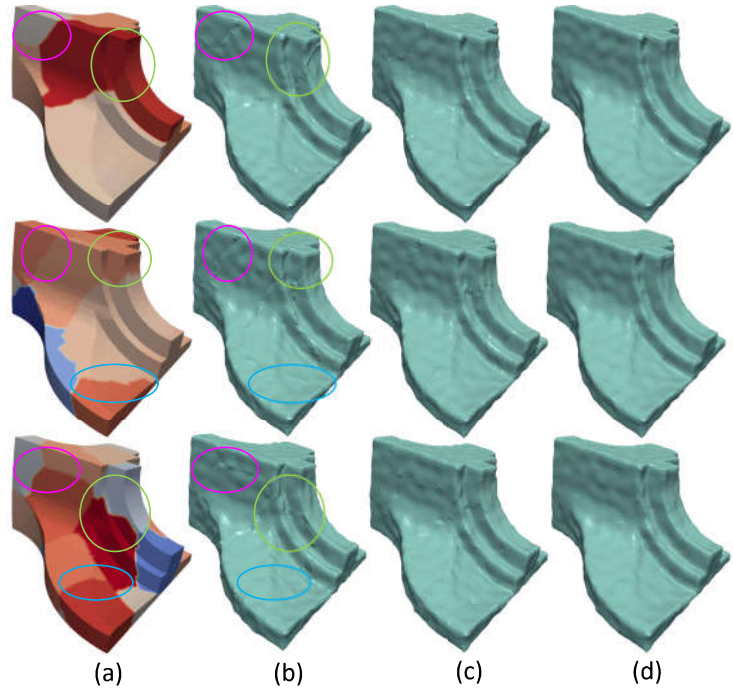

Figure 5.52: (**a**) The model separated in different number of parts (10, 15 and 20, respectively). Additionally, indicative areas have been selected where two or more submeshes are connected; (**b**) Non Overlapping case, the edge effect is apparent in areas where submeshes are connected; (**c**) Overlapping case, the edge effect have been mitigated but have not been eliminated yet. The bigger the number of the partitioning the more intense the problem of the effect; (**d**) Weighted Overlapping case, the results seem to be independent and unaffected of the partitioning (Fandisk $\sigma^2 = 0.2$).

**Spatial Coherence between Submeshes of the Same Mesh**

The previously presented approach, using OI for the estimation of matrices $\mathbf{U}_{n_l}[i] \; \forall \; i = 1, \ldots, n_t$, strongly depends on the assumption that there is a spatial coherence between submeshes of the same mesh. Supposing the correctness of this assumption, the matrix $\mathbf{U}_{n_l}[i-1]$, which is used for initializing Algorithm 11, is the best-related approximation meaning that its form is very close to the real solution. The best-provided initialization matrix has as a result a faster con-vergence, providing at the same time the most reliable results. In this approach, the proposed initialization strategy suggests using as an initial estimation the solution of the previous adjacent submesh. In the following, we will study the validity of this assumption via extensive simulations using different models. Our study is based on the observation that the surface's form of a mesh follows the same pattern, which means that neighboring parts of the same mesh have:

1. Similar connectivity properties (degree and distance).

2. Same geometrical characteristics which are shared between connected points (curvature, small-scale features, texture pattern etc.).

Fig. 5.53 presents colored images representing the Laplacian matrices $\mathbf{T}[i] \: \forall \: i = \{1,2,3\}$ of different submeshes for several 3D models. Providing an easier comparison between the images, we have created matrices of submeshes with the same size $100 \times 100$ so that $\mathbf{T} \in \mathbb{R}^{100 \times 100}$. Each pixel $(x,y)$ of an image represents the corresponding color-coded value of $\mathbf{T}(x,y)$. Additionally, a color bar is also provided showing the range of colors between the lowest and the highest value of each matrix $\mathbf{T}$, where, the deep blue represents the lowest value of each matrix while the bright yellow represents the highest value. We can observe that different submeshes of the same model follow a similar form while they are totally different in comparison with submeshes of different meshes.

Similar conclusions could be perceived by observing Table 5.7. Each row of this table presents the Mean Squared Error (MSE) estimated by the comparison between the random matrix $\mathbf{T}$ of a model, represented as $\mathbf{T}[1]$, and the mean matrix $\tilde{\mathbf{T}}$ of any other model which appears in Fig. 5.53, including the mean matrix of the same model. This comparison is repeated using different models (other rows of this table). For the sake of simplicity, we used only one random matrix $\mathbf{T}[1]$. However, similar results are extracted using any other random matrix of a model.

**Number of Submeshes**

The ideal selected number of submeshes depends on the total number of points of the mesh. Large submeshes create large matrices increasing significantly the processing time since the number of edge points increases. On the other hand, using small submeshes the final results are negatively affected by the edge effects. Table 5.8 shows how the number of segments affects the metric of Mean Normal Difference for both averaging cases (simple and weighted average), where MND represents the average distance from the resulting mesh normals to the ground truth mesh surface.

Table 5.7: Mean squared error between the $\mathbf{T}[1]$ of different models and the mean $\tilde{\mathbf{T}}$ of each model. The lowest value per row is highlighted in bold.

|  | Armadillo $\tilde{\mathbf{T}}$ | Fandisk $\tilde{\mathbf{T}}$ | Sphere $\tilde{\mathbf{T}}$ | Trim star $\tilde{\mathbf{T}}$ | Twelve $\tilde{\mathbf{T}}$ |
|---|---|---|---|---|---|
| **Armadillo** $\mathbf{T}[1]$ | **0.0606** | 13.9720 | 10.0905 | 1.2347 | 37.4199 |
| **Fandisk** $\mathbf{T}[1]$ | 15.6144 | **1.4120** | 11.1582 | 8.4506 | 29.8815 |
| **Sphere** $\mathbf{T}[1]$ | 10.4615 | 11.4700 | **0.8857** | 3.9065 | 26.4125 |
| **Trim star** $\mathbf{T}[1]$ | 1.3122 | 8.5019 | 3.919 | **0.6095** | 29.6996 |
| **Twelve** $\mathbf{T}[1]$ | 37.8481 | 30.2103 | 26.6648 | 30.0618 | **4.5641** |

Table 5.8: Mean Normal Difference using different number of segments (Bunny Model with 34,817 vertices). We also compare the mean normal difference by using normal average and weighted average based on the number of the connected vertices.

| Number of Submeshes | Number of Vertices per Segment | MND Using Simple Average | MND Using Weighted Average |
|---|---|---|---|
| 25 | 1392 | 0.0921 | 0.0915 |
| 40 | ~870 | 0.0931 | 0.0925 |
| 50 | ~696 | 0.0941 | 0.0934 |
| 70 | ~497 | 0.0960 | 0.0952 |
| 100 | ~348 | 0.0988 | 0.0980 |
| 200 | ~174 | 0.1039 | 0.1028 |
| 500 | ~69 | 0.1163 | 0.1150 |

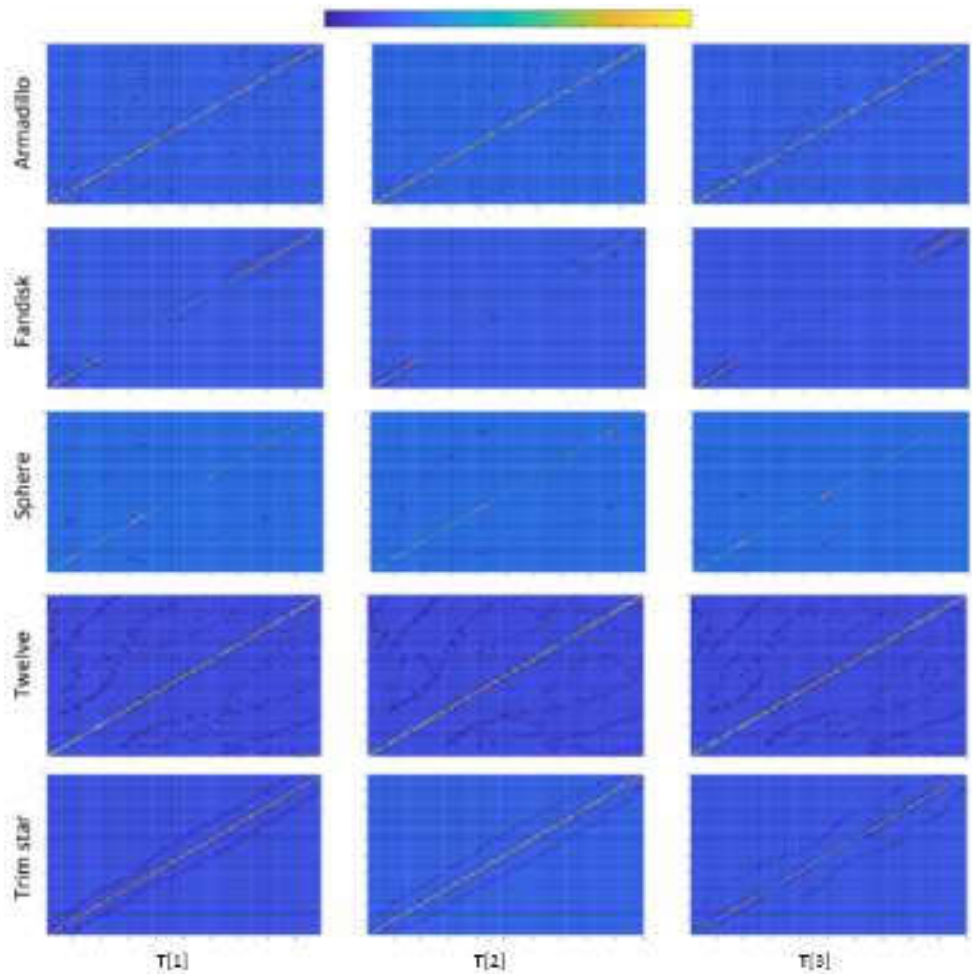

Figure 5.53: Laplacian matrices of different submeshes for different models in color based on the values of their cells. It can be easily observed that different submeshes of the same model follow a similar form while they are totally different in comparison with submeshes of different meshes.

Table 5.9: Mean normal difference using different size of equal-sized overlapped submeshes (Julio Model with 36,201 vertices 70 segments).

| Type of Overlapping | Number of Vertices per Segment | Coarse Denoising MND | Fine Denoising MND |
|:---:|:---:|:---:|:---:|
| max | 532 | 0.1248 | 0.1176 |
| 1.05 · max | 558 | 0.1228 | 0.1173 |
| 1.10 · max | 585 | 0.1203 | 0.1172 |
| 1.15 · max | 611 | 0.1188 | 0.1169 |
| 1.20 · max | 638 | 0.1174 | 0.1169 |
| 1.25 · max | 665 | 0.1164 | 0.1164 |

Table 5.10: Mean normal difference using different size of equal-sized overlapped submeshes (Julio Model with 36,201 vertices 100 segments).

| Type of Overlapping | Number of Vertices per Segment | Coarse Denoising MND | Fine Denoising MND |
|:---:|:---:|:---:|:---:|
| max | 372 | 0.1276 | 0.1189 |
| 1.05 · max | 390 | 0.1248 | 0.1187 |
| 1.10 · max | 409 | 0.1228 | 0.1185 |
| 1.15 · max | 427 | 0.1208 | 0.1183 |
| 1.20 · max | 446 | 0.1184 | 0.1174 |
| 1.25 · max | 465 | 0.1175 | 0.1168 |

**Size of Overlapped Submeshes**

The real motivation behind the processing in parts is strongly supported by the existence of a great amount of state-of-the-art applications in which large 3D models cannot be scanned at once using portable 3D scanners. As a result, the output of the sequential scanning would be a sequence of submeshes that arrive sequentially in time. An extensive evaluation study carried out using different overlapped sizes (Tables 5.9–5.11) showed that the reconstruction quality is strongly affected by the size of the submeshes themselves rather than the number of overlapped vertices.

Regarding the ideal size of the overlapped patches, we investigated the effect of using different sizes of overlapped submeshes in a range from 5 to 25% of the maximum submeshes length, in the quality of the reconstructed model. More specifically, as shown in Tables 5.10 and 5.11 and in Fig. 5.54, the mean normal difference and the visual smoothed results have not significant differences between the different case studies, especially for percentages up to 10% of the max segment. Additionally, if we consider the fact that this process takes place in the coarse denoising step we can conceive the negligible contribution of the overlapped submeshes size to the final denoising results.

Table 5.11: Mean normal difference using different size of equal-sized overlapped submeshes (Julio Model with 36,201 vertices 50 segments).

| Type of Overlapping | Number of Vertices per Segment | Coarse Denoising MND | Fine Denoising MND |
|---|---|---|---|
| max | 741 | 0.1229 | 0.1167 |
| 1.05 · max | 778 | 0.1207 | 0.1166 |
| 1.10 · max | 815 | 0.1188 | 0.1163 |
| 1.15 · max | 852 | 0.1172 | 0.1160 |
| 1.20 · max | 889 | 0.1160 | 0.1159 |
| 1.25 · max | 926 | 0.1159 | 0.1158 |

By inspecting the results, we can definitely state that the number and size of segments are much more important than the size of the overlapped patches. The overlapping process mainly contributes in the case of on-the-edge points helping for a more accurate estimation of their position by creating full-connected points. A sufficient overlapping size corresponds to the 15% of the total points in the submesh.

Fig. 5.54 illustrates the reconstruction results of the coarse denoising step using 70 overlapped submeshes consisting of a different number of vertices in each case. As we can observe, in cases where the number of overlapping vertices is higher than 15% of the total number of submesh points then the reconstructed results are almost identical with the 15% case.

**Block-Based Spectral Compression and Experimental Analysis**

The spectral compression methods utilize the subspace of the eigenvector matrix $\mathbf{U}_{n_l}[i]$ for encoding the geometry of a 3D mesh. This matrix can be computed by a direct SVD implementation or by executing a number of orthogonal iterations on $\mathbf{T}^z[i]$, and it is used as the encoding dictionary to provide a compact representation of the vertices of each submesh.

- At the encoder: The coordinates $\mathbf{v}_x[i] \in \mathbb{R}^{n_d \times 1}$ are projected to the dictionary and we finally take the feature vector $\mathbf{U}_{n_l}^T[i]\mathbf{v}[i]$, where $n_l << n_{d_i}$.

- At the decoder: The inverse process takes place, the vertices of the original 3D mesh are reconstructed from the feature vector and the dictionary $\mathbf{U}_{n_l}[i]$ according to:
$$\tilde{\mathbf{v}}[i] = \mathbf{U}_{n_l}[i]\mathbf{U}_{n_l}^T[i]\mathbf{v}[i] \tag{5.45}$$

The sender transmits only the connectivity of the mesh and the $n_l$ respective spectral coefficients of each block. On the other hand, the receiver evaluates the dictionary $\mathbf{U}_{n_l}[i]$, based on the received connectivity, and uses the spectral coefficients to retrieve the coordinates of the original mesh $\hat{\mathbf{x}}, \hat{\mathbf{y}}, \hat{\mathbf{z}}$ [155]. To mention here that the subspace size $n_l$ has a fixed value in the case of OI, providing fast execution times but having a lack of reconstruction accuracy. On the other hand, the DOI approach provides reconstructed results with high and stable reconstruction quality, since it searches for the "ideal" subspace size, but as a result, it adds an extra computational cost.

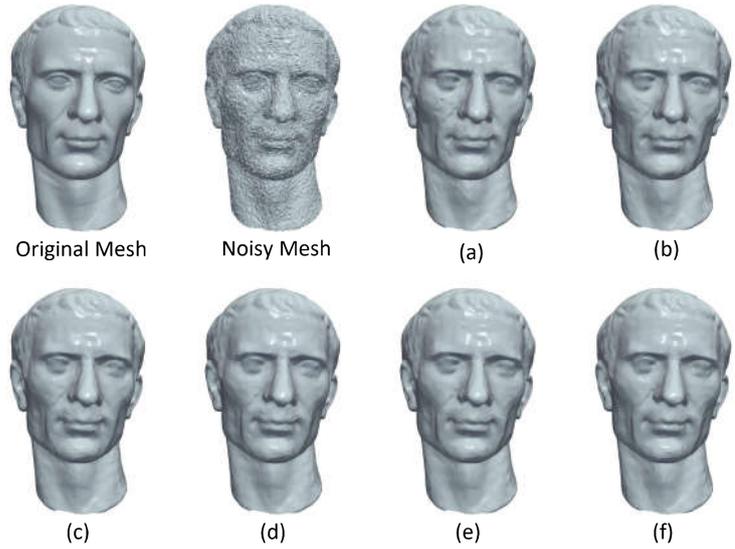

Figure 5.54: Coarse denoising meshes with 70 equal-sized overlapped submeshes consisting of (**a**) 532 vertices (max), (**b**) 558 vertices (1.05 · max), (**c**) 585 vertices (1.10 · max), (**d**) 611 vertices (1.15 · max), (**e**) 638 vertices (1.20 · max), (**f**) 665 vertices (1.25 · max).

Fig. 5.55 shows how the selected rate of bpv affects the metric NMSVE for different compared approaches. We also provide the execution times, next to each line, that encapsulates the respective time needed to run each method (e.g., to construct the matrix $\mathbf{T}^\zeta, \zeta \geq 1$ and execute the OI). As we can also conclude by observing this figure, OI performs almost the same reconstructed quality with the SVD method, in considerably less time, since it can be executed up to 20 times faster. It is obvious that the more the number of the iterations of OI, the better the reconstructed accuracy of the model, converging towards the (optimal) SVD result. Obviously, this strategy increases the total execution time due to the more iterations, however, the total execution of OI still remains much faster than this one of the direct SVD.

For the case of DOI, there is an increase in the execution time, in comparison to OI. However, it still is significantly faster than the SVD (it needs almost half time). On the other hand, there is a significant increase in the final compression rate bpv, which is captured as a right shifting of the plot in Fig. 5.55. The shifting is more obvious when the initial value of $n_l$ is small which means that more iterations are necessary for achieving the satisfying accuracy. The theoretical complexity of the proposed schemes is in tandem with the measured time. More specifically, the OI approach for the "Bunny" model can be executed much faster than the direct SVD approach. While running more OI iterations yields a better NMSVE, converging towards the (optimal) SVD result, it comes at the cost of a linear increase in the decoding time. On the other hand, one iteration of $\mathbf{T}^4$ achieves the same visual error as executing four OI, in considerably less time. Fig. 5.56 plots the squared error per each submesh of the Bunny model (70 submeshes in total). Each presented approach has a different reconstruction performance in different submeshes except for the DOI that provides a stable reconstruction accuracy due to the "ideal" value of subspace size that is required to satisfy a predefined reconstruction quality threshold. Fig. 5.57 presents the heatmap visualization of the normals' difference between the ground truth and the reconstructed models for different OI approaches and SVD.

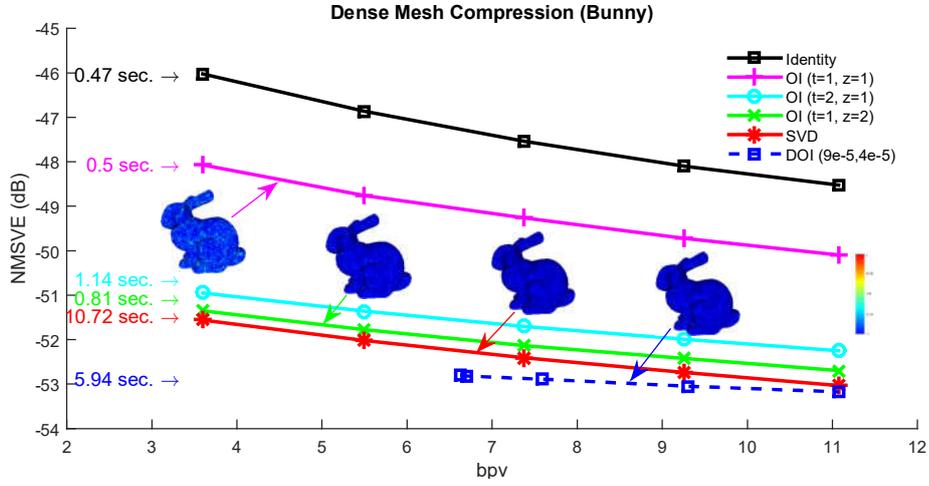

Figure 5.55: Normalized mean square visual error (NMSVE) of the reconstructed models per different bpv for different compared approaches.

The plot of bpv vs. NMSVE for the "Dragon" model is shown in Fig. 5.58(a). Note that the execution times shown next to each line encapsulate the respective time needed to construct $\mathbf{T}^\zeta, \zeta \geq 1$, and to run the respective number of OI, with the speed-up as compared to SVD shown in parenthesis. By inspecting the figure, it can be easily concluded that the quality of the OI method performs almost the same as with SVD, especially when the number of iterations increases. Additionally, in Fig. 5.58(b), we provide the heatmap visualization of the normals' difference between the ground truth and the reconstructed models for different OI approaches and SVD.

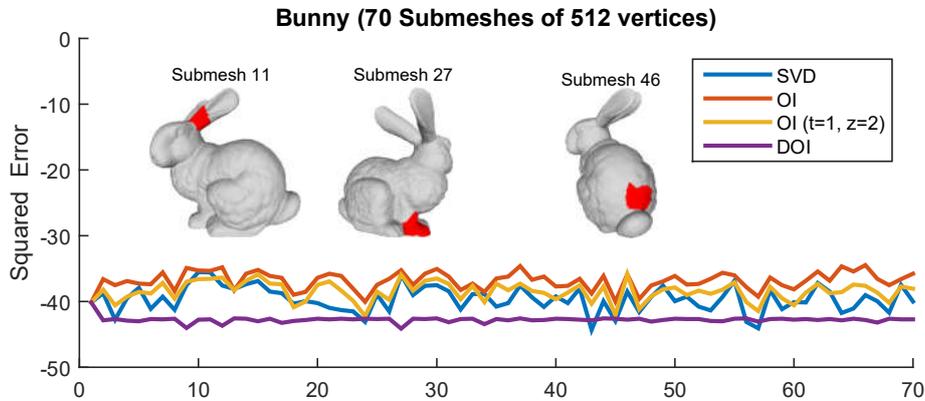

Figure 5.56: Squared error per each submesh for different approaches.

**Comparisons of the Execution Times with a Relative Method**

In this paragraph, we present the execution time effectiveness of our method in comparison with the relevant method of Vallet and Levy [340]. The main contribution of Vallet and Levy's method is the development of an efficient numerical mechanism that computes the eigenfunctions of the Laplacian. The eigenfunctions are computed band by band based on spectral transforms and an efficient eigensolver, and also using an out-of-core implementation that can compute thousands of eigenvectors for meshes with up to a million vertices. They also propose a limited-memory filtering algorithm, that does not need to store the eigenvectors. Vallet and Levy's method is very fast, especially in comparison with the traditional SVD decomposition and it also shares a lot of common ideas with our method, trying to solve a similar problem. The main similarity between Vallet and Levy's method and our approach is that both of them can be used as low-pass filtering. Nevertheless, Vallet and Levy's method has some limitations that our method can efficiently handle and overcome. More specifically:

- Their method is not able to preserve the creases and as a future extension, they suggested the use of eigenfunctions of an anisotropic version of the Laplace operator that could improve the frequency localization of the creases and therefore to better preserve them when filtering. We overcome this limitation by using an extra stage of processing (called as a fine step) that handles each area with an anisotropic way taking into account the different geometrical characteristics of small surfaces (e.g., creases, corners, edges, etc.)

- Another limitation is the fact that Vallet and Levy's method cannot be directly applied to mesh compression since they took particular care, making their Laplacian geometry dependent. On the other hand, our method

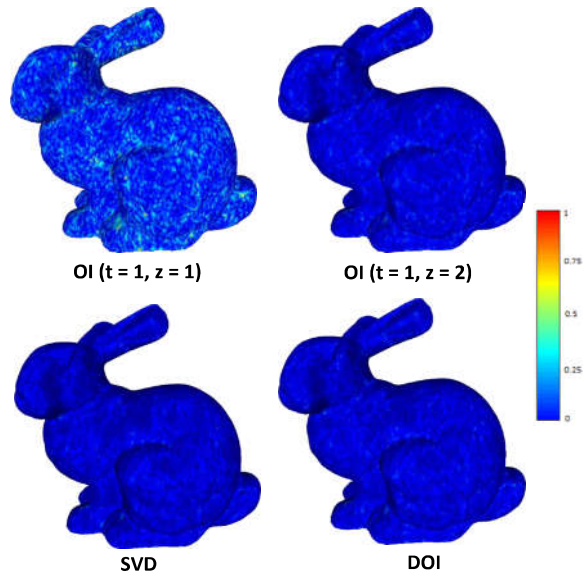

Figure 5.57: Heatmap visualazation of normal difference with bpv 5.64 for different reconstructed approaches.

- can be efficiently used for mesh compression as also many experimental results can verify.

- Regarding the performance of the computational complexity, Vallet and Levy's forecast that partitioning also partially fixes the problem of spatial localization at the expense of losing continuity (this is also why JPEG uses small blocks instead of applying the DCT to the whole image). Their suggestion can be verified by our implementation since we achieve tremendously faster execution times by participating in patches the whole 3D mesh and proceed them separately.

Fig. 5.59 depicts plots that show the execution times of two OI approaches (i.e., OI ($t = 1, \zeta = 1$) and OI ($t = 1, \zeta = 4$)) in comparison with the execution times of the Manifold Harmonics Basis and Limited-memory Filtering, as presented in [340]. The main reason why our method is much faster than other decomposition methods is due to the fact that it handles many but much smaller matrices (of submeshes) than the large Laplacian matrix of the initial whole mesh. The execution time to decompose a matrix exponentially increases as the dimension of the matrix also increases. On the other hand, the cumulative time to decompose many but small matrices is significantly lower.

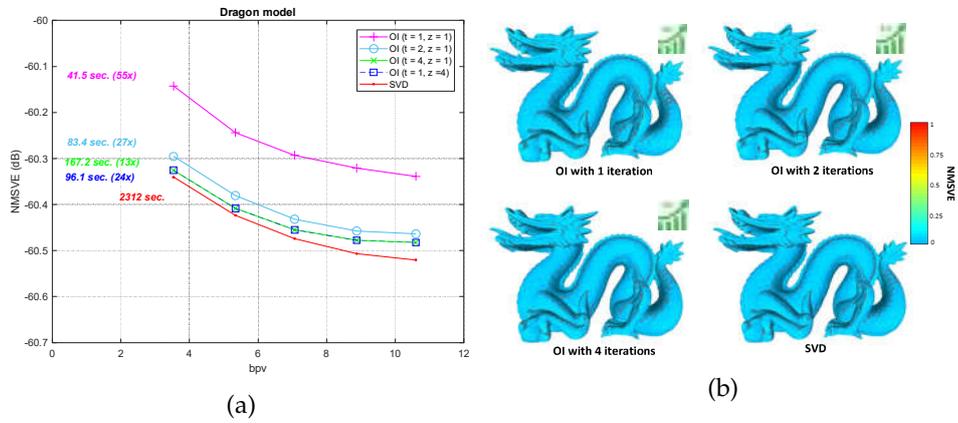

Figure 5.58: (**a**) NMSVE vs bpv plot of the Dragon model (437,645 vertices); (**b**) normal difference with bpv 7.06. The arrow represents the speed-up in execution time compared to SVD.

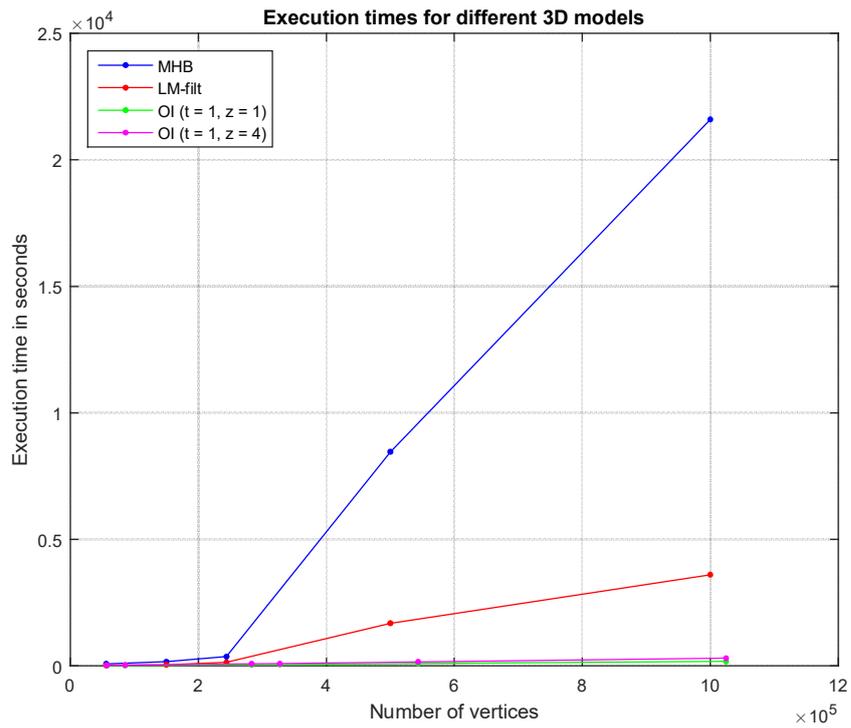

Figure 5.59: Execution times for different 3D models.

### 5.2.2 Spatio-temporal Compression

In this subsection, an approach for dynamic 3D mesh compression is presented, that effectively exploits the spatio-temporal coherence of animated sequences, achieving low compression ratios without noticeably affecting the visual quality of the animation. We show that, on contrary to mainstream approaches that either exploit spatial (e.g., spectral coding) or temporal redundancies (e.g., PCA-based method), the proposed scheme, achieves increased efficiency, by projecting the differential coordinates sequence to the subspace of the covariance of the point trajectories. The method formulates a spatio-temporal matrix and takes advantage of the geometric and temporal coherences that consecutive differential frame coordinates of the same animation have, by projecting the spatio-temporal matrix of the $\delta$ coordinates on the eigen-trajectories, we can derive superior spatio-temporal compression. The pipeline of the proposed method is presented in Fig. 5.60.

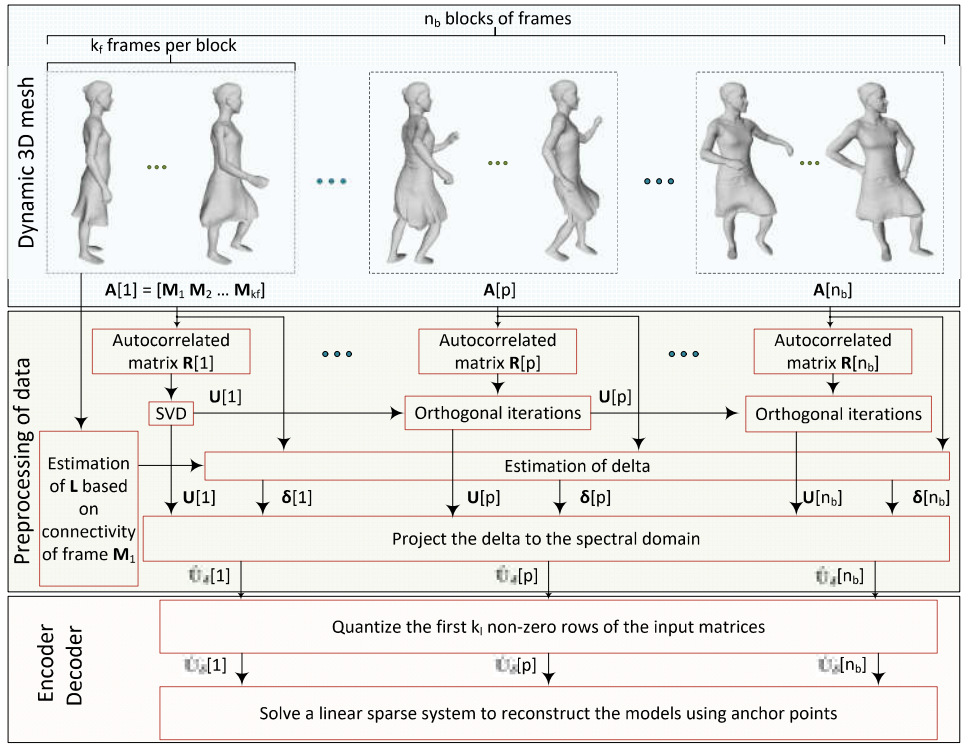

Figure 5.60: Pipeline of the proposed method for the compression of dynamic 3D meshes using spatio-temporal information.

We start by formulating the spatio-temporal matrices $\mathbf{A}_x, \mathbf{A}_y, \mathbf{A}_z \in \mathbb{R}^{n_s \times n}$, where the matrix $\mathbf{A}_x$ consists of the $x$ coordinates of the $n$ vertices of the $n_s$ frames of the animated mesh. To mention here that the following process is applied three times, one for each coordinate matrix, however, for the sake of simplicity, we will present it here once, only for the $x$ coordinate. Firstly, we estimate the autocorrelation matrix $\mathbf{R}_x \in \mathbb{R}^{n_s \times n_s}$

$$\mathbf{R}_x = \mathbf{A}_x \mathbf{A}_x^T \tag{5.46}$$

and we decompose it, via Eq. (4.3), in order to estimate the orthonormal matrix $\mathbf{U}_x \in \mathbb{R}^{n_s \times n_s}$ of the eigenvectors. We also estimate the delta coordinates of the spatio-temporal matrix $\mathbf{A}_x$:

$$\boldsymbol{\delta}_x = \mathbf{L}\mathbf{A}_x^T \in \mathbb{R}^{n \times n_s} \tag{5.47}$$

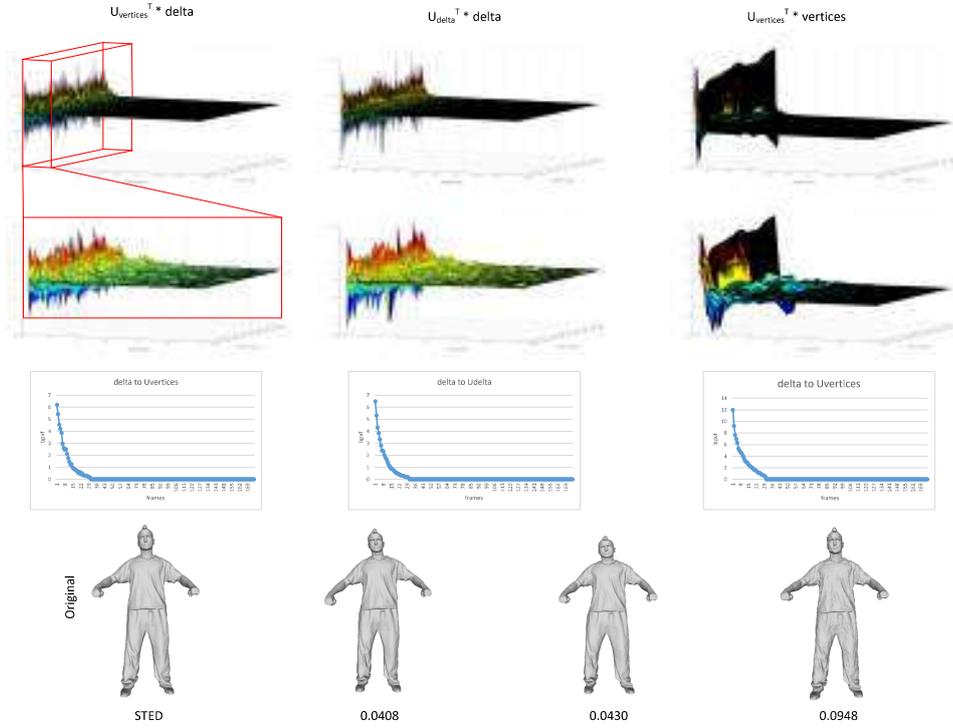

Figure 5.61: Visual representation of matrix $\hat{\mathbf{U}}_{\delta_x}$, and enlarged detail showing how sparse it is.

We project the delta coordinates to the matrix of the eigenvectors $\mathbf{U}_x$, using the GFT function:

$$\hat{\mathbf{U}}_{\delta_x} = \mathcal{T}(\boldsymbol{\delta}_x^T) = \mathbf{U}_x^T \boldsymbol{\delta}_x^T \in \mathbb{R}^{n_s \times n} \tag{5.48}$$

**Encoder:** The matrix $\hat{\mathbf{U}}_{\delta_x}$ represents the information that we want to compress (e.g., for storage, transmission or any other purposes). We take advantage of the observation that the visual representation of matrix $\hat{\mathbf{U}}_{\delta_x}$ is very sparse (i.e., a lot of values are close to zero), as shown in Fig. 5.61. So, we assume that we can encode only the values of its first $k_l$ rows without losing a lot of information, where $k_l < n_s$. The values of the rest rows are replaced with zeros (or in other words, we encode them using 0 bits).

$$\dot{\mathbf{U}}_{\delta_x i} = \begin{cases} Q(\hat{\mathbf{U}}_{\delta_x i}), & \text{if row } i < k_l \\ 0 & \text{otherwise} \end{cases} \quad (5.49)$$

where $Q(.)$ denotes the quantization function. Fig. 5.62 shows how the selection of the $k_l$ value affects the reconstructed results. Comparing the Figs. 5.62-(d) and 5.62-(e), we can conclude that after a current value of $k_l$ (e.g., $> 20$), maintaining more information does not mean better visual results.

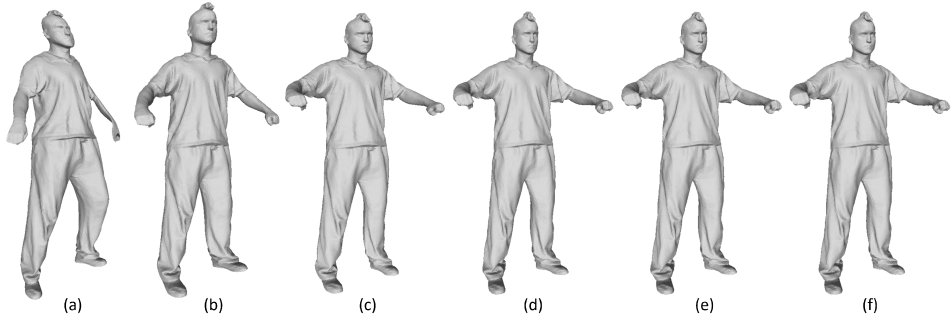

(a) (b) (c) (d) (e) (f)

Figure 5.62: Reconstructed results of "Handstand" model, by encoding only the first (a) 2, (b) 5, (c) 10, (d) 20, (e) 30 out of 175 rows of the matrix $\hat{\mathbf{U}}_{\delta_x}$, (f) original 3D mesh.

**Decoder:** The decoder has to solve the following Eq. (5.50) for the reconstruction of matrix $\mathbf{A}_x$.

$$\tilde{\mathbf{A}}_x = \mathbf{L}^{-1}(\mathbf{U}_x \dot{\mathbf{U}}_{\delta_x})^T \in \mathbb{R}^{n \times n_s} \quad (5.50)$$

where $\tilde{\mathbf{A}}_x$ denotes the reconstructed matrix, $\mathbf{L}$ can be estimated by the known connectivity, which is also the same for all frames, and $\mathbf{U}_x$ is a dictionary matrix which is assumed as known.

Solving the Eq. (5.50), we take an acceptable solution. However, to further increase the level of detail of the reconstructed 3D meshes, we can also quantize a set of $n_a$ uniformly distributed vertices, known as anchor points, where $n_a$ usually corresponds to the 1% of the total number of vertices $n$. The indices of these vertices represented by the set $\mathcal{N}$. The reconstruction of the 3D mesh vertices is performed as in [147], by solving the following sparse linear system:

$$\begin{bmatrix} \mathbf{L} \\ \mathbf{I}_{n_a} \end{bmatrix} \tilde{\mathbf{A}}_x = \begin{bmatrix} \mathbf{U}_x \dot{\mathbf{U}}_{\delta_x} \\ \mathbf{I}_{n_a} Q(\mathbf{A}_x^T) \end{bmatrix} \quad (5.51)$$

where $\mathbf{I}_{n_a} \in \mathbb{R}^{n_a \times n}$ is a subset of the identity matrix $\mathbf{I} \in \mathbb{R}^{n \times n}$, since its rows has been constructed by those $i$ rows of $\mathbf{I}$ where $i \in \mathcal{N}$.

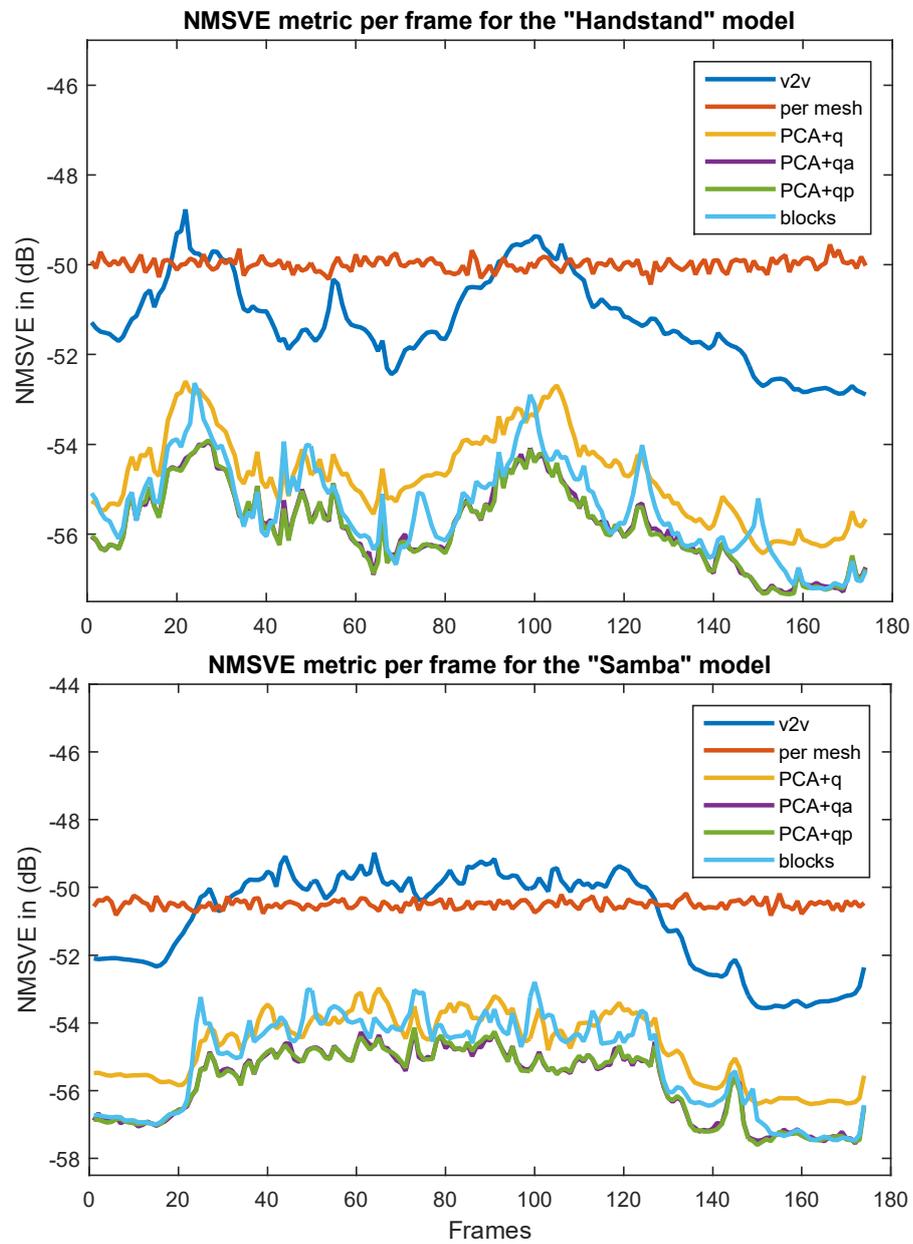

Figure 5.63: NMSVE metric per each frame for different approaches and models.

**Speed-up Process using OI for Online Procedure**

To make the process more computationally light, we suggest separating the dynamic 3D mesh into $n_b$ equal-sized blocks of $k_f = n_s/n_b$ consecutive meshes and then using OI. In this case, the dimensions of the matrix $\mathbf{U}_x[1]$, which is the only matrix that is decomposed via SVD, is $k_f \times k_f \ll n_s \times n_s$, making the execution of the process to be very fast. Additionally, the reconstruction of the animation in blocks is also faster due to the lower dimensions of the proceeding matrices.

The assumption, concerning the coherence, is based on the observation that blocks of meshes of the same 3D animation maintain both geometric (same connectivity and geometrical features) and temporal characteristics (similar motion between consecutive frames). This coherence leads to a very fast solution since the initialization of the OI starts to a subspace which is very close (i.e., coherent) to the real one. It is worth mentioning here that the OI is a solution that also can be used in real-time applications since both matrix multiplications and QR factorizations have been highly optimized for maximum efficiency on modern serial and parallel architectures.

Nevertheless, despite the fast execution of this approach, the accuracy of the reconstructed results is inferior to these of using the whole sequence of meshes in one block. The advantages of implementing this approach are more apparent when the animated 3D model consists of many frames $> 1000$ or when the application priority is a fast implementation instead of a very accurate reconstruction. The case of using the whole animated mesh as input could be assumed as a sub-case of this pipeline (Fig. 5.60) where $n_b = 1$.

**Experimental Analysis and Results**

For the comparisons, we use different variants that either exploit temporal and/or spatial coherence's. More specifically, we present the results by using: (i) Projection of the delta coordinates to the eigen-trajectories (PCA), (ii) Projection of the quantized delta coordinates to the eigen-trajectories (PCA+q), (iii) PCA and then quantization to the projected vertices (v2v) [341], (iv) (PCA+q) and serial reconstruction using anchor points (PCA+qs), (v) (PCA+q) and parallel reconstruction using anchor points (PCA+qp), (vi) OI into blocks of frames and parallel reconstruction using anchor points (blocks), (vii) per mesh GFT without taking advantage of the temporal information (per mesh) [332].

In Fig. 5.63, we present the NMSVE metric [155] per each frame for different approaches and models. PCA+qs and PCA+qp approaches provide the best results. We can also see that the NMSVEs, for these methods that exploit the temporal coherence's, follow a similar pattern. This means that the quality of the reconstruction depends on the type of motion of the dynamic model. On the other hand, the "per mesh" approaches provide reconstructed results with almost the same NMSVE value per each frame.

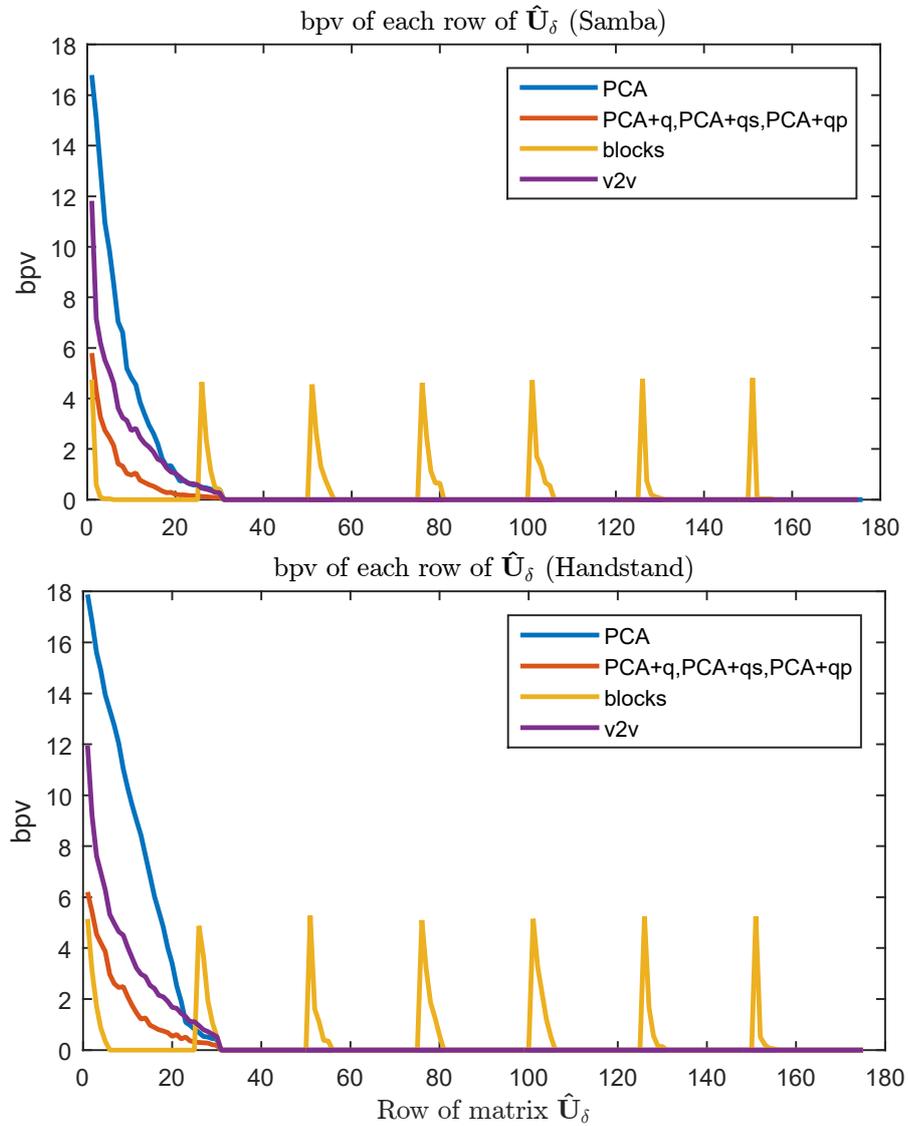

Figure 5.64: bpv for different approaches.

Fig. 5.64 shows the bpv for each row of the matrix $\hat{\mathbf{U}}_\delta$ for different approaches. For most of the methods, only the first $k_l$ rows of the matrix are quantized with bpv $> 0$. In the case of the "block"-based approach, we need to quantize the first rows of each block separately, increasing however in this way the final bvpf. Additionally, we can observe that in blocks, consisting of similar frames (e.g., the last blocks of these animations where the frames are significantly similar due to a static motion), the required bits in order to achieve the same reconstruction error are less than in other blocks capturing fast motions.

| Model | Q | PCA | PCA+q | v2v | blocks | PCA+qs PCA+qp |
|---|---|---|---|---|---|---|
| Hand-stand | $q$ | 1.2215 | 0.2893 | 0.5783 | 0.4006 | 0.2893 |
| | $q_a$ | - | - | 0.1600 | 0.1600 | 0.1600 |
| | $q_d$ | 0.2799 | 0.2799 | 0.2799 | 0.1847 | 0.2799 |
| | $q_s$ | 1.5014 | 0.5692 | 1.0182 | 0.7453 | 0.7292 |
| Samba | $q$ | 0.7289 | 0.1827 | 0.4362 | 0.3057 | 0.1827 |
| | $q_a$ | - | - | 0.1595 | 0.1595 | 0.1595 |
| | $q_d$ | 0.2808 | 0.2808 | 0.2808 | 0.1498 | 0.2808 |
| | $q_s$ | 1.0088 | 0.4635 | 0.8761 | 0.6155 | 0.6226 |

Table 5.12: Total $q_s$ bpvf for different approaches.

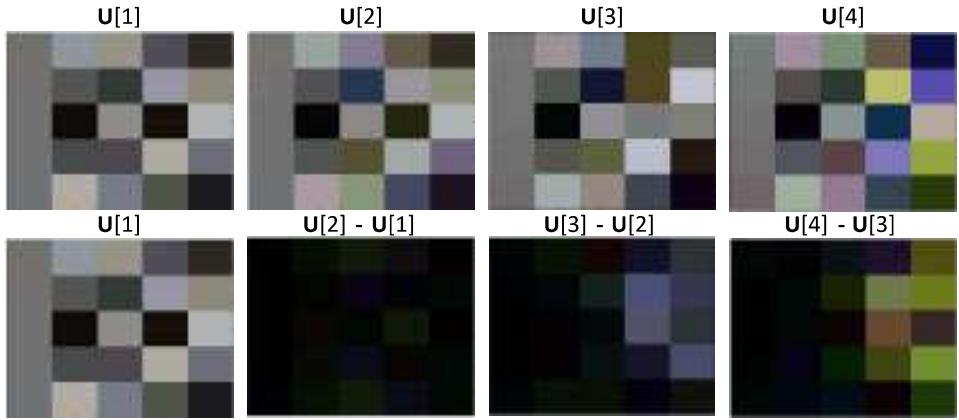

Figure 5.65: Consecutive dictionary matrices of handstand model and the corresponding difference matrices.

In Table 5.12, we present the bpvf for the quantization bits $q$ of the matrix elements $\hat{\mathbf{U}}_\delta$ according to Eq. (5.49), or based on the average estimation of Fig. 5.64. We also present the bpvf for the quantization of the anchor points $q_a$, the dictionary matrix $q_d$, as well as the total bpvf $q_s$ which represent the summary of all the quantization bits, as presented in Eq. (5.52). We assume that for a lossless quantization, both each anchor point and each element of a dictionary matrix is represented using 16 bits.

$$q = \frac{bits \cdot n \cdot n_s}{n \cdot n_s}, \; q_a = \frac{bits \cdot n_a \cdot n_s}{n \cdot n_s}, \; q_d = \frac{bits \cdot n_b^2 \cdot k_f \cdot k_f}{n \cdot n_s} \quad (5.52)$$
$$q_s = q + q_a + q_d$$

At this point it should be mentioned, that for the "block"-based approach, we encode the difference $\mathbf{U}[i+1] - \mathbf{U}[i]$ between two consecutive matrices (second row of Fig. 5.65), instead of the original dictionary matrices $\mathbf{U}[i]$, taking advantage of the spatial coherency between the consecutive dictionaries. This information is easier to be quantized using fewer bits.

Table 5.13 presents the execution times during the data processing and encoding phase (first row) and the reconstruction phase (second row). The "blocks" approach provides the fastest solution in both cases. For meshes with more than $> 20,000$ vertices the "per mesh" approach can not be implemented due to the high computational cost for estimating the per frame SVD decomposition.

| Models (n/$n_s$) | Per mesh | PCA+q | PCA+qs | PCA+qp | blocks |
|---|---|---|---|---|---|
| **Handstand** [317] | 39557.42 | 0.89 | 0.89 | 0.89 | 0.07 |
| (10002/175) | 4736.25 | 26.47 | 4736.25 | 28.67 | 22.12 |
| **Dinosaur** [316] | - | 0.69 | 0.69 | 0.69 | 0.06 |
| (20218/152) | - | 40.63 | 6920.56 | 45.53 | 37.19 |
| **Samba** [317] | 37314.55 | 0.87 | 0.87 | 0.87 | 0.07 |
| (9971/175) | 4214.55 | 24.86 | 4214.55 | 26.17 | 20.89 |
| **Chinchilla** [316] | 22.49 | 0.17 | 0.17 | 0.10 | 0.07 |
| (4307/84) | 476.04 | 4.41 | 476.04 | 4.68 | 4.22 |
| **Camel-g.** [342] | - | 0.12 | 0.12 | 0.12 | 0.06 |
| (21887/48) | - | 54.27 | 2730.24 | 56.88 | 51.27 |
| **Flag** | 7062.21 | 2.38 | 2.38 | 2.38 | 0.07 |
| (2704/1000) | 2633.36 | 2.22 | 2633.36 | 2.37 | 1.84 |

Table 5.13: Execution times for the compression and reconstruction of different models.

Fig. 5.66 depicts the reconstructed results and additional enlarged details of the models for easier visual comparison among the methods. We also provide the STED [343] metric that is ideal for evaluating the quality of the reconstructed animation and a heatmap visualization of the angle $\theta$ representing the difference between surface normals of the reconstructed and the original models. The PCA+qs and the PCA+qp approaches have very similar reconstructed results, as we can see in Figs. 5.63 and 5.66, however the PCA+qs method is much slower (Table 5.13) since it needs to solve the Eq. (5.51) in $n_s$ times. Additionally, it appears temporal artifacts that are apparent only in the animation mode (non in a static figure).

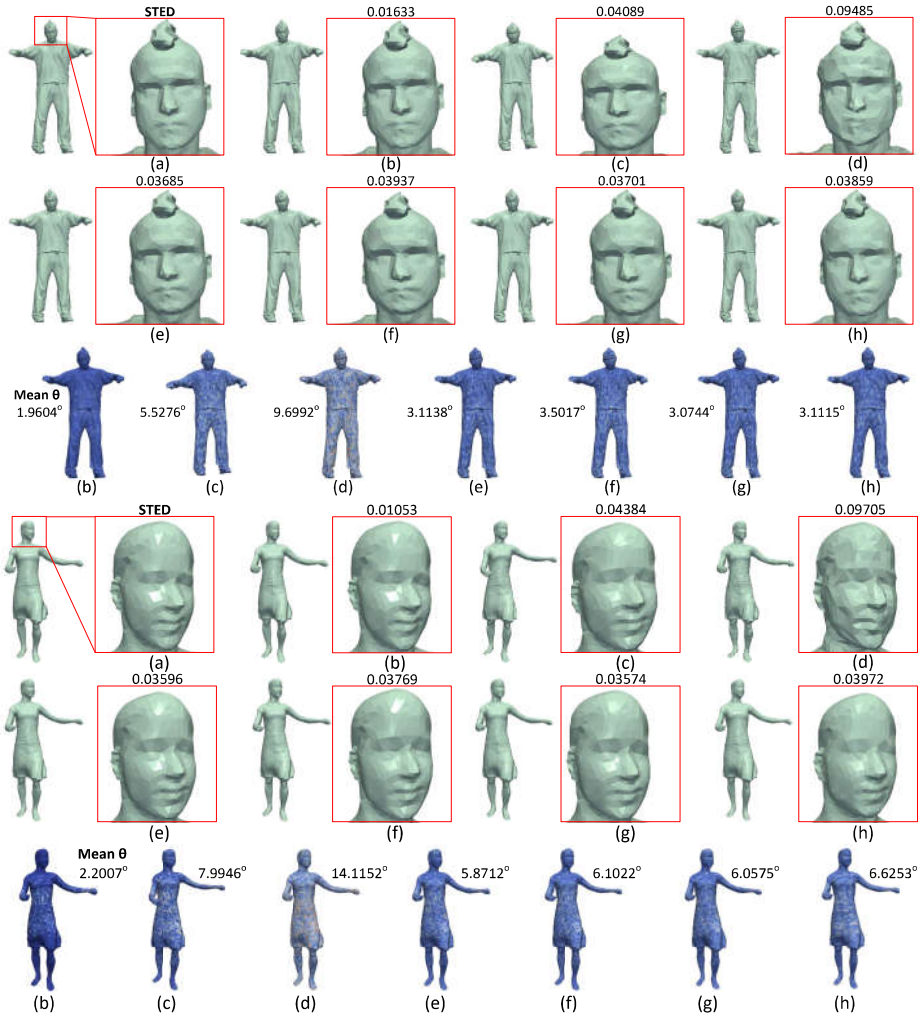

Figure 5.66: (a) Original models and reconstructed in (b) PCA, (c) PCA+q, (d) v2v, (e) PCA+qs, (f) PCA+qp using 0.5% of anchor points, (g) PCA+qp using 1% of anchor points, (h) "blocks" approach using 1% of anchor points.

### 5.2.3 Feature Aware 3D Mesh Compression using Robust Principal Component Analysis

Feature identification is a very important task for distinguishing high-frequency features, such as sharp edges and corners from surface noise. Although there are a number of works on detecting noise and sharp features in static meshes, their robustness is significantly deteriorated in the presence of noise, as it is also shown in Fig. 5.67. A progressive compression scheme can enable aggressive compression ratios, by successfully identifying and encoding sharp and small-scale geometric features. The accurate identification of the features can be achieved by exploiting the low-rank property of the captured geometry and the sparsity of the features in the Laplacian domain, permeating benefits from RPCA. Due to the visual importance of the identified geometric features, the geometry coding process is optimized for preserving the geometric features at extremely low bit rates. The proposed feature-aware high pass quantization method achieves extremely high compression ratios, offering at the same time meaningful approximations of the given surfaces.

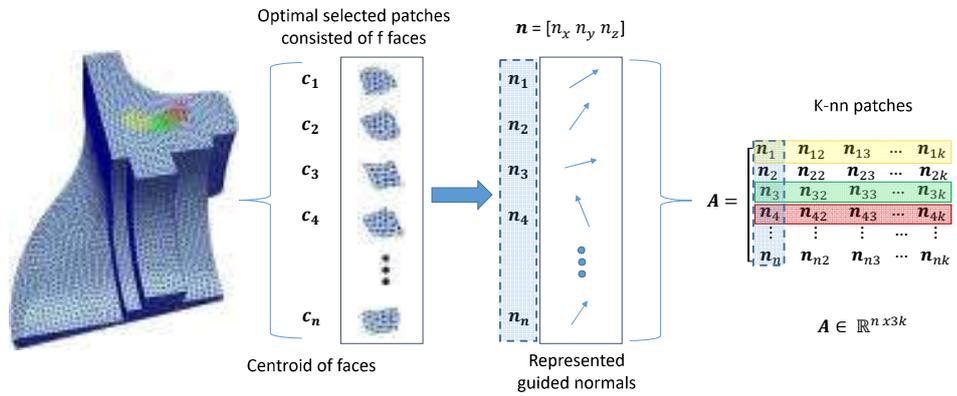

Figure 5.67: Construction of a low-rank matrix using guided normals.

**RPCA Based Identification**

To be able, to identify accurately features by exploiting their sparsity in the Laplacian domain, we construct a matrix $\mathbf{M} \in \mathbb{R}^{n_f \times 3n}$, where $n_f$ is the number of faces, as shown in Fig. 5.67. To be more specific, we construct a low-rank geometry matrix by following the approach of [17] and building on the assumption that a triangle mesh can be decomposed into many small patches, consisting of faces with similar normal directions. For each face, we initially search for such a small patch that contains $f_i$ and choose its average normal as the guidance normal of $f_i$. For each guidance normal we choose a set of $k$ surrounding faces, which are selected using either a topological or a geometric criterion. In the first case, the set consists of the faces, that share at least one vertex with $f_i$, while in the geometrical neighborhood case, the set consists of the nearest faces (centroids).

The approach that we follow, is based on the following observations: i) an underlying surface is a piecewise $C^2$-smooth surface with sharp features ii) the sharp features can be sparsely represented in some coherent dictionary which is constructed by the pseudo-inverse matrix of the Laplacian of the shape [175] and iii) the direction of the Laplacian of the feature signal approximates the local normal direction of the features. Based on the aforementioned observations, we then apply the RPCA approach directly on the matrix with the normals of the meshes, presented in Fig. 5.67. Each column of S provides the indices of the identified features. Therefore, the diversity offered by the different estimations results in a robust identification, even in the presence of noise as it is also shown in Fig. 5.68.

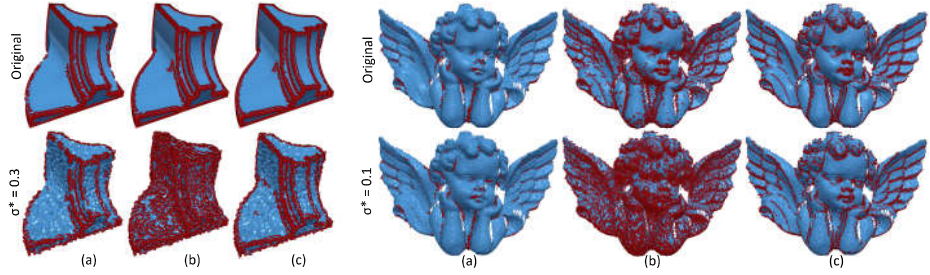

Figure 5.68: Identification of features using the fandisk and angel models: (a) using the conventional approach of sec. 3 (b) by applying RPCA on normals and their nearest neighbors (c) by applying RPCA on the guided normals and their nearest neighbors

**Progressive Compression of 3D Meshes**

In the proposed 3D mesh compression scheme, we are able to progressively encode geometry information allowing the generation of a sequence of levels of detail. Motivated by the fact that geometric features like high curvature regions, e.g., corners and edges, usually convey important visual information we initially provide a compact representation of the model by generating the quantized $\delta$ coordinates of the geometric features and assigning zeros to the delta coordinates of the non-feature points generating the vector $\boldsymbol{\delta} = [\delta_1, \ldots, \delta_n]$, where:

$$\bar{\delta}_i = \begin{cases} Q(\delta_i), & \mathbf{v}_i \text{ is feature} \\ 0 & \text{otherwise} \end{cases} \quad (5.53)$$

and $Q(\cdot)$ is a scalar quantization function. To further increase the level of detail, we progressively generate the delta coordinates of the non feature points. Together, with the vector $\boldsymbol{\delta}$ we also quantize a set of known vertices which are also known as control or anchor points $\mathbf{v}_a$, that are uniformly distributed on the model surface $\mathbf{v}_a = Q\left(\left[\mathbf{v}_{i_1}, \ldots, \mathbf{v}_{i_{n_a}}\right]\right)$ where $i_{n_a}$ is the vertex index and $n_a$ correspond to the 1% of the total number of vertices $n$. The reconstruction of the 3D mesh vertices is performed as in [147], by solving the following sparse linear system:

$$\begin{bmatrix} \mathbf{L} \\ \mathbf{I} \end{bmatrix} \mathbf{v} = \begin{bmatrix} \bar{\boldsymbol{\delta}} \\ \mathbf{I}\mathbf{v}_a \end{bmatrix}, \quad \mathbf{L} = \mathbf{D} - \mathbf{C}, \quad (5.54)$$

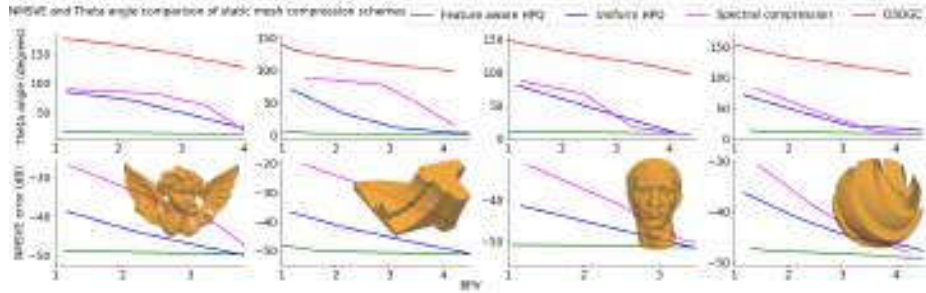

Figure 5.69: Reconstruction results ($\delta$, $NMSVE$) vs Compression efficiency using different 3D models.

---

**Algorithm 9:** Feature aware high pass quantization
---
1. *Construct a low rank matrix with normals or guided normals as shown in Fig.* 5.67
2. *Evaluate Features by executing RPCA*
3. *Evaluate the delta coordinates of feature points using eq.* (4.12).
4. *Perform uniform quantization of the delta features, as shown in eq.* (5.53)
5. *Evaluate the delta coordinates of a sparse set of non feature vectors in order and perform quantization- This step is executed for progressively increasing the level of detail*
6. *At the decoder site solve a sparse linear system defined in* (5.54)

---

In the proposed 3D mesh compression scheme, we are able to encode geometry information allowing the generation of a sequence of levels of detail. Geometric features like high curvature regions, that convey important visual information, bare a high saliency value. Algorithm 10 presents briefly the multiscale feature aware compression strategy.

**Evaluation Results**

In the following, we show reconstruction results evaluated using Algorithm 9, and compare them against a feature unaware (uniform) high pass quantization scheme in order to highlight the compression benefits when exploiting the knowledge of geometric features. In the feature aware case we assign a number of bits for quantizing only the feature points while in the second case the same number of bits is used to encode the delta coordinates of the whole 3D mesh. The compression efficiency is measured in bpv encapsulating the information related to the delta coordinates and the quantized anchor points. As conventional schemes we consider:

1. The approaches that provide compact representation of 3D meshes in the GFT basis (Spectral approach executed in overlapped parts as described in [152] due to complexity constraints) [155], [152].

---
**Algorithm 10:** Multiscale feature aware compression
---
**Input** : 3D model $\mathcal{M}$, desired compression ratio, $N$ number of classes, saliency map of vertices
**Output:** Compressed representation $\dot{\mathcal{M}}$
`// Anchor points sampling`
1 Sample uniformly 10% of the the vertices to be used as anchor points.
`// Delta coordinates sampling`
2 Sort the vertices based on the saliency values of the saliency map provided by the CNN approach;
3 Separate the vertices into $N$ salient classes based on their salient values;
4 Depending on the compression scenario, different ratios are selected from each class, giving emphasis to the perceptually salient vertices;
5 The final number of the remaining vertices must satisfy the initial percentage scenario;
6 The delta coordinates of selected vertices are quantized with 12 bits.
7 The delta coordinates of the rest of the vertices are set to zero
`// Connectivity coding`
8 Employ edge breaker strategy [344]
---

2. The uniform high pass quantization approach presented in [345].

3. The OD3GC method [20] that performs direct quantization to the 3-space coordinates.

Figs. 5.69 and 5.70 presents the reconstruction quality versus the compression efficiency of the feature aware high pass quantization approach and the aforementioned conventional approaches. By inspecting these figures, it is obvious that the robust identification of features at the encoder enables aggressive compression ratios, that correspond to even 1.5 bpv, without introducing significant loss on the visual quality. Finally, all the other methods introduce high-frequency errors into the model that modify significantly the appearance of the surface, resulting in a model with a blocky or noisy structure, where the reconstruction errors are highly noticeable even in compression ratios that are greater than 2.5 bpv.

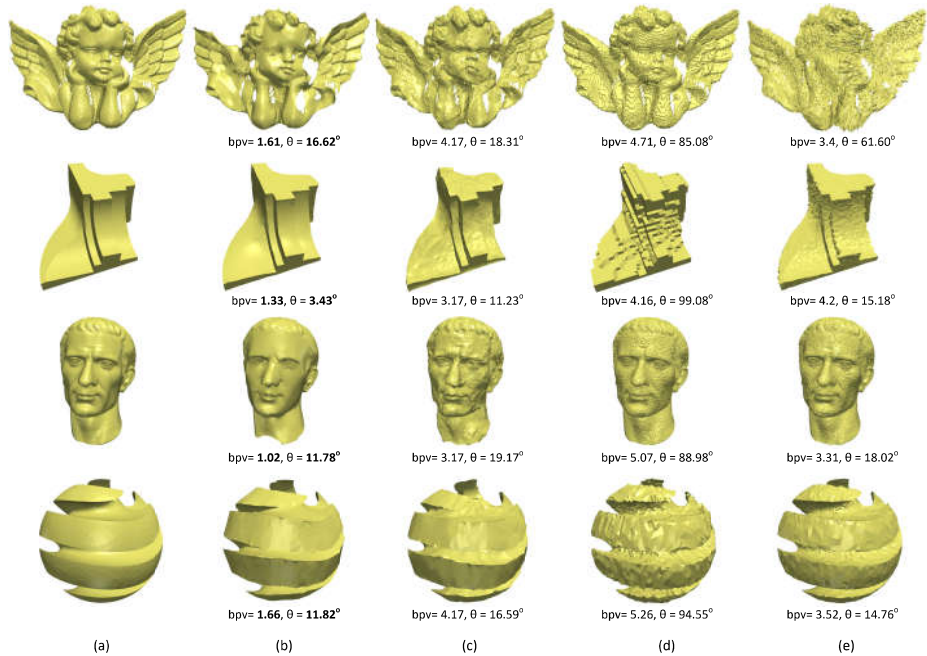

Figure 5.70: (a) Original Models vs Reconstruction results using: (b) feature aware & (c) uniform high pass quantization of different 3D models, (d) the OD3GC approach of [20] and (e) spectral compression in parts ($\sim$ 600 vertices per part).

To avoid over-smoothing effects and reduce the complexity of identifying patches consisting of faces with similar normal directions, one could potentially use the individual normal instead of the guidance normals directions assigned to each centroid, since the exploitation of spatio-temporal coherence via RPCA provides a robust identification of features in the presence of noise. After the identification of the normal directions of the basis frames and the features that correspond to each frame, we evaluate the corresponding mesh vertices by executing a conventional bilateral approach [21]. By inspecting the dynamic mesh denoising results in Fig. 5.71, it is clearly shown that the use of the individual normals does not smooth as much the sharp features, as compared to the use of the guided normals.

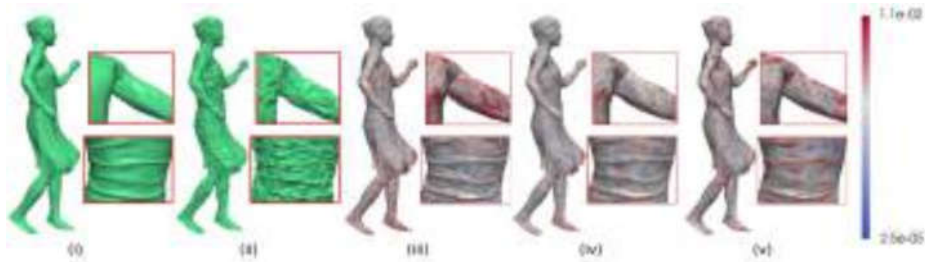

Figure 5.71: Denoising of Dynamic Mesh using Robust RPCA on normals - Samba model, (i) original frame, (ii) noisy frame, (iii) Output of RPCA on normals (iv) Output of RPCA on normals using nearest neighbors (v) Bilateral normal filetring [21].

### 5.2.4 Scalable Coding of Dynamic 3D Meshes for Low-latency Applications

This subsection presents an efficient approach for online scalable coding of dynamic 3D meshes. This method is totally parameter-free and can be used without further changes or extra parameterization. A significant advantage of the proposed method is its ability to transmit different bpv per frame depending on the instant network's capability. Additionally, the selection of the transmitted vertices is optimized, taking into account both the spatial and temporal information. This gives an extra benefit to the reconstruction process to handle more efficiently the received vertices providing more accurate results. The main objective of our method is to provide the most efficient solution combining both the low execution time and the high perception quality of the reconstructed result.

In Fig. 5.72, the proposed framework is briefly presented, highlighting the most important procedures of our approach. In a nutshell, we start with the layer decomposition process taking into account the spatial and temporal information of the dynamic mesh. The output of this process corresponds to the active points at the end of the removal process[2]. After the transmission, each reduced frame is reconstructed. Firstly, we use a weighted Laplacian interpolation approach, as a coarse reconstruction process, in order to estimate the position of the removed vertices. Then, we perform a fine estimation step by tracking the normals subspace deviation between different layers using ISVD, significantly reducing the required complexity as compared to a conventional SVD based approach. The convergence of this approach is significantly accelerated using RPCA as an initialization procedure. Finally, each frame is fine-reconstructed penalizing displacement of the vertices over a tangent plane per-pendicular to the local surface normal.

---

[2] At this point, it should be noted that the number of the removed vertices depends on the network's capability per frame.

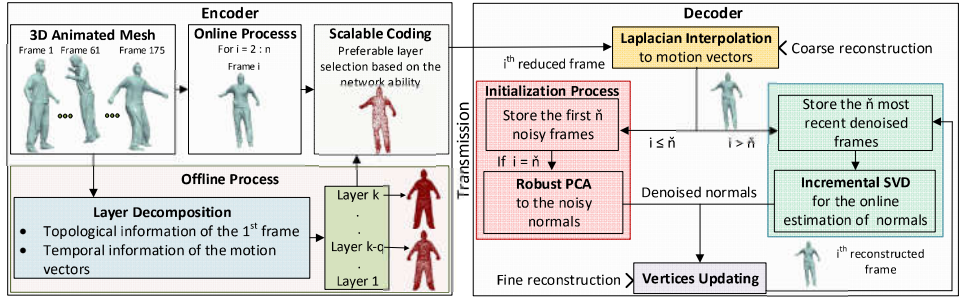

Figure 5.72: The pipeline of the scalable coding approach.

**Layer Decomposition**

Regarding the layer decomposition part, we propose a spatiotemporal algorithm for scalable coding of mesh sequences, similar to [22], which removes the vertices of a mesh taking into account both topological and temporal criteria. The vertex removal order is a vital process because it could affect the visual quality of each frame/mesh reconstruction influencing the prediction accuracy and hence the compression performance. We assume that only one vertex is removed at each layer so that $n$ spatial layers are created. We notate as $\mathbf{M}_1$ the set of points of the layer 1 which has only one vertex while in the highest layer $n$ there is the set of points $\mathbf{M}_n$ consisting of $n$ vertices. The relation between the set of points $\mathbf{M}_l$ and the exactly previous set $\mathbf{M}_{l-1}$ can be described as:

$$\mathbf{M}_l = \mathbf{M}_{l-1} \cup \{\mathbf{v}\} \tag{5.55}$$

It is obvious that each layer has one more vertex in comparison to its exactly previous layer, and always $\mathbf{M}_l \subseteq \mathbf{M} \ \forall \ l = 1 \cdots n$. We need to mention here that the layer decomposition process is an offline procedure taking place before the start of the transmission.

**Vertex Removal Using Topology Information**

We use an iterative process that removes one single vertex per iteration. Firstly, the removal cost for any candidate vertex is estimated and then the vertex with the lowest value is eliminated. The process is repeated $n$ times until only one vertex will have remained in layer 1. The proposed removal cost function consists of two terms, namely the spatial $C_s$ and the temporal $C_t$. The spatial factor is related to the geometry of the first frame and it is estimated as follows. We define as $R_j(i)$ the set of the $j$-ring neighbors of the $i$ vertex and the

$\hat{R}_j(i, l)$ as a subset of $R_j(i)$ with the remaining $j$-ring neighbors of $i$ vertex in the $l$ level. We also define as $b$ the topological distance between any vertex $v$ and its neighboring vertices. Topological distance shows the minimum number of edges with which two vertices are connected to each other. The selected $j$ value of $b_j \ \forall \ j = 1 \cdots \infty$ is an important variable to determine the spatial prediction accuracy. We suggest the maximum value of $j$ to be equal to 3 otherwise the process becomes time-consuming without providing any reconstruction benefit. At each layer, we remove this specific vertex which can be efficiently predicted by the reconstruction process. This is the reason why the vertex with the lowest value is selected. Finally, the removed vertex $v_l$ at layer $l$ is given by:

$$v_l = \arg\min_{v \in \mathbf{M}_l} C(i, l) \tag{5.56}$$

where $C(i, l)$ is the removal cost of vertex $i$ at layer $l$. We define the prediction inaccuracy $C_s(i, l)$ (spatial term) for vertex $i$ at layer $l$ as:

$$C_s(i, l) = \sum_{j=1}^{3} \frac{|R_j(i)| - |\hat{R}_j(i, l)|}{|R_j(i)|} \rho^j \tag{5.57}$$

where $|.|$ operator returns the number of elements in a set and $\rho$ is a positive parameter. In the experiments, the fixed $\rho = 0.6$ is used for all the models. Note that $(|R_j(i)| - |\hat{R}_j(i, l)|)/|R_j(i)|$ becomes equal to 0 when $|\hat{R}_j(i, l)| = |R_j(i)|$, and it is equal to 1 when $|\hat{R}_j(i, l)| = 0$. In any case $|R_j(i)| \geq |\hat{R}_j(i, l)|$. The authors of [22] suggested the use of an extra factor which evaluates how each candidate vertex will affect its neighboring vertices if it is removed. However, its estimation is very time-consuming and the provided results do not appear any significant advantage. Instead of this factor, we propose the use of temporal information term $C_t$ as described in the next paragraph.

**Vertex Removal Using Temporal Information**

Generally, a stationary or slow-motioned point is more likely to be accurately predicted. On the other hand, highly deformable surface patches are less accurately predictable. According to this observation, we propose a new factor that exploits the temporal information from frame to frame, taking into account the mean motion vector of each point, as shown below:

$$C_t(i) = \frac{\sum_{t=2}^{n_s} \|\mathbf{v}_i(s) - \mathbf{v}_i(s-1)\|_2}{n_s} \ \forall \ i = 1 \cdots n \tag{5.58}$$

where $(s)$ represents the current frame while the $(s-1)$ represents the previous frame. Then, the final removal cost is estimated as:

$$C(i, l) = C_s(i, l) + \eta C_t(i) \tag{5.59}$$

where $\eta = 0.1$. In Fig. 5.73, a frame of the animated model Samba is presented under different removal layers and using different removal cost functions. The difference between the two removal cost functions is more apparent in case Fig. 5.73-(a), wherein the second case more vertices of the hand are selected as active since they experience large deviations within frames as compared to other slowly varying vertices and consequently are expected to be predicted less accurately. One of the strongest benefits of our approach is the fact that the layer decomposition algorithm sorts vertices based on their contribution to the accurate prediction of their real position, using both spatial and temporal criteria. In each layer decomposition step, the easiest predicted vertex is removed (i.e., this one that takes the lowest value of the cost function). This means that if time-variant wrinkles are apparent somewhere, then our algorithm will keep more vertices from this area and will remove vertices for other "more static" areas.

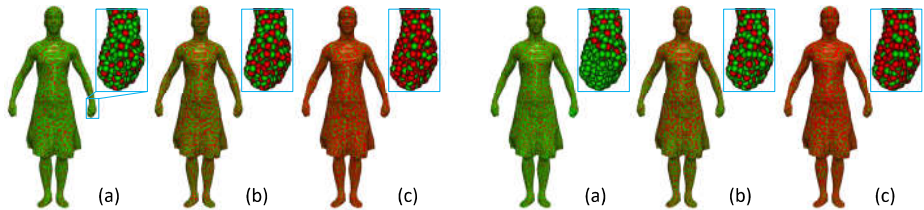

(a)　　　　(b)　　　　(c)　　　　　(a)　　　　(b)　　　　(c)

Figure 5.73: [Left] Only topological information, [Right] Topological and temporal information (Samba of 9971 vertices). (a) 3000 vertices have been removed (layer $P_{6971}$), (b) 5000 vertices have been removed (layer $P_{4971}$), (c) 7000 vertices have been removed (layer $P_{2971}$). The green color represents the remaining vertices while the red color represents the removed vertices.

**Laplacian Interpolation to the Motion Vectors for Coarse Reconstruction**

In this paragraph, we describe the process for the online coarse reconstruction. For the reconstruction of each frame, the topological and geometrical information of the previously reconstructed frame is used. We follow the main idea of the method, proposed by [346], but we take into account that the number of vertices, for any appeared incomplete frame, is variable and depends on the network's transmission ability.

**Weighted Graph Laplacian Matrix**

Generally, a binary Laplacian matrix provides information regarding the connectivity of vertices. However, weighted Laplacian matrices are able to provide additional information which can efficiently be utilized by a variety of other processes. In order to set the preferable constraints, we construct a modified weighted Laplacian matrix, similar to [328], that takes into account the two following factors $\mathbf{P} \in \mathbb{R}^{n \times n}$ and $\mathbf{O} \in \mathbb{R}^{n \times n}$. Parameter $\mathbf{H}$ is related to the distance between connected vertices. The value of this parameter represents the inverse $L^2$ norm between two vertices $\mathbf{v}_i$, $\mathbf{v}_j$ provided that $\mathbf{v}_i$ and $\mathbf{v}_j$ are connected to each other. It is estimated according to:

$$\mathbf{P}_{ij} = \begin{cases} \frac{1}{\|\mathbf{v}_i(t-1) - \mathbf{v}_j(t-1)\|_2 + \epsilon} & \text{if } j \in R_1(i) \\ 0 & \text{otherwise} \end{cases} \quad \forall\, i = 1 \cdots n \quad (5.60)$$

where $\epsilon$ is a very small positive number. For the estimation of this parameter, the values of the vertices from the previously reconstructed frame $(t-1)$ are used. Parameter $\mathbf{O}$ is related to the connecting proximity $b$ (degree or topological distance) of an unknown vertex with an already known vertex. The initial known vertices have a value equal to 4 (reinforcing the contribution of the known values), while the value of the unknown vertices depends on their connectivity degree $b$, as shown in the following equation:

$$\mathbf{O}_{ij} = \begin{cases} 4\mathbf{C}_{ij} & \text{if } \mathbf{v}_i \text{ is known} \\ \frac{\mathbf{C}_{ij}}{b+1} & \text{otherwise} \end{cases} \quad \forall\, i,j = 1 \cdots n \quad (5.61)$$

where $\mathbf{C}$ is the adjacency matrix. Finally, the weighted adjacency matrix $\mathbf{C}_w$ is created using the estimated parameters $\mathbf{P}$ of Eq. (5.60) and $\mathbf{O}$ Eq. (5.61) according to:

$$\mathbf{C}_w = \mathbf{P} \circ \mathbf{O} \circ \mathbf{C} \quad (5.62)$$

where $\circ$ denotes the Hadamard product. Then, the weighted Laplacian matrix is estimated according to the following equation:

$$\mathbf{L}_w = \mathbf{D} - \mathbf{C}_w \quad (5.63)$$

**Weighted Laplacian Interpolation**

We use the estimated weighted Laplacian matrix $\mathbf{L}_w$ of Eq. (5.63) which encloses all the necessary constraints for an efficient weighted Laplacian interpolation. The Laplacian of $\mathbf{B}$ is written as: $\Delta \mathbf{m} = \mathbf{L}_w \mathbf{B}$.

For the fast solving of the Eq. (4.26), we follow the same line of thought as presented by [347]. We start by simplifying the representation of the parameters $\mathbf{P}$ and $\mathbf{O}$. We can say that the correspondingly matrices could be written as a set of $n$ vectors of parameters with variance size depending on the first-ring area size: $\mathbf{P} = \{\mathbf{p}_1; \mathbf{p}_2; \cdots; \mathbf{p}_n\}$, $\mathbf{O} = \{\mathbf{o}_1; \mathbf{o}_2; \cdots; \mathbf{o}_n\}$ where $\mathbf{p}_i, \mathbf{o}_i \in \mathbb{R}^{|\hat{R}_1(i)| \times 1} \; \forall\, i = 1 \cdots n$ and $\mathbf{B}_{\mathbf{k}i}$ consist of the known motion vectors of the neighbor vertices belonging on $\hat{R}_1(i)$. The solution is estimated based on an iterative process which is executed until $|\hat{R}_1(i)| = |R_1(i)| \; \forall\, i = 1 \cdots n$.

$$\mathbf{B}_{\mathbf{u}i} = \frac{\sum_{\forall\, j \in \hat{R}_1(i)} \mathbf{h}_{ij} \mathbf{o}_{ij} \mathbf{B}_{\mathbf{k}i}}{\sum_{\forall\, j \in \hat{R}_1(i)} |\mathbf{h}_{ij} \mathbf{o}_{ij}|} \quad \forall\, i = 1 \cdots n \quad (5.64)$$

The closer the value of $|\hat{R}_1(i)|$ to $|R_1(i)|$ the faster the reconstruction process. For this reason the process is faster when the used decomposition layer is higher. Despite the fact that the geometric approach solves the same problem, described of our mathematical background in Eqs. (5.60)-(4.26), the execution is much faster. The coordinates of the missing vertices are estimated by updating their position of the previous frame using the motion vectors of Eq. (4.26).

$$\mathbf{v_u}(t) = \mathbf{v_u}(t-1) + \mathbf{B}_u, \ \forall\, t = 2 \cdots n \tag{5.65}$$

Finally, all vertices of the incomplete frame (t) are known $\mathbf{v}(t) = \mathbf{v_k}(t) \cup \mathbf{v_u}(t)$ where $\mathbf{v_k} = [\mathbf{v_{k1}} \cdots \mathbf{v_{k}}_{n_k}]$ and $\mathbf{v_u} = [\mathbf{v_u}_{n_k+1} \cdots \mathbf{v_u}_n]$. Please note that for the estimation of the motion vectors of the second frame, the first frame has to be known.

Despite the extremely good results that this step provides, especially when the decomposition layer is high, there are some misaligned points affecting the visual quality of the final results. We refine these abnormalities following the procedure described below.

**Online Scalable Coding using Fine Reconstruction**

The coarse reconstructed method, provided by the previously presented step, demonstrates impressive performance. However, in cases where the scalable coding is responsible for highly incomplete frames > 60% then noise appears in specific areas (e.g., nonrigid areas, areas with high motion between consecutive frames). To remove these abnormalities, a fine reconstruction step is utilized. Firstly, we estimate the ideal (denoised) centroid normals of all faces and then we use them to update the position of any vertex of the mesh using an iterative process. Although a deformation method, like this one presented by [348], could be used as the main method for the fine reconstruction step, providing very plausible visual results, it would be a very time-consuming process for the final reconstruction of the whole dynamic 3D mesh.

**Initialization Strategy by Exploiting Outliers via RPCA**

For the denoising of the first $\bar{n}_t$ frames, we follow a batch approach in order to exploit more effectively their coherence using RPCA. As a result, we initially create a spatiotemporal matrix $\mathbf{M} \in \mathbb{R}^{n_f \times 3\bar{n}_t}$ according to:

$$\mathbf{M} = \begin{bmatrix} \mathbf{n_{c1}}(1) & \mathbf{n_{c1}}(2) & \cdots & \mathbf{n_{c1}}(\bar{n}_t) \\ \mathbf{n_{c2}}(1) & \mathbf{n_{c2}}(2) & \cdots & \mathbf{n_{c2}}(\bar{n}_t) \\ \vdots & \vdots & \ddots & \vdots \\ \mathbf{n_{c}}_{n_f}(1) & \mathbf{n_{c}}_{n_f}(2) & \cdots & \mathbf{n_{c}}_{n_f}(\bar{n}_t) \end{bmatrix} \tag{5.66}$$

where $\bar{n} \ll n$ and $\mathbf{n}_{ci}(\bar{n}_t) = [\mathbf{n}_{\mathbf{cx}_i}(\bar{n}_t); \mathbf{n}_{\mathbf{cy}_i}(\bar{n}_t); \mathbf{n}_{\mathbf{cz}_i}(\bar{n}_t)]$ represents the $i^{th}$ centroid normal of $\bar{n}_t^{th}$ frame. To note here that the experimental process has shown that a patch with size $\bar{n}_t = [5 - 10]$ is enough.

The motivation for using RPCA, as an initialization strategy, counts on the observation that the noisy (misaligned) normals of the coarse reconstructed mesh are also affected by outliers. This impulsive noise structure is attributed to the fact that most of the vertices keep their original position or have been already well reconstructed due to the weighted Laplacian interpolation step. Because of this observation, we assume that the presented type of noise is more like sparse outliers than noise with normal distribution.

Then, we use the elements of the low-rank matrix $\mathbf{E}$ to refine the $\bar{n}_t$ meshes updating the positions of their vertices. The low-rank matrix $\mathbf{E}$, consisting of the denoised normals represented as $\dot{\mathbf{n}}$. However, the process does not return unit normals so we need to normalized them, as shown in Eq. (5.67).

$$\mathbf{E} = \begin{bmatrix} \dot{\mathbf{n}}_{c1}(1) & \dot{\mathbf{n}}_{c1}(2) & \ldots & \dot{\mathbf{n}}_{c1}(\bar{n}_t) \\ \dot{\mathbf{n}}_{c2}(1) & \dot{\mathbf{n}}_{c2}(2) & \ldots & \dot{\mathbf{n}}_{c2}(\bar{n}_t) \\ \vdots & \vdots & \ddots & \vdots \\ \dot{\mathbf{n}}_{cn_f}(1) & \dot{\mathbf{n}}_{cn_f}(2) & \ldots & \dot{\mathbf{n}}_{cn_f}(\bar{n}_t) \end{bmatrix} \quad (5.67)$$

$$\dot{\mathbf{n}}_{ci}(j) = [\frac{\dot{\mathbf{n}}_{\mathbf{cx}_i}(j)}{\|\dot{\mathbf{n}}_{ci}(j)\|_2}; \frac{\dot{\mathbf{n}}_{\mathbf{cy}_i}(j)}{\|\dot{\mathbf{n}}_{ci}(j)\|_2}; \frac{\dot{\mathbf{n}}_{\mathbf{cz}_i}(j)}{\|\dot{\mathbf{n}}_{ci}(j)\|_2}] \ \forall \ j = 1 \cdots \bar{n}_t \quad (5.68)$$

**Online Refining using ISVD**

The previously mentioned step (initialization strategy) is applied once, only for the patch of the first $\bar{n}_t$ noisy frames. After that, we use the knowledge of the reconstructed frames in order to estimate the denoised normals of any new presented frame. In the literature, many of the proposed methods are trying to estimate the ideal normals. However, none of them are fast enough for real-time or online applications. To overcome this limitation, we suggest the use of an incremental approach. The SVD updating algorithm, described in detail in [349], [350], provides an efficient way to carry out the SVD of a larger matrix $[\mathbf{E}_{n_f \times 3\bar{n}_t}, \mathbf{F}_{n_f \times 3k}]$, where $\mathbf{F}$ is an $n_f \times 3k$ matrix consisting of the $n_f$ centroid normals of the $k$ additional frames. At this point, it should be noted that the matrix $\mathbf{E}$ is an already low-rank matrix consisting of the denoised normals of the $\bar{n}_t$ previous frames. Specifically, for the normal's estimation of the $\bar{n}_t + 1$ frame, the matrix $\mathbf{E}$, as described in Eq. (5.67), is used, while for any frame $> \bar{n}_t + 2$ the matrix $\mathbf{E}$ is updated as we will show later in Eq. (5.72). The $k \leq \bar{n}_t$ represents the number of the observed noisy frames where we want to estimate the denoise normals. Typically, $k = 1$ because each new frame appears online, however, if a buffer could store a sequence of coarse reconstructed meshes then more frames (block of frames) could be used, increasing the benefits that ISVD method provides regarding the execution times. By exploiting the orthonormal properties and block structure, the SVD computation of $[\mathbf{E}, \mathbf{F}]$ can be efficiently carried out by using the smaller matrices, $\mathbf{U}_q, \mathbf{V}_q$, and the SVD of the smaller matrix $\begin{bmatrix} \mathbf{\Lambda}_q & \mathbf{U}_q^T\mathbf{F} \\ \mathbf{0} & \mathbf{R} \end{bmatrix}$. The computational complexity analysis and details of the SVD updating algorithm are described in [349]. The steps of the ISVD method that we use are described in the next Eqs. (5.69) - (5.71). Firstly, we apply a qr(.) decomposition of the $(\mathbf{I} - \mathbf{U}_q\mathbf{U}_q^T)\mathbf{F}$ in order to estimate the matrices $\mathbf{Q}$ and $\mathbf{R}$:

$$\mathbf{QR} = \text{qr}((\mathbf{I} - \mathbf{U}_q\mathbf{U}_q^T)\mathbf{F}) \tag{5.69}$$

Next, we obtain the $q$-rank SVD of the $(q + k) \times (q + k)$ matrix:

$$\begin{bmatrix} \mathbf{\Lambda}_q & \mathbf{U}_q^T\mathbf{F} \\ \mathbf{0} & \mathbf{R} \end{bmatrix} = \hat{\mathbf{U}}\hat{\mathbf{\Lambda}}\hat{\mathbf{V}}^T \tag{5.70}$$

Then, the best $q$-rank approximation of $[\mathbf{E}, \mathbf{F}]$ is:

$$[\mathbf{E}, \mathbf{F}] = ([\mathbf{U}_q, \mathbf{Q}]\hat{\mathbf{U}})\hat{\mathbf{\Lambda}}(\begin{bmatrix} \mathbf{V}_q & \mathbf{0} \\ \mathbf{0} & \mathbf{I} \end{bmatrix}\hat{\mathbf{V}})^T \tag{5.71}$$

Finally, the matrix $\mathbf{F}$ obtains the denoised normals of the new frame which will be used to update the vertices. The matrix $\mathbf{E}$ will be updated, by a left shifting operation denoting as $\mapsto$, in order to obtain the most recent information for more efficient online estimation of the denoised normals of the next frame, according to:

$$\mathbf{E} : (\dot{\mathbf{n}}_{c1}, \dot{\mathbf{n}}_{c2}, \cdots, \dot{\mathbf{n}}_{c(\bar{n}_t-1)}, \dot{\mathbf{n}}_{c\bar{n}_t}) \mapsto (\dot{\mathbf{n}}_{c2}, \dot{\mathbf{n}}_{c3}, \cdots, \dot{\mathbf{n}}_{c\bar{n}_t}, \mathbf{F}) \tag{5.72}$$

where $\dot{\mathbf{n}}_{ci}$ represents the $i^{th}$ row of matrix $\mathbf{E}$. Please note that similar to the previous step using RPCA, the centroid normals must be normalized again so that any normal to be equal to the unit vector.

**Ideal Normals for the Vertex Updating**

As we referred to earlier, the apriori knowledge of the denoised normals of a mesh could be effectively used for fine denoising. The next step is to use the centroid normals, estimated in the previous steps, for achieving online fine reconstruction. In the past, a lot of researchers have adopted the same vertex updating algorithm, firstly described in [16], mainly because of its robustness and the very good provided results. We follow the same line of thought but we give different reliability to different points. More specifically, we assume that some points are more reliable than others. Two parameters affect mostly the classification of a point as trustable or not. The first one is the rigidness of the area (first-ring) where a point lies (Please see Fig. 5.74-[Right]) and the second is the total distance that a point covers through the frames (Please see Fig. 5.74-[Left]). More specifically, regarding the first parameter, we suggest that only vertices with a significant change ($> 1\%$) of their first-ring area between two frames will be updated, otherwise, we assume that they have been correctly reconstructed under the Laplacian interpolation step. The criterion that is used for the characterization of a point as rigid or not depends on the percentage change of its first-ring area between two consecutive frames and it is described in the Eq. (5.73):

$$\Phi(i) = \begin{cases} 1 & \text{if } \frac{|\text{area}(i,l) - \text{area}(i,l-1)|}{\max(\text{area}(i,l), \text{area}(i,l-1))} > 1\% \\ 0 & \text{otherwise} \end{cases} \quad (5.73)$$

where $\text{area}(i,l)$ is the first-ring area of point $i$ as it appears in the $l$ frame and $\Phi(i) = 1$ means that the $i$ point needs to be updated for more accurate results. An example of two areas (rigid and no-rigid) is shown in Fig. 5.75. Rigid area (e.g., head of camel) means better reconstruction using Laplacian interpolation regardless of the value of the motion vectors. Regarding the second parameter, the temporal factor $C_t$ is used, as it has been defined in Eq. (5.58). Then the updated weights are estimated according to Eq. (5.75), emphasizing the known and less moving points. Additionally, in order to make the process more time-efficient, the known points included in the $P_l$ set, are excluded from the updating process. The fine-tuned normals are then used to update the vertices accord-ing to [16]. The stability and the robustness of this approach has made it very popular and it has been used in a lot of other papers [9], [10], [165], [351], [12].

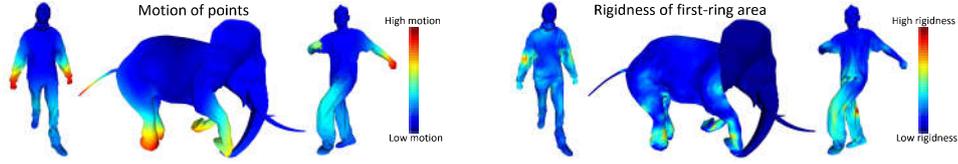

Figure 5.74: Heat map visualization showing: [Left] the total distance of each centroid that travels through the frames, [Right] the total change of the first-ring area of each centroid. The heat maps have been applied in a random snapshot of each 3D animation models (Crane, Elephant gallop and March 2 respectively). The deep blue color represents low changes and the dark red represents high changes, as also shown in the provided colormaps.

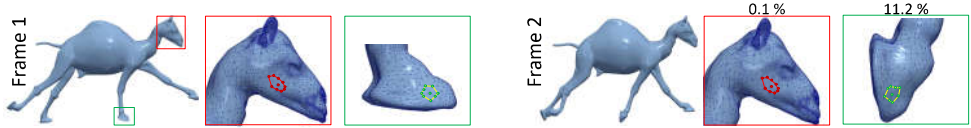

Figure 5.75: Two consecutive frames of the animated mesh Camel gallop. Different areas have different behavior, with respect to the change of their shape from frame to frame, depending on the type of motion that each area manifests.

$$\mathbf{v}_i^{(t+1)} = \mathbf{v}_i^{(t)} + \frac{\sum_{j \in \Psi_i} \nu_j \bar{\mathbf{n}}_{cj}(\langle \bar{\mathbf{n}}_{cj}, (\mathbf{c}_j^{(t)} - \mathbf{v}_i^{(t)}) \rangle)}{|\Psi_i|} \ \forall \ i \notin \mathbf{M}_l, \ \Phi(i) = 1 \quad (5.74)$$

$$\nu_i = \begin{cases} 4C_t(i) & \text{if } \mathbf{v}_i \text{ is known} \\ C_t(i) & \text{otherwise} \end{cases} \quad (5.75)$$

$$\mathbf{c}_j^{(t+1)} = (\mathbf{v}_{j1}^{(t+1)} + \mathbf{v}_{j2}^{(t+1)} + \mathbf{v}_{j3}^{(t+1)})/3 \ \forall \ j \in \Psi_i \quad (5.76)$$

($t$) represents the number of iteration and matrix $\Psi_i$ is the cell of vertices that are directly connected with the vertex $\mathbf{v}_i$. This iterative process can be considered as a gradient descent process that is executed for minimizing the energy term $\sum_{j \in \Psi_i} \|\bar{\mathbf{n}}_{cj}(\mathbf{c}_j^{(t)} - \mathbf{v}_i^{(t)})\|_2$ across all faces. Finally, we need to mention here that for any of the used animated models, the ideal number of iterations $t$ for an efficient fine reconstruction, was only 1-2. This is due to the fact that the coarse reconstructed frames are already denoised and need just a refinement through the aforementioned approach so as to remove outliers. Algorithm 11 briefly presents the steps of the proposed process.

**Impact of using ISVD**

ISVD is used not only because it allows an online solution but also because it is a very fast process especially when it is used in blocks of frames. Specifically, the larger the size of a block (e.g., frames per block (F/B)) the more apparent the benefits of the ISVD, regarding the execution time. ISVD is a vital step for this research because it allows the online reconstruction of frames in acceptable rates. Table 5.15 shows the execution times of incremental and traditional SVD, for different animated models and for different blocks of frames. Observing the table, it is obvious that ISVD is much faster than the traditional SVD (from 8 to 35.91 times faster). The bold values represent the execution times per block of frames while the values in the parenthesis represent the execution times per frame. To mention here that an extra benefit of ISVD is that it can be used for large matrices too, like in the case of Elephant model (last lines of Table 5.15). On the other hand, SVD cannot be directly applied without extra modification (e.g., separation in parts). Table 5.14 presents the execution times of the coarse reconstruction step using two different approaches, mentioned as "linear" and "geometric". More specifically, the "linear" approach directly solves the linear system of Eq. (4.26) while the "geometric" approach uses the iterative process of Eq. (5.64) for the solution of the system.

**Algorithm 11:** Scalable Coding of Dynamic 3D Meshes

```
// Offline and Online Processes in the Encoder Side
```
**Input**  : Dynamic 3D mesh
**Output:** $n$ different Layers (set of vertices) and reduced frames
```
// Offline Process (Layer Decomposition)
```
1 **for** $l = 1 \cdots n$ **do**
2   | Estimate the set of vertices $\mathbf{M}_l$ based on topological and temporal information via Eq. (5.59);
3 **end**
```
// Online Process (Scalable Coding)
```
4 **for** $i = 1 \cdots n_t$ **do**
5   | Choose which layer $\mathbf{M}_l$ (set of vertices) you would send based on network ability
6 **end**
```
// Online Process in the Decoder Side (Reconstruction)
```
**Input**  : Reduced frames of the dynamic 3D mesh
**Output:** Reconstructed dynamic 3D mesh
7 **for** $i = 1 \cdots n_t$ **do**
8   | Coarse reconstruction based on Laplacian interpolation to the motion vectors via Eqs. (5.60)-(5.65);
    | `// Denoised normals and ouliers removal`
9   | Creation of spatiotemporal matrix $\mathbf{M}$ Eq. (5.66);
10  | **if** $i < \bar{n}_t$ **then**
11  |   | Estimation of low-rank matrix based on Robust PCA;
12  | **else**
13  |   | Online refining using Incremental SVD via Eqs. (5.69)-(5.72);
14  | **end**
15  | Fine reconstruction using the vertices updating Eqs. (5.73)-(4.29);
16 **end**

**Impact of using RPCA as Initialization Strategy**

As mentioned earlier, for efficiently applying the ISVD step, some initial knowledge is required. More specifically, the low-rank matrix of any new entrance is estimated using the updated matrix $\mathbf{E}$ which contains the values of the $\bar{n}_t$ previous frames, as described in Eq. (5.72). However, for the first $\bar{n}_t$ frames, this knowledge is missing. The proposed method fails if the initial matrix $\mathbf{S}$ is not already low-rank (e.g., using the noisy frames without further processing). Alternative approaches, returning a low-rank matrix of the first $\bar{n}_t$ frames, could also be used (e.g., SVD, PCA) as an initialization strategy. However, RPCA satisfies simultaneously both the reconstruction quality and the lower computational complexity, as shown in Fig. 5.76.

| Name of Model | Decomposition layer | Execution time per frame in (sec.) (Linear Approach) | Execution time per frame in (sec.) & Speed up (Geometric Approach) |
|---|---|---|---|
| Bouncing | $P_{8000}$ | 1.392561 | 0.040240 (34.61x) |
| | $P_{6000}$ | 2.739982 | 0.066960 (40.92x) |
| | $P_{4000}$ | 3.691753 | 0.085024 (43.42x) |
| | $P_{2000}$ | 8.080768 | 0.111490 (72.48x) |
| Crane | $P_{8000}$ | 1.422852 | 0.041686 (34.13x) |
| | $P_{6000}$ | 2.665091 | 0.065021 (40.99x) |
| | $P_{4000}$ | 3.682691 | 0.082462 (44.66x) |
| | $P_{2000}$ | 8.132842 | 0.144083 (56.44x) |
| Handstand | $P_{8000}$ | 1.401952 | 0.039392 (35.59x) |
| | $P_{6000}$ | 2.604363 | 0.060651 (42.94x) |
| | $P_{4000}$ | 3.714263 | 0.078702 (47.19x) |
| | $P_{2000}$ | 8.293463 | 0.125693 (65.98x) |
| Jumping | $P_{8000}$ | 1.396755 | 0.041670 (33.52x) |
| | $P_{6000}$ | 2.798114 | 0.065335 (42.83x) |
| | $P_{4000}$ | 3.721518 | 0.083198 (44.73x) |
| | $P_{2000}$ | 8.378740 | 0.122232 (68.55x) |
| Swing | $P_{8000}$ | 1.171689 | 0.034310 (34.15x) |
| | $P_{6000}$ | 2.403907 | 0.059041 (40.72x) |
| | $P_{4000}$ | 3.481418 | 0.076251 (45.66x) |
| | $P_{2000}$ | 7.726545 | 0.122875 (62.88x) |

Table 5.14: Execution times per each frame for different decomposition layers and different models.

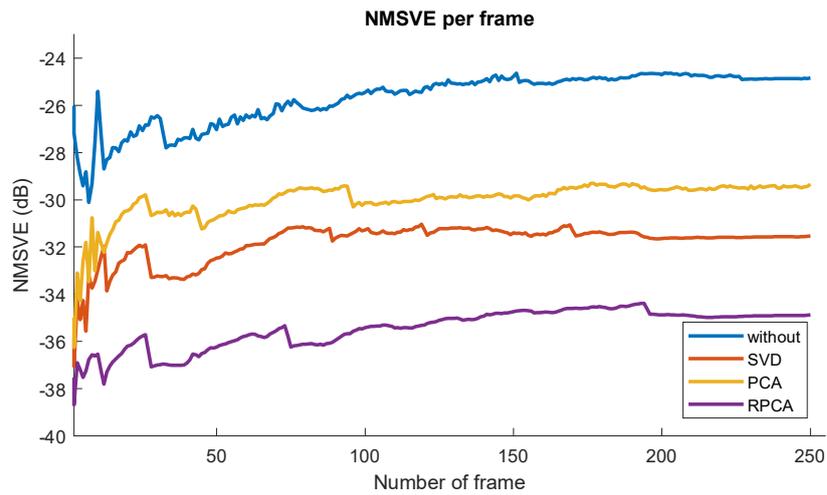

Figure 5.76: NMSVE error per frame in (dB) for the dynamic 3D model "March 2". Each line represents a different initialization strategy.

|  | F/B | ISVD | SVD | Speed up |
|---|---|---|---|---|
| Jumping | 1 | **0.10** (0.100) | **0.98** (0.980) | 9.8x |
|  | 10 | **0.24** (0.024) | **4.15** (0.415) | 17.2x |
|  | 50 | **1.10** (0.022) | **9.45** (0.189) | 8.59x |
|  | 75 | **1.43** (0.019) | **11.52** (0.153) | 8.05x |
| Camel | 1 | **0.28** (0.280) | **4.61** (4.610) | 16.46x |
|  | 8 | **0.48** (0.060) | **17.24** (2.155) | 35.91x |
|  | 12 | **0.61** (0.050) | **24.20** (2.016) | 14.76x |
|  | 24 | **1.30** (0.054) | **45.84** (1.910) | 35.26x |
| March 2 | 1 | **0.11** (0.110) | **0.88** (0.880) | 8x |
|  | 25 | **0.57** (0.022) | **10.11** (0.404) | 17.73x |
|  | 50 | **1.15** (0.023) | **12.12** (0.242) | 10.53x |
|  | 125 | **2.59** (0.020) | **16.51** (0.132) | 6.37x |
| Elephant | 1 | **0.52** (0.520) | - | - |
|  | 8 | **1.61** (0.201) | - | - |
|  | 12 | **1.73** (0.144) | - | - |
|  | 24 | **2.90** (0.120) | - | - |

Table 5.15: Execution times by using ISVD Vs. SVD. The numerical results are expressed in seconds.

**Benefits of Using Temporal Information for the Layer Decomposition**

The proposed layer decomposition algorithm sorts vertices based on their contribution to the accurate prediction of their real position, using both spatial and temporal criteria. In each layer decomposition step, the easiest predicted vertex is removed (i.e., this one that takes the lowest value of the cost function). This means that if time-variant wrinkles are apparent somewhere, then our algorithm will keep more vertices from this area and will remove vertices for other "more static" areas. The cost function takes into account the temporal information related to vertices variance with respect to the motion vectors. In other words, the temporal information benefits the preserving of more of these vertices that are moved more.

In Fig. 5.77, we present some examples of animated 3D models in different motion scenarios for the decomposition layer $P_{7000}$ (7000 vertices remain). The first column of Fig. 5.77-(a) illustrates, in different colors, which vertices are chosen by our algorithm to be kept (highlighted in red color) while the remaining vertices are removed (highlighted in green color). In Fig. 5.77-(b), we present some arbitrary frames of each 3D animated sequence, indicating in which areas of the moving 3D object, the motion is taking mostly place. We also highlight areas with different densities of remaining vertices using boxes of different colors. More specifically, areas with many coded vertices (i.e., very dense red points), because of their intense motion, are highlighted in red boxes. Correspondingly for medium intense motion, we used cyan boxes. A very representative example is the Jumping model in which the 3D object is moving mostly at the torso (i.e., upper the pelvis) so the density of the remaining vertices there is higher than this of the legs.

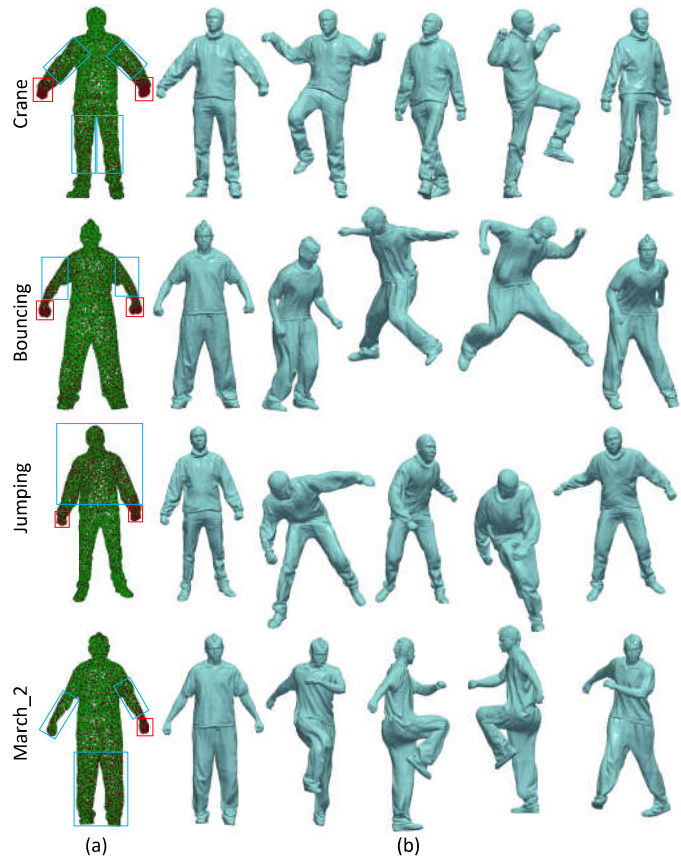

Figure 5.77: (a) Remaining (in red color) and removed (in green color) vertices for the decomposition layer $P_{7000}$, (b) Arbitrary frames of different 3D models in different motion scenarios.

**Performance of Augmented Lagrange Multiplier (ALM) Algorithm**

In Fig. 5.78, we present plots of different models showing (i) how the error changes per each iteration and (ii) how many iterations are required until the convergence to a steady state. As we can observe, all models seem to have very similar behavior in terms of convergence rate.

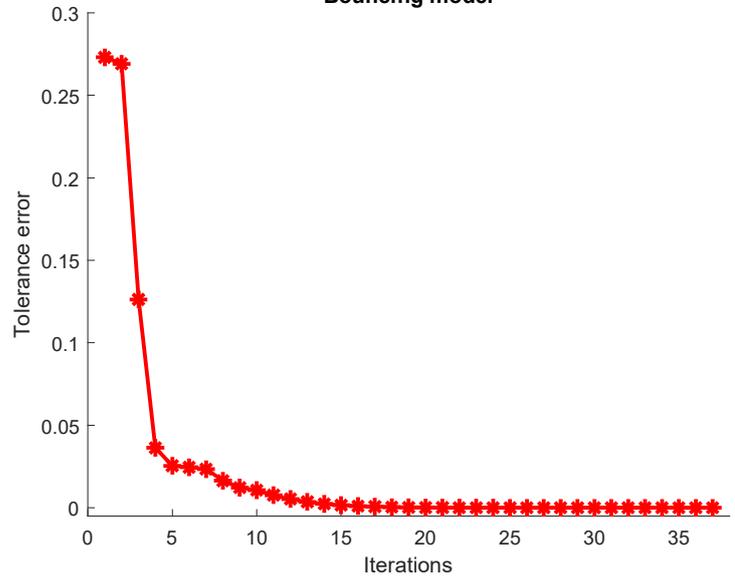

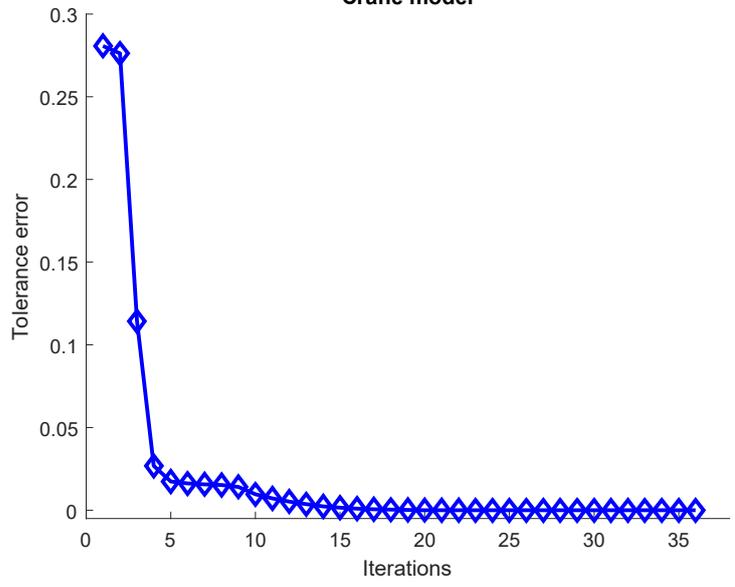

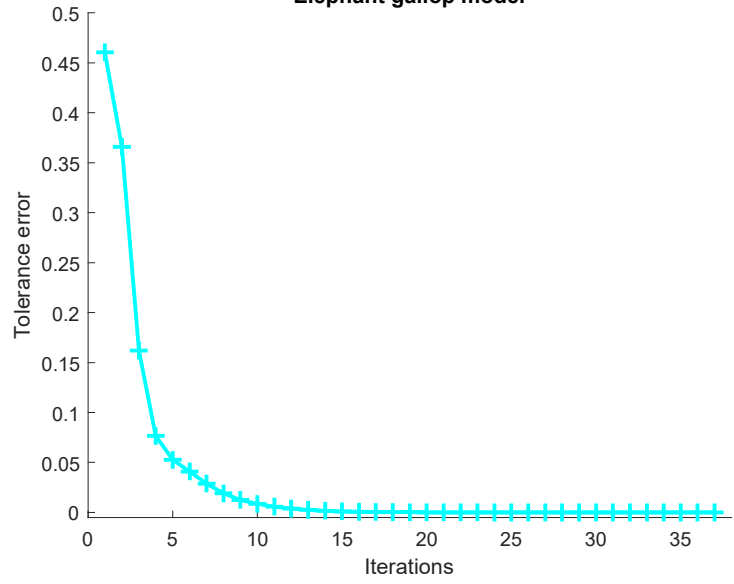
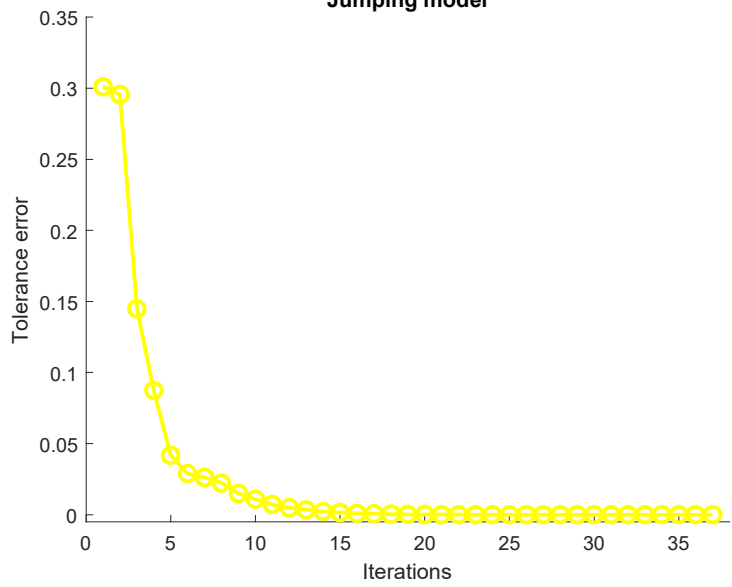

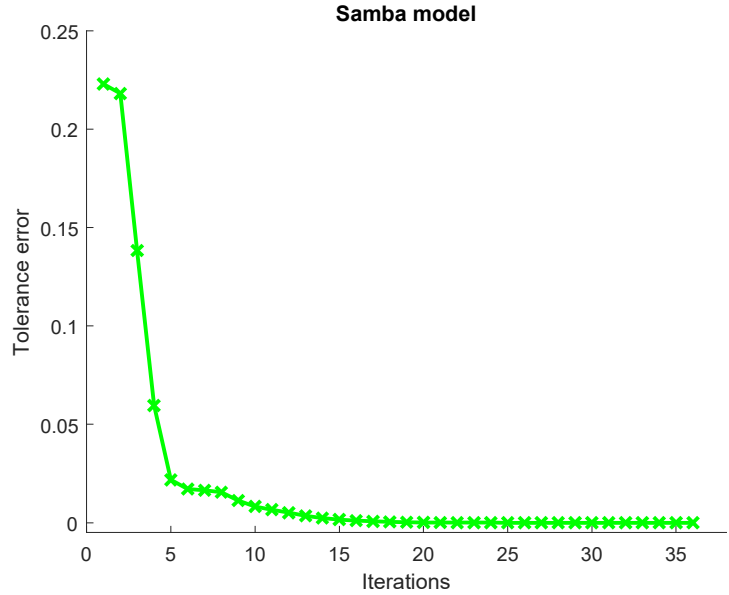

Figure 5.78: Tolerance error per each iteration for different 3D models.

Additionally, we provide the following table, as a summary. More specifically, Table 5.16 presents (i) the maximum number of iterations, (ii) the final error and (iii) the execution time per each model until the convergence of ALM.

| Model | Iterations | Tolerance error | Execution time |
|---|---|---|---|
| Bouncing | 37 | $6.89 \cdot 10^{-8}$ | 0.485894 |
| Crane | 36 | $9.48 \cdot 10^{-8}$ | 0.40282 |
| Elephant gallop | 37 | $7.87 \cdot 10^{-8}$ | 2.485669 |
| Jumping | 37 | $7.10 \cdot 10^{-8}$ | 0.470657 |
| Samba | 36 | $9.36 \cdot 10^{-8}$ | 0.456937 |

Table 5.16: Summary values per model.

**Relation Between Number of Decomposition Layers and Other Variables**

In Fig. 5.79, we present the relationship among the decomposition layer and the bpvf for different models. As we can see, all models have very similar behavior. By inspecting Fig. 5.80, we can easily observe the relationship between the decomposition layer and the KG error for different models. Apart from the number of the decomposition layer, KG error is also affected by the characteristics of the 3D animation (e.g., features of each 3D frame model, motion among the frames).

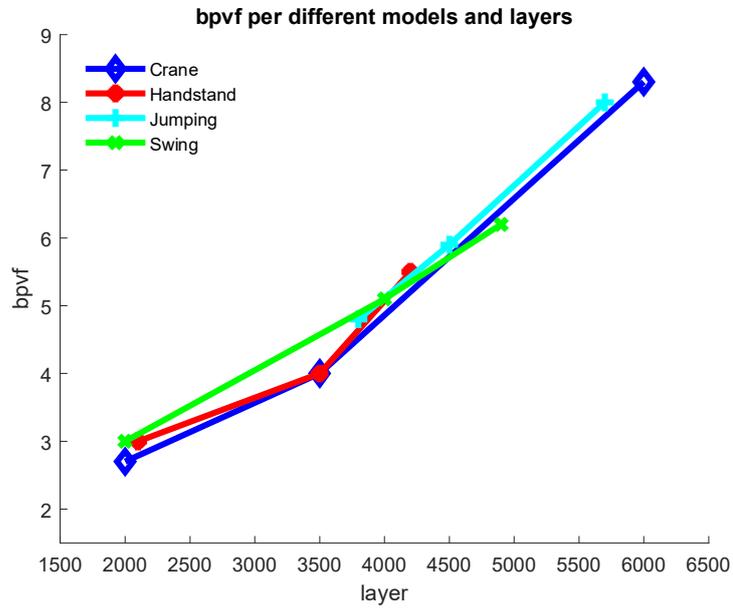

Figure 5.79: bpvf for different 3D models and decomposition layers.

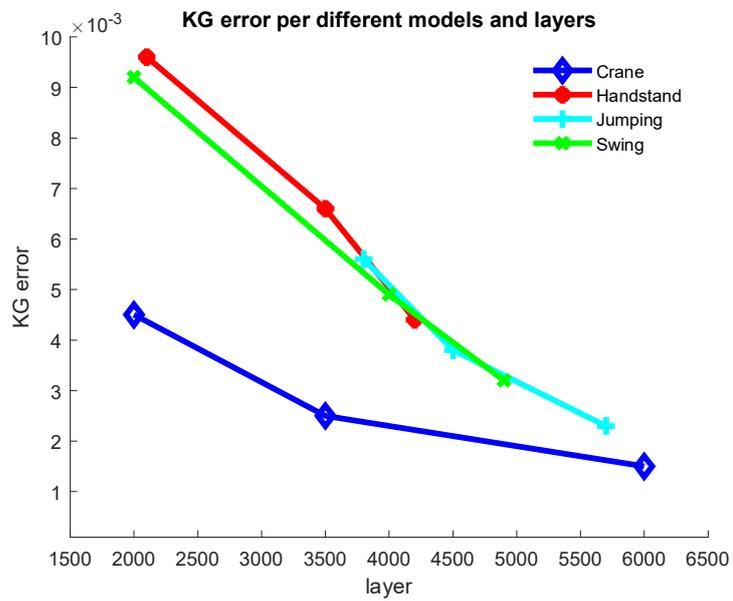

Figure 5.80: KG error for different 3D models and decomposition layers.

| Name of Model | Decomposition layer | Execution time per frame in (sec.) |
|---|---|---|
| Bouncing | $P_{8000}$ | 0.040240 |
| | $P_{6000}$ | 0.066960 |
| | $P_{4000}$ | 0.085024 |
| | $P_{2000}$ | 0.111490 |
| Crane | $P_{8000}$ | 0.041686 |
| | $P_{6000}$ | 0.065021 |
| | $P_{4000}$ | 0.082462 |
| | $P_{2000}$ | 0.144083 |
| Handstand | $P_{8000}$ | 0.039392 |
| | $P_{6000}$ | 0.060651 |
| | $P_{4000}$ | 0.078702 |
| | $P_{2000}$ | 0.125693 |
| Jumping | $P_{8000}$ | 0.041670 |
| | $P_{6000}$ | 0.065335 |
| | $P_{4000}$ | 0.083198 |
| | $P_{2000}$ | 0.122232 |
| Swing | $P_{8000}$ | 0.034310 |
| | $P_{6000}$ | 0.059041 |
| | $P_{4000}$ | 0.076251 |
| | $P_{2000}$ | 0.122875 |

Table 5.17: Execution times per each frame for different decomposition layers and different models.

The performance of the algorithm, in relation to the number of layers, is presented in Table 5.17. As we can see, incomplete models with more vertices (e.g. $P_{8000}$) require less execution time for their reconstruction during the completion process, in comparison with more sparse incomplete models (e.g. $P_{2000}$).

**Experimental Analysis and Comparisons**

The results of the proposed method are compared with: (a) A layer decomposition approach using an Efficient Fine-granular Scalable Coding Algorithm (EFSCA) of 3D mesh sequences for low-latency streaming applications, as described in details in [22]. (b) The Frame-based Animated Mesh Compression (FAMC) method which is promoted within the MPEG-4 standard [2]. This method uses different types of transforms to encode the residual motion compensation error, namely (1) the Discrete Cosine Transform (DCT) and (2) the integer to integer lifting-based bi-orthogonal wavelet transform. In Fig. 5.81, we present the reconstructed results using the pipeline of our proposed method for different frames (50, 100, 150) of the same dynamic mesh (Samba). In this example, we have assumed that any frame of the animation has been compressed with the same compression rate, through the same compression scenario ($\sim 30\%, 50\%, 70\%, 90\%$), showing how the reconstructed dynamic mesh is affected because of a consistent compression rate.

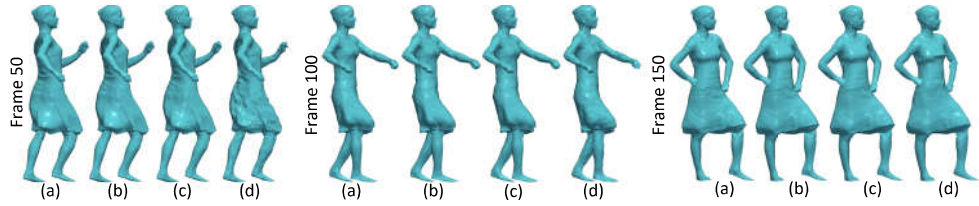

Figure 5.81: Dynamic 3D mesh Samba (9971 points each frame). (a) layer $P_{6971}$ (3000 points have been removed $\sim$ 30%), (b) layer $P_{4971}$ (5000 points have been removed $\sim$ 50%), (c) layer $P_{2971}$ (7000 points have been removed $\sim$ 70%), (d) layer $P_{971}$ (9000 points have been removed $\sim$ 90%).

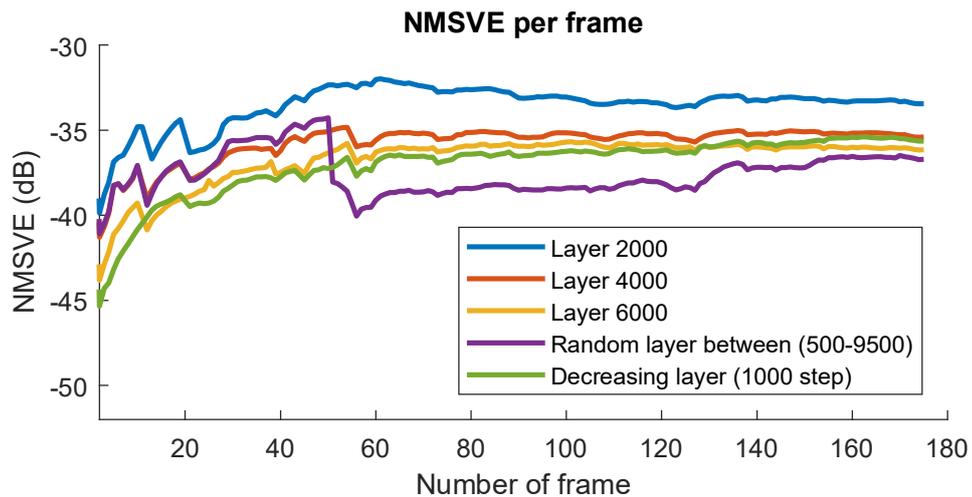

Figure 5.82: NMSVE error expressed in (dB) per frame for the dynamic 3D model Bouncing. Each line of this plot represents a different experimental scenario.

Fig. 5.82 shows how the NMSVE metric changes during the frames of the dynamic mesh under of different experimental scenarios. Targeting a more realistic scenario, we assume that the change of layers happens in blocks of frames, each block consisting of 50 consecutive mesh instances. Specifically, the experimental scenarios that we studied are: (i) Stable coding: the points of any frame of the dynamic mesh is reduced based on the same layer ($P_{2000}$, $P_{4000}$, $P_{6000}$ correspondingly). (ii) Random coding: the points of each block of frames is reduced based on a random layer in a range of ($P_{500}$-$P_{9500}$). (iii) Scalable coding: means that the points of each block of frames are reduced based on a layer which is lower of the layer of the previous block of frames using an incremental step of 1000. The initial layer of the first block of frames is $P_{8000}$ (2002 points have been removed). We can observe that the NMSVE error is much lower in the first frames, where the motion of the object has not started yet (relatively stationary frames). However, after that, the error is slightly increased but without being affected by the type of the motion of the object. As shown in this figure, the random coding, which typically is more close to a realistic scenario (network with variable transmission capability), has overall better results. It should be noticed that in this scenario it is possible to remove 9502 points (in layer $P_{500}$) from a frame which is equal to 95% of the dynamic mesh (for the Bouncing model consisting of 10,002 points).

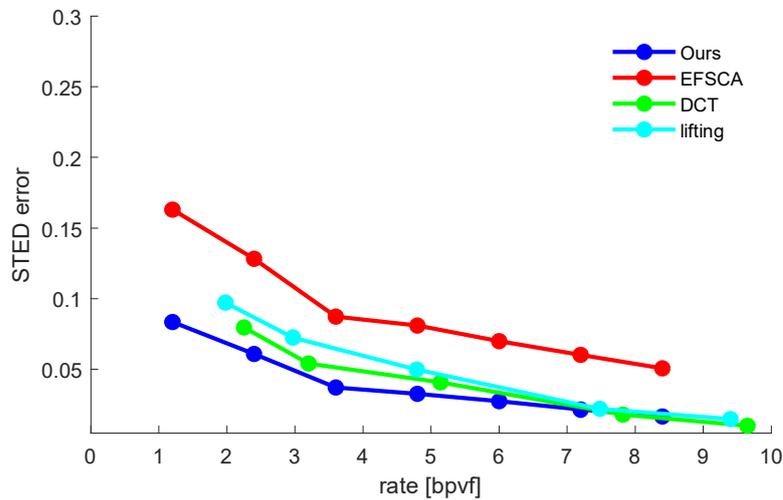

Figure 5.83: STED error of the reconstructed results using different approaches and different rates of bpvf. (Jumping Model).

Fig. 5.83 illustrates the STED error through the change of bpvf rate for different approaches. We can see that in high rates of bpvf, our proposed method has a very similar response to the two approaches of FAMC, however, when the bpvf is decreased our approach seems performs better. In Fig. 5.84, we present comparisons between our method and others using different rates of bpvf. For the evaluation we use the KG error metric. In Fig. 5.85, we present the heatmap visualization of $\theta$ metric for different reconstructed models. Additionally, we provide the mean $\theta$ of each frame for the different approaches, as well as enlarged detail of the reconstructed models for easier comparison.

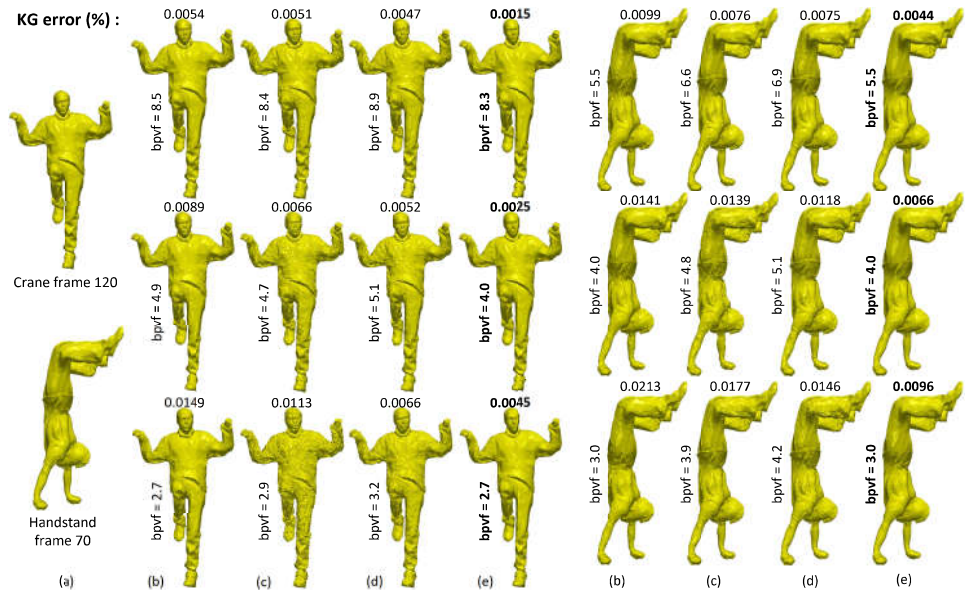

Figure 5.84: KG error of the reconstructed results using different approaches and different rates of bpvf. (a) Original mesh and reconstructed using: (b) EF-SCA [22], (c) the lifting approach of the FAMC method [2], (d) the DCT approach of the FAMC method [2], (e) our approach.

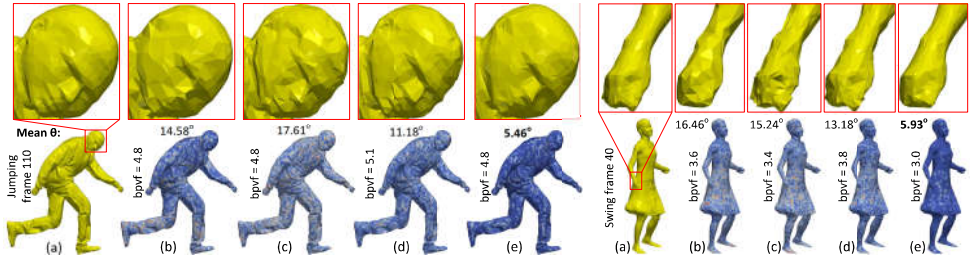

Figure 5.85: Heatmap visualizing the $\theta$ metric per face in different colors. The mean $\theta$ of the reconstructed mesh is also provided. (a) Original mesh and reconstructed using: (b) EFSCA [22], (c) the lifting approach of the FAMC method [2], (d) the DCT approach of the FAMC method [2], (e) our approach.

Besides the good results that the approaches of the FAMC method provide, a significant disadvantage of this method is the fact that it compresses the whole animation in a file using the same rate of bit-per-vertex for any frame. However, this approach is inappropriate for online applications where the bandwidth is unstable. To simulate variable bandwidth capabilities and adjust rates for chunks we separate the whole animation in blocks-of-frames (e.g., 10 frames per block) and at any block, a different bpvf is used. Note, that even if our method does not theoretically guarantee non-occurrence of tangled meshes, neverthe-less we use centroid normals with always positive direction (outside of the 3D object) so as to practically never face this problem.

**Applications in Virtual Reality/ Augmented Reality (VR/AR)**

In Figs. 5.86, 5.87, 5.88, we present some practical implementations of the proposed framework which have been designed using the Unity platform[3], Vuforia engine[4], Android Studio[5] and the SDK of Meta glasses[6]. The scripts have been written in the language C#. In the following examples, we show different reconstructed dynamic 3D meshes which are displayed online in different devices using AR and VR technologies.

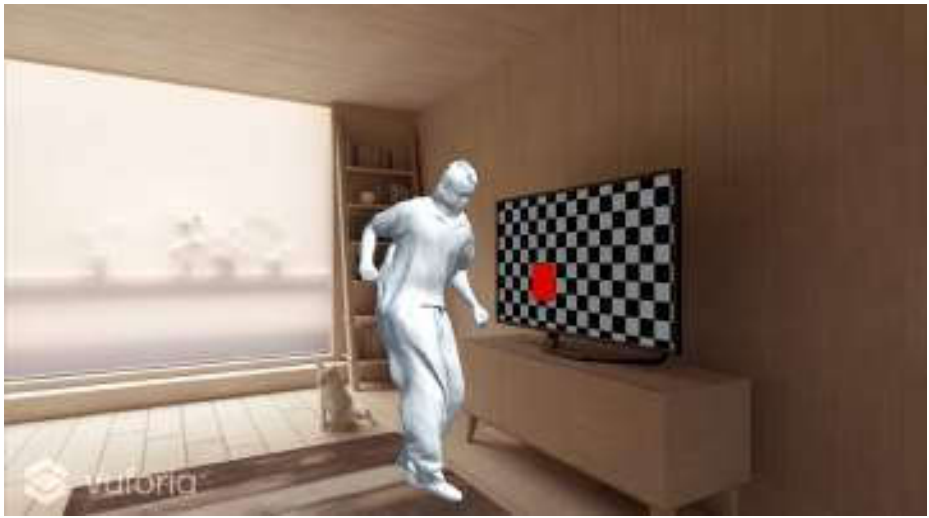

Figure 5.86: Online displaying of the reconstructed 3D animation (Handstand model) in a LG Nexus 5X Emulator of Android Studio. We used the virtual camera in a simulated environment that Android Virtual Device (AVD) provides.

---

[3]Unity. https://unity3d.com/
[4]Vuforia. https://www.vuforia.com/
[5]Android Studio. https://developer.android.com/studio/
[6]Meta. https://www.metavision.com/

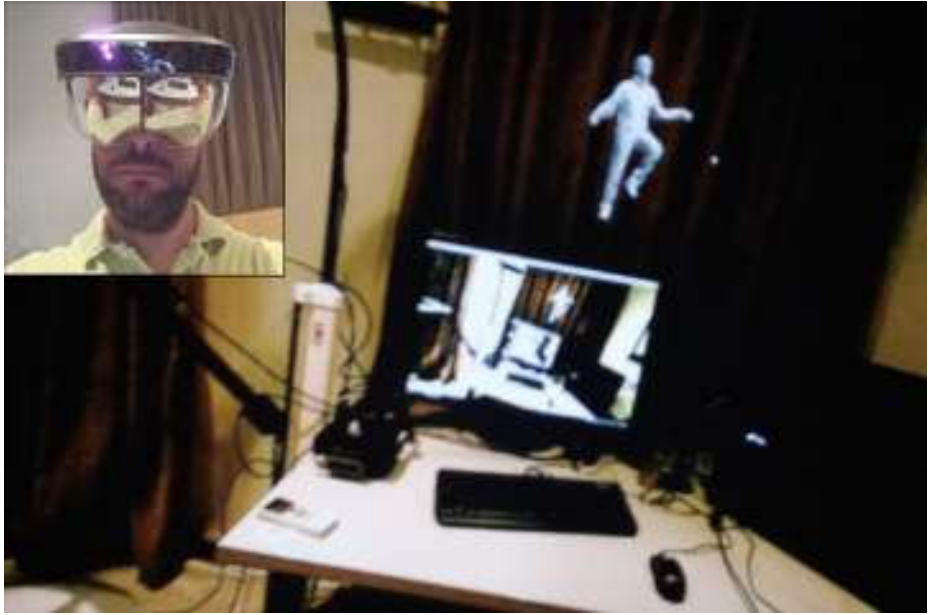

Figure 5.87: Online displaying of the reconstructed 3D animation (Crane model) using Meta AR glasses.

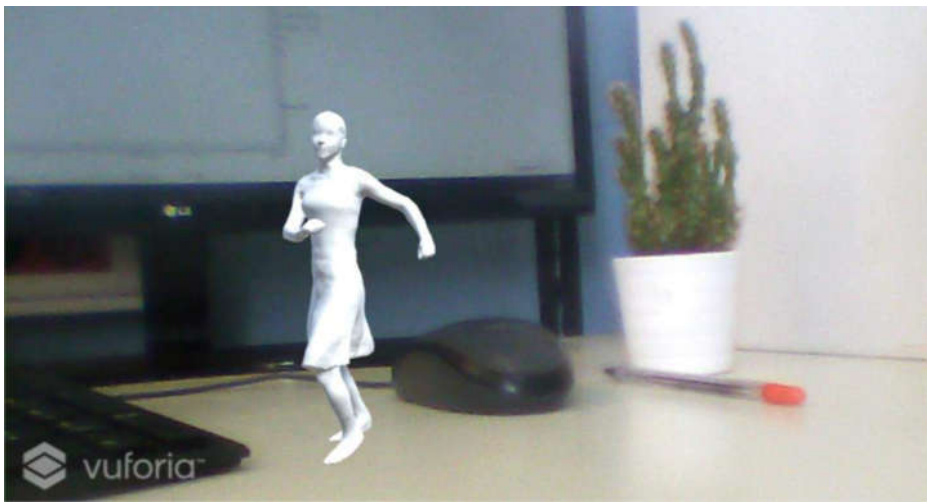

Figure 5.88: Online displaying of the reconstructed 3D animation (Samba model) using external camera connected to a Personal Computer (PC).

> **Publications that have contributed to this section:**
>
> 1. Content of the subsection 5.2.1 has been presented in **J5**, **J4** and **C7**
> 2. Content of the subsection 5.2.2 has been presented in **C4**
> 3. Content of the subsection 5.2.3 has been presented in **C5** and **C6**
> 4. Content of the subsection 5.2.4 has been presented in **J6**

## 5.3 Completion and Reconstruction of Incomplete Static and Dynamic 3D Models

This section refers to the problem of completion and reconstruction of incomplete 3D structures where some missing areas appear through their surface. Three different approaches are presented here with respect to the type of the used 3D model. More specifically, this section is divided into three subsections regarding the type of 3D object that we handle:

1. Static 3D meshes (subsection 5.3.1)

2. Dynamic 3D meshes (subsection 5.3.2)

3. Dynamic point clouds (subsection 5.3.3)

### 5.3.1 Completion and Inpainting of Static 3D Meshes

The purpose of the surface inpainting is the completion or the recovery of missing or damaged regions taking into account the spatial coherences, identified in the observed data. In real scenarios, there are a lot of reasons which can cause the creation of incomplete areas in a 3D object, such as:

1. limited quality of captured scanning devices

2. a relative motion between captured device and target which can create a limited observation

3. obstacles which can be located in front of the object of interest

4. on purpose removing of areas in order to dislodge deformed areas

Typically, surface reconstruction of real objects is achieved by combining the outcome of objects scanned from multiple directions. Nevertheless, due to limitations on the number of 3D sensors (hardware limitations) enclosing the scene, occlusions inevitably occur, causing missing areas to appear on the surface [352]. Additionally, multiple scanning is not always feasible to be applied due to the lack of time or expertise of the user.

So, there is a need for using algorithms that are able to overcome the problem of missing information. The objective is the implementation of a fast approach that efficiently recovers missing areas of a mesh taking advantage both of global and local coherence. A solution, which has been proposed in the literature and has been used a lot specifically in the area of image processing, is the matrix completion approach that takes advantage of the low-rank property attributed to spatial coherence. A problem, in this case, is how the initially observed matrix will be created.

For the creation of this matrix, we suggest the use of a shifting window that passes over the overlapped areas of connected points. For the points that have been observed, their coordinates values are assigned in the matrix otherwise if there is no information about the point then it is assumed as a missing point and the value of the matrix for this position sets equal to zero. Two different realistic

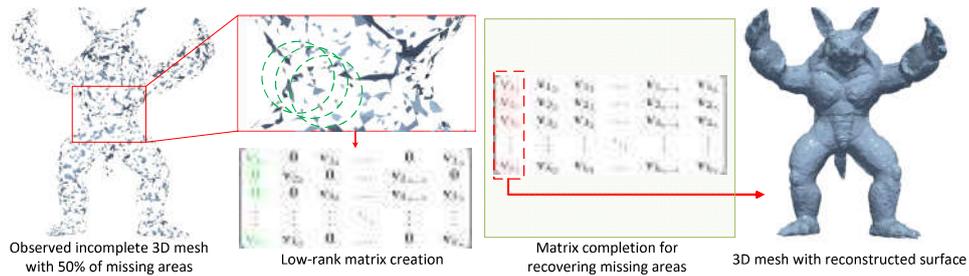

Observed incomplete 3D mesh with 50% of missing areas | Low-rank matrix creation | Matrix completion for recovering missing areas | 3D mesh with reconstructed surface

Figure 5.89: Pipeline of 3D mesh surface inpanting via matrix completion approach.

scenarios are studied and presented: (i) the repairing of partial-observed 3D meshes consisting of incomplete surfaces, (ii) the recovery of 3D objects with holes which are created by the intentional removal of deformed areas. In Fig. 5.89, we briefly illustrate all the steps of the proposed schema starting from the incomplete observation of a mesh to the finally reconstructed result. In some applications (e.g., repairing damaged surface regions) the mesh connectivity is known in advance before the reconstruction process. On the other hand, for hole filling applications, the connectivity is completely unknown without any further information. In these cases, a different approach seems more relevant to be followed. Firstly, we need to implant a set of points in random positions inside the hole, and then to use a triangulation algorithm, like this proposed in [353], before using our method for the final refinement.

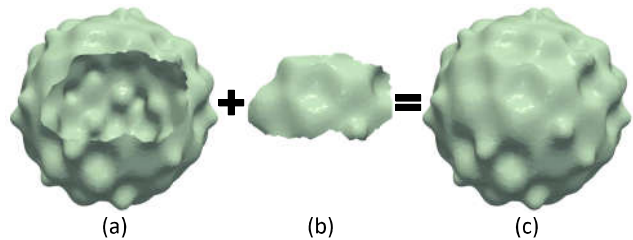

(a) (b) (c)

Figure 5.90: (a) Partial-observed mesh $\tilde{\mathcal{V}}$, (b) unknown or removed part of mesh $\dot{\mathcal{V}}$, (c) the entire original mesh $\mathcal{V} = \tilde{\mathcal{V}} + \dot{\mathcal{V}}$ (Bumpy sphere model).

It is assumed that only a subset $\tilde{\mathcal{V}}$ of the set $\mathcal{V}$ has been already observed and it is considered as known, where $\tilde{\mathcal{V}} \subset \mathcal{V}$. The main goal of the inpainting process is to estimate the unknown values of missing vertices, represented as $\dot{\mathcal{V}} = \mathcal{V} \backslash \tilde{\mathcal{V}}$, based on the information of the remaining set, where the operator $\backslash$ denotes the difference between the two sets (Fig. 5.90).

**Creation of the Low-Rank Matrix**

Despite the lack of knowledge about the real position of the missing surfaces, the initial connectivity of points it is assumed as known. Firstly, the matrix $\mathbf{A} \in \mathbb{R}^{n \times k}$ is estimated, each row of which contains the indices of the $i^{th}$ point and its $k$ nearest connected points $\forall\ i = 1, n$, as shown below:

$$\mathbf{A} = \begin{bmatrix} \mathcal{Q}(\mathbf{v}_{1_1}) & \mathcal{Q}(\mathbf{v}_{1_2}) & \cdots & \mathcal{Q}(\mathbf{v}_{1_k}) \\ \mathcal{Q}(\mathbf{v}_{2_1}) & \mathcal{Q}(\mathbf{v}_{2_2}) & \cdots & \mathcal{Q}(v_{2_k}) \\ \vdots & \vdots & \ddots & \vdots \\ \mathcal{Q}(\mathbf{v}_{n_1}) & \mathcal{Q}(\mathbf{v}_{n_2}) & \cdots & \mathcal{Q}(\mathbf{v}_{n_k}) \end{bmatrix} \quad (5.77)$$

where $\mathcal{Q}(\mathbf{v})$ is a function which returns the index of point $\mathbf{v}$. The formulation of $\mathbf{A}$ starts by gathering points of the $i^{th}$ first-ring area, $\forall\ i = 1, n$, until $k$ neighbor points have been selected. If $k$ is bigger than the sum of points that lie in the first-ring area then the process will continue by gathering points of a higher ring level (e.g., second-ring, etc.). Although the order of the selected points is arbitrary, it does not affect the final results because all points of the same ring can provide similar geometric constraints. The Fig. 5.91 presents an example in which $k$ neighbor points ($k = 10$) have been arbitrary selected (starting from the first-ring area). The selected points have been marked with the $\otimes$ symbol.

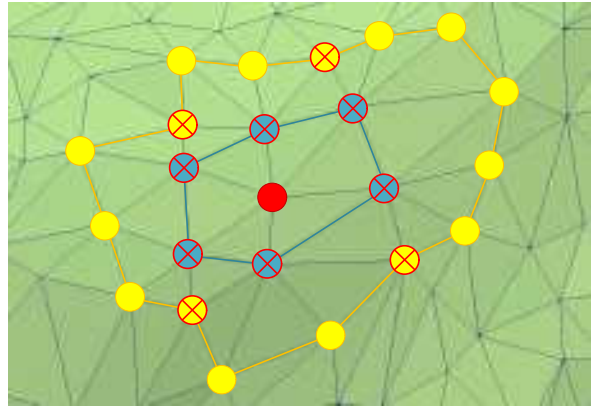

Figure 5.91: Blue contour represents the first-ring area of red point, while the yellow contour represents the second-ring area.

Once the matrix $\mathbf{A}$ has been completed, the matrix $\mathbf{M}$ is estimated according to the following assumption. If a point is unknown then we set its coordinates equal to the zero vector $\mathbf{0} = [0\ 0\ 0]$, otherwise, if it has been observed, we set its real values. An example of a low-rank matrix $\mathbf{M} \in \mathbb{R}^{n \times k}$ is shown below:

$$\mathbf{M} = \begin{bmatrix} \mathbf{v}_{1_1} & \mathbf{0} & \mathbf{v}_{1_3} & \ldots & \mathbf{0} & \mathbf{v}_{1_k} \\ \mathbf{0} & \mathbf{v}_{2_2} & \mathbf{0} & \ldots & \mathbf{v}_{2_{k-1}} & \mathbf{0} \\ \mathbf{0} & \mathbf{0} & \mathbf{v}_{3_3} & \ldots & \mathbf{v}_{3_{k-1}} & \mathbf{v}_{3_k} \\ \vdots & \vdots & \vdots & \ddots & \vdots & \vdots \\ \mathbf{v}_{n_1} & \mathbf{v}_{n_2} & \mathbf{0} & \ldots & \mathbf{0} & \mathbf{v}_{n_k} \end{bmatrix} \quad (5.78)$$

The value of the parameter $k$ depends on the percentage of the observed mesh. More specifically, the larger the percentage of missing areas (or the size of the removed surface), the bigger the value of $k$ for ensuring that each row of the matrix $\mathbf{M}$ has at least one non-zero element.

**Recovering of Missing Areas Using Matrix Completion**

The matrix completion can be considered as a subcategory of the matrix recovery problem, which has to recover the missing entries of a matrix, given a limited number of known entries, as described in [303]. The authors in [104] proved that a matrix $\mathbf{M}$ of rank $r$ can be recovered by solving the following optimization problem:

$$\min_{\mathbf{E}} \|\mathbf{E}\|_*, \text{ subject to } \mathbf{E} = \mathbf{M} \quad (5.79)$$

which can be formulated using a similar approach as this one described in [23]:

$$\min_{\mathbf{E}} \|\mathbf{E}\|_*, \text{ subject to } \mathbf{E} + \mathbf{S} = \mathbf{M},\ P_\omega(\mathbf{S}) = 0 \quad (5.80)$$

where $P_\omega$ is an operator keeping unchanged the known elements which belong to $\omega$ and sets equal to zeros the rest unknown elements which belong to $\bar{\omega}$. We denote as $\omega = \{\mathbf{M}(\mathcal{Q}(\mathbf{v}_{ij})) > 0\},\ \forall\, i = 1\ldots n,\ j = 1\ldots k\}$ the set of indices with nonzero elements of matrix $\mathbf{M}$, where $\omega \cup \bar{\omega} = \Omega$ is the sample space. For the estimation of Eq. (5.80) we use the augmented Lagrange multiplier algorithm:

$$l(\mathbf{E}, \mathbf{S}, \mathbf{Y}, \mu) = \|\mathbf{E}\|_* + \langle \mathbf{Y}, \mathbf{M} - \mathbf{E} - \mathbf{S} \rangle + \frac{\mu}{2}\|\mathbf{M} - \mathbf{E} - \mathbf{S}\|_F^2 \quad (5.81)$$

which is solved by repeatedly setting:

$$\mathbf{E}^{(t+1)} = \arg\min_{\mathbf{E}} l(\mathbf{E}, \mathbf{S}^{(t)}, \mathbf{Y}^{(t)}, \mu^{(t)}) \quad (5.82)$$

$$\mathbf{S}^{(t+1)} = \arg\min_{P_\omega(S)=0} l(\mathbf{E}^{(t+1)}, \mathbf{S}, \mathbf{Y}^{(t)}, \mu^{(t)}) \quad (5.83)$$

$$\mathbf{Y}^{(t+1)} = \mathbf{Y}^{(t)} + \mu^{(t)}(\mathbf{M} - \mathbf{E}^{(t+1)} - \mathbf{S}^{(t+1)}),\ \mu^{(t+1)} = \rho\mu^{(t)} \quad (5.84)$$

For the estimation of matrix **E**, we perform singular value decomposition according to:

$$(\mathbf{U}, \mathbf{\Lambda}, \mathbf{V}) = \text{SVD}(\mathbf{M} - \mathbf{S}^{(t)} + (\mu^{(t)})^{-1}\mathbf{Y}^{(t)}) \tag{5.85}$$

$$\mathbf{E}^{(t+1)} = \mathbf{U}\mathcal{S}_{(\mu^{(t)})^{-1}}[\mathbf{\Lambda}]\mathbf{V}^T, \ \mathbf{\Lambda} = diag(\lambda_i) \ \forall \ i = 1, k \tag{5.86}$$

where the function $\mathcal{S}_m[.] = \text{sgn}(.)\max(\|.\| - 1/m, 0)$ is a soft thresholding for a given value $m$ and $\lambda_i$ are the positive singular values with $\lambda_1 > \cdots > \lambda_k$. Correspondingly, the matrix **S** is estimated based on:

$$\mathbf{S}^{(t+1)} = P_{\bar{\omega}}[\mathbf{M} - \mathbf{E}^{(t+1)} + (\mu^{(t)})^{-1}\mathbf{Y}^{(t)}] \tag{5.87}$$

which can be easily written as:

$$\mathbf{S}^{(t+1)} = P_{\bar{\omega}}(\mathbf{E}^{(t+1)}) - \mathbf{E}^{(t+1)} \tag{5.88}$$

When the converge criterion $\|\mathbf{M} - \mathbf{E}^{(t)} - \mathbf{S}^{(t)}\|_F / \|\mathbf{M}\|_F < \epsilon$ has been satisfied, where $\epsilon$ is a very small positive threshold, then the iterative process stops and the matrix $\mathbf{E}^{(t)}$ are returned.

For the mesh reconstruction, we take into account the first column of the low-rank matrix $\mathbf{E}_{1i}$, as shown in Fig. 5.89, containing the estimated values of each $i = 1, n$ vertex. However, for taking advantage of the ground truth knowledge of the observed areas, the estimated values of matrix $\mathbf{E}_{1i}$ can be replaced with the corresponding observed values of $\mathbf{M}_{1i}$, such as $\mathbf{E}_{1i} = (i \in \omega \rightarrow \mathbf{M}_{1i}) \wedge (\neg i \in \omega \rightarrow \mathbf{E}_{1i})$.

**Progressive Scheme**

To accelerate the iterative process, we suggest a progressive scheme in which, fewer than $k$, singular values are selected. The number $n_\lambda < k$ of the remaining singular values is estimated in each iteration separately, according to:

$$n_\lambda^{(t)} = \begin{cases} n_\lambda^{(t)} & \text{if } \frac{\lambda_{n_\lambda^{(t)}}}{\lambda_{n_\lambda^{(t)}-1}} > 0.995 \\ \min(n_\lambda^{(t)} + 1, \lfloor 0.75k \rfloor) & \text{otherwise} \end{cases} \tag{5.89}$$

The value of $n_\lambda$ is initially defined equal to $\lfloor 0.5k \rfloor$, avoiding the over-smoothing of using few singular values. On the other hand, a value bigger than $\lfloor 0.75k \rfloor$ increases the execution time without providing any further corresponding quality. Giving border values for $n_\lambda$, we can ensure the best performance of the algorithm in terms of both reconstruction quality and computational complexity. In each iteration, the value $k$ appeared in Eq. (5.86), is replaced with the estimated $n_\lambda^{(t)} < k$.

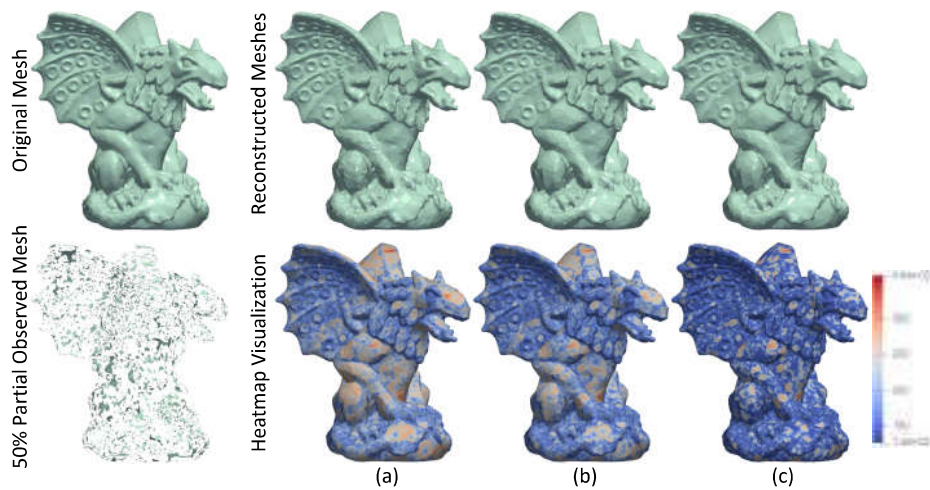

Figure 5.92: Reconstructed results of (a) Lin approach [23], (b) Our approach using the estimated low-rank $\mathbf{E}_1$, (c) Our approach replacing the estimated low-rank $\mathbf{E}_1$ with the ground truth values (Gargoyle model).

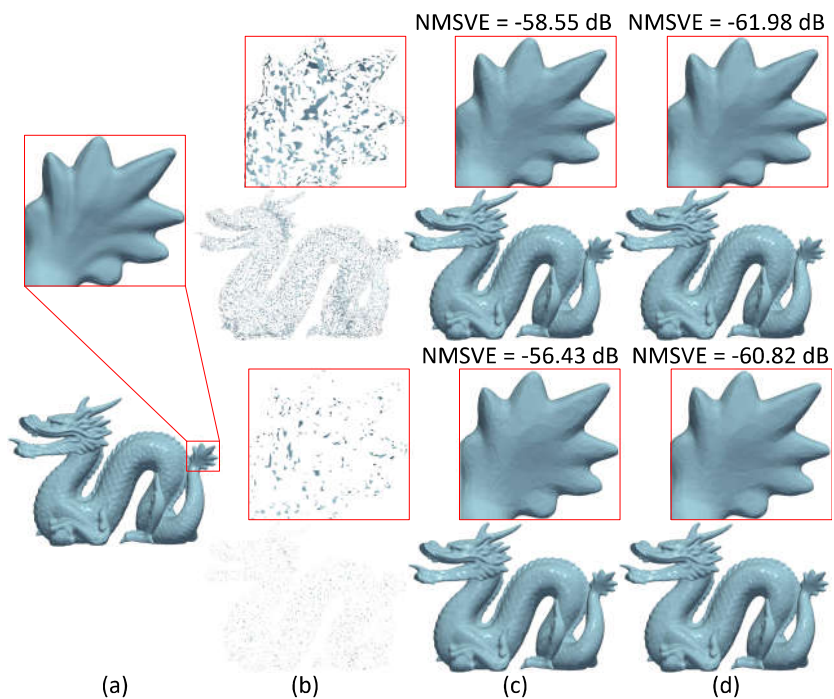

Figure 5.93: (a) Original mesh, (b) [up] 50% partial-observed mesh and [down] 30% partial-observed mesh, (c) reconstructed results using the approach in [23], (d) reconstructed results using our approach (Dragon model).

**Experimental Results of Inpainting in Static 3D Meshes**

In Fig. 5.92, the reconstructed results using different approaches are presented. Moreover, the heatmap visualizations of each reconstructed model are illustrated providing an easily-observed comparison with the ground truth model, making more apparent the superiority of our approach. Fig. 5.93 illustrates the reconstructed results of different approaches under different percentages of observed surfaces. NMSVE metric is also provided for easier comparison of the reconstructed models. In Table 5.18, we present the execution times for a full reconstruction using different dense and very dense models with different percentages of observed surfaces, so that the reconstruction error to be $< 10^{-5}$. It seems that the smaller the percentage of the observed mesh is, the more time-consuming the process is. The reason is that a larger value of $n$ is required, as mentioned above, result in the dimensionality of the matrix **M**.

| Model | Number of Vertices / Faces | Percentage % of observation | Execution time (s) | Rec. error |
|---|---|---|---|---|
| Buddha | 543,524/ 1,087,451 | 50 | 222.85 | $10^{-5}$ |
|  |  | 70 | 177.17 | $10^{-5}$ |
| Dragon | 437,645/ 871,414 | 50 | 91.58 | $10^{-5}$ |
|  |  | 70 | 72.29 | $10^{-5}$ |
| Gargoyle | 85,558/ 171,112 | 50 | 22.89 | $10^{-5}$ |
|  |  | 70 | 16.98 | $10^{-5}$ |

Table 5.18: Execution time for a full reconstruction of incomplete observed meshes.

Table 5.19 presents the execution times of different reconstruction approaches. We can easily observe that Laplacian Interpolation (LI) [25], [354] and Least-Squares Meshes (LSM) [24] are much slower in comparison with our approach.

| Model | Vertices | LSM [24] | LI [25] | Our Approach |
|---|---|---|---|---|
| Elephant | 24,955 | 186.74 s | 53.74 s | **4.62** s |
| Armadillo | 20,002 | 135.83 s | 34.83 s | **3.78** s |
| Tyra | 20,002 | 134.01 s | 32.50 s | **3.89** s |
| Samba | 9,971 | 20.23 s | 6.87 s | **2.16** s |
| Fandisk | 6.475 | 8.54 s | 2.99 s | **1.75** s |

Table 5.19: Execution time (in seconds) for a full reconstruction of 50% partial-observed meshes.

Additionally, for meshes with vertices $> 3 \cdot 10^4$, LSM and LI are unable to be used without further modification (e.g., separation in submeshes), because of the large dimensions of the used Laplacian matrix. On the other hand, our method is able to efficiently handle the completion of very dense meshes (e.g., Buddha: 543,524 vertices in 222.85 s, Dragon: 437,645 vertices in 91.58 s, Gargoyle: 85,558 vertices in 22.89 s). Fig. 5.94 illustrates the reconstructed result of our approach in comparison with LSM and LI, which also take advantage of the initial connectivity of the meshes. It is obvious that our method is not only the fastest (see Table 5.19) but it actually provides better reconstruction results avoiding the extra smoothing of areas because of the used low-rank matrix which encloses geometrical constraints as described in the previous subsection. In Fig. 5.95, the reconstructed results of a different case study are presented. In this case, we assume that a deformation has appeared in an arbitrary area of the mesh because of (e.g., bad illumination, movement obstacle), so we need to remove it on purpose inevitably creating, however, a missing area.

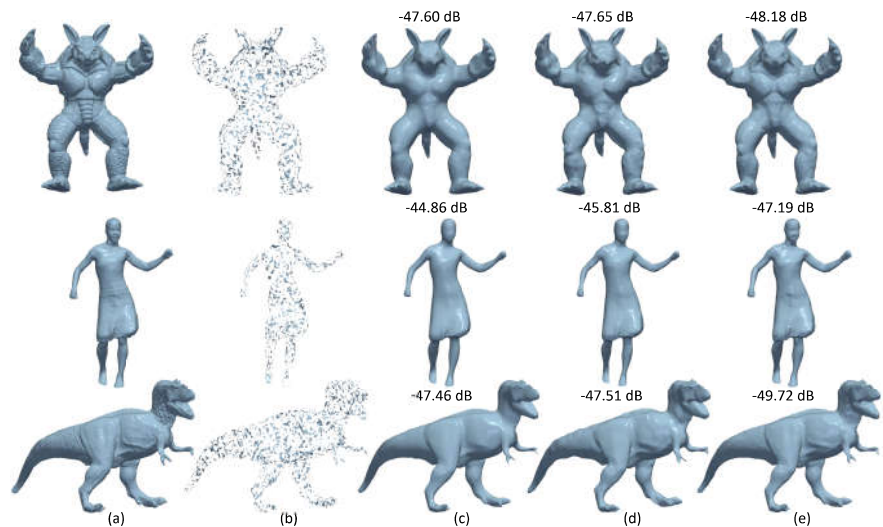

Figure 5.94: (a) Original models (Armadillo, frame 78 of dynamic mesh Samba and Tyra), (b) 50% partial-observed meshes and reconstructed models using: (c) Least-Squares Meshes [24], (d) Laplacian Interpolation [25], (e) our approach.

### 5.3.2 Dynamic Mesh Completion

The reconstruction of incomplete dynamic 3D meshes is a task that demands different handling than this of using static 3D meshes. This can be achieved by exploiting the known connectivity and a subset of motion vectors correspond-ing to the known points in an online setting. The dynamic mesh sequence is not known, a priori, and in each iteration, the used algorithm has to exploit the information that has been presented so far.

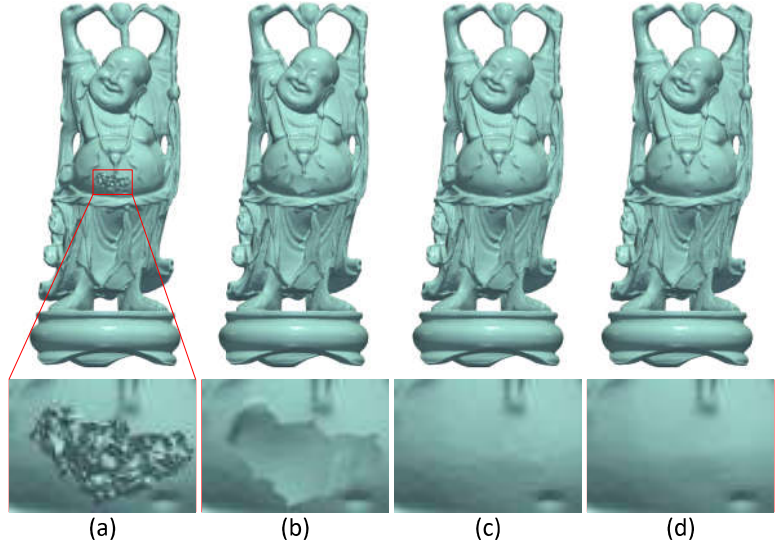

|     |     |     |     |
| --- | --- | --- | --- |
| (a) | (b) | (c) | (d) |

Figure 5.95: (a) Partial-damaged mesh, (b) mesh with hole created by the intentional removal of the damaged part, (c) reconstructed mesh using our approach, (d) original mesh (Buddha model).

Let us assume that $\mathcal{A}'$ is a highly incomplete dynamic mesh. In other words, each mesh of the animation has only a subset of known vertices while the rest have been removed. The incomplete animated model is represented by a sequence of $n_s$ incomplete meshes $\mathcal{A}' = [\mathcal{M}_1; \mathcal{M}'_2; \ldots \mathcal{M}'_{n_s}]$ where $\mathcal{M}'_i \subset \mathcal{M}_i \; \forall \; i = 2, n_s$. Each incomplete dynamic mesh is described by a matrix of dimension $3n \times n_s$:

$$\mathcal{A}' = \begin{bmatrix} \mathcal{M}_1 \\ \mathcal{M}'_2 \\ \mathcal{M}'_3 \\ \mathcal{M}'_4 \\ \vdots \\ \mathcal{M}'_{n_s} \end{bmatrix} = \begin{bmatrix} \mathbf{v}_{11} & \mathbf{v}_{12} & \mathbf{v}_{13} & \mathbf{v}_{14} & \ldots & \mathbf{v}_{1(n-1)} & \mathbf{v}_{1n} \\ 0 & \mathbf{v}_{22} & 0 & 0 & \ldots & 0 & 0 \\ \mathbf{v}_{31} & 0 & 0 & \mathbf{v}_{34} & \ldots & 0 & 0 \\ 0 & 0 & \mathbf{v}_{43} & 0 & \ldots & \mathbf{v}_{4(n-1)} & \mathbf{v}_{4n} \\ \vdots & \vdots & \vdots & \vdots & \vdots & \vdots & \vdots \\ 0 & \mathbf{v}_{n_s 2} & 0 & 0 & \ldots & \mathbf{v}_{n_s(n-1)} & 0 \end{bmatrix} \quad (5.90)$$

The incomplete meshes are created by randomly removing points from the original ones. Fig. 5.96 depicts some indicative frames of an animated mesh assuming different densities of known points, creating missing areas.

To estimate the coordinates of the missing points, we initially use the prescribed connectivity information of the adjacency matrix. Despite the fact that the position of vertices is changing, the adjacency matrix remains fixed over time, since each mesh of a dynamic 3D mesh has the same connectivity [355]. This observation allows us to estimate the adjacency matrix only once and use it repeatedly for any subsequent mesh of the same model. Moreover, we assume that we have full knowledge of the first mesh $\mathcal{M}_1$ of the sequence.

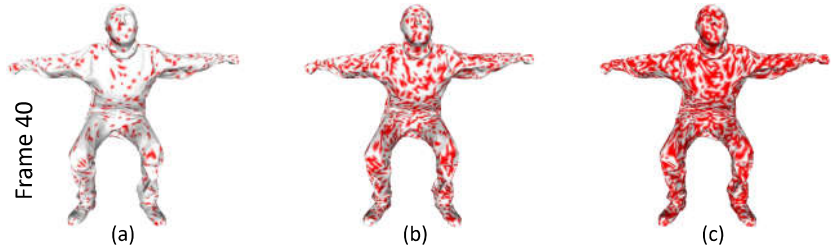

Figure 5.96: Missing areas are represented with white color while the known are represented with red. The percentage of considered known points are: (a) 10%, (b) 30%, (c) 50% (Squat model).

**Spatial Classification of Each Vertex in Each Frame**

For each one of the $n$ vertices of the mesh $\mathcal{M}_1$, we create a cell of nodes $c_i \ \forall \ i = 1, \cdots, n$ using the knowledge of the adjacency matrix $\mathbf{C}$. Each cell $c_i = [\dot{c}_{i1}, \ \dot{c}_{i2}, \ \cdots \ \dot{c}_{in_{v_i}}]$ consists of the $n_{v_i}$ indices of the vertices that belongs to the first-ring area $\Psi_{i1}$ of each vertex $i$. However, it is worth mentioning here that each $i$ cell has different connectivity valence $n_{v_i}$. The element $\dot{c}_{ij}$ represents the index of $j$-th connected vertex with the vertex $i$. The set of cells remains fixed for every frame (mesh) and it is used for recovering the missing points. Fig. 5.97 illustrates an example of two connected cells $c_1, c_2$ with their related connections, cell $c_1$ has $n_{v_1} = 6$ connected neighbors (valence), while $c_2$ has a valence $n_{v_2} = 5$.

Then, we classify the vertices for each frame $\mathcal{M}'_i \ \forall \ i = 2, \cdots, n_s$. This process is executed in a sequential manner for each mesh starting from the second mesh until the end of the animation sequence. We assume three different classes for each vertex:

- **Anchor vertices** which are assumed as the known vertices of the mesh.

- **Satellites vertices** are the vertices that belong to a cell of an already known vertex. More specifically, if a vertex belongs to a cell where an anchor vertex exists, then it is defined as first generation satellite. Following the same line of thought, if a vertex belongs to a cell of an $t$ generation satellite vertex, then it is defined as $t-1$ generation satellite vertex.

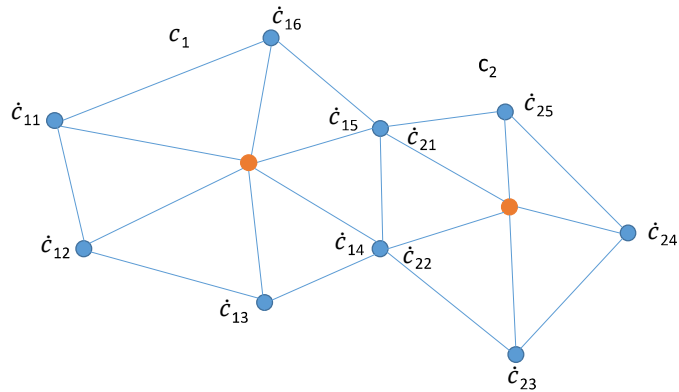

Figure 5.97: Representation of two connected cells

- **Unknown vertices** are the vertices that do not belong to a cell of an already known vertex (i.e., anchor or satellites). These vertices will be recovered and placed in the mesh at a future iteration.

The classification procedure is an iterative process that tries to eliminate any remaining unknown vertex so that the 3D mesh will be composed of only anchor or satellite vertices. This means that the coordinates of an unknown vertex are evaluated in a following iteration. This method has been proved to be robust and all the missing vertices are always recovered.

The method starts (Iteration 1) with the assumption that if the position of a vertex is known, then it is referred to as anchor point. In Iteration 2, all the vertices that belong to the cell of an anchor point are classified as satellite vertices (first generation satellite), while the rest vertices are classified as unknown. At the following iteration (Iteration 3), the position of the satellite points are taken into account and their cells are used for identifying new satellites (second generation satellites). The process stops when only known points exist.

Fig. 5.98 illustrates the aforementioned classification procedure. Red points represent the known vertices (anchor points). Blue points represent the satellites (first generation) which are connected with the anchor points, and correspondingly green point represent the satellites (second generation) that are connected with the first generation satellites. A vertex can be a satellite point for more than one anchor point while an anchor point can be a satellite point for another neighbor anchor point. A cell $c_i$ can indicate the existence of satellite vertices in the first-ring area of vertex $i$, nevertheless, it does not specify their real position. The classification procedure is used for assigning weights and then estimating the unknown vertex position using a weighted filtering approach based on their previous position and the motion vector of the anchor points. A detailed description of the method is provided in the following sub-section.

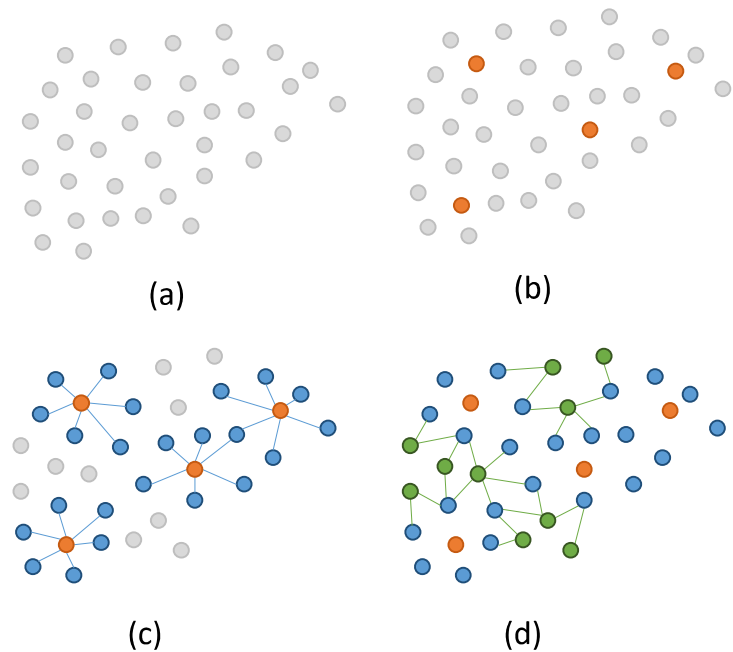

Figure 5.98: (a) Part of mesh with unknown points, (b) Anchors are indicated with red (Iteration 1), (c) First generation satellites identified based on anchors cells (Iteration 2), (d) Second generation satellites are set based on first generation satellites cells (Iteration 3).

**Geometry Completion Based on Topological Characteristics**

The interpolation procedure, which is proposed for the geometry completion, is divided into two stages. In the first stage, a spatial iterative process is executed for reconstructing the mesh using only the known vertices of the current mesh. In the second stage, a temporal process tries to reconstruct the entire animated mesh using knowledge of the previous frame. Fig. 5.99 presents the process while the reconstruction takes place gradually, starting from the second mesh and continues until the end of the animation sequence rendering the method appropriate for online setups.

We assume that the satellite vertices are expected to move towards the direction of an anchor keeping their common topological characteristics (e.g., distances between each other) unchanged. However, some satellites are connected with more than one anchors, meaning that their new position will be affected by the motion vectors of every connected anchor. For estimating coordinates more accurately we suggest a weighted reconstruction function which is de-fined by exploiting the following observations. As it was mentioned earlier, when a satellite is connected with more than one anchors then its new position is affected by the motion vectors of all the anchors. However, each anchor contributes with a different weight that is related to the relative distance between anchors and satellites. Smaller cells are more rigid so that their satellite points are expected to follow the motion vectors of the closest anchor, as shown in Fig. 5.100.

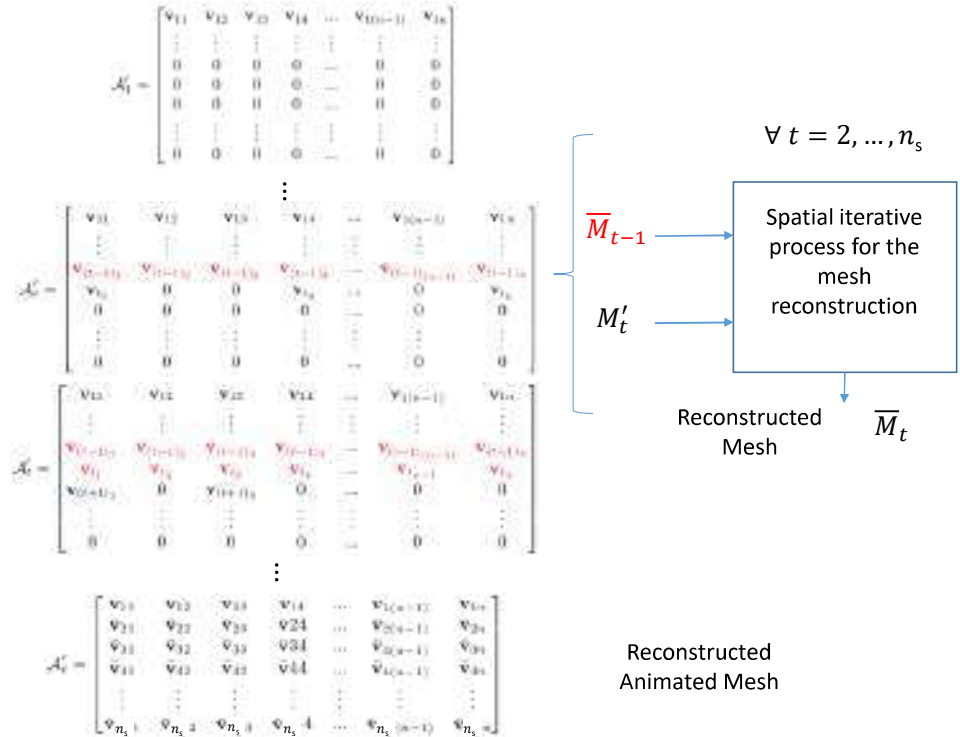

Figure 5.99: The animation is reconstructed frame by frame (temporal process) taking into account the previous reconstructed mesh and the current incomplete (spatial iterative process).

The second rule that we apply, is based on the fact that some points are more trustworthy than others. In other words, we give more emphasis on the anchor points instead of the satellites due to their known position. Subsequently, we give more emphasis on the first generation satellites rather than the next generation because they are connected directly with the anchor points so that their estimated position is expected to be more accurate. According to the aforementioned observations, we distinguish two different weighted factors:

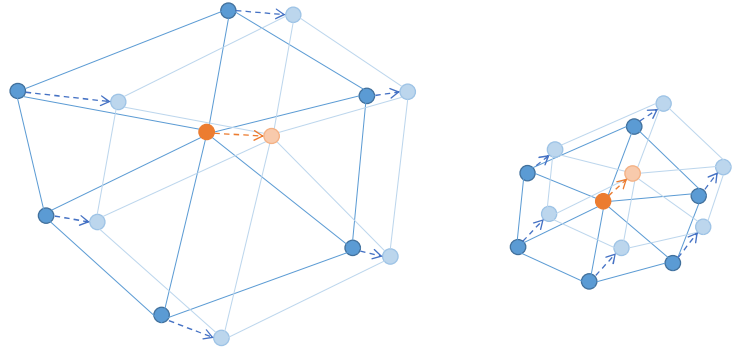

Figure 5.100: Example of large and small variations in cell movements.

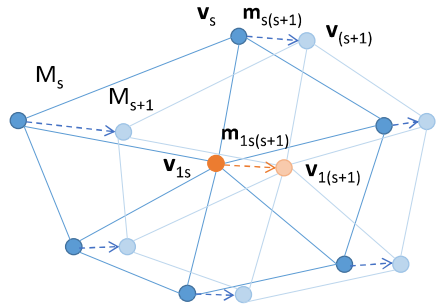

Figure 5.101: Anchor (in red color) and satellites (in blue color) movement in a cell of two sequential meshes (s) and (s+1).

1. The factor $w_{d_{ij}}$ that represents the inverse distance between the vertex $i$ and the known connected vertex $j$, $\forall\, j = 1, \cdots, n_{v_i}$, such as $w_{d_{ij}} = 1/||\mathbf{v}_i - \mathbf{v}_j||_2^2$.

2. The factor $w_{p_i}$ that represents the prioritization weight of vertex $i$, $\forall\, i = 1, \cdots, n$. Anchors prioritization weights have the highest value (equal to the number of iterations) while the last generation satellites have the smallest one (equal to one). More specifically, in each iteration of the classification procedure the weights of the known vertices increase by 1.

For each satellite vertex $i$, we estimate its new coordinates that it has into the $(s+1)$ mesh, by updating the coordinates that it has in the previous $(s)$ mesh, based on the following equation:

$$\mathbf{v}_{(s+1)i} = \mathbf{v}_{(s)i} + \mathbf{m}_{s(s+1)i} \tag{5.91}$$

where

$$\mathbf{m}_{s(s+1)i} = \frac{\sum_{j=1}^{n_{v_i}} w_{d_{ij}} w_{p_i} \mathbf{m}_{s(s+1)j}}{\sum_{j=1}^{n_{v_i}} |w_{d_{ij}} w_{p_i}|} \tag{5.92}$$

**Algorithm 12:** Reconstruction of 3D animation model

    **Function:** 3D mesh reconstruction based on the previous complete mesh
    **Input**   : Animated 3D model $\mathcal{A}'$ with missing data.
    **Output** : A reconstructed animated model $\bar{\mathcal{A}}$.
1 Find the connectivity **C** of $\mathcal{M}_1$;
2 **for** $i \leq n_s$ **do**
3     **while** *Number of known vertices < n* **do**
4        Search for satellite points using the connectivity of the cell $c_i$;
5        Estimate the weighted distance via Eq. (5.92);
6        Update vertex based on its previous frame coordinates via Eq. (5.91);
7     **end**
8 **end**
9 **return** Reconstructed 3D animation models $\bar{\mathcal{A}}$;

The $\mathbf{m}_{s(s+1)ij}$ represents the motion vector of vertex $\mathbf{v}_i$ from $(s)$-th mesh to $(s+1)$-th mesh (see Fig. 5.101). An overview of the proposed method is briefly presented in the following Algorithm 12.

**Experimental Results in Completion of Dynamic 3D Meshes**

A big advantage of this approach is its fast execution times, especially when it is compared with matrix completion methods. Nevertheless, the processing time is related to the number of initial known vertices. Specifically, the execution time increases linearly with the number of unknown vertices. In Fig. 5.102, we present the processing time for the reconstruction of each mesh of the animated sequence using different initialization schemes. The required number of iterations for a complete mesh reconstruction depends on the percentage of known vertices. Additionally, we can observe that the two presented models have similar behavior.

In Fig. 5.103 the missing vertices are visualized with red color and the known vertices with blue (1st and 3rd row). A heatmap visualization is also offered to present the squared difference between original and reconstructed mesh for different numbers of known vertices (2nd and 4th row). For comparison purposes, we have also employed conventional techniques for the reconstruction of the animation models, namely the LSM algorithm [100] and the LIA [356]. The compared results of our approach and LIA are presented in Fig. 5.104 showing that our method outperforms LIA in both reconstruction quality and execution time. A major disadvantage of LIA is the smoothed results even in cases with a high percentage of remaining points. In terms of execution time, our method becomes faster when more remaining points are used, because of the fewer iterations that are required, contrary to LIA where the execution time increases because of the larger matrix operations.

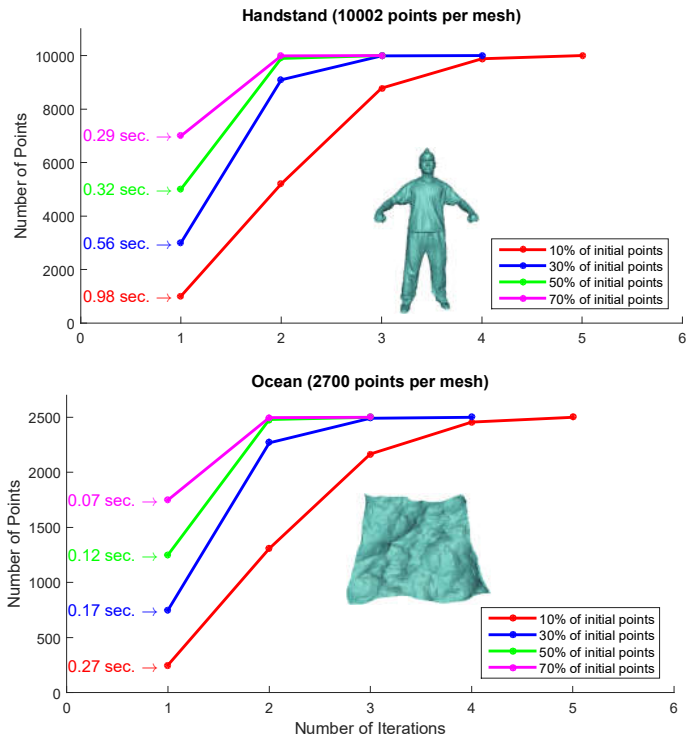

Figure 5.102: Number of iterations and processing time for a full mesh reconstruction.

Fig. 5.106 presents some indicative reconstructed frames of different animated models. For the sake of completeness, the NMSVE values are also illustrated under each reconstructed mesh. Finally, it should be noted that despite the high motion variance of animated trajectories, the perceptual quality of the reconstructed dynamic meshes when the density of the known point is higher than $> 30\%$ is considerably high. This totally satisfies the main goal of this work which is the design and implementation of fast and effective dynamic mesh reconstruction approaches. Fig. 5.105 illustrates some indicative reconstructed frames of the Handstand model after using the aforementioned approaches.

The evaluation of both the execution time and the reconstruction quality highlights the effectiveness of this method even in complex motion scenario that include rapid changes between sequential frames or in cases with a small percentage of known points.

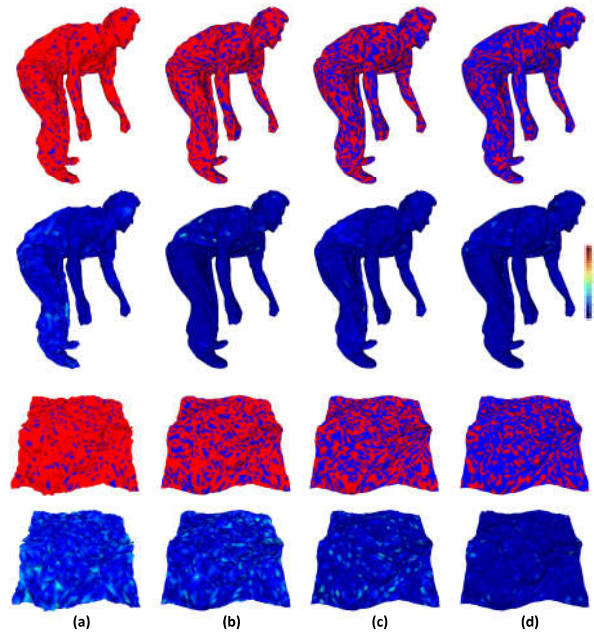

Figure 5.103: Visualized missing data and heatmap visualization for different density of points (a) 10% of original points, (b) 30% of original points, (c) 50% of original points, (d) 70% of original points. (Handstand frame 110, Ocean frame 1500).

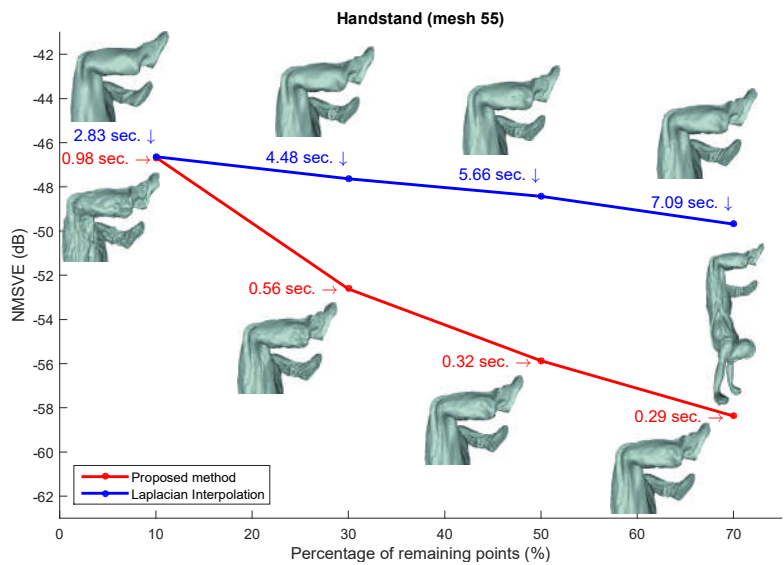

Figure 5.104: NMSVE and processing time results for the two compared methods.

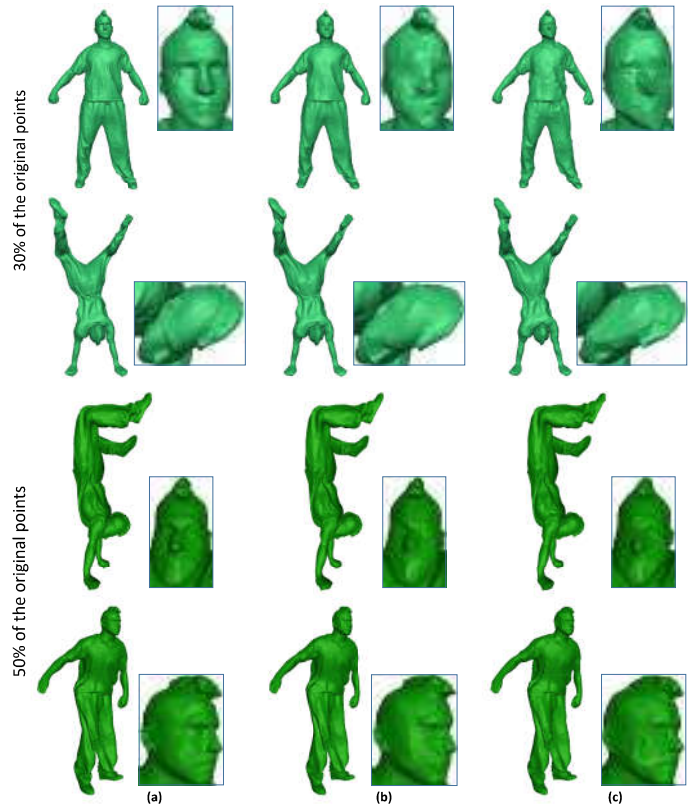

Figure 5.105: Handstand with 30% of original points (Frames 150 & 80) and with 50% of the original points (Frames 55 & 120) (a) our method, (b) LIA, (c) LSM.

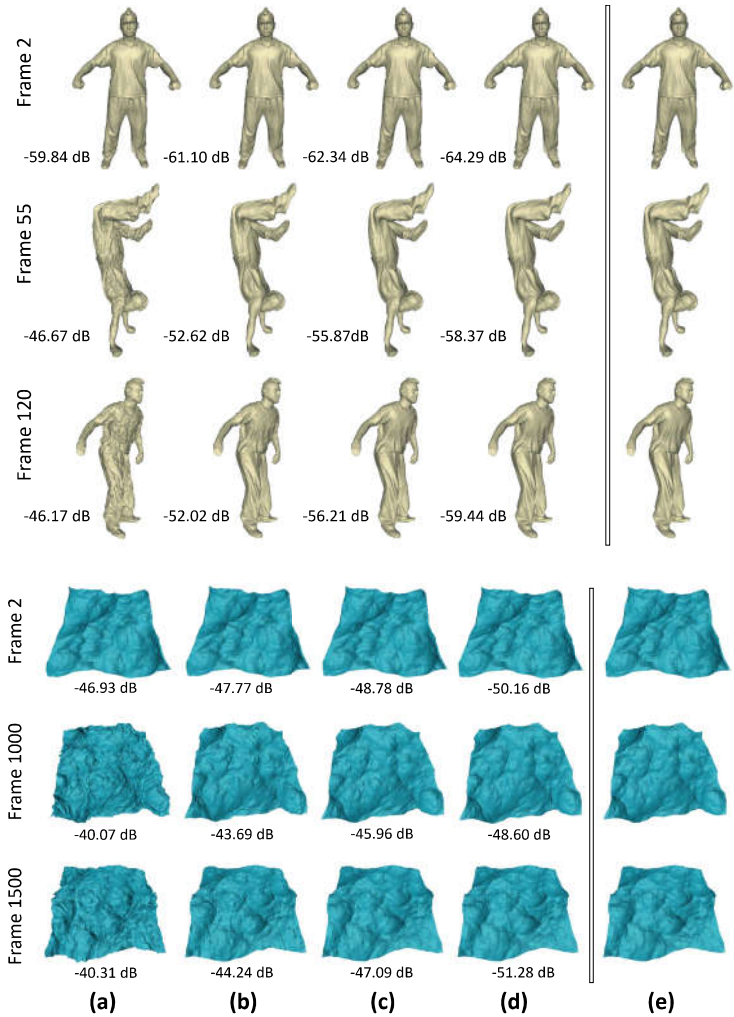

Figure 5.106: Reconstructed meshes for different density of remaining points (a) 10% of original points, (b) 30% of original points, (c) 50% of original points, (d) 70% of original points, (e) original mesh. (Handstand & Ocean)

### 5.3.3 Completion of Dynamic Point Clouds

Here, we present an extension of the previously presented subsection. More specifically, this extension makes also use of the motion vectors but it is applied to dynamic point cloud instead of dynamic 3D meshes. Additionally, it uses a weighted regularized Laplacian interpolation scheme providing more robust reconstruction results. Fig. 5.107 depicts an example of point clouds of a highly incomplete model assuming different numbers of known points.

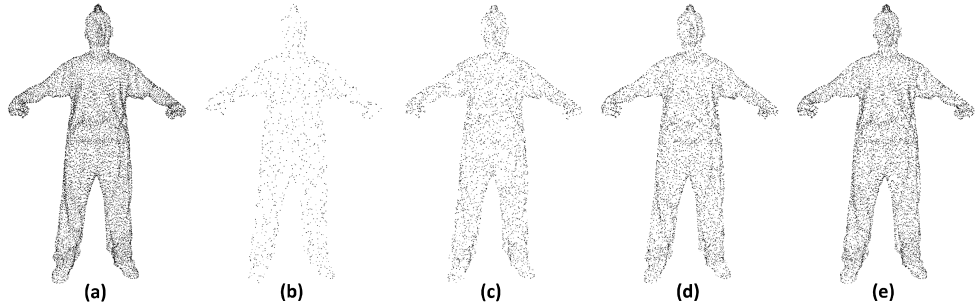

Figure 5.107: Indicative incomplete frames of the animated sequence: (a) original mesh (10,002 points), (b) 10% of original points, (c) 30% of original points, (d) 50% of original points, (e) 70% of original points.

The connectivity between vertices in a static point cloud is initially unknown. For this reason, we have to approximately construct it, based on the following two techniques:

- the $\epsilon \in \mathbb{R}$ neighborhoods ($\epsilon - N$)
- the $k \in \mathbb{N}$ nearest neighbors ($k - NN$)

It is known that $\epsilon - N$ graphs are symmetric and are more geometric meaningful, although depending on the choice of the parameter $\epsilon$ they could lead to heavy or disconnected graphs. On the other hand, the choice of the parameter for $k \in \mathbb{N}$ graphs is more straightforward, usually leading to connected graphs. Even though the position of vertices changes from frame to frame, we assume that the topology is global, meaning that every static point cloud of the same sequence has the same topology over time [357]. Similar to Eq. (3.6), we define as $\mathbf{C} \in \mathbb{R}^{n \times n}$ the binary adjacency matrix with the following elements:

$$\mathbf{C}_{ij} = \begin{cases} 1 & \text{if } i,j \in \Psi^k \\ 0 & \text{otherwise} \end{cases} \quad (5.93)$$

where $\Psi^k$ is a matrix estimated by $k - NN$ or $\epsilon - N$ process and consists of the neighbors of each vertex. The adjacency matrix is estimated once and it can be used during the consolidation of any other frame.

The point cloud consolidation process takes place sequentially starting from the $\mathcal{M}'_1$ and continuing until all point clouds of the sequence have been consolidated. Only the previous (consolidated) and the currently incomplete frame are required for the process, making the proposed schema ideal for real-time applications. In this way, our approach can be directly applied in real time 3D scanning scenarios. On the contrary, other methods require the entire knowledge of the incomplete point cloud sequence before the execution of the consolidated process. Fig. 5.108 briefly illustrates the proposed schema for a real-time consolidation of a dynamic point cloud.

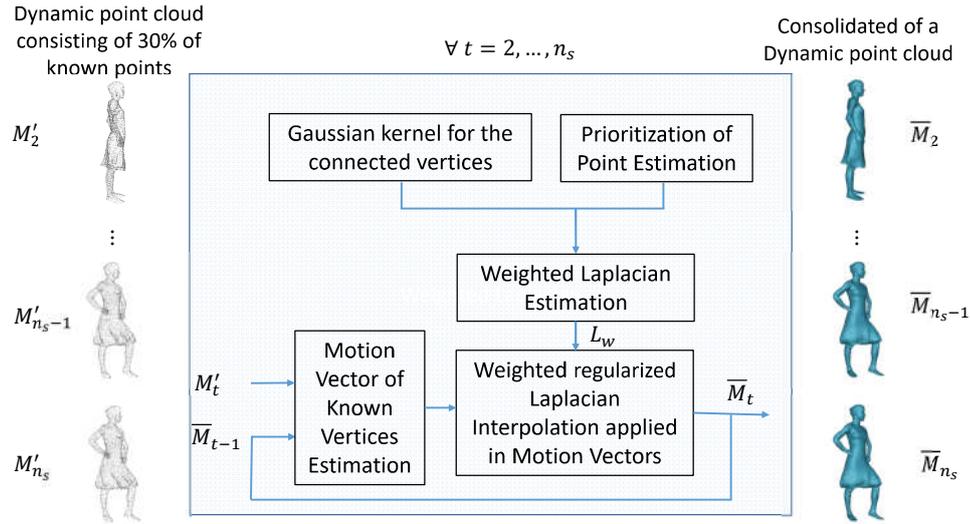

Figure 5.108: Dynamic point cloud consolidation schema.

**Weighted Laplacian Matrix**

The binary Laplacian matrix provides information about the connectivity of vertices. Nevertheless, weighted Laplacian matrices are able to provide additional geometric information that can be efficiently exploited by a variety of processes. In this work, we suggest constructing a modified weighted Laplacian matrix that takes into account:

1. The spatial coherence between connected vertices
2. A prioritization ranking of the vertices

The prioritization of a vertex $i$ is performed by assigning an integer factor $w_{p_i} \in [1, N]$ expressing the proximity to an already known vertex. Using this annotation rule, the highest value, let's denote as $N$, is assigned to known vertices, the value $N-1$ is assigned to vertices that are directly connected with known vertices while 1 is assigned to the most remote vertices. Fig. 5.109 presents an example of vertices with different prioritization values. We assume that adjacent vertices exhibit similar behavior because they share common topological characteristics. In this way, we use a Gaussian kernel for expressing this spatial relation. The weighted adjacency matrix takes into account the two mentioned parameters which are applied to the adjacency matrix according to:

$$\mathbf{C_w}_{ij(t)} = \begin{cases} w_{p_i} \cdot e^{-\|\mathbf{v}_{i(t-1)} - \mathbf{v}_{j(t-1)}\|_2^2} \cdot c_{ij} & if\ c_{ij} = 1 \\ 0 & otherwise \end{cases} \quad (5.94)$$

where $(t)$ and $(t-1)$ denotes the current and previous time indices (frames). The proposed weighted Laplacian matrix is estimated by:

$$\mathbf{L_w} = \mathbf{D} - \mathbf{C_w} \quad (5.95)$$

where $\mathbf{D} = diag\{D_1, \ldots, D_k\}$ is a diagonal matrix with $D_i = \sum_{j=1}^{k} c_{wij}$.

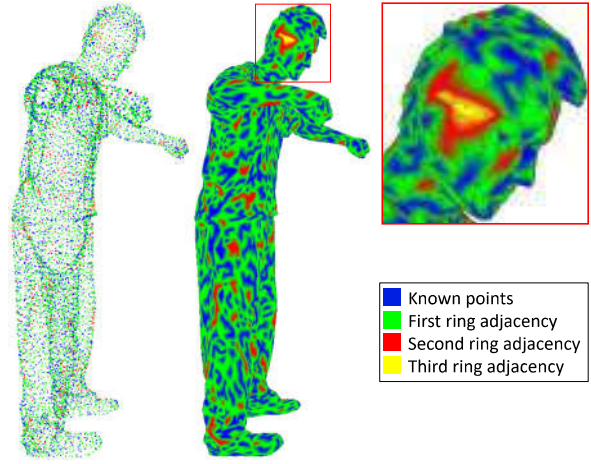

Figure 5.109: Colorized points with different prioritization values.

**Weighted Regularized Laplacian Interpolation (WRLI)**

According to [25], a way to interpolate a triangulated 3D model with a curved surface in a three-dimensional space is by putting constraints on the Laplacian $\mathbf{L}$. The proposed weighted Laplacian matrix, presented in Eq. (5.94), encloses all the necessary constraints in order to be efficiently used by WRLI. The process is applied to the motion vector of known vertices managing to overcome the smoothness limitation of the original algorithm, according to the process described in subsection 4.6.

To add further constraints to the problem, given that is overdetermined, we suggest using the $l_1$ regularizer, which efficiently exploits the fact that there are small variations in the motion vectors of the given vertex and the mean motion vector of its neighbors. Thus, we suggest estimating the unknown motion vectors $\mathbf{B_u}$ by solving the following Lasso problem:

$$\arg\min_{\mathbf{z}} 0.5\|\mathbf{P} - \mathbf{Rz}\|_2^2 + \gamma\|\mathbf{z}\|_{l_1} \tag{5.96}$$

where $\mathbf{P} = \mathbf{QB_k} - \mathbf{R\hat{B}_u}$, $\mathbf{Q} = \begin{pmatrix} \mathbf{L}_{w_{11}} \\ \mathbf{L}_{w_{21}} \end{pmatrix}$ and $\mathbf{R} = \begin{pmatrix} \mathbf{L}_{w_{12}} \\ \mathbf{L}_{w_{22}} \end{pmatrix}$, $\mathbf{z} = \mathbf{B_u} - \mathbf{\hat{B}_u}$ and $\mathbf{\hat{B}}_i = \sum_{\Psi_1} \mathbf{d}$ represents the mean motion vector of those vertices which are contained into the first ring area $\Psi_1$ of vertex $i$. Algorithmically, the convex optimization problem in Eq. (5.96), known also as LASSO problem, can be tackled by any generic SOCP solver, while in our case we have used the ADMM as described in [358]. The coherence between the motion vector of neighboring vertices enhances the recovery efficiency of $\mathbf{z}$ by efficiently exploiting its sparsity. The evaluated motion vectors are then used for estimating the coordinates of the missing vertices based on their previous position according to:

$$\mathbf{v_{u(t)}} = \mathbf{v_{u(t-1)}} + \mathbf{\hat{B}_u} \tag{5.97}$$

where $\mathbf{v_k} = [\mathbf{v_{k1}} \cdots \mathbf{v_{kn_k}}]$ and $\mathbf{v_u} = [\mathbf{v_{u_{n_k+1}}} \cdots \mathbf{v_{un}}]$. Fig. 5.110 correspond to the known motion vectors and the estimated ones based on WRLI. The described process is briefly presented in the following Algorithm 13.

---

**Algorithm 13:** Completion of dynamic point cloud

**Input** : Incomplete dynamic point cloud $A'$.
**Output:** A consolidated dynamic point cloud $\bar{A}$.
1 Find the adjacency matrix $\mathbf{C}$ of $M_1$ based on $\kappa - NN$ or $\epsilon - N$;
2 **while** $i \leq n$ **do**
3     Estimate the Gaussian Kernel of connected vertices;
4     Estimate the prioritization factor of each point;
5     Estimate the weighted Laplacian Matrix via Eq. (5.95);
6     Estimate the motion vectors of known vertices according to subsection 4.6;
7     Estimate the motion vectors of the unknown vertices based on WRLI by solving Eq. (5.96);
8     Estimate the coordinates of vertices via Eq. (5.97)
9 **end**

---

**Experimental Results in Consolidation of Dynamic Point Clouds**

In this paragraph, we present experimental results using a variety of initial conditions for different incomplete dynamic point cloud like complex motion scenarios with rapid changes in the point trajectories of sequential frames or using highly incomplete dynamic point clouds where only a small percentage of points are known. The quality performance of the proposed technique is evaluated in comparison with other SoA techniques, namely:

- **LSM** [24]: is described as the solution of an extended system of equations $[L^T I_{n_f \times n_k}^T]^T x_t = [0^T, x_{n_k,n}]$, for $i = 1, \cdots, n_s$, for $n_k$ known vertices in the $i$-th frame.

- **MC** [359]: which works on the animation matrix $A \in \mathbb{R}^{3n \times n_s}$. This technique represents the geometry-myopic approach for the consolidation of $A$, where no prior information is required. It belongs to the offline case, where all the available data has already been captured.

- **GMC** [359]: which exploits the spatial static point cloud geometry, represented by the Laplacian matrix **L**. It also belongs to the offline case.

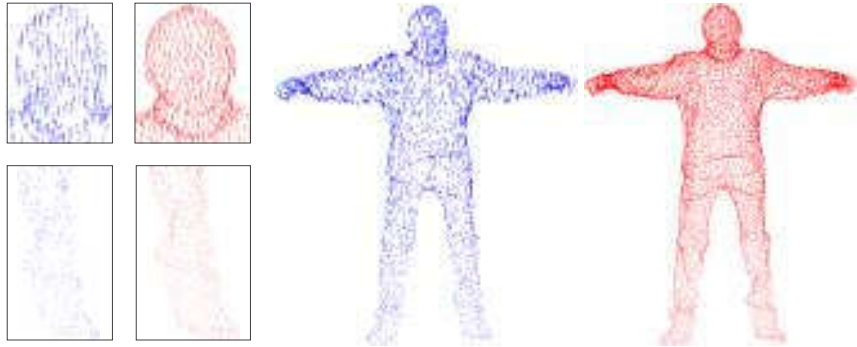

Figure 5.110: Motion vectors of known vertices (blue) and motion vectors estimated via WRLI (red).

Fig. 5.111 presents the NMSVE per frame for each compared method. According to this figure, the proposed method provides better results than any other method. LMS and GMC always have a constant NMSVE value independently of the frame.

On the other hand, results of MC and WRLI strongly depend on the temporal coherence between consecutive frames. More specifically, the NMSVE value of the proposed method is $\approx -62\ dB$ when the temporal coherence is high. Theoretically, we can guarantee that if the sampling frequency of a captured dynamic point cloud is high then the consolidated result using our method does not have a difference that can be easily perceived, even in highly incomplete cases. In Fig. 5.112, the consolidated results for the Handstand model are presented. The experiment is executed after assuming different numbers of known vertices (30% and 50%). For the sake of completeness, the NMSVE values of the consolidated 3D static point cloud are also shown.

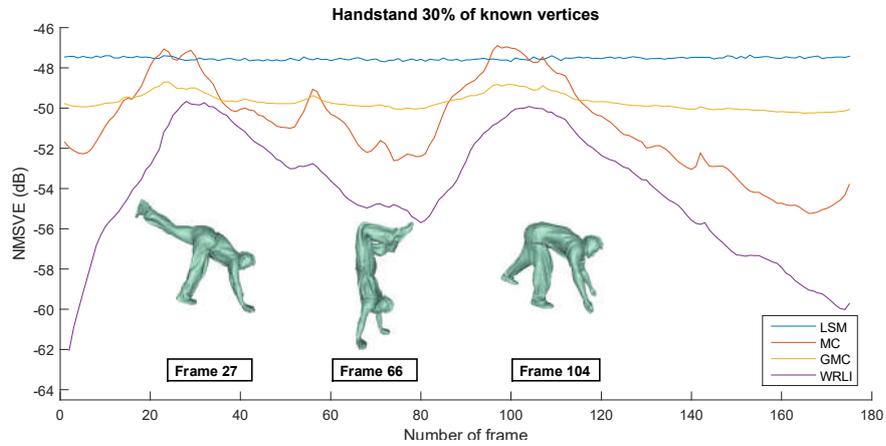

Figure 5.111: NMSVE values per frame for different methods.

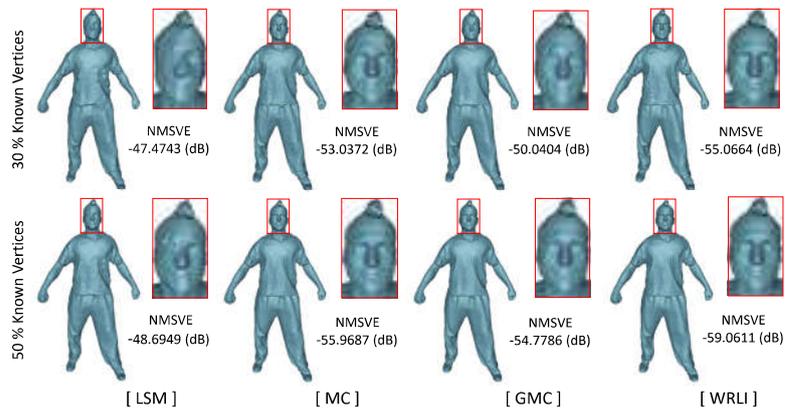

Figure 5.112: Reconstructed results using different percentages of know vertices (Handstand Frame 140).

In Fig. 5.113, a heatmap visualization is finally presented highlighting the difference between the original and the consolidated 3D static point cloud for each one of the considered approaches. The results show that the proposed approach efficiently exploits the coherence between motion vectors of known vertices and is ideally suited for online settings where the point cloud sequence is not known a priori and is dynamically generated.

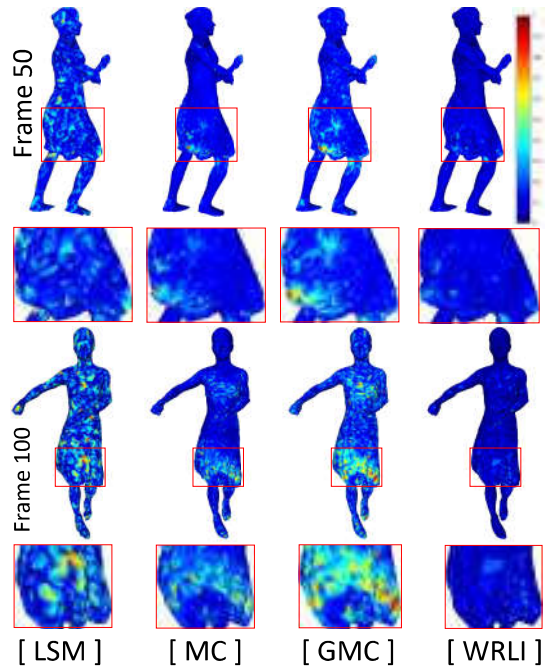

Figure 5.113: Heatmap visualization of consolidated results using different methods (Samba model with 30% of known vertices).

**Publications that have contributed to this section:**

1. Content of the subsection 5.3.1 has been presented in **C8**
2. Content of the subsection 5.3.2 has been presented in **C10**
3. Content of the subsection 5.3.3 has been presented in **C9**

## 5.4 Outliers Removal and Consolidation of Static and Dynamic 3D Meshes

Nowadays, there is a tremendous interest in the processing of unorganized static or dynamic point clouds, generated using a variety of 3D scanning technologies such as structured light and LIDAR systems, ideally suited for challenging applications such as immersive tele-presence systems and gaming. These captured 3D point clouds are usually noisy with a perceptive percentage of outliers, stressing the need for an approach with low computational requirements which will be able to automatically remove the outliers and create a consolidated point cloud. Additionally, robots are able to carry out a complex series of actions, to make decisions, to interact with their environment and generally to perform plausible reactions. The visual ability of the robots plays an important role in their behavior, helping them to efficiently manage the received information. To effectively address the aforementioned issues, we present a novel method, that detects accurately and efficiently the outliers by exploiting the spatial coherence in the object geometry and the sparsity of the outliers in the spatial domain. This is achieved by solving a convenient convex method called RPCA.

This section presents the results of the outliers removal approach applied in static and dynamic 3D point clouds under three different use cases, as shown below:

- Outliers removal in highly dense and unorganized data, acquired by laser scanner (subsection 5.4.1).

- Real-time outliers removal and consolidation of data, received by robot's depth camera (subsection 5.4.2).

- Outliers removal and consolidation of dynamic point clouds (subsection 5.4.3).

Each one of these use cases has its own challenges that need to be addressed.

### 5.4.1 Outliers Removal of Highly Dense and Unorganized Point Clouds

Point clouds exhibit common characteristics that can be effectively exploited by recent advances in low-rank matrix analysis, which has been proposed as a method to decompose a given data matrix into a low-rank and a sparse component [302]. More specifically, benefiting from the sparsity of the outliers in the spatial domain, we provide an approach that applied to remove outliers from a captured unorganized point cloud [131]. To demonstrate the effectiveness of RPCA as the proposed technique, we evaluate it by using real scanned point clouds of urban environments which are extremely dense consisting of millions of points. The proposed processing pipeline is briefly presented in Fig. 5.114. This figure pertains to outliers removal on unorganized point clouds after their accurate identification using as an indicator the sparse matrix of an RPCA approach.

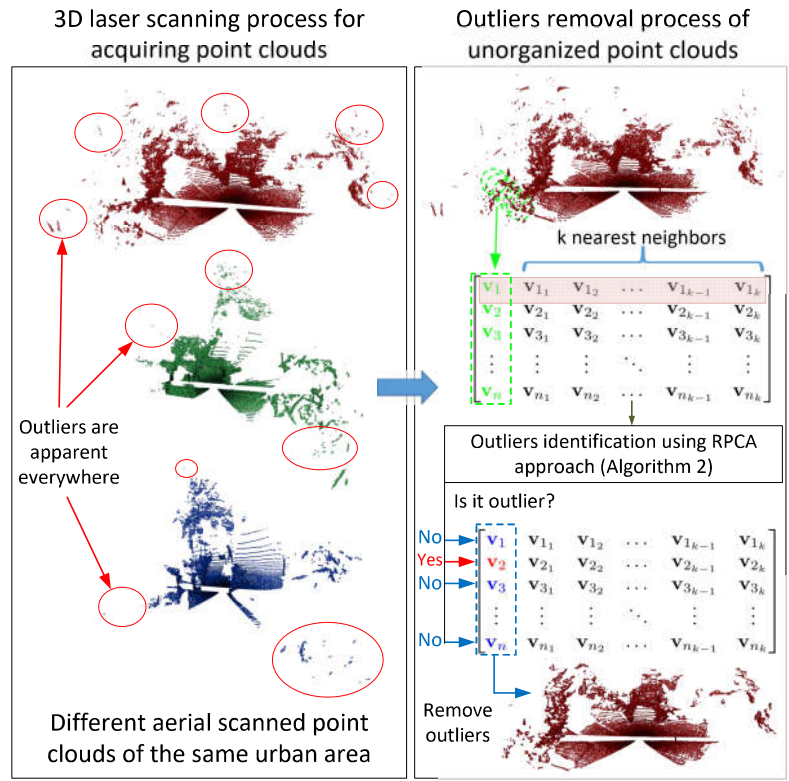

Figure 5.114: Pipeline of the outliers removal approach.

**Outliers Identification and Removal**

We assume the existence of an unorganized point cloud $\mathcal{M}$ consisting of $n$ points, each one of them represented by a vector $\mathbf{v} = [x, y, z] \in \mathbb{R}^{3 \times 1}$ in a 3D coordinate system. Inspired by the observation that neighboring vertices, which lie in a small cluster, they are sharing similar geometrical and topological information, we create a spatial coherence matrix $\mathbf{M}$ which is used as input for the RPCA analysis. For each point $\mathbf{v}_i \; \forall \; i = 1, ..., n$ of the point cloud, we create a cluster consisting of its $k$ nearest neighbors. The coordinates of the cluster points $\mathbf{v}_{i_j}$ may then be organized by placing the $j$ point $\forall \; j = 1, ..., k$ of the $i$ cluster into the matrix $\mathbf{M} \in \mathbb{R}^{3n \times k+1}$, as shown in Eq. (5.98).

$$\mathbf{M} = \begin{bmatrix} \mathbf{v}_1 & \mathbf{v}_{1_1} & \cdots & \mathbf{v}_{1_k} \\ \mathbf{v}_2 & \mathbf{v}_{2_1} & \cdots & \mathbf{v}_{2_k} \\ \vdots & \vdots & \ddots & \vdots \\ \mathbf{v}_n & \mathbf{v}_{n_1} & \cdots & \mathbf{v}_{n_k} \end{bmatrix} = \begin{bmatrix} x_1 & x_{1_1} & \cdots & x_{1_k} \\ y_1 & y_{1_1} & \cdots & y_{1_k} \\ z_1 & z_{1_1} & \cdots & z_{1_k} \\ \vdots & \vdots & \ddots & \vdots \\ x_n & x_{n_1} & \cdots & x_{n_k} \\ y_n & y_{n_1} & \cdots & y_{n_k} \\ z_n & z_{n_1} & \cdots & z_{n_k} \end{bmatrix} \quad (5.98)$$

**Identifying and Removing Outliers**

We do not use the RPCA in the conventional way, taking advantage of the low-rank matrix $\mathbf{E}$, but we identify outliers using the sparse matrix $\mathbf{S}$. More specifically, for specifying which point should be considered as an outlier, only the first column of the sparse matrix $\mathbf{S}$ is used. The $k$ next columns, representing values of the neighbor points, are ignored.

$$\mathbf{S} = \begin{bmatrix} \mathbf{s}_1 & \mathbf{s}_{1_1} & \mathbf{s}_{1_2} & \cdots & \mathbf{s}_{1_{k-1}} & \mathbf{s}_{1_k} \\ \mathbf{s}_2 & \mathbf{s}_{2_1} & \mathbf{s}_{2_2} & \cdots & \mathbf{s}_{2_{k-1}} & \mathbf{s}_{2_k} \\ \vdots & \vdots & \vdots & \ddots & \vdots & \vdots \\ \mathbf{s}_n & \mathbf{s}_{n_1} & \mathbf{s}_{n_2} & \cdots & \mathbf{s}_{n_{k-1}} & \mathbf{s}_{n_k} \end{bmatrix} \quad (5.99)$$

Firstly, we estimate the scalar value of each $i$ element $||\mathbf{s}_i||_2 = \sqrt{(x_i^2 + y_i^2 + z_i^2)}$. The number of the removed outliers depends on the value of the selected thresh-

---

**Algorithm 14:** Outliers identification and removal process using RPCA

**Input** : A highly dense and unorganized collection of points $\mathcal{M}$, threshold $\tau$
**Output:** A point cloud $\mathcal{M}' \subseteq \mathcal{M}$ without outliers
1 Decompose $\mathbf{M}$ into a low-rank approximative $\mathbf{E}$ and a sparse $\mathbf{S}$ component using RPCA;
2 **for** $i = 1$ *to* $n$ **do**
3     **if** $||\mathbf{s}_i||_2 < \tau$ **then**
4         Keep the point $i$;
5     **else**
6         Remove the point $i$; // It is outlier
7     **end**
8 **end**

---

old $\tau$. In other words, we can change the value of $\tau$ in order to take more or fewer outliers. Specifically, the higher this value the fewer the identified outliers. Considering the above, we can conclude that the proposed approach is completely adaptable and can be easily modified for the individual needs of different applications. Additionally, we can mention here that the matrix **E** (low-rank matrix) could be possibly used for denoising purposes since it provides a smoothed version of the input data **M** [360]. The steps of the proposed process are described in Algorithm 14.

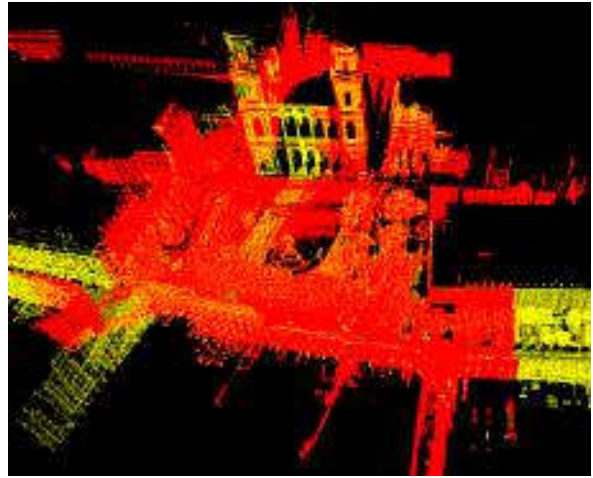

Figure 5.115: Registration of two point clouds after the removal of outliers.

**Outliers Removal of Highly Dense Point Cloud**

In Fig. 5.116, we present the results of the outliers removal process using, as input, two real-scanned and unorganized point clouds containing an unknown percentage of outliers. Fig. 5.116-(c) & 5.116-(d) includes zoomed regions of these highly dense point clouds providing an easier evaluation of our method's performance. Two different types of outliers seem to be apparent: (i) the large-scale outliers (identified in red circles) which lie away from the point cloud and (ii) the small-scale outliers (identified in yellow circles) which are tangled with the useful information and mistakenly could be recognized as points. The proposed approach can efficiently remove both of these abnormalities, as shown in Fig. 5.116-(b) & 5.116-(d). Additionally, Fig. 5.115 presents a registration example of two different point clouds, located in overlapped areas, after the removal of outliers.

The experimental process shows that the proposed method is capable to efficiently handle both large and small-scale outliers as well as providing a smoothed version, i.e., **E**, which could be used in denoising applications.

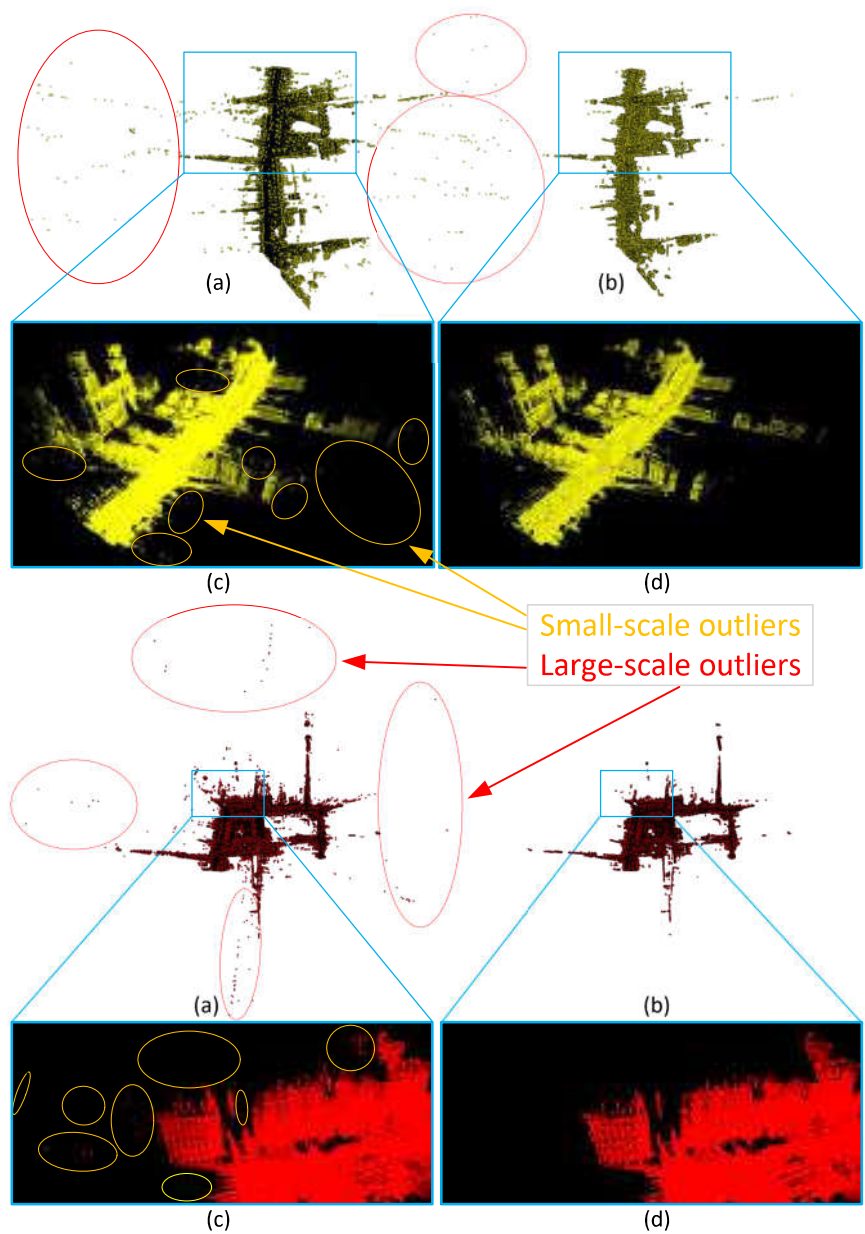

Figure 5.116: [First-Third line] Point clouds from above. [Second-Forth line] Zoomed region of the point clouds from a different view. (a) & (c) Real-scanned urban environment with a great amount of outliers, (b) & (d) point clouds without outliers after using the proposed approach.

### 5.4.2 Removing of Outliers in Real Time Robotic Applications

A step further than the previous subsection, we present here the method for removing outliers applied in 3D point clouds which are captured by the optical system of robots having depth camera at their disposal. We investigated real scenarios where the robot moves while it acquires the point cloud in a natural light environment so that unpleasant noise and outliers become apparent.

More specifically, the robot captures visual information by collecting points using its depth camera. However, in the general case, the precision of cameras is limited or a relative motion between robot and target exists, and as a result, the acquired 3D point cloud suffers from noise, outliers and incomplete surfaces. These problems must be solved before the robot uses this inappropriate information. In Fig. 5.117 the basic steps of our approach are briefly presented.

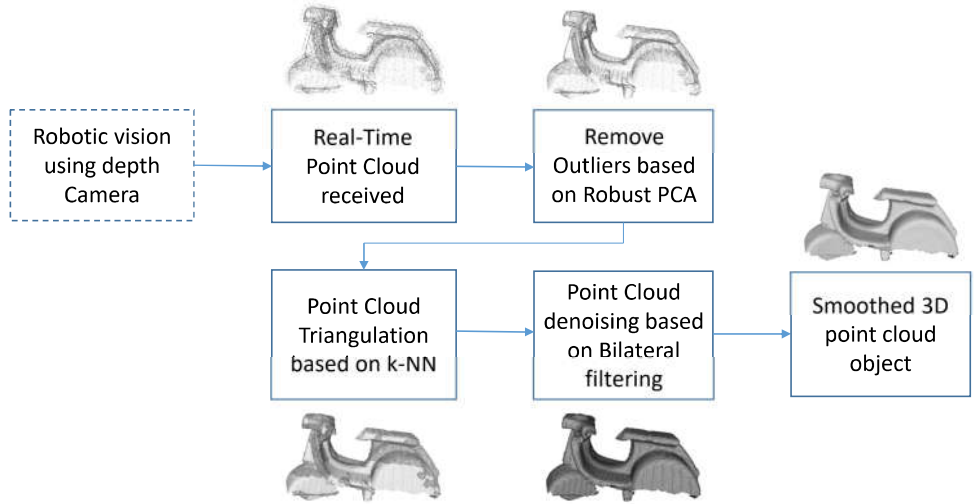

Figure 5.117: Framework of our proposed method.

Firstly, the robotic optical sensor starts capturing points of objects that exist in the robot's line of vision. The acquired 3D point clouds are usually noisy and they also have misaligned outliers. We use RPCA for removing the outliers and consequently, we create a triangulated model based on the k-NN algorithm. The triangulation process helps us to specify the neighbors of each point so that the bilateral filtering method can efficiently be used as a denoising technique. In the end, a smoothed 3D mesh is created which has a suitable form for being used by other applications or processing tasks; namely navigation, object recognition and segmentation.

We assume that the captured 3D point cloud $\mathcal{M}$ consists of $n$ points represented as a vector $\mathbf{v} = [\ \mathbf{x},\ \mathbf{y},\ \mathbf{z}\ ]$ in a 3D coordinate space $\mathbf{x}, \mathbf{y}, \mathbf{z} \in \mathbb{R}^{n \times 1}$ and $\mathbf{v} \in \mathbb{R}^{n \times 3}$. Some of these points are considered as outliers and they must be removed.

**Triangulation Based on k-NN**

The acquired 3D point cloud is unorganized, meaning that the connectivity of its points is unknown thus we are unable to take advantage of the geometric relation between neighboring vertices. To overcome this limitation we estimate the connectivity between points based on the k-NN algorithm using $k = 7$ as the length of each cell, as a matter of fact, each point is connected with the 7 nearest neighbor points. The defined connectivity helps us to create the binary adjacency matrix $\mathbf{C} \in \mathbb{R}^{n \times n}$. The 1s of this sparse matrix represents edges; by connecting 3 corresponding vertices a face is created such that $f_i = [\mathbf{v}_{i1}, \mathbf{v}_{i2}, \mathbf{v}_{i3}] \ \forall \ i = 1, n_f$ where $n_f > n$.

**3D Mesh Denoising Using Bilateral Filtering**

The final step of our approach is the mesh denoising using the bilateral filtering algorithm directly applied to the triangulated 3D point cloud. The used approach of the bilateral algorithm, as described in [7] [4], is a fast method and it can be easily implemented in real-time applications. The denoising process occurs by firstly estimating the *d* factor according to Eq. (5.100) and then updating the noisy vertices according to Eq. (5.101).

$$w_i = \frac{\sum_{j \in \Psi^k} W_1 W_2 \langle \mathbf{n}_i, \mathbf{v}_i - \mathbf{v}_j \rangle}{\sum_{j \in \Psi^k} |W_1 W_2|} \tag{5.100}$$

where $W_1 = e^{\frac{-(\|\mathbf{v}_i - \mathbf{v}_j\|)^2}{2\sigma_1^2}}$, $W_2 = e^{\frac{-(\langle \mathbf{n}, \mathbf{v}_i - \mathbf{v}_j \rangle)^2}{2\sigma_2^2}}$ and $\mathbf{n}_i = \frac{\sum_{j \in \Psi^k} \mathbf{n}_j}{k} \ \forall \ i = 1, n, \ \forall \ j \in \Psi^k$. Once the $\mathbf{w} = [w_1 \ w_2 \ \cdots \ w_n]$ is estimated then each vertex is updated according to:

$$\hat{\mathbf{v}} = \mathbf{v} + \mathbf{n} \cdot \mathbf{w}^T \tag{5.101}$$

where $\hat{\mathbf{v}} \in \mathbb{R}^{n \times 3}$ represents the vector of smoothed vertices. This method manages to preserve the features of the object without further shrinking of the already smoothed areas.

**Real Time Implementation and Results**

Two models of two different geometrical categories are used for the evaluation. The first model consists of flat surfaces (Moto) while the second has a lot of features and details (Cactus), as presented in Fig. 5.118. To establish experimental conditions as close as possible to the real cases, the experiments have taken place using only natural lights, no artificial lights or other additional light sources are used. Fig. 5.119 depicts the moment when the robot's vision starts, while in Fig. 5.120, snapshots of the capturing process are illustrated for the two models.

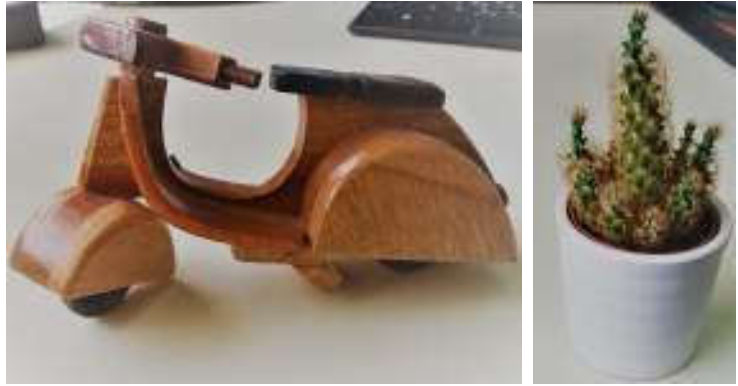

(a) (b)

Figure 5.118: Images of the original objects used as models for the evaluation process, (a) Moto (b) Cactus.

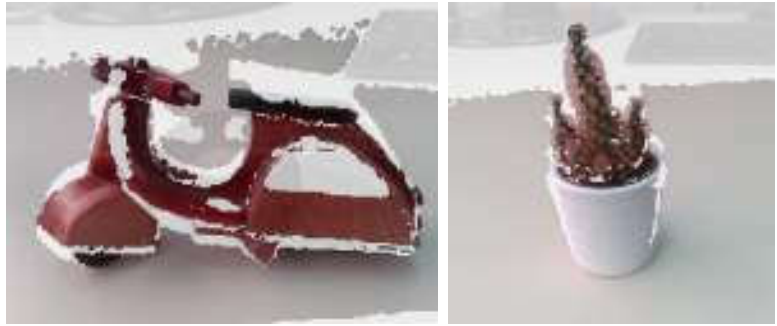

Figure 5.119: The moment when the capturing point process starts.

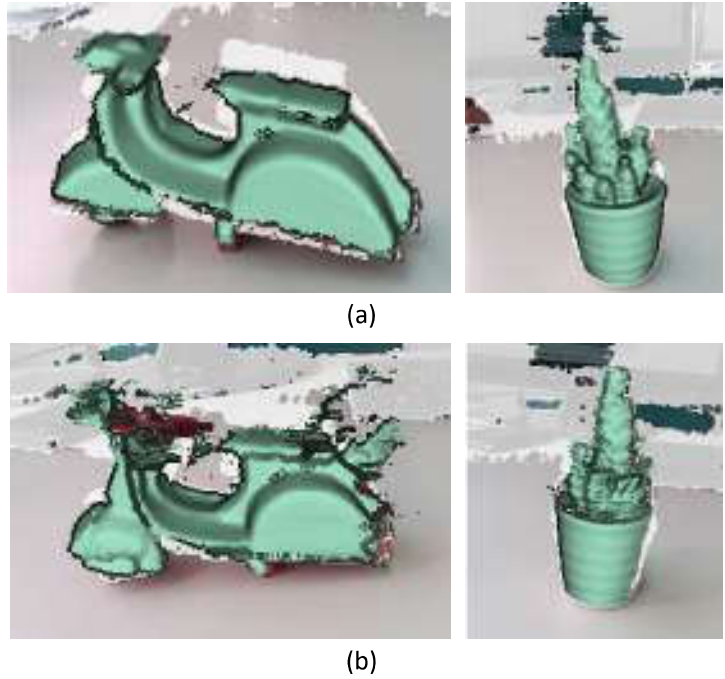

Figure 5.120: (a) Stable collection of points, (b) Collection of points with a sudden robot's movement.

Fig. 5.120 depicts two different case studies of acquiring points. In the first case of Fig. 5.120 (a), a careful gathering takes place while in the second case of Fig. 5.120 (b), a sudden movement of the robot causes noise and outliers. Nevertheless, observing carefully the figures, we can see that even in the first case there are imperfections that need to be covered.

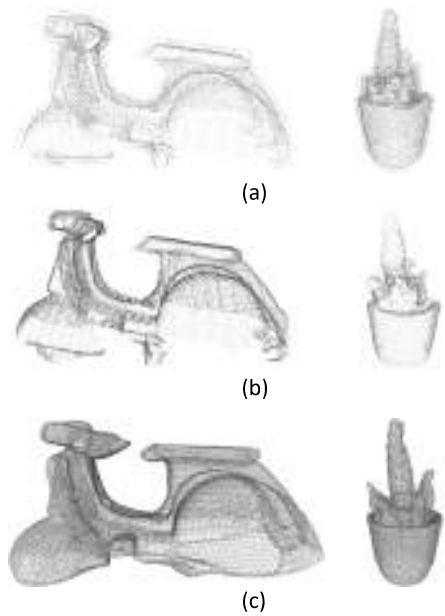

(a)

(b)

(c)

Figure 5.121: (a) 3D point clouds with noise and outliers, (b) outliers have been removed but noisy parts still remain, (c) triangulated and smoothed results.

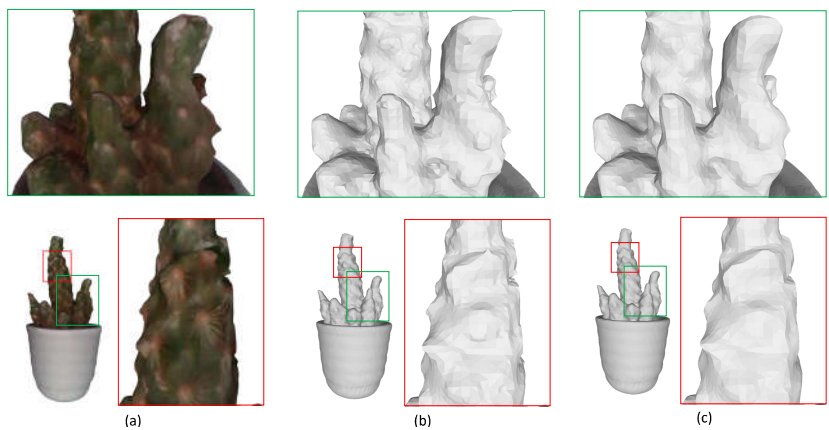

Figure 5.122: (a) Cactus model is presented having its original texture, (b) noisy surfaces after outliers' removing and triangulation process, (c) the 3D denoised mesh based on bilateral filtering.

The step-by-step extracted results of the proposed method are presented in Fig. 5.121. More specifically, Fig. 5.121 (a) presents the point clouds having noise and outliers, as received by the depth camera sensor. In Fig. 5.121 (b) the outliers of the point clouds have been removed, however, abnormalities of noise still remain. In Fig. 5.121 (c) the final triangulated and smoothed objects are presented. In Fig. 5.122, the smoothed results using bilateral filtering for the Cactus model are presented. In addition, an enlarged presentation of details is also shown for easier comparison between the 3D objects.

### 5.4.3 Outliers Removal and Consolidation of Dynamic Point Cloud

RPCA can also be used, in a similar way, to remove outliers from a Dynamic Point Cloud Sequence (DPCS). Then, a weighted laplacian interpolation can be used to fill the holes and gaps created during the outliers removal stage. Fig. 5.123 briefly illustrates the basic steps of the proposed schema.

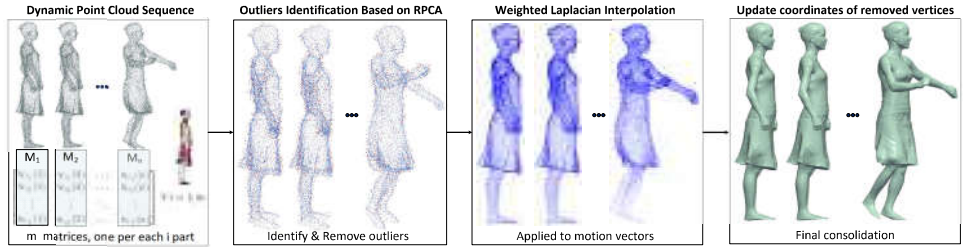

Figure 5.123: Outliers removal pipeline of an unorganized dynamic point cloud sequence.

**Creating the Spatial-Temporal Matrices**

Each frame of the point cloud sequence is separated into $n_t$ parts such as $\mathcal{M}_i = \mathcal{M}_i[1] \cup \mathcal{M}_i[2] \cup \cdots \cup \mathcal{M}_i[n_t] \ \forall \ i = 1, \cdots, n_s$, using the METIS algorithm [325]. However, METIS does not separate the mesh into equal-sized parts. To overcome this problem we create overlapped patches by extending each part adding neighbor connected points, so that each part has exactly the same number of points $n_m$. Then we create the matrices $\mathbf{M}[i] \in \mathbb{R}^{3n_m \times n_s}$ according to:

$$\mathbf{M}[i] = \begin{bmatrix} \mathbf{v}_{1_1}[i] & \mathbf{v}_{2_1}[i] & \cdots & \mathbf{v}_{n_{s_1}}[i] \\ \mathbf{v}_{1_2}[i] & \mathbf{v}_{2_2}[i] & \cdots & \mathbf{v}_{n_{s_2}}[i] \\ \vdots & \vdots & \ddots & \vdots \\ \mathbf{v}_{1_{n_m}}[i] & \mathbf{v}_{2_{n_m}}[i] & \cdots & \mathbf{v}_{n_{s_{n_m}}}[i] \end{bmatrix} \ \forall \ i = 1, \ldots, n_t \qquad (5.102)$$

where $\mathbf{v}_{n_{s_{n_m}}}[n_t] = [x_{n_{s_{n_m}}}[n_t]; y_{n_{s_{n_m}}}[n_t]; z_{n_{s_{n_m}}}[n_t]$ represents the $n_m{}^{th}$ vertex of $n_t{}^{th}$ part of $n_s{}^{th}$ frame.

The reason why we separate each frame into parts, instead of using the entire object, is because different parts of a mesh have different types of motion as well as different spatial information. The main purpose is to take advantage of both spatial and temporal information which a dynamic point cloud sequence has.

**Identifying and Removing Outliers**

The sparse matrices $\mathbf{S}[i]$, the formulation of which is presented below, is used for specifying which vertices should be considered as outliers.

$$\mathbf{S}[i] = \begin{bmatrix} \mathbf{s}_{1_1}[i] & \mathbf{s}_{2_1}[i] & \cdots & \mathbf{s}_{n_{s_1}}[i] \\ \mathbf{s}_{1_2}[i] & \mathbf{s}_{2_2}[i] & \cdots & \mathbf{s}_{n_{s_2}}[i] \\ \vdots & \vdots & \ddots & \vdots \\ \mathbf{s}_{1_{n_m}}[i] & \mathbf{s}_{2_{n_m}}[i] & \cdots & \mathbf{s}_{n_{s_{n_m}}}[i] \end{bmatrix} \quad \forall \, i = 1, \ldots, n_t \quad (5.103)$$

More specifically, we estimate the scalar value of each element $|\mathbf{s}|$ and we assume that its magnitude is proportional to the probability of the corresponding vertex being an outlier. In general, the elements representing the original areas have values $< 10^{-7}$, while the outlier values are usually $> 10^{-3}$. We can accurately identify outliers using an empirical threshold $\tau = 10^{-4}$ that provides reliable results. Then, the inevitably created gaps and holes are filled using the method described in the following section.

**Weighted Graph Laplacian Matrix**

The binary Laplacian matrix provides information about the connectivity of vertices. On the other hand, a weighted Laplacian matrix is able to provide additional geometric information which can be efficiently exploited by a variety of processes. We suggest the construction of a modified weighted Laplacian matrix that takes into account the three following factors:

1. The spatial coherence between connected vertices, expressed as the invert absolute value of their distance:

$$\mathbf{W}_{1_{ij}} = \begin{cases} \frac{1}{\|\mathbf{v}_{i(t-1)} - \mathbf{v}_{j(t-1)}\|_2^2} & \text{if } j \in \Psi_i^k \\ 0 & \text{otherwise} \end{cases} \quad \forall \, i = 1, n \quad (5.104)$$

2. The similarity of the motion vectors between connected vertices, expressed by the cosine of their angle $\widehat{\theta_{\mathbf{m}_i \mathbf{m}_j}}$:

$$\mathbf{W}_{2_{ij}} = \begin{cases} \frac{\mathbf{m}_i \cdot \mathbf{m}_j}{|\mathbf{m}_i| \cdot |\mathbf{m}_j|} & \text{if } j \in \Psi_i^k \\ 0 & \text{otherwise} \end{cases} \quad \forall \, i = 1, n \quad (5.105)$$

where $\mathbf{m}_i = \mathbf{v}_i(t-1) - \mathbf{v}_i(t)$ is the motion vector which represents the distance of a vertex $\mathbf{v}_i$ between two sequential frames (time indices). We need to keep the elements of $\mathbf{W}_2$ always non-negative, for this reason, we add +1 to any value so that $\mathbf{W}_2 \in [0,2]$.

3. A prioritization ranking which expresses the connecting proximity of an unknown vertex with an already known vertex. The known vertices have the highest value 4, while the value of the unknown vertices depends on the approximation with the nearest known vertex based on their connectivity degree $n_v$.

$$\mathbf{W}_{3_{ij}} = \begin{cases} 4w_{ij} & \text{if } \mathbf{v}_i \text{ is known} \\ \frac{w_{ij}}{n_v+1} & \text{otherwise} \end{cases} \quad \forall\, i,j = 1,n \quad (5.106)$$

where $w \in [0,1]$ represents the corresponding value of the binary adjacency matrix.

Fig. 5.124 illustrates examples of points with different prioritization ranking.

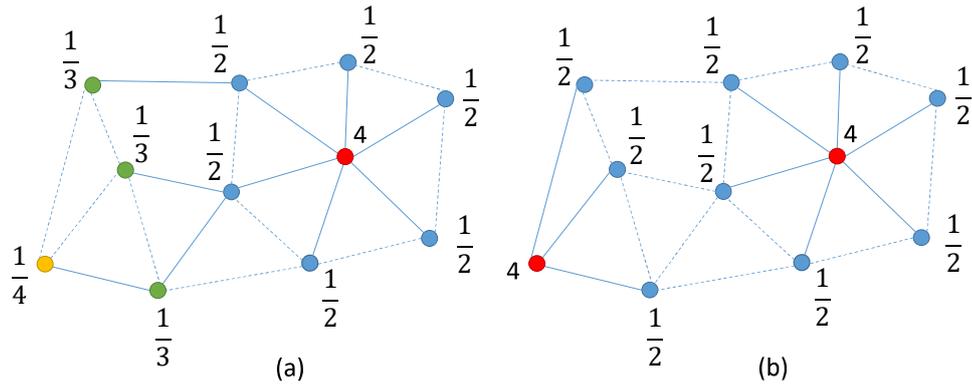

Figure 5.124: Different values of prioritization factor per each point. The factor is defined based on the number of the known points (red color) and their relative degree of connectivity. Example with (a) one known point, (b) two known points.

The weighted adjacency matrix takes into account all the aforementioned parameters which are applied according to:

$$\mathbf{C}_w = \mathbf{W}_1 \circ \mathbf{W}_2 \circ \mathbf{W}_3 \circ \mathbf{C} \quad (5.107)$$

where $\circ$ denotes the Hadamard product. Finally, the proposed weighted Laplacian matrix is estimated according to:

$$\mathbf{L}_w = \mathbf{D} - \mathbf{C}_w \quad (5.108)$$

where $\mathbf{D} = diag\{D_1, \ldots, D_n\}$ is a diagonal matrix with $D_i = \sum_{j=1}^{n} c_{w_{ij}}$.

**Weighted Laplacian Interpolation**

Then, the coordinates of the missing vertices are estimated by updating their position of the previous frame using the motion vectors of Eq. (4.27), as presented in the following equation:

$$\mathbf{v_u}(t) = \mathbf{v_u}(t-1) + \mathbf{B}_u \ \forall \ t = 2, n_s \qquad (5.109)$$

Finally, all vertices of the incomplete frame $t$ are known $\mathbf{v}(t) = \mathbf{v_k}(t) \cup \mathbf{v_u}(t)$ where $\mathbf{v_k} = [\mathbf{v_{k1}} \cdots \mathbf{v_{k n_k}}]$ and $\mathbf{v_u} = [\mathbf{v_{u n_k+1}} \cdots \mathbf{v_{u n}}]$. We need to mention here that the first frame is necessary for the estimation of motion vector between the second and the first frame. For this reason, for the first frame only, we utilize the low-rank matrix $\mathbf{E}$.

**Outliers Removal and Consolidation Results**

The quality performance of the proposed technique is compared with other state-of-the-art techniques: (i) the RPCA method using the returned low-rank matrix and (ii) the method described in [26], which is used for the removal of outliers in combination with laplacian interpolation to points in order to estimate the missing points.

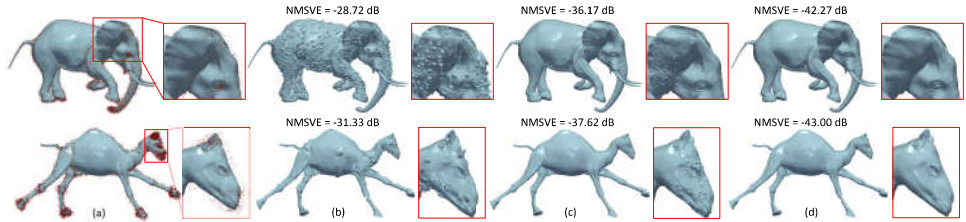

Figure 5.125: (a) Original mesh and point cloud with 20% of outliers, reconstructed models using: (b) Liu-Chan method in [26], (c) the low-rank matrix of RPCA, (d) our approach. (Elephant frame 33, Camel frame 36 [27]).

In Fig. 5.125, we present the reconstructed results of our approach in comparison with the aforementioned approaches. As we can observe, RPCA seems to have very good results, although it fails to remove outliers that appear close to the original data. In these cases, our method outperforms the conventional one because it does not only remove the identified outliers but can be modified to remove some controversial points, without losing information as the second step is able to qualitatively reconstruct the object. In Fig. 5.126, a heatmap visualization is presented highlighting the difference between the original and reconstructed 3D point cloud for each compared method. The superiority of our method is obvious in Fig. 5.127 where the value of metric $\theta$ is presented per each frame of the reconstructed animation and for any compared method. The robustness of our method is apparent in Fig. 5.128, in which the consolidated results of different cases are almost the same in terms of metric $\theta$, regardless of the different initial percentage of outliers. Specifically, the difference in degree between the results of cases with 10% and 30% of initial outliers is $\sim 1^o$, which is negligible.

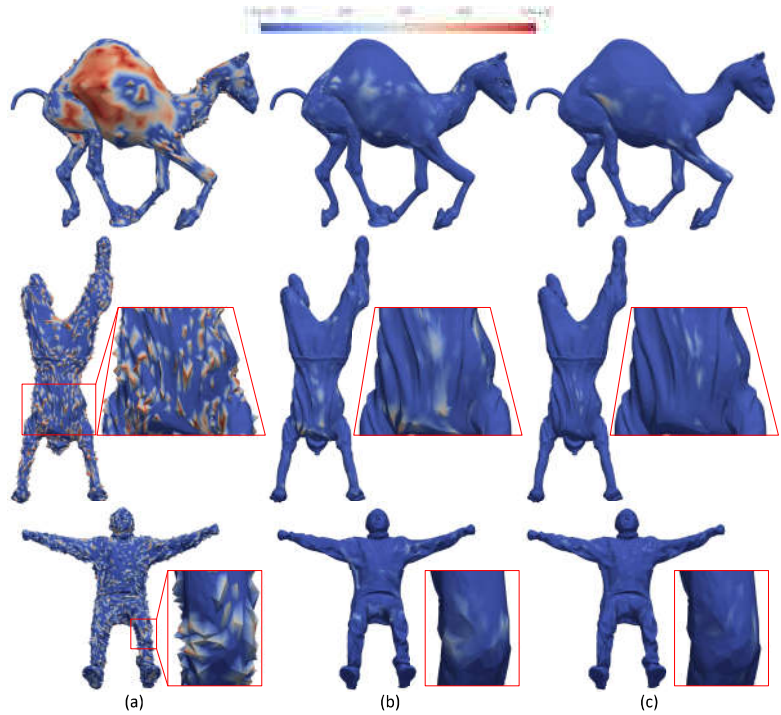

Figure 5.126: Heatmap visualization of reconstructed models, using: (a) Liu-Chan method in [26], (c) the low-rank matrix of RPCA, (c) our approach (Camel frame 18 of 48 and 15% outliers, Handstand frame 48 of 175 and 20% outliers, Squat frame 115 of 120 and 25 % outliers [28]).

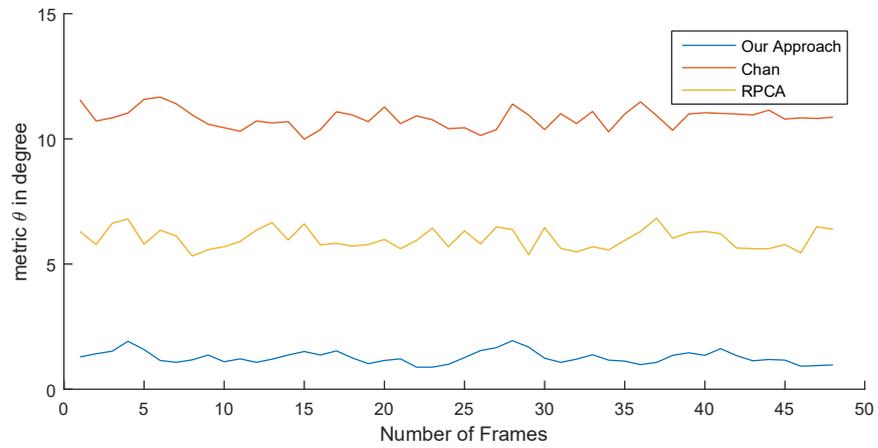

Figure 5.127: Value of metric $\theta$ per each frame of the animated model (Camel), presented for the compared methods.

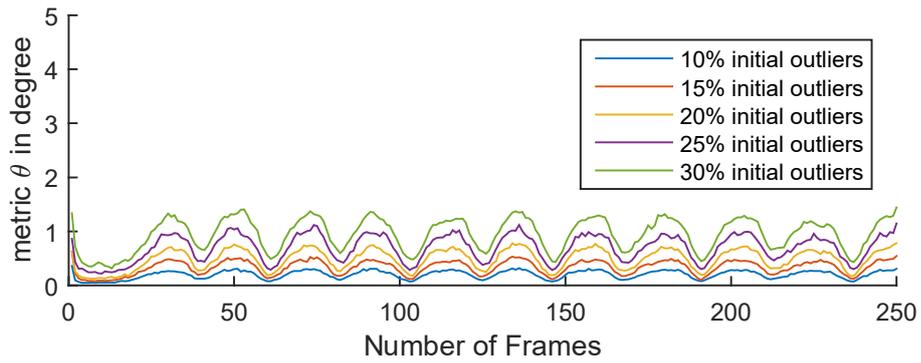

Figure 5.128: Value of metric $\theta$ per each frame of the animated model (Squat), presented for different percentages of initial outliers.

**Publications that have contributed to this section:**

1. Content of the subsection 5.4.1 has been presented in **C11**
2. Content of the subsection 5.4.2 has been presented in **C12**
3. Content of the subsection 5.4.3 has been presented in **C13**

CHAPTER 6

# 3D Geometry Processing and Pattern Recognition

In this chapter, we will present the contribution of this thesis in 3D geometric processing and pattern recognition. In contrary to the previous Chapter 5, the works of the following sections do not try to solve critical issues but they can be used to facilitate other tasks and high-level applications. However, a part of tools of the same mathematical framework, as has been presented in Chapter 4, is also used in the following sections, as briefly illustrated in Fig. 6.1. More specifically, the areas that we investigate can be separated into the following sections:

- Saliency map extraction for features identification (Section 6.1)

- Registration and identification of 3D objects in cluttered scenes (Section 6.2 )

In the following, we will discuss in detail our contribution to these areas.

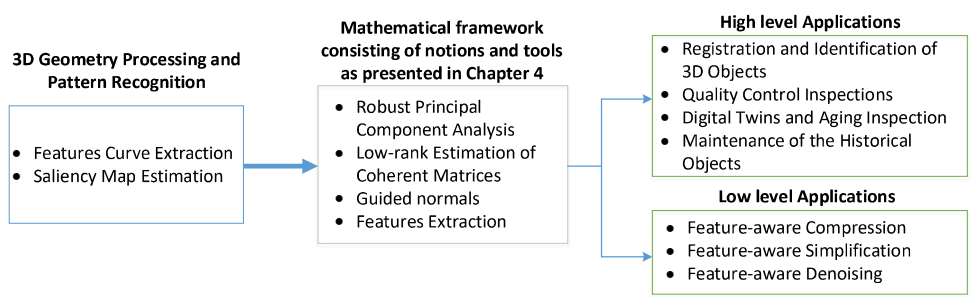

Figure 6.1: 3D geometry processing and pattern recognition tasks to facilitate low and high-level applications.

## 6.1 Saliency Mapping Applications and Features Identification

Saliency maps can be used as a unique signature of a 3D model to describe the geometry of its shape by providing different values to vertices with different geometric characteristics. In this way, the saliency mapping procedure could be used as a shape descriptor to evaluate the similarity between different models' representation of the same 3D object, which have been reconstructed under different conditions. Saliency mapping estimation can be used, as a pre-processing step facilitating many other applications like registration, partial matching, tracking, object recognition, etc. More specifically, the extracted saliency map can evaluate the similarity among different models of the same 3D object under different conditions like scale, resolution quality, pose, and level of noise.

Feature extraction from 3D triangle meshes is a prevalent and important task that could contribute to many scientific fields such as computer vision, pattern recognition, medical 3D modeling, etc. However, there is no universal generic descriptor of 3D shapes that can be used for real case scenarios under challenging real-life situations. In industrial environments, for example, the same type of a 3D object (e.g., gears) may appear in different scales and random poses (i.e., rotation and translation) making a real-time processing task (like the recognition and matching between a "target" and a "query" object) to be very challenging. Additionally, the compared 3D models may have been acquired by using different scanner devices. This variety of devices can affect the pattern of noise and the quality resolution of the digitized models introducing extra difficulties for an efficient processing algorithm. In the following subsections, we will present our contribution to the area of feature extraction, saliency mapping estimation and use of the extracted saliency mapping in applications.

- Feature curve extraction and retrieval of 3D objects (subsection 6.1.1)
- Saliency mapping estimation of 3D objects (subsection 6.1.2).
- Utilization of saliency mapping to facilitate the performance of high level applications (subsection 6.1.3).

### 6.1.1 Feature Curve Extraction on Triangle 3D Meshes

Feature extraction from 3D triangle meshes is a very popular and important task that could contribute to many scientific fields such as computer vision, pattern recognition, medical 3D modeling, etc. However, the main challenge is not just finding corners and edges of 3D models but to automatically extract connected clusters of vertices that jointly represent a feature curve. Here, we present an approach for feature curve extraction and similarity evaluation among feature curves of the same or other models robust to differences in scale, resolution quality, pose, or partial observation.

The method for the feature curve extraction on 3D meshes is separated into two basic steps. At the first step, we estimate the saliency of each vertex using spectral analysis. The magnitude of the estimated saliency identifies if a vertex is a feature or not. Based on the geometry, we can say that the feature vertices represent the edges of a feature curve (both crests and valleys) or corners. In the second step, we estimate the mean curvature of the extracted features and we use it to classify the different feature curves (if exist). Additionally, we use the information related to the mean curvature and the saliency of each feature curve in order to find similarities with the feature curves of other models.

**Spectral Analysis for Estimating the Saliency of Each Vertex (Feature Vertex Extraction)**

In order to identify the vertices, which represent features, we firstly follow a spectral analysis process for estimating the saliency of each vertex. Then, we use the magnitude of the saliency to classify the vertices into features (big values) and non-features (small values). The used covariance matrix is created according to Section 4.4 and then it is decomposed:

$$\text{eig}(\mathbf{R}_i) = \mathbf{U}_i \mathbf{\Lambda}_i \; \forall \; i = 1, \cdots, n \tag{6.1}$$

where $\mathbf{U}_i \in \mathbb{R}^{3\times 3}$ denotes the matrix whose columns are the corresponding right eigenvectors and $\mathbf{\Lambda}_i = \text{diag}(\lambda_{i1}, \lambda_{i2}, \lambda_{i3})$ denotes a diagonal matrix with the corresponding eigenvalues $\lambda_{ij}$, $\forall \; j = 1, \cdots, 3$, so that:

$$\mathbf{R}_i * \mathbf{U}_i = \mathbf{U}_i * \mathbf{\Lambda}_i \tag{6.2}$$

The saliency $s_i$ of a vertex $\mathbf{v}_i$ is defined as the value given by the inverse norm 2 of the corresponding eigenvalues:

$$s_i = \frac{1}{\sqrt{\lambda_{i1}^2 + \lambda_{i2}^2 + \lambda_{i3}^2}} \; \forall \; i = 1, \cdots, n \tag{6.3}$$

We also normalized the values in order to be in the range of [0-1], according to:

$$\bar{s}_i = \frac{s_i - \min(s_i)}{\max(s_i) - \min(s_i)} \; \forall \; i = 1, \cdots, n \tag{6.4}$$

We assume that a small value of saliency means that the vertex lies in a flat area while a big value means that the vertex lies in an edge or corner. Observing the Eq. (6.3), we can see that a large value of $\sqrt{\lambda_{i1}^2 + \lambda_{i2}^2 + \lambda_{i3}^2}$ corresponds to small saliency. This observation can be explained under the aspect of spectral analysis, having used the steps of Eqs. (6.30)-(4.15). The normal of a vertex, lying in a flat area, is represented by one dominant eigenvector, the corresponding eigenvalue of which has a very big value $\lambda_1 \gg \lambda_2 \cong \lambda_3$. On the other hand, the normal of a vertex lying in a corner is represented by three eigenvectors, the corresponding eigenvalues of which have small but almost equal values $\lambda_1 \cong \lambda_2 \cong \lambda_3$.

For the identification of which vertices represent features, we use the k-means algorithm for separating the normalized values of the saliency into 5 different classes. We assume that the first two classes consist of non-features vertices, while the three next classes consist of features. In Fig. 6.2, we present the model "14.ply" in 5 different colors. Each color represents one of the 5 classes. The most salient vertices are those with the highest value, represented with red color, while vertices in flat areas are represented with blue color.

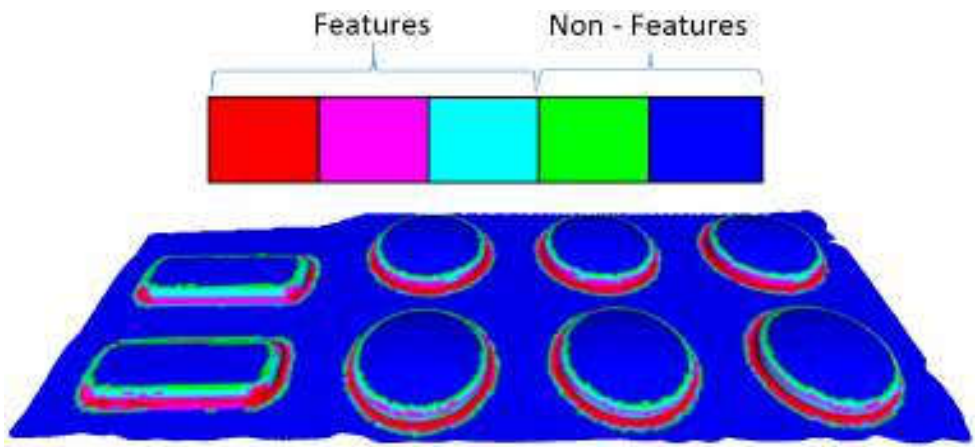

Figure 6.2: Colored vertices classified in different classes based on their saliency.

Finally, in Fig. 6.3 we present the features extraction (red vertices) based on the grouping of the three high salient classes.

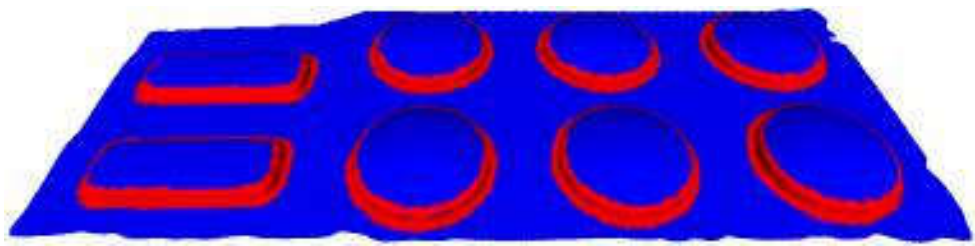

Figure 6.3: Corresponding .obj output with features representation of the model "14.ply" with connectivity.

To note here that the execution time of the algorithm depends on: (i) the size of the mesh and (ii) the size of the patches, but generally, it is very fast.

**Feature Curve Identification Based on Mean Curvature**

Once we have estimated the features of a mesh, we use the mean curvature $m_c$ for identifying the feature curves based on the clustering of the $m_c$ values. In Fig. 6.4, we present examples of feature curves of different models. The initial number of the feature curves is unknown for each model hence we evaluate the optimal number of clusters, in a range of [1-5], using the Calinski-Harabasz clustering evaluation criterion, and then we use the k-means algorithm for the clustering.

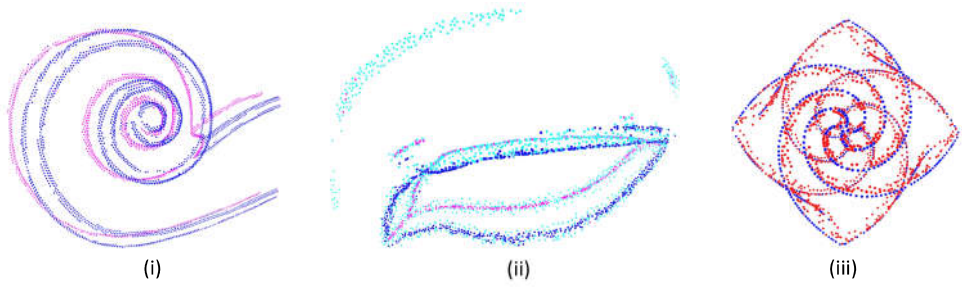

Figure 6.4: Feature curves of different models: (i) 7.ply, (ii) 12.ply, (iii) 13.ply.

**Feature Curve Similarities Between Different Models**

The normalized mean curvature $\bar{m}_c$ and salient values $\bar{s}$ are additionally used in order to compare feature curves of different models and find possible similarities between them. More specifically, we estimate the vectors $\dot{s} \in \mathbb{R}^{10 \times 1}$ and $\dot{m} \in \mathbb{R}^{10 \times 1}$ which consist of the values of the corresponding histograms, according to:

$$\dot{s} = \text{histogram}(\bar{s})$$
$$\dot{m} = \text{histogram}(\bar{m}_c) \tag{6.5}$$

and we create the vector $q \in \mathbb{R}^{20 \times 1}$ by concatenating the vectors $\dot{s}$ and $\dot{m}$:

$$q = [\dot{s}\ \dot{m}] \tag{6.6}$$

To investigate the similarity between two feature curves A and B, we estimate the correlation coefficient using the vectors $q_A$ and $q_B$, according to:

$$r = \frac{\sum_{i=1}^{20}(q_{A_i} - \bar{q}_A)(q_{B_i} - \bar{q}_B)}{\sqrt{(\sum_{i=1}^{20}(q_{A_i} - \bar{q}_A))(\sum_{i=1}^{20}(q_{B_i} - \bar{q}_B))}} \tag{6.7}$$

where $\bar{q}_A$ and $\bar{q}_B$ indicate the mean values.

In Fig. 6.5, we present a table 15 × 15 showing the similarities between all models, following the aforementioned pipeline. A value equal to 0 means that the feature curves of two different models are identical. Inspecting the numerical results of this figure, we can see that the models: 11 and 12, 7 and 8, 6 and 9, 4 and 10 are highly related to each other, which is obviously true. On the other hand, models 3 and 15 are also highly related to each other however without an apparent similarity. Models 13 and 14 are totally unrelated to any other model.

To note here that the results may differ a little by run to run since we use k-means for the clustering which does not give exactly the same values in each execution, however, this does not negatively affect the quality of the results. The technical manual of this application is presented in Appendix C.

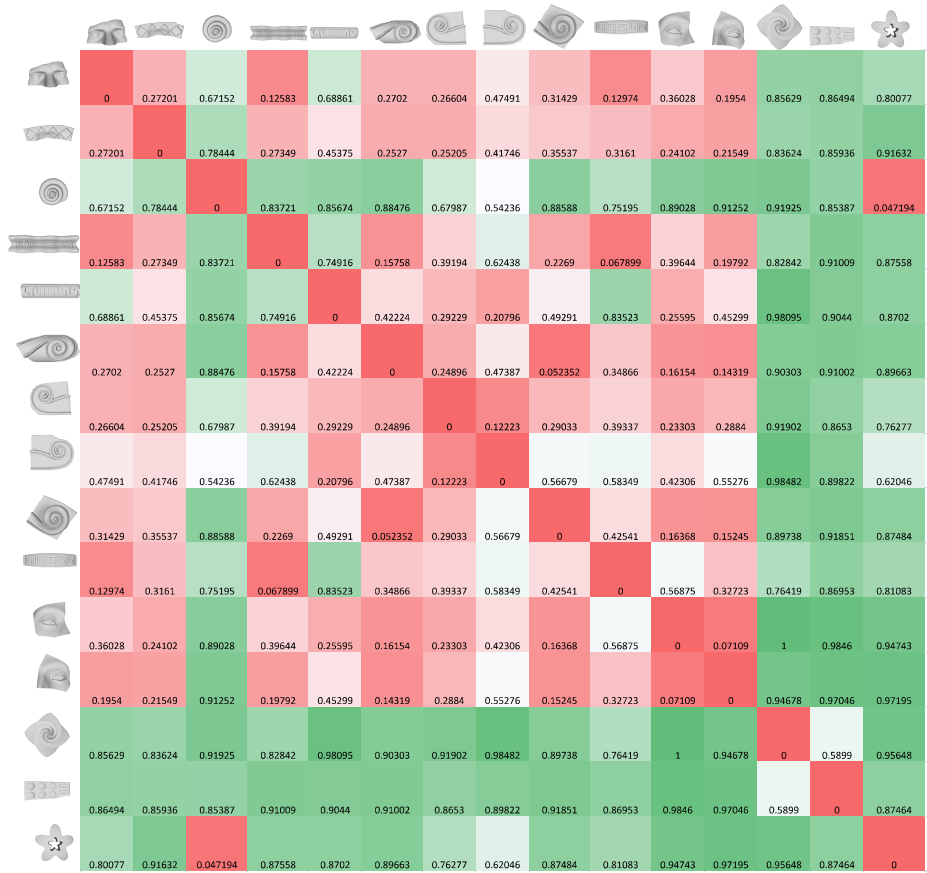

Figure 6.5: Correlation matrix with values ranging from 0 to 1 showing the similarity of feature curves between different models. The models are ordered by the increased number which indicates their name.

### 6.1.2 Robust and Fast 3D Saliency Mapping Estimation

In this subsection, we present a 3D feature-aware saliency estimation approach, taking into account both spectral and geometrical information of a 3D object. The main purpose is to provide a meaningful 3D saliency mapping which could be beneficial for industrial applications. For the processing of the 3D saliency mapping estimation, we start by separating the whole mesh into $n_f$ (i.e., equal to the number of centroids) overlapped and equally-sized patches. Then, we estimate the spectral and geometrical saliency and finally, we combine these two values. Once the saliency mapping of a mesh has been estimated, it can be used in several different industrial applications, facilitating several processes in manufacturing, maintenance, inspection and repairing. Fig. 6.6 briefly presents the pipeline of our approach.

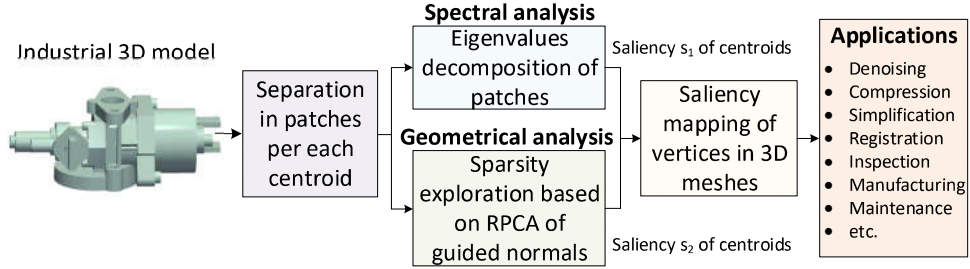

Figure 6.6: Pipeline of the proposed method for the estimation of the saliency mapping of 3D meshes based on spectral and geometrical analysis.

**Geometry-Based Saliency Analysis**

The geometrical saliency features are estimated by exploiting the sparsity of the guided normals. Centroid normals $\mathbf{n}_c$ can be also used in this analysis, however, guided normals have a more robust behavior in the presence of scanning noise [332]. The estimated patches $\mathcal{P}_i$ are used for the construction of matrix $\mathbf{M} \in \mathbb{R}^{3n_f \times (k+1)}$:

$$\mathbf{M} = \begin{bmatrix} \mathbf{g}_1 & \mathbf{g}_{11} & \mathbf{g}_{12} & \cdots & \mathbf{g}_{1k} \\ \mathbf{g}_2 & \mathbf{g}_{21} & \mathbf{g}_{22} & \cdots & \mathbf{g}_{2k} \\ \vdots & \vdots & \vdots & \ddots & \vdots \\ \mathbf{g}_{n_f} & \mathbf{g}_{n_f 1} & \mathbf{g}_{n_f 2} & \cdots & \mathbf{g}_{n_f k} \end{bmatrix} \quad (6.8)$$

where $\mathbf{g}_i = [g_{i_x}, g_{i_y}, g_{i_z}]^T$. Then, we apply the RPCA approach to this matrix, taking advantage of the geometrical coherence between neighboring guided normals. By the decomposition, the low-rank $\mathbf{E}$ and sparse $\mathbf{S}$ matrices are estimated. However, the estimation of the geometric saliency feature $s_{1i}$ of the centroid $\mathbf{c}_i$ requires only the values of the first column of the sparse matrix,

according to:
$$s_{1i} = \sqrt{S_{i1_x}^2 + S_{i1_y}^2 + S_{i1_z}^2} \quad \forall\, i = 1, \cdots, n_f \tag{6.9}$$

where $S_{i1_x}$ denotes the scalar value of the $x$ coordinate, of the $i^{th}$ row, of the $1^{st}$ column, of the **S** matrix. For the sake of completeness, the saliency mapping estimation using other RPCA approaches is provided in the Appendix E.

The motivation for exploiting the sparsity of the guided normals is based on the observation that the similarity of the normals between neighboring triangles is an index of the geometrical coherence of the triangles. Low values of the sparse matrix mean that the normals of a triangle and its neighbors are similar (low-rank), so if all triangles of a neighboring area have similar geometrical behavior this means that this patch represents a flat area. On the other hand, if there is a big dissimilarity this means that the surface has an abnormal shape.

**Spectral-Based Saliency Analysis**

For each face $f_i$ of the mesh, we use $\mathbf{M}_i \in \mathbb{R}^{3 \times (k+1)}$, representing the $i$ row of the matrix **E** in Eq. (6.8):

$$\mathbf{M}_i = \begin{bmatrix} g_{ix} & g_{ix_1} & g_{ix_2} & \cdots & g_{ix_k} \\ g_{iy} & g_{iy_1} & g_{iy_2} & \cdots & g_{iy_k} \\ g_{iz} & g_{iz_1} & g_{iz_2} & \cdots & g_{iz_k} \end{bmatrix} \quad \forall\, i = 1, \cdots, n_f \tag{6.10}$$

$\mathbf{M}_i$ is used for the estimation of the covariance matrices $\mathbf{R}_i$:

$$\mathbf{R}_i = \mathbf{M}_i \mathbf{M}_i^T \in \mathbb{R}^{3 \times 3} \tag{6.11}$$

Then, it is followed a similar approach like this that has been presented in the paragraph "Spectral Analysis for Estimating the Saliency of Each Vertex" of the previous subsection 6.1.1.

**Estimating the Saliency of Vertices**

We then normalize spectral and geometric saliency in a range [0-1], according to:

$$\bar{s}_{ji} = \frac{s_{ji} - \min(s_{ji})}{\max(s_{ji}) - \min(s_{ji})} \quad \forall\, i = 1, \cdots, n_f,\ j \in \{1,2\} \tag{6.12}$$

For the sake of completeness, we denote the saliency mapping as the weighted combination of the normalized geometrical $\bar{s}_1$ and spectral $\bar{s}_2$ saliency features, according to:

$$s_{ci} = \frac{w_1 \bar{s}_{1i} + w_2 \bar{s}_{2i}}{w_1 + w_2} \quad \forall\, i = 1, \cdots, n_f \tag{6.13}$$

where $w_1$ and $w_2$ are the corresponding weights which can be tuned for giving emphasis to the one or the other approach. However, we suggest the use of $w_1 = w_2 = 1$ which are also used in all of our experiments.

The proposed method is robust, even when we assume complex surfaces with different geometrical characteristics, since it exploits spectral characteristic (i.e., over-sensitivity in the variation of neighboring centroid normals) and geometrical characteristics (i.e., sparsity property of intense features). For the estimation of the saliency value of each vertex we use the following equation:

$$s_i = \frac{\sum_{\forall \mathbf{c}_j \in \Psi_{1_i}} s_{cj}}{|\Psi_{1_i}|} \ \forall \ i = 1, \cdots, n \tag{6.14}$$

where $\Psi_{1_i}$ represents the first-ring area of the vertex $\mathbf{v}_i$.

The saliency mapping of our method does not just detect defects, but it highlights areas with high-frequency spatial components (which means sharp features, noise, and abnormalities) and areas where a neighborhood of normals have a random variance behavior (normals of this neighborhood do not have a prevailing direction). In other words, our approach highlights areas where the range distribution of normals is high. In the following Fig. 6.7, we present boxplots showing the standard deviation of the normals for all overlapped patches which have been categorized (into 8 categories) based on the salient values of each vertex. More specifically, we estimate the standard deviation of the normals of each patch $\mathbf{M}_i$:

$$\sigma_i = \text{std}(\mathbf{M}_i) \ \forall \ i = 1, \cdots, n_f \tag{6.15}$$

where std(A) represents the equation that estimates the standard deviation of the values consisting of A. Then, we categorize each vertex based on the saliency values of our method in 8 categories.

As we can see, categories consisting of less salient vertices, representing flat areas, have smaller mean values of standard deviation since all the normals of a neighborhood have a common direction and form. On the other hand, as we move to categories including more salient vertices, the mean value of the standard deviation increases, meaning that the directions of the normals of a neighborhood become more irregular.

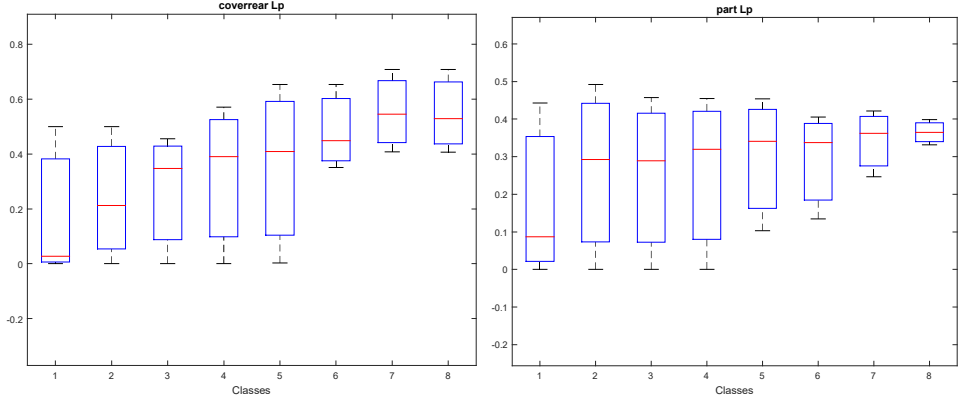

Figure 6.7: Boxplot of standard deviation of normals per different classes.

**Speed up Process - Sampling Matrix M**

Besides the fact that we use a very fast variant of RPCA, the most time-consuming step still is the decomposition of the coherent matrix **M** into a low-rank and a sparse matrix. The computational complexity of this method is related to the size of the data, so an approach to decrease the execution time is to use without of course reducing the detection accuracy. To achieve this, we firstly use the saliency mapping results of the spectral method. The salient map helps us to make a first coarse estimation about where the sharp features and the flat areas exist.

We start by assuming that each vertex can be categorized into a saliency class, based on its salient value that has been extracted by the spectral method only. We use 64 classes in total which is equal to the number of different colors of the "jet" colormap that is also used for the visualization of the saliency mapping. Class 1 consists of the least salient vertices, while class 64 consists of the most salient vertices. Then, we are based on the observation that a large quantity of vertices belongs to class 1 (as we can also see in the Histograms of Fig. 6.41) and we exclude these vertices. Finally we create a smaller dimension matrix $\mathbf{M}' \in \mathbb{R}^{3n'_f \times (k+1)}$ using all the vertices of the rest of the classes [2-64] where $n'_f < n_f$.

The final execution times of the fast approach are further improved and the speed up of the new algorithm is up to $\sim 85\%$, as we can see in the Table 6.1. Fig 6.9 also presents the results of the saliency mapping in different models using the two presented approaches, namely the original RPCA and the faster approach applied only to the salient vertices of the mesh. As we can observe, there is no big perceptual difference between the two results.

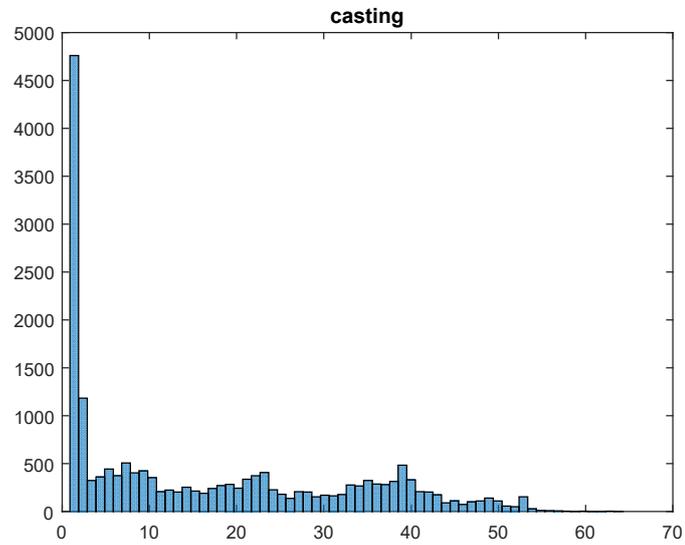

Figure 6.8: Histograms showing the number of vertices per each class. We have used 64 bins equal to the number of colors of the "jet" colormap.

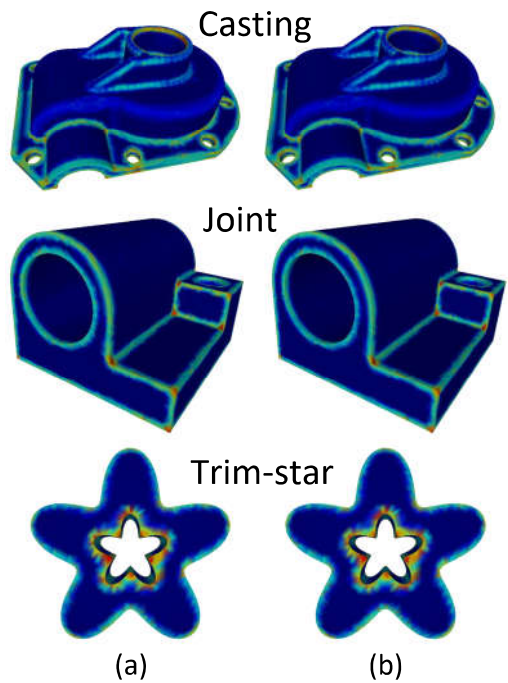

Figure 6.9: Heat map visualization of the salient map extracted by the (a) Original approach, (b) faster approach applied only to the salient vertices of the mesh.

| Model | Number of vertices belongs to category 1 | Total number of Vertices | Execution times using the original approach (sec) | Execution times using the fast approach (sec) | Speed up |
|---|---|---|---|---|---|
| fandisk | 2782 | 6475 | 0.46 | 0.19 | 59.7 % |
| cad | 8735 | 19398 | 1.22 | 0.42 | 65.2 % |
| block | 2062 | 8771 | 0.38 | 0.22 | 40.3 % |
| joint | 15848 | 20902 | 1.40 | 0.34 | 76.1 % |
| part_Lp | 898 | 4261 | 0.31 | 0.15 | 50.1 % |
| coverrear_Lp | 4902 | 7872 | 0.86 | 0.13 | 85.2 % |
| rockerarm | 1782 | 9413 | 0.53 | 0.30 | 43.1 % |
| casting | 4760 | 18410 | 1.14 | 0.21 | 81.8 % |
| trim-star | 1806 | 5192 | 0.23 | 0.13 | 42.0 % |

Table 6.1: Execution times for the original approach and the fast approach using a smaller coherency matrix.

**Selected Value of Parameter** $k$

It is generally true that the value of parameters affects the results of a method. The selection of the ideal value plays an important role in the final outcome. However, it is not always an easy process and in some cases, it needs an extensive evaluation study for identifying the ideal values. The problem becomes harder when we deal with more parameters. On the other hand, pre-defined parameters can be used in some applications, overcoming this problem, especially in cases where the input data (e.g., 3D models) have some similar behavior (e.g., industrial objects). Our purpose, in this research, was to present an easily-tuned method in which the user does not need to exhaustively search for ideal values per each model separately. For this reason, we suggest a recommended value that provides good results in most of the cases. The proposed value of $k$ is empirical and it is selected after performing a thorough experimental study. A patch, consisting of $k = 25$ faces, seems to provide the best results taking into account both feature identification accuracy and complexity, in the majority of 3D models. Furthermore, we have to say that the ideal selection of $k$ is not crucial and for this reason, an exhaustive search is not necessary. Additionally, the ideal value for the same model may be different depending also on the application in which the saliency mapping is applied. Of course, very high $> 50$ or very low values $< 10$ are not acceptable but in a range of $[15 - 35]$ our method can find small and large-scale features of 3D models.

In Figs. 6.10, 6.11, 6.12, we present heatmaps visualization showing how the saliency mapping of different 3D models changes while the value of $k$ also changes in a range of $[5 - 55]$ with a step of 10.

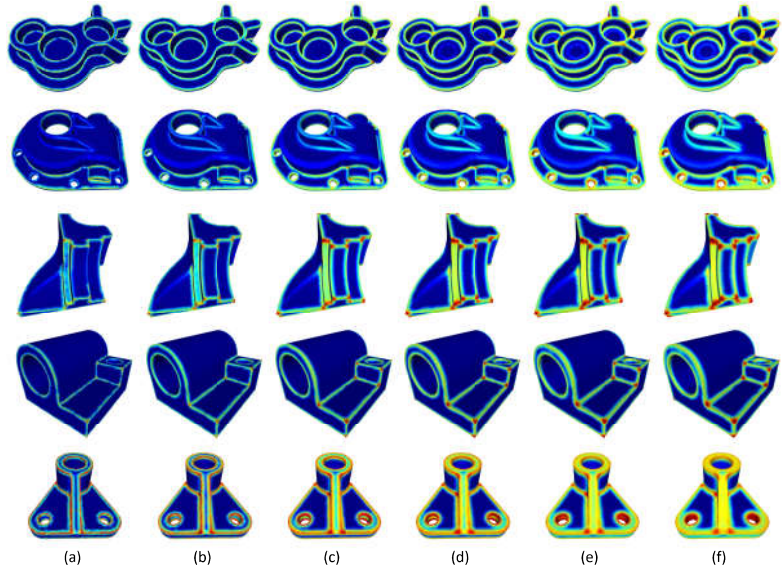

Figure 6.10: Heatmap visualization of saliency mapping (only spectral analysis) for different models (Cad, Casting, Fandisk, Joint and Part Lp respectively) using k equal to: (a) 5, (b) 15, (c) 25, (d) 35, (e) 45, (f) 55.

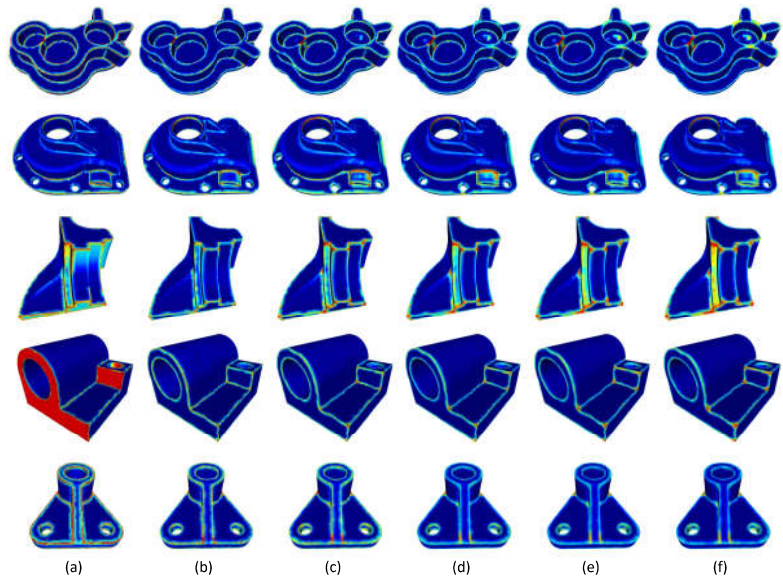

Figure 6.11: Heatmap visualization of saliency mapping (only geometrical analysis) for different models (Cad, Casting, Fandisk, Joint and Part Lp respectively) using k equal to: (a) 5, (b) 15, (c) 25, (d) 35, (e) 45, (f) 55.

Fig. 6.10 presents the heatmap visualization of saliency mapping for a variety of models using only the step of the spectral analysis of the proposed method for different values of parameter $k$. Fig. 6.11 presents the same as in the previous figure but using only the step of the geometrical analysis of the proposed method. Finally, Fig. 6.12 present the heatmap visualization of saliency mapping of the proposed method.

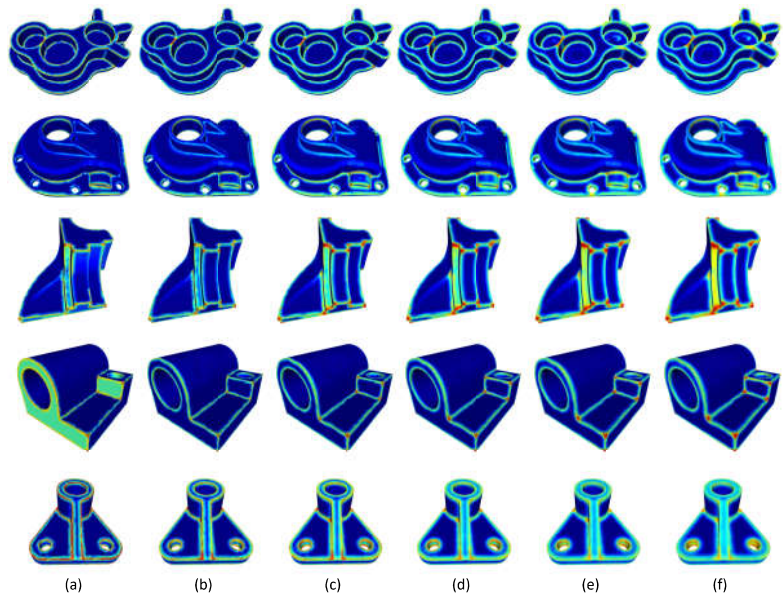

Figure 6.12: Heatmap visualization of saliency mapping (the proposed method) for different models (Cad, Casting, Fandisk, Joint and Part Lp respectively) using k equal to: (a) 5, (b) 15, (c) 25, (d) 35, (e) 45, (f) 55.

Another very important factor that we need to take into account for the selection of the value $k$ is the total execution time of the process which is affected by the value of $k$. As we can see in the next tables, while the length of the patches increases, the execution time also increases, both for the spectral and geometrical analysis. In some applications where the processing time is important, we may consider that a smaller-sized patch could be a better choice.

Additionally, we show how the value of $k$ affects the results of applications like the denoising of noisy 3D models. In Fig. 6.13, we present the reconstructed denoised results using our method for different values of $k$. In this experiment, we used a variety of different models and we also provide enlarged details (in red boxes) for easier comparison among the different cases. In Fig. 6.14, we present plots of different models showing how the Hausdorff distance error between the denoised reconstructed and the original mesh changes while the value of $k$ changes. This analysis showed that there is no common ideal value for all models, however, there is a range $[15-35]$ in which the results are acceptable.

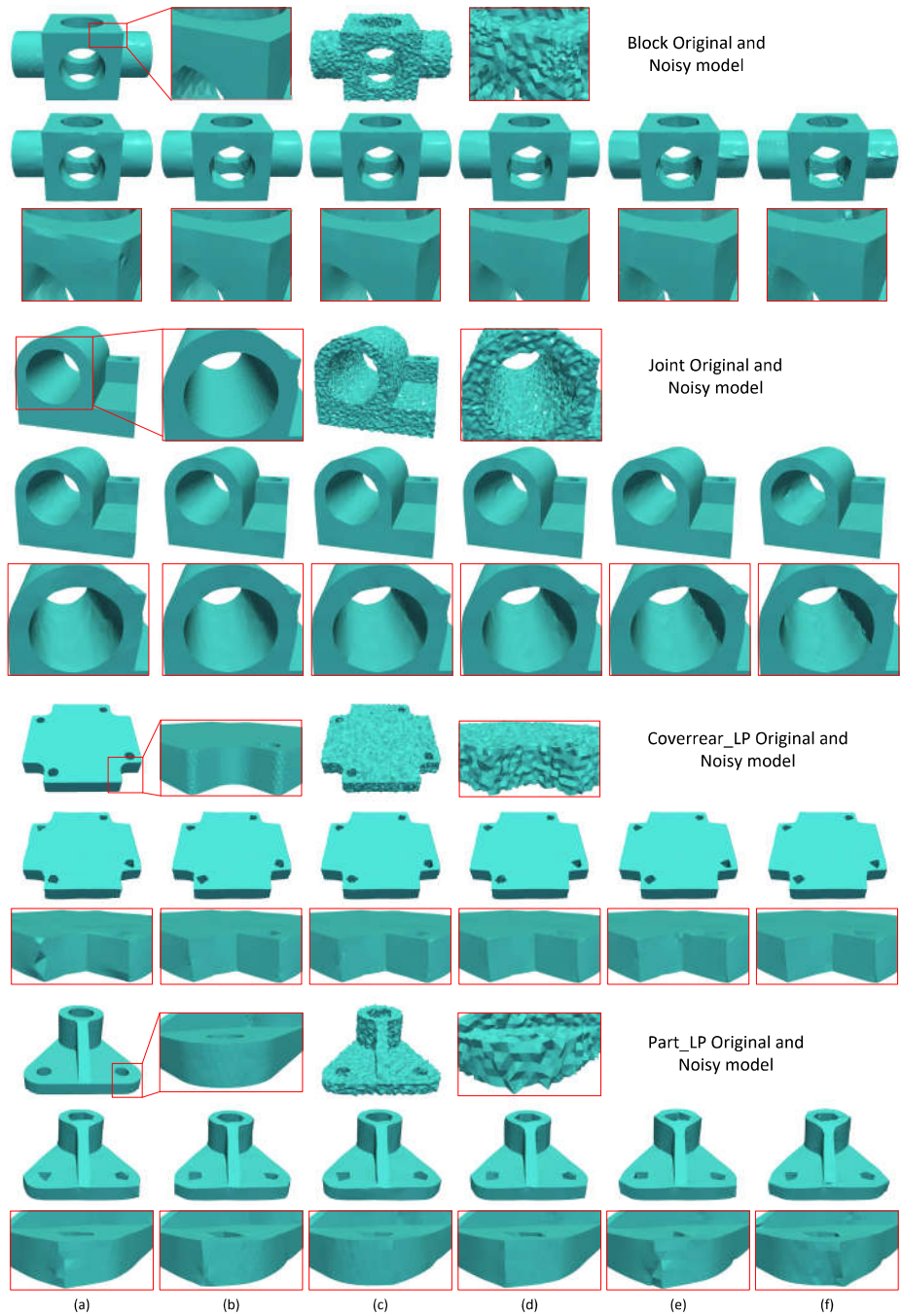

Figure 6.13: Denoised results using k: (a) 5, (b) 15, (c) 25, (d) 35, (e) 45, (f) 55.

To remind here that our method is not a denoising technique but we use this application only to evaluate our method. The results of these examples verify our assumptions and show the meaningful characteristics of our saliency mapping approach.

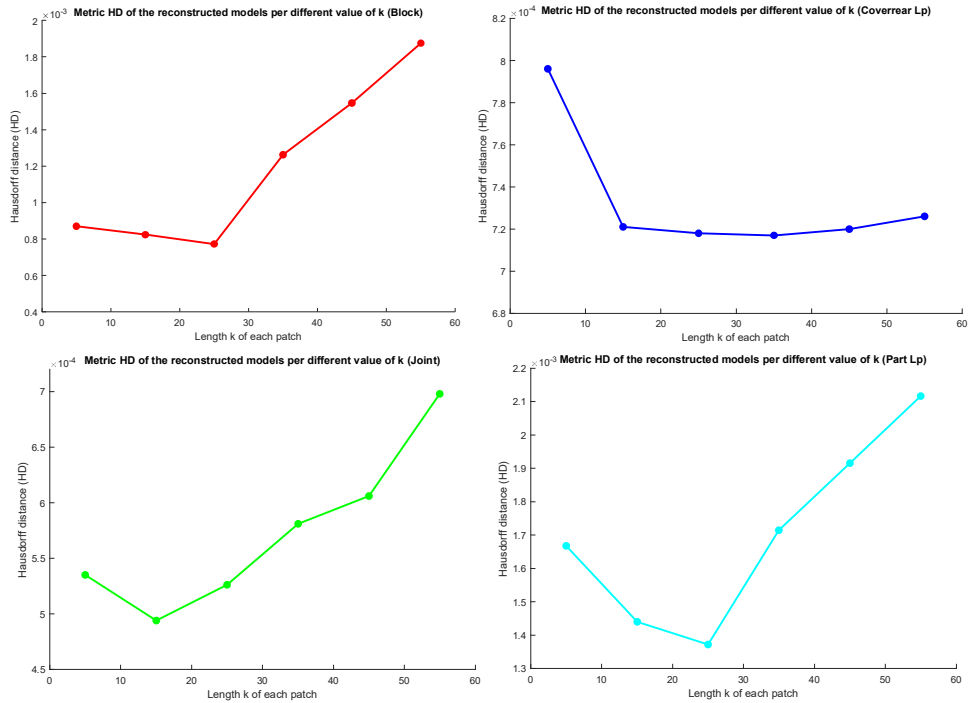

Figure 6.14: Plots of different models presenting the Hausdorff Distance (HD) error per different value of *k* for the denoised reconstructed models.

We decided to use a fixed value for *k* in order to avoid the complexity of finding the ideal value for each model. The experimental analysis shows that $k = 25$ gives acceptable results, in most of the cases. A patch consisting of 25 faces is a very good representative area which covers the area of a geometric feature of a model, as shown in Fig. 6.15-(b). Small areas of patches (in Fig. 6.15-(c)) are not able to cover a feature (like a corner), while big areas of patches (in Fig. 6.15-(d)) may overlap both sharp features and flat areas. This observation depends of course on the model. When the vertices of the model are not dense then the *k* could take a smaller number. However, our method is totally adaptive and the value of *k* can easily change (if necessary), depending on the requirements of an application or the special geometric characteristics of a model.

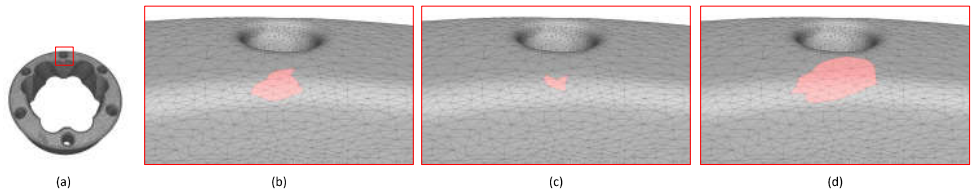

Figure 6.15: (a) "Rolling stage" model, and patches of different $k$ (b) $k = 25$, (c) $k = 5$, (d) $k = 55$.

**Rotation Invariant Saliency Mapping and Robustness in Noise**

Our purpose is to evaluate the consistency of the proposed saliency map under different conditions that could appear in a real-case scenario. The experimental analysis shows that our method is totally scale- and pose-invariant and very robust to several other constraints like different resolution quality and pattern of noise. In Figs. 6.16-6.17 we present the saliency mapping of different models under different conditions. In both of these experiments, we evaluate the consistency of the saliency maps in models with different scale and pose. As we can see, in all cases, the results of the extracted saliency maps are identical to those of the original models.

Additionally, Fig. 6.16 shows how the saliency map of a model changes when the resolution quality of a model also changes. As we see, even in cases where the resolution quality of a model is 50%, lower than the original model, the saliency maps of the same objects are very relevant to each other. Similar conclusions we can make observing the Fig. 6.17. However, in this case, it seems that a high level of noise negatively affects the saliency map in comparison with the corresponding of the original model, even though it also provides rational results. More specifically, when the level of noise is smaller than 0.1, the extracted saliency maps are very similar to the original. After this threshold, the saliency maps can not be used for comparisons between the two objects. Nevertheless, we can say that in real case scenarios, the contemporary scanner devices do not introduce so high level of noise to the reconstructed models.

**Speed-up Process for the Estimation of Saliency Maps in Dynamic 3D Meshes**

Motivated by the observation that the shape of a non-rigid dynamic 3D model does not significantly change frame by frame, we make the assumption that it is not necessary to estimate the saliency map of whole vertices all over again for each frame, but we can take advantage of the saliency information of the previously estimated frame and use it "as is" in the corresponding areas that geometrically remain the same.

In other words, we suggest estimating the saliency values only of these ver-tices that have a significant change regarding the normalized difference **d** of their first-ring area, as compared between two consecutive frames. Otherwise, we assume that they have preserved their geometry result in the same saliency value as the corresponding vertices of the previous frame, so this information is just transferred to the vertices of the next frame. The criterion that is used to define if we have to estimate again the saliency value of a vertex is presented in the following Eq. (6.16):

$$\Phi(i,l) = \begin{cases} 1 & \text{if } \mathbf{d} = \frac{|\mathbf{A}(i,l) - \mathbf{A}(i,l-1)|}{\max(\mathbf{A}(i,l), \mathbf{A}(i,l-1))} \geq 0.1 \\ 0 & \text{otherwise} \end{cases} \qquad (6.16)$$

where $\mathbf{A}(i,l)$ is the first-ring area of point $i$ as it appears in the $l$ frame and $\Phi(i,l) = 1$ means that we have to estimate again the saliency value of the $i$ point that lies in frame $l$.

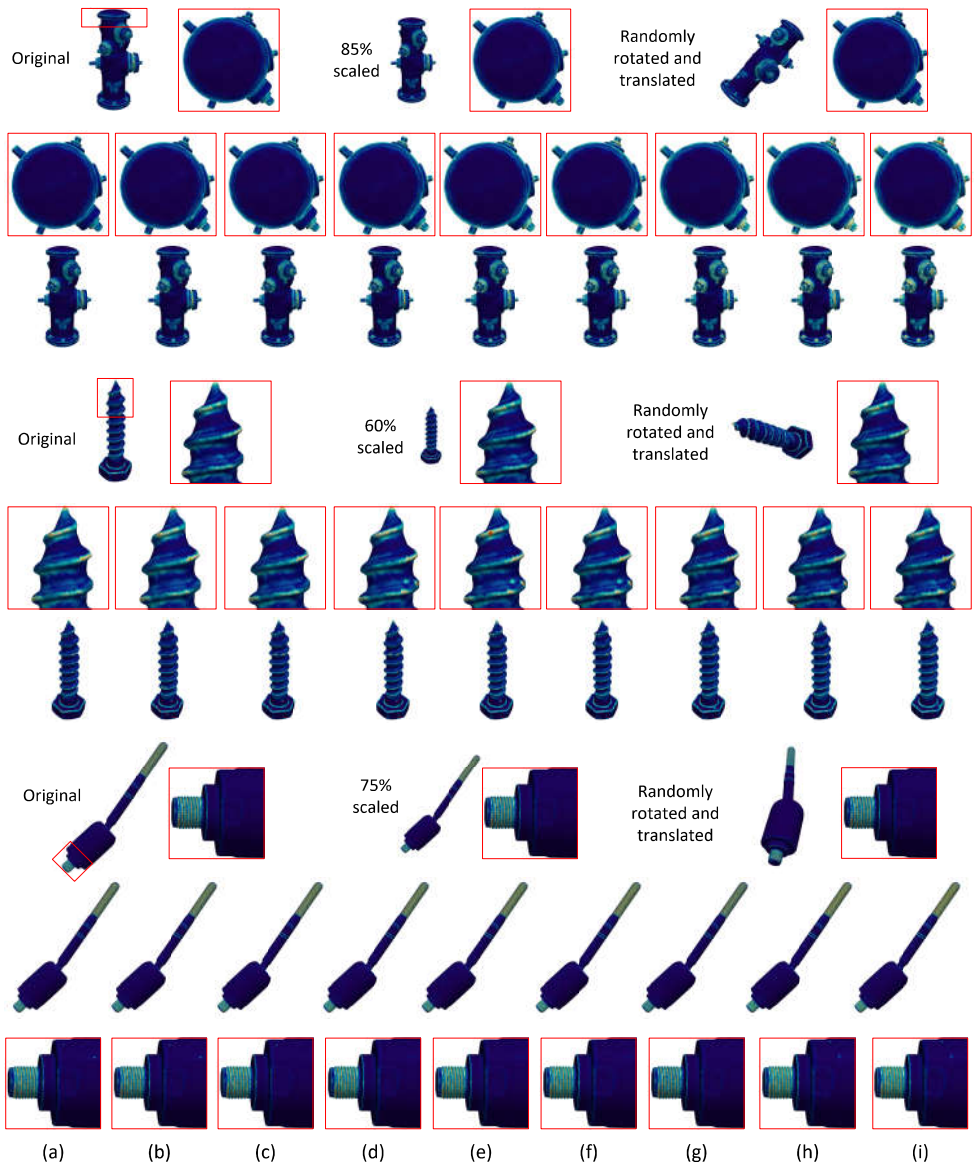

Figure 6.16: Saliency mapping of models with different resolution quality. (a) 10%, (b) 15%, (c) 20%, (d) 25%, (e) 30%, (f) 35%, (g) 40%, (h) 45%, (i) 50%, less vertices in comparison with the ground truth model.

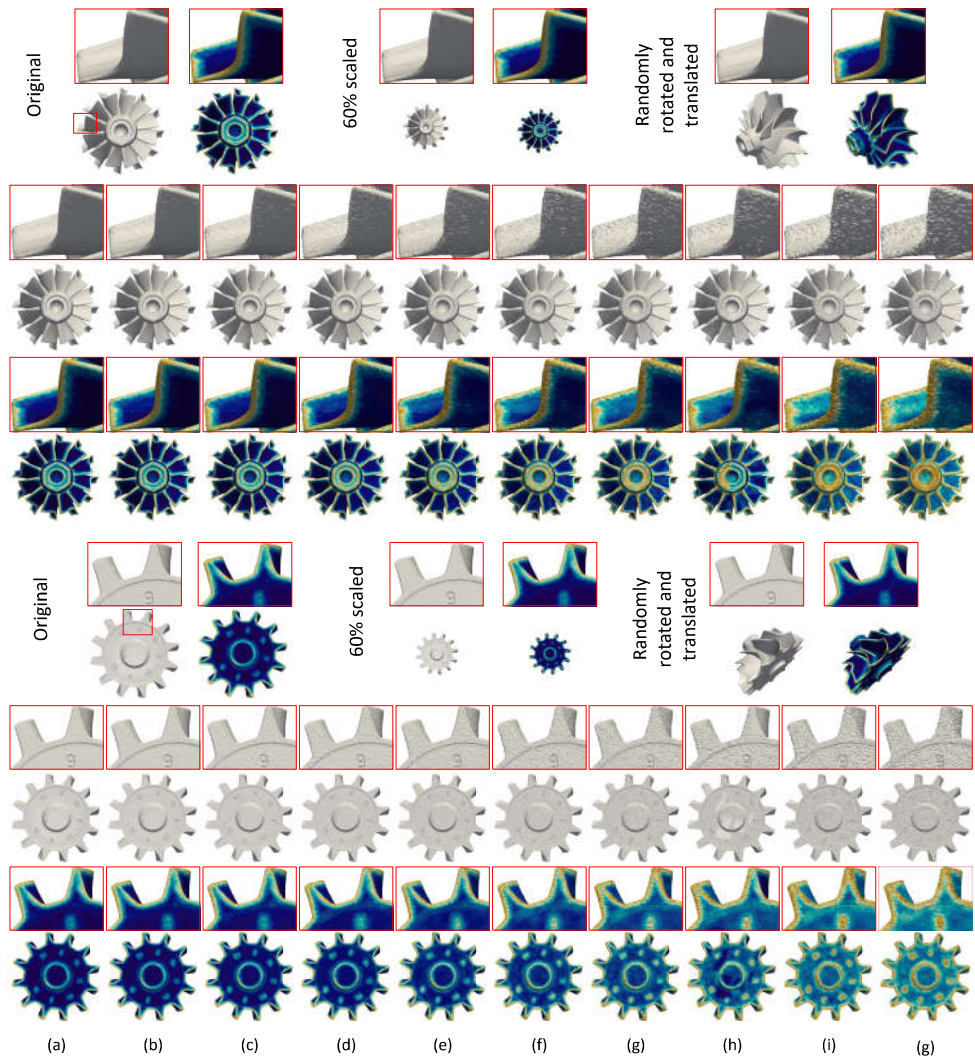

Figure 6.17: Saliency mapping of models with different level of noise: (a) 0.02, (b) 0.04, (c) 0.06, (d) 0.08, (e) 0.1, (f) 0.12, (g) 0.14, (h) 0.16, (i) 0.18, (g) 0.2.

Observing the Fig. 6.18, we can see that most of the vertices, of the dynamic mesh Handstand (10,002 vertices and 175 frames), have a small value of first-ring area difference (i.e., $\mathbf{d} < 0.1$). The results of this figure show that the first-ring area of only a few vertices has been significantly changed, so the saliency values of only these vertices have to be estimated in this frame.

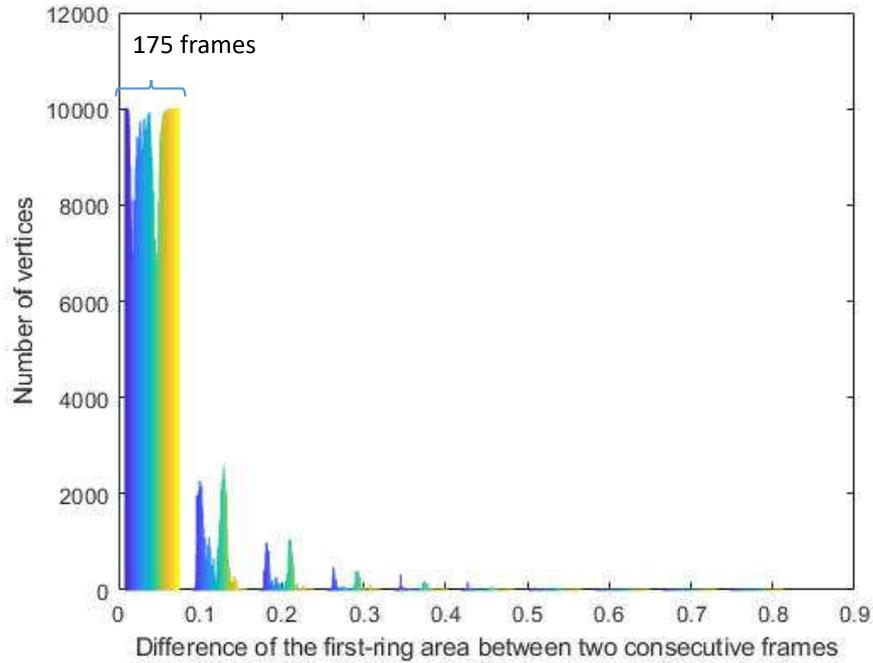

Figure 6.18: Histogram representing the number of vertices per frame that has a specific value of first-ring area difference **d**.

Additionally, we can see that the value of **d** is affected by the motion of the animation. More specifically, the value of **d** for the first and last frames is very small $< 0.1$ for almost all the vertices ($\sim 10,000$) of the model. In Fig. 6.19, we present two small sequences of the animation (frames 2-10 and 78-86). The first row of each part presents the heatmap visualization of saliency maps for nine consecutive frames. The second row highlights, in red color, the vertices of each model that correspond to $d \geq 0.1$ and the rest vertices are highlighted in blue color. The third row depicts the heatmap visualization of the value **d** per each model. The frames of the first sequence change slowly having, as a result, the saliency map of any model to remain almost the same to each other. On the other hand, in the second sequence, the movement of the mode is more intense affecting more the difference between the first-ring area of each frame.

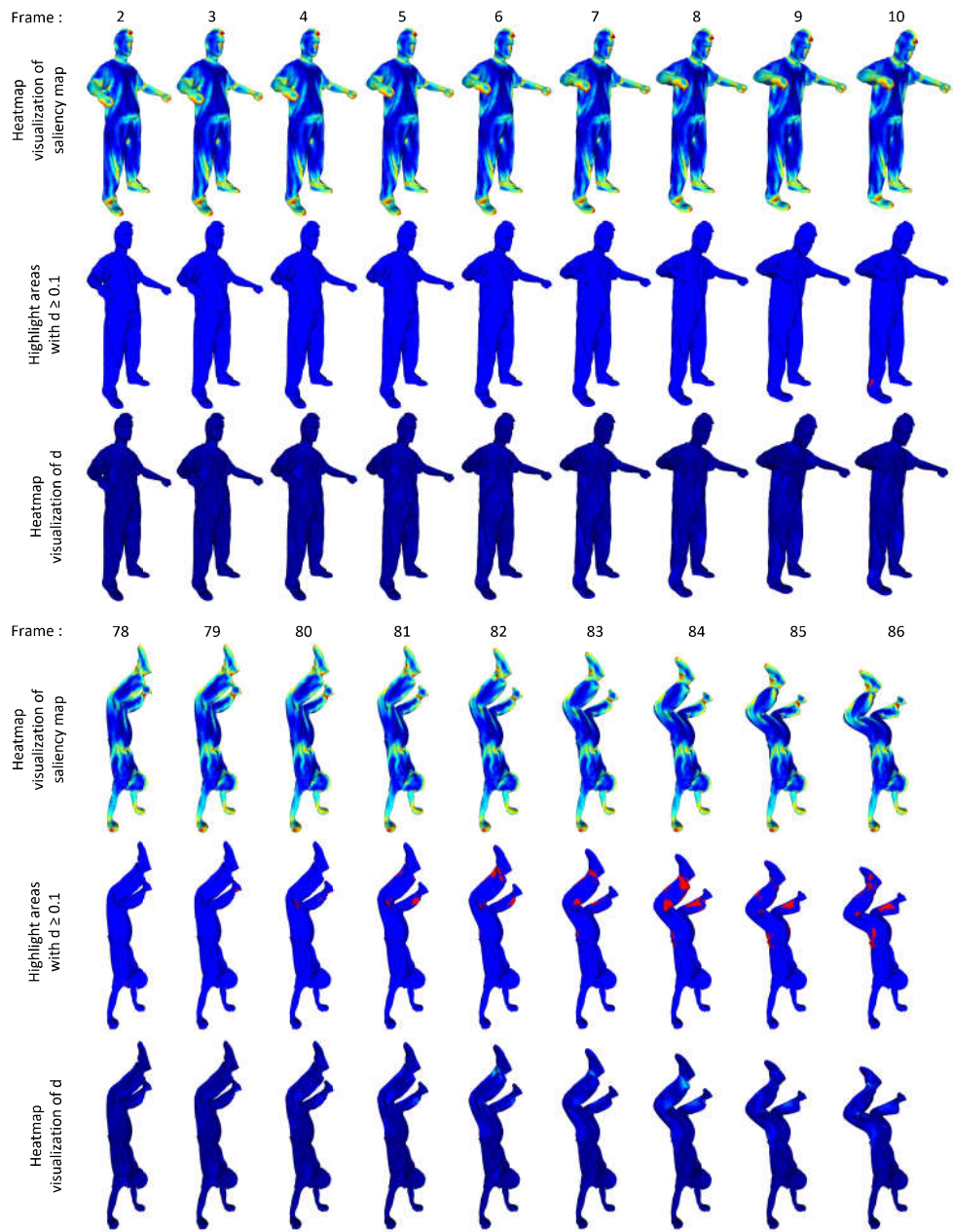

Figure 6.19: Heatmap visualization of different consecutive frames of the dynamic 3D mesh (Handstand) and identification of the vertices (highlighted in red color) that have a first-ring area difference $\mathbf{d} \geq 0.1$.

**RPCA Analysis for Geometrical Saliency Estimation**

Generally, RPCA has been used for identifying outliers in large datasets including images and videos while very recently have been applied in the area of 3D meshes and point clouds. More specifically, RPCA is able to decompose the original observed raw data **M** into two categories: (i) the low-rank data matrix **E** and (ii) and the matrix **S** with the outliers. The features are the residual of observed data minus the low-rank data (i.e., $\mathbf{S} = \mathbf{M} - \mathbf{E}$). The higher the value of **S** is, the more intense the corresponding feature is.

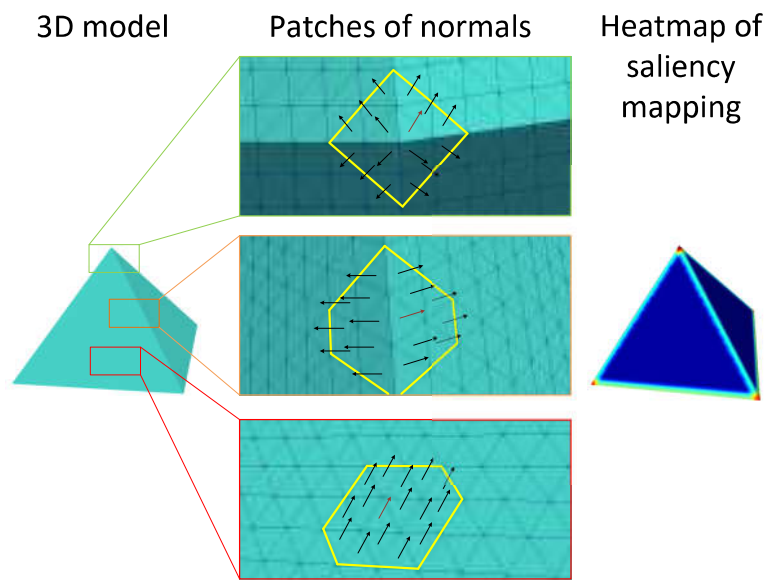

Figure 6.20: Normals' direction in three different type of patches (i.e., corner, edge and flat area).

The low-rank data represent a new estimated version of the original data where the abnormal values (i.e., outliers) have been "smoothed" so that they have similar behavior and distribution with the other "original" data. Following this line of thought, RPCA creates two matrices: (i) the low-rank matrix with the smoothed geometry and (ii) the sparse matrix, composed of few non zero elements corresponding to outliers, represents geometric features of different scale. In this sparse matrix, the higher the value of an entry is, the more possible the corresponding datum to be an outlier is. The high values correspondingly show how much a datum differs from its other neighboring data. In this application, we do not use RPCA to find outliers, with the conventional way of finding points lying out of the normal surface of the 3D model, but we use the values of the sparse matrix to evaluate its perceptual importance, i.e., its saliency property. More specifically, we search the similarity of all guided normals with their neighborhood for small patch area of the mesh. The similarity of the normals between neighboring triangles shows also the geometrical coherence of these triangles. Low values of such a sparse matrix characterize the similarity between the normals of a triangle and its neighbors (low-rank). The low-rank property is a feature that characterizes flat areas, meaning that if all triangles of a neighboring area have similar geometrical behavior this means that this patch represents a flat area. On the other hand, if there is a significant variation this means that the surface of the patch has a non-flat structure. In Fig. 6.20, we present a simplified example of normals' direction in three different types of patches. In the flat area, all normals have the same direction while in the corner, normals follow four different directions. One big drawback of other geometry-based saliency approaches is the fact that they are susceptible to the presence of noise since the geometry of the surface is super affected by the noise. However, at this point we would like also to highlight as an additional benefit, the fact that our proposed method can be efficiently used under of the presence of noise since we use guided normals (i.e., filtered) instead of centroid normals, which is a very important characteristic especially for using it in industrial application where the 3D objects, representing the data, are usually noisy due to the scanning process.

### 6.1.3 Utilization of Saliency Mapping in Applications

The extracted saliency mapping by the method described above is evaluated in a variety of actual industrial applications (i.e., manufacturing inspection of scaled object, inspection of aging mechanical parts, facilitation of heritage repairing/maintenance) in which our method can be successfully utilized, in different industrial areas (i.e., manufacturing, heritage, medical).

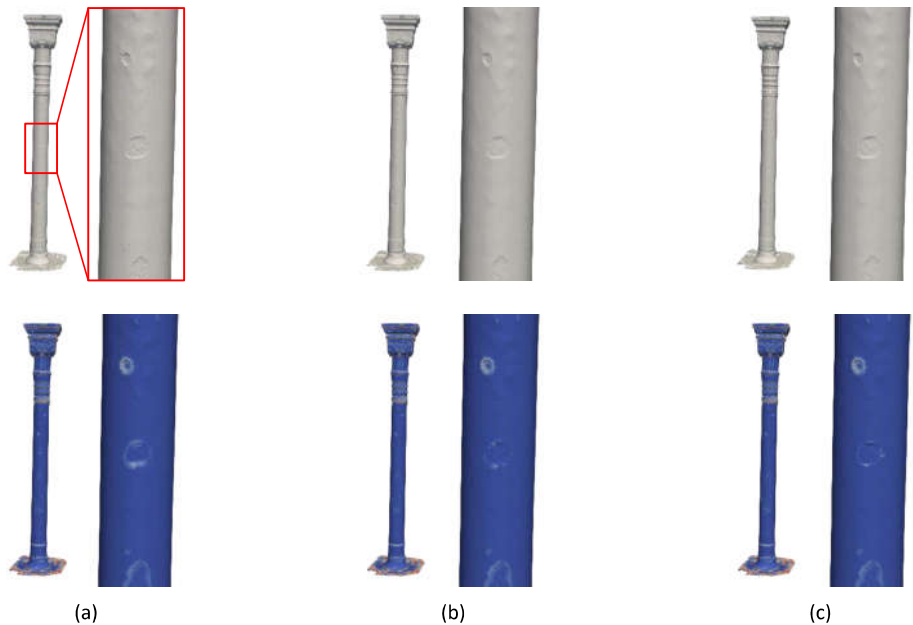

(a) (b) (c)

Figure 6.21: Heat map visualization of the saliency map for the same pillar in three different temporal phases (a) temporal phase 1, (b) temporal phase 2, (c) temporal phase 3.

**Spatial and Temporal Abnormalities**

Saliency mapping can be used not only for identifying spatial abnormalities but also for the temporal differences of a model that may be caused by the slow and inevitable aging processes due to: (1) material deterioration, (2) modifications by humans (e.g., in the urban environment), and (3) climate and environmental changes. In Fig. 6.21, we present the Heat map visualization of the saliency map for the same pillar in three different temporal phases, while in Fig. 6.22, we present temporal differences of the pillar in different phases based on their saliency maps.

**Utilization in the Manufacturing Industry for Quality Control Inspections**

It is very common, in the manufacturing industry, objects to be produced in different sizes, retaining however the same form with the prototype model. Nonetheless, to assure quality, the reconstructed objects must satisfy a range of statutory and contractual obligations. In this case, inspection is used to verify and certify that the new scaled object has been manufactured in full compliance with all specified requirements and constraints. In Fig. 6.23, we present examples of inspection between real-scanned industrial objects, denoted as pro-

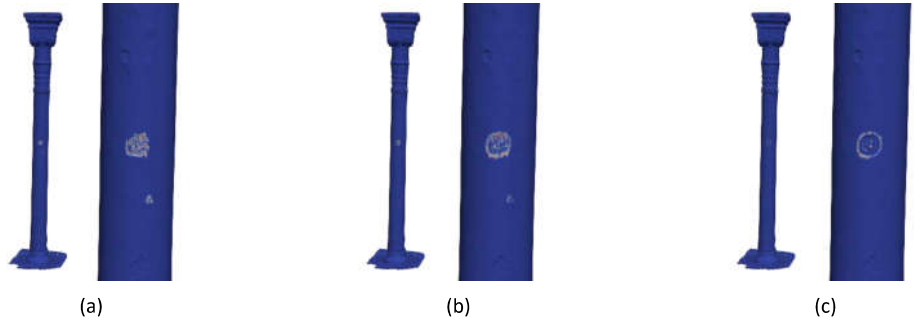

(a)            (b)            (c)

Figure 6.22: Saliency mapping comparison between the models of (a) temporal phase 1 and temporal phase 2, (b) temporal phase 1 and temporal phase 3, (c) temporal phase 2 and temporal phase 3.

totype models[1] [2] [3] (Fig. 6.23-(a)), and their corresponding scaled and deformed 3D objects (Fig. 6.23-(b)). Our purpose is to inspect if the new manufactured 3D object has the exact same design details as the original (regarding the fidelity of its form) and also to ensure that it has not been affected by irregularities encountered during the manufacturing processes. In Fig. 6.23-(c), we present an enlarged representation of the scaled model, presented in Fig. 6.23-(b), with red cycles that specify the deformed areas. The purpose of this application is to automatically identify deformations or other abnormalities from the surface of the manufactured 3D object in comparison with the original model. For easier comparison, we provide a heatmap visualization of the difference between the original and the constructed model. Blue color means that there is no difference between the compared models while red color indicates a big difference. Our method is able to find and highlight possible differences between two objects with similar shapes comparing the saliency values of their surface. In this way, it is capable to automatically inspect degradations of the surface standards of manufactured objects despite the constraints posed by scaled manufactured objects or objects created by different materials. For the comparisons between the original and the reconstructed models, we used two different approaches. In the first approach, we deployed the Hausdorff distance (HD) (Fig. 6.23-(d)) of the normalized models (with values in the range [0-1]) while in the second approach, we used both HD and the salient values (Fig. 6.23-(e)) according to Eq. (6.14). We assume that we have two normalized 3D models $\mathcal{M}_1 \in \mathbb{R}^{n_1 \times 3}$ and $\mathcal{M}_2 \in \mathbb{R}^{n_2 \times 3}$, where $n_1 \neq n_2$ (e.g., original and compared respectively). Firstly, for each vertex of the these two models, we create a representative vector con-

---

[1] "Aeronautics actuator casting" model
[2] "Automobile Hubcap" model
[3] "Oil pump" model

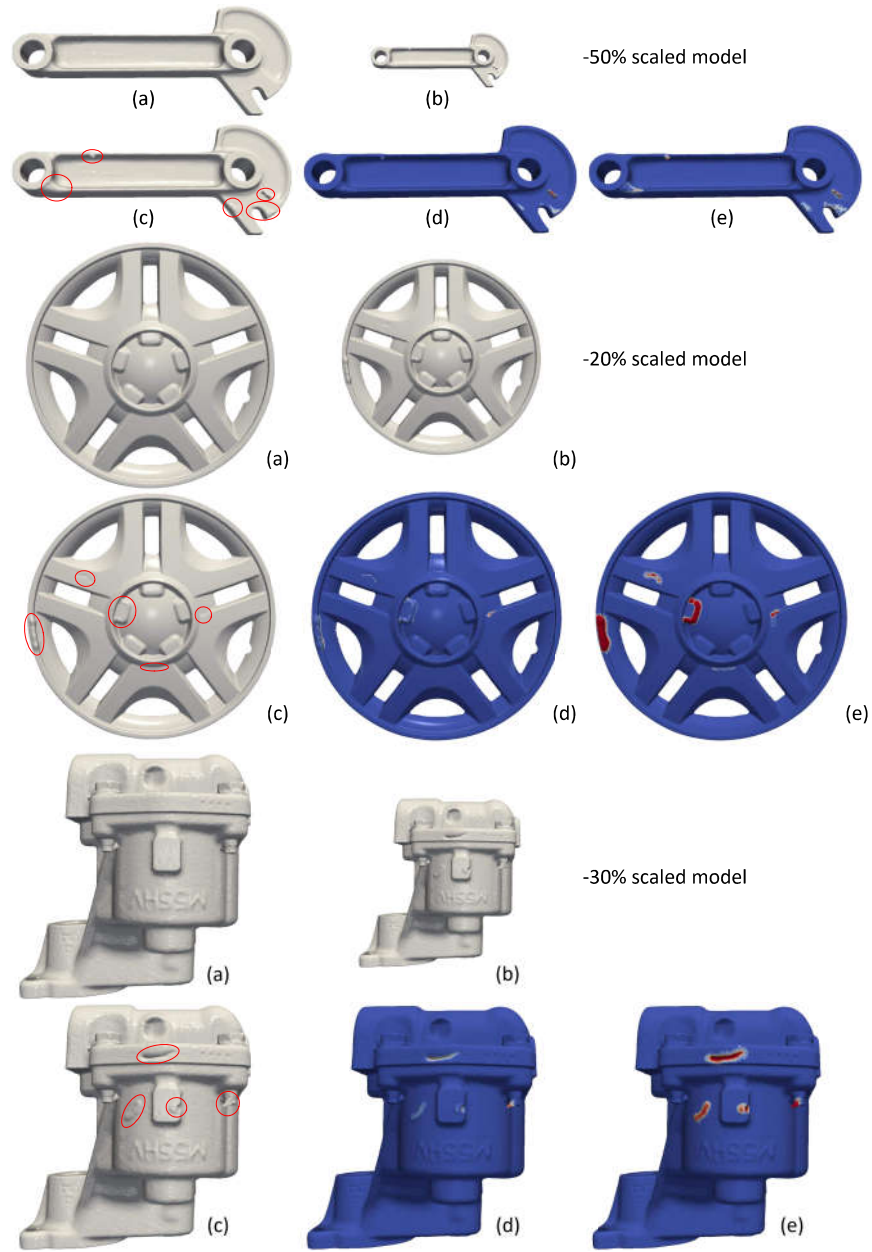

Figure 6.23: **[First line]** (a) Original models, (b) scaled and deformed models. **[Second line]** Enlarged representations of (b) with: (c) red cycles for highlighting the deformed areas, (d) heatmap visualization of HD applied to vertices, (e) heatmap visualization of HD applied both to vertices and salient values.

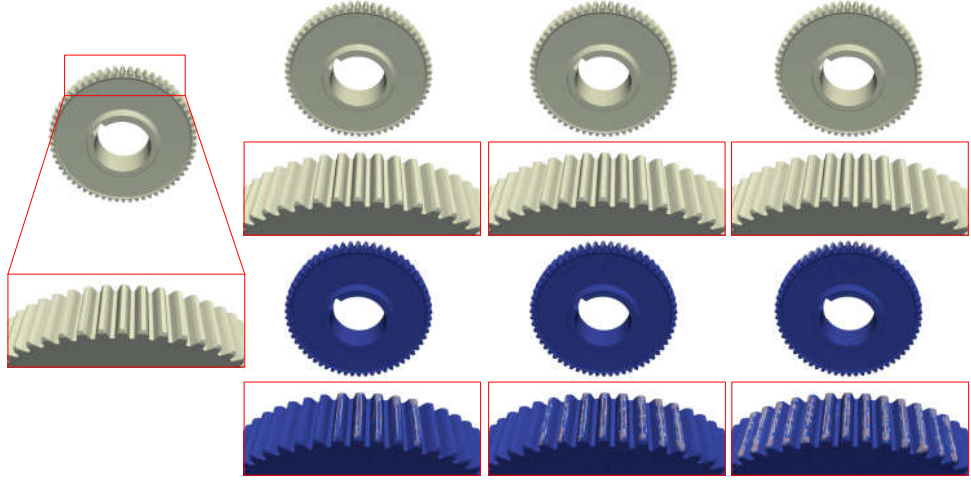

Figure 6.24: Digital twin of gear model in a early stage and after 3 consecutive temporal moments.

sisting of the values of its coordinates and its saliency:

$$\mathbf{q}_{ji} = [\mathbf{v}_{ji}\ s_{ji}] \in \mathbb{R}^{4\times 1} \tag{6.17}$$

$\forall\ j \in \{1,2\}, i \in [1,\cdots,n_1],\ [1,\cdots,n_2]$. Then, for each $i$ vertex of $\mathcal{M}_2$ we find the closest $d_i^*$ vertex of $\mathcal{M}_1$ based on their norm 2 difference:

$$d_i^* = \arg\min_d D(\mathbf{q}_{2i}, \mathbf{q}_{1d})\ \forall\ d \in [1,\cdots,n_1] \tag{6.18}$$

$\forall\ i \in [1,\cdots,n_2])$, where:

$$D(\mathbf{q}_2, \mathbf{q}_1) = \sqrt{(x_2 - x_1)^2 + (y_2 - y_1)^2 + (z_2 - z_1)^2 + (s_2 - s_1)^2} \tag{6.19}$$

Then for these indices $d_i^*$, we estimate the saliency difference $(s_{2i} - s_{1d_i^*})\ \forall\ i = 1,\cdots,n_2$ and we visualize it. The experiments verifies that the proposed approach can identify deviations easily and with great detail.

**Utilization for the Creation of Digital Twins and Aging Inspection**

The proposed method supports detecting changes that can be caused by aging, comparing the saliency mapping of a 3D object having been acquired in two or more different temporal moments. In this way, our approach could be used to identify surface differences of the same object, affected by mechanical stress (e.g., a gear of a machine) or deteriorated due to environmental conditions (e.g., an ancient statue or columns). In Fig. 6.24, we present visual representations of the same gear in 4 different occasions (i.e., in a early stage and after 3 consecutive temporal moments). This figure shows that our method is able to capture differences due to aging, so indiscernible, that even the human eye could not easily notice.

**Utilization in the Heritage Industry for the Maintenance of the Historical Objects**

In Fig. 6.25, we present examples of real-scanned historical objects [4] [5] in which the proposed method of the saliency mapping can be utilized so as to automatically detect cracks and other defects on their surfaces. The presented figures verify that the proposed 3D saliency mapping approach is very useful since it can be used for identifying areas of the original model that need to be repaired, facilitating the work of the experts during the maintenance process. Our method uses small patches of neighboring vertices, thus if an abnormality appears somewhere in the surface, then our algorithm is able to recognize it and highlight this specific area. The higher the abnormality, the higher the value of salience, so it is ideal for inspection of cracks, damages, etc,. In this way, this method could be used as a pre-processing step for the creation of a digital replica of the original cultural object, without imperfections, since it highlights the areas that need repair (i.e., digital repairing is also available). The recent trend for digitalization and creation of digital twin models has a lot of historical interest in the heritage industry. A digital repaired 3D model can be used for the VR/AR representation of a heritage object (e.g., for educational purposes) showing how it looked like originally and additionally giving to the visitors the opportunity to see reconstructed views of the object.

**Utilization in Medical Industry for the Creation of Patient-Specific Models**

The ability to produce patient-specific parts directly from scan data is an obvious benefit for the medical industry. Nowadays, it can be achieved using conventional techniques [361], however, the reconstructed 3D models need refinement before their use in other applications, due to the staircase effect caused by the 3D mesh reconstruction of MRI/CT images). The 3D reconstructed results using the CTs images are noisy, with a unique usually noise pattern (i.e., staircase effect), the distribution of which is totally different than the typical Gaussian noise distribution. The traditional methods of the literature are usually fail to handle noise patterns different from Gaussian noise and other more sophisticated approaches, like non-isotropic feature-aware denoising methods, are required. Our method is capable to provide the saliency mapping (i.e., a

---

[4]"Calcite Vase" model
[5]"Epichysis 26332 b" model

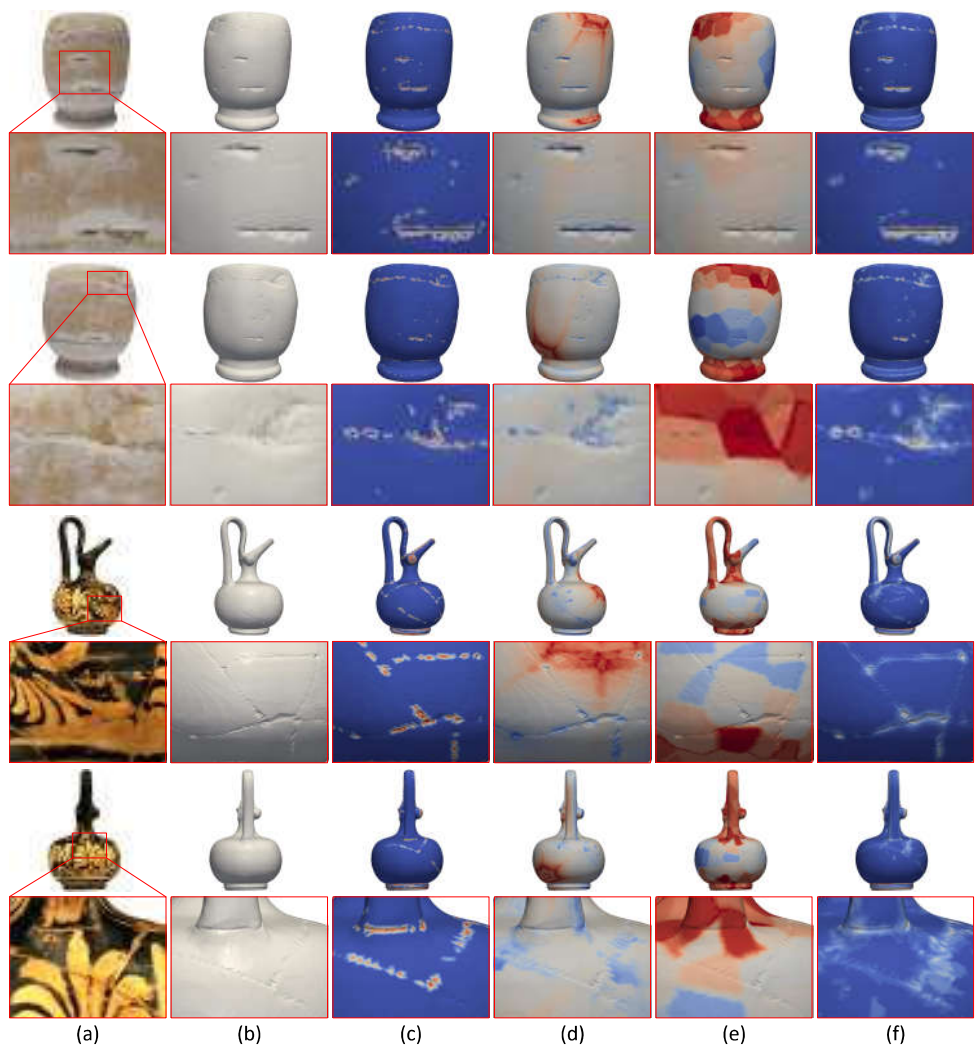

Figure 6.25: Original model (a) with texture, (b) with no texture. Heatmap visualization of saliency mapping using: (c) curvature co-occurrence histogram [29], (d) entropy based salient model [30], (e) mesh saliency via spectral processing [31], (f) the proposed method.

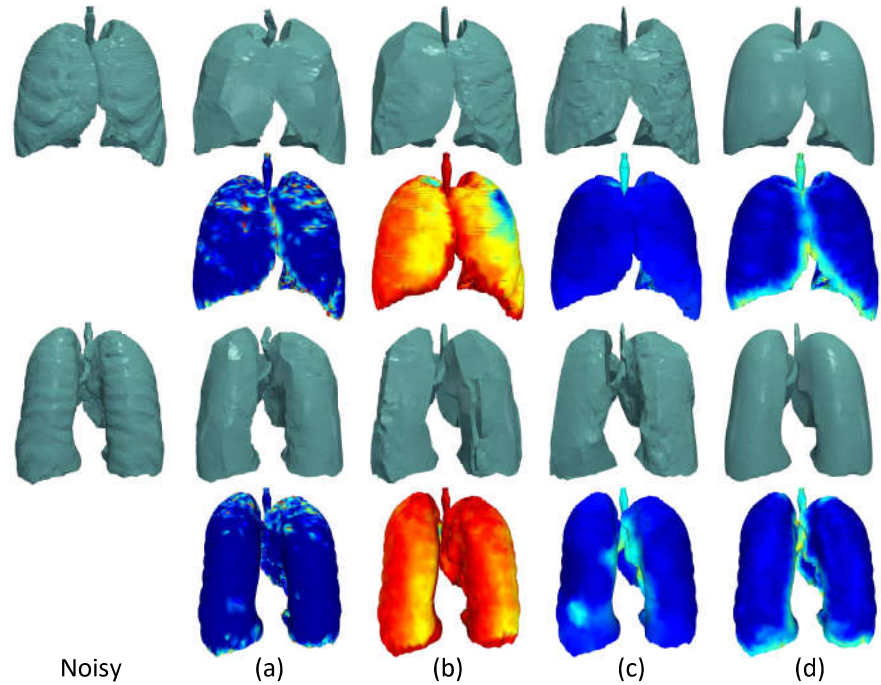

Noisy      (a)      (b)      (c)      (d)

Figure 6.26: Smoothing results and the corresponding saliency mapping using: (a) curvature co-occurrence histogram [29], (b) entropy based salient model [30], (c) mesh saliency via spectral processing [31], (d) our approach.

saliency value per vertex) that can be used as weights for feature-aware denoising giving more emphasis to the vertices related to high salient values, since they represent more intense features. Fig. 6.26 presents the smoothing results of noisy 3D lung models affected by the staircase effect. Our approach perfectly smooths the noisy 3D meshes, creating a reliable patient-specific model which can be used for personalized testing and simulations.

**Heatmap Visualization of Saliency Mapping**

Fig. 6.27 presents the heatmap visualization of the 3D saliency mapping applied in different industrial 3D models. For easier comparison, all the results are normalized, taking values [0-1] Eq. (6.12). The used colormap for the visualization is the "jet" contenting 64 colors (deep blue = 0, deep red = 1). Saliency mapping of a 3D object must provide visual information that can be easily recognizable. This means that different areas with different characteristics will be highlighted with a different color. On the other hand, different areas with the same characteristics will be highlighted with the same color. The experimental results show that our method (in Fig. 6.27-(e)) successfully follows this direction providing more robust and meaningful results than the other approaches. More specifically, the highest values (red colors) represents very distinctive vertices (e.g., corners), while the lowest values (blue colors) represents flat areas.

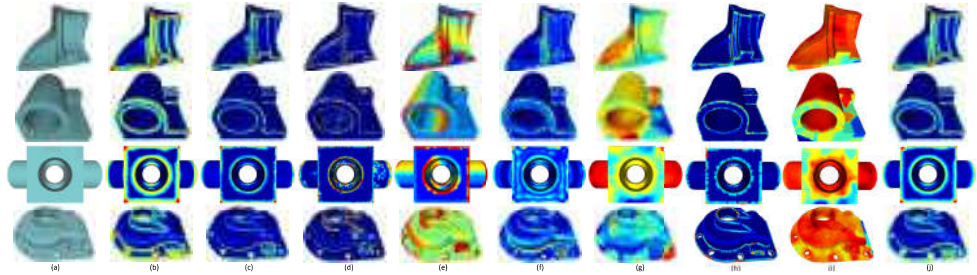

Figure 6.27: (a) Original model, and heatmaps visualization of saliency mapping based on: (b) the eigenvalues of small patches (spectral analysis), as described in paragraph 6.2.3, (c) the RPCA approach (geometrical analysis), as described in paragraph 6.1.2, (d) Wei *et al.* [29], (e) Tao *et al.* [30], (f) Lee *et al.* [32], (g) Song *et al.* [31], (h) Guo *et al.* [33], (i) Song *et al.* (CNN) [34], (j) our approach.

**Simplification Based on the Saliency of Vertices**

Due to the easiness of creating digital 3D content nowadays, a great amount of information can be captured and stored. The information, acquired by 3D scanners, is usually huge creating dense 3D models that are very difficult to be efficiently handled by other applications (i.e., high computational complexity). This information must be simplified, keeping only of the most representative information, and removing least important information. Simplification is a low-level application that focus on representing an object using a lower resolution mesh without errors or with errors that cannot be easily perceived. The main objective of a successful simplification approach is to remove only those vertices which do not offer significant geometric information to the simplified 3D object and their removal will not change significantly the shape or perceptual details of the 3D object. Following this line of thought, we suggest to remove the least perceptually important vertices, preserving only the most salient vertices for the reconstruction of the new simplified 3D model. More specifically, the steps of the suggested simplification process are: (i) all vertices are sorted based on their salient values. (ii) The $K$-th vertices with the higher salient values remain. (iii) the rest $n - K$ less salient vertices are removed and the $k$-nn algorithm is used for the recreation of the new connectivity (triangulation). Fig. 6.28 presents simplified meshes under different simplification scenarios.

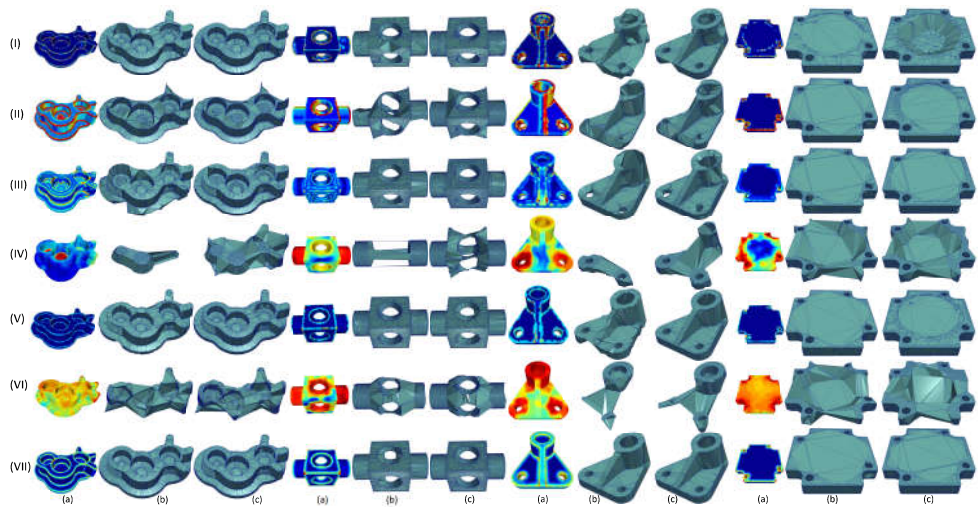

Figure 6.28: Simplification of 3D models using the saliency mapping of different methods, namely: (I) Wei *et al.* [29], (II) Tao *et al.* [30], (III) Lee *et al.* [32], (IV) Song *et al.* [31], (V) Guo *et al.* [33], (VI) Song *et al.* (CNN) [34] and (VII) our approach respectively (from up to down). (a) Heatmap visualization of the saliency mapping and simplified results using different simplification approaches, **Cad** (19,398 points): (b) 2000 ($\sim$ 10.3%), (c) 4000 ($\sim$ 20.6%), **Block** (8,771 points): (b) 2000 ($\sim$ 22.8%), (c) 4000 ($\sim$ 45.6%), **Part Lp** (4,261points): (b) 500 ($\sim$ 11.7%), (c) 1000 ($\sim$ 23.4%), **Coverrear Lp** (7,872 points): (b) 2000 ($\sim$ 25.4%), (c) 3000 ($\sim$ 38.1%).

**Simplification in the Field of Heritage Culture Preservation**

The simplification process in the field of heritage culture preservation requires a different type of handling with respect to the conventional process of simplification in other domains.

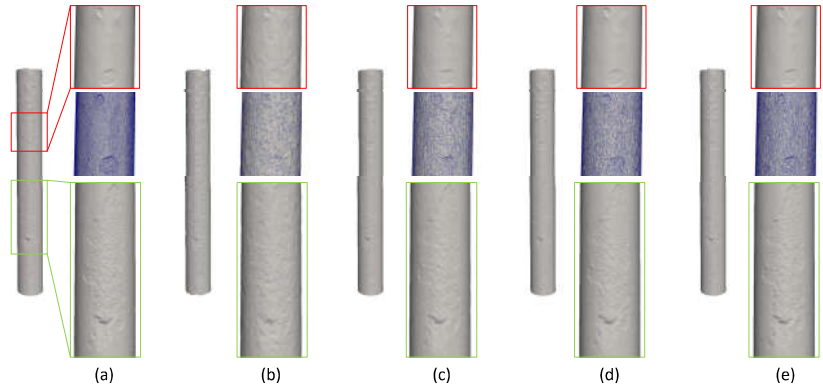

Figure 6.29: (a) Original 3D model (pillar 66,113 vertices) and reconstructed simplified models preserving the: (b) 90% simplification (6,640 vertices have been remained), (c) 70% simplification (19,314 vertices have been remained), (d) 50% simplification (33,095 vertices have been remained), (e) 30% simplification (45,652 vertices have been remained).

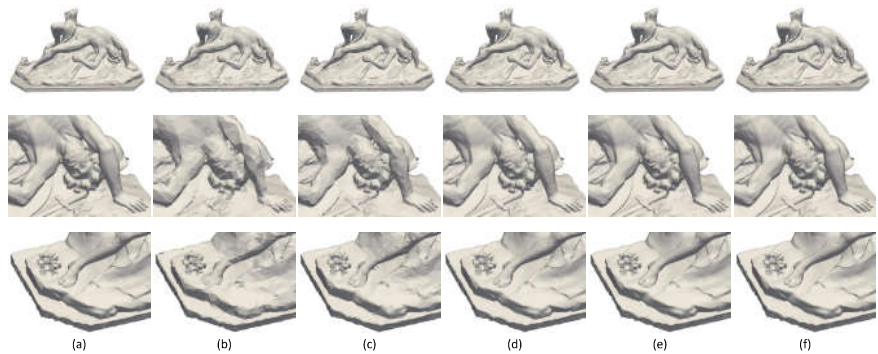

Figure 6.30: (a) Original model (Gladiator 249,964 vertices) and simplified models in different simplification scenarios: (b) 95% simplification (12,488 vertices have been remained) (c) 90% simplification (24,427 vertices have been remained), (d) 70% simplification (75,634 vertices have been remained), (e) 50% simplification (125,360 vertices have been remained), (f) 30% simplification (175,281 vertices have been remained).

More specifically, we do not aim to just create a simplified model, but also to emphasize some specific details with significant meaning such as the geometric features of the model as well as the defected areas. Thanks to our approach, the salient areas are identified and highlighted with higher values. In Fig. 6.29 and 6.30, we present examples of simplified reconstructed results of different heritage 3D models under different simplification scenarios from a simplification range 30% to 95% percentage. To note here that the simplification percentage denotes the number of vertices that have to be removed. More specifically, in a hypothetical scenario where we apply to a model a simplification ratio equal to 90%, it means that 90% of its initial vertices should be removed and only 10% of them will be used for the reconstruction. As we can observe, the simplified models have been reconstructed concerning the most salient vertices which represent high-frequency features (i.e., perception details and defects). More specifically, the presented simplified models in Fig. 6.29 seem to preserve the defect information even in very high percentages of simplification.

In Tables 6.2 and 6.3, we present in details the percentages of vertices that we preserve per each class {c1, c2, c3, c4} and per each simplification scenario {30%, 50%, 70%, 90%}, where $ci \ \forall \ i = 1, \cdots, 4$ represents the $i$-th class. The vertices of a higher class are more important than the vertices of a lower class. This is the reason why we keep a higher percentage of vertices from higher classes.

Table 6.2: Simplification for "Gladiator" model.

| \multicolumn{6}{c}{Gladiator model 249,964 vertices} |
|---|---|---|---|---|---|
|  | vertices | c1 | c2 | c3 | c4 |
| ~30% | 175,281 | 0.61 | 0.76 | 0.85 | 0.99 |
| ~50% | 125,360 | 0.4 | 0.57 | 0.65 | 0.95 |
| ~70% | 75,634 | 0.20 | 0.39 | 0.41 | 0.85 |
| ~90% | 24,427 | 0.033 | 0.1 | 0.3 | 0.6 |
| ~95% | 12,488 | 0.02 | 0.04 | 0.14 | 0.4 |

Table 6.3: Simplification for "Pillar14" model.

| \multicolumn{6}{c}{Pillar14 model 66,113 vertices} |
|---|---|---|---|---|---|
| simplification | vertices | c1 | c2 | c3 | c4 |
| ~30% | 45,652 | 0.69 | 0.8 | 0.85 | 0.9 |
| ~50% | 33,095 | 0.5 | 0.6 | 0.85 | 0.9 |
| ~70% | 19,314 | 0.29 | 0.75 | 0.8 | 0.85 |
| ~90% | 6,640 | 0.033 | 0.1 | 0.55 | 0.7 |

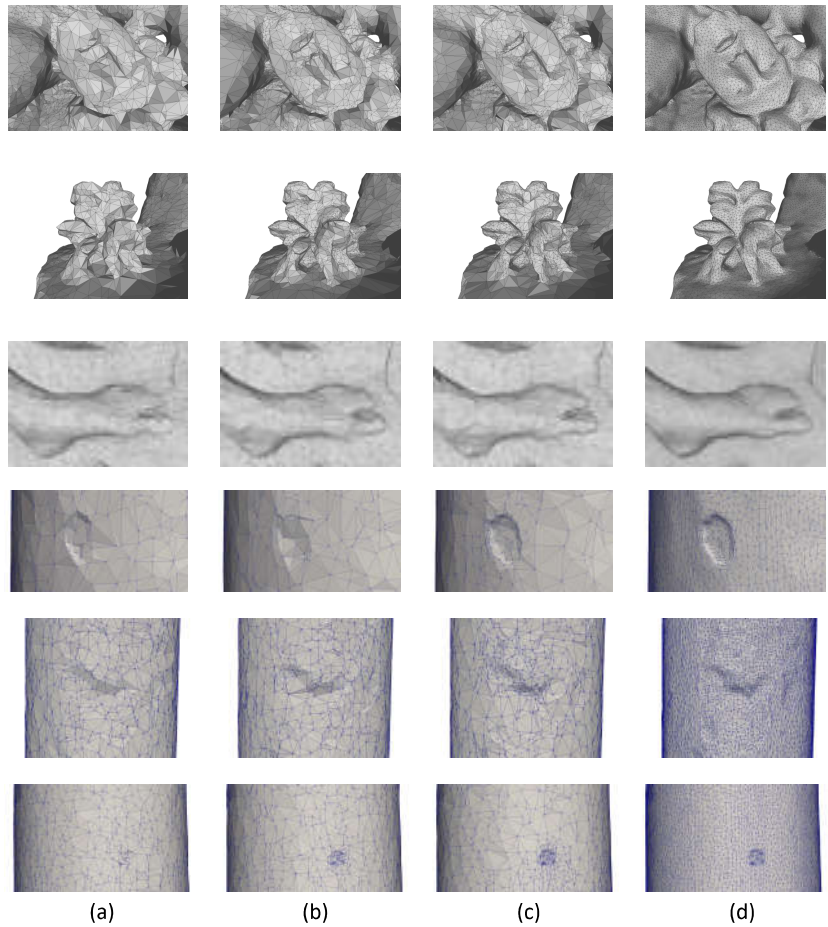

(a) (b) (c) (d)

Figure 6.31: (a) simplification without using saliency information, (b) using 4 saliency classes, (c) using 8 saliency classes, (d) original mesh.

Table 6.4: Percentages of remaining vertices per different classes.

|  | c1 | c2 | c3 | c4 | c5 | c6 | c7 | c8 |
|---|---|---|---|---|---|---|---|---|
| Gladiator model | | | | | | | | |
| 4 classes | 0.1 | 0.25 | 0.39 | 0.76 | - | - | - | - |
| 8 classes | 0.09 | 0.1 | 0.15 | 0.25 | 0.37 | 0.67 | 0.7 | 1 |
| Pillar14 model | | | | | | | | |
| 4 classes | 0.2 | 0.6 | 0.7 | 0.8 | - | - | - | - |
| 8 classes | 0.17 | 0.5 | 0.95 | 0.95 | 0.97 | 0.97 | 1 | 1 |

Fig. 6.31 presents the simplification results for two heritage models (i.e.,

gladiator and pillar14). For an easier comparison among the used approaches, we present enlarged details and we also highlight the remain vertices and the connectivity information between them. In both models and for all approaches, we used a 80% simplification scenario. The results showed that the traditional approach of the simplification does not handle differently the perceptually salient vertices (Fig. 6.31-a) in contrast to our approach that gives emphasis on the preservation of these vertices. The percentages of vertices that we preserve per each class, for the two different approaches (i.e., 4 classes, 8 classes) are presented in Table 6.4. To study the effect of granularity in simplification cases, we compared the simplified and reconstructed models in the case of 8 classes, 4 classes and saliency agnostic uniform sampling. Tables 6.2, 6.3 and 6.4 record the portion of vertices kept per class during the simplification of "gladiator" and "pillar14" models. As Fig. 6.31 visualizes the area around defects is better reconstructed when employing more classes and using an anisotropic strategy for the selection of vertices per class.

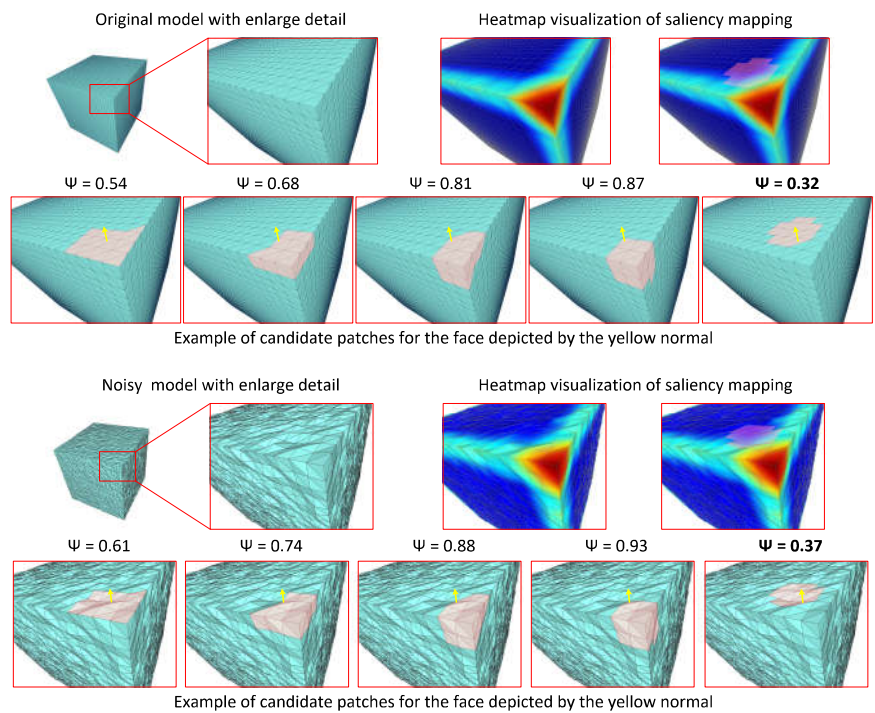

Figure 6.32: Ideal patch selection based on the proposed saliency mapping.

The reconstructed 3D models, which represent real objects, may be very dense, consisting of millions of vertices. This highly dense information is usually overwhelming, and the original 3D models can be efficiently simplified, keeping only the most representative details, and removing the least important information representing vertices lying in flat or smoothed areas. Simplification is a low-level application that focuses on representing an object using a lower resolution mesh without errors or with errors that cannot be easily perceived. The main objective of a successful simplification approach is to remove only those vertices which do not offer significant geometric information to the simplified 3D object, and their removal will not change the shape or perceptual details of the 3D object significantly. Most importantly, in the area of heritage maintenance any perceptual detail, both for features and defects, must be preserved. Following this line of thought, we suggest removing the least perceptually important vertices, preserving only the most salient vertices for the reconstruction of the new simplified 3D model. Nevertheless, to avoid the loss of information on areas with low salient vertices, we suggest the selection of different portions for vertices belonging in different salient classes. Algorithm 15 briefly describes the step that we follow for the proposed multiscale feature aware simplification.

---

**Algorithm 15:** Multiscale feature aware simplification

**Input** : 3D model $\mathcal{M}$, desired simplification percentage, $N$ number of classes, saliency map of vertices

**Output:** Simplified model $\hat{\mathcal{M}}$

`// Vertex simplification`

1 Sort the vertices based on the saliency values of the saliency map provided by the CNN approach;
2 Separate the vertices into $N$ salient classes based on their salient values;
3 Depending on the simplification scenario, different percentages of vertices are selected from each class, giving emphasis to the perceptually salient vertices;
4 The final number of the remaining vertices must satisfy the initial percentage scenario;

`// Connectivity simplification`

5 **for** *each i removed vertex* **do**
6     Find the nearest vertex that remains;
7     Replace any edge connection of the *i* removed vertex with the nearest remaining vertex;
8 **end**

---

**Feature-aware Denoising Based on the Saliency of Vertices**

Guided normals filtering has been used in past works [362], [12] providing excellent denoising results. In [362], the saliency is estimated by using the difference between the normals of the two incident faces. We follow the same line of thought but we use a different way for the estimation of the ideal patch.

More specifically, we select the patch that has the smallest value of $\Psi$, according to Eqs. (6.20)-(6.21), since it consists of "less salient" faces (flat areas that are depicted with deep blue color).

$$\mathcal{P}_i^* = (\mathcal{P}_{ij} \mid \min(\Xi_{ij})) \ \forall \ i = 1, \cdots, n, j = 1, \cdots, n_p \tag{6.20}$$

$$\text{where} \ \ \Xi_{ij} = \frac{\sum_{\forall g \in \mathcal{P}_{ij}} s_g}{|\mathcal{P}_{ij}|} \ \forall \ g = 1, \cdots, k \tag{6.21}$$

In Fig. 6.32, we present an example of five candidate patches (for the face which is depicted by the yellow normal). In these examples, we show that the selected ideal patch is this one with the lowest value of $\Xi$ (i.e., $\Xi = 0.32$ and $\Xi = 0.37$), representing the area with the less salient features. As we can observe, both the first and the last patches represent totally flat areas, however they do not have the same $\Xi$ value since the first patch consists of more salient triangles in comparison to the last patch, so the last area is more preferable to represent the ideal patch. We also can observe that our method provides reliable results of saliency mapping even under the presence of noise, which makes it ideal for use in applications with noisy 3D models. The purpose of this example is the estimation of the most representative centroid normal (i.e., guided normal) in order to use it for a more efficient bilateral filtering [15]:

$$\mathbf{W}_{1_{ij}} = \exp(\frac{-\|\mathbf{c}_i - \mathbf{c}_j\|^2}{2\sigma_1^2}), \ \mathbf{W}_{2_{ij}} = \exp(\frac{-\|\mathbf{g}_i - \mathbf{g}_j\|^2}{2\sigma_2^2}) \tag{6.22}$$

$$\hat{\mathbf{n}}_c = \frac{\sum_{f_j \in \mathcal{P}_i} A_j \mathbf{W}_{1_{ij}} \mathbf{W}_{2_{ij}} \mathbf{n}_{cj}}{\|A_j \mathbf{W}_{1_{ij}} \mathbf{W}_{2_{ij}} \mathbf{n}_{cj}\|_2} \tag{6.23}$$

Finally, the denoised normals $\hat{\mathbf{n}}_c$ are used to update the vertices according to [6]: Note here that we do not search for ideal parameters per each 3D model or method. Instead, in all the experiments and for any of the approaches, we use exactly the same values for each parameter. Specifically, we define $\sigma_2 = 0.25$, $\sigma_{1_{ij}} = \sum_{\forall \mathbf{c}_j \in \Psi_{1_i}} \|\mathbf{c}_i - \mathbf{c}_j\|_2^2 / |\Psi_{1_i}|$ [15], 15 iterations for the bilateral filtering Eqs. (6.22)-(6.23) and 20 iterations for the vertex updating. The ideal selected patch must consist of normals with similar direction (in order to satisfy the normals' consistency). The patches that have a lot of corners or edges must be banned (i.e., high salient values in our case) since they consist of normals lying in different directions. As a result, the value of $\Xi$ would be totally misleading since it would not represent a specific planar area. Fig. 6.33 presents the denoising results with enlarged regions for easier comparisons. The quality of the reconstructed models is evaluated using the metrics: (i) $\theta$ representing the mean angle between the normals of the ground truth and the reconstructed faces and (ii) the Hausdorff distance (HD).

It should be emphasized that, in most cases, there is no ground truth saliency map or a reliable metric that can be used for benchmarking purposes. The typical way to evaluate a saliency map is via subjective evaluation. The subjective evaluation can clearly show if a specific saliency mapping has achieved its purpose, applied in a specific application, and provides a fair comparison with the results of other salient mapping methods.

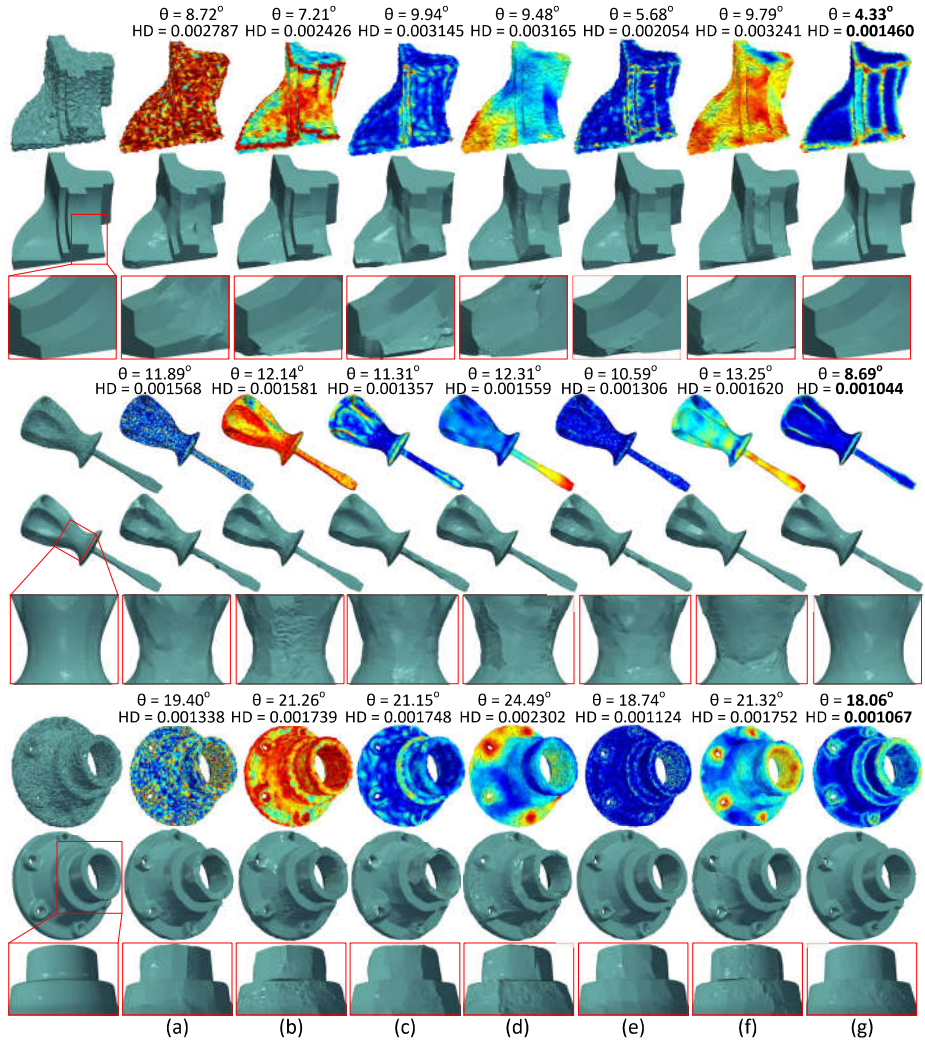

Figure 6.33: Heatmaps of saliency mapping and denoising results using the methods: (a) curvature co-occurrence histogram [29], (b) entropy based salient model [30], (c) a mesh saliency [32], (d) mesh saliency via spectral processing [31], (e) Point-wise saliency detection [33], (f) mesh saliency via CNN [34], (g) our approach.

> **Publications that have contributed to this section:**
>
> 1. Content of the subsection 6.1.1 has been presented in **C15** and **C16**
> 2. Content of the subsection 6.1.2 has been presented in **C14** and **J7**
> 3. Content of the subsection 6.1.3 has been presented in **J7**, **J8** and **J9**

## 6.2 Registration and Identification of Partially-observed 3D Objects in Cluttered Scenes

In this section, we present a methodology for identifying and registering partially-scanned and noisy 3D objects, lying in arbitrary positions in a 3D scene, with corresponding high-quality models. The methodology is assessed on point cloud scenes with multiple objects with large missing parts. The proposed approach does not require connectivity information and is thus generic and computationally efficient, thereby facilitating computationally demanding applications, like augmented reality. The main contributions of this work are the introduction of a layered joint registration and indexing scheme of cluttered partial point clouds using a novel multi-scale saliency extraction technique to identify distinctive regions, and an enhanced similarity criterion for object-to-model matching. The processing time of the process is also accelerated through 3D scene segmentation. The objective of this work is to match and replace these objects with the corresponding high-quality 3D objects that are assumed to be available beforehand.

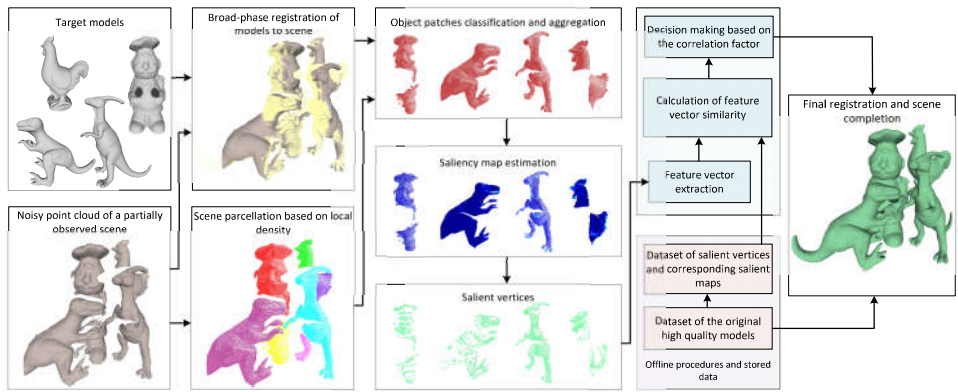

Figure 6.34: Pipeline of the proposed methodology. For the sake of simplicity, we illustrate here only the *target* models that are also present in the selected partially observed scene.

### 6.2.1 Work-flow of the proposed method

The irrelevant objects and the noise seriously affect the optimization process [363–371]. Our methodology extends previous work on point cloud registration [291] to overcome its limitations in the case of noisy point clouds. Our idea is that the obtained alignment solution can be improved if registration is guided by the most relevant salient part of the scene. The subsequent steps after the broad-phase registration include feature extraction, similarity assessment and saliency estimation. These steps are more computationally efficient when a part of the point cloud is used rather than the whole scene, as in some cases the scene might be very big. In the next sections, we describe how we robustify and accelerate computations at the same time by identifying and focusing only on salient parts of the scene that potentially correspond to the *target* model. This data selection step not only accelerates the calculations as it allows to search for pairs of landmarks (necessary for the final registration step) in a reduced space, but also leads to a reduction of the impact of the outliers.

An outline of our methodology is illustrated in Fig. 6.34 and can be summarized as follows:

1. **Broad-phase registration**: First, a fast global registration technique [291] is applied, which helps both for the decision of the matching and the final fine registration process, providing a better initial alignment between the *query* and the *target* object (subsection 6.2.2).

2. **Segmentation, robustification, feature extraction and matching**: In parallel, the whole scene is divided into clusters using our parameter-free implementation of the popular density-based clustering algorithm [372]. Scene clusters, which are geometrically more similar to the registered point cloud of the previous step, are merged to create the *query* object (subsection 6.2.3). A robustification step is applied to facilitate the identification and removal of spurious point sets (obtained by imperfect scanning) that might blur the object boundaries affecting the registration and the execution time (subsection 6.2.5). The proposed feature vectors, combining pose with local multi-scale geometric information (subsection 6.2.3), are then extracted and used as descriptors for model to object correspondence assessment (subsection 6.2.3). Finally, based on the defined point similarity criterion, the best-related pairs of vertices between the matched (complete and partial) objects are identified (subsection 6.2.3).

3. **Narrow-phase registration**: The final step includes the calculation of a rigid transformation that brings the previously identified pairs of corresponding points into alignment (subsection 6.2.4).

### 6.2.2 Broad-Phase Registration

The first step of the matching process is to align each *target* model to the scanned scene by global registration, without incorporating knowledge of the model class. We have selected a recently proposed global registration algorithm [291] that has shown very good performance in different realistic datasets. For completeness, we present here an overview of the algorithm, while details can be found in [291]. The algorithm finds a number of candidate transformations by matching pairs in a roughly uniformly distributed subset of vertices of the input objects based on local shape properties (i.e., principal curvatures and the first principal direction). The optimal transformation is selected by localizing a density peak in the space of candidate rigid transformations. In order to find the density peak, a metric $d(T_1, T_2)$ is needed, which measures the distance of a transformation $T_1$ from a transformation $T_2$. A density estimation function can be defined using some kernel function $F$ as:

$$\rho(x) = \sum_i F(d(T_i, x)). \tag{6.24}$$

Various kernel functions can be used for $F$, however, mostly for efficiency reasons, a simple Gaussian $F(r) = e^{-(br)^2}$ is preferred, where $b$ is a spread parameter. Instead of looking for the general location of the true global maximum of $\rho$, $\rho$ is only evaluated at each candidate position and the maximum among them is chosen as the result. Given the spread parameter $b$ and some small threshold $\phi$, only samples within the radius $r = \sqrt{-ln(\phi)}/b$ contribute significantly ($> \phi$) to the density. Therefore, the task is to find, for each candidate, a set of candidates up to the distance $r$. Because of the non-Euclidean topology of the search space, a KD-tree cannot be used for this purpose, however, a more general acceleration structure - the Vantage Point Tree [373] - can be used. Measuring the distance between two transformations commonly involves relating their rotation and translation components, which is notoriously difficult. Instead, embracing the inherent dependence of such relation on the character of the input data, the metric can be derived from the difference of the effect the two transformations have on the vertices of the input objects:

$$d(T_1, T_2) = \sum_i ||\mathbf{R}_1 \mathbf{v}_i + \mathbf{t}_1 - \mathbf{R}_2 \mathbf{v}_i - \mathbf{t}_2|| \tag{6.25}$$

where $\mathbf{R}_1$ and $\mathbf{R}_2$ are the rotations of $T_1$ and $T_2$ respectively and $\mathbf{t}_1$ and $\mathbf{t}_2$ are the translation vectors of $T_1$ and $T_2$ respectively. Since the sampling density of the input objects may be quite irregular, a more robust option is to integrate distance over triangles instead of summing over vertices. The value of the corresponding integral over a single triangle $t$ can be expressed as:

$$d_t(T_1, T_2) = \int_t ||\mathbf{R}_1 \mathbf{v} + \mathbf{t}_1 - \mathbf{R}_2 \mathbf{v} - \mathbf{t}_2|| da. \tag{6.26}$$

where a(t) represents the simplest geometric area (i.e., consisting of three vertices) represented by a triangle t. Since this work is designed for point clouds and does not request actual connectivity information, triangles are defined based on the point's closest neighbors. Finally, the full rigid transformation metric is obtained by summing over all triangles:

$$d(T_1, T_2)^2 = \sum_{i=1} d_{t_i}(T_1, T_2)^2 \tag{6.27}$$

A remarkable property of both expressions for transformation distance, i.e. the vertex sum and the triangle integral, is that with linear pre-processing, they can be evaluated in constant time, i.e. independently on the sampling density of the input objects.

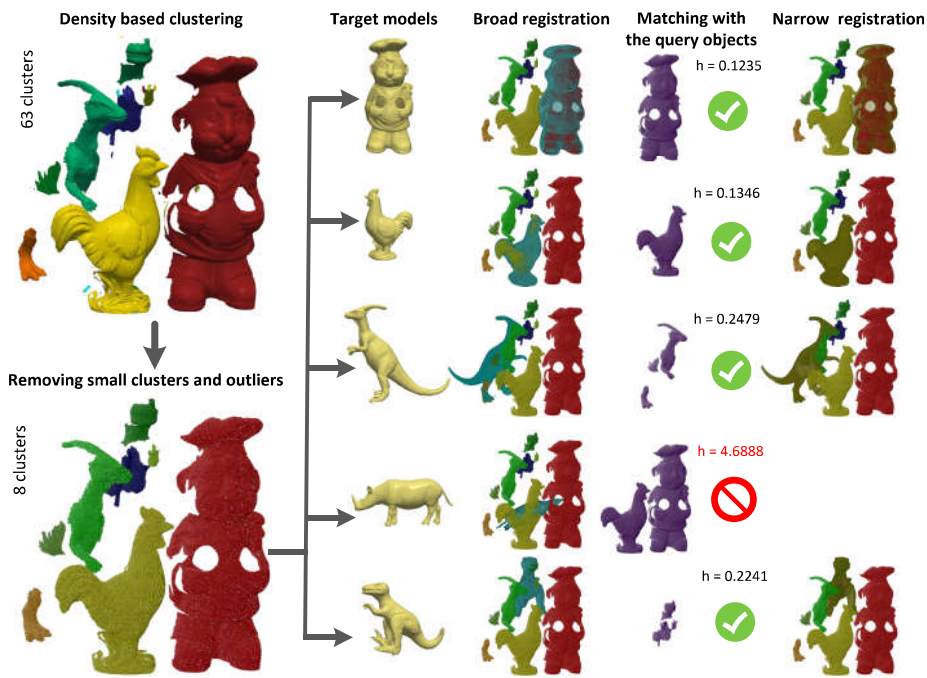

Figure 6.35: Broad-to-narrow phase registration. Each target model is registered to the partial scene. The clusters that are geometrically close to the registered model are selected to create the query object for the matching comparison step.

It is important to note here that the *target* object may not appear in the partially observed scene, as presented in the forth line (i.e., rhino model) of Fig. 6.35. The global registration step is not required to identify if there is a correlation between the *target* model and the *query* model, but only to provide the best possible solution. The final decision on the correspondence of the two models will be made in the narrow-phase registration step according to subsection 6.2.3. After the broad-phase registration, all the high-resolution models have been mapped to the scene using global alignment. Then the solution is further refined to become more robust to outliers.

### 6.2.3 Scene Segmentation and Model-to-Object Matching

**Point Cloud Segmentation by Density-based Clustering**

The semantic segmentation of the scene is often challenging, as the 3D objects lying in the scene might appear tangled with each other, due to abnormalities created by imperfect scanning. Let's also note that supervised [374] or semi-supervised [375] learning techniques that exploit prior knowledge in the form of shape priors or large training datasets with semantic annotations cannot be applied here to facilitate segmentation, because such large scale annotations are not always available. Our method is based on the assumption that, even if different objects overlap, (i.e. their distance is small in some regions), the local point density within each object is larger than these across different objects in the scene. Therefore for parcellation of the scene, we formulate a density-based algorithm, i.e the Density-Based Spatial Clustering of Applications with Noise (DBSCAN) [372], and implement a parameter-free approach, as explained in the sequel. More specifically, DBSCAN is used for the automated segmentation of a point cloud scene into separate clusters, which can be potentially used for matching and registration, reducing the total execution time. The number of clusters is not required beforehand. Considering point cloud $\mathbf{M} \in \mathbb{R}^{n \times 3}$, the clustering result is represented by an vector $\mathbf{k} \in \mathbb{Z}^n$ [376] with elements:

$$\mathbf{k}_i = \begin{cases} j & \text{if } \mathbf{v}_i \text{ belongs to cluster } j \in \{1, \cdots, n_c\} \\ -1 & \text{if } \mathbf{v}_i \text{ is an outlier} \end{cases}, \qquad (6.28)$$

where $n_c$ represents the number of clusters. Each scene may consist of a different number of 3D objects and a 3D object may be represented by more than one clusters due to imperfect scanning, occlusions, etc. One parameter of the DBSCAN algorithm which needs to be predefined is the neighborhood radius $\epsilon$. Our contribution is that we allow the threshold $\epsilon$ to be spatially adapted and be larger in sparse regions in order to retain sufficient neighbors everywhere. The method becomes more robust with this adaptation, especially in cases where the 3D objects have different density of points in different areas of their surface. More specifically, we assign a value that is inversely proportional to the local density, such as twice the average distance of the $K_a$ nearest neighbors.

$$\epsilon_i = 2 \sum_{j=1}^{K_a} \frac{D_{ij}}{K_a} \ \forall \ i = 1, \ldots, n \qquad (6.29)$$

where $D_{ij} \equiv D(i,j)$ is the Euclidean distance of the $i$ vertex to its $j^{th}$ nearest vertex. The spatially adaptive value of $\epsilon$ allows us to differentiate which points belong to a cluster, and which are large-scale outliers or noise points. After the broad-phase registration step, all clusters being geometrically close to the registered target model (according to a predefined distance threshold) are merged to form a new point cloud (i.e., *query* object), denoted as $\mathbf{Q}$.

**Salient Points Detection**

Our purpose in this step is to identify if each high-quality *target* model $\mathbf{T} \in \mathbb{R}^{n_t \times 3}$ and each segmented *query* object $\mathbf{Q} \in \mathbb{R}^{n_q \times 3}$ (where $n_t \geq n_q$ due to occlusion, low-quality, etc), represents the same structure. To define similarity between each set of point clouds we propose descriptors that encode spectral saliency. In the following we describe the proposed features, and how they are used to extract point-to-point correspondences, necessary for the final registration step.

The feature descriptors that we use are related to the saliency map of the point cloud. Saliency is a value assigned to each vertex of a point cloud that represents its perceived importance. In the case of raw point clouds without context information, saliency characterizes the geometric properties. High values of saliency represent more perceptually protruding vertices. In this work, we assume that geometric lines, corners, and edges are more distinctive perceptually than flat areas, according to the theory of visual saliency of sight.

To estimate the saliency map, we use a similar pipeline, as the one described in [377], but we extract the saliency map using only spectral analysis, avoiding the computationally complex geometric analysis. For each point $\mathbf{v}_i$ of the point cloud $\mathbf{Q} \in \mathbb{R}^{n_q \times 3}$, we construct a matrix $\mathbf{N}_i \in \mathbb{R}^{(k+1) \times 3}$ comprising of the normals of $\mathbf{v}_i$ and the normals of the k-nearest neighboring points of $\mathbf{v}_i$ (generally, we set $k = 20$):

$$\mathbf{N}_i = [\mathbf{n}_i, \mathbf{n}_{i_1}, \mathbf{n}_{i_2}, \cdots, \mathbf{n}_{i_k}]^T \; \forall \; i = 1, \cdots, n_q \tag{6.30}$$

For the estimation of the point normals, a plane is approximated based on the set of closest neighboring points, according to [378]:

$$\mathbf{n}_i = \frac{1}{|\Psi_i|} \sum_{\mathbf{v} \in \Psi_i} (\mathbf{v} - \mathbf{v}_i)(\mathbf{v} - \mathbf{v}_i)^T \tag{6.31}$$

The eigenvector corresponding to the smallest eigenvalue of $\mathbf{n}_i$, is the best estimation of its normal vector. The matrices $\mathbf{N}_i$ are also used for the computation of covariance matrices $\mathbf{C}_i$:

$$\mathbf{C}_i = \mathbf{N}_i^T \mathbf{N}_i \in \mathbb{R}^{3 \times 3} \tag{6.32}$$

The matrix $\mathbf{C}_i = \mathbf{U} \mathbf{\Lambda} \mathbf{U}^T$ is decomposed into a matrix $\mathbf{U}$, consisting of the eigenvectors, and a diagonal matrix $\mathbf{\Lambda} = \text{diag}(\lambda_{i1}, \lambda_{i2}, \lambda_{i3})$, consisting of the corresponding eigenvalues. Finally, the saliency value $s_i$ of a vertex $\mathbf{v}_i$ is determined as the value given by the inverse $l_2$-norm of the corresponding eigenvalues:

$$s_i = \frac{1}{\sqrt{\lambda_{i1}^2 + \lambda_{i2}^2 + \lambda_{i3}^2}} \tag{6.33}$$

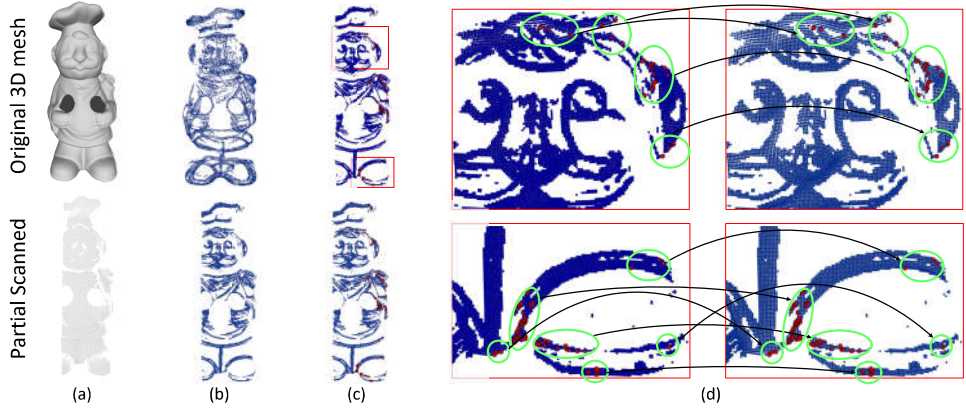

Figure 6.36: (a) [Up] Original high-quality 3D mesh (Chef model) (consisting of 176,912 vertices), [Down] partially scanned point cloud object of the same model (consisting of 45,811 vertices), (b) [Up] salient vertices of the original model (consisting of 55,935 vertices), [Down] salient vertices of the partial scanned object (consisting of 16,324), (c) [Up] remaining salient vertices of the high-quality model that creates unique pairs with the salient vertices of partially-scanned model (consisting of 11,162 vertices) and [Down] vice versa, (d) Enlarged regions illustrating the identified correspondences in red color.

Based on this equation, flat areas producing high eigenvalues correspond to small saliency values, while the most salient vertices are those that represent high-frequency features (i.e., edges and corners) and thereby produce small eigenvalues. These features are more recognizable by the human brain, since they are perceptually more distinctive. The same process for extracting the saliency maps is followed for partial objects as well as the *target* models. The only difference is that computations for the *target* models are performed only once (offline) and a small set of salient points is stored to be used for the subsequent matching process.

**Multi-Scale Feature Extraction**

First, the saliency values of the two compared models are normalized according to:

$$s_{t_i} = 1 - \frac{e^{-K_s * s_{t_i}}}{s_{\max}} \ \forall \ i = 1, \cdots, n_t$$
$$s_{q_i} = 1 - \frac{e^{-K_s * s_{q_i}}}{s_{\max}} \ \forall \ i = 1, \cdots, n_q$$
(6.34)

where $s_{\max} = \max(\max(s_t), \max(s_q))$. Then, we perform spatial smoothing of the saliency map with a uniform kernel of increasing size and use the obtained values to form a feature vector with the multi-scale saliency values. The neighborhood size is selected as $\Psi^{k \cdot K}$ with $K = 5$ and $k = 1,...,5$, although these parameters may vary. Smaller scales increase feature vector specificity, while larger scales smooth out noise and irregularities making the shape descriptor more robust. The use of multiple scales allows us to combine both properties and leads to unique and accurate correspondences. This process is applied for each point cloud in **T** as well as **Q**. Specifically, for each vertex $i$, we create a corresponding vector $\mathbf{a}_i \in \mathbb{R}^5$, according to:

$$\mathbf{a}_i = [\frac{\sum\limits_{j \in \Psi_i^K} s_j}{K} \quad \frac{\sum\limits_{j \in \Psi_i^{2K}} s_j}{2K} \quad ... \quad \frac{\sum\limits_{j \in \Psi_i^{5K}} s_j}{5K}]^T \tag{6.35}$$

Then, we concatenate the multi-scale saliency values with the vertex coordinates to obtain the final feature representation. Finally, for each one vertex, we create the augmented feature vector $\mathbf{f} \in \mathbb{R}^8$, consisting of the vertex coordinates and the corresponding values of the vector $\mathbf{a}$:

$$\mathbf{f} = \mathbf{v} \cup \mathbf{a} = \begin{bmatrix} \mathbf{v}^T \\ \mathbf{a}^T \end{bmatrix} \tag{6.36}$$

**Model-to-Object Correspondence Estimation**

The feature vectors $\mathbf{f}$, calculated using Eq. (6.36), are used for the evaluation of similarity between the *query* and *target* point clouds, looking for their unique pairs of vertices $\mathbf{p}$ which exhibit the smallest feature vector distance. It is expressed through the $l_2$-norm:

$$\mathbf{p} = (\mathbf{v}_t, \mathbf{v}_q) = \underset{\mathbf{v}_t, \mathbf{v}_q}{\arg\min} ||\mathbf{f}_t - \mathbf{f}_q||_2 \tag{6.37}$$

Finally, we keep only the first $K_p$ pairs having the highest feature vector similarity. An example is shown in Fig. 6.36c-6.36d, where feature vector correspondences are illustrated with red color. These $K_p$ pairs are the best-identified correspondences between model **T** and object **Q**. Let's note here that we use the augmented feature vectors $\mathbf{f}$, which include multi-scale geometric descriptors in addition to 3D location, to avoid erroneous surface mapping obtained by chance due to accidentally good local geometric fit (small Euclidean distance) of the partial point cloud. These augmented feature vectors can ensure not only spatial proximity, but also local shape similarity. At this point of the methodology, the obtained pairs of corresponding vertices include matches for each *target* model **T** to each *query* object $\mathbf{Q}^j, j \in \{1, \cdots, n_{c_q}\}$, where the $n_{c_q}$ denotes the number of the *query* objects. To identify the correspondences, we introduce a dissimilarity factor $h_j$ which is defined as the mean distance of the $K_p$ pairs of the best-related vertices between model **T** and each object $\mathbf{Q}^j$.

$$h_j = \frac{1}{K_p} \sum_{i=1}^{K_p} ||\mathbf{f}_{t_i} - \mathbf{f}_{q_i^j}||_2 \tag{6.38}$$

where $\mathbf{f}_{q_i^j} \in \mathbf{Q}^j$ is the feature vector matched to $\mathbf{f}_{t_i} \in \mathbf{T}$. The lower the value of $h$, the more similar are the two point clouds.

### 6.2.4 Narrow-Phase Registration

Having identified and matched the target and query object pairs in the scene, the fine registration is achieved by identifying a set of corresponding points and then finding the optimal transformation that brings those pairs of points (control points) into alignment. In this step, we initialize the registration with the solution obtained from the global initial alignment and refine it using a weighted ICP approach. The objective is, given a set of control points $\mathbf{p} = (\mathbf{v}_{t_i}, \mathbf{v}_{q_i})$ with $\mathbf{v}_t \in \mathbf{T}$ and $\mathbf{v}_q \in \mathbf{Q}$, to estimate a rigid transformation $T$ that minimizes a distance (or more general an error) function.

As the partial object contains only a subset of the shape represented by the high resolution model, point matching is performed starting with a set of control points in **Q** and identifying the corresponding points in **T** as described in the previous section. Generally, in weighted ICP approaches [379], [380] the error function is composed of a feature vector distance term and a weighting term used to downgrade the contribution of pairs that have high likelihood to be outliers or wrong correspondences. The optimal transformation is obtained by solving a weighted least squares minimization problem. For rigid transformations expressed by a rotation matrix **R** and a translation vector **t**, it can be written as:

$$\arg\min_{\mathbf{R},\mathbf{t}} \sum_i \psi(D(\mathbf{R}\mathbf{v}_{t_i} + \mathbf{t}, \mathbf{v}_{q_i})) \tag{6.39}$$

where $D$ is the Euclidean distance and $\psi(.)$ is an even, $C^1$-continuous on $\mathbb{R}$ and monotonically increasing function on $[0, \infty)$. The function $\psi(r)$ that we use is the Tukey's bi-weight function formulated as:

$$\psi(r) = \begin{cases} \frac{\gamma^2}{6}\{1 - (1 - \frac{r^2}{\gamma^2})^3\} & \text{if } |r| \leq \gamma \\ \frac{\gamma^2}{6} & \text{if } |r| > \gamma \end{cases} \tag{6.40}$$

and we use as weights $w(r)$, the first order derivative of $\psi(r)$ function, denoted $w(r) = \psi'(r)$:

$$w(r) = \begin{cases} r(1 - \frac{r^2}{\gamma^2})^2 & \text{if } |r| \leq \gamma \\ 0 & \text{if } |r| > \gamma \end{cases} \tag{6.41}$$

### 6.2.5 Robustification by Outliers Removal

Outliers and other surface abnormalities of the scanning procedure might be interpreted as salient points, leading to wrong matching and registration results. So, we have to ignore both the outliers and the open edges, which form the boundary of the segmented point cloud, because they do not represent characteristic discriminative features to guide the registration process.

**Small-Scale Outliers Removal**

Scanned objects or scenes usually include noisy parts represented by vertices that do not belong to the geometry of the real object. Two different types of outliers occur in scanned point clouds; (a) the large-scale outliers that lie away from the point cloud and (b) the small-scale outliers which are tangled with the useful information and could be mistakenly recognized as points [381]. As we mentioned earlier, the large-scale outliers could be removed through the application of the clustering method presented in subsection 6.2.3. For the small-scale outliers removal process, we use a Robust Principal Component Analysis (RPCA) approach, as described in Section 4.5.

We finally estimate the value $m_i$ that quantifies the probability of vertex $i$ to be an outlier. The sparse matrix is usually full of zeros (representing vertices on flat areas), and only some high non-zero values exist that correspond to outliers.

$$m_i = \sqrt{\tilde{x}_i^2 + \tilde{y}_i^2 + \tilde{z}_i^2} \tag{6.42}$$

**Boundary Edges Identification and Removal**

Similarly to the outliers, the boundary edges of a partially-scanned object may mistakenly be recognized as edge features, so they must be identified and removed as well. It is necessary to differentiate salient points on edges and corners, useful for guiding the registration process, from misleading boundary edges around holes and missing parts, caused by imperfect scanning. To identify such boundary edges, we estimate the mean distance $d_{\Psi_i} = \frac{\sum_{\mathbf{v} \in \Psi_i^{K_D}} \|\mathbf{v} - \mathbf{v}_i\|}{K_D}$ between each $i$ vertex and its $K_D$ nearest neighbors, and characterize each point as internal or boundary point according to:

$$\mathbf{v}_i = \begin{cases} \text{is an internal vertex, if } d_{\Psi_i} \leq 2\bar{d} \\ \text{is a boundary vertex, if } d_{\Psi_i} \geq 2\bar{d} \end{cases} \tag{6.43}$$

Table 6.5: Speed up of our approach (Narrow-phase Registration)

| Target to scene | O. Registration (in sec.) | | | Our Approach (in sec.) | | | | Speed up |
|---|---|---|---|---|---|---|---|---|
| | Pre-proce-cessing | Regi-stra-tion | Total | Segme-nta-tion | Pre-proce-ssing | Regi-stra-tion | Total | |
| Chef to Scene 5 | 26.08 | 9.82 | 35.90 | 0.81 | 3.78 | 1.81 | 6.40 | **x 5.6** |
| Chic. to Scene 12 | 27.91 | 12.80 | 40.71 | 0.29 | 2.29 | 2.15 | 4.73 | **x 8.6** |
| Par. to Scene 27 | 29.36 | 26.82 | 56.18 | 0.31 | 4.20 | 5.08 | 9.59 | **x 5.8** |
| T-rex to Scene 32 | 26.72 | 18.48 | 45.20 | 0.16 | 2.99 | 3.12 | 6.27 | **x 7.2** |
| Rhino to Scene 43 | 31.06 | 8.76 | 39.82 | 0.46 | 1.41 | 0.71 | 2.58 | **x 15.4** |

where the $\bar{d}$ represents the mean distance of all mean distances $d_{\Psi_j}$ in the *query object* **Q**,

$$\bar{d} = \frac{\sum_{i=1}^{n_q} d_{\Psi_j}}{n_q} \qquad (6.44)$$

### 6.2.6 Parameter Adjustment

In this paragraph, we will present and justify the selection of parameter values that are fixed through the steps of the proposed methodology in order to provide reproducible results. The chosen number of neighboring vertices in Eq. (6.29) is equal to $K_a = 5$, but in any case, we have observed that the algorithm is not sensitive to this value. In fact, the results are very similar for the range of $K_a \in [5, 8]$. The only important fact is that those vertices should be retrieved from the geometric area, which is very close to the reference vertex, such as the first ring area.

In Eq. (6.42), we estimate the quantity $m_i$ for each $i$ vertex that is used to identify if a vertex is an outlier or not. However, we first need to specify a threshold for this identification. We interpret as outliers those vertices that have a value of $m_i$ bigger than 0.6, and then we remove them. Obviously, this threshold can be adjusted; The experimental analysis demonstrates that the lower the threshold, the more vertices are considered as outliers and are therefore removed. The user can adjust the value of this parameter; however, we suggest the threshold to be equal to 0.6.

Table 6.6: Speed up of our approach (Clustering)

| Scene | Number of Points | Original DBSCAN | Our Approach | Speed up |
|---|---|---|---|---|
| Scene 3 | 101,644 | 18.75 | 0.46 | **x 41.7** |
| Scene 24 | 144,130 | 61.25 | 0.64 | **x 95.7** |
| Scene 33 | 112,755 | 22.53 | 0.53 | **x 42.5** |
| Scene 42 | 153,486 | 68.42 | 0.70 | **x 97.7** |
| Scene 50 | 124,464 | 29.11 | 0.57 | **x 51** |

Table 6.7: Default values for parameters.

| Variable | Default Value | Short description |
|---|---|---|
| $K_a$ | 5 | Number of neighboring vertices in Eq. (6.29) |
| $K_D$ | 10 | Number of neighboring vertices in Eq. (6.43) |
| $K_s$ | 1000 | Value of kernel in Eq. (6.34) |
| $K_p$ | 200 | Number of compared vertices in Eq. (6.38) |
| $\zeta$ | 0.4 | Threshold that denotes the dissimilarity |

Eq. (6.43) is used to determine if a vertex is considered as a boundary point or not. The used number of neighboring vertices in this case is $K_D = 10$. The selection of this value is not critical and it is advised to be selected in the range of $K_D \in [8, 12]$. These values provide enough, but not too many, instances. In subsection 6.2.3, we described the use of feature vectors **f** for finding unique pairs. However, to make the process more computationally efficient, we do not compare the two full-sized point clouds, vertex by vertex, but we keep only the highest saliency vertices (as presented in Fig. 6.36-(b)), corresponding to saliency values higher than a threshold set to $\tau = 0.4$.

Eq. (6.38) expresses the dissimilarity factor $h$. If the value of $h$ is lower than a threshold then we assume that the two examined point clouds are related, representing the same 3D object. The experimental analysis (please refer to subsection 6.2.8) has shown that a threshold $\zeta = 0.4$ allows to differentiate intra- from inter-class object pairs. We also define the number of the compared vertices to be equal to $K_p = 200$.

We would like to mention here that the selected values are invariant to affine transformations and sampling density. Table 6.7 summarizes the default values that we used and a short description.

### 6.2.7 Computational Efficiency

In this paragraph, we present a time complexity analysis showing how our contribution can speed up the clustering process. Table 6.5 assesses the execution time of the narrow-phase registration phase and Table 6.6 assesses the modified clustering algorithm.

More specifically, in Table 6.5, we present the speed up of our approach in comparison with the one-shot registration (i.e, O. Registration in Table 6.5) (without scene segmentation) applied for different *target* to scene registration examples. In this case, the effectiveness of our approach depends mainly on the size of the *query* point cloud, however for all tested examples the speed up is more than 5x. We would like to note here that the pre-processing step consists of the outliers' removal process and feature extraction.

In Table 6.6, we present the execution times of the original DBSCAN algorithm and our approach for some random scenes of the used dataset (UWAOR). As we can see, our approach is up to 97.7 times faster. Also, we can observe that the effectiveness of our approach is more apparent with increasing number of points. The main reason why our implementation is faster than the original DBSCAN is that our approach does not exhaustively estimate distances between each vertex and all the other vertices of the point cloud, but searches only within a small spatial area consisting of a specific and predefined number of neighbors equal to $K_a$. In this case the time complexity of the algorithm is not $O(n^2)$ but $O(K_a^2)$, where $K_a \ll n$.

### 6.2.8 Performance Evaluation

In the following subsection, we present and evaluate the performance and accuracy of the proposed matching and registration process.

**Evaluation of Matching in Partial Scenes**

For this experiment, we used the 50 partially scanned scenes of the UWAOR dataset (Fig. 4.5) and the five *target* models (i.e., Chef, Chicken, Parasaurodophus, T-rex, Rhino). Fig. 6.37 presents the boxplots of the dissimilarity factor between the *query* and the *target* models. The first five boxplots present the value of the dissimilarity factor when the *target* model is registered through the broad phase in a scene where a partial representation of the same object also exists (intra-class). On the opposite, the last boxplot presents the values of the dissimilarity factor when the *target* model is globally registered in a scene which does not include a partial representation of the same object (inter-class), like in the case of the rhino model as shown in Fig. 6.35. As we can see, the dissimilarity factor of the first five cases is smaller than 0.5. This means that the global registration provides a correct initial estimate, regardless of the noise, amount of cluttering, and type of object. On the other hand, as we expected, the dissimilarity factor between *query* and *target* models is very high, when they represent different objects.

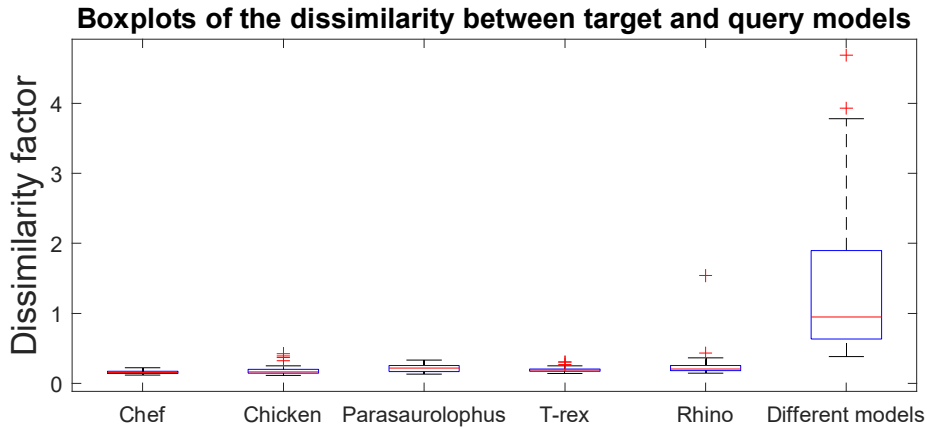

Figure 6.37: Range of the dissimilarity factor calculated between each target and query (partial) model when they represent the same object (intra-class) or different objects (inter-class). The first 5 boxplots show intra-class comparisons for every individual object, whereas the last boxplot summarizes the values of all inter-class comparisons.

Fig. 6.38-[Right] depicts how the occlusion of a model (in percentage %) affects the dissimilarity factor (in case of intra-class matching). In other words, this figure shows that the value of the dissimilarity factor depends on the percentage of the occlusion of the query object. The bigger the occlusion, the higher the possible values of the dissimilarity factor, as observed from the slight shift of the distribution towards higher values. The colors of the heatmap represent the number of the instances per occlusion and dissimilarity factor as presented in the corresponding axes. Fig. 6.38-[Left] presents a histogram showing the number of instances that in certain range of the dissimilarity factor. The distribution shows that most of the cases have a dissimilarity factor between $[0.1 - 0.3]$. Each bar of the histogram is the total sum of each row of the right figure.

Moreover, we calculated the Precision-Recall (PR) curve which is one of the most common indicators used in the literature for the evaluation of a descriptor or algorithm in retrieval tasks [275]. Precision denotes the number of correct

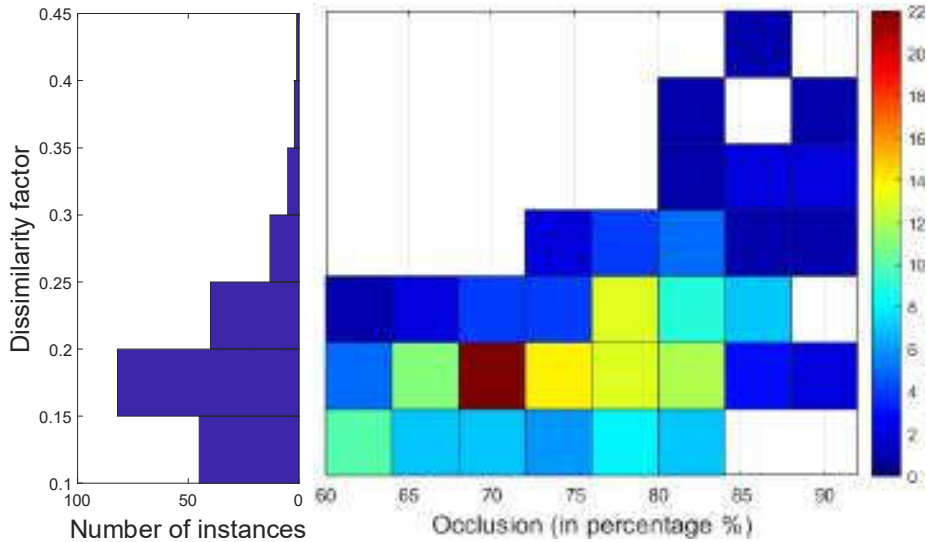

Figure 6.38: Effect of occlusion on dissimilarity factor. [Left] Marginal distribution of dissimilarity factor. [Right] Joint distribution of dissimilarity factor and amount of occlusion in the case of intra-class object matching.

matches to the total number of matches. Recall denotes the number of correct matches to the total number of possible correct matches. In Fig. 6.39, the PR curve has been created by changing the threshold of the dissimilarity factor (which defines the similarity between two models) in a range of $[0.05 - 1]$ with a step of 0.05. Our method provides a binary decision (if the *query* and

Table 6.8: Results of matching the five target models to the 50 partial scan scenes [UWAOR].

| Name of model | Chef | Chicken | Parasaur. | T-rex | Rhino |
|---|---|---|---|---|---|
| TP | 50/(50) | 47/(48) | 45/(45) | 45/(45) | 28/(29) |
| TN | 0/(0) | 2/(2) | 4/(5) | 5/(5) | 21/(21) |
| FP | 0 | 0 | 1 | 0 | 0 |
| FN | 0 | 1 | 0 | 0 | 1 |

*target* models represent the same object or not), based on the value of the predefined threshold. The broad-phase registration only identifies the area of the point cloud scene in which the query model may lie, but the value of the dissimilarity factor is what ultimately defines if there is an actual match. Table 6.8 presents the results of our matching process. The numbers in parenthesis present the correct (existing or non-existing) matches which we aim to identify. For example, the "Chicken" model appears in 48 of the 50 scenes and the method has found 47 true positives (TP) and 2 true negatives (TN) cases. As can be seen, the "Chef" and the "T-rex" models are correctly matched in all of the 50 and 45 scenes, in which they correspondingly appear. Also, all models except the "Parasaurolophus" (having one false positive (FP)) were correctly identified as missing in all scenes in which they do not appear. Statistical measures of performance are shown in Table 6.9 for each model separately as well as averaged across all models. Finally, in Fig. 6.40, we present the recognition rate, i.e.

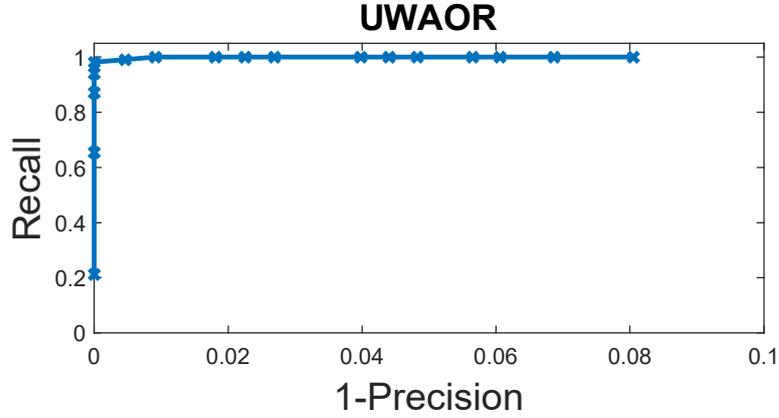

Figure 6.39: PR curve under different threshold of the dissimilarity factor.

Table 6.9: Statistical measures for the evaluation of our method.

|          | Spec. | Prec. | Recall | Acc.  | FPR   | F1    |
|----------|-------|-------|--------|-------|-------|-------|
| Chef     | -     | 1     | 1      | 1     | -     | 1     |
| Chicken  | 1     | 1     | 0.979  | 0.980 | 0     | 0.989 |
| Parasaur.| 0.800 | 0.978 | 1      | 0.98  | 0.200 | 0.989 |
| Tyra     | 1     | 1     | 1      | 1     | 0     | 1     |
| Rhino    | 1     | 1     | 0.965  | 0.980 | 0     | 0.982 |
| **Total**| **0.969** | **0.995** | **0.991** | **0.988** | **0.030** | **0.993** |

the fraction of TP over the total number of correct matches, of our method in comparison with other well-known methods of the literature for different occlusion rates. The recognition rate of our method is less than 100% only when the occlusion is higher than $> 85\%$.

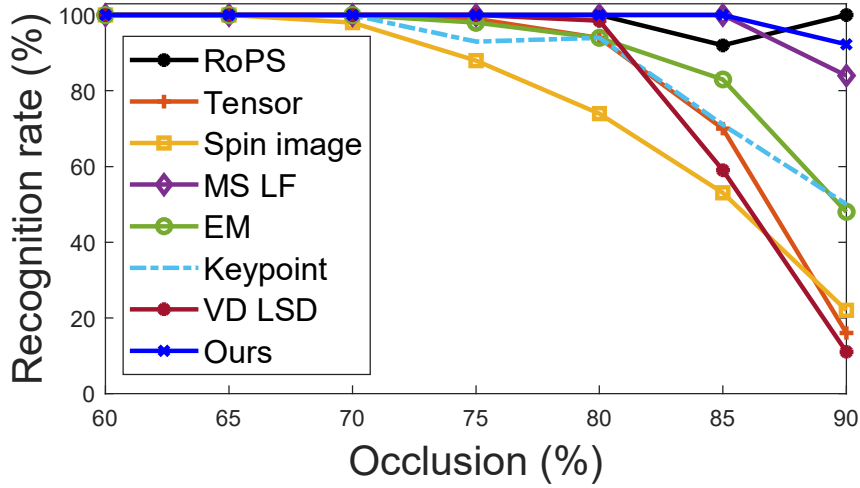

Figure 6.40: Recognition rates per different percentage of occlusion, in comparison with other methods, namely tensor [35], spin image [36], keypoint [37], VD-LSD [38], EM based [39], RoPS [40] and MS_LF [41].

**Evaluation of Matching with Noisy Query Models**

In this experimental case, we compare the *target* models with *query* objects that represent a partial scan of the *target* model, under different levels of noise. For the creation of the noisy models, we added, to each *query* model, different levels of Gaussian noise with intensity $\sigma_E = \{0.1, 0.3, 0.5\}$ to each of the original representations [12]. An example of a noisy scene with different levels of Gaussian noise is presented in Fig. 6.48.

Table 6.10: True positive and false negative matches using the original and the noisy datasets [UWA3M].

|  | Chef | Chicken | Parasaur. | T-rex |
|---|---|---|---|---|
| Original | TP = 22<br>FN = 0 | TP = 16<br>FN = 0 | TP = 16<br>FN = 0 | TP = 21<br>FN = 0 |
| Noise $\sigma_E = 0.1$ | TP = 22<br>FN = 0 | TP = 16<br>FN = 0 | TP = 16<br>FN = 0 | TP = 21<br>FN = 0 |
| Noise $\sigma_E = 0.3$ | TP = 21<br>FN = 1 | TP = 15<br>FN = 1 | TP = 15<br>FN = 1 | TP = 16<br>FN = 5 |
| Noise $\sigma_E = 0.5$ | TP = 20<br>FN = 2 | TP = 10<br>FN = 6 | TP = 13<br>FN = 3 | TP = 13<br>FN = 8 |

In Table 6.10, we present the TP and the false negative (FN) matches for each of the different models of the dataset. The TP matches are 100% for the original dataset and under the presence of noise with $\sigma_E = 0.1$. When the level of noise starts to increase ($> 0.1$), the first false negatives start to appear. Similar conclusions can be observed in Fig. 6.43. The recognition rate decreases as the level of noise increases, but nevertheless does not drop a lot, as even for a high amount of noise (i.e., $\sigma_E = 0.5$), the mean recognition rate remains high (75%).

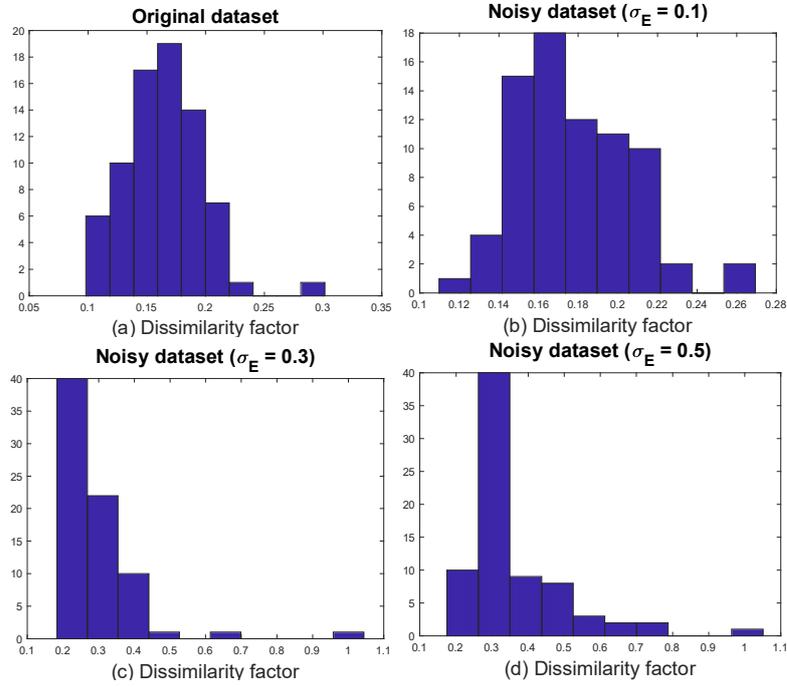

Figure 6.41: Effect of noise ($\sigma_E$) on dissimilarity factor. (a) original dataset, (b) $\sigma_E = 0.1$, (c) $\sigma_E = 0.3$, (d) $\sigma_E = 0.5$.

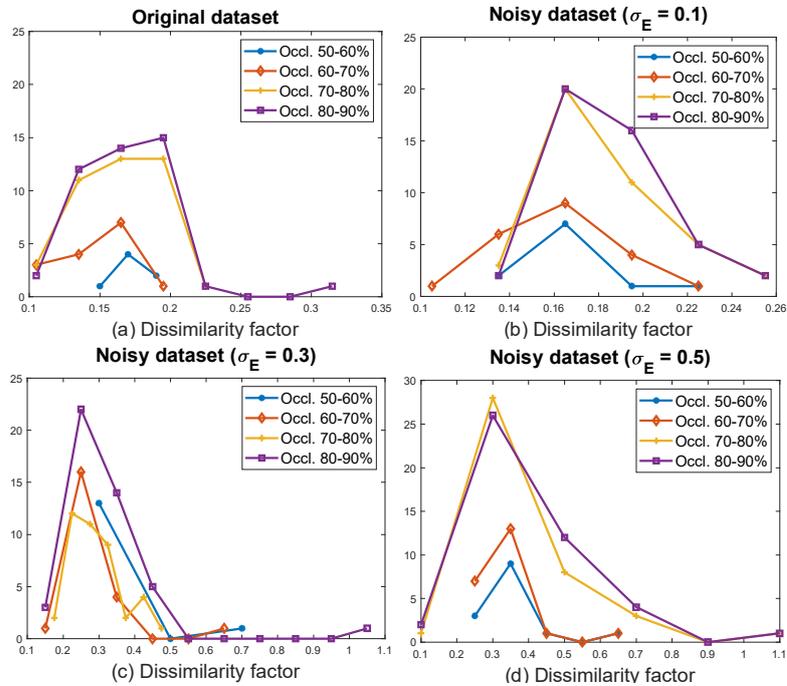

Figure 6.42: Distribution of dissimilarity factor between target and query models for different occlusion rate of the query model and noise levels ($\sigma_E$) affecting the dataset. (a) original dataset, (b) $\sigma_E = 0.1$, (c) $\sigma_E = 0.3$, (d) $\sigma_E = 0.5$.

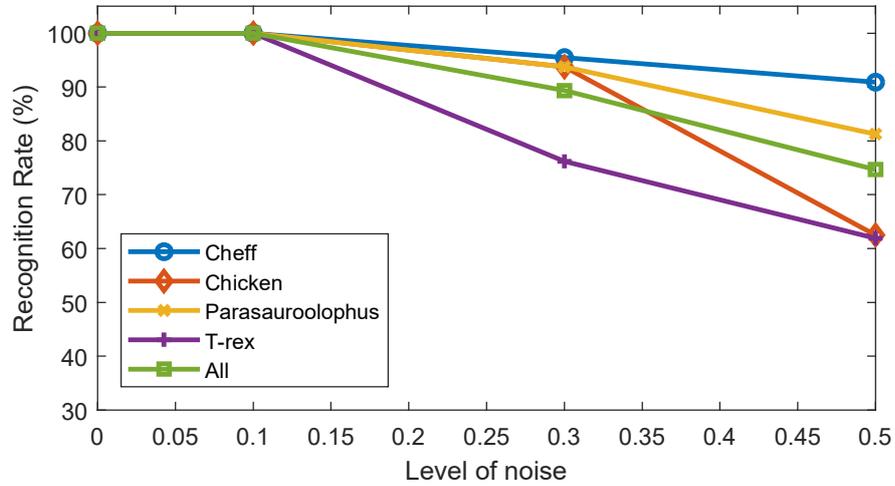

Figure 6.43: Recognition rates per different levels of noise for the each model separately and in total.

Fig. 6.41 shows the histograms representing the number of intra-classes comparisons that have a certain value of dissimilarity (x-axis). As we can observe, in the original dataset and in the dataset that has been affected by $\sigma_E = 0.1$ noise, the dissimilarity factors are less than 0.4. As the amount of noise increases, the dissimilarity factor also increases, affecting the accuracy of object identification. Additionally, in Fig. 6.42, we present the dissimilarity factor for different percentages of occlusion for the different noisy datasets. As expected, the dissimilarity factor depends on the percentage of the occlusion of the query object, as well as on the level of noise.

**Evaluation of Segmentation and Registration in Noisy and Low Quality Scenes**

Besides the fact that the UWAOR dataset consists of models that have been affected by real noise due to the limitations of the scanner device, we further investigate more challenging situations of noisy and low quality scenes. In Fig. 6.46, we present how the clustering method works under different levels of Gaussian noise (0.1 - 0.5) and different levels of visual quality (10% - 50% simplification).

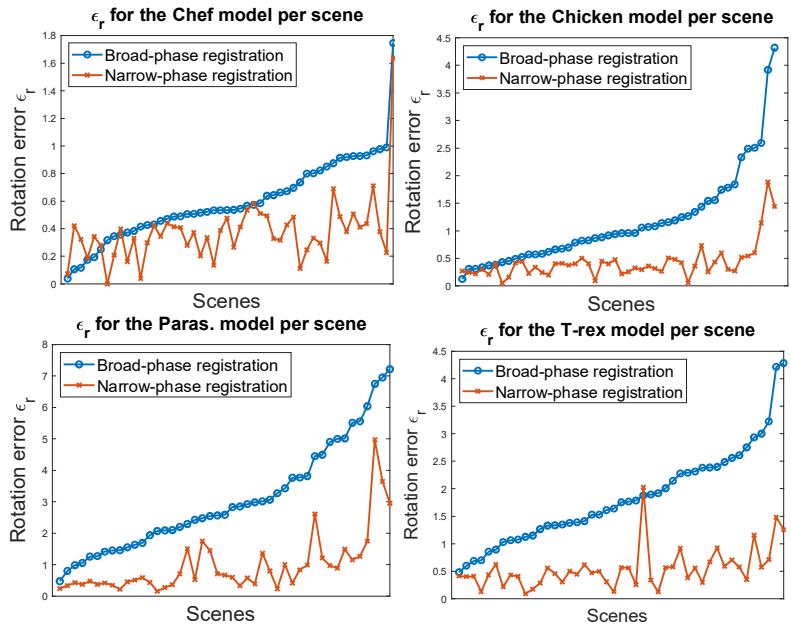

Figure 6.44: Rotation error $\epsilon_r$ for the broad-phase and the narrow-phase registration for each of the 50 scenes. The scenes are sorted by increasing value of broad-phase registration error.

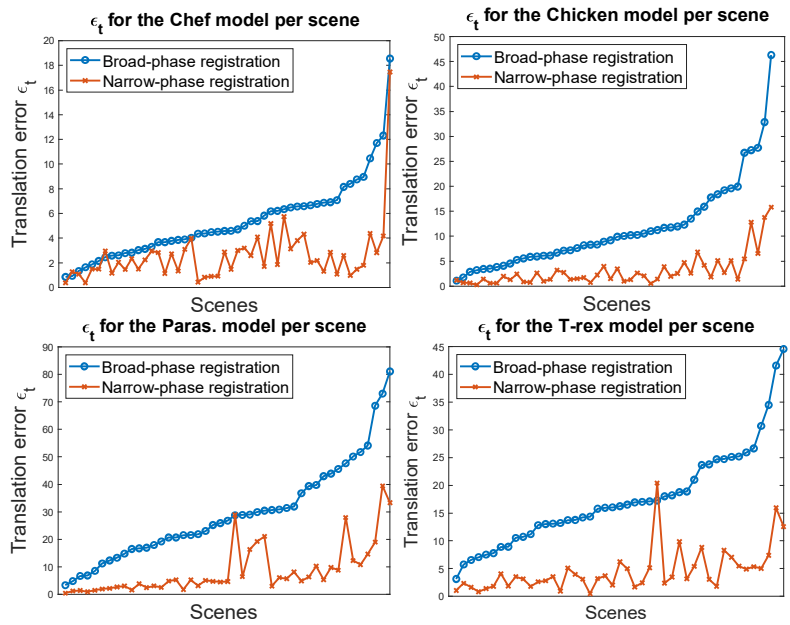

Figure 6.45: Translation error $\epsilon_t$ for the broad-phase and the narrow-phase registration for each of the 50 scenes. The scenes are sorted by increasing value of broad-phase registration error.

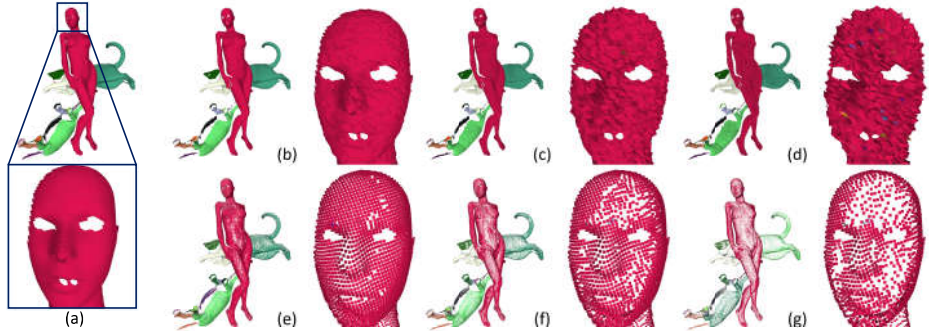

Figure 6.46: (a) Original scene clustering and clustering under (b) $\sigma_E = 0.1$, (c) $\sigma_E = 0.3$, (d) $\sigma_E = 0.5$ Gaussian noise, and (e) 10%, (f) 30%, (g) 50% of total points simplification.

Finally, in Fig. 6.50, we present the rotation $\epsilon_r$ and translation $\epsilon_t$ error for each model presented in the 50 scenes of the UWAOR dataset and in the 150 scenes of the Rodolà et al. dataset [321], for different competing methods. In Figs. 6.48, 6.49, we present some experimental results that show how the performance of the broad-to-narrow registration is affected by different levels of noise and different resolution. More specifically, we applied different levels of Gaussian noise [$\sigma_E = 0.1, 0.3, 0.5$] to the models, and then we followed the same steps as in the original implementation of our approach. Additionally, in Figs. 6.48, 6.49, we present how the low resolution quality of a scene could also affect the performance of the alignment. In this experiment, we downsampled the scene about 10%, 30%, and 50% of the original points (i.e., sim. 10%, sim. 30%, sim. 50%). The experiments show that the alignment of noisy or low-resolution models is deteriorated, as was anticipated. However, the figures and the results show that the performance of the whole proposed pipeline is not significantly affected (in a way to hinder correct matching and registration).

**Evaluation of the Narrow-Phase Registration**

Besides the evaluation of the matching process, which results to object identification, we also evaluated the robustness of the registration process that allows the high-quality 3D model to accurately replace the partially scanned model. Even if object identification is always correct, accurate model-to-object registration is not granted.

Figs. 6.44-6.45 present the rotation ($\epsilon_r$) and translation ($\epsilon_t$) errors after the broad-phase and the narrow-phase registration of the models in each scene. In the majority of the cases, the narrow-phase registration reduces significantly the error of broad-phase registration. Additionally, the broad-phase registration,

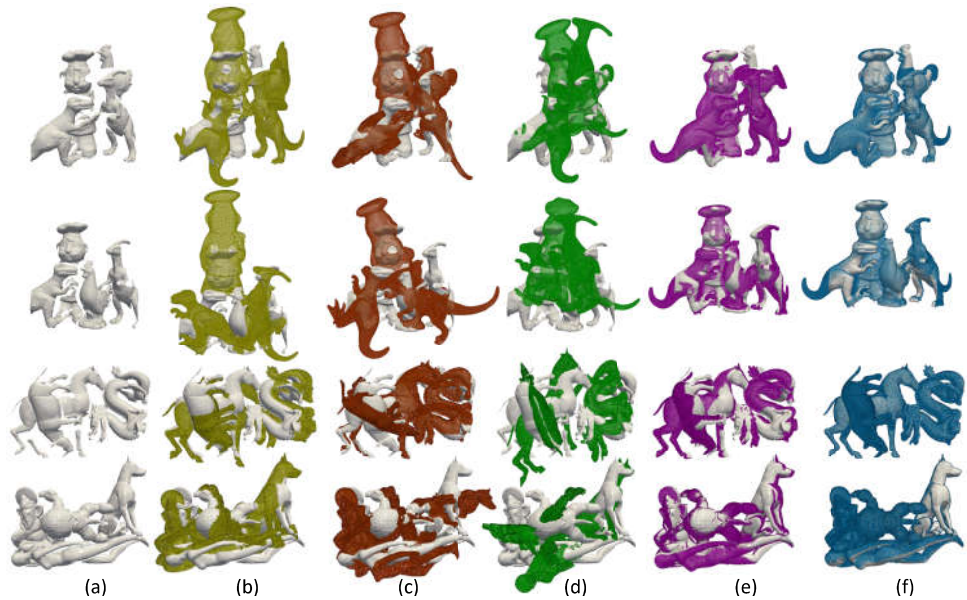

Figure 6.47: Visualization of registration results of different methods for 4 random partial scenes. (a) partial scenes consisting of different models in arbitrary positions, registration results using: (b) Robust low-overlap 3-D point cloud registration approach [42], (c) Discriminative Optimization approach [43], (d) Density Adaptive Point Set Registration [44], (e) the Super 4PCS approach [45], (f) the proposed method.

used as an initialization step, affects the results of the narrow-phase registration, however the upper bound of the latter remains limited even for inaccurate initializations.

In Fig. 6.51 we present the results of the broad-phase and the narrow-phase registration for the "Chef" model of the $7^{th}$ scene and the "Chicken" model of the $49^{th}$ scene. We also provide enlarged details for an easier comparison between the approaches and heatmaps that visualize the mean squared error between the position of the registered and the original models.

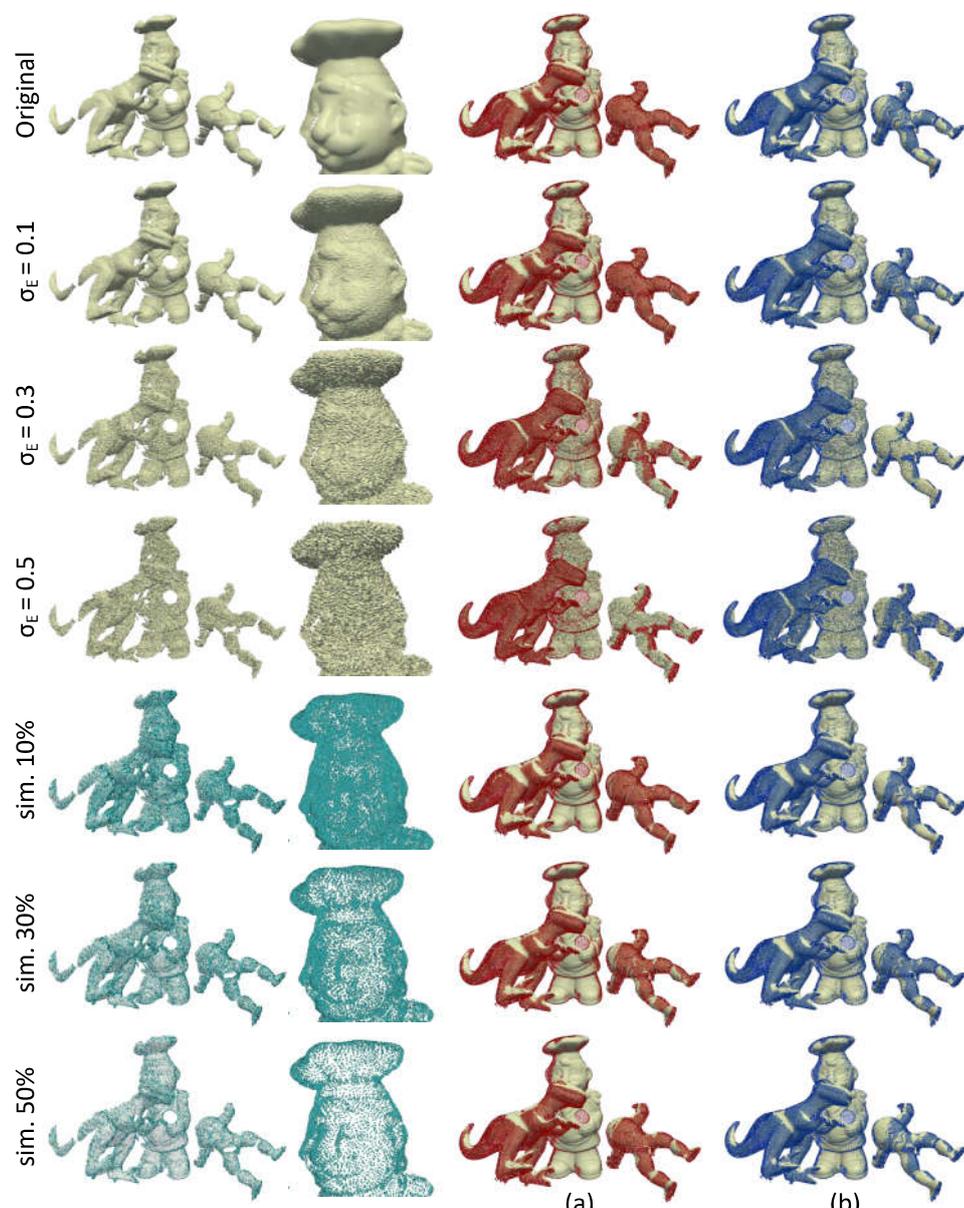

Figure 6.48: (a) broad-phase registration, (b) narrow-phase registration, under different conditions.

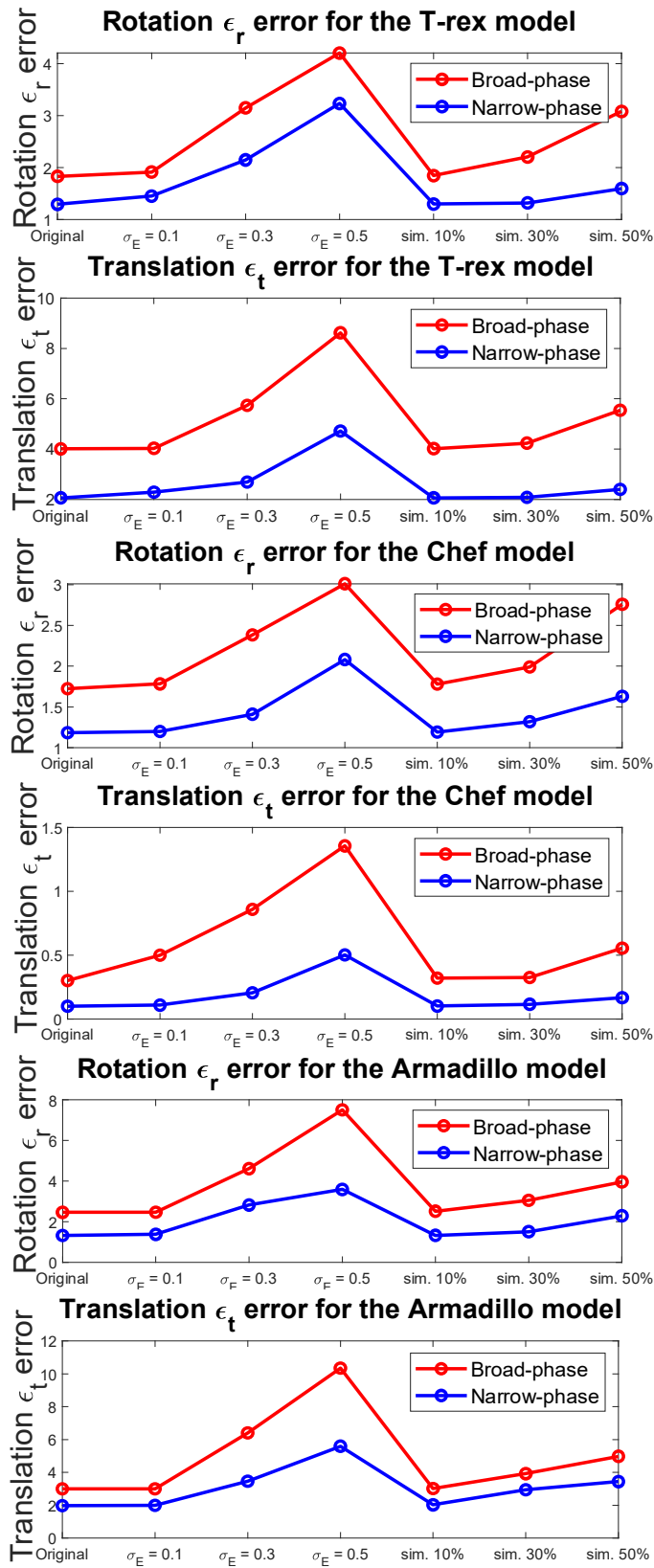

Figure 6.49: Rotation and translation error for each model of the scene.

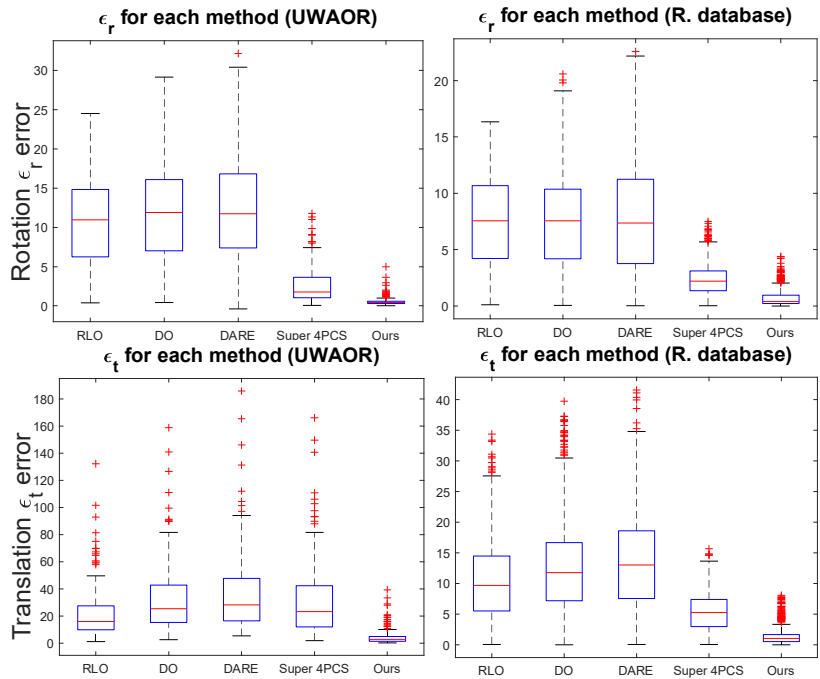

Figure 6.50: Rotation $\epsilon_r$ and translation $\epsilon_t$ error for each method applied to the two datasets (i.e, UWAOR and Rodolà's dataset).

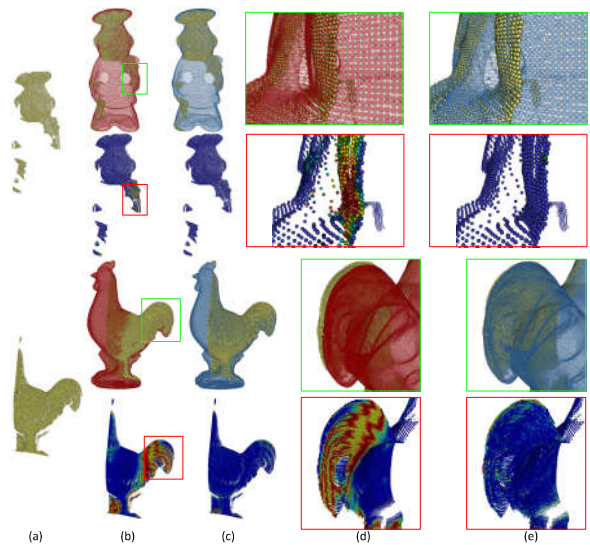

Figure 6.51: (a) Partially scanned segments of a scene (b) broad-phase registration and heatmap visualization of mean squared error, (c) narrow-phase registration and heatmap visualization of mean squared error, (d) enlarge details of broad-phase registration, (e) enlarge details of narrow-phase registration.

Many methods obtain high registration accuracy when applied to full objects, however, they achieve a much lower accuracy when used with partial objects [382]. Many approaches provide good results only if there is a big overlap between the input models. Other approaches have some constraints, like equally sized models with respect to their number of vertices. In Fig. 6.47, we present the registration results of different state-of-the-art methods as applied in some partially-scanned scenes of the dataset. More specifically, the used methods for the registration comparisons are: (i) Robust low-overlap 3-D point cloud registration for outlier rejection (RLO) [42], (ii) Discriminative Optimization: Theory and Applications to Point Cloud Registration (DO) [43], (iii) Density Adaptive Point Set Registration (DARE) [44] and (iv) Super 4PCS Fast Global Point cloud Registration via Smart Indexing [45]. To note here that for a fair comparison with all aforementioned approaches, we firstly applied the model-to-object matching results obtained by our method, and then used them for the narrow-phase registration step. The experimental process shows that our robustified weighted ICP-based method provides the most accurate registration for all the models of the partial scenes.

> **Publications that have contributed to this section:**
>
> 1. Content of the section 6.2 has been presented in **J10**

CHAPTER 7

# Conclusion

In the previous chapters, we presented and discussed some common geometrical-related problems and challenges, covering a large area of different 3D geometric processing applications. Then, we presented in detail our contribution (i.e., assumptions and general introduction, the workflow of the methods, detailed description of the processes, experimental analysis and results) regarding the implementation of suggested solutions in order to overcome these situations. More specifically, the scientific areas that we have been involved in this thesis are: (i) the completion and reconstruction of incomplete static and dynamic 3D models, (ii) the outliers removal and consolidation of static and dynamic 3D meshes, (iii) the denoising of static and dynamic 3D meshes, (iv) the compression of static and dynamic 3D meshes, (v) the feature extraction from 3D objects, (vi) the saliency mapping estimation and (vii) the registration and identification of partially-observed 3D objects in cluttered scenes.

To mention also here that for each area, we provided several approaches taking into our consideration different aspects (e.g., structure type of the model, real-time restriction, etc.) of the problem that we intend to solve. Additionally, we compared the results of our research with a variety of existing state-of-the-art methods of each specific scientific field, showing how the proposed approaches outperform the results of other methods, taking also into account their implementation in realistic scenarios.

The above work presents our approaches to some of the most important tasks in 3D geometry processing. Many successful applications are presented with still a lot more left to be discovered. Limited to the scope of this dissertation, this chapter will make a conclusion and leave some openings for future work. Future directions would be the use of Divide-and-Conquer Matrix Factorization techniques.

## 7.1 Publications

**Publications in journals related to this Thesis:**

J1  G. Arvanitis, A. S. Lalos, K. Moustakas and N. Fakotakis, "Feature Preserving Mesh Denoising Based on Graph Spectral Processing," in IEEE Transactions on Visualization and Computer Graphics, vol. 25, no. 3, pp. 1513-1527, 1 March 2019, doi: 10.1109/TVCG.2018.2802926 [12].

J2  G. Arvanitis, A. S. Lalos, K. Moustakas, "Denoising of dynamic 3D meshes via low-rank spectral analysis", Computers and Graphics, Volume 82, 2019, Pages 140-151, ISSN 0097-8493 [383].

J3  S. Nousias, G. Arvanitis, A. S. Lalos and K. Moustakas, "Fast Mesh Denoising With Data Driven Normal Filtering Using Deep Variational Autoencoders," in IEEE Transactions on Industrial Informatics, vol. 17, no. 2, pp. 980-990, Feb. 2021, doi: 10.1109/TII.2020.3000491 [384].

J4  A. S. Lalos, E. Vlachos, G. Arvanitis, K. Moustakas and K. Berberidis, "Signal Processing on Static and Dynamic 3D Meshes: Sparse Representations and Applications," in IEEE Access, vol. 7, pp. 15779-15803, 2019, doi: 10.1109/ACCESS.2019.2894533 [305].

J5  G. Arvanitis, A. S. Lalos, K. Moustakas, "Spectral Processing for Denoising and Compression of 3D Meshes Using Dynamic Orthogonal Iterations". Journal of Imaging. 2020; 6(6):55. https://doi.org/10.3390/jimaging6060055 [385].

J6  G. Arvanitis, A. S. Lalos, K. Moustakas, "Adaptive representation of dynamic 3D meshes for low-latency applications", Computer Aided Geometric Design, Volume 73, 2019, Pages 70-85, ISSN 0167-8396, https://doi.org/10.1016/j.cagd.2 [386].

J7  G. Arvanitis, A. S. Lalos and K. Moustakas, "Robust and Fast 3-D Saliency Mapping for Industrial Modeling Applications," in IEEE Transactions on Industrial Informatics, vol. 17, no. 2, pp. 1307-1317, Feb. 2021, doi: 10.1109/TII.2020.3003455 [387].

J8  S. Nousias et al., "A Saliency Aware CNN-Based 3D Model Simplification and Compression Framework for Remote Inspection of Heritage Sites," in IEEE Access, vol. 8, pp. 169982-170001, 2020, doi: 10.1109/ACCESS.2020.3023167 [388].

J9  S. Nousias, G. Arvanitis, A. S. Lalos and K. Moustakas, "Deep Saliency Mapping for 3D Meshes and Applications," will be submitted in ACM Transactions on Multimedia Computing, Communications, and Applications.

J10 G. Arvanitis, E. Zacharaki, L. Vasa and K. Moustakas, "Broad-to-Narrow Registration and Identification of 3D Objects in Partially Scanned and Cluttered Point Clouds," in IEEE Transactions on Multimedia, doi: 10.1109/TMM.2021.3089838 [389].

**Publications in conferences related to this Thesis:**

C1 G. Arvanitis, A. Lalos and K. Moustakas, "Feature-Aware and Content-wise Denoising of 3D Static and Dynamic Meshes using Deep Autoencoders," 2019 IEEE International Conference on Multimedia and Expo (ICME), Shanghai, China, 2019, pp. 97-102, doi: 10.1109/ICME.2019.00025. [19].

C2 G. Arvanitis, A. S. Lalos and K. Moustakas, "Image-Based 3D MESH Denoising Through A Block Matching 3D Convolutional Neural Network Filtering Approach," 2020 IEEE International Conference on Multimedia and Expo (ICME), London, UK, 2020, pp. 1-6, doi: 10.1109/ICME46284.2020.9102938 [390].

C3 S. Nousias, G. Arvanitis, A. S. Lalos and K. Moustakas, "Fast mesh denoising with data driven normal filtering using deep autoencoders," 2019 IEEE 17th International Conference on Industrial Informatics (INDIN), Helsinki, Finland, 2019, pp. 260-263, doi: 10.1109/INDIN41052.2019.8972221 [391].

C4 G. Arvanitis, A. S. Lalos and K. Moustakas, "Fast Spatio-temporal Compression of Dynamic 3D Meshes" it has been submitted to IEEE MMSP 2021

C5 A. S. Lalos, G. Arvanitis, A. Spathis-Papadiotis and K. Moustakas, "Feature Aware 3D Mesh Compression Using Robust Principal Component Analysis," 2018 IEEE International Conference on Multimedia and Expo (ICME), San Diego, CA, USA, 2018, pp. 1-6, doi: 10.1109/ICME.2018.8486541 [332].

C6 A. Lalos, G. Arvanitis, E. Vlachos and K. Moustakas, "Energy Efficient Transmission of 3D Meshes Over MMWave-Based Massive MIMO Systems," 2019 IEEE International Conference on Multimedia and Expo (ICME), Shanghai, China, 2019, pp. 1714-1719, doi: 10.1109/ICME.2019.00295 [392].

C7 A. Lalos, G. Arvanitis, A. Dimas, K. Moustakas, "Block based Spectral Processing of Dense 3D Meshes using Orthogonal Iterations". In Proceedings of the 13th International Joint Conference on Computer Vision, Imaging and Computer Graphics Theory and Applications - Volume 1: GRAPP, ISBN 978-989-758-287-5, pages 122-132. DOI: 10.5220/0006611401220132 [393].

C16  I. Romanelis, G. Arvanitis, K. Moustakas, "Fast Feature Curve Extraction for Similarity Estimation of 3D Meshes". Eurographics Workshop on 3D Object Retrieval, 2020 [396].

APPENDIX A

# Fast Spectral Smoothing Using Orthogonal Iterations

The Orthogonal Iteration is an iterative procedure for evaluating the singular vectors corresponding to the $m$ dominant singular values of a symmetric, non-negative definite matrix, at a lower cost than the direct evaluation of SVD. A common way to reduce the number of iterations required for convergence is to replace $\mathbf{R}$ in Eq.(A.1) with $\mathbf{R}^\beta[i]$. We can directly apply Lemma 1 to matrix $\mathbf{R}[i]$.

*Lemma* 1. Consider a symmetric, positive definite matrix $\mathbf{R}$ of size $n$ where $\lambda_1 \geq \ldots \geq \lambda_c \geq \ldots \lambda_n > 0$ are its singular values and $\mathbf{u}_1, \ldots, \mathbf{u}_n$ the corresponding singular vectors. Consider the sequence of matrices $\{\mathbf{U_m}(i)\}$ of dimensions $n \times g$, defined by the iteration:

$$\mathbf{U}_m[i] = Orthonorm\{\mathbf{R}\mathbf{U}_m[i-1]\}, i = 1, 2, \ldots \quad (A.1)$$

where *Orthonorm* stands for orthonormalization. Then,

$$\lim_{t \to \infty} \mathbf{U}_m(t) = [u_1, \ldots, u_m] \quad (A.2)$$

provided that the matrix $\mathbf{U}_m^T(0) = [u_1, \ldots, u_m]$ is not singular. In the following, we will adapt the proposed scheme to the problems at hand.

To preserve orthonormality, it is important that the initial subspace $\widetilde{\mathbf{U}}_m[0]$ is orthonormal. For that reason, $\widetilde{\mathbf{U}}_m[0]$ is estimated by applying SVD directly, while the following subspaces $\widetilde{\mathbf{U}}_m[i], i = 2, \ldots, k$ are adjusted using Eqs. (A.1). There are a number of different choices that can be used for the orthonormalization of the estimated subspace in (A.1) that affect both complexity and performance, with the QR decomposition being one of the most widely adopted. The orthonormalization step using QR can be described by:

$$\begin{aligned}\mathbf{R}^\beta[i]\,\widetilde{\mathbf{U}}_m[i] &\Rightarrow \mathbf{Q}_{qr}[i]\,\mathbf{R}_{qr}[i] \\ \widetilde{\mathbf{U}}_m[i] &= \widetilde{\mathbf{Q}}_{qr}[i] = \left[\mathbf{Q}_{qr}[i]_{(:,1)}, \ldots, \mathbf{Q}_{qr}[i]_{(:,m)}\right]\end{aligned} \quad (A.3)$$

where matrix $\tilde{\mathbf{Q}}_{qr}[i]$ is evaluated by applying $c$ sequential Householder (HH) reflections. Therefore, $\tilde{\mathbf{Q}}_{qr}[i]$ is the submatrix that corresponds to the first $c$ columns of:

$$\mathbf{Q}_{qr}[i] = \mathbf{H_1}^T \cdot \mathbf{H_2}^T \cdot \ldots \cdot \mathbf{H_m}^T \tag{A.4}$$

APPENDIX B

# Estimation of the clean projected vertices and the noise variance

The probability density of the observations given $\mathbf{U}_{\mathbf{m}_{ij}}, \hat{\mathbf{v}}_{ij}\ \forall\ i = 1, s\ ,\ j = \{x, y, z\}$ can be factored due to the noise independence assumption to give:

$$p\left(\mathbf{v}_{ij} \mid \mathbf{U}_{\mathbf{m}_{ij}}, \hat{\mathbf{v}}_{ij}\right) = \prod_{n=1}^{k_i} p\left(\mathbf{v}_{ij} \mid (\mathbf{U}_{\mathbf{m}_{ij}})_{(:,n)}, \hat{\mathbf{v}}_{ij}\right)$$

$$= \frac{exp\left(-\frac{1}{2\sigma_{\mathbf{z}_{ij}}} \left\|\mathbf{v}_{ij} - \mathbf{U}_{\mathbf{m}_{ij}}\hat{\mathbf{v}}_{ij}\right\|_2^2\right)}{\left(2\pi\sigma_{\mathbf{z}_{ij}}\right)^{q/2}} \sim \mathcal{N}\left(\mathbf{U}_{\mathbf{m}_{ij}}\hat{\mathbf{v}}_{ij}, \sigma_{\mathbf{z}_{ij}}\mathbf{I}_{k_i}\right) \quad (B.1)$$

The posterior distribution can be computed based on Bayes' rule as:

$$p\left(\hat{\mathbf{v}}_{ij} \mid \mathbf{v}_{ij}, \mathbf{U}_{\mathbf{m}_{ij}}\right) = \frac{p\left(\mathbf{v}_{ij} \mid \mathbf{U}_{\mathbf{m}_{ij}}, \hat{\mathbf{v}}_{ij}\right) p\left(\hat{\mathbf{v}}_{ij}\right)}{p\left(\mathbf{v}_{ij} \mid \mathbf{U}_{\mathbf{m}_{ij}}\right)} \quad (B.2)$$

where the normalizing constant, also known as the marginal likelihood is independent of the graph Fourier coefficients and is given by $p\left(\mathbf{v}_{ij} \mid \mathbf{U}_{\mathbf{m}_{ij}}\right) = \int p\left(\mathbf{v}_{ij} \mid \mathbf{U}_{\mathbf{m}_{ij}}, \hat{\mathbf{v}}_{ij}\right) p\left(\hat{\mathbf{v}}_{ij}\right) d\hat{\mathbf{v}}_{ij}$.

By writing only the terms of the posterior in Eq. (B.2) that depends on $\hat{\mathbf{v}}_{ij}$ and "completing the square" we obtain:

$$p\left(\hat{\mathbf{v}}_{ij} \mid \mathbf{v}_{ij}, \mathbf{U}_{\mathbf{m}_{ij}}\right) \propto exp\left(-\frac{\left(\hat{\mathbf{v}}_{ij} - \mathbf{H}_{ij}\right)^T \left(\sigma_{\mathbf{z}_i}^{-1}\mathbf{U}_{\mathbf{m}_{ij}}^T\mathbf{U}_{\mathbf{m}_{ij}} + \mathbf{\Pi}_{0ij}^{-1}\right)\left(\hat{\mathbf{v}}_{ij} - \mathbf{H}_{ij}\right)}{2}\right) \quad (B.3)$$

where

$$\mathbf{H}_{ij} = \sigma_{z_i}^{-1} \left( \sigma_{z_i}^{-1} \mathbf{U}_{\mathbf{m}_{ij}}^T \mathbf{U}_{\mathbf{m}_{ij}} + \mathbf{\Pi}_{0ij}^{-1} \right)^{-1} \mathbf{U}_{\mathbf{m}_{ij}} \mathbf{v}_{ij} \quad (B.4)$$

Thus, we recognize the form of the posterior distribution as Gaussian, i.e. $p\left(\hat{\mathbf{v}}_{ij} \mid \mathbf{v}_{ij}, \mathbf{U}_{\mathbf{m}_{ij}}\right) \sim \mathcal{N}\left(\frac{1}{\sigma_{z_{ij}}} \mathbf{\Pi}_{ij} \mathbf{U}_{\mathbf{m}_{ij}} \mathbf{v}_{ij}, \mathbf{\Pi}_{ij}\right)$, where

$$\begin{align}
\mathbf{\Pi}_{ij} &= \left( \sigma_{z_i}^{-1} \mathbf{U}_{\mathbf{m}_{ij}}^T \mathbf{U}_{\mathbf{m}_{ij}} + \mathbf{\Pi}_{0ij}^{-1} \right)^{-1} \quad (B.5) \\
&= \mathbf{\Pi}_{0ij} - \mathbf{\Pi}_{0ij} \mathbf{U}_{\mathbf{m}_{ij}}^T \left( \sigma_{z_{ij}} \mathbf{I}_{k_i} + \mathbf{U}_{\mathbf{m}_{ij}} \mathbf{\Pi}_{0ij} \mathbf{U}_{\mathbf{m}_{ij}}^T \right)^{-1} \mathbf{U}_{\mathbf{m}_{ij}} \mathbf{\Pi}_{0ij}. \quad (B.6)
\end{align}$$

So given the parameters $\sigma_{z_{ij}}, \mathbf{\Pi}_{0ij}$ the MAP estimate is the mean of the posterior distribution $p\left(\hat{\mathbf{v}}_{ij} \mid \mathbf{v}_{ij}, \mathbf{U}_{\mathbf{m}_{ij}}\right)$ Eq. (B.4) may be written as:

$$\begin{align}
\hat{\mathbf{v}}_{\mathbf{p}_{ij}} &= \left( \sigma_{z_{ij}} \mathbf{\Pi}_{0ij}^{-1} + \mathbf{U}_{\mathbf{m}_{ij}}^T \mathbf{U}_{\mathbf{m}_{ij}} \right)^{-1} \mathbf{U}_{\mathbf{m}_{ij}}^T \mathbf{v}_{ij} \\
&= \mathbf{\Pi}_{0ij} \mathbf{U}_{\mathbf{m}_{ij}}^T \left( \sigma_{z_{ij}} \mathbf{I}_{k_i} + \mathbf{U}_{\mathbf{m}_{ij}} \mathbf{\Pi}_{0ij} \mathbf{U}_{\mathbf{m}_{ij}}^T \right)^{-1} \mathbf{v}_{ij} \quad (B.7)
\end{align}$$

where the last equation follows the matrix identity $(\mathbf{I} + \mathbf{AB})^{-1} \mathbf{A} \equiv \mathbf{A} (\mathbf{I} + \mathbf{BA})^{-1}$.

The EM algorithm seeks to find the maximum likelihood estimate of the marginal likelihood by iteratively applying the following two steps (E & M step):

**E:** Calculate the expected value of the likelihood function:

$$\begin{align}
Q\left(\boldsymbol{\theta} \mid \boldsymbol{\theta}^{(t)}\right) &= E_{\hat{\mathbf{v}}_{ij} \mid \mathbf{v}_{ij}; \boldsymbol{\theta}^{(t)}} \left[ \log p\left(\mathbf{v}_{ij}, \hat{\mathbf{v}}_{ij}; \boldsymbol{\theta}^{(t)}\right) \right] \\
&+ E_{\hat{\mathbf{v}}_{ij} \mid \mathbf{v}_{ij}; \boldsymbol{\theta}^{(t)}} \left[ \log p\left(\hat{\mathbf{v}}_{ij}; \boldsymbol{\gamma}_{ij}, \boldsymbol{\Sigma}_{ij}\right) \right] \quad (B.8)
\end{align}$$

**M:** Find the parameters $\boldsymbol{\theta} = \left\{ \sigma_{z_{ij}}, \boldsymbol{\gamma}_{ij}, \boldsymbol{\Sigma}_{ij} \right\}$ that maximize

$$\boldsymbol{\theta}^{(t+1)} := \arg\max_{\boldsymbol{\theta}} Q\left(\boldsymbol{\theta} \mid \boldsymbol{\theta}^{(t)}\right) \quad (B.9)$$

To estimate $\sigma_{z_{ij}}$ we simplify the function in Eq. (B.8) by dropping the terms that do not include $\sigma_{z_{ij}}$:

$$Q\left(\sigma_{\mathbf{z}_{ij}} \mid \boldsymbol{\theta}^{(t)}\right) \propto E_{\hat{\mathbf{v}}_{ij}\mid \mathbf{v}_{ij};\boldsymbol{\theta}^{(t)}}\left[\log p\left(\mathbf{v}_{ij} \mid \hat{\mathbf{v}}_{ij}; \sigma_{\mathbf{z}_{ij}}\right)\right]$$
$$= -\frac{k_i}{2}\log \sigma_{\mathbf{z}_{ij}} - \frac{1}{2\sigma_{\mathbf{z}_{ij}}}\left\|\mathbf{v}_{ij} - \mathbf{U}_{\mathbf{m}_{ij}}\mathbf{H}_{ij}\right\|_2^2$$
$$-\frac{1}{2\sigma_{\mathbf{z}_{ij}}}E_{\hat{\mathbf{v}}_{ij}\mid\mathbf{v}_{ij};\boldsymbol{\theta}^{(t)}}\left[\left\|\mathbf{U}_{\mathbf{m}_{ij}}\left(\hat{\mathbf{v}}_{ij}-\mathbf{H}_{ij}\right)\right\|_2^2\right] \quad \text{(B.10)}$$

with

$$E_{\hat{\mathbf{v}}_{ij}\mid\mathbf{v}_{ij};\boldsymbol{\theta}^{(t)}}\left[\left\|\mathbf{U}_{\mathbf{m}_{ij}}\left(\hat{\mathbf{v}}_{ij}-\mathbf{H}_{ij}\right)\right\|_2^2\right] = Tr\left(\mathbf{\Pi}_{ij}\mathbf{U}_{\mathbf{m}_{ij}}^T\mathbf{U}_{\mathbf{m}_{ij}}\right) \quad \text{(B.11)}$$
$$= \sigma_{\mathbf{z}_{ij}}^{(t)} Tr\left(\mathbf{\Pi}_{ij}\left(\mathbf{\Pi}_{ij}^{-1}-\mathbf{\Pi}_{0ij}^{-1}\right)\right) \quad \text{(B.12)}$$
$$= \sigma_{\mathbf{z}_{ij}}^{(t)}\left(\mathbf{m}_{ij} - Tr\left(\mathbf{\Pi}_{ij}\mathbf{\Pi}_{0ij}^{-1}\right)\right) \quad \text{(B.13)}$$

where Eq. (B.12) can be directly derived by substituting Eq. (B.5) in Eq. (B.11). The learning rule for $\sigma_{\mathbf{z}_{ij}}$ is obtained by setting the derivative of $Q\left(\sigma_{\mathbf{z}_{ij}} \mid \boldsymbol{\theta}^{(t)}\right)$ to zero leading to $\sigma_{\mathbf{z}_{ij}}^{(t+1)} = \frac{\left\|\mathbf{v}_{ij}-\mathbf{U}_{\mathbf{m}_{ij}}\mathbf{H}_{ij}\right\|_2^2 + \sigma_{\mathbf{z}_{ij}}^{(t)}\left(\mathbf{m}_{ij}-Tr\left(\mathbf{\Pi}_{ij}\mathbf{\Pi}_{0ij}^{-1}\right)\right)}{k_i}$. To estimate $\gamma_{ij}$ and $\mathbf{\Sigma}_{ij}$ we notice that the first term in Eq. (B.8) is unrelated to the aforementioned parameters and thus it can be simplified to:

$$Q\left(\gamma_{ij},\mathbf{\Sigma}_{ij} \mid \boldsymbol{\theta}^{(t)}\right) \propto E_{\hat{\mathbf{v}}_{ij}\mid\mathbf{v}_{ij};\boldsymbol{\theta}^{(t)}}\left[\log p\left(\hat{\mathbf{v}}_{ij};\gamma_{ij},\mathbf{\Sigma}_{ij}\right)\right] \quad \text{(B.14)}$$

where it can be easily shown that $\log p\left(\hat{\mathbf{v}}_{ij};\gamma_{ij},\mathbf{\Sigma}_{ij}\right) \propto -\frac{\mathbf{m}_{ij}}{2}\log\left|\gamma_{ij}\mathbf{\Sigma}_{ij}\right| - \frac{\hat{\mathbf{v}}_{ij}^T\left(\gamma_{ij}\mathbf{\Sigma}_{ij}\right)^{-1}\hat{\mathbf{v}}_{ij}}{2}$. Therefore Eq. (B.14) may be written as:

$$\begin{aligned}Q\left(\gamma_{ij},\mathbf{\Sigma}_{ij} \mid \boldsymbol{\theta}^{(t)}\right) \propto \;& -\frac{\mathbf{m}_{ij}}{2}\log\left|\gamma_{ij}\right| - \frac{\mathbf{m}_{ij}}{2}\log\left|\mathbf{\Sigma}_{ij}\right| \\ & -\frac{Tr\left(\left(\gamma_{ij}\mathbf{\Sigma}_{ij}\right)^{-1}\left(\mathbf{\Pi}_{ij}+\mathbf{H}_{ij}\mathbf{H}_{ij}^T\right)\right)}{2}\end{aligned} \quad \text{(B.15)}$$

where $\mathbf{H}_{ij}$, $\mathbf{\Pi}_{ij}$ are evaluated according to:

$$\text{(B.16)}$$

$$mathbf{f}H_{ij} = \mathbf{\Pi}_{0ij}\mathbf{U}_{\mathbf{m}_{ij}}^T\left(\mathbf{U}_{\mathbf{m}_{ij}}\mathbf{\Pi}_{0ij}\mathbf{U}_{\mathbf{m}_{ij}}^T + \sigma_{\mathbf{z}_{ij}}\mathbf{I}_{k_i}\right)^{-1}\mathbf{v}_{ij} \quad \text{(B.17)}$$

$$\mathbf{\Pi}_{ij} = \mathbf{\Pi}_{0ij} - \mathbf{\Pi}_{0ij}\mathbf{U}_{\mathbf{m}_{ij}}^T \left(\mathbf{U}_{\mathbf{m}_{ij}}\mathbf{\Pi}_{0ij}\mathbf{U}_{\mathbf{m}_{ij}}^T + \sigma_{\mathbf{z}_{ij}}\mathbf{I}_{k_i}\right)^{-1}\mathbf{U}_{\mathbf{m}_{ij}}\mathbf{\Pi}_{0ij} \qquad (B.18)$$

The derivative of Eq. (B.15) with respect to $\gamma_{ij}$ is given by $\frac{\partial Q(\gamma_{ij},\Sigma_{ij}|\boldsymbol{\theta}^{(t)})}{\partial \gamma_{ij}} = -\frac{\mathbf{m}_{ij}}{2\gamma_{ij}} - \frac{1}{2\gamma_i^2}Tr\left(\Sigma_{ij}^{-1}\left(\mathbf{\Pi}_{ij} + \mathbf{H}_{ij}\mathbf{H}_{ij}^T\right)\right)$, and the learning rule for $\gamma_{ij}$ will be $\gamma_{ij} = \frac{Tr\left(\Sigma_{ij}^{-1}\left(\mathbf{\Pi}_{ij}+\mathbf{H}_{ij}\mathbf{H}_{ij}^T\right)\right)}{\mathbf{m}_{ij}}$.

While the gradient of (B.15) with respect to $\Sigma_{ij}$ will be $\frac{\partial Q(\gamma_{ij},\Sigma_{ij}|\boldsymbol{\theta}^{(t)})}{\partial \Sigma_{ij}} = -\frac{\mathbf{m}_{ij}\Sigma_{ij}^{-1}}{2} + \frac{\Sigma_{ij}^{-1}\left(\mathbf{\Pi}_{ij}+\mathbf{H}_{ij}\mathbf{H}_{ij}^T\right)\Sigma_{ij}^{-1}}{2\gamma_{ij}}$. By setting the derivative to zero we obtain the learning rule for $\Sigma_{ij}$.

Appendix C

# Technical Manual of the Application

In this section, we present and explain every functionality of the application for helping the first time user. Fig. C.1 shows the general view of the application when a user opens it for the first time.

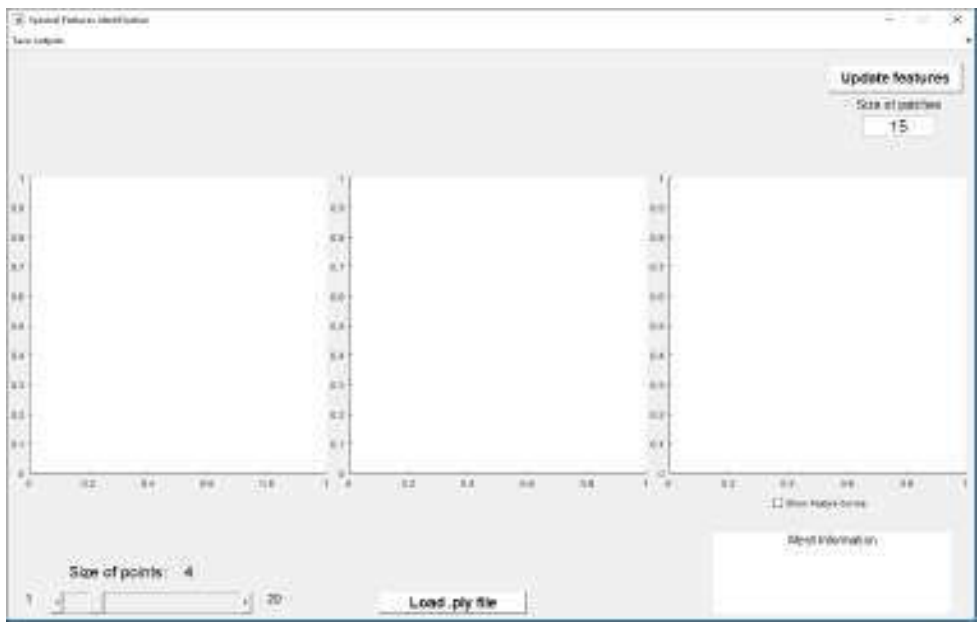

Figure C.1: General view of the application when a user opens it for the first time.

Firstly, the user has to load a .ply file. The "Load .ply file" button (Fig. C.2 highlighted in red) is used for this purpose. When it is pushed, a new window opens for searching the preferable file.

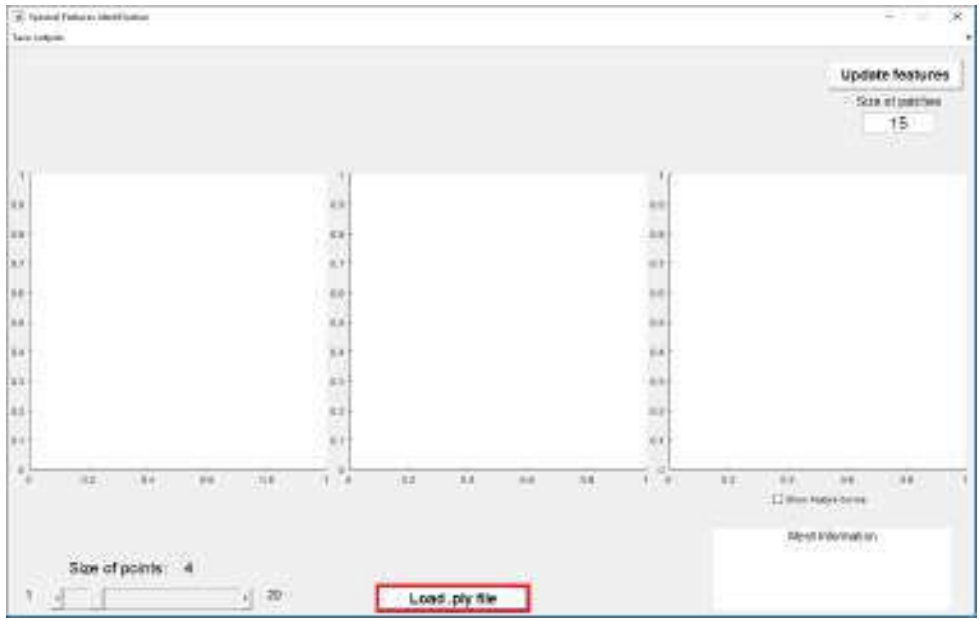

Figure C.2: The "Load .ply file" button for loading a new .ply file.

When a .ply file has been selected, two point clouds are presented in the corresponding axes areas, as presented in Fig. C.3.

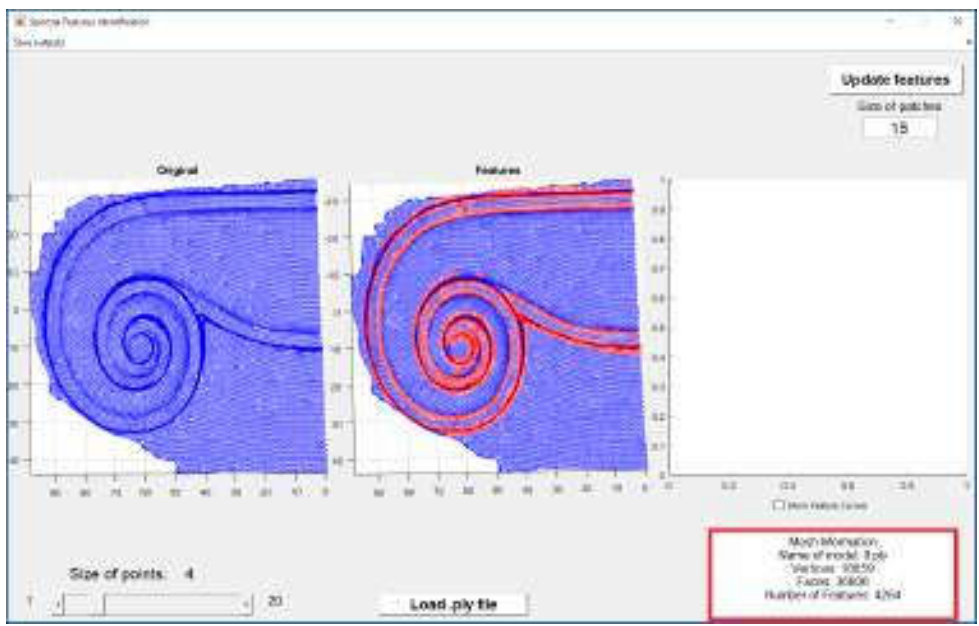

Figure C.3: Displaying on the selected model and features identification.

In the left area, the original mesh, without any feature identification, is presented while in the right area the same 3D model is presented but its features are highlighted in red color. Additionally, the "Mesh Information" box area is filled with information related to the mesh. More specifically:

- The name of the loaded model (e.g., 8.ply)
- The number of vertices that the model has (e.g., 18659)
- The number of faces that the model has (e.g., 36806)
- The number of vertices that have been identified as features (e.g., 4264)

The feature identification is a very fast procedure so it runs automatically when a new model is loaded. On the other hand, the feature curve identification is not so fast, especially for large models, so it is disabled by default. However, the user can easily enable it by checking the "Show Feature Curves" checkbox (Fig. C.4 box highlighted in red color).

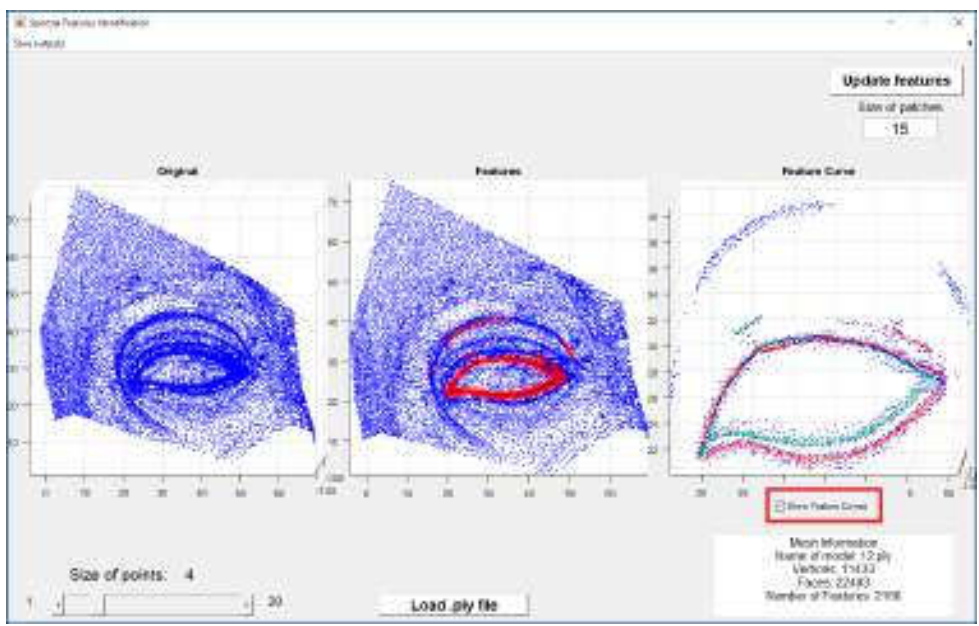

Figure C.4: Enabling the feature curve identification functionality.

To note here that when a model is chosen, the application automatically estimates the features of a model using the default pre-defined value of patches size equal to 15. However, this value is only optional and the user can change it for better results. Specifically, the user can change the number, which indicates the size of patches, using the editable area, as presented in Fig. C.5 (highlighted in red color), and then pressing the "Update features" button. The ideal patch size depends on the geometric characteristics (small or large-scale features) of each model. Instinctively, we could say that the larger the patch size, the more the number of the identified feature vertices. Despite the fact that it is generally true, it is not perfectly correct since larger patch size means the identification of large-scale features (low-spacial frequencies features) while smaller patch size means the identification of small-scale features.

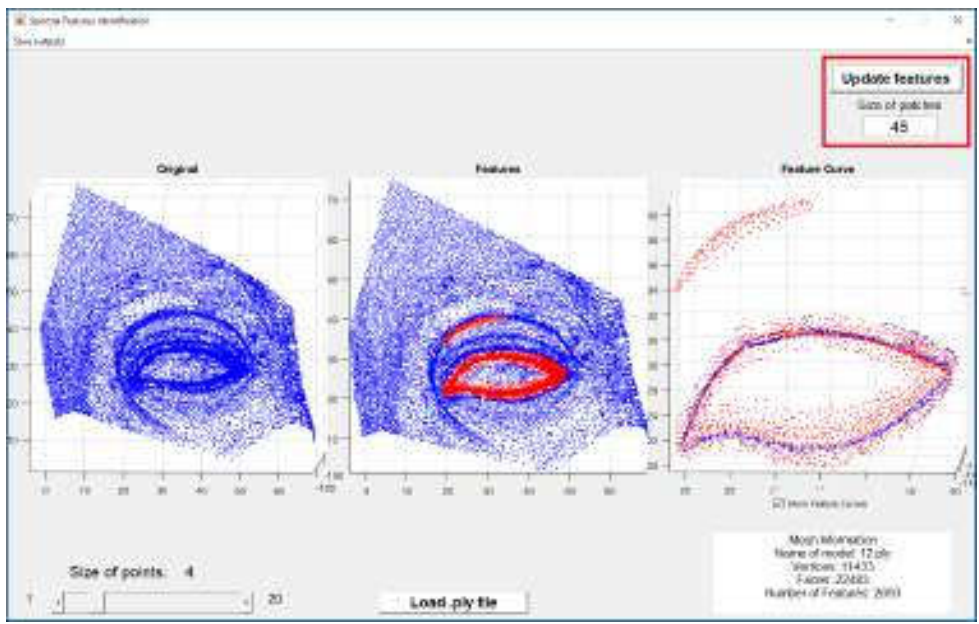

Figure C.5: Features identification of model "12.ply" using a patches size equal to 45.

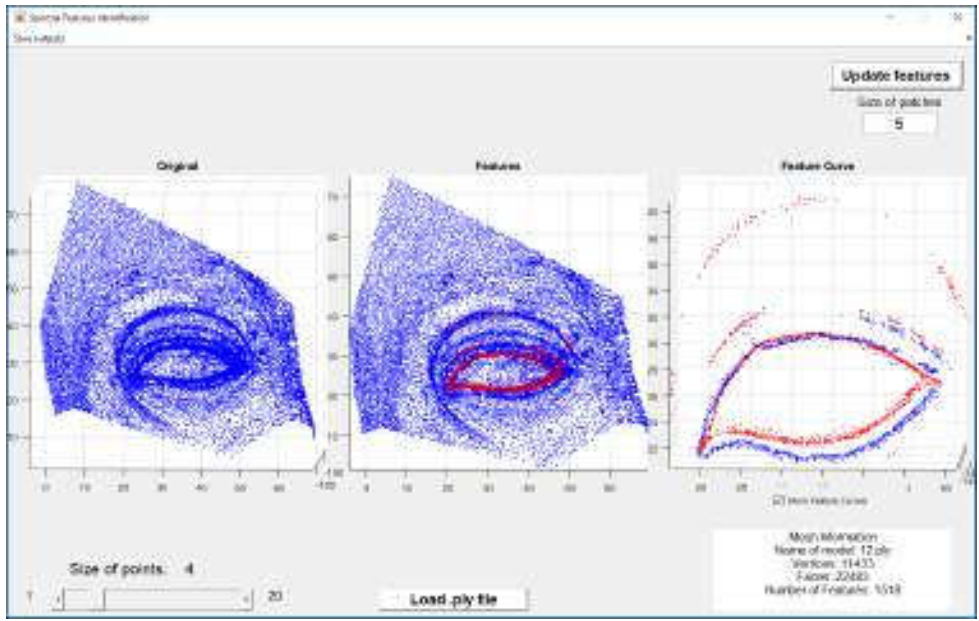

Figure C.6: Features identification of model "12.ply" using a patches size equal to 5.

In Figs. C.5 and C.6, we present two examples in which a larger (e.g., 45) and a smaller (e.g., 5) patch size area is selected for the same model. It is obvious that a different number of features are identified in each case, as also shown in the information box. To note here that in all of our experiments, we use the same patch size, equal to 15, for any model and we do not search for ideal patch size per each model individually. Some models are denser than others. For this reason, we have added a slider which defines the size of the points, as presented in Fig. C.7 (highlighted in red box).

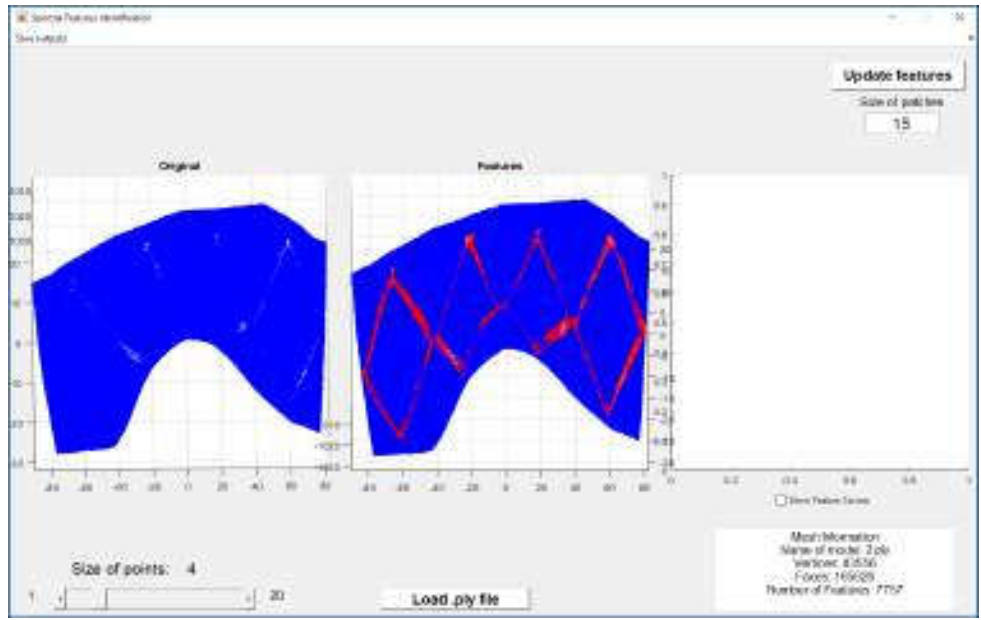

Figure C.7: Displaying point clouds with point size equal to 4.

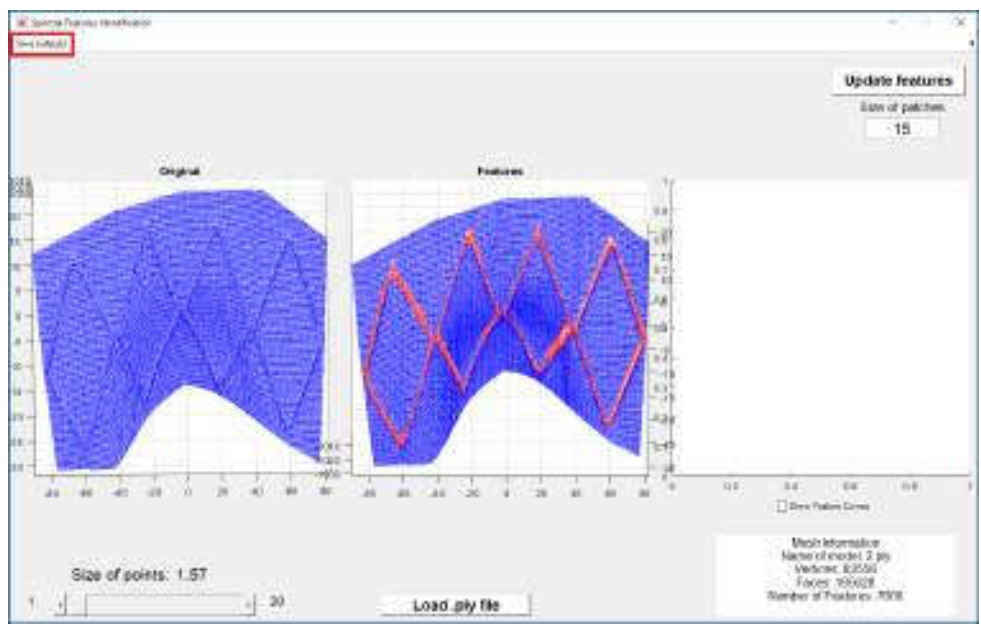

Figure C.8: Displaying point clouds with point size equal to 1.57.

APPENDIX D

# Symbolic System

In the followings, we present the symbols that we use and a brief description of them. More specifically, lower and upper-case letters (i.e., x) represent scalar variables, lower-case bold letters (i.e., **x**) represent vectors, upper-case bold letters (i.e., **X**) represent matrices, and calligraphic letters (i.e., $\mathcal{X}$) represent sets or (i.e., $\mathcal{X}(.)$) functions and operators.

| Lower-case symbol | Brief Description of the Symbol |
|---|---|
| $a,b$ | Auxiliary variables |
| $c$ | Cell of vertices. It represents the first-ring area |
| $\dot{c}$ | Elements of the cell $c$. It represents the index of a vertex that lies in the the cell $c$ |
| $d$ | Elements of **D** matrix |
| $e$ | Scalar variable representing the smoothing factor of an area |
| $g, i, j$ | pointers |
| $h$ | Dissimilarity factor |
| $k$ | Number of neighbors |
| $k_f$ | number of frames per block |
| $k_l$ | number of rows of the spatiotemporal matrix that are preserved |
| $l$ | Layer |
| $m$ | Parts of submeshes |
| $n$ | Number of verices |
| $n_a$ | Number of anchor points |
| $n_b$ | number of blocks |
| $n_c$ | Number of clusters |
| $n_d$ | Number of vertices in equalsized submeshe |
| $n_e$ | Number of edges |
| $n_f$ | Number of faces |
| $n_h$ | Number of low dimensional space (removed components) |
| $n_k$ | Number of known motion vectors |
| $n_l$ | Number of low dimensional space (remained components) |
| $n_m$ | Number of points in each submesh |
| $n_p$ | Number of candidate patches |
| $n_r$ | Number of remaining vertices |
| $n_s$ | Number of meshes |
| $n_t$ | Number of submeshes, Number of eigenvalues |
| $n_v$ | Valence of the first-ring area |
| $o$ | Elements of matrix **O** |
| $p$ | Elements of matrix **P** |
| $q, q0, q1$ | Scalar variables used as parameters of a Toeplitz symmetric structure |
| $r$ | The rank of a matrix, radius |
| $s$ | The current frame of a dynamic sequence |
| $t$ | Number of iterations |
| $w$ | Weights |
| $w_d$ | The inverse distance weight |
| $w_p$ | The prioritization weight |
| $x, y, z$ | Scalar variable representing the euclidean coordinates of a vertex |

Table D.1: Lower-case symbols and their brief description.

| Upper-case symbol | Brief Description of the Symbol |
|---|---|
| $A$ | Area of a triangular face |
| $C$ | Curvature |
| $C_t$ | Temporal removal cost |
| $C_s$ | Spatial removal cost |
| $E_s$ | The total energy |
| $F(.)$ | Kernel function |
| $K$ | The vertices with the highest salient values |
| $P$ | Candidate patches |
| $R$ | Ring areas per point and level |
| $T$ | Transformation for registration |

Table D.2: Upper-case symbols and their brief description.

| Lower and upper-case greek symbol | Brief Description of the Symbol |
|---|---|
| $\alpha$ | Auxiliary variable |
| $\beta$ | A small threshold |
| $\gamma$ | Auxiliary variable |
| $\delta$ | The delta coordinates |
| $\epsilon$ | A small positive scalar |
| $\zeta$ | The power in OI, denoising factor |
| $\eta$ | Auxiliary variable |
| $\theta$ | The metric of angle difference |
| $\lambda$ | Eigenvalues |
| $\mu$ | A small threshold |
| $\xi$ | Auxiliary variable |
| $\rho$ | A positive parameter |
| $\rho(.)$ | A density estimation function |
| $\sigma$ | The standard deviation |
| $\tau, \phi$ | A small threshold |
| $\omega$ | Auxiliary variable |
| $\psi(.)$ | Tukey's bi-weight function |
| $\Psi_1$ | The first-ring area of a vertex |
| $\Psi^k$ | The area of the k-nn of a vertex |

Table D.3: Lower and upper-case greek symbols and their brief description.

| Lower-case bold symbol | Brief Description of the Symbol |
|---|---|
| **a** | Vector of kernels |
| **c** | The vector of centroids |
| **e** | The vectors of elements of a low-rank matrix |
| **f** | The vector of the faces |
| **g** | The vector of the guided normals |
| **k** | vectors identicate in which cluster each vertex belongs |
| **m** | The vector of the motion vectors |
| **n** | The vector of the normals |
| **ṅ** | The vector of the denoised normals |
| $\mathbf{n}_c$ | The vector of the centroid normals |
| **p** | Vector with the pairs of vertices between two point clouds |
| **q** | An augmented auxiliary vector |
| **r** | The mean residual vector |
| **s** | The vector of the elements of a sparse matrix |
| **t** | Translation vector |
| **v** | The vector of the vertices |
| $\mathbf{v}_k$ | The vector of the known vertices |
| $\mathbf{v}_u$ | The vector of the unknown vertices |
| $\mathbf{v}_c$ | The vector of the clear vertices |
| $\hat{\mathbf{v}}$ | The vector of the smoothed vertices |
| $\mathbf{w}$ | The vector of the weights |
| $\mathbf{z}$ | The vector of the noise |

Table D.4: Lower-case bold symbols and their brief description.

| Upper-case bold symbol | Brief Description of the Symbol |
|---|---|
| $\mathbf{A}$ | Auxiliary matrix |
| $\mathbf{B}$ | The matrix of motion vectors |
| $\mathbf{B}_k$ | The matrix of known motion vectors |
| $\mathbf{B}_u$ | The matrix of unknown motion vectors |
| $\mathbf{C}$ | The adjacency matrix, convenience matrix |
| $\mathbf{D}$ | The Diagonal matrix |
| $\mathbf{E}$ | The low-rank matrix |
| $\mathbf{F}$ | An augmented auxiliary matrix |
| $\mathbf{G}$ | The Noisy matrix in RPCA |
| $\mathbf{H}$ | An parameter matrix |
| $\mathbf{I}$ | The identity matrix |
| $\mathbf{L}$ | The Laplacian matrix |
| $\mathbf{L}_w$ | The weighted Laplacian matrix |
| $\mathbf{M}$ | The matrix of observed data for decomposition |
| $\mathbf{M}'$ | A smaller-dimensions matrix of $\mathbf{M}$ |
| $\mathbf{N}$ | The matrix of the normals |
| $\mathbf{O}$ | The connectivity proximity matrix |
| $\mathbf{P}$ | An Auxiliary matrix related to distance |
| $\mathbf{Q}$ | The matrix Q from QR decomposition, Query model |
| $\mathbf{R}$ | The autocorrelated matrix, matrix R from QR, Rotation matrix for registration |
| $\mathbf{S}$ | The Sparse matrix |
| $\mathbf{T}$ | The formulated Laplacian matrix $L + I * \epsilon$, Target model |
| $\mathbf{U}$ | The matrix of the eigenvectors |
| $\mathbf{V}$ | The matrix of the vertices |
| $\mathbf{W}$ | The matrices of weights |
| $\mathbf{Y}$ | An auxiliary matrix |

Table D.5: Upper-case bold symbols and their brief description.

| Lower and upper-case bold greek symbol | Brief Description of the Symbol |
|---|---|
| $\boldsymbol{\mu}$ | The matrix of the mean values |
| $\boldsymbol{\Lambda}$ | The matrix of the eigenvalues |
| $\boldsymbol{\Pi}$ | A parameter auxiliary matrix |
| $\boldsymbol{\Sigma}$ | A parameter auxiliary matrix |

Table D.6: Lower and upper-case bold greek symbols and their brief description.

| Calligraphic symbol | Brief Description of the Symbol |
|---|---|
| $\mathcal{A}$ | The animated mesh |
| $\mathcal{A}'$ | The incomplete 3D animation |
| $\bar{\mathcal{A}}$ | The reconstructed 3D animation |
| $\mathcal{B}$ | The set of the candidate patches P |
| $\mathcal{D}$ | The singular value thresholding operator RPCA |
| $\mathcal{E}$ | The set of the edges |
| $\mathcal{F}$ | The set of the faces |
| $\mathcal{G}$ | A graph |
| $\mathcal{I}(.)$ | The augmented Lagrangian operator |
| $\mathcal{L}$ | The normalized Laplacian matrix |
| $\mathcal{M}$ | A 3D mesh or a point cloud |
| $\tilde{\mathcal{M}}$ | A noisy Mesh |
| $\mathcal{M}'$ | An incomplete 3D mesh |
| $\mathcal{N}(.)$ | The distribution of noise |
| $\mathcal{P}_\omega$ | An operator keeping unchanged the known elements |
| $\mathcal{Q}(.)$ | The shrinkage operator in RPCA |
| $\mathcal{T}(.)$ | The GFT function |
| $\mathcal{V}$ | The set of vertices |
| $\tilde{\mathcal{V}}$ | The known subset of vertices |
| $\check{\mathcal{V}}$ | The unknown subset of vertices |

Table D.7: Calligraphic symbols and their brief description.

APPENDIX E

# Saliency Maps and Execution Times and Results using alternative RPCA Approaches

We exhaustively searched various RPCA versions that could be successfully used in our problem. In the following table E.1, we present results using different approaches for decomposing the matrix $\mathbf{E}$ into a low-rank and a sparse matrix, as well the execution time for the decomposition of an example ("block" model). Finally to visually evaluate the efficiency of the considered methods we provide the heat map visualization of the salient map that each approach provides.

| RPCA method | Decomposition, Minimization, Constraints, Solver | Execution time in (sec) | Visualized Results of salient mapping |
|---|---|---|---|
| (ADM) Alternating Direction Method  Yuan and Yang [397] | $\mathbf{E} = \mathbf{L} + \mathbf{S}$ $\min_{\mathbf{L},\mathbf{S}} \|\mathbf{L}\|_* + \lambda \|\mathbf{S}\|_1$ s.t. $\mathbf{L} + \mathbf{S} = \mathbf{E}$ ADM solver | 480.609 | 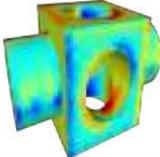 |
| **(ALM)** Augmented Lagrange Multiplier<br><br>Tang and Nehorai [398] | $\mathbf{E} = \mathbf{L} + \mathbf{S}$<br>$\min_{\mathbf{L},\mathbf{S}} \|\mathbf{L}\|_* + \kappa(1-\lambda)\|\mathbf{L}\|_{2,1} + \kappa\lambda\|\mathbf{S}\|_{2,1}$<br>s.t. $\mathbf{L} + \mathbf{S} = \mathbf{E}$<br>ALM solver | 17.724 | 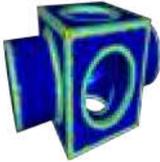 |
| **(APG)** Accelerated Proximal Gradient<br><br>Lin et al. [399] | $\mathbf{E} = \mathbf{L} + \mathbf{S}$<br>$\min_{\mathbf{L},\mathbf{S}} \|\mathbf{S}\|_*$<br>s.t. $rank(L) < r$<br>APG solver | 2.741 | 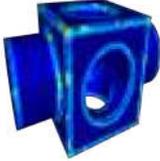 |
| **(AS-RPCA)** Active Subspace: Towards Scalable Low-Rank Learning<br><br>Liu and Yan [400] | $\mathbf{E} = \mathbf{L} + \mathbf{S}$<br>$\min_{\mathbf{L},\mathbf{S}} \|\mathbf{L}\|_* + \lambda\|\mathbf{S}\|_1$<br>s.t. $\mathcal{Q}(\mathbf{L}) + \mathbf{S} = \mathbf{E}$ | 1.276 | 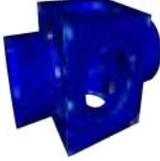 |
| **(DECOLOR)** Contiguous Outliers in the Low-Rank Representation<br><br>Zhou et al. [401] | $\mathbf{E} = \mathbf{L} + \mathbf{S} + \mathbf{N}$<br>$\min_{\mathbf{L},\mathbf{S}} \alpha\|\mathbf{L}\|_* + \beta\|\mathbf{F}\|_1 + \gamma\|C_{vec}(\mathbf{F})\|_1$<br>$+ \frac{1}{2}\|P_F(E-L)\|_F^2$<br>s.t. $rank(\mathbf{L}) < K$ | 1.877 | 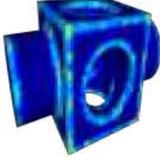 |
| **(DUAL)** Dual RPCA<br><br>Lin et al. [399] | $\mathbf{E} = \mathbf{L} + \mathbf{S}$<br>$\min_{\mathbf{L},\mathbf{S}} \|\mathbf{S}\|_*$<br>s.t. $rank(L) < r$<br>DUAL Approach | 277.080 | 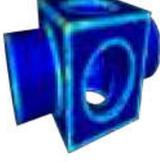 |
| **(EALM)**<br>Exact ALM<br><br>Lin et al. [402] | $\mathbf{E} = \mathbf{L} + \mathbf{S}$<br>$\min_{\mathbf{L},\mathbf{S}} \|\mathbf{L}\|_* + \lambda\|\mathbf{S}\|_1$<br>s.t. $\mathbf{L} + \mathbf{S} = \mathbf{E}$<br>EALM solver | 19.594 | 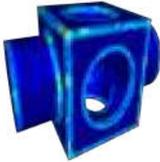 |
| **(SPCP max QN)**<br>Flip-Flop version<br>of Stable PCP-max<br>solved by<br>Quasi-Newton<br><br>Aravkin et al. [403] | $\mathbf{E} = \mathbf{L} + \mathbf{S} + \mathbf{N}$<br>$\min \Phi(\mathbf{L}, \mathbf{S})$<br>$\rho(\mathbf{L} + \mathbf{S} - \mathbf{Y}) < \epsilon$<br>flip - flop version | 58.277 | 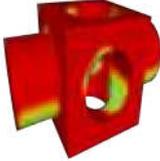 |
| **(FPCP)**<br>Fast PCP<br><br>Rodriguez<br>and Wohlberg [404] | $\mathbf{E} = \mathbf{L} + \mathbf{S}$<br>$\arg\min_{\mathbf{L},\mathbf{S}} \frac{1}{2}\|\mathbf{L} + \mathbf{S} - \mathbf{E}\|_F + \lambda\|\mathbf{S}\|_1$<br>s.t. $rank(\mathbf{L}) = t$<br>alternating minimization | 0.197 | 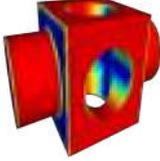 |
| **(FW-T)**<br>SPCP solved<br>by Frank-Wolfe<br>method<br><br>Mu et al. [405] | $\mathbf{E} = \mathbf{L} + \mathbf{S} + \mathbf{N}$<br>$\min_{\mathbf{L},\mathbf{S}} \frac{1}{2}\|P_F(\mathbf{L} + \mathbf{S} - \mathbf{E})\|_F^2$<br>$\|\mathbf{L}\|_* \leq \tau_L, \|\mathbf{S}\|_1 \leq \tau_S$<br>Frank-Wolfe method | 4.221 | 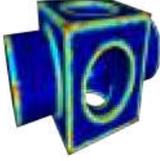 |
| **(GA)**<br>Grassmann<br>Average<br><br>Hauberg et al. [406] | $q = \arg\min_{\mathbf{u} \in S^{D-1}} \sum_{n=1}^{N}$<br>$w_n dist^2_{S^{D-1}}(u_n, \mathbf{u})$<br>$\|q_i - q_{i-1}\| \leq e$<br>Grassmann Average | 0.019 | 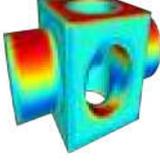 |
| **(GoDec)** <br> Go Decomposition <br><br> Zhou and Tao [407] | $\mathbf{E} = \mathbf{L} + \mathbf{S} + \mathbf{N}$ <br> $\min_{\mathbf{L},\mathbf{S}} \|\mathbf{E} - \mathbf{L} - \mathbf{S})\|_F^2$ <br> $rank(\mathbf{L}) \leq e, \, card(\mathbf{S}) \leq k$ | 0.060 | 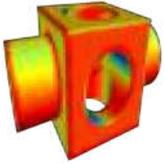 |
| **(GreGoDec)** <br> Greedy Semi-Soft GoDec Algotithm <br><br> Zhou and Tao [329] | $\mathbf{E} = \mathbf{L} + \mathbf{S} + \mathbf{N}$ <br> $\min_{\mathbf{L},\mathbf{S}} \|\mathbf{E} - \mathbf{L} - \mathbf{S})\|_F^2$ <br> $rank(\mathbf{L}) \leq e, \, card(\mathbf{S}) \leq k$ <br> via greedy semi-soft approach | 0.210 | 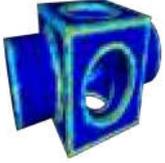 |
| **(IALM)** <br> Inexact ALM <br><br> Lin et al. [402] | $\mathbf{E} = \mathbf{L} + \mathbf{S}$ <br> $\min_{\mathbf{L},\mathbf{S}} \|\mathbf{L}\|_* + \lambda \|\mathbf{S}\|_1$ <br> s.t. $\mathbf{L} + \mathbf{S} = \mathbf{E}$ <br> Inexact ALM solver | 0.455 | 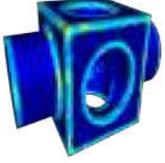 |
| **(IALM BLWS)** <br> IALM with Block Lanczos with Warm Start <br><br> Lin and Wei [408] | $\mathbf{E} = \mathbf{L} + \mathbf{S}$ <br> $\min_{\mathbf{L},\mathbf{S}} \|\mathbf{L}\|_* + \lambda \|\mathbf{S}\|_1$ <br> s.t. $\mathbf{L} + \mathbf{S} = \mathbf{E}$ <br> BLWS approach | 3.700 | 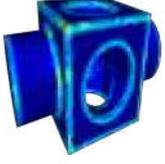 |
| **(L1F)** <br> $L_1$ Filtering <br><br> Liu et al. [409] | $\mathbf{E} = \mathbf{L} + \mathbf{S}$ <br> $\tilde{\mathbf{L}}^s = \mathbf{L}^r (\mathbf{L}^s)^\dagger \mathbf{L}^c$, <br> $\min_{\mathbf{S}^c, \tilde{\mathbf{Q}}} \|\mathbf{S}^c\|_1$ s.t. $\mathbf{E}^c + \mathbf{U}^s \tilde{\mathbf{Q}} + \mathbf{S}^c$ <br> $\min_{\mathbf{S}^r, \tilde{\mathbf{P}}} \|\mathbf{S}^r\|_1$ s.t. $\mathbf{E}^r + \tilde{\mathbf{P}}^T (\mathbf{V}^s)^T + \mathbf{S}^r$ | 1.832 | 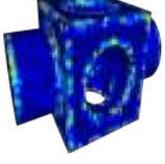 |
| **(Lag SPCP QN)** Lagrangian SPCP solved by Quasi-Newton  Aravkin et al. [403] | $\mathbf{E} = \mathbf{L} + \mathbf{S} + \mathbf{N}$ $\min \Phi(\mathbf{L}, \mathbf{S})$ $\rho(\mathbf{L} + \mathbf{S} - \mathbf{Y}) < \epsilon$ Lagrangian version | 3.126 | 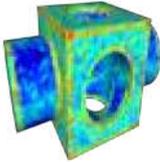 |
| **(LSADM)** Linearized Symmetric ADM  Goldfarb et al. [410] | $\mathbf{E} = \mathbf{L} + \mathbf{S}$ $\min_{\mathbf{L},\mathbf{S}} \|\mathbf{L}\|_* + \lambda \|\mathbf{S}\|_1$ s.t. $\mathbf{L} + \mathbf{S} = \mathbf{E}$ LS ADM solver | 0.846 | 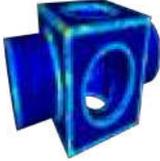 |
| **(MBRMF)** Markov BRMF  Wang and Yeung [411] | $\mathbf{E} = \mathbf{U}\mathbf{V}^T + \mathbf{S}$ $log(p(\mathbf{U}, \mathbf{V}|\mathbf{E}, \lambda, \lambda_U, \lambda_V))$ Bayesian distribution constraints | 56.200 | 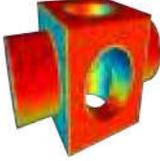 |
| **(MoG-RPCA)** Mixture of Gaussians RPCA  Zhao et al. [412] | $\mathbf{E} = \mathbf{L} + \mathbf{S}$ $min$ **KL** divergence MOG distribution constraints for **S** | 21.062 | 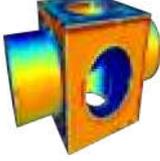 |
| **(noncvxRPCA)** Robust PCA via Nonconvex Rank Approximation  Kang et al. [413] | $\mathbf{E} = \mathbf{L} + \mathbf{S}$ $\min_{\mathbf{L},\mathbf{S}} \|\mathbf{L}\|_* + \lambda \|\mathbf{S}\|_1$ s.t. $\mathbf{L} + \mathbf{S} = \mathbf{E}$ via Nonconvex Rank Approximation | 0.070 | 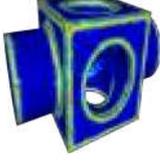 |
| **(NSA)** Non-Smooth Augmented Lagrangian  Aybat et al. [414] | $\mathbf{E} = \mathbf{L} + \mathbf{S}$ $\min_{\mathbf{L},\mathbf{S}} \|\mathbf{L}\|_* + \lambda\|\mathbf{S}\|_1$ s.t. $\mathbf{L} + \mathbf{S} = \mathbf{E}$ via Non-Smooth Augmented Lagrangian | 1.079 | 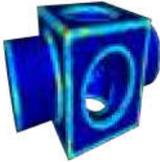 |
| **(PCP)** Principal Component Pursuit  Candes et al. [304] | $\mathbf{E} = \mathbf{L} + \mathbf{S}$ $\min_{\mathbf{L},\mathbf{S}} \|\mathbf{L}\|_* + \lambda\|\mathbf{S}\|_1$ s.t. $\mathbf{E} - \mathbf{L} - \mathbf{S} = 0$ | 13.112 | 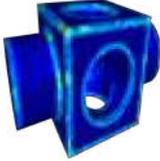 |
| **(PRMF)** Probabilistic Robust Matrix Factorization  Wang et al. [415] | $\mathbf{E} = \mathbf{U}\mathbf{V}^T + \mathbf{S}$ $log(p(\mathbf{U},\mathbf{V}|\mathbf{E},\lambda,\lambda_U,\lambda_V))$ distribution constraints | 18.626 | 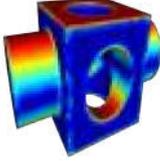 |
| **(PSPG)** Partially Smooth Proximal Gradient  Aybat et al. [416] | $\mathbf{E} = \mathbf{L} + \mathbf{S}$ $\min_{\mathbf{L},\mathbf{S}} \|\mathbf{L}\|_* + \lambda\|\mathbf{S}\|_1$ s.t. $\mathbf{L} + \mathbf{S} = \mathbf{E}$ via Partially Smooth Proximal Gradient | 0.474 | 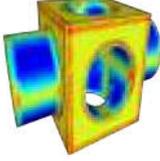 |
| **(RegL1-ALM)** Low-Rank Matrix Approximation under Robust L1-Norm  Zheng et al. [417] | $\mathbf{E} = \mathbf{U}\mathbf{V} + \mathbf{S}$ $\min_{\mathbf{U},\mathbf{V}} \|W_5 \odot (\mathbf{E} - \mathbf{U}\mathbf{V})\|_1 + \lambda\|\mathbf{V}\|_*$ s.t. $\mathbf{U}\mathbf{U} = \mathbf{I}_r$ | 2.148 | 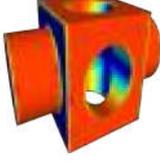 |
| **(SSGoDec)** Semi-Soft GoDec  Zhou and Tao [407] | $\mathbf{E} = \mathbf{L} + \mathbf{S} + \mathbf{N}$ $\min_{\mathbf{L},\mathbf{S}} \|\mathbf{E} - \mathbf{L} - \mathbf{S})\|_F^2$ $rank(\mathbf{L}) \leq e$, $card(\mathbf{S}) \leq \tau$ $\tau$ is a soft threshold | 0.636 | 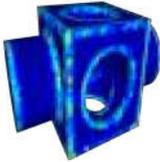 |
| **(STOC-RPCA)** Online Robust PCA via Stochastic Optimization  Feng et al. [418] | $\mathbf{E} = \mathbf{LR} + \mathbf{S} + \mathbf{N}$ $\min_{\mathbf{L}\in\mathbb{R}^{n\times p},\mathbf{R}\in\mathbb{R}^{n\times r}} \frac{1}{2}\|\mathbf{E} - \mathbf{LR}^T - \mathbf{S})\|_F^2 +$ $\frac{\lambda_1}{2}(\|\mathbf{L}\|_F^2 + \|\mathbf{S}\|_F^2) + \lambda_2 \|\mathbf{S}\|_1$ $\mathbf{E} - \mathbf{LR} - \mathbf{S} = 0$ | 1.354 | 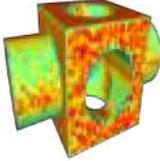 |
| **(SVT)** A singular value thresholding algorithm for matrix completion  Cai et al. [419] | $\mathbf{E} = \mathbf{L} + \mathbf{S}$ $\min \|\mathbf{L}\|_*$ s.t. $\mathcal{Q}(\mathbf{L}) = b$ via singular value thresholding | 474.073 | 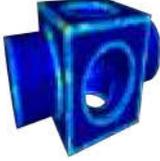 |
| **(VBRPCA)** Variational Bayesian RPCA  Babacan et al. [420] | $\mathbf{E} = \mathbf{D}\mathbf{B}^T + \mathbf{S} + \mathbf{N}$ $p(\mathbf{E}, \mathbf{D}, \mathbf{B}, \mathbf{S})$ Distribution constraints | 1.701 | 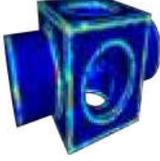 |
| Ours | $\mathbf{E} = \mathbf{L} + \mathbf{S}$ $\min_{\mathbf{L},\mathbf{S}} \frac{1}{2}\|\mathbf{L} + \mathbf{S} - \mathbf{E}\|_F + \lambda \|\mathbf{S}\|_1$ s.t. $rank(\mathbf{L}) = K$ augmented Lagrange multiplier | 0.25 | 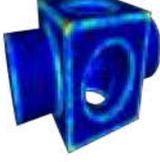 |

Table E.1: Execution times for the decomposition of the corresponding matrix $\mathbf{E} \in \mathbb{R}^{n \times 3 \cdot 15}$ of the model "block" and heat map visualization of the saliency map for different RPCA approaches.